\documentclass[final,5p,times,twocolumn]{elsarticle}
\usepackage{amssymb}
\usepackage{amsmath}
\usepackage{array}
\usepackage{color}

\newcommand{\newtext}[1]{#1}
\newcommand{\newcoloringtext}[1]{#1}

\journal{Computers and Fluids}
\date{June 17, 2024}

\begin{document}

\graphicspath{{./figs_compressed/}, {./figs/pdfs/}}

\begin{frontmatter}

\title{%
	Space-time adaptive ADER-DG finite element method with LST-DG predictor and\\
	\textit{a posteriori} sub-cell ADER-WENO finite-volume limiting for multidimensional detonation waves simulation
}
\author{Popov I.S.}
\ead{diphosgen@mail.ru, popovis@omsu.ru}

\affiliation{organization={Department of Theoretical Physics, Dostoevsky Omsk State University}, city={Omsk}, country={Russia}}

\begin{abstract}
The space-time adaptive ADER-DG finite element method with LST-DG predictor and a posteriori sub-cell ADER-WENO finite-volume limiting was used for simulation of multidimensional reacting flows with detonation waves. The presented numerical method does not use any ideas of splitting or fractional time steps methods. The modification of the LST-DG predictor has been developed, based on a local partition of the time step in cells in which strong reactivity of the medium is observed. This approach made it possible to obtain solutions to classical problems of flows with detonation waves and strong stiffness, without significantly decreasing the time step. The results obtained show the very high applicability and efficiency of using the ADER-DG-$\mathbb{P}_{N}$ method with a posteriori sub-cell limiting for simulating reactive flows with detonation waves. The numerical solution shows the correct formation and propagation of ZND detonation waves. The structure of detonation waves is resolved by this numerical method with subcell resolution even on coarse spatial meshes. The smooth components of the numerical solution are correctly and very accurately reproduced by the numerical method. Non-physical artifacts of the numerical solution, typical for problems with detonation waves, \newcoloringtext{such as the propagation of non-physical shock waves and weak detonation fronts ahead of the main detonation front,} did not arise in the results obtained. The results of simulating rather complex problems associated with the propagation of detonation waves in significantly inhomogeneous domains are presented, which show that all the main features of detonation flows are correctly reproduced by this numerical method. It can be concluded that the space-time adaptive ADER-DG-$\mathbb{P}_{N}$ method with LST-DG predictor and a posteriori sub-cell ADER-WENO finite-volume limiting is perfectly applicable to simulating fairly complex reacting flows with detonation waves.
\end{abstract}

%


\end{frontmatter}

\section*{Introduction}
\label{sec:intro}

Modern problems of the movement of liquids, gases and plasmas are accompanied by problems of simulating complex multi-scale flows~\cite{Fortov_ESM, Fortov_HEDP, Zeldovich_Raizer_2002, Drake_2018}. Typical problems are characterized by the formation in flows of non-stationary shock waves, contact discontinuities, rarefaction waves, vortex generation domains, coherent structures and a wide range of hydrodynamic instabilities. These peculiarities of the solution place strict conditions on the possibilities and accuracy of numerical schemes for simulating hydrodynamic flows.

Non-stationary flows that arise in problems of hydrodynamics and gas dynamics are characterized by the formation of flow discontinuities, even in the case of smooth initial conditions~\cite{Kulikovskii, Rozhdestvenskii_Janenko}. This places strong and often contradictory requirements~\cite{Handbook_NMHP_BF_2016, Handbook_NMHP_AM_2017} on numerical methods designed to simulate such flows. On the one hand, numerical methods must be able to correctly resolve discontinuities in hydrodynamic flow and preserve the properties of monotonicity of the numerical solution, as a result of which they must have no more than first order, which is explained by the well-known Godunov theorem. On the other hand, to describe non-stationary processes, a sufficiently accurate description of the formation and propagation of small disturbances against the general background of the flow is necessary, as a result of which numerical methods must be of high order and characterized by weak numerical dissipation.

Non-stationary compressible flows of multicomponent reactive flows are characterized by some significant additional difficulties that are not encountered in classical gas-dynamic problems~\cite{Oran_Boris_2005, Lunev_2017, Nagnibeda_2009, Anderson_1989}. The kinetics of reactions in a multicomponent gas medium is described by the equations of chemical kinetics, such as the law of mass action. The reaction rates and the local energy yield of the reactions strongly nonlinearly depend on local density, temperature and composition fields. This imposes significant restrictions on the resolution accuracy of discontinuous components and sharp solution gradients. The strong dependence of the reaction parameters on the local temperature fields and concentrations of the medium components and the strong dependence of the fields of hydrodynamic quantities on the reaction fields (especially on the energy yield) lead to the formation of a complex feedback structure. This imposes significant restrictions on the accuracy of reproduction in the numerical solution of small perturbations, which is especially important in the study of non-stationary processes~\cite{Oran_Boris_2005}.

A special place among the problems of hydrodynamics and gas dynamics is occupied by the problems of simulating flows containing detonation waves~\cite{Oran_Boris_2005}. Detonation waves are usually formed in reactive flows in which chemical or physical reactions occur with a positive release of energy, while certain conditions are imposed on the resulting feedbacks in the system and the geometric characteristics of the problem~\cite{Lee_2008}. Detonation waves are characterized by characteristic time and spatial scales that are significantly smaller than the characteristic time and spatial scales of the hydrodynamic flow~\cite{frac_steps_detwave_sim_2000}. This leads to the appearance of strong stiffness in the solution describing the flow with detonation waves. It should be noted that detonation waves are characterized by a complex flow structure, when well-consistent and self-organized patterns are observed, one of which is the detonation wave itself -- a structure consisting of a shock wave and a zone of reactions occurring in the medium, directly adjacent to the wave front~\cite{Oran_Boris_2005, frac_steps_detwave_sim_2000, chem_kin_hrs_rev_1, chem_kin_hrs_rev_2, chem_kin_hrs_weno}. The mismatch of processes that can be introduced by the numerical method can lead to a physically incorrect solution when the detonation front splits into a shock wave and an independent front of weak detonation~\cite{correct_det_wave_speed_2017}. For this reason, numerical methods for simulating flows with detonation waves must be able to correctly resolve the spatial and temporal scales of flow with detonation waves, organizing a consistent description of the hydrodynamic and reaction properties of the medium~\cite{DG_FR_2023}.

Among the numerical methods of computational fluid dynamics for simulating detonation flows, the most widely used are methods of splitting into physical processes~\cite{Oran_Boris_2005}, including the Marchuk-Strang methods and methods of fractional steps~\cite{frac_steps_detwave_sim_2000, chem_kin_hrs_weno}. However, as the practice of using splitting in the case of flows with detonation waves shows, obtaining a physically correct numerical solution is only possible using native additional procedures~\cite{frac_steps_detwave_sim_2000}. Otherwise, in the case of the formation of detonation waves during the flow, nonphysical shock and weak detonation fronts may arise~\cite{correct_det_wave_speed_2017}, significantly separated in space. It is clear that the use of the idea of splitting is poorly compatible with the use of high-order and high-resolution numerical methods, because usually splitting by physical processes reduces the order of the numerical method to the first. However, this does not prevent us from combining high-precision WENO schemes with splitting according to physical processes~\cite{chem_kin_hrs_weno}.

Over the past quarter century, many types of high-order and high-resolution numerical methods have been applied to numerically simulate detonation phenomena. It should be noted that the correct description of hydrodynamic flows with detonation waves using numerical methods without explicit identification of discontinuities causes difficulties even in one-dimensional problems. This usually requires the inclusion of additional non-trivial and native procedures in the general-purpose numerical method. A review of works on simulating detonation flows using various high-precision numerical methods is presented in the works~\cite{chem_kin_hrs_rev_1, chem_kin_hrs_rev_2, RTS_split_2019}. Without pretending to be a systematic review, some works were noted in which the simulation of detonation flows was carried out using numerical methods of the second and higher orders. In work~\cite{Gao_2016}, the RKDG method with $h$-adaptivity was used to simulate one-dimensional detonation waves. The work used the second-order Godunov method to simulate one-dimensional flows with detonation waves in a medium with a three-stage reaction. The work~\cite{Henrick_2006} describes the possibilities of using fifth-order finite-difference WENO schemes to simulate various modes in one-dimensional detonation. The paper presented a new approach to the use of the DG method with flux reconstruction to simulate various modes in one-dimensional detonation. In the last two works, rather complex features associated with the formation of a stable and unstable detonation wave were studied. The works~\cite{Gao_2011, Hu_2017, Shu_Wang_2012} developed finite-volume and discontinuous Galerkin methods for simulating two-dimensional detonation flows. In work~\cite{Shu_Wang_2015} discontinuous Galerkin method has been developed for simulation three-dimensional reactive flows with detonation waves.

\newcoloringtext{Adding source terms associated with reactions in a multicomponent medium, which can lead to the formation of detonation, to classical numerical fluid dynamics methods can lead to the formation of the nonphysical artifacts~\cite{frac_steps_detwave_sim_2000, correct_det_wave_speed_2017} described above, such as weak detonation fronts. Further in the text of this article, non-physical artifacts of detonation modeling will be understood as precisely the propagation of non-physical shock waves and weak detonation fronts ahead of the main detonation front.} One of the main problems in modeling reacting flows with detonation waves is the requirement for correct resolution of very small time and spatial scales\newcoloringtext{, and numerical methods ``out of the box'' are required to be able to use strongly stiff systems of equations for solving}. Numerical methods based on the ADER paradigm, in particular the numerical methods ADER-DG and finite-volume ADER-WENO methods, are well suited to this criterion. \newcoloringtext{Therefore, it is expected that numerical methods of this family will make it possible to correctly describe the development and propagation of detonation waves without significant modification or adaptation of the original methods.} The ADER paradigm originates from basic works~\cite{ader_init_1, ader_init_2}. The space-time adaptive ADER finite element DG method with \textit{a posteriori} correction technique of solutions on sub-cells by the finite-volume ADER-WENO limiter using adaptive mesh refinement~\cite{ader_dg_ideal_flows} demonstrate unprecedented accuracy of numerical solution and resolution of discontinuities, and may be considered as a new generation of shock capturing schemes for computational fluid dynamics. This numerical method has arisen as a result of the development of special schemes~\cite{mood_par} using a paradigm MOOD, using high-precision ADER-DG methods~\cite{ader_stiff_1, ader_stiff_2, ader_dg_dev_1, ader_dg_dev_2} and adaptive correction with a sufficiently accurate and stable ADER-WENO-FV limiter~\cite{ader_weno_lstdg_ideal, ader_weno_lstdg_diss}. This method was extended for application in simulation of dissipative GD and MHD flows in~\cite{ader_dg_diss_flows} \newcoloringtext{and low Mach number flow~\cite{ader_dg_semiexpl}}; \newcoloringtext{in simulation of the deformable solids dynamics~\cite{ader_dg_hyperelastic}, the propagation of seismic waves~\cite{ader_dg_seiemic, ader_dg_seiemic_underwater}, the blood flow~\cite{ader_eno_fv_blood_2022};} to the case of unstructured mesh, with application of direct ALE schemes in~\cite{ader_dg_ale}; it was applied for solution of GRGD and GRMHD problems in~\cite{ader_dg_grmhd, ader_dg_gr_prd}. \newcoloringtext{In~\cite{ader_dg_gr_z4_2024} Dumbser \textit{et al} developed \textit{well-balanced} ADER-DG method for GRGD problems, and in~\cite{ader_dg_wb_shwater_2022} \textit{well-balanced} ADER-DG was constructed for the equations of the theory of multilayer shallow water. In~\cite{ader_stiff_1, PNPM_DG_2010, ader_dg_ode_ivp} ADER-DG methods were used to solve ODE systems.} In~\cite{ader_dg_simple_mod} in the Appendix A it was described the main actions that must be taken to transform a given DG code into a DG code that is stabilized with an a posteriori finite volume sub-cell limiter. In~\cite{ader_dg_PNPM}, a generalization of the ADER-DG method with a posteriori correction technique of solutions on sub-cells by the finite-volume ADER-WENO limiter using adaptive mesh refinement to the $P_{N} P_{M}$-schemes~\cite{PNPM_DG_2009, PNPM_DG_2010} was constructed. Peculiarities of mathematical formulation and efficiency possibility of implementation are discussed in~\cite{ader_dg_eff_impl, fron_phys} and~\cite{exahype, ader_dg_hpc_impl_1, ader_dg_hpc_impl_2, ader_dg_hpc_impl_3, ader_dg_hpc_impl_4, ader_eff}. \newcoloringtext{In~\cite{dg_entropy} Gaburro \textit{et al} developed entropy preserving ADER-DG schemes (see also~\cite{dg_entropy_add}). In~\cite{ader_weno_sph}, based on the ADER paradigm, a SPH numerical method was created, which demonstrated a very high quality of modeling hydrodynamic and magnetohydrodynamic flows with discontinuities, compared to classical SPH methods.} The modern state of development of the space-time adaptive ADER finite element DG method with a posteriori correction technique of solutions on sub-cells by the finite-volume limiter using AMR for use on unstructured meshes and using the ALE approach is presented in~\cite{ader_dg_mod_1, ader_dg_mod_2} \newcoloringtext{(see also~\cite{ader_rev_2024})}. \newcoloringtext{The papers~\cite{dec_vs_ader_2021, dec_vs_ader_2023} discuss the features of the relationship between the ADER paradigm and the DeC paradigm, which also allows one to obtain numerical methods of an arbitrary high order.}

An important feature of the space-time adaptive ADER finite element DG method with a posteriori correction technique of solutions on sub-cells by the finite-volume ADER-WENO limiter, as a new generation of high-resolution shock capturing numerical schemes, is the property of high-precision discontinuity resolution in the numerical solution -- shock waves and contact discontinuities are resolved within a single finite element cell, while contact discontinuities do not spread over time; at the same time with high-precision resolution of small components of the solution. Therefore, it is of interest to use numerical methods based on the ADER paradigm to simulation reactive flows with detonation waves. The work~\cite{ader_dg_dev_2} involved modeling one-dimensional flows with detonation waves using the finite-volume ADER-WENO method. The work~\cite{popov_j_sci_comp_2023} explored the possibilities of using ADER-DG methods with a posteriori correction of the solution for modeling one-dimensional flows with detonation waves, while using the method of adaptive time step correction, which was used globally in the computing domain. In these cases, physically correct solutions were obtained even in the case of coarse coordinate meshes.

In this work, an attempt is made to use a numerical method, without using additional procedures that significantly change the numerical method, to simulate two-dimensional and three-dimensional reacting flows with detonation waves.

\section{General description of the numerical method}
\label{sec:gen_descript}

\subsection{Mathematical formulation of the problem}
\label{sec:gen_descript:math_framework}

The system of non-stationary Euler equations and the system of non-stationary convection-reaction equations formed a quasilinear system of hyperbolic equations for describing compressible reacting flows, which takes the following form:
\begin{equation}\label{eq:system_of_equations}
\frac{\partial\mathbf{U}}{\partial t} + \nabla\cdot\mathbf{F} = \mathbf{S};
\end{equation}
where $\mathbf{U}$ is the vector of conserved values, $\mathbf{F}$ is the flux term and $\mathbf{S}$ is the source term:
\begin{equation}
\mathbf{U} = \left[
\begin{array}{c}
\rho\\
\rho\mathbf{v}\\
\varepsilon\\
\rho\mathbf{c}
\end{array}
\right];\quad
\mathbf{F} = \left[
\begin{array}{c}
\rho\mathbf{v}\\
\rho\mathbf{v}\otimes\mathbf{v} + p\mathbf{I}\\
(\varepsilon + p) \mathbf{v}\\
\rho\mathbf{c}\otimes\mathbf{v}
\end{array}
\right];\quad
\mathbf{S} = \left[
\begin{array}{c}
0\\
\mathbf{S}_{f}\\
\mathrm{S}_{e}\\
\mathbf{S}_{r}
\end{array}
\right];
\end{equation}
where $\rho$ is the mass density; $\mathbf{v} = (u, v, w)$ is the velocity; $p$ is the pressure; $\varepsilon$ is the total energy density including the thermal $e$ and the kinetic contributions $\varepsilon = e + \frac{1}{2} \rho v^{2}$; $\mathbf{c}^{T} = [c_{1}, \ldots, c_{R}]$ is a vector of mass concentrations $c_{k}$ of the component of the reacting medium, which determines the mass fraction of the $k$-th component in the mixture: the density of the $k$-th component can be obtained by the formula $\rho_{k} = \rho c_{k}$, and thus the relation $\sum_{k} c_{k} = 1$ follows from $\rho = \sum_{k} \rho_{k}$; $R$ is the amount of components in the mixture; $\mathbf{S}_{f}$ is the external forces density; $\mathrm{S}_{e} = \mathrm{S}_{e, r} + \mathrm{S}_{e, f}$ which is determined by the sum of energy yield associated with reactions occurring in the medium $\mathrm{S}_{e, r}$ and the external forces $\mathrm{S}_{e, f}$; $\mathbf{S}_{r}$ is the source term that determines the rates of reactions; $\mathbf{v}\otimes\mathbf{v} \equiv \mathbf{v}\mathbf{v}^{T}$ and $\mathbf{c}\otimes\mathbf{v} \equiv \mathbf{c}\mathbf{v}^{T}$ are the tensor products.

The source term $\mathbf{S} = \mathbf{S}(\mathbf{U};\, \mathbf{r},\, t)$ allows arbitrary dependence on conserved variables $\mathbf{U}$, spatial coordinates $\mathbf{r} = (x, y, z)$ and time $t$.

The multicomponent and reaction properties of the flow are considered in the form of non-stationary convection-reaction equations
\begin{equation}\label{eq:convection_reaction_equations}
\frac{\partial \left(\rho\mathbf{c}\right)}{\partial t} + \nabla\left(\rho\mathbf{c}\otimes\mathbf{v}\right) = \mathbf{S}_{r},
\end{equation}
where, as can be seen from the system of equations (\ref{eq:system_of_equations}), the transfer velocity $\mathbf{v}$ of the mixture components and the total mass density $\rho$ are determined from the Euler equations.

The reaction rates included in the term $\mathbf{S}_{r}$ were chosen in the form of the law of mass action, where the reaction rate constants could have an arbitrary dependence on the temperature $T$ of the medium. The equation of state of the multicomponent mixture was chosen in the form of the perfect gas equation. The caloric equation of state was specified in the form $p = (\gamma - 1) e$, where $\gamma$ can be calculated as the ratio of specific heat capacities of components averaged over mass concentrations $c_{k}$. Empirical values of the effective gas constant and $\gamma$, which are used in the local approximation of the wide-range equations of state by the equation of state of a perfect gas~\cite{Zeldovich_Raizer_2002, Lunev_2017, Anderson_1989}, can be used instead of the classical values. 

Thus, in the present work, a computational scheme was used for the simplest mathematical model of unsteady compressible reacting flows, however, it is quite in demand in solving applied problems~\cite{Oran_Boris_2005, Lunev_2017, Nagnibeda_2009, Anderson_1989}.

\subsection{Formulation of the numerical method}
\label{sec:gen_descript:numeric_framework}

The numerical method used in this work is based on the space-time adaptive ADER finite-element DG method, which is characterized by a very high accuracy and resolution of the smooth components of the solution. Non-physical anomalies of the numerical solution arising due to the fundamental linearity of the ADER-DG method, which are explained by the well-known Godunov theorem, are corrected by an a posteriori limiter, for which the high-precision finite-volume ADER-WENO method was chosen. An excellent detailed description of this computational scheme is given in the basic works~\cite{ader_dg_ideal_flows, ader_dg_dev_1, ader_dg_dev_2, ader_weno_lstdg_ideal, ader_weno_lstdg_diss, ader_dg_diss_flows, ader_dg_ale, ader_dg_grmhd, ader_dg_gr_prd, ader_dg_PNPM, PNPM_DG_2009, PNPM_DG_2010} of the developers of this method. Details of the internal structure of the ADER-DG are presented in the works~\cite{ader_dg_dev_1, ader_dg_dev_2, PNPM_DG_2009, PNPM_DG_2010}. Details of the internal structure of the finite-element ADER-DG and finite-volume ADER-WENO methods, in which the LST-DG prediction method is used, are presented in the work~\cite{ader_weno_lstdg_ideal, ader_weno_lstdg_diss}. Peculiarities of mathematical formulation and efficient software implementation are discussed in~\cite{ader_dg_eff_impl, fron_phys} and~\cite{exahype, ader_dg_hpc_impl_1, ader_dg_hpc_impl_2, ader_dg_hpc_impl_3, ader_dg_hpc_impl_4}. The modern state of development of the space-time adaptive ADER finite element DG method with a posteriori correction technique of solutions on sub-cells by the finite-volume limiter using AMR for use on unstructured meshes and using the ALE approach is presented in the works~\cite{ader_dg_mod_1, ader_dg_mod_2}. For this reason, the description of the computational method in this paragraph will be given only briefly enough to understand the general structure of the method used in this particular case.

The space-time adaptive ADER-DG finite element method with LST-DG predictor and a posteriori sub-cell ADER-WENO finite-volume limiting involves a sequence of steps~\cite{ader_dg_ideal_flows, ader_dg_dev_1, ader_dg_dev_2, ader_weno_lstdg_ideal, ader_weno_lstdg_diss, ader_dg_diss_flows, ader_dg_ale, ader_dg_grmhd, ader_dg_gr_prd, ader_dg_PNPM, PNPM_DG_2009, PNPM_DG_2010}:
\begin{itemize}
	\item a LST-DG predictor, using which a local discrete space-time solution in the small is obtained;
	\item a pure ADER discontinuous Galerkin $\mathbb{P}_{N}\mathbb{P}_{N}$ scheme, using which a candidate high accuracy solution is obtained;
	\item a determination of the admissibility of the obtained high accuracy candidate solution and identification of ``troubled'' cells;
	\item a recalculation of the solution in ``troubled'' cells by a stable ADER-WENO finite-volume limiter.
\end{itemize}

The coordinate computational domain $\Omega$ in this work was chosen as $\Omega = [x_{L}, x_{R}]\times[y_{L}, y_{R}]\times[z_{L}, z_{R}] \subset R^{3}$. The computational domain $\Omega$ was represented by the mesh $\Omega = \cup_{i_{1}i_{2}i_{3}} \Omega_{i_{1}i_{2}i_{3}}$, where $\Omega_{i_{1}i_{2}i_{3}} = \left[x_{i_{1}}, x_{i_{1}+1}\right]\times\left[y_{i_{2}}, y_{i_{2}+1}\right]\times\left[z_{i_{3}}, z_{i_{3}}\right]$ is the mesh cell, and is characterized by mesh coordinate steps $(\Delta x, \Delta y, \Delta z)$ for individual coordinate directions.

The system of quasilinear hyperbolic equations of gas dynamics of reacting flows (\ref{eq:system_of_equations}) was formulated in the local $4$-coordinate system $(\tau, \boldsymbol{\xi}) \equiv (\tau, \xi, \eta, \zeta)$ of the reference space-time finite element $\omega_{4} = [0, 1]^{4}$, which corresponds to the space-time finite element of the mesh $[t^{n}, t^{n+1}]\times\Omega_{i_{1}i_{2}i_{3}}$:
\begin{equation}\label{eq:rescaled_system_of_equations}
\frac{\partial\mathbf{u}}{\partial \tau} + \nabla_{\boldsymbol{\xi}}\cdot\mathbf{f} = \mathbf{s};
\end{equation}
where $\mathbf{f} = (\mathbf{f}_{\xi}, \mathbf{f}_{\eta}, \mathbf{f}_{\zeta})$ and $\mathbf{s}$ are the rescaled flux terms and the source term, $\nabla_{\boldsymbol{\xi}} \equiv (\partial_{\xi}, \partial_{\eta}, \partial_{\zeta})$, $t^{n+1} = t^{n} + \Delta t^{n}$ and $\Delta t^{n}$ is the time step. The functional dependence of the source terms $\mathbf{s}(\mathbf{u}; \tau, \xi, \eta, \zeta) = \Delta t^{n}\cdot\mathbf{S}(\mathbf{u}; \mathbf{r}(\xi, \eta, \zeta), t(\tau))$ retains an explicit dependence on the coordinates $(x, y, z)$ and time $t$.

Functional representations of solutions within the framework of the space-time adaptive ADER-DG finite element method with LST-DG predictor and
a posteriori sub-cell ADER-WENO finite-volume limiting were constructed on the functional basis of tensor products of one-dimensional basis functions $\varphi_{k} = \varphi_{k}(\xi)$, for which the Legendre interpolation polynomials of degree $N$ were chosen, the interpolation nodes of which were the nodal points of the Gauss-Legendre quadrature formula, which are the roots of the shifted Legendre polynomials $P_{N+1}(\xi)$. 

The piecewise polynomials DG-representation $\mathbf{u}_{h}(\mathbf{r}, t^{n})$ on each time layer $t^{n}$ specified on the space cell $\Omega_{i_{1}i_{2}i_{3}}$ (mapped into the space reference element $\omega_{3} = [0, 1]^{3}$):
\begin{equation}\label{DG_representation}
\begin{split}
\mathbf{u}_{h}(\mathbf{r}, t^{n}) = \sum\limits_{\mathbf{k}} \hat{\mathbf{u}}_{\mathbf{k}}^{n} \cdot \Phi_{\mathbf{k}}\big(\boldsymbol{\xi}(\mathbf{r})\big),
\end{split}
\end{equation}
where $\hat{\mathbf{u}}_{\mathbf{k}}^{n} = \hat{\mathbf{u}}_{k_{1}k_{2}k_{3}}^{n}$ are the coefficients of the piecewise polynomials DG-representation; $\Phi_{\mathbf{k}}(\boldsymbol{\xi}) = \varphi_{k_{1}}\left(\xi\right) \varphi_{k_{2}}\left(\eta\right) \varphi_{k_{3}}\left(\zeta\right)$ are the basis functions represented as tensor products representation; $\varphi_{k} = \varphi_{k}(\xi)$ are the one-dimensional basis functions; and multiindex $\mathbf{k} = (k_{1}, k_{2}, k_{3}) \in [0, N]^{3}$.

The ADER paradigm for solving the generalized Riemann problem is based on the use of a local solution in the small, which, within the framework of the development of the paradigm, will be obtained in the form of a local discrete space-time solution specified on the space-time finite element $[t^{n}, t^{n+1}]\times\Omega_{i_{1}i_{2}i_{3}}$ (mapped into the space-time reference element $\omega_{4} = [0, 1]^{4}$):
\begin{equation}\label{q_solution}
\begin{split}
\mathbf{q}_{h}(\mathbf{r}, t) = 
	\sum\limits_{\mathbf{p}} \hat{\mathbf{q}}_{\mathbf{p}}^{n} \cdot 
	\Theta_{\mathbf{p}}\big(\tau(t), \boldsymbol{\xi}(\mathbf{r})\big),
\end{split}
\end{equation}
where $\hat{\mathbf{q}}_{\mathbf{p}}^{n} = \hat{\mathbf{q}}_{p_{0}p_{1}p_{2}p_{3}}^{n}$ are the coefficients of the local discrete space-time solution; $\Theta_{\mathbf{p}}(\tau, \boldsymbol{\xi}) = \varphi_{p_{0}}\left(\tau\right) \varphi_{p_{1}}\left(\xi\right) \varphi_{p_{2}}\left(\eta\right) \varphi_{p_{3}}\left(\zeta\right)$ are the basis functions represented as tensor products representation; and multiindex $\mathbf{p} = (p_{0}, p_{1}, p_{2}, p_{3}) \in [0, N]^{4}$. It should be noted that the discrete space-time solution contains $(N+1)^{4}$ degrees of freedom (\texttt{DOF}), so it contains quite a lot of information about the local solution.

The local discrete space-time solution $\mathbf{q}_{h}(\mathbf{r}, t)$ is obtained as a result of a local space-time predictor. The LST-DG predictor allows you to obtain a local solution in the small, representing the structure of the solution within the coordinate constraints of one mesh cell, without taking into account the influence of other mesh cells. The LST-DG predictor represents the classical DG method, in which a discrete scheme is obtained by taking the $L_{2}$-projection of the residual of the system (\ref{eq:rescaled_system_of_equations}) onto the basis functions $\Theta_{\mathbf{p}}(\tau, \boldsymbol{\xi})$ of the solution representation:
\begin{equation}
\begin{split}
\int\limits_{\omega_{4}} d\tau d\boldsymbol{\xi} \cdot \Theta_{\mathbf{p}}(\tau, \boldsymbol{\xi}) \cdot \left[
	\frac{\partial\mathbf{u}}{\partial \tau} + \nabla_{\boldsymbol{\xi}}\cdot\mathbf{f} - \mathbf{s}
\right] = 0;
\end{split}
\end{equation}
which, as a result of substituting the representation of the discrete space-time solution $\mathbf{u}\mapsto\mathbf{q}_{h}$ in the form (\ref{q_solution}), subsequent integration by parts in time $\tau$, the use solution at the previous time step $\mathbf{u}_{h}(\mathbf{r}, t^{n})$ as initial condition -- $\mathbf{q}_{h}(\mathbf{r}, t^{n}) = \mathbf{u}_{h}^{n}(\mathbf{r}, t^{n})$, and the use of point-wise evaluation of the physical fluxes when using Gauss-Legendre quadrature formulas, takes on the following form:
\begin{equation}
\begin{split}
\sum_{\mathbf{q}}\left\{
	\mathbb{K}_{\mathbf{p}\mathbf{q}}^{\tau} \hat{\mathbf{q}}_{\mathbf{q}} +
	\mathbb{K}_{\mathbf{p}\mathbf{q}}^{\boldsymbol{\xi}} \hat{\mathbf{f}}_{\mathbf{q}} -
	\mathbb{M}_{\mathbf{p}\mathbf{q}} \hat{\mathbf{s}}_{\mathbf{q}}
\right\} = \sum_{\mathbf{k}} \mathbb{F}_{\mathbf{p}\mathbf{k}} \hat{\mathbf{u}}_{\mathbf{k}}^{n},
\end{split}
\end{equation}
where $\hat{\mathbf{f}}_{\mathbf{q}} = \mathbf{f}(\hat{\mathbf{q}}_{\mathbf{q}})$ and $\hat{\mathbf{s}}_{\mathbf{q}} = \mathbf{s}(\hat{\mathbf{q}}_{\mathbf{q}})$ were obtained as a result of the use of point-wise evaluation of the physical fluxes; $\mathbb{K}_{\mathbf{p}\mathbf{q}}^{\tau}$, $\mathbb{K}_{\mathbf{p}\mathbf{q}}^{\boldsymbol{\xi}} = \left(\mathbb{K}_{\mathbf{p}\mathbf{q}}^{\xi}, \mathbb{K}_{\mathbf{p}\mathbf{q}}^{\eta}, \mathbb{K}_{\mathbf{p}\mathbf{q}}^{\zeta}\right)$, $\mathbb{M}_{\mathbf{p}\mathbf{q}}$ and $\mathbb{F}_{\mathbf{p}\mathbf{k}}$ are multiindex matrices depending on integrals containing basis functions $\Theta_{\mathbf{p}}(\tau, \boldsymbol{\xi})$ and their derivatives; $\hat{\mathbf{u}}_{\mathbf{k}}^{n}$ are the coefficients of the DG representation of the solution at the previous time step $t^{n}$. All these multiindex matrices can be expressed as Kronecker products of two-index matrices depending on integrals containing basis functions $\varphi_{k}(\xi)$ and their derivatives. The resulting expression for the discrete numerical scheme was reformulated in the following form, convenient for writing a system of nonlinear algebraic equations:
\begin{equation}\label{eq:lst_dg_predictor_final}
\begin{split}
\hat{\mathbf{q}}_{\mathbf{p}} &-
\sum_{\mathbf{q}, \mathbf{r}}
\left[\left(\mathbb{K}^{\tau}\right)^{-1}\right]_{\mathbf{p}\mathbf{q}} \mathbb{M}_{\mathbf{q}\mathbf{r}} \hat{\mathbf{s}}_{\mathbf{r}}\\
&=
\sum_{\mathbf{q}, \mathbf{r}} \left[\left(\mathbb{K}^{\tau}\right)^{-1}\right]_{\mathbf{p}\mathbf{q}}
\mathbb{K}_{\mathbf{q}\mathbf{r}}^{\boldsymbol{\xi}} \hat{\mathbf{f}}_{\mathbf{r}} -
\sum_{\mathbf{k}, \mathbf{q}} 
\left[\left(\mathbb{K}^{\tau}\right)^{-1}\right]_{\mathbf{p}\mathbf{q}}\mathbb{F}_{\mathbf{q}\mathbf{k}} \hat{\mathbf{u}}_{\mathbf{k}}^{n},
\end{split}
\end{equation}
where all matrix constructions can be precomputed in code, summation over the internal silent multiindex $\mathbf{q}$ can also be precomputed. It should be noted that the mass matrix $\mathbb{M}_{\mathbf{p}\mathbf{q}}$ is completely diagonal in all pairs of multi-index components, which is associated with the orthogonality of the basis functions $\varphi_{k}(\xi)$, and for the matrix $\mathbb{F}_{\mathbf{p}\mathbf{k}}$ a very remarkable property is satisfied
\begin{equation}
\left\{\sum_{\mathbf{q}}\left[
	\left(\mathbb{K}^{\tau}\right)^{-1}
\right]_{\mathbf{p}\mathbf{q}} \mathbb{F}_{\mathbf{q}\mathbf{k}}
\right\}_{p_{0}p_{1}p_{2}p_{3}, k_{1}k_{2}k_{3}} = \mathbb{I}_{p_{0}} \delta_{p_{1}k_{1}} \delta_{p_{2}k_{2}} \delta_{p_{3}k_{3}},
\end{equation}
where $\delta_{p_{1}, k}$ is the Kronecker delta symbol and $\mathbb{I}_{p_{0}} = 1$ is just a one; which is a consequence of the correspondence principle: in the case of null fluxes terms $\mathbf{F} \equiv 0$ and null sources terms $\mathbf{S} \equiv 0$, the solution $\mathbf{q}(\mathbf{r}, t)$ should not depend on time $t$ and for the representation coefficients the solution $\hat{\mathbf{q}}_{p_{0}\mathbf{k}} = \hat{\mathbf{u}}_{\mathbf{k}}^{n}$ can be obtained. The matrices $[(\mathbb{K}^{\tau})^{-1}]_{\mathbf{p}\mathbf{q}} \mathbb{K}_{\mathbf{q}\mathbf{r}}^{\boldsymbol{\xi}}$ have an important property -- all their eigenvalues are strictly equal to zero~\cite{Jackson_2017}. Therefore, the Picard interaction process strictly converges due to Banach's fixed point theorem~\cite{Zanotti_lectures_2016}. The resulting system of nonlinear algebraic equations of the LST-DG predictor in the case of zero source terms was solved using the Picard iterative process. In the case of non-zero source terms, an internal iterative process was used~\cite{Zanotti_lectures_2016}, which in the case of continuous functions in the source terms was the Newton iterative method~\cite{Jackson_2017}, and in the case of discontinuous functions, the additional Picard iterative process. 

The presence in the source terms of terms $\mathbf{S}$ related to the kinetics of reactions in a multicomponent reacting medium leads to the occurrence of high and anomalously high stiffness. In this case, it is possible to use adaptive time step correction or adaptive mesh refinement (AMR), an additional refinement criterion in which will be the relative rate of reactions occurring in the reacting environment. In this work, a new approach was proposed based on modification of the LST-DG predictor, which allows obtaining a conditioned local discrete space-time solution $\mathbf{q}_{h}(\mathbf{r}, t)$ without using adaptive time step correction. The proposed approach is presented in detail in Subsection~\ref{sec:detonation_waves:lst_dg_predictor} ``LST-DG predictor for stiffness reactive flows'' in Section~\ref{sec:detonation_waves} ``Detonation waves simulation''. A more detailed description of the design and the implementation of the LST-DG predictor and the possibilities of implementing iterative processes for obtaining a solution to a system of nonlinear algebraic equations, that arise within the LST-DG predictor, is presented in the basic works~\cite{ader_dg_ideal_flows, ader_dg_dev_1, ader_dg_dev_2, ader_weno_lstdg_ideal, ader_weno_lstdg_diss, ader_dg_diss_flows, ader_dg_ale, ader_dg_grmhd, ader_dg_gr_prd, ader_dg_PNPM, PNPM_DG_2009, PNPM_DG_2010}, as well as in the works~\cite{ader_dg_eff_impl, fron_phys} and~\cite{exahype, ader_dg_hpc_impl_1, ader_dg_hpc_impl_2, ader_dg_hpc_impl_3, ader_dg_hpc_impl_4}.

The local discrete space-time solution $\mathbf{q}_{h}$ describes the dynamics of the flow locally in the small in one spatial cell $\Omega_{i}$, while the fluxes across the cell boundaries $\partial\Omega_{i}$ are not taken into account. The discrete solution $\mathbf{q}_{h}$ is used in a high-order space-time one-step discrete ADER-DG scheme to solve the generalized Riemann problem (GRP) when calculating the boundary flux terms and calculating the internal integrals over the cell volume, which determine the fluxes in the cell and the source terms. 

The one-step discrete ADER-DG scheme is obtained~\cite{ader_dg_dev_2, ader_dg_ideal_flows} by $L_{2}$-projecting the residual of the system of equations (\ref{eq:system_of_equations}) onto the basis functions $\Phi_{\mathbf{k}}(\boldsymbol{\xi})$ that are used to DG representation of the solution $\mathbf{u}_{h}$~\cite{dg_base_1, dg_base_2, dg_base_3, dg_base_4}, and integrating over the full time step $[t^{n}, t^{n+1}]$:
\begin{equation}
\begin{split}
\int\limits_{t^{n}}^{t^{n+1}} dt \int\limits_{\Omega_{i}} d\mathbf{r} \cdot \Phi_{\mathbf{k}}(\boldsymbol{\xi}(\mathbf{r})) \cdot \left[
	\frac{\partial\mathbf{U}}{\partial t} + \nabla\cdot\mathbf{F} - \mathbf{S}
\right] = 0;
\end{split}
\end{equation}
which, as a result of the transition to the reference space-time finite element $(\tau, \boldsymbol{\xi})$, substituting the representation $\mathbf{u}_{h}$ in the form (\ref{DG_representation}) into the term with the time derivative, subsequent integration by parts in space coordinates $\boldsymbol{\xi}$, and the use discrete space-time solution $\mathbf{u}\mapsto\mathbf{q}_{h}$ in the form (\ref{q_solution}) in volume and surface integrals, takes on the following form:
\begin{equation}
\begin{split}
\int\limits_{\omega_{3}} d\boldsymbol{\xi} \cdot \Phi_{\mathbf{k}} \cdot
\Bigg(\mathbf{u}_{h}^{n+1} &- \mathbf{u}_{h}^{n}\Bigg)
- \int\limits_{0}^{1} d\tau \int\limits_{\omega_{3}} d\boldsymbol{\xi} \cdot
\nabla_{\boldsymbol{\xi}} \Phi_{\mathbf{k}} \cdot \mathbf{f}(\mathbf{q}_{h})\\
&+ \int\limits_{0}^{1} d\tau \oint\limits_{\partial\omega_{3}} d\boldsymbol{\sigma}_{\boldsymbol{\xi}} \cdot
\Phi_{\mathbf{k}} \cdot \boldsymbol{\mathfrak{G}}\left(\mathbf{q}_{h}^{(-)}, \mathbf{q}_{h}^{(+)}\right)\\
&=\int\limits_{0}^{1} d\tau \int\limits_{\omega_{3}} d\boldsymbol{\xi} \cdot
\Phi_{\mathbf{k}} \cdot \mathbf{s}(\mathbf{q}_{h}; \mathbf{r}, t),
\end{split}
\end{equation}
where $\mathbf{u}_{h}^{n}$ and $\mathbf{u}_{h}^{n+1}$ is the DG representations of the solution $\mathbf{u}_{h}$ at the time steps $t^{n}$ and $t^{n+1}$, respectively, $\boldsymbol{\mathfrak{G}}$ is the Riemann solver, which rescaled on the reference space-time finite element $\omega_{4}$, $\mathbf{q}_{h}^{(-)}$ and $\mathbf{q}_{h}^{(+)}$ are the discrete solutions inside and outside the normalized surface element $d\boldsymbol{\sigma}_{\boldsymbol{\xi}}$ of the reference finite volume $\omega_{3}$ of the cell $\Omega_{i}$, respectively, $\mathbf{q}_{h}$ is the discrete solution in the volume $\omega_{3}$ of the cell $\Omega_{i}$. In the implementation of the ADER-DG scheme, all integrals included in the presented expression for one-step discrete ADER-DG scheme were calculated using Gauss-Legendre quadrature formulas of degree $N$. In the resulting expressions, matrix-matrix operations were identified, multiindex matrices for which were presented in the form of Kronecker products of two-index matrices containing values and integrals of basis functions $\varphi_{k}$ and their derivatives. The Rusanov solver~\cite{Rusanov_solver}, sometimes referred to as the local Lax-Friedrichs flux~\cite{ader_dg_ideal_flows}, and the HLLE solver~\cite{Toro_solvers_2009} were used as the Riemann solver in this work.

The one-step discrete ADER-DG scheme is explicit, despite the presence of a locally implicit LST-DG predictor. The Courant-Friedrichs-Lewy stability criterion is imposed on the time step $\Delta t^{n}$~\cite{ADER_DG_time_step_1, ADER_DG_time_step_2}:
\begin{equation}
\Delta t^{n} = \mathtt{CFL} \cdot \frac{1}{d}\cdot\frac{1}{2N+1} \cdot \min\limits_{k = 1,\ldots,d}\left[\frac{h_{k}}{\left|\lambda_{k}^{max}\right|}\right],
\end{equation}
where $\mathtt{CFL} \leqslant 1$ is the Courant number, $d$ is the spatial dimension of the problem, $N$ is the degrees of polynomials used in the DG representation (\ref{DG_representation}), $h_{k}$ is the spatial mesh step in the $k$-direction, $|\lambda_{k}^{max}|$ is the maximum signal speed in the $k$-direction for which the expression $|\lambda_{k}^{max}| = |u_{k}|+c$ was used, where $u_{k}$ is the flow velocity in the $k$-direction, $c$ is the sound speed. This form differs from the classical form of the Courant-Friedrichs-Lewy stability criterion, used in finite-volume numerical methods, by the presence of the expression $2N+1$ in the denominator. This is used to choose effectively the spatial step of the subgrid in which the a posteriori limiting of the solution is performed.

A more detailed description of the one-step discrete ADER-DG scheme is presented in the basic works~\cite{ader_dg_ideal_flows, ader_dg_dev_1, ader_dg_dev_2, ader_weno_lstdg_ideal, ader_weno_lstdg_diss, ader_dg_diss_flows, ader_dg_ale, ader_dg_grmhd, ader_dg_gr_prd, ader_dg_PNPM, PNPM_DG_2009, PNPM_DG_2010}, as well as in the works~\cite{ader_dg_eff_impl, fron_phys} and~\cite{exahype, ader_dg_hpc_impl_1, ader_dg_hpc_impl_2, ader_dg_hpc_impl_3, ader_dg_hpc_impl_4}.

The one-step discrete ADER-DG scheme is fundamentally a linear scheme of arbitrarily high order, so the numerical solution $\mathbf{u}_{h}^{n+1}$ may contain a violation of monotonicity, which is determined by the well-known Godunov theorem, which shows that there are no linear monotonic numerical schemes above first order for hyperbolic equations. Violation of the monotonicity of the numerical solution $\mathbf{u}_{h}^{n+1}$ leads to the inadmissibility of the numerical solution, in particular, non-physical oscillations of the numerical solution arise, leading to negative values of density $\rho$, internal energy density $e$ and pressure $p$. Therefore, the numerical solution $\mathbf{u}_{h}^{n+1}$ obtained as a result of using the one-step discrete ADER-DG scheme is only preliminary -- the so-called candidate solution $\mathbf{u}_{h}^{*}$.

Within the framework of the space-time adaptive ADER-DG finite element method with LST-DG predictor and a posteriori sub-cell ADER-WENO finite-volume limiting, the resulting numerical candidate solution $\mathbf{u}_{h}^{*}$ is checked for admissibility, for which admissibility criteria are used. In this work, two main admissibility criteria were used: the physical admissibility detector (PAD) and the numerical admissibility detector (NAD), which are widely used~\cite{ader_dg_ideal_flows, ader_dg_dev_2, ader_dg_diss_flows, ader_dg_ale, ader_dg_grmhd, ader_dg_gr_prd, ader_dg_PNPM, PNPM_DG_2009, PNPM_DG_2010, ader_dg_eff_impl, fron_phys} in numerical methods for solving quasi-linear equations with a posteriori correction of the numerical solution. The physical admissibility detector checks the candidate numerical solution for the admissibility of the main physical assumptions of the problem; in this work, these were the condition of positivity of the density $\rho$ and internal energy density $e$, as well as the non-negativity of the mass concentration $c_{k}$ of the components of a multicomponent medium. The positiveness of the pressure $p$ and the non-negativity of the density of the components $\rho_{k}$ of the medium automatically follow from the fulfillment of these admissibility conditions. The numerical admissibility detector in this work is chosen in a cell representation, and is based on the use of the relaxed discrete maximum principle (DMP) in the polynomial sense, and is expressed by the following inequality~\cite{ader_dg_ideal_flows, ader_dg_dev_2, ader_dg_diss_flows}:
\begin{equation}\label{eq:nad_ineq}
\begin{split}
\min\limits_{\mathbf{r}' \in V_{i}} \left(\mathbf{u}_{h}(\mathbf{r}', t^{n})\right) - \boldsymbol{\delta}
\leqslant \mathbf{u}_{h}^{*}(\mathbf{r}, t^{n+1}) \leqslant
\max\limits_{\mathbf{r}' \in V_{i}} \left(\mathbf{u}_{h}(\mathbf{r}', t^{n})\right) + \boldsymbol{\delta},\\
\forall \mathbf{r} \in \Omega_{i},
\end{split}
\end{equation}
where the maximum and minimum are taken over the set $V_{i}$ that contains this cell $\Omega_{i}$ and its Voronov neighboring cells. An additional small vector quantity $\boldsymbol{\delta}$, which is given by the expression
\begin{equation}
\boldsymbol{\delta} = \max\left[\mathbf{\delta}_{0}, \epsilon_{0} \cdot \left(
	\max\limits_{\mathbf{r}' \in V_{i}} \left(\mathbf{u}_{h}(\mathbf{r}', t^{n})\right) -
	\min\limits_{\mathbf{r}' \in V_{i}} \left(\mathbf{u}_{h}(\mathbf{r}', t^{n})\right)
\right)\right],
\end{equation}
determines the tolerance of the criterion -- the real calculation in software implementation is carried out not on the polynomial representation $\mathbf{u}_{h}$ (\ref{DG_representation}) of the solution, but on the basis of a finite-volume sub-cell representation $\mathbf{v}_{j}$ (\ref{eq:u_to_v}), which is formed in a subgrid of the cell; the use of exact extrema of the solution representation would require significant computational costs, especially in the case of high degrees of $N$ and two- and three-dimensional problems, so an approach was chosen with the analysis of the finite-volume representation $\mathbf{v}_{j}$ and a small expansion of the criterion admissibility window by $\boldsymbol{\delta}$. The values $\delta_{0} = 10^{-4}$ and $\epsilon = 10^{-3}$ were chosen in accordance with the recommendations of the works~\cite{ader_dg_ideal_flows, ader_dg_dev_2}. The inequality (\ref{eq:nad_ineq}) is stated in vector form -- feasibility is checked for each individual component of the vector of conservative variables, and the final conclusion of feasibility is determined by the $\land$ logical operation for all components of the vector. Subcell forms of the numerical admissibility detector (the so-called SubNAD), used in particular in methods with flux reconstruction~\cite{DG_FR_2019, DG_FR_2023}, were not used in this work. In general, the choice of criteria for the admissibility of a candidate numerical solution corresponds to the approaches proposed in the works~\cite{ader_dg_ideal_flows, ader_dg_dev_2, ader_dg_diss_flows}, with the addition of a physical admissibility detector with a condition on the concentration of components of a multicomponent medium.

Based on the fulfillment of the admissibility criterion, a troubled cells indicator $\beta$ is calculated, which marks $\beta = 0$ cells in which the candidate solution $\mathbf{u}_{h}^{*}$ is admissible, and marks $\beta = 1$ cells in which the candidate solution is not admissible -- the so-called troubled cells. The numerical solution $\mathbf{u}_{h}^{n+1}$ in troubled cells is recalculated using a high-precision finite-volume ADER-WENO scheme with high stability. To recalculate the solution using a finite-volume numerical scheme, a subgrid $\{\Omega_{i, j}\}$ is created in troubled cells $\Omega_{i} = \cup_{j}\Omega_{i, j}$, in the sub-cells of which an alternative finite-volume sub-cell representation $\mathbf{v}_{j}$ of the solution is determined
\begin{equation}\label{eq:u_to_v}
\mathbf{v}_{j} = \frac{1}{\left|\Omega_{i, j}\right|} \int\limits_{\Omega_{i, j}} \mathbf{u}_{h} dV,
\end{equation}
where $\Omega_{i, j}$ is the subcell, $\left|\Omega_{i, j}\right|$ is the volume of the subcell $\Omega_{i, j}$. The spatial subgrid step $h_{s}$ was chosen to be $h/N_{s}$, which assumed a subgrid size of $N_{s}^{d}$ in each troubled cell. This corresponds to the spatial subgrid step $h_{s} = h/(2N+1)$, which made it possible to use the same time step $\Delta t^{n}$ at the same value of the Courant number \texttt{CFL} as in the case of the one-step discrete ADER-DG scheme.

The finite-volume representation $\mathbf{v}^{n}$ of the solution at the previous time step $t^{n}$ is calculated based on the DG representation $\mathbf{u}_{h}(\mathbf{r}, t^{n})$ of the solution at the previous time step $t^{n}$ using a conservative suitable projector operator $\hat{\mathbb{P}}$:
\begin{equation}\label{eq:p_operator}
\mathbf{v}^{n} = \hat{\mathbb{P}}\cdot\mathbf{u}_{h}^{n},
\end{equation}
which is defined by an integral expression (\ref{eq:u_to_v}). The finite-volume scheme, based on the representation at the previous time step, calculates the solution at the new time step. This is the essence of a posteriori limiting of the solution -- the limiter uses the solution from the previous time step $t^{n}$ to obtain a solution at the new time step $t^{n+1}$ based on information already received from the candidate solution about a possible solution at the new time step $t^{n+1}$. The solution $\mathbf{v}^{n+1}$ at the next time step $t^{n+1}$ obtained using the finite-volume ADER-WENO limiter is converted into a DG representation of the solution $\mathbf{u}_{h}(\mathbf{r}, t^{n+1})$ at the new time step using the suitable high order accurate reconstruction operator $\hat{\mathbb{R}}$:
\begin{equation}\label{eq:r_operator}
\mathbf{u}_{h}^{n+1} = \hat{\mathbb{R}}\cdot\mathbf{v}^{n+1},
\end{equation}
which implements a least squares approximation of the solution $\mathbf{v}^{n+1}$ by DG representation (\ref{DG_representation}) using a matrix representation of the operator $\hat{\mathbb{P}}$ and a pseudo-inverse matrix. The operators $\hat{\mathbb{P}}$ and $\hat{\mathbb{R}}$ have one-side reversibility: $\hat{\mathbb{R}}\circ\hat{\mathbb{P}} = 1$. However, reversibility in a different order of action of the operators does not occur --  $\hat{\mathbb{P}}\circ\hat{\mathbb{R}} \neq 1$, in the chosen case $N_{s} = 2N+1 > N+1$ for $N \geqslant 1$. Due to this property, it is necessary to make an important note -- if the cell $\Omega_{i}$ is a troubled cell for several sequential time steps, then at each new time step it is necessary to use the original finite-volume representation $\mathbf{v}$ of the solution as a solution at the previous time step, and not the representation obtained from the DG representation $\mathbf{u}_{h}$ of previous time step. A more detailed description of the suitable projector operator and high order accurate reconstruction operator is presented in the works~\cite{ader_dg_ideal_flows, ader_dg_dev_2, ader_dg_diss_flows, ader_dg_ale, ader_dg_grmhd, ader_dg_gr_prd, ader_dg_PNPM}.

The high order finite-volume ADER-WENO scheme proposed in the work~\cite{ader_weno_lstdg_ideal, ader_weno_lstdg_diss} was chosen as a limiter. This scheme is also based on the use of the ADER paradigm, and the solution of the generalized Riemann problem for calculating flux terms and calculating contributions from source terms in this scheme is carried out using a local discrete space-time solution obtained using a predictor. The expression used to calculate the finite-volume representation of the solution $\mathbf{v}^{n+1}$ at the next time step $t^{n+1}$ in the coordinates of the reference space-time finite element (mapped $\Omega_{i, j}\times[t^{n}, t^{n+1}]$ into $\omega_{4} = [0, 1]^{4}$), was chosen as
\begin{equation}
\begin{split}
\mathbf{v}^{n+1} = \mathbf{v}^{n} & - 
	\int\limits_{0}^{1} d\tau \oint\limits_{\partial\omega_{3}} d\boldsymbol{\sigma}_{\boldsymbol{\xi}} \cdot
	\boldsymbol{\mathfrak{G}}\left(\mathbf{q}_{h}^{(-)}, \mathbf{q}_{h}^{(+)}\right)\\
	& +\int\limits_{0}^{1} d\tau \int\limits_{\omega_{3}} d\boldsymbol{\xi} \cdot
	\mathbf{s}(\mathbf{q}_{h}; \mathbf{r}, t),
\end{split}
\end{equation}
where $\mathbf{v}^{n+1}$ and $\mathbf{v}^{n}$ is the finite-volume representations of the solution $\mathbf{v}$ at the time steps $t^{n}$ and $t^{n+1}$, respectively, $\boldsymbol{\mathfrak{G}}$ is the Riemann solver, which rescaled on the reference space-time finite element $\omega_{4}$, $\mathbf{q}_{h}^{(-)}$ and $\mathbf{q}_{h}^{(+)}$ are the discrete solutions inside and outside the normalized surface element $d\boldsymbol{\sigma}_{\boldsymbol{\xi}}$ of the subcell reference finite volume $\omega_{3}$ of the subcell $\Omega_{i, j}$, respectively, $\mathbf{q}_{h}$ is the discrete solution in the volume $\omega_{3}$ of the subcell $\Omega_{i, j}$. In the implementation of the finite-volume ADER-WENO scheme, all integrals included in the presented expression for one-step discrete ADER-DG scheme were calculated using Gauss-Legendre quadrature formulas of degree $N_{\rm WENO}$. In the resulting expressions, matrix-matrix operations were identified, multiindex matrices for which were presented in the form of Kronecker products of two-index matrices containing values and integrals of basis functions $\varphi_{k}$ and their derivatives.

The expression for the LST-DG predictor for finite-volume ADER-WENO scheme, which was used to calculate the discrete solution $\mathbf{q}_{h}$, has a form similar to the expression for the ADER-DG scheme (\ref{eq:lst_dg_predictor_final}), with the exception of the expression on the right side -- the expression $\hat{\mathbf{u}}_{k}$ is replaced on the expression $\hat{\mathbf{w}}_{k}$ for the coefficients of a conservative WENO-reconstruction, which was chosen in the following form:
\begin{equation}\label{WENO_reconstruction}
\begin{split}
\mathbf{w}_{h}(\mathbf{r}) = \sum\limits_{\mathbf{k}} \hat{\mathbf{w}}_{\mathbf{k}} \cdot \Phi_{\mathbf{k}}\big(\boldsymbol{\xi}(\mathbf{r})\big),
\end{split}
\end{equation}
where $\hat{\mathbf{w}}_{\mathbf{k}} = \hat{\mathbf{w}}_{k_{1}k_{2}k_{3}}$ are the coefficients of the WENO-reconstruction; $\Phi_{\mathbf{k}}(\boldsymbol{\xi}) = \varphi_{k_{1}}\left(\xi\right) \varphi_{k_{2}}\left(\eta\right) \varphi_{k_{3}}\left(\zeta\right)$ is the basis functions represented as tensor products representation; $\varphi_{k} = \varphi_{k}(\xi)$ are the one-dimensional basis functions; and multiindex $\mathbf{k} = (k_{1}, k_{2}, k_{3}) \in [0, N_{\rm WENO}]^{3}$. The procedure for obtaining reconstruction in one-dimensional and multidimensional cases was chosen in the form proposed in the work: a system of equations for reconstruction coefficients $\hat{\mathbf{w}}_{\mathbf{k}}$ was obtained from conservative conditions; in the two-dimensional and three-dimensional cases, a dimension-by-dimension technique was used. 

The order of the polynomials $N_{\rm WENO}$ used in the WENO-reconstruction in the finite-volume scheme was chosen to be different from the order of the polynomials $N$ used in the DG representation. All results presented in this work were obtained using $N_{\rm WENO} = 1$, which determined the second order for the limiter -- finite-volume ADER-WENO2 method. It will be shown below that this choice did not in any way reduce the accuracy of the space-time adaptive ADER-DG finite element method with LST-DG predictor and a posteriori sub-cell ADER-WENO finite-volume limiting. A more detailed description of the finite-volume ADER-WENO scheme is presented in the works~\cite{ader_stiff_1, ader_stiff_2, ader_weno_lstdg_ideal, ader_weno_lstdg_diss}.

It should be noted that the use of the finite-volume ADER-WENO method as a limiter is not strictly necessary. In the works~\cite{ader_dg_diss_flows, ader_dg_ale}, finite-volume TVD limiters were used, and a solution with high accuracy was obtained. In this work, the classical Godunov method of the first order as a limiter was also implemented, however, the results obtained demonstrated a significantly greater numerical diffusion than when using ADER-WENO2 -- the width of shock and detonation waves increased by $2$-$5$ times. Therefore, all the results presented later in the text of this work were obtained using finite-volume ADER-WENO2 method as a limiter.

The software implementation of the method was developed using the \texttt{C++} programming language with partial implementation of \texttt{x86} intrinsics and \texttt{x86} assembly language into the code to use the \texttt{avx-512} instruction set. Multithreading and multiprocessing execution was organized using the \texttt{OpenMP} and \texttt{MPI} interfaces. All main matrix-matrix operations were implemented within the BLAS interface. Software implementations were carried out separately for one-dimensional, two-dimensional and three-dimensional problems. All calculations were carried out on the HEDT class workstation with Intel i9-10980xe processor and 256 GB RAM.

\section{Applications of the numerical method to the classical gas dynamics problems}
\label{sec:apps_cgd_problems}

\subsection{Accuracy and convergence}
\label{sec:apps_cgd_problems:acc_conv}

The accuracy and convergence of the space-time adaptive ADER finite-element DG numerical method with a posteriori sub-cell ADER-WENO finite-volume limiting were tested based on a numerical solution of the three-dimensional problem of the advection of an isentropic vortex in a periodic cubic spatial domain~\cite{ader_dg_ideal_flows}. The chosen problem relates to sufficiently complex problems for classical numerical methods of gas dynamics and allows one to adequately study the dissipative and dispersive properties of the numerical method~\cite{Shu_weno_nasa_rep_1997}. The use of an isentropic vortex advection problem is standard in the field of research of high-order numerical methods.

The initial conditions are presented in the form of a superposition of stationary ambient gas flow with parameters $\rho_{\infty} = 1$, $(u_{\infty}, v_{\infty}, w_{\infty}) = (1, 1, 0)$, $p_{\infty} = 1$, $T_{\infty} = 1$ and a local perturbation, which is an isentropic vortex $\delta\rho$, $(\delta u, \delta v, \delta w)$, $\delta p$, $\delta T$:
\begin{equation}
\left[
\begin{array}{c}
\rho\\ 
u\\ 
v\\ 
w\\ 
p
\end{array}
\right] = 
\left[
\begin{array}{c}
\rho_{\infty} + \delta\rho\\ 
u_{\infty} + \delta u\\ 
v_{\infty} + \delta v\\ 
w_{\infty} + \delta w\\ 
p_{\infty} + \delta p
\end{array}
\right].
\end{equation}
The expression for the coordinate dependencies of the initial conditions was presented in the form:
\begin{equation}
\begin{split}
&\delta p = \left(1 + \delta T\right)^{\frac{\gamma}{\gamma-1}} - 1;\\
&\delta u = - (y - y_{0})\cdot\cfrac{\varepsilon}{2\pi}\cdot\exp\left[\cfrac{1-r^{2}}{2}\right];\\
&\delta v = + (x - x_{0})\cdot\cfrac{\varepsilon}{2\pi}\cdot\exp\left[\cfrac{1-r^{2}}{2}\right];\\
&\delta w = 0;\\
&\delta T = -\frac{\varepsilon^{2}(\gamma-1)}{8\gamma\pi^{2}}\cdot\exp\left[1-r^{2}\right];
\end{split}
\end{equation}
where $r^{2} = (x-x_{0})^{2} + (y-y_{0})^{2}$ determines the distance to the initial position of the vortex core $(x_{0}, y_{0}) = (5, 5)$ in the $xy$-plane; the vortex strength $\varepsilon = 5$, the adiabatic index $\gamma = 1.4$. The initial conditions were chosen to be independent of the $z$ coordinate, therefore, from the coordinate dependence point of view, this problem is two-dimensional. The mass density of entropy $s = p/\rho^{\gamma}$ is constant everywhere in the flow.

\begin{figure}[h!]
\centering
\includegraphics[width=0.49\textwidth]{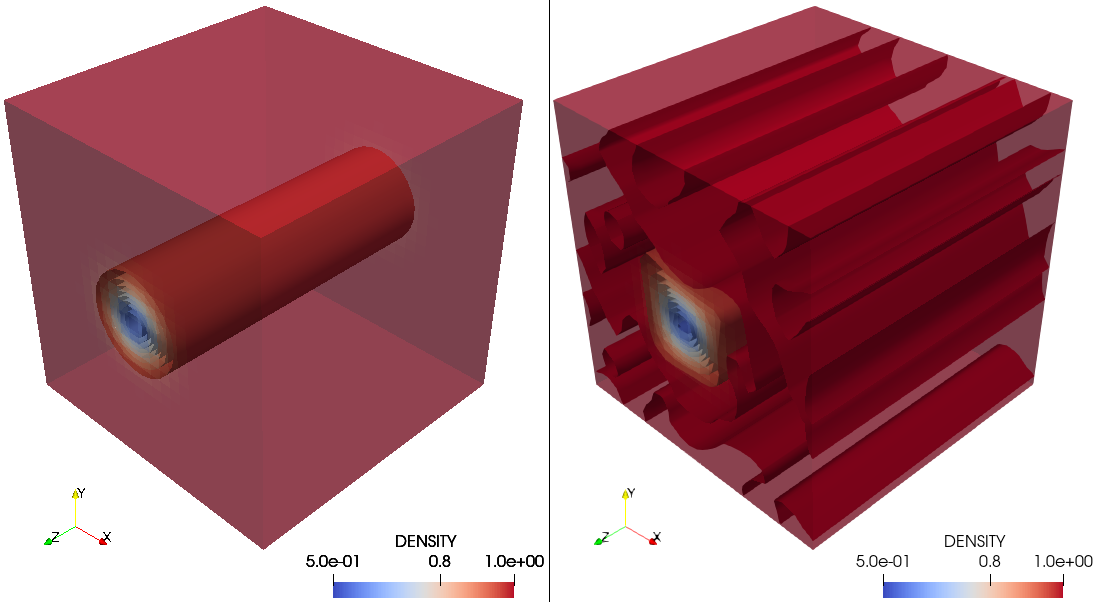}
\includegraphics[width=0.49\textwidth]{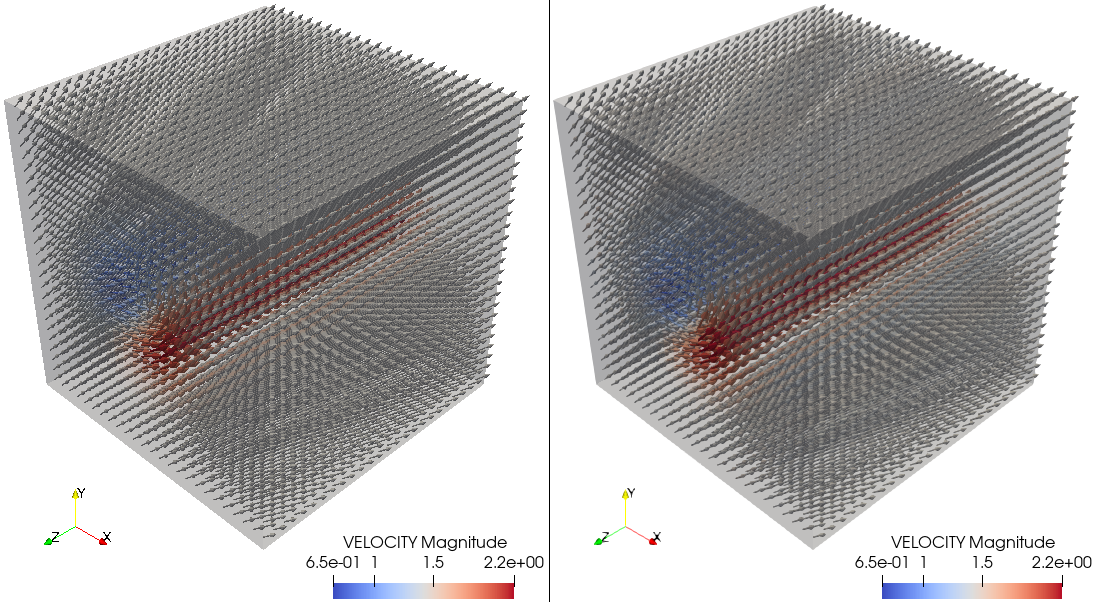}
\caption{\label{fig:vortex_adv_dg_12_1x1x1}
Numerical solution of the three-dimensional problem of an isentropic vortex advection 
(a detailed statement of the problem is presented in the text)
obtained using the ADER-DG-$\mathbb{P}_{12}$ method on single cell mesh $1 \times 1 \times 1$
at the final time $t_{\rm final} = 10.0$:
subcells finite-volume representation of exact (left) and numerical (right) solutions
for density $\rho$ (top) and velocity $\mathbf{v}$ (bottom).
}
\end{figure}

The coordinate domain was chosen in the form of a cube $\Omega = [0, 10]\times[0, 10]\times[0, 10]$ with periodic boundary conditions. The exact analytical solution of the problem represents the process of simple advection of an isentropic vortex, which can be expressed by a function $\phi = \phi(x_{0} - u_{\infty}t, y_{0} - v_{\infty}t, z_{0} - w_{\infty}t)$ for the main hydrodynamic variables. The use of the selected initial conditions, coordinate domain and periodic boundary conditions leads to the fact that after a period of time $\Delta t = 10$ the vortex returns to its original coordinate position, and the exact solution returns to the initial conditions, therefore the final time $t_{\rm final} = 10.0$ was chosen. The initial conditions and the exact solution are shown in Figure~\ref{fig:vortex_adv_dg_12_1x1x1} (left). 

\begin{table*}[h!]
\begin{center}
\caption{\label{tab:conv_orders}
$L_{1}$, $L_{2}$ and $L_{\infty}$ norms of errors $\epsilon$, by density $\rho$,
and convergence orders $p$ for ADER-DG-$\mathbb{P}_{N}$ method with a posteriori 
limitation of the solution by a ADER-WENO2 finite volume limiter;
$p_{\rm theor.} = N+1$ is the theoretical accuracy order of the ADER-DG-$\mathbb{P}_{N}$ method.
}
\begin{tabular}{||l|l|ccc|ccc|l||}
\hline
& cells 
& $\qquad \epsilon_{L_{1}} \qquad$ & $\qquad \epsilon_{L_{2}} \qquad$ & $\qquad \epsilon_{L_{\infty}} \qquad$
& $\qquad p_{L_{1}} \qquad$ & $\qquad p_{L_{2}} \qquad$ & $\qquad p_{L_{\infty}} \qquad$ & theor. \\
\hline
DG-$\mathbb{P}_{1}$		&	$10^{3}$	&	1.05E+01	&	8.74E--01	&	2.21E--01	&	--		&	--		&	--		&	2\\
						&	$20^{3}$	&	2.63E+00	&	1.98E--01	&	4.87E--02	&	1.99	&	2.14	&	2.18	&	\\
						&	$30^{3}$	&	1.05E+00	&	8.02E--02	&	3.46E--02	&	2.28	&	2.23	&	0.85	&	\\
						&	$40^{3}$	&	5.10E--01	&	3.94E--02	&	1.80E--02	&	2.50	&	2.47	&	2.27	&	\\[1mm]
DG-$\mathbb{P}_{2}$		&	$10^{3}$	&	1.56E+00	&	1.13E--01	&	8.54E--02	&	--		&	--		&	--		&	3\\
						&	$20^{3}$	&	1.65E--01	&	1.10E--02	&	4.32E--03	&	3.24	&	3.36	&	4.30	&	\\
						&	$30^{3}$	&	3.49E--02	&	2.58E--03	&	1.55E--03	&	3.83	&	3.58	&	2.53	&	\\
						&	$40^{3}$	&	1.21E--02	&	9.55E--04	&	6.70E--04	&	3.68	&	3.46	&	2.92	&	\\[1mm]
DG-$\mathbb{P}_{3}$		&	$10^{3}$	&	3.10E--01	&	1.75E--02	&	6.37E--03	&	--		&	--		&	--		&	4\\
						&	$20^{3}$	&	1.62E--02	&	8.31E--04	&	3.33E--04	&	4.26	&	4.39	&	4.26	&	\\
						&	$30^{3}$	&	2.72E--03	&	1.35E--04	&	7.44E--05	&	4.40	&	4.49	&	3.70	&	\\
						&	$40^{3}$	&	7.73E--04	&	3.87E--05	&	2.57E--05	&	4.37	&	4.33	&	3.70	&	\\[1mm]
DG-$\mathbb{P}_{4}$		&	$10^{3}$	&	5.42E--02	&	2.88E--03	&	1.07E--03	&	--		&	--		&	--		&	5\\
						&	$20^{3}$	&	1.48E--03	&	6.65E--05	&	3.02E--05	&	5.19	&	5.43	&	5.15	&	\\
						&	$30^{3}$	&	1.76E--04	&	7.87E--06	&	5.12E--06	&	5.26	&	5.26	&	4.38	&	\\
						&	$40^{3}$	&	3.75E--05	&	1.73E--06	&	9.74E--07	&	5.38	&	5.27	&	5.77	&	\\[1mm]
DG-$\mathbb{P}_{5}$		&	$ 5^{3}$	&	4.26E--01	&	2.04E--02	&	5.95E--03	&	--		&	--		&	--		&	6\\
						&	$10^{3}$	&	8.16E--03	&	4.27E--04	&	2.04E--04	&	5.71	&	5.58	&	4.86	&	\\
						&	$15^{3}$	&	8.15E--04	&	3.70E--05	&	1.46E--05	&	5.68	&	6.03	&	6.51	&	\\
						&	$20^{3}$	&	1.32E--04	&	7.12E--06	&	2.32E--06	&	6.33	&	5.73	&	6.38	&	\\[1mm]
DG-$\mathbb{P}_{6}$		&	$ 5^{3}$	&	1.33E--01	&	6.51E--03	&	2.29E--03	&	--		&	--		&	--		&	7\\
						&	$10^{3}$	&	1.88E--03	&	9.31E--05	&	4.68E--05	&	6.15	&	6.13	&	5.61	&	\\
						&	$15^{3}$	&	1.22E--04	&	5.36E--06	&	2.53E--06	&	6.76	&	7.04	&	7.19	&	\\
						&	$20^{3}$	&	1.54E--05	&	6.69E--07	&	3.31E--07	&	7.18	&	7.24	&	7.07	&	\\[1mm]
DG-$\mathbb{P}_{7}$		&	$ 5^{3}$	&	5.15E--02	&	3.73E--03	&	1.50E--03	&	--		&	--		&	--		&	8\\
						&	$10^{3}$	&	3.72E--04	&	1.60E--05	&	7.98E--06	&	7.11	&	7.86	&	7.56	&	\\
						&	$15^{3}$	&	1.69E--05	&	7.18E--07	&	3.65E--07	&	7.63	&	7.66	&	7.61	&	\\
						&	$20^{3}$	&	1.61E--06	&	6.92E--08	&	3.19E--08	&	8.17	&	8.13	&	8.47	&	\\[1mm]
DG-$\mathbb{P}_{8}$		&	$ 5^{3}$	&	1.98E--02	&	1.35E--03	&	5.62E--04	&	--		&	--		&	--		&	9\\
						&	$10^{3}$	&	7.41E--05	&	3.34E--06	&	1.44E--06	&	8.06	&	8.66	&	8.61	&	\\
						&	$15^{3}$	&	1.89E--06	&	8.22E--08	&	4.06E--08	&	9.04	&	9.14	&	8.80	&	\\
						&	$20^{3}$	&	1.39E--07	&	6.62E--09	&	3.25E--09	&	9.08	&	8.76	&	8.77	&	\\[1mm]
DG-$\mathbb{P}_{9}$		&	$ 2^{3}$	&	2.55E+00	&	1.26E--01	&	5.63E--02	&	--		&	--		&	--		&	10\\
						&	$ 4^{3}$	&	3.86E--02	&	2.50E--03	&	1.50E--03	&	6.05	&	5.65	&	5.23	&	\\
						&	$ 6^{3}$	&	1.38E--03	&	7.99E--05	&	3.50E--05	&	8.22	&	8.49	&	9.27	&	\\
						&	$ 8^{3}$	&	9.00E--05	&	4.20E--06	&	2.21E--06	&	9.49	&	10.24	&	9.60	&	\\[1mm]
DG-$\mathbb{P}_{10}$	&	$ 2^{3}$	&	1.40E+00	&	6.44E--02	&	2.31E--02	&	--		&	--		&	--		&	11\\
						&	$ 4^{3}$	&	1.47E--02	&	8.94E--04	&	4.73E--04	&	6.58	&	6.17	&	5.61	&	\\
						&	$ 6^{3}$	&	3.35E--04	&	1.88E--05	&	1.15E--05	&	9.32	&	9.52	&	9.16	&	\\
						&	$ 8^{3}$	&	2.07E--05	&	1.02E--06	&	4.62E--07	&	9.69	&	10.14	&	11.18	&	\\[1mm]
DG-$\mathbb{P}_{11}$	&	$ 2^{3}$	&	8.40E--01	&	4.00E--02	&	8.58E--03	&	--		&	--		&	--		&	12\\
						&	$ 4^{3}$	&	5.63E--03	&	3.30E--04	&	1.54E--04	&	7.22	&	6.92	&	5.80	&	\\
						&	$ 6^{3}$	&	9.28E--05	&	5.11E--06	&	2.55E--06	&	10.12	&	10.28	&	10.12	&	\\
						&	$ 8^{3}$	&	3.24E--06	&	2.12E--07	&	9.70E--08	&	11.66	&	11.06	&	11.36	&	\\[1mm]
DG-$\mathbb{P}_{12}$	&	$ 2^{3}$	&	5.72E--01	&	2.79E--02	&	6.75E--03	&	--		&	--		&	--		&	13\\
						&	$ 4^{3}$	&	2.09E--03	&	1.15E--04	&	4.96E--05	&	8.10	&	7.92	&	7.09	&	\\
						&	$ 6^{3}$	&	2.89E--05	&	1.55E--06	&	6.78E--07	&	10.55	&	10.63	&	10.58	&	\\
						&	$ 8^{3}$	&	9.57E--07	&	5.64E--08	&	2.24E--08	&	11.85	&	11.51	&	11.86	&	\\
\hline
\end{tabular}
\end{center}
\end{table*}

Figure~\ref{fig:vortex_adv_dg_12_1x1x1} shows numerical solution to this problem obtained using the ADER-DG-$\mathbb{P}_{12}$ method on single cell mesh $1 \times 1 \times 1$. The presented results show that the numerical solution is visually different from the exact analytical solution. However, when the single cell size of the spatial mesh is taken into account, this result demonstrates the unusually high accuracy of the ADER-DG-$\mathbb{P}_{N}$ method.

The error $\epsilon$ of the numerical solution was calculated in three functional norms $L_{1}$, $L_{2}$, $L_{\infty}$ for the density $\rho$:
\begin{equation}
\begin{split}
&\epsilon_{L_{1}} = \int\limits_{\Omega} \left|\rho(\mathbf{r}, t_{\rm final}) - \rho_{\rm exact}(\mathbf{r})\right| dV;\\
&\epsilon_{L_{2}}^{2} = \int\limits_{\Omega} \left[\rho(\mathbf{r}, t_{\rm final}) - \rho_{\rm exact}(\mathbf{r})\right]^{2} dV;\\
&\epsilon_{L_{\infty}} = \operatorname{ess}\sup\limits_{\hspace{-5mm}\mathbf{r}\in\Omega} 
						 \left|\rho(\mathbf{r}, t_{\rm final}) - \rho_{\rm exact}(\mathbf{r})\right|;
\end{split}
\end{equation}
where the integrals were calculated as the sum of the integrals for each finite-element cell $\Omega_{n}$ over the function of the DG representation of the solution $\mathbf{u}$, and the calculation was carried out using the Gauss-Legendre quadrature formula based on polynomials of degree $40$ for each coordinate direction; the calculation of the supremum to determine the norm of error $\epsilon_{L_{\infty}}$ was performed using a finite-volume subcell representation of the solution. The errors $\epsilon$ were obtained for a set of mesh coordinate steps $h$, from which the empirical values of the convergence orders $p$ were calculated: $\epsilon \sim h^{p}$, therefore $p = \ln(\epsilon(h_{1})/\epsilon(h_{2}))/\ln(h_{1}/h_{2})$.

The calculated empirical convergence orders $p_{1}$, $p_{2}$, $p_{\infty}$ for errors $\epsilon_{L_{1}}$, $\epsilon_{L_{2}}$, $\epsilon_{L_{\infty}}$ for ADER-DG-$\mathbb{P}_{N}$ method with a posteriori limitation of the solution by a ADER-WENO2 finite volume limiter are presented in Table~\ref{tab:conv_orders}. The results are obtained for degrees $N = 1, \ldots, 12$. It should be noted that the solution to this test problem is smooth, and the limiter was called only for the central cell in the case $N = 1$; in all other cases the solution was obtained from the use of a limiter -- the admissibility criteria were not activated in the case $N \geqslant 2$. In the case $N = 1$ when several troubled cells were formed in the central region of the coordinate domain, the DW representation obtained from the finite-volume subcell representation was also used to calculate the errors $\epsilon_{L_{1}}$ and $\epsilon_{L_{2}}$.

The expected theoretical values of the convergence orders $p_{\rm theor.} = N+1$~\cite{ader_dg_dev_1} are presented in Table~\ref{tab:conv_orders} for comparison. The presented results show that in the case of orders of polynomials $N \leqslant 9$ in the DG-representation, there is a good correspondence between the empirical $p$ and theoretical $p_{\rm theor.}$ convergence orders. It should be noted, in order to obtain correct values of empirical orders of convergence in the case of large degrees $N = 8, \ldots, 10$, a special choice of the mesh coordinate step $h$ is necessary -- in the region of small $h$ there is a significant increase in round-off errors, which leads to an increase in error $\epsilon$. However, in the case of degrees of polynomials $N \geqslant 11$, there is a decrease in the empirical convergence orders compared to the theoretical convergence orders $p_{\rm theor.}$, which is also associated with round-off calculation errors. The case $N = 10$ is marginal in this sense -- the consistency between the empirical and theoretical orders is reached only for the error norm $\epsilon_{L_{\infty}}$, for which the calculation was carried out not using the exact DG-representation, but using the finite-volume subcell representation of the solution. ``Superconvergence'' is observed in some cases, what is expressed in larger values of empirical convergence orders compared to theoretical orders $p_{\rm theor.}$; however, this is characteristic of crossover effects in the transition to a step $h \rightarrow 0_{+}$.

The presented results on the convergence orders for ADER-DG-$\mathbb{P}_{N}$ method with a posteriori limitation of the solution by finite volume limiter are in good agreement with the results of the basic work~\cite{ader_dg_dev_1, ader_dg_ideal_flows}. This allows us to conclude that the software implementation of the ADER-DG-$\mathbb{P}_{N}$ method is correct.

\subsection{One-dimensional gas dynamics problems}
\label{sec:apps_cgd_problems:problems_1d}

Obtaining a correct solution to the problem of advection of an isentropic vortex made it possible to determine the correctness of the software implementation of the ADER-DG-$\mathbb{P}_{N}$ method; however, numerical simulation of detonation waves requires clarification of the correctness of the full software implementation of the ADER-DG-$\mathbb{P}_{N}$ method with a posteriori sub-cell ADER-WENO finite-volume limiting for simulation of flows with discontinuities in the solution.

\begin{table}[h!]
\caption{%
Data for one-dimensional Riemann problem tests.
The parameter values $(\rho_{L}, u_{L}, p_{L})$ correspond to the state of the flow to the left of the discontinuity; 
the parameter values $(\rho_{R}, u_{R}, p_{R})$ correspond to the state of the flow to the right of the discontinuity.
}
\label{tab:classical_tests_1d}
\centering
\begin{tabular}{|c||c|c|c||c|c|c|}
\hline
Test & $\rho_{L}$ & $u_{L}$ & $p_{L}$ & $\rho_{R}$ & $u_{R}$ & $p_{R}$ \\
\hline
1 & $1.000$ & $0.000$ & $1.000$ & $0.125$ & $0.000$ & $0.100$ \\
\hline
2 & $0.445$ & $0.698$ & $3.528$ & $0.500$ & $0.000$ & $0.571$ \\
\hline
3 & $1.000$ & $-1.000$ & $1.000$ & $1.000$ & $+1.000$ & $1.000$ \\
\hline
4 & $1.000$ & $+1.000$ & $1.000$ & $1.000$ & $-1.000$ & $1.000$ \\
\hline
\end{tabular}
\end{table}

\begin{figure*}[h!]
\centering
\includegraphics[width=0.245\textwidth]{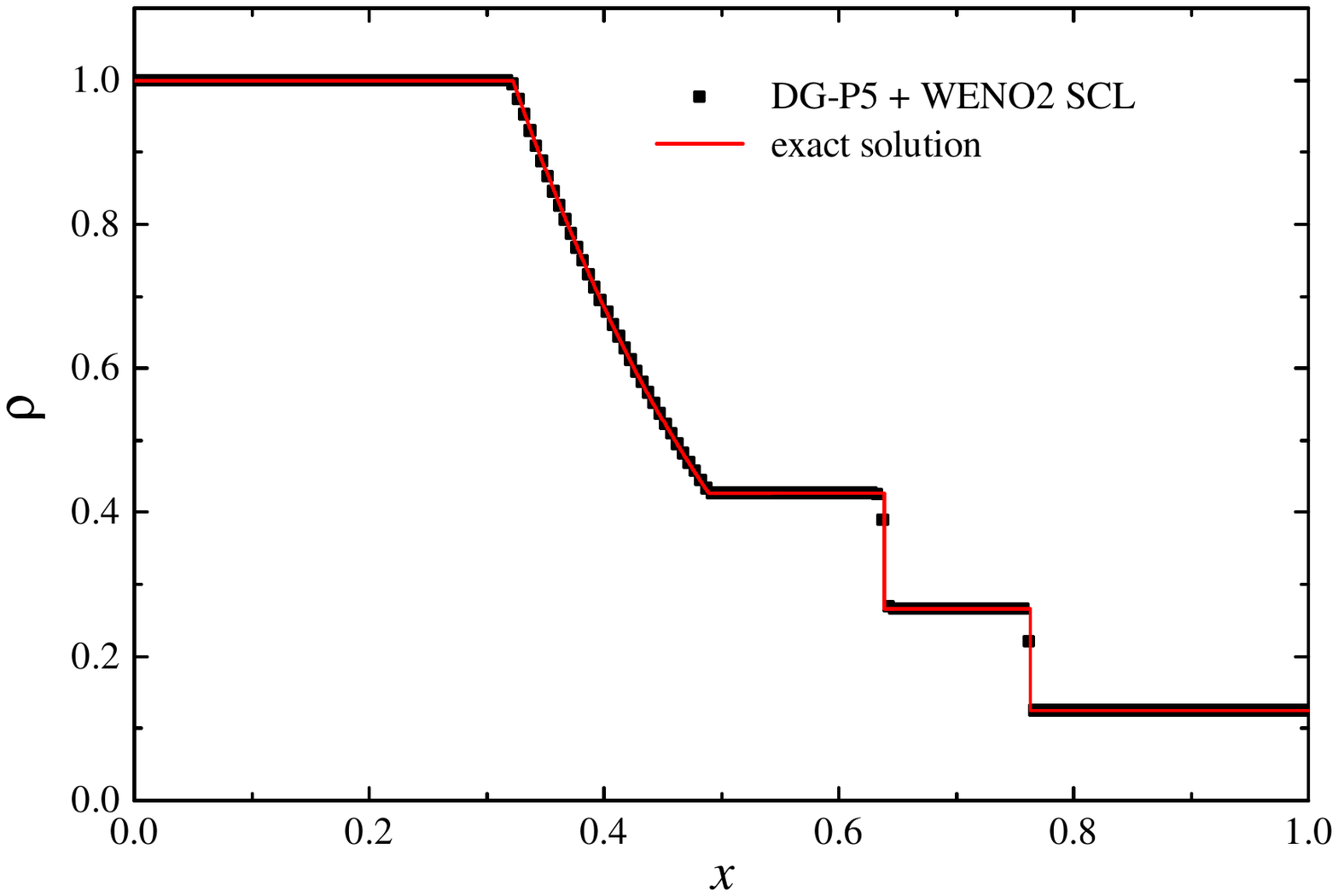}
\includegraphics[width=0.245\textwidth]{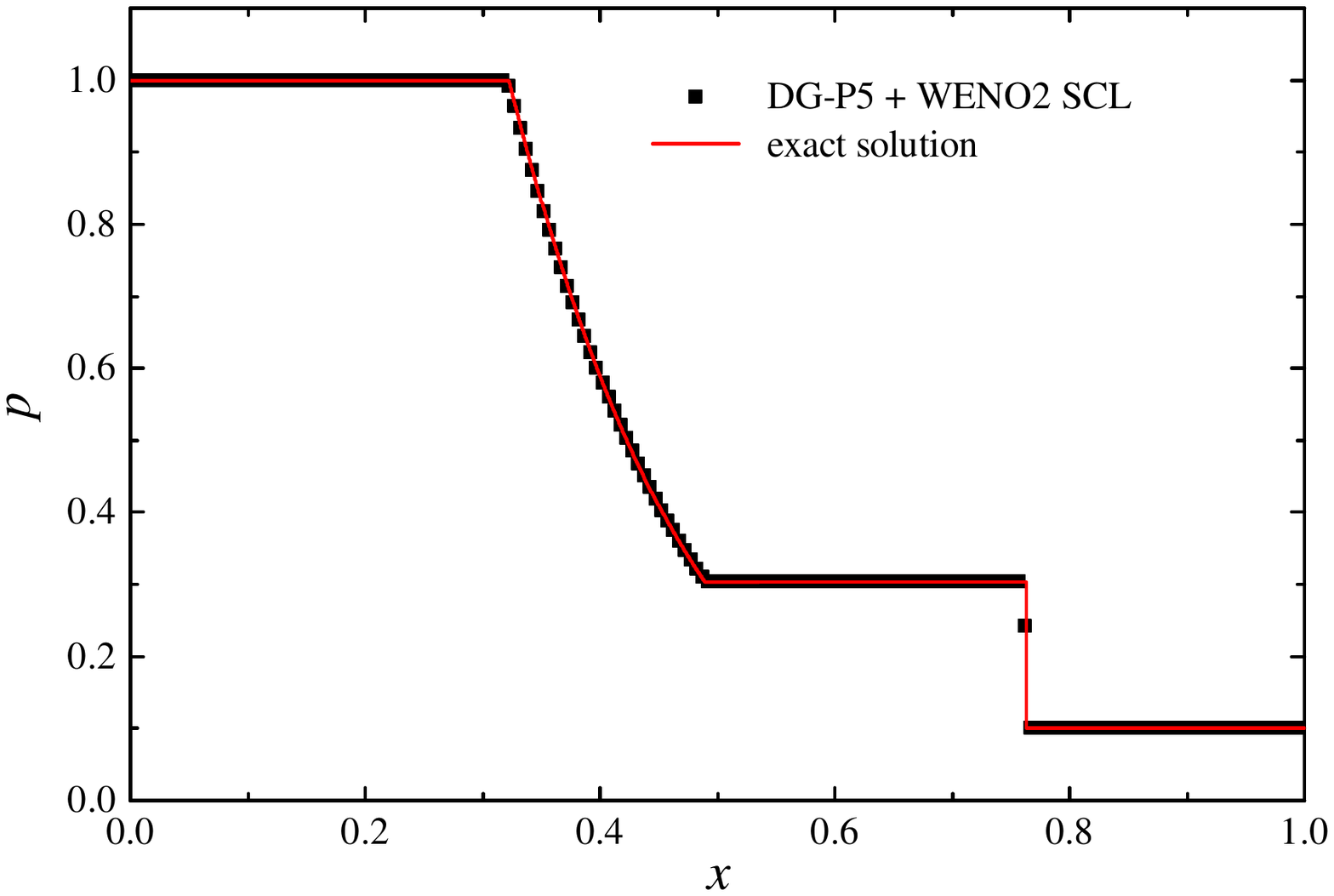}
\includegraphics[width=0.245\textwidth]{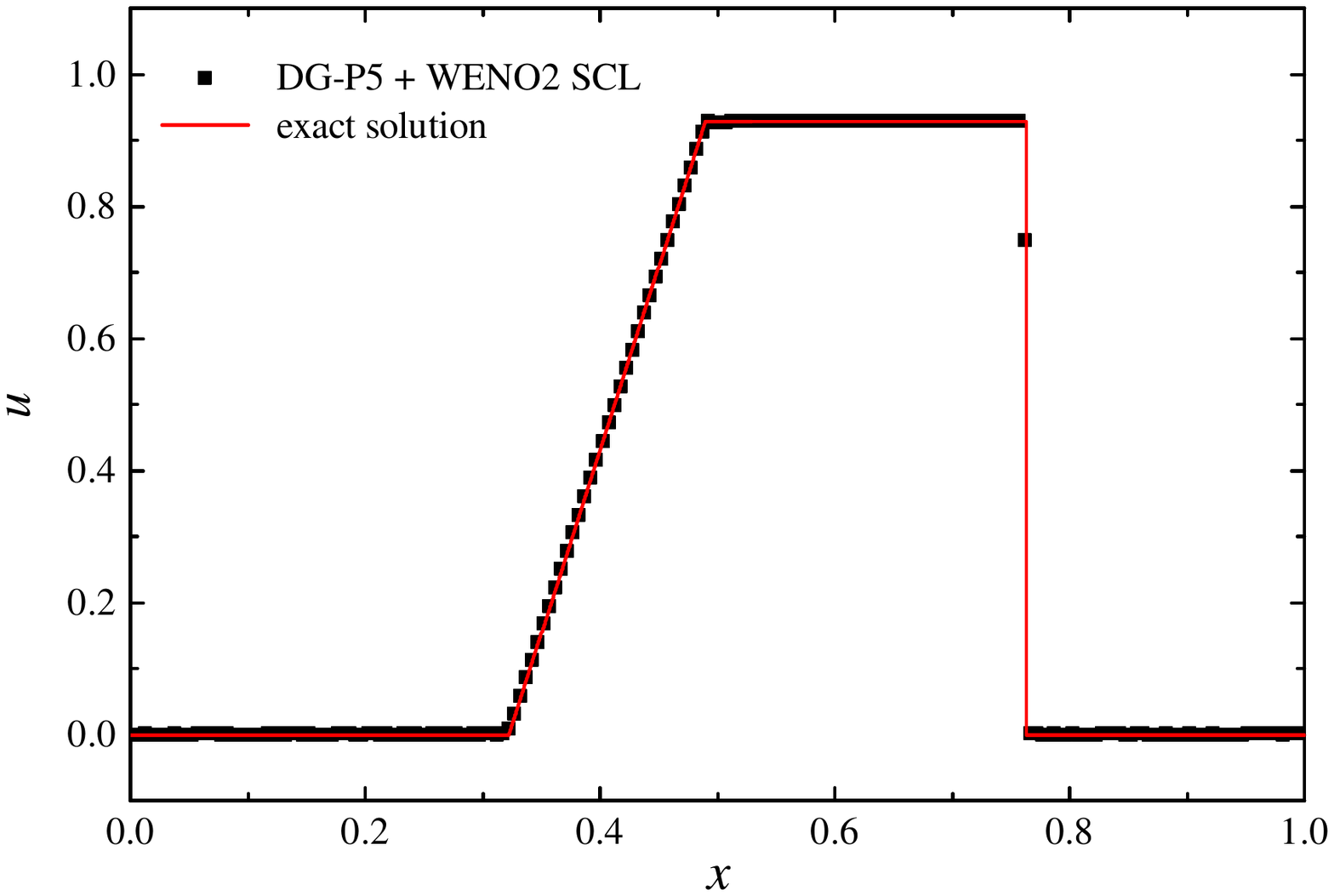}
\includegraphics[width=0.245\textwidth]{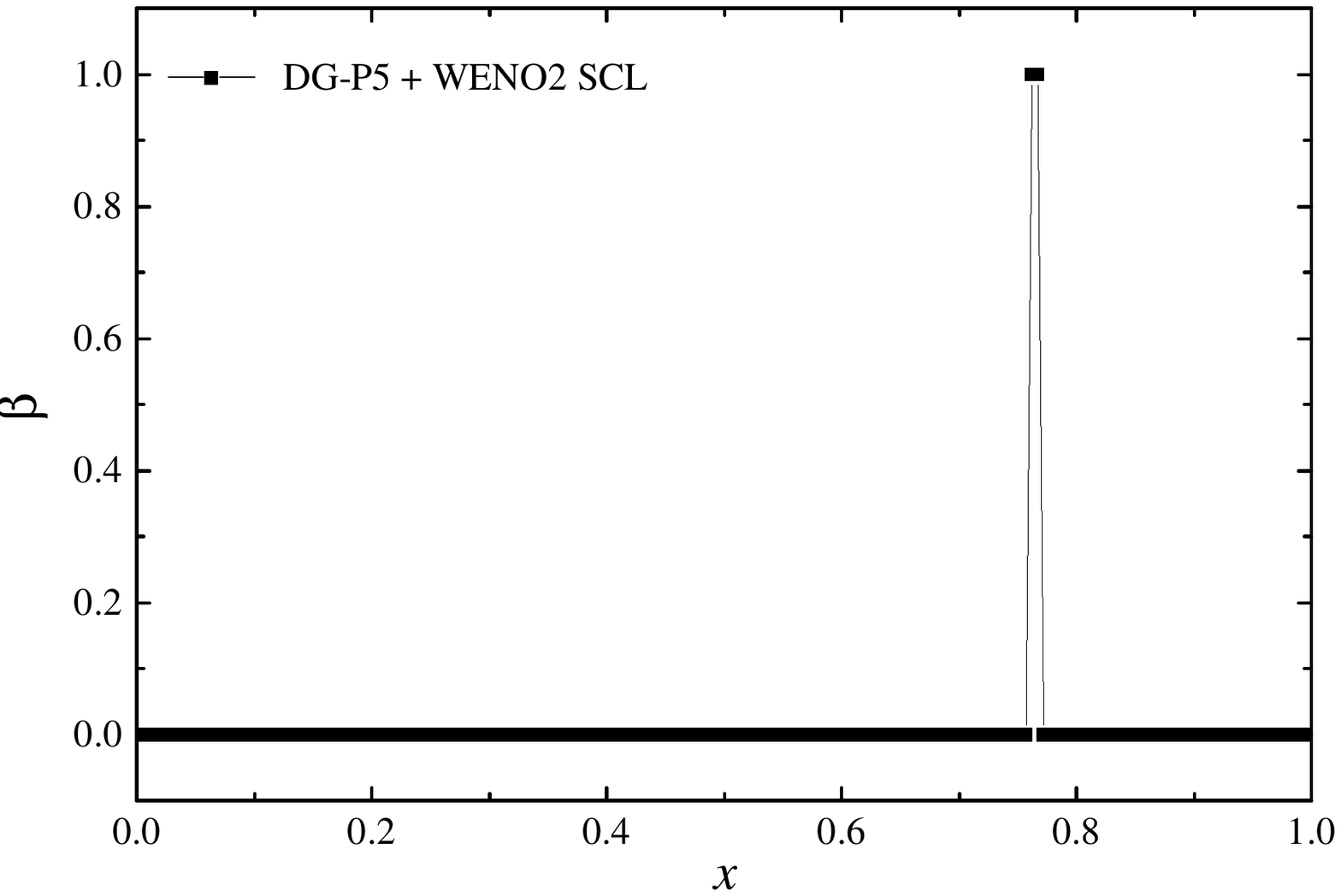}\\
\includegraphics[width=0.245\textwidth]{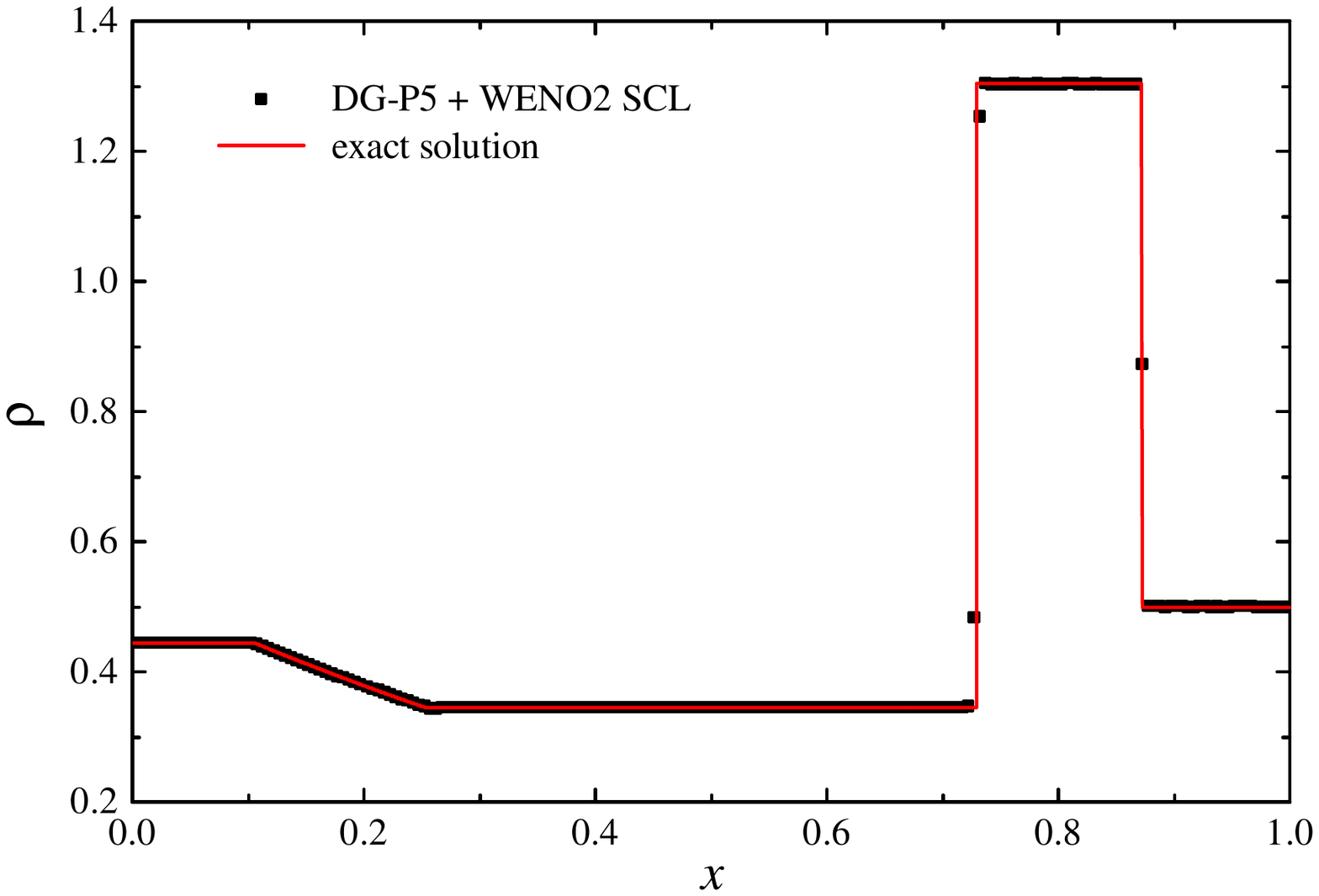}
\includegraphics[width=0.245\textwidth]{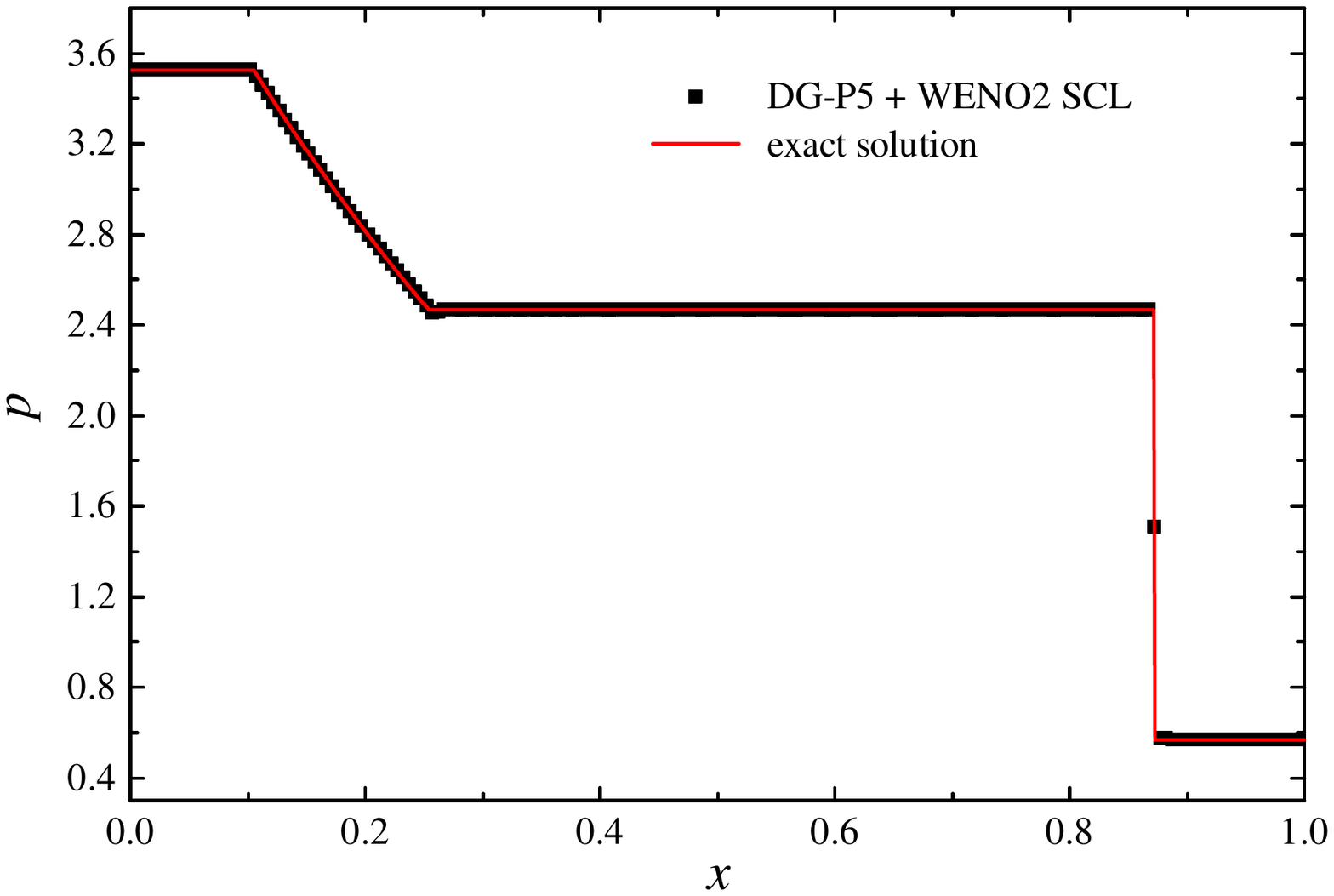}
\includegraphics[width=0.245\textwidth]{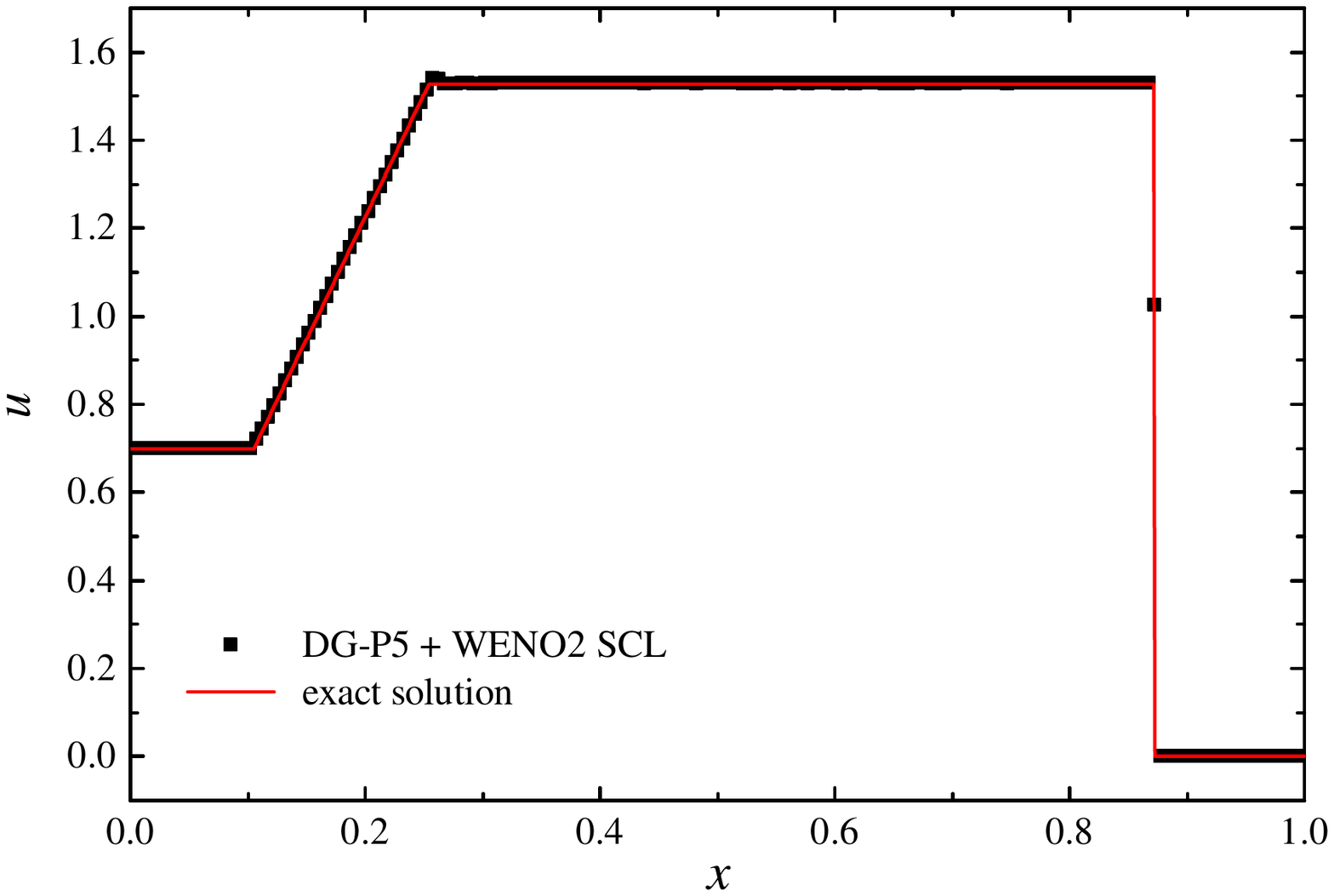}
\includegraphics[width=0.245\textwidth]{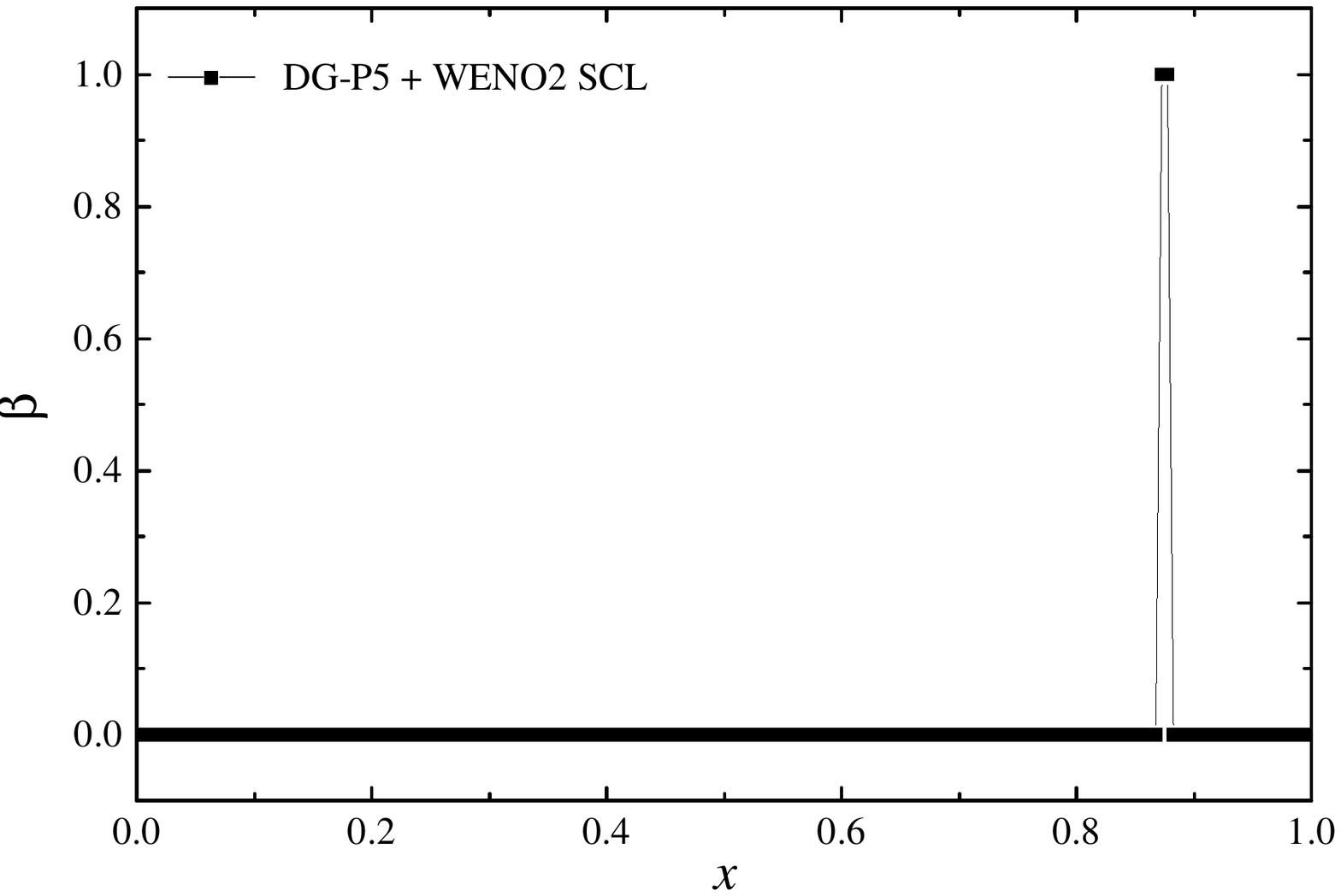}\\
\includegraphics[width=0.245\textwidth]{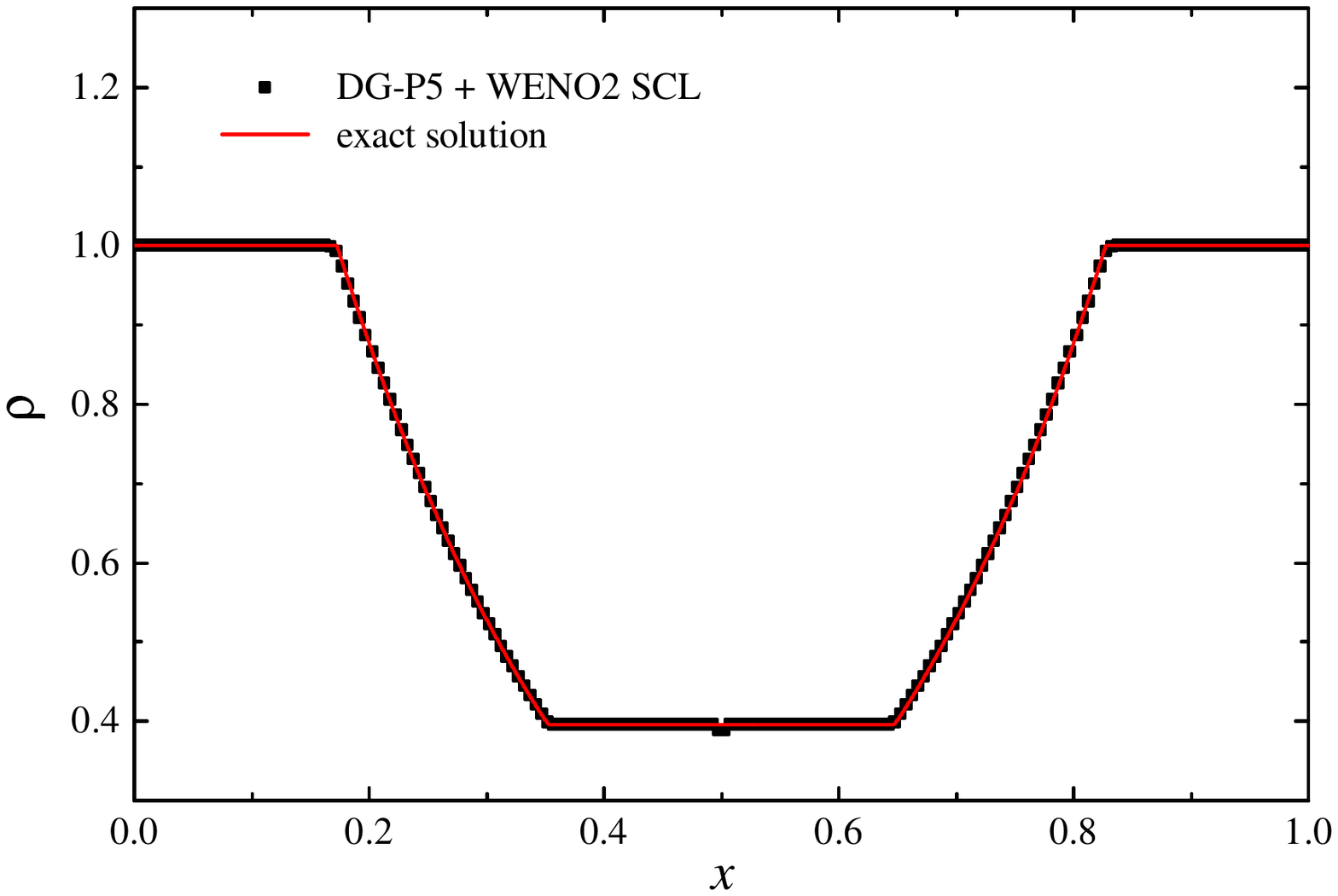}
\includegraphics[width=0.245\textwidth]{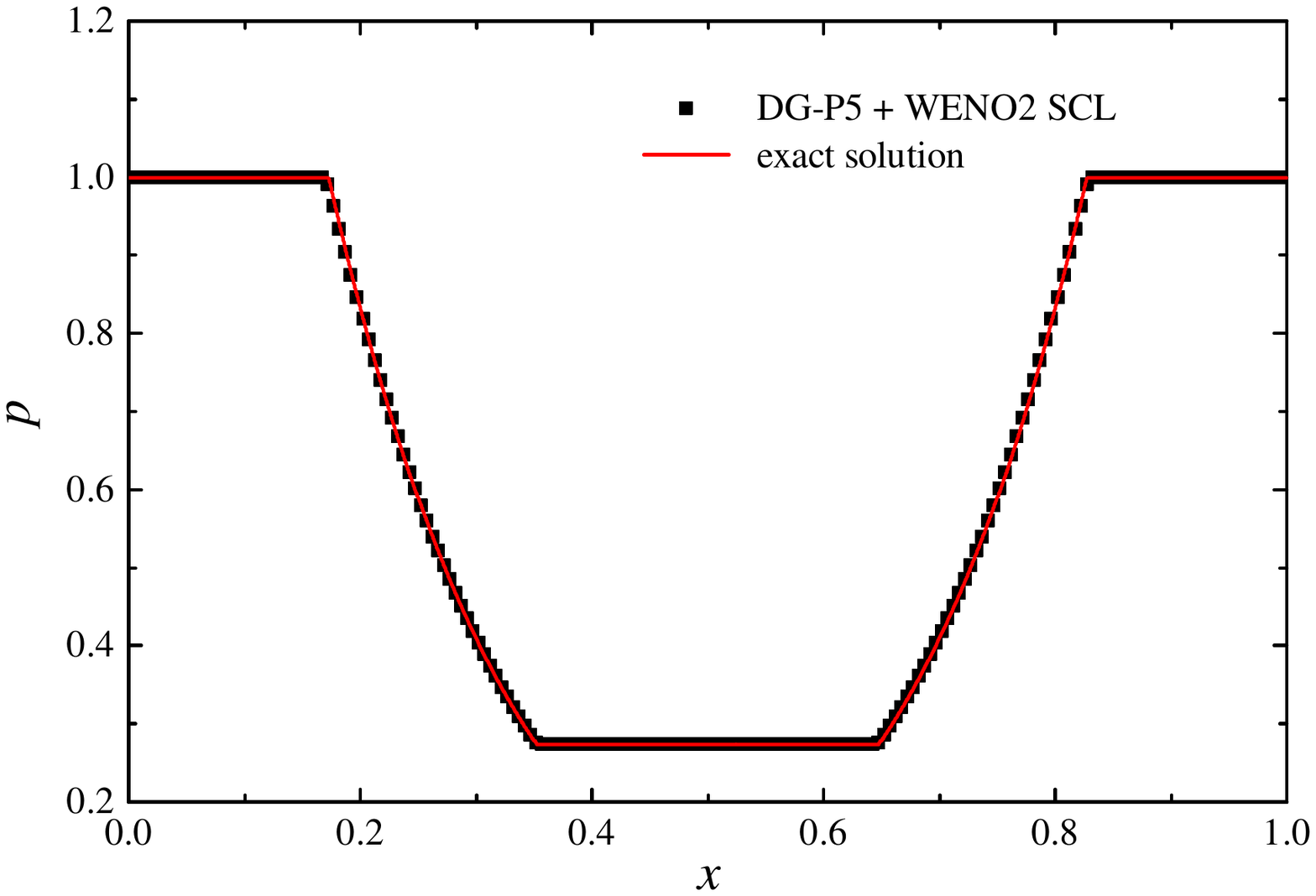}
\includegraphics[width=0.245\textwidth]{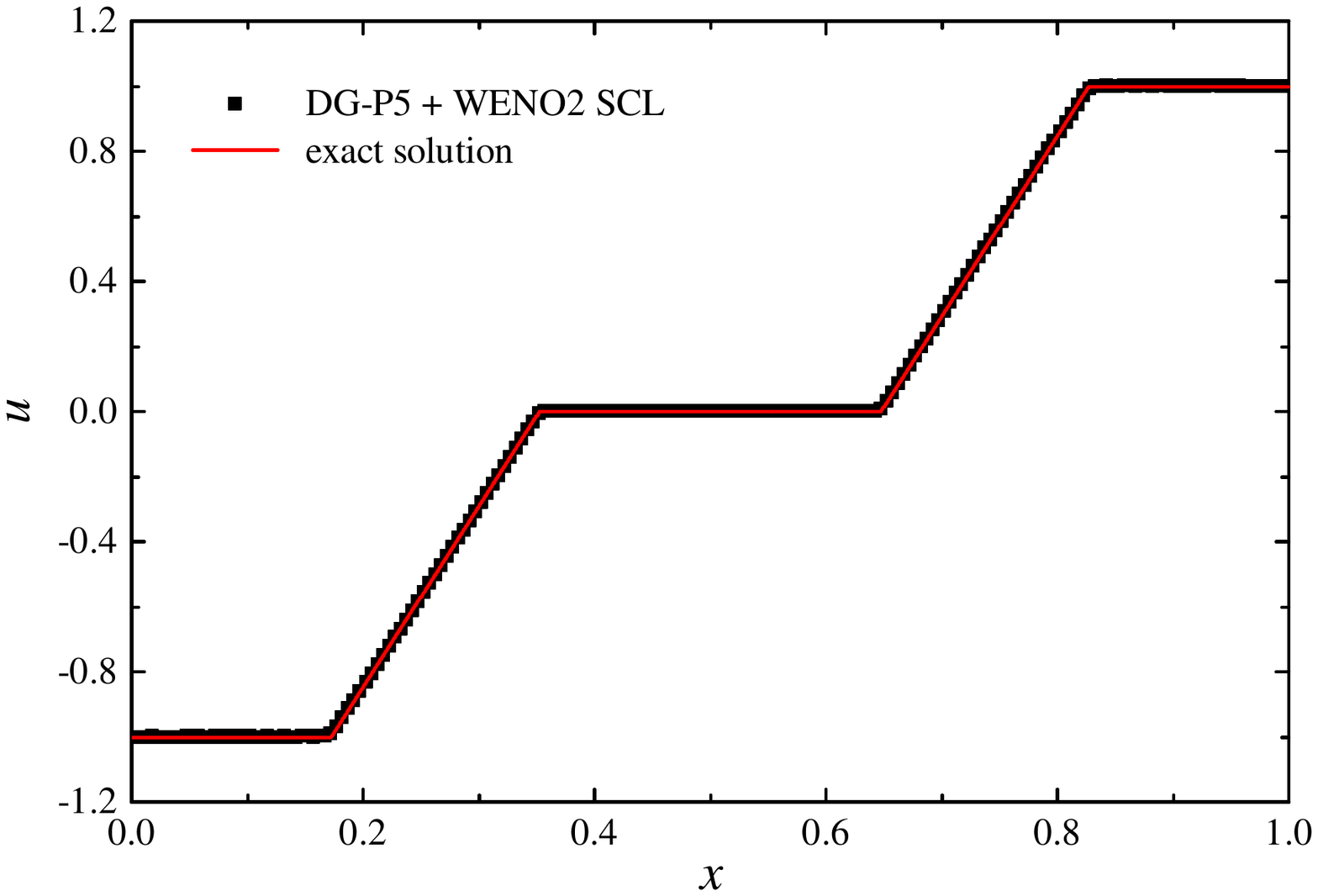}
\includegraphics[width=0.245\textwidth]{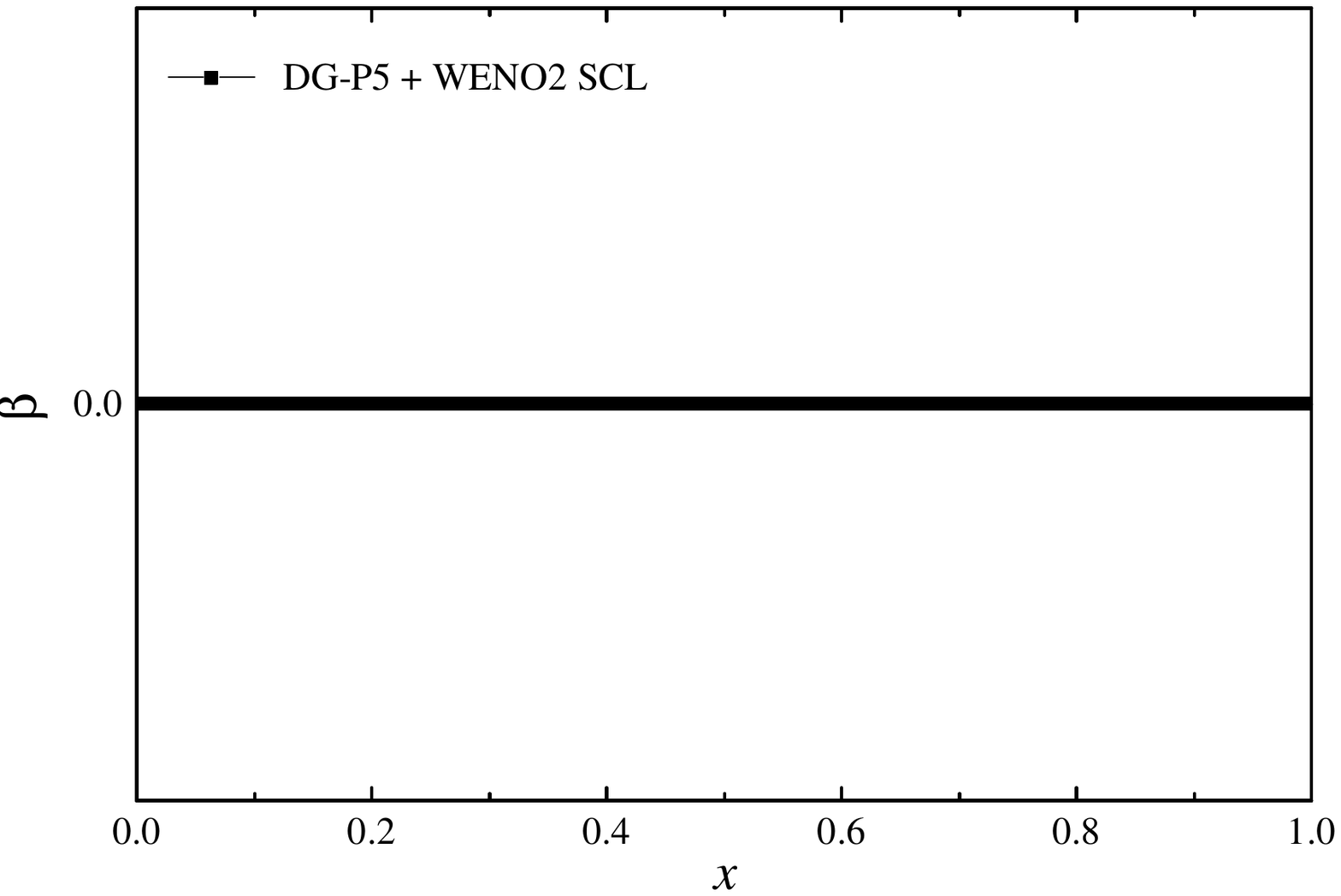}\\
\includegraphics[width=0.245\textwidth]{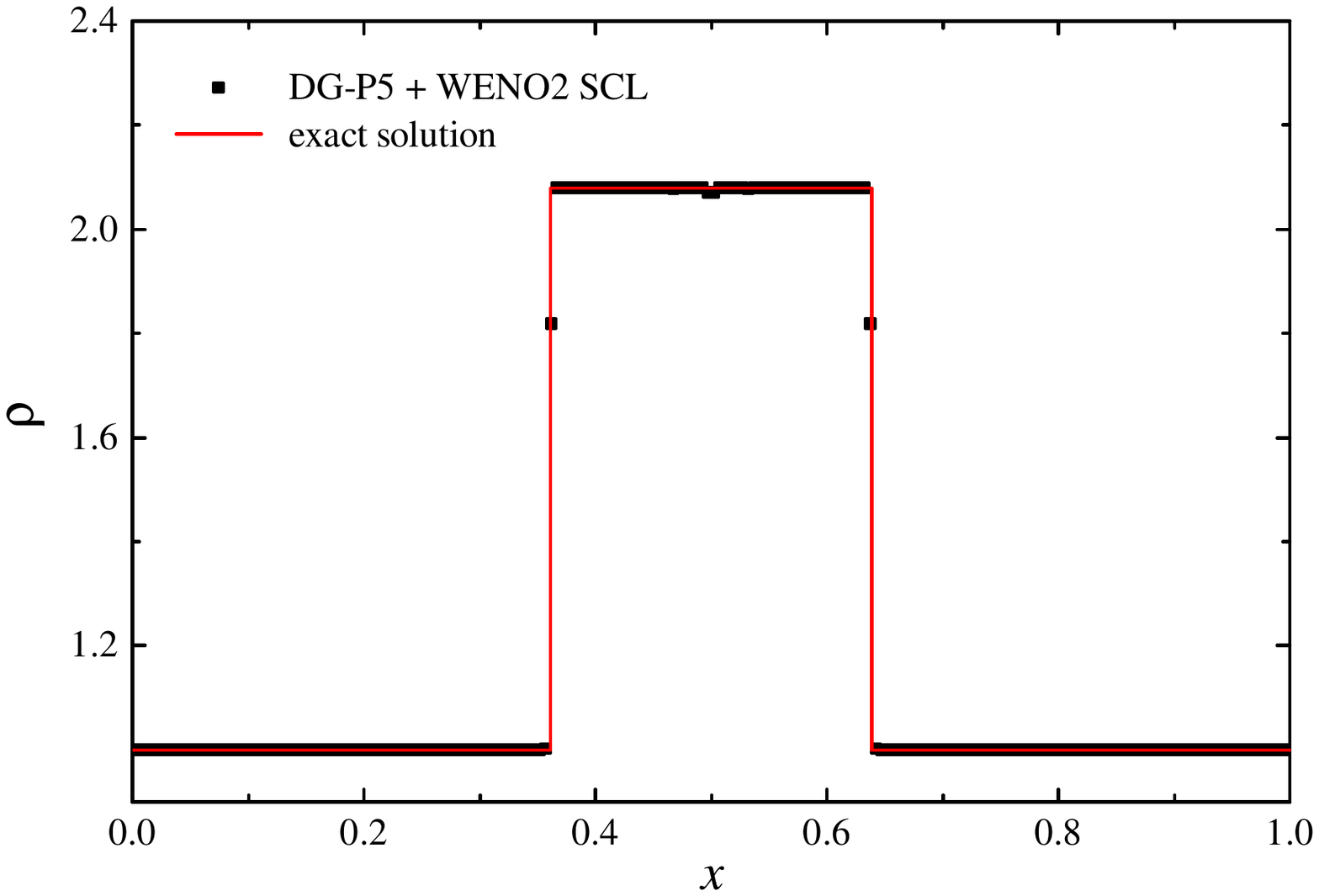}
\includegraphics[width=0.245\textwidth]{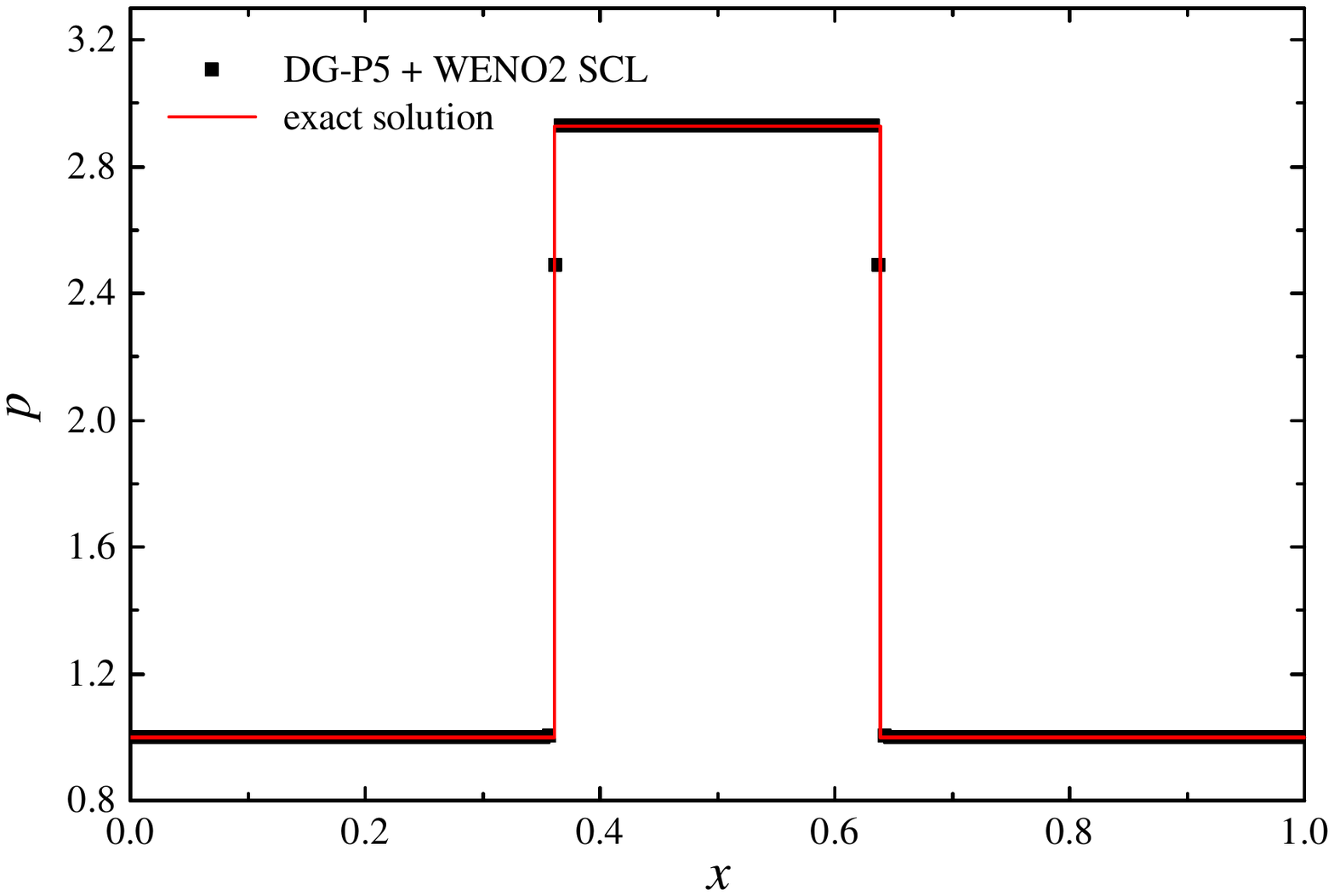}
\includegraphics[width=0.245\textwidth]{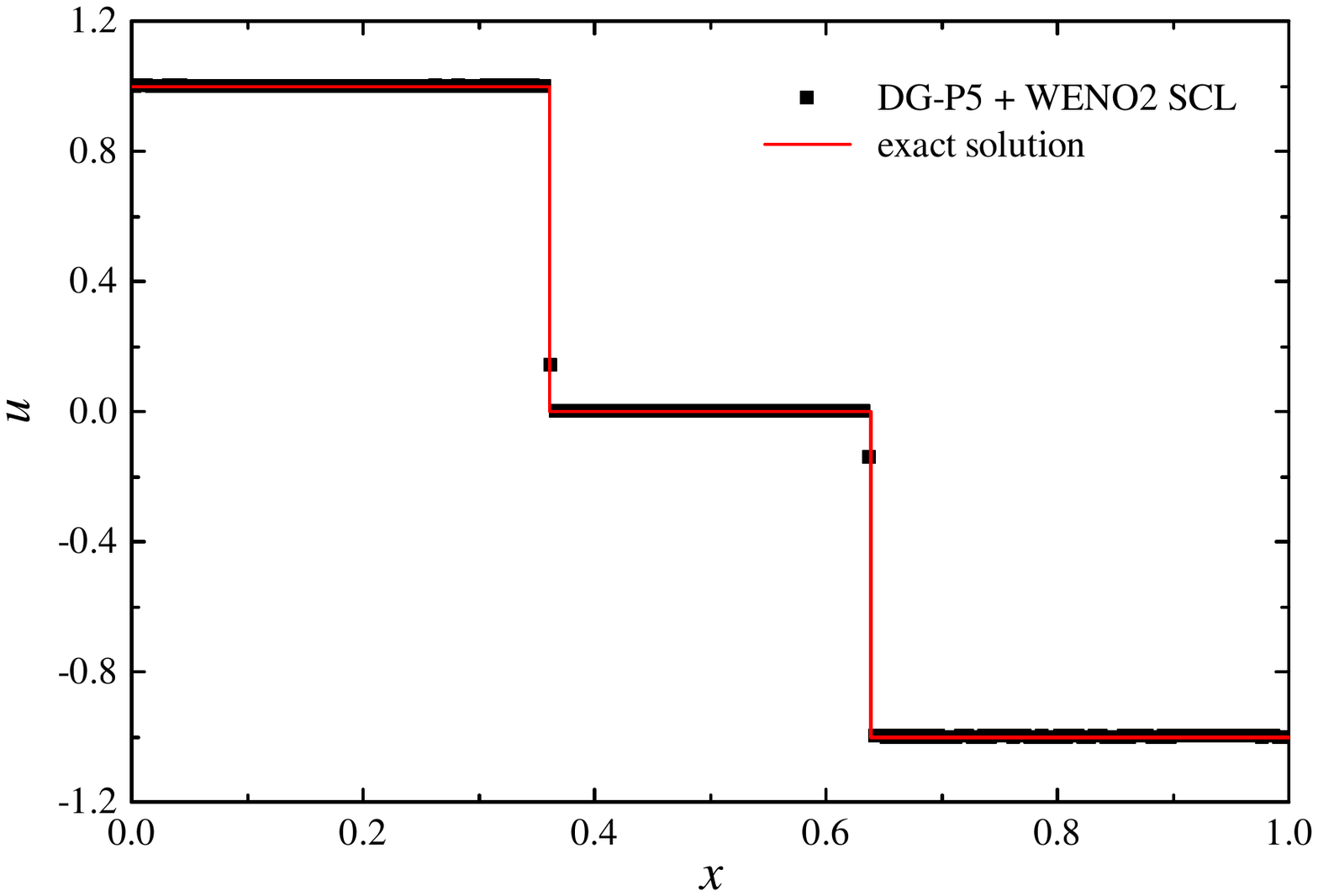}
\includegraphics[width=0.245\textwidth]{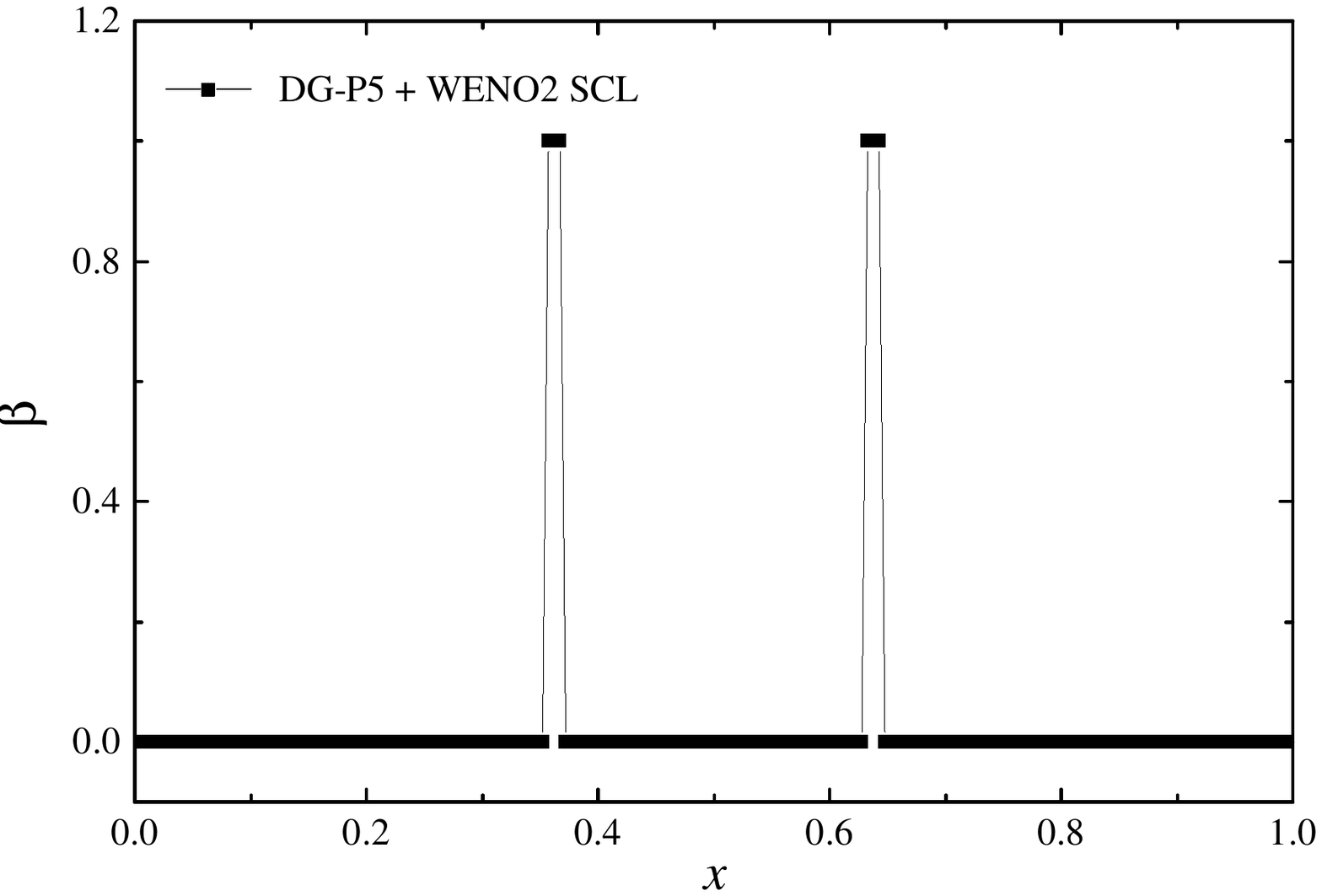}
\caption{\label{fig:classical_tests_1d}
Numerical solution of the classical Sod, Lax, two rarefaction waves and two shock waves problems (from top to bottom)
obtained using the ADER-DG-$\mathbb{P}_{5}$ method with a posteriori limitation of the solution by a ADER-WENO2 finite volume limiter
on mesh with $200$ cells (a detailed statement of the problem is presented in the text).
The graphs show the coordinate dependencies of density $\rho$, pressure $p$, flow velocity $u$ 
and troubled cells indicator $\beta$ (from left to right) at the final time $t_{\rm final} = 0.15$.
The black square symbols represent the numerical solution; 
the red solid lines represents the exact analytical solution of the problem.
}
\end{figure*} 

This subsection presents the results of calculating classical test cases based on exactly solvable classical one-dimensional Riemann problems~\cite{Toro_solvers_2009}. Verification and testing of the developed software implementation in this work was carried out on four gas-dynamic tests~\cite{Toro_solvers_2009}: the classical Sod and Lax problems ($1$ and $2$ tests), the problem with two strong rarefaction waves ($3$ test), and the problem with two shock waves ($4$ test). The spatial domain of the flow was chosen as $\Omega = [0, 1]$. The initial discontinuity was located at the coordinate $x_{\rm c} = 0.5$. The initial conditions in the two-dimensional Riemann problems were chosen in the following form:
\begin{equation}\label{eq:rp_1d_init}
(\rho, u, p)(x, t = 0) = \left\{
\begin{array}{ll}
(\rho_{L}, u_{L}, p_{L}), & \mathrm{if}\ x \leqslant 0.5; \\
(\rho_{R}, u_{R}, p_{R}), & \mathrm{if}\ x >\, 0.5;
\end{array}
\right.
\end{equation}
where the parameter values $(\rho_{L}, u_{L}, p_{L})$ and $(\rho_{R}, u_{R}, p_{R})$ correspond to the state of the flow to the left and right of the discontinuity. Data for the parameter values $(\rho, u, p)$ of these four Riemann problem tests are presented in Table~\ref{tab:classical_tests_1d}. The boundary conditions were chosen as free outflow conditions. The final time of the simulation was chosen as $t_{\rm final} = 0.15$ for all four tests. The adiabatic index $\gamma = 1.4$. The Courant number $\mathtt{CFL} = 0.4$.

Numerical solution of the classical Sod, Lax, two rarefaction waves and two shock waves problems obtained using the ADER-DG-$\mathbb{P}_{5}$ method with a posteriori limitation of the solution by a ADER-WENO2 finite volume limiter on mesh with $200$ cells is presented in Figure~\ref{fig:classical_tests_1d}.

The one-dimensional flow that occurs in the Sod problem contains a shock wave, a contact discontinuity, and a rarefaction wave. The obtained coordinate dependencies show that the shock front and contact discontinuity are resolved in the solution with an accuracy of one mesh cell. In this case, the width of the contact discontinuity does not increase over time, which is often characteristic of first- and second-order methods. The contact discontinuity does not appear in any way in the coordinate dependence of pressure $p$ in the form of non-physical artifacts. The one-dimensional flow that arises in the Lax problem is also characterized by a shock wave, a contact discontinuity and a rarefaction wave, however, compared to the Sod problem, it is much more difficult to solve this problem by high-order methods. The obtained coordinate dependencies demonstrate properties of the numerical solution similar to the solution to the Sod problem -- the shock wave and contact discontinuity are resolved with an accuracy of one mesh cell. An important feature of the numerical solution of the Sod and Lax problems is that in the solution only two troubled cells arise at each time step grid, as can be seen from the presented results for the troubled cells indicator $\beta$. Troubled cells are formed only in the coordinate vicinity of the shock wave front, while in the vicinity of the contact discontinuity, as well as the boundary characteristics of the rarefaction wave, troubled cells are not formed.

The one-dimensional flow that occurs in Test 3 contains two symmetrical rarefaction waves. The presented results show that the main features of a flow with two rarefaction waves are resolved quite accurately in the numerical solution. It should also be noted that in this test case the limiter was not called -- there are no grid cells, the troubled cells indicator $\beta = 0$ is everywhere. It can also be noted that on the coordinate dependence of the density at the point of the initial position of the discontinuity, one can observe a small area of convexity downwards, the size of which is two cells. The features of the resolution of rarefaction waves in Test 3, in general, are no different from the resolution of rarefaction waves in the Sod and Lax problems. The one-dimensional flow that occurs in Test 4 contains two symmetrical shock waves. Shock waves are resolved in a numerical solution with an accuracy of one mesh cell, while two troubled cells are formed in the vicinity of the shock wave fronts. The features of shock wave resolution in Test 4 are, in general, not different from the shock wave resolution in the Sod and Lax problems.

As a result of the analysis, we can conclude that the developed software implementation of the ADER-DG-$\mathbb{P}_{N}$ method with a posteriori limitation of the solution by a ADER-WENO finite volume limiter makes it possible to obtain numerical solutions to one-dimensional problems that are characterized by the appearance of discontinuous components in the solution. The properties of the numerical solution correspond to the results presented in the basic works~\cite{ader_dg_ideal_flows, ader_dg_dev_1}.

\subsection{Kelvin-Helmholtz instability problem}
\label{sec:apps_cgd_problems:khi_2d}

The evolution of the Kelvin-Helmholtz instability was considered as the first two-dimensional problem. The formation of a vortex street in a shear layer significantly depends on the dissipative properties of the numerical method used. In the one-dimensional Riemann problems discussed above, it was shown that contact discontinuities are resolved very well by the method, while the contact discontinuity region does not form troubles for mesh cells. Two-dimensional tangential discontinuities in gas dynamics exhibit instability to small perturbations of the discontinuity, accompanied by the generation of vortices that form an irregular structure of multiple vortices with fractal properties, which in dynamic evolution form a regular structure of large vortices in the form of a shifted sequence~\cite{Springel_mnras_2010}. 

In this work, the problem statement was chosen in the form proposed in the work~\cite{Springel_mnras_2010} and used in the work~\cite{Springel_mnras_2015}. The coordinate domain was chosen in the form of a square $\Omega = [0, 1]\times[0, 1]$ with periodic boundary conditions. The initial conditions in the two-dimensional Kelvin-Helmholtz instability problem were chosen in the form~\cite{Springel_mnras_2015}:
\begin{equation}
\begin{split}
&p(x, y, t = 0) = 2.5;\\[5mm]
&\rho(x, y, t = 0) = \left\{
\begin{array}{ll}
2, & \mathrm{if}\ 0.25 < y < 0.75;\\
1, & \mathrm{if}\ y \leqslant 0.25 \lor y \geqslant 0.75;
\end{array}
\right.\\[5mm]
&u(x, y, t = 0) = \left\{
\begin{array}{ll}
-0.5, & \mathrm{if}\ 0.25 < y < 0.75;\\
+0.5, & \mathrm{if}\ y \leqslant 0.25 \lor y \geqslant 0.75;
\end{array}
\right.\\[5mm]
&v(x, y, t = 0) = v_{0} \sin(4 \pi x) \Bigg\{ 
	\exp\left(-\frac{(y - 0.25)^{2}}{2\sigma^{2}}\right)\\
&\qquad\qquad\qquad\qquad\qquad\,\,+ \exp\left(-\frac{(y - 0.75)^{2}}{2\sigma^{2}}\right)
\Bigg\};
\end{split}
\end{equation}
where $v_{0} = 0.1$ and $\sigma = 0.05/\sqrt{2}$, the adiabatic index $\gamma = 1.4$. The Courant number $\mathtt{CFL} = 0.4$.

\begin{figure}[h!]
\centering
\includegraphics[width=0.239\textwidth]{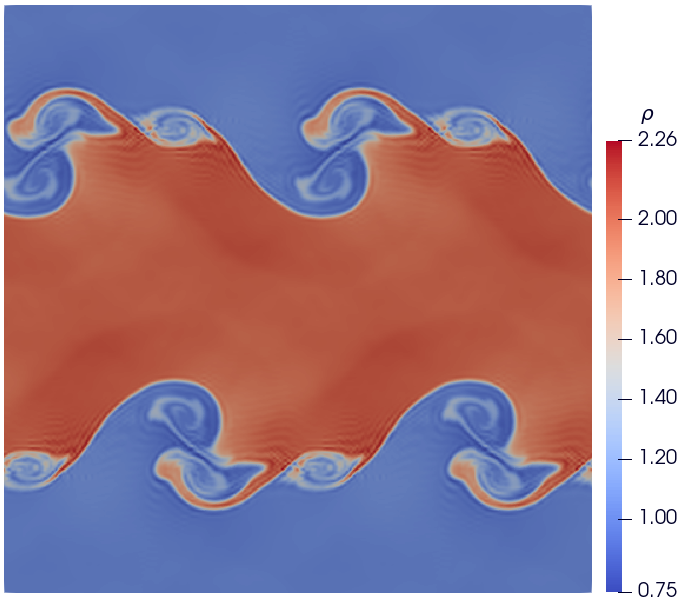}
\includegraphics[width=0.239\textwidth]{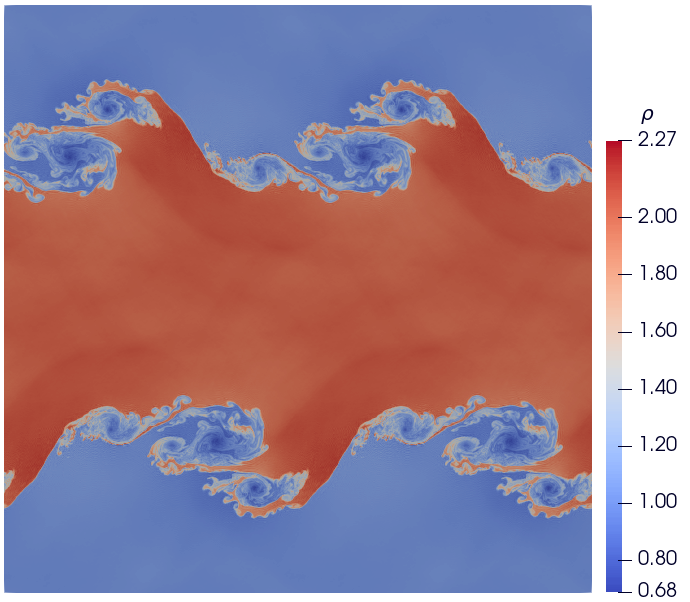}\\
\includegraphics[width=0.239\textwidth]{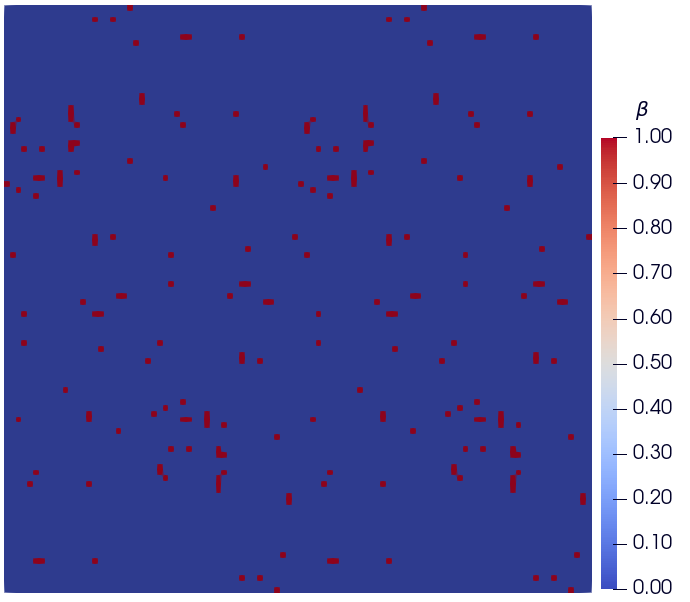}
\includegraphics[width=0.239\textwidth]{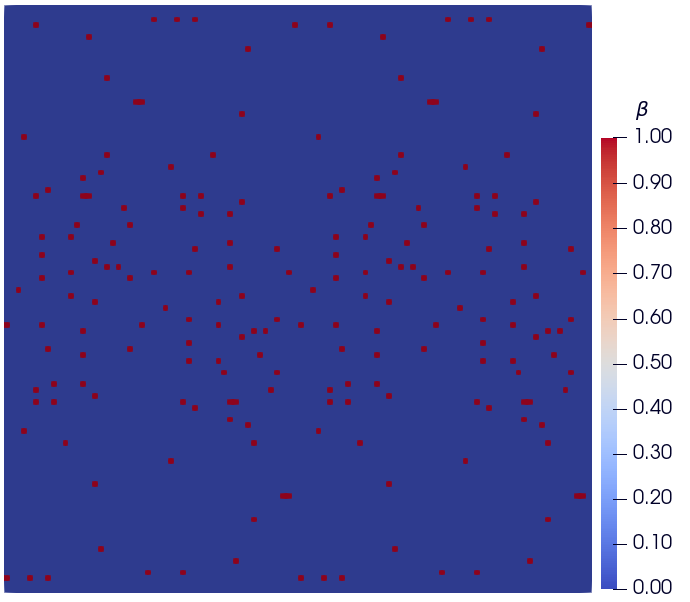}
\caption{\label{fig:khi_2d_08}
Numerical solution of the two-dimensional Kelvin-Helmholtz instability problem (a detailed statement of the problem is presented in the text)
obtained using the ADER-DG-$\mathbb{P}_{2}$ (left) and ADER-DG-$\mathbb{P}_{9}$ (right) methods on mesh with $100 \times 100$ cells 
at the time $t = 0.8$. 
The graphs show the coordinate dependencies of the subcells finite-volume representation of density $\rho$ (top) and troubled cells indicator $\beta$ (bottom).
}
\end{figure}
\begin{figure}[h!]
\centering
\includegraphics[width=0.239\textwidth]{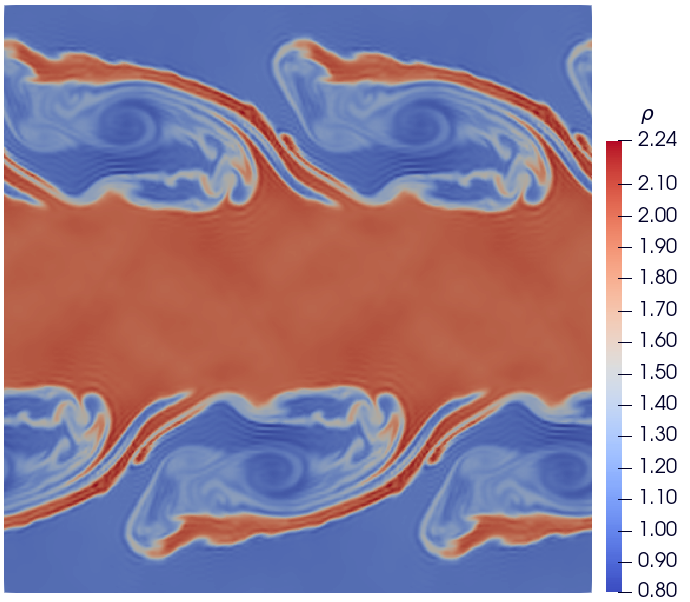}
\includegraphics[width=0.239\textwidth]{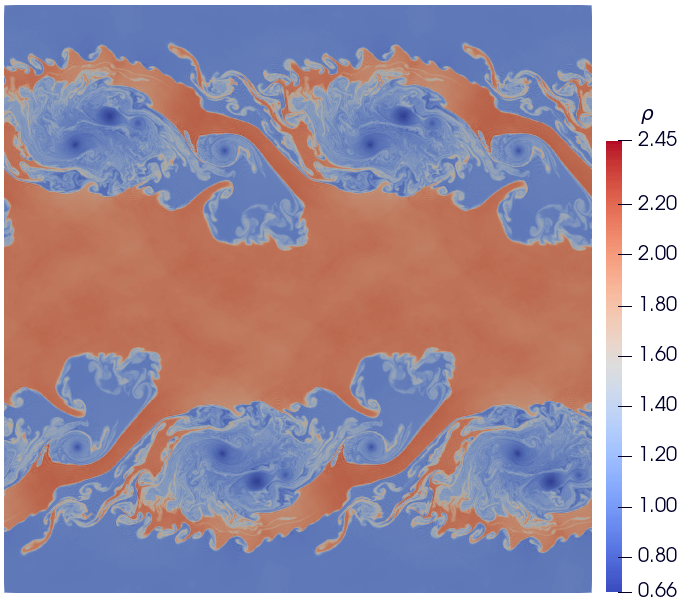}\\
\includegraphics[width=0.239\textwidth]{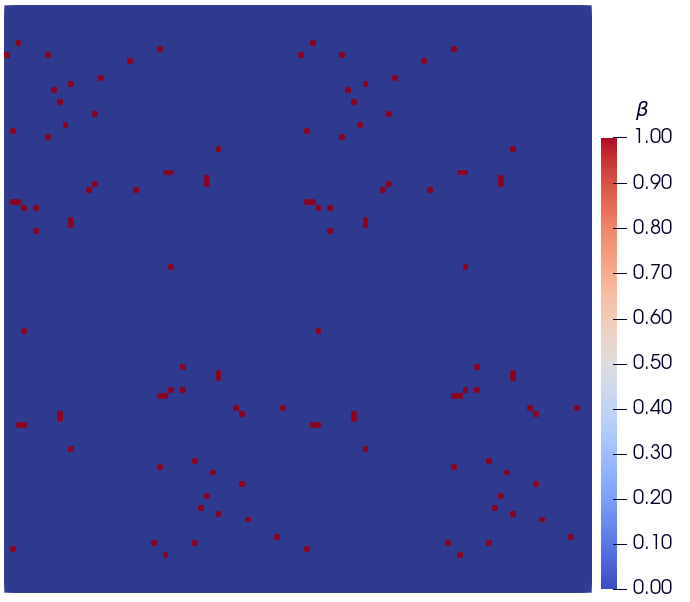}
\includegraphics[width=0.239\textwidth]{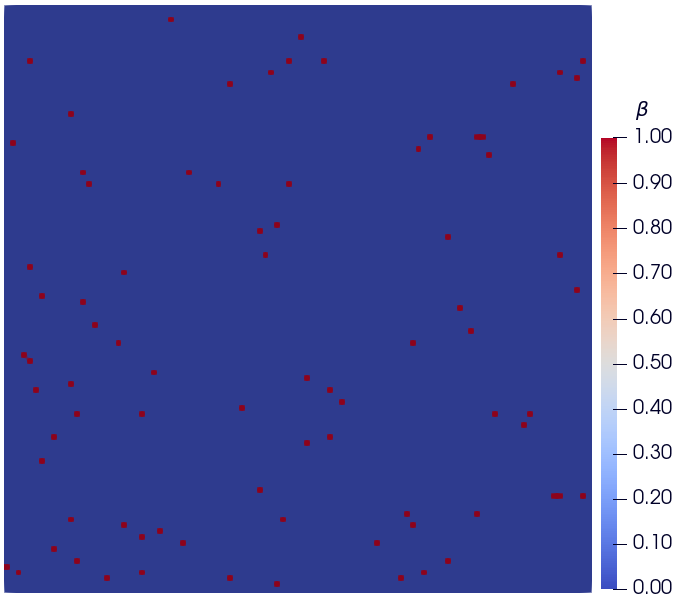}
\caption{\label{fig:khi_2d_20}
Numerical solution of the two-dimensional Kelvin-Helmholtz instability problem (a detailed statement of the problem is presented in the text)
obtained using the ADER-DG-$\mathbb{P}_{2}$ (left) and ADER-DG-$\mathbb{P}_{9}$ (right) methods on mesh with $100 \times 100$ cells 
at the time $t = 2.0$. 
The graphs show the coordinate dependencies of the subcells finite-volume representation of density $\rho$ (top) and troubled cells indicator $\beta$ (bottom).
}
\end{figure}

\begin{table*}[h!]
\caption{%
\label{tab:rp_2d}
Data for two-dimensional Riemann problem tests in the square domain $\Omega = [0, 1]\times[0, 1]$.
The values of density $\rho$, pressure $p$ and velocity projections $(u, v)$ in individual domains of initial conditions are presented. These cases correspond to Configurations $3$, $4$, $6$, $8$ and $12$ in~\cite{Riemann_2d_catalog_2002} and RP1, RP2, RP3, RP4 and RP5 in~\cite{ader_dg_dev_1, ader_weno_lstdg_ideal}.
}
\centering
\begin{tabular}{|c|c|cccc|c|cccc|}
\hline
Test	& domain	& $\rho$	& $u$		& $v$		& $p$	& domain	& $\rho$	& $u$		& $v$		& $p$\\
\hline
RP1	& $x \leqslant 0.5 \land y >\, 0.5$	& $0.5323$	& $1.206$	& $0.0$	& $0.3$	
	& $x > 0.5 \land y >\, 0.5$	& $1.5$	& $0.0$	& $0.0$	& $1.5$\\
	& $x \leqslant 0.5 \land y \leqslant 0.5$	& $0.138$	& $1.206$	& $1.206$	& $0.029$
	& $x > 0.5 \land y \leqslant 0.5$	& $0.5323$	& $0.0$	& $1.206$	& $0.3$\\
\hline
RP2	& $x \leqslant 0.5 \land y >\, 0.5$	& $0.5065$	& $0.8939$	& $0.0$	& $0.35$
	& $x > 0.5 \land y >\, 0.5$		& $1.1$		& $0.0$		& $0.0$	& $1.1$\\
	& $x \leqslant 0.5 \land y \leqslant 0.5$	& $1.1$		& $0.8939$	& $0.8939$	& $1.1$
	& $x > 0.5 \land y \leqslant 0.5$			& $0.5065$	& $0.0$	& $0.8939$	& $0.35$\\
\hline
RP3	& $x \leqslant 0.5 \land y >\, 0.5$	& $2.0$	& $0.75$	& $0.5$		& $1.0$	
	& $x > 0.5 \land y >\, 0.5$		& $1.0$	& $0.75$	& $-0.5$	& $1.0$\\
	& $x \leqslant 0.5 \land y \leqslant 0.5$	& $1.0$		& $-0.75$	& $0.5$	& $1.0$
	& $x > 0.5 \land y \leqslant 0.5$	& $3.0$	& $-0.75$	& $-0.5$	& $1.0$\\
\hline
RP4	& $x \leqslant 0.5 \land y >\, 0.5$	& $1.0$		& $-0.6259$	& $0.1$		& $1.0$		
	& $x > 0.5 \land y >\, 0.5$		& $0.5197$	& $0.1$		& $0.1$		& $0.4$\\
	& $x \leqslant 0.5 \land y \leqslant 0.5$	& $0.8$		& $0.1$		& $0.1$		& $1.0$
	& $x > 0.5 \land y \leqslant 0.5$	& $1.0$		& $0.1$		& $-0.6259$	& $1.0$\\
\hline
RP5	& $x \leqslant 0.5 \land y >\, 0.5$	& $1.0$		& $0.7276$	& $0.0$		& $1.0$		
	& $x > 0.5 \land y >\, 0.5$		& $0.5313$	& $0.0$		& $0.0$		& $0.4$\\
	& $x \leqslant 0.5 \land y \leqslant 0.5$	& $0.8$		& $0.0$		& $0.0$		& $1.0$
	& $x > 0.5 \land y \leqslant 0.5$	& $1.0$		& $0.0$		& $0.7276$	& $1.0$\\
\hline
\end{tabular}
\end{table*}

Numerical solutions were obtained on a mesh $100 \times 100$ by the ADER-DG-$\mathbb{P}_{9}$ method. The ADER-WENO2 finite volume method was used as a posteriori limiter. The main results obtained are presented in Figure~\ref{fig:khi_2d_08} at the time $t = 0.8$ and  in Figure~\ref{fig:khi_2d_20} at the time $t = 2.0$. To quantify the accuracy of the numerical solution, the results in the case of using $N = 2$ are also presented for comparison.

The presented results allow us to conclude that the ADER-DG-$\mathbb{P}_{9}$ method with a posteriori ADER-WENO2 finite volume limitation allows us to obtain a solution of very high accuracy for the selected size of the spatial mesh $100 \times 100$. A direct comparison with the solution presented in the work~\cite{Springel_mnras_2015} seems pointless, because it, of course, has much greater accuracy, which is associated with the use of AMR and a much larger spatial mesh size. However, taking into account the differences between spatial meshes $100 \times 100$ and $4096 \times 4096$, it can be said that the solution in this work agrees well with the results presented in the work~\cite{Springel_mnras_2015}, and the solution agrees well with the movie referenced in the work~\cite{Springel_mnras_2015} [Fig.~11].

In the numerical solution obtained using the ADER-DG-$\mathbb{P}_{9}$ method, a high accuracy of resolution of small-scale eddies, significantly smaller than the coordinate step of the spatial mesh, is observed. In dynamics, an evolutionary process of the formation of a small-scale vortex structure is observed, followed by the coarsening of small-scale disturbances into large vortex structures.

The comparison with the results obtained using the ADER-DG-$\mathbb{P}_{2}$ shows that the use of $N = 2$ leads to a significantly less accurate resolution of the vortex generation process, which, in principle, was initially expected. However, it should be noted that the numerical solution in the case of $N = 2$ correctly resolves large-scale vortex structures -- a direct comparison of the coordinate dependencies of the density shows that large-scale vortex structures are located in approximately the same coordinates in the numerical solution and demonstrate similar dynamics as in the case of the solution obtained with using the ADER-DG-$\mathbb{P}_{9}$ method.

It should be noted that there are very few troubled cells in the solution and all of them are irregularly distributed over the spatial mesh. This is due to the fact that in solving the problem, shock waves of significant amplitude are not formed, the front of which could be distinguished against the general background of the flow. This result applies to both method ADER-DG-$\mathbb{P}_{9}$ and method ADER-DG-$\mathbb{P}_{2}$. However, in the case of method ADER-DG-$\mathbb{P}_{2}$, there are usually slightly more troubled cells than in the case of method ADER-DG-$\mathbb{P}_{9}$.

\begin{figure*}[h!]
\centering
\includegraphics[width=0.245\textwidth]{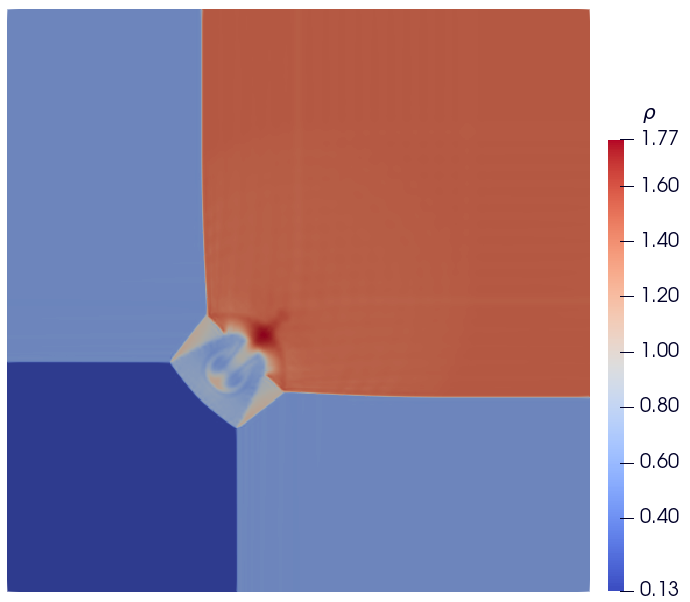}
\includegraphics[width=0.245\textwidth]{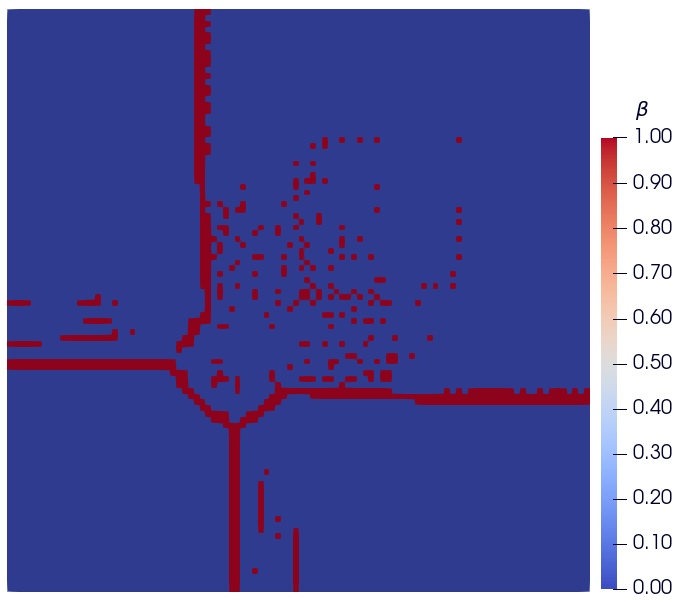}
\includegraphics[width=0.245\textwidth]{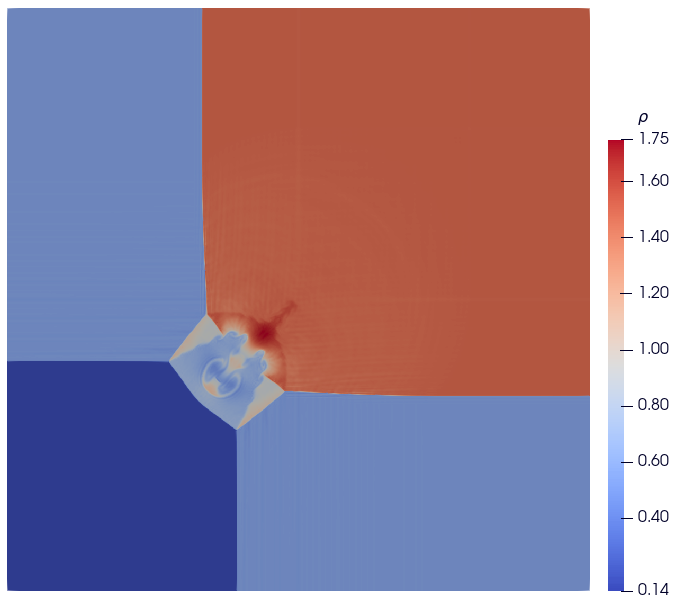}
\includegraphics[width=0.245\textwidth]{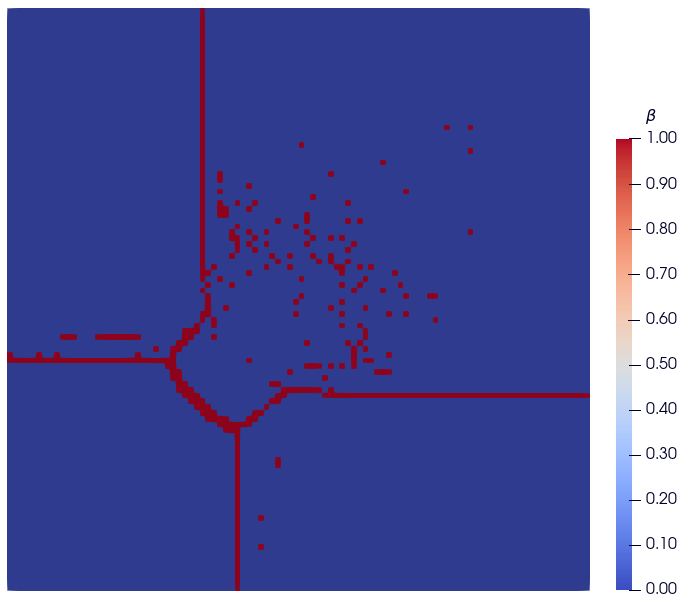}\\
\includegraphics[width=0.245\textwidth]{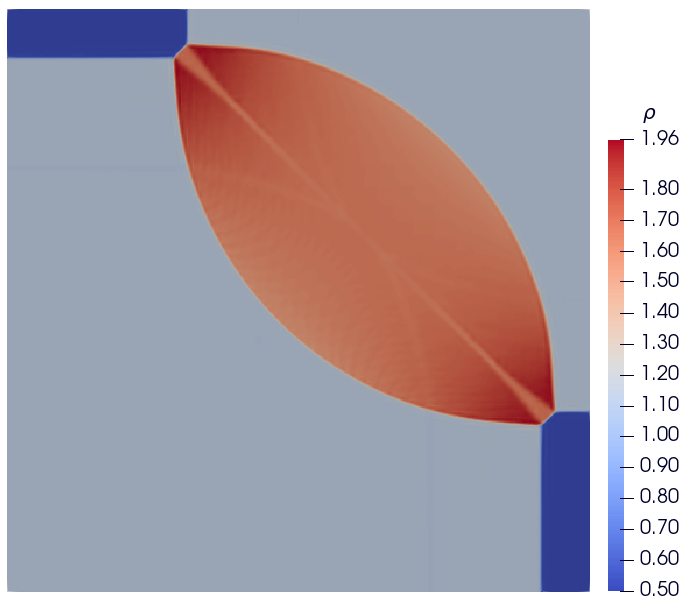}
\includegraphics[width=0.245\textwidth]{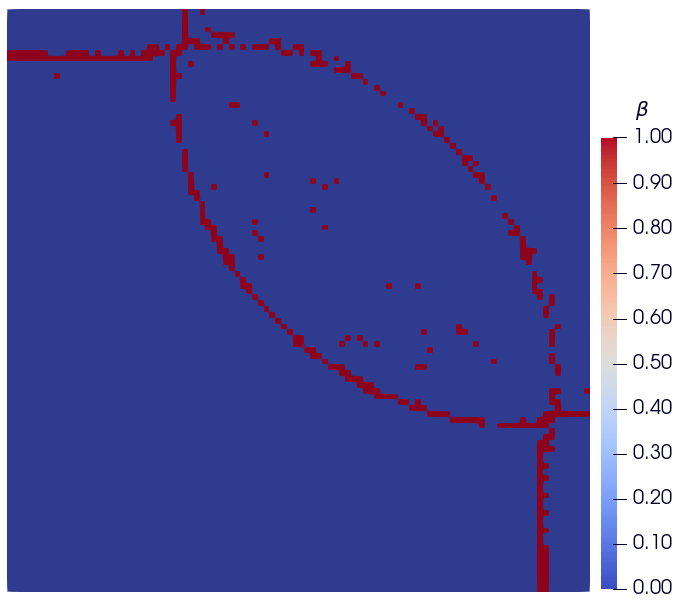}
\includegraphics[width=0.245\textwidth]{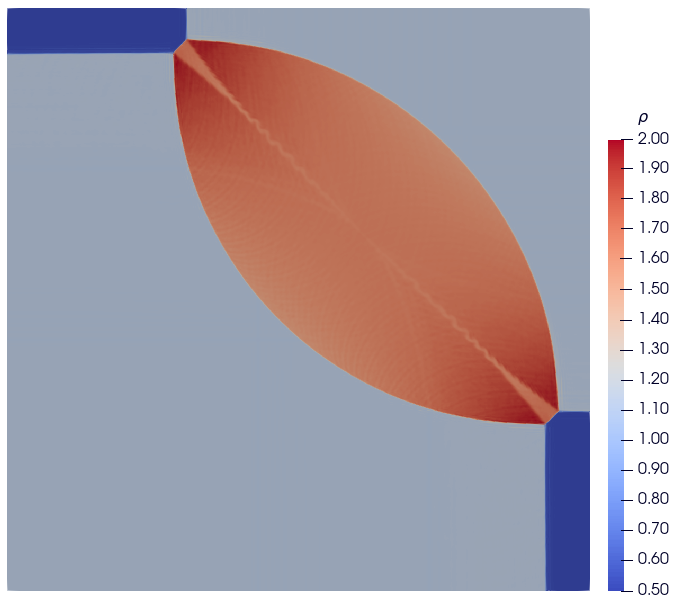}
\includegraphics[width=0.245\textwidth]{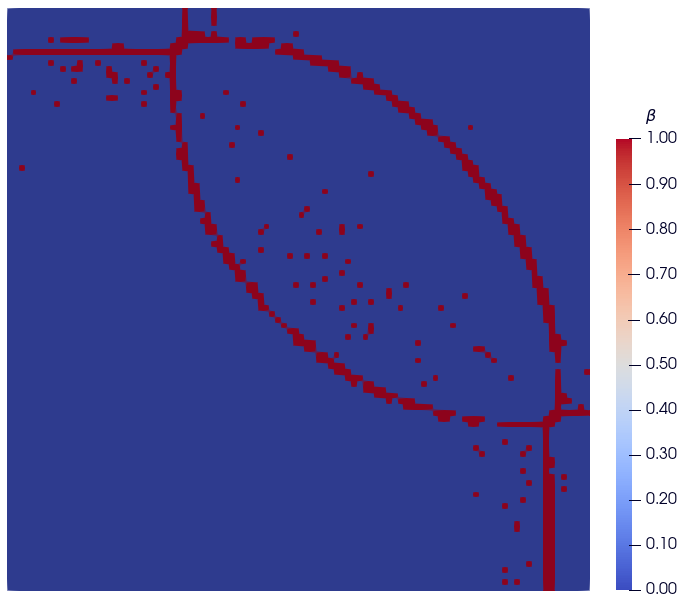}\\
\includegraphics[width=0.245\textwidth]{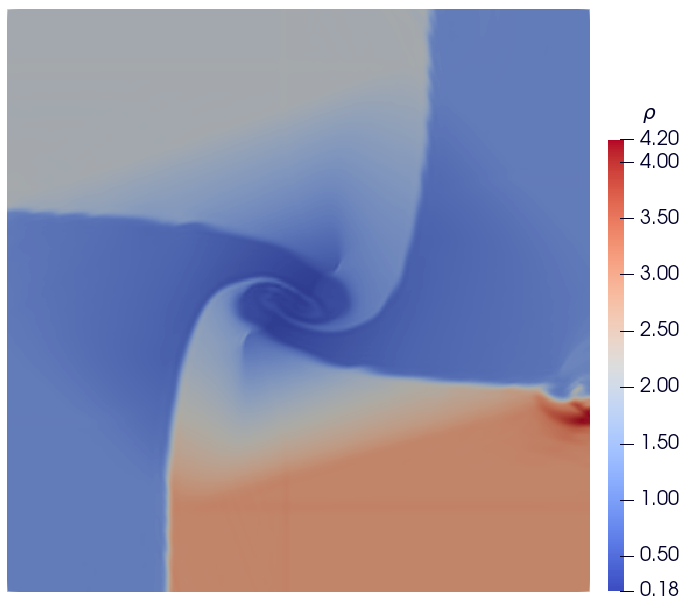}
\includegraphics[width=0.245\textwidth]{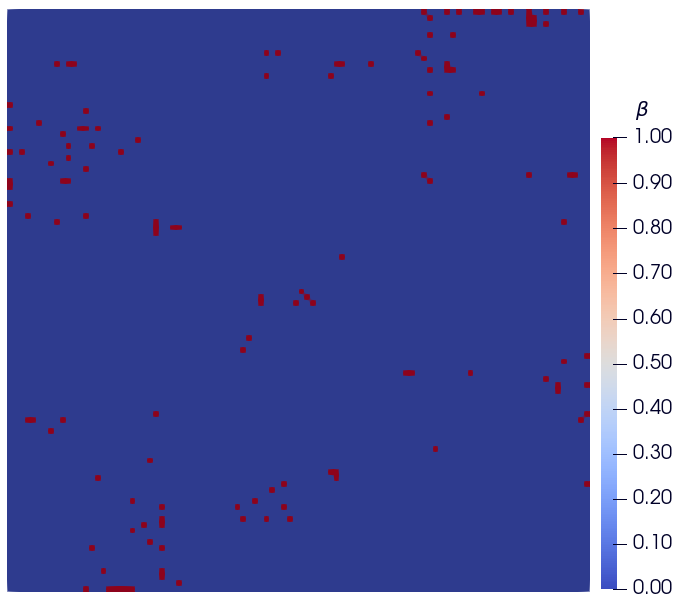}
\includegraphics[width=0.245\textwidth]{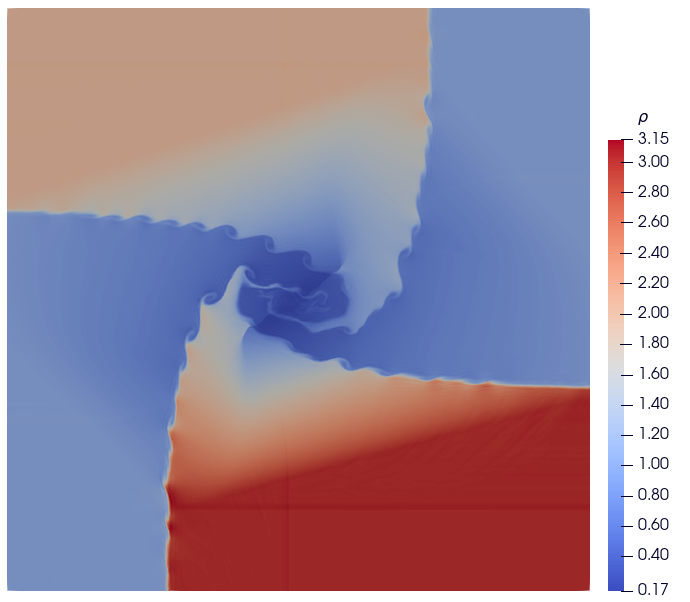}
\includegraphics[width=0.245\textwidth]{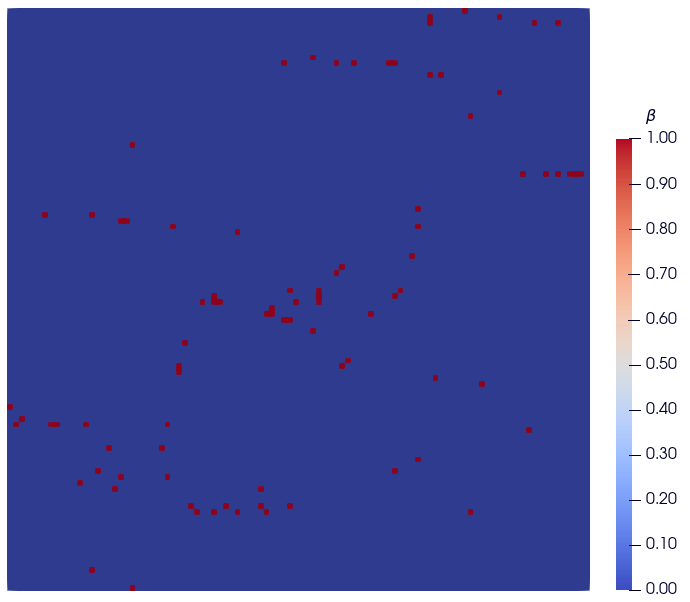}\\
\includegraphics[width=0.245\textwidth]{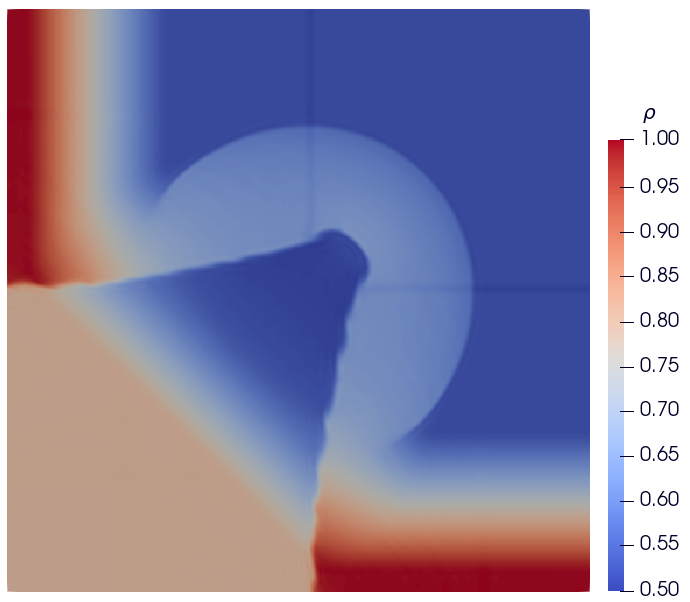}
\includegraphics[width=0.245\textwidth]{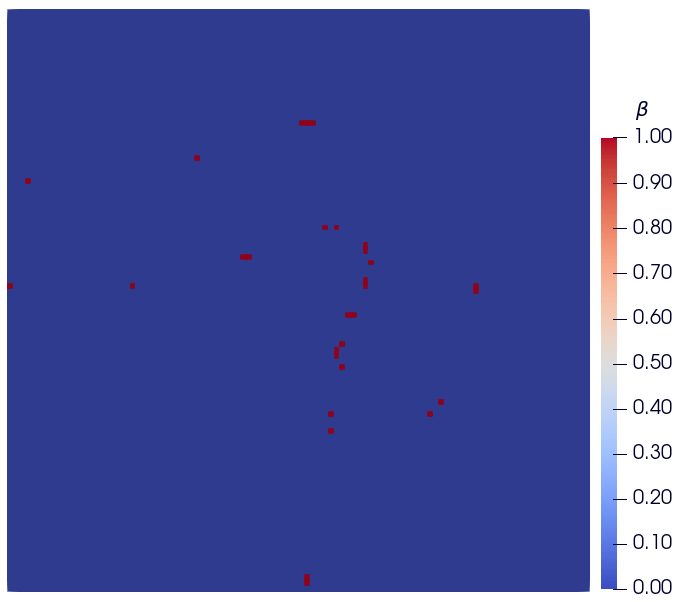}
\includegraphics[width=0.245\textwidth]{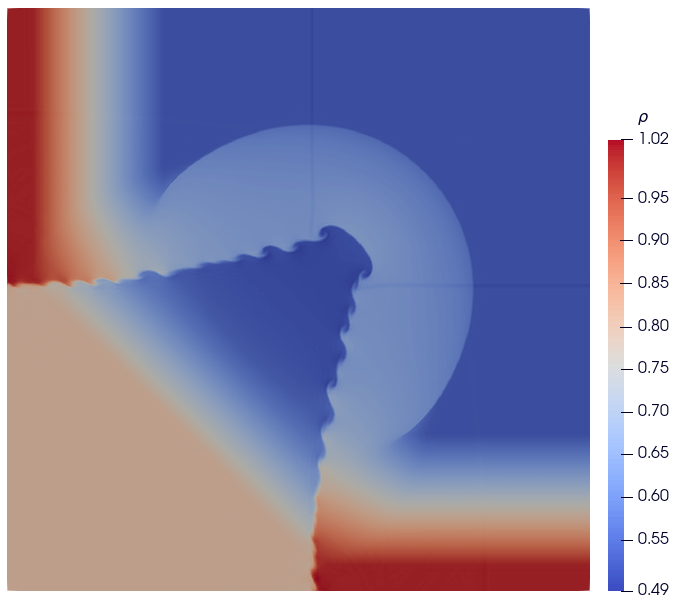}
\includegraphics[width=0.245\textwidth]{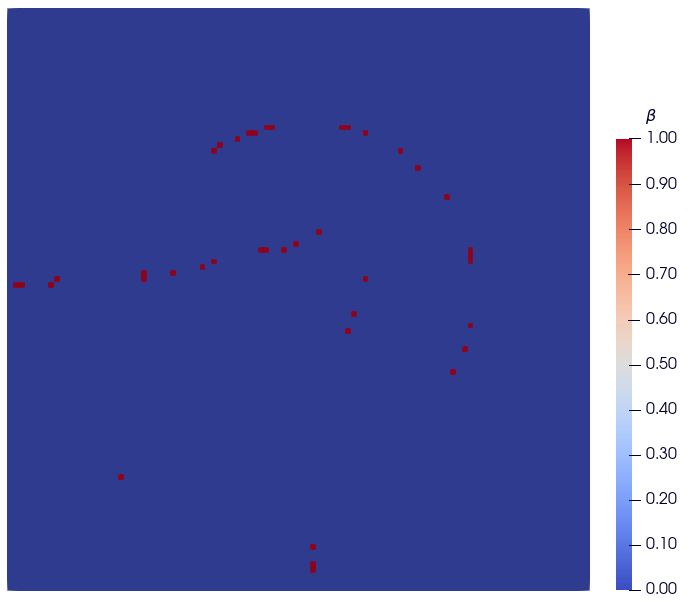}\\
\includegraphics[width=0.245\textwidth]{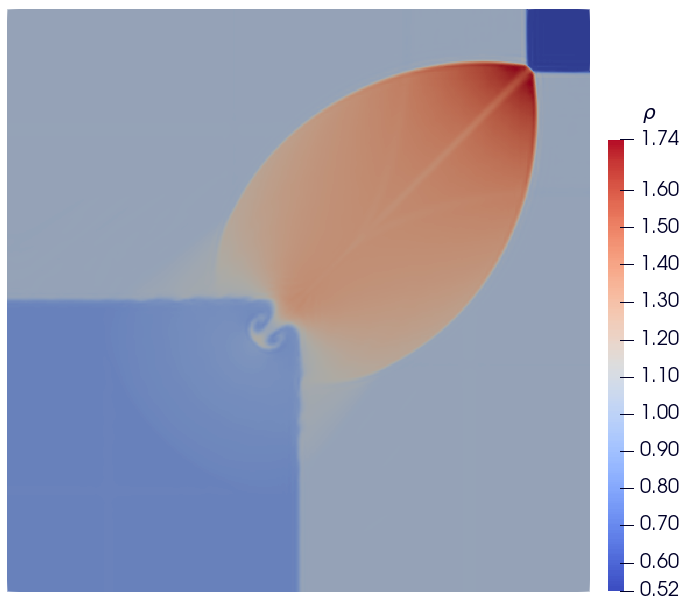}
\includegraphics[width=0.245\textwidth]{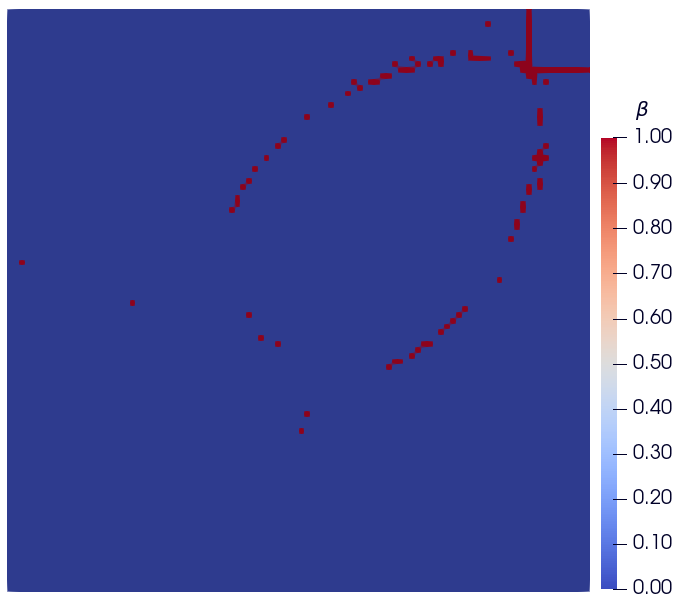}
\includegraphics[width=0.245\textwidth]{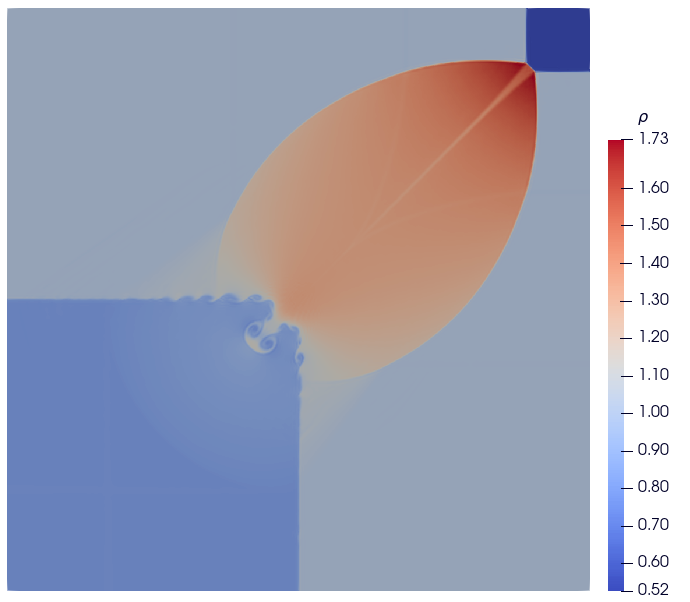}
\includegraphics[width=0.245\textwidth]{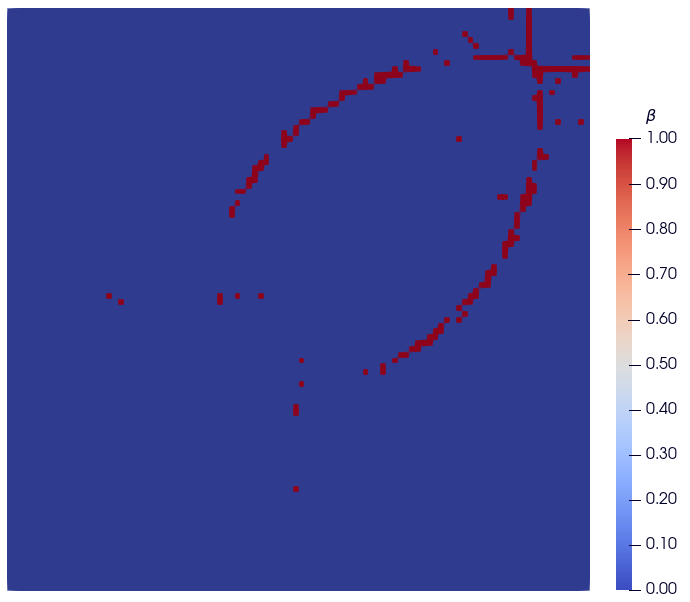}
\caption{\label{fig:rp_2d_dg_2_5_100x100}
Numerical solution of the classical two-dimensional Riemann problems RP1, RP2, RP3, RP4 and RP5 (from top to bottom)
obtained using the ADER-DG-$\mathbb{P}_{2}$ method (left two columns) and the ADER-DG-$\mathbb{P}_{5}$ method (right two columns) methods
with a posteriori limitation of the solution by a ADER-WENO2 finite volume limiter
on mesh with $100 \times 100$ cells (a detailed statement of the problem is presented in the text).
The graphs show the coordinate dependencies of density $\rho$ (first and third columns) 
and troubled cells indicator $\beta$ (second and fourth columns)
at the final time $t_{\rm final} = 0.25$ (RP1, RP2, RP4, RP5) and $0.30$ (RP3).
}
\end{figure*}

\begin{figure}[h!]
\centering
\includegraphics[width=0.239\textwidth]{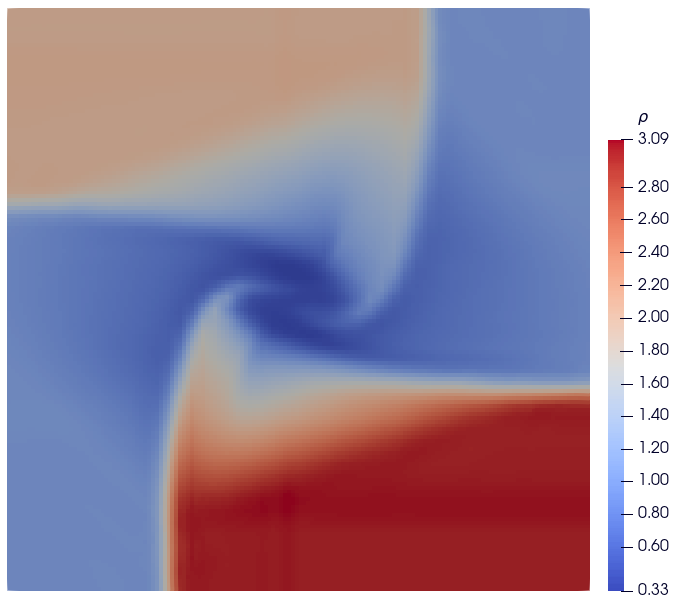}
\includegraphics[width=0.239\textwidth]{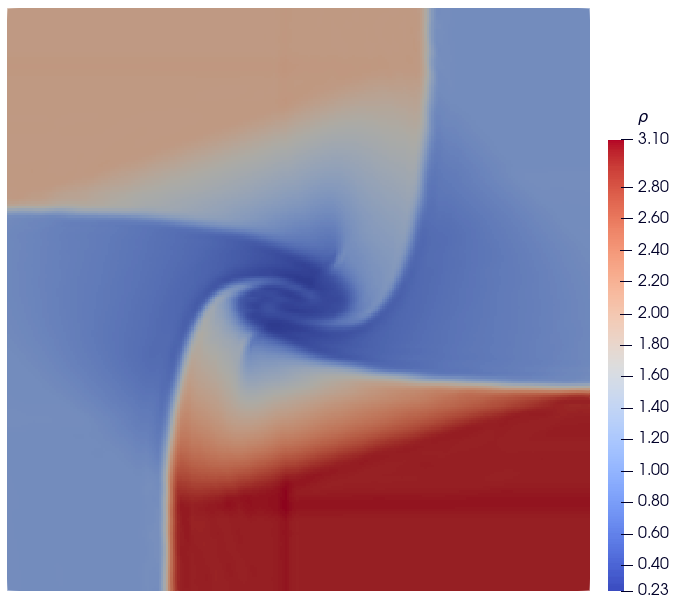}\\
\includegraphics[width=0.239\textwidth]{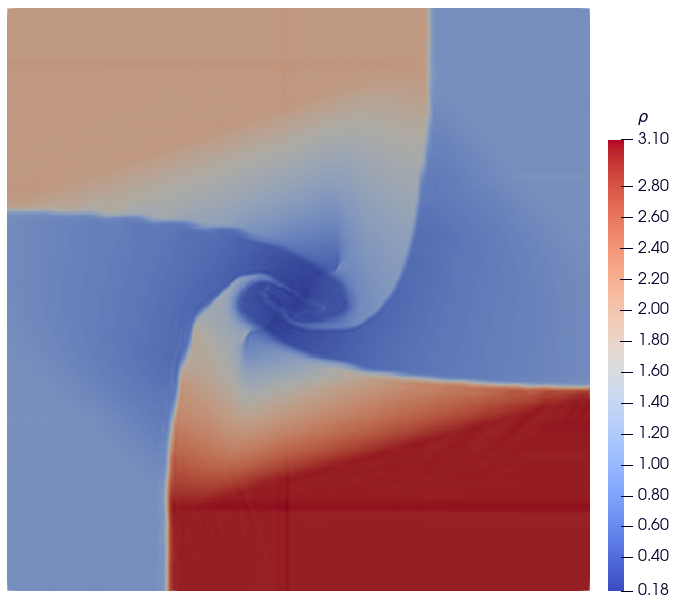}
\includegraphics[width=0.239\textwidth]{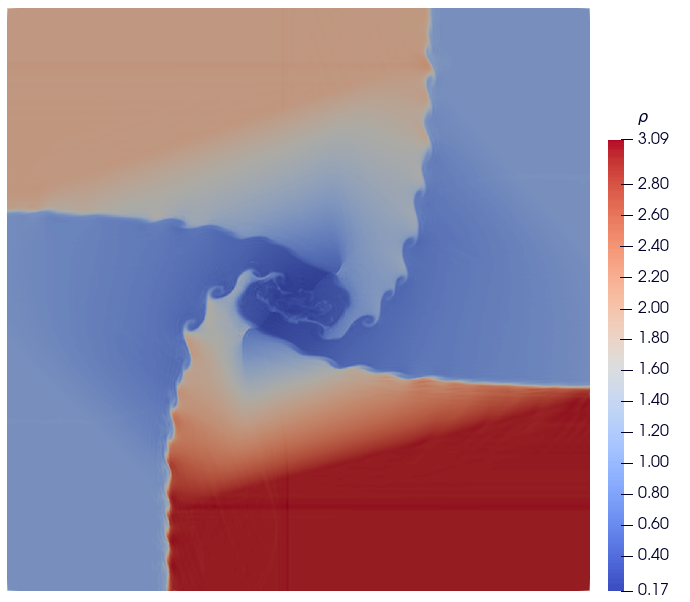}
\caption{\label{fig:rp_2d_dg_9_50x50}
The coordinate dependency of density $\rho$ of the numerical solution of the RP3 (a detailed statement of the problem is presented in the text) 
at the final time $t_{\rm final} = 0.30$ obtained using the ADER-DG-$\mathbb{P}_{1}$ (top left), ADER-DG-$\mathbb{P}_{2}$ (top right), ADER-DG-$\mathbb{P}_{5}$ (bottom left) and ADER-DG-$\mathbb{P}_{9}$ (bottom right) methods with a posteriori limitation of the solution by a ADER-WENO2 finite volume limiter 
on mesh with $50 \times 50$ cells.
}
\end{figure}

\subsection{Two-dimensional Riemann problems}
\label{sec:apps_cgd_problems:rp_2d}

The correctness of the software implementation of the numerical method was checked for a set of five two-dimensional Riemann problems. There is a wide catalog of classical two-dimensional Riemann problems presented in the work~\cite{Riemann_2d_catalog_2002}. In this work, we selected problems $3$, $4$, $6$, $8$ and $12$ that are commonly used~\cite{ader_dg_ideal_flows, ader_dg_dev_1, ader_weno_lstdg_ideal} to test high-order numerical methods.

The coordinate domain was chosen in the form of a square $\Omega = [0, 1]\times[0, 1]$ with free outflow boundary conditions. The initial conditions in the two-dimensional Riemann problems were chosen in the following form~\cite{ader_dg_dev_1}:
\begin{equation}\label{eq:rp_2d_init}
(\rho, u, v, p)(x, y, t = 0) = \left\{\hspace{-1mm}
\begin{array}{ll}
(\rho_{1}, u_{1}, v_{1}, p_{1}), & \hspace{-2mm}\mathrm{if}\ x >\, 0.5 \land y >\, 0.5; \\
(\rho_{2}, u_{2}, v_{2}, p_{2}), & \hspace{-2mm}\mathrm{if}\ x \leqslant 0.5 \land y >\, 0.5; \\
(\rho_{3}, u_{3}, v_{3}, p_{3}), & \hspace{-2mm}\mathrm{if}\ x \leqslant 0.5 \land y \leqslant 0.5; \\
(\rho_{4}, u_{4}, v_{4}, p_{4}), & \hspace{-2mm}\mathrm{if}\ x >\, 0.5 \land y \leqslant 0.5; \\
\end{array}
\right.
\end{equation}
where the lines $x = 0.5$ and $y = 0.5$ are chosen as discontinuity lines in the initial conditions. Detailed information about the initial conditions is presented in Table~\ref{tab:rp_2d}. The final time has been chosen $t_{\rm final} = 0.25$ for RP1, RP2, RP4, RP5 cases and $0.30$  for RP3 case. More detailed information about the structure of the solution in two-dimensional Riemann problems and information about other configurations of the problem are presented in the works~\cite{Riemann_2d_catalog_2002, Riemann_2d_descr_1993}. The adiabatic index $\gamma = 1.4$. The Courant number $\mathtt{CFL} = 0.4$.

Numerical solutions for all considered two-dimensional Riemann problems were obtained on a mesh $100 \times 100$. The ADER-DG-$\mathbb{P}_{N}$ method was used. The ADER-WENO2 finite volume method was used as a posteriori limiter. The best accuracy of the numerical solution was obtained using the degree $N = 5$ in the DG representation, which is consistent with the results of the basic work~\cite{ader_dg_dev_1}. The main results obtained are presented in Figure~\ref{fig:rp_2d_dg_2_5_100x100}. To quantify the accuracy of the numerical solution, the results in the case of using $N = 2$ are also presented for comparison. 

The obtained results, and in particular the generation of the main structures in all these two-dimensional Riemann problems, are in good agreement with the literature~\cite{Riemann_2d_catalog_2002}. All the main non-stationary compressible flow patterns observed in these problems are correctly identified in the numerical solution. At the same time, the features of the generation of the main flow structures are in accordance with the results of the work~\cite{ader_dg_dev_1}.

The comparison of the numerical solution for cases $N = 2$ and $5$ shows that in case $N = 5$ the processes of generation and propagation of direct and oblique shock waves and contact discontinuities are reproduced in the flow much more accurately than in case $N = 2$. This is especially well observed in the evolution of vortex streets in the region of shear wave propagation and the evolution of the Kelvin-Helmholtz instability in all five Riemann problems considered. It should be noted that, in comparison with one-dimensional problems, troubled cells are formed not only in areas of shock waves, but also in some areas of instability near contact discontinuities -- this is observed in the troubled cells indicator $\beta$ in Figure~\ref{fig:rp_2d_dg_2_5_100x100}. Moreover, in the cases of RP1 and RP2, the troubled cells completely cover the vicinity of the fronts of direct and oblique shock waves, and in the case RP5, not completely cover -- in the areas of intersection of vortex streets and shock waves (where the intensity of the shock waves becomes weak in the reference frame associated with the local flow velocity), there are not so many troubled cells.

Of particular interest is the case of RP3, where the interaction of four regions of the evolution of the Kelvin-Helmholtz instability is observed and quite complex dynamics of vortex structures in the flow are generated. To determine the possibilities of the ADER-DG-$\mathbb{P}_{N}$ method, simulations were also carried out for the coarse mesh $50 \times 50$, and the results are presented in Figure~\ref{fig:rp_2d_dg_9_50x50} for cases $N = 1$, $2$, $5$ and $9$. From the presented results we can conclude that in case $N = 1$ the vortex street is not resolved in the numerical solution -- the tangential discontinuities diffusely expand and preserve their structure at the final time $t_{\rm final}$. Similar results are observed in case $N = 2$, however, in this case the discontinuities are resolved more sharply in the numerical solution. In case $N = 5$, in the numerical solution there is some slight regular ``ripple'' at the discontinuities, however, practically no vortex generation process is observed. In the numerical solution for case $N = 5$, which is presented in Figure~\ref{fig:rp_2d_dg_2_5_100x100} obtained on a more refined mesh $100 \times 100$, the evolution of the Kelvin-Helmholtz instability was resolved more accurately. In case $N = 9$, the numerical solution on a coarse mesh turned out to be quite accurate -- all the main vortex generation processes are observed correctly. The results obtained in case $N = 9$ are comparable in accuracy to the results presented in the work~\cite{ader_dg_dev_1} [Fig.~11] where the sixth order ADER-WENO6 finite volume method with AMR (equivalent resolution: $1250 \times 1250$) was used. This can be considered expected, taking into account the capabilities of the subgrid resolution of the solution by the DG method -- if for simple estimates we assume that the resolution of the ADER-DG-$\mathbb{P}_{N}$ method of subcell structures of a smooth solution is ``equivalent'' to $2N+1$ subcells, then in case $N = 9$ the $50 \times 50$ grid approximately corresponds to the ADER-WENO6 finite volume method with AMR in the work~\cite{ader_dg_dev_1}.

\subsection{Cylindrical and spherical explosion problems}
\label{sec:apps_cgd_problems:sod_md}

\paragraph{Formulation of the problem}
Explosion problems occupy an interesting place among two-dimensional and three-dimensional problems. On the one hand, these are conceptually one-dimensional problems, where the solution depends on only one spatial coordinate -- the distance $r$ to the center of the explosion, so these problems may seem quite simple. On the other hand, the use of two-dimensional and three-dimensional computational codes to solve these problems can reveal certain problems of numerical methods and their implementations associated with maintaining the spatial symmetry of the original problem -- axial in the two-dimensional case and spherical in the three-dimensional case. Therefore, it seems interesting to obtain a numerical solution to the two-dimensional cylindrical and three-dimensional spherical explosion problem.

The formulation of the explosion problem was chosen in the form of a multidimensional Sod problem:
\begin{equation}\label{eq:sod_2d_init}
(\rho, |\mathbf{v}|, p)(\mathbf{r}, t = 0) = \left\{
\begin{array}{ll}
(1.000, 0.0, 1.0), & \mathrm{if}\ r \leqslant 0.5; \\
(0.125, 0.0, 0.1), & \mathrm{if}\ r >\, 0.5; \\
\end{array}
\right.
\end{equation}
where $r$ is the distance to the center of the explosion, $|\mathbf{v}|$ is the absolute value of flow velocity: $r^{2} = x^{2} + y^{2}$ and $|\mathbf{v}|^{2} = u^{2} + v^{2}$ in two-dimensional case, $r^{2} = x^{2} + y^{2} + z^{2}$ and $|\mathbf{v}|^{2} = u^{2} + v^{2} + w^{2}$ in three-dimensional case. The computational coordinate domain was chosen in the form of a $d$-dimensional cube $\Omega = [-1, +1]^{d}$, where $d = 2$ in two-dimensional case and $d = 3$ in three-dimensional case. The boundary conditions were specified in the form of free outflow conditions. The final time has been chosen $t_{\rm final} = 0.25$ for two-dimensional and three-dimensional explosion problems. The adiabatic index $\gamma = 1.4$. The Courant number $\mathtt{CFL} = 0.4$.

To obtain a reference solution, a one-dimensional problem with a geometric source term was used:
\begin{equation}
\frac{\partial}{\partial t}\left[
\begin{array}{c}
\rho\\
\rho \mathrm{v}\\
\varepsilon
\end{array}
\right] + 
\frac{\partial}{\partial r}\left[
\begin{array}{c}
\rho \mathrm{v}\\
\rho \mathrm{v}^{2} + p\\
(\varepsilon + p) \mathrm{v}
\end{array}
\right] =
-\frac{d - 1}{r} \left[
\begin{array}{c}
\rho \mathrm{v}\\
\rho \mathrm{v}^{2}\\
(\varepsilon + p) \mathrm{v}
\end{array}
\right];
\end{equation}
where $\mathrm{v} = |\mathbf{v}|$ is the absolute value of the flow velocity, the remaining designations coincide with those already introduced above. The coordinate domain for obtaining the reference solution was chosen in the form of a range $\Omega_{r} = [0, 1]$. The initial conditions were chosen according to the initial conditions of the original problem, which corresponds to the Sod problem. A solid wall condition was specified at the left boundary, and a free outflow condition -- at the right boundary. During the calculations, the point $r = 0$ was not directly involved in the calculations of the source term. The reference solution was obtained for cylindrical and spherical explosion problems using the ADER-WENO2 finite volume method on a mesh with $6000$ finite-volume cells. The resulting standard solution will be further used in this work for comparison with solutions to the explosion problem in the full two-dimensional and three-dimensional formulations of the problems.

\begin{figure*}[h!]
\begin{minipage}{1.0\textwidth}
\centering
\includegraphics[width=0.245\textwidth]{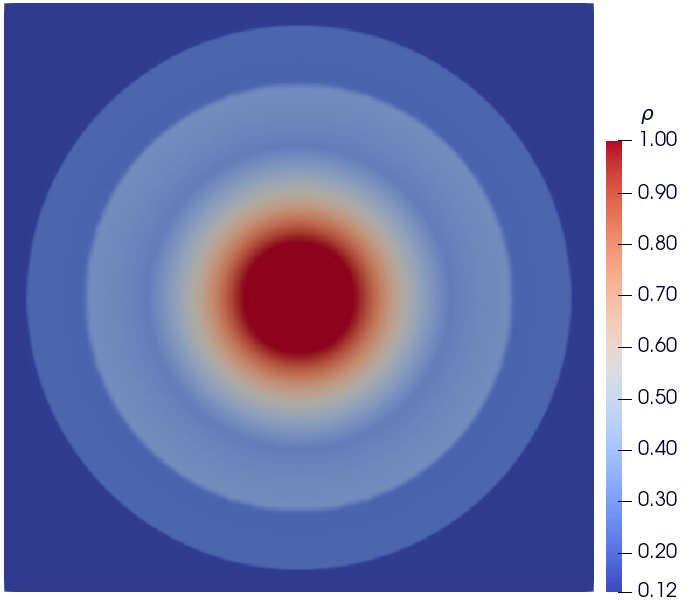}
\includegraphics[width=0.245\textwidth]{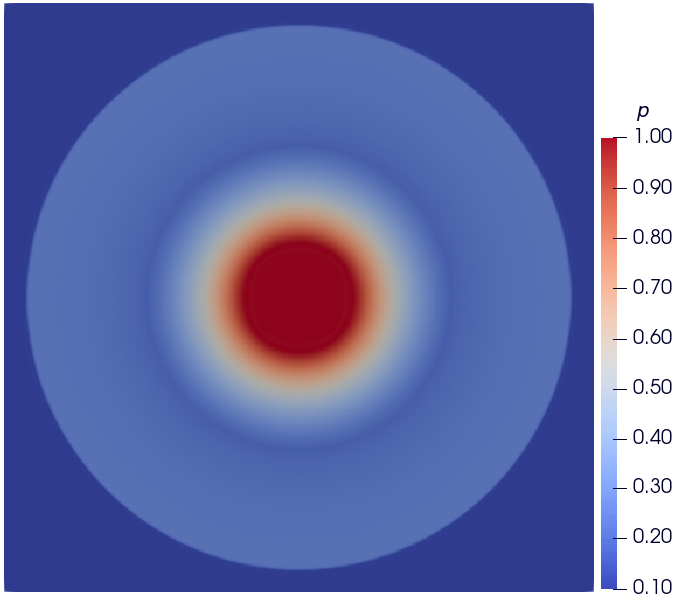}
\includegraphics[width=0.245\textwidth]{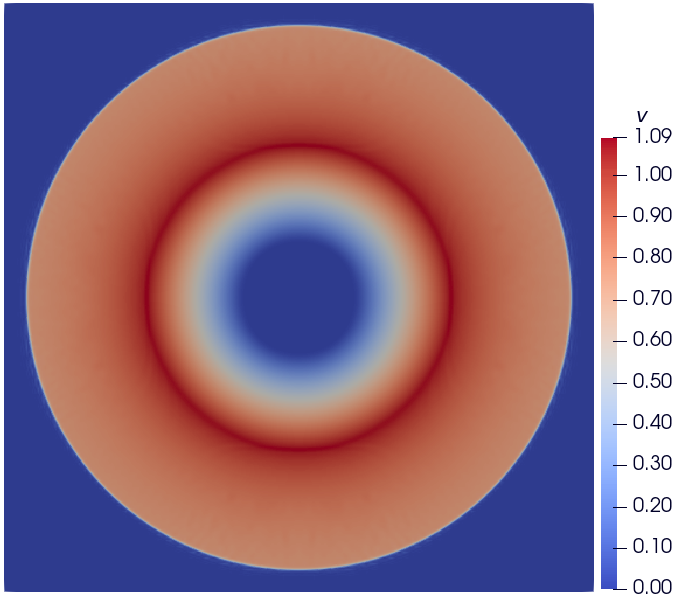}
\includegraphics[width=0.245\textwidth]{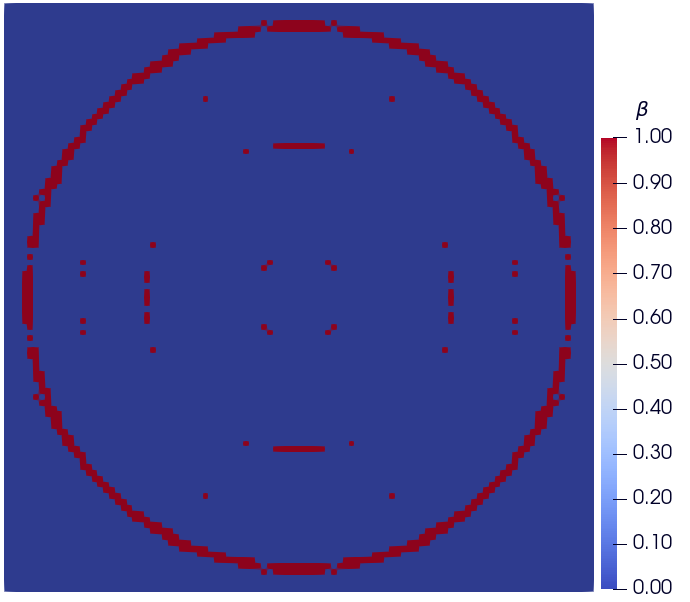}\\
\includegraphics[width=0.245\textwidth]{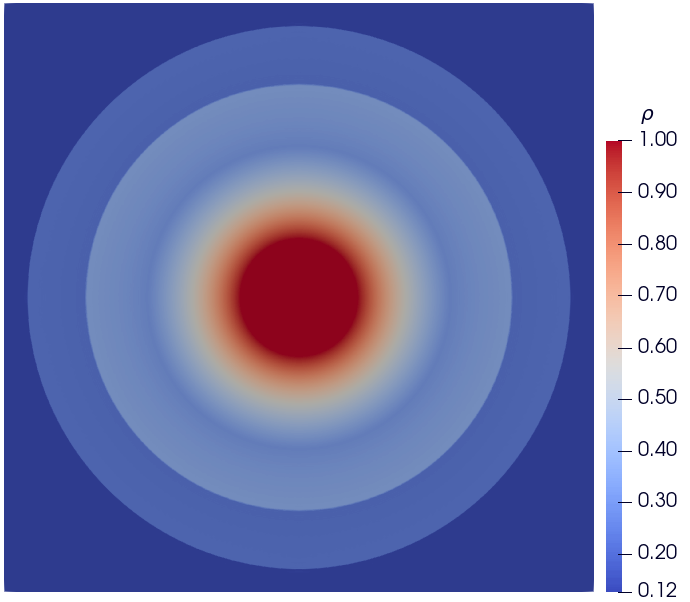}
\includegraphics[width=0.245\textwidth]{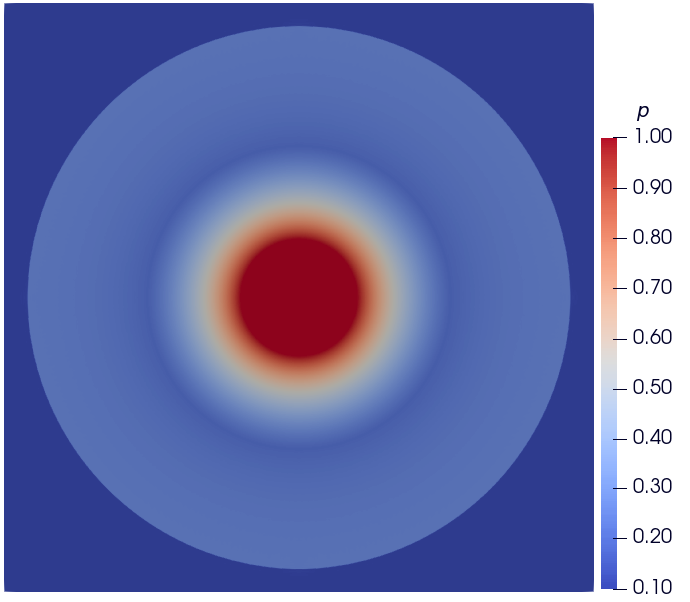}
\includegraphics[width=0.245\textwidth]{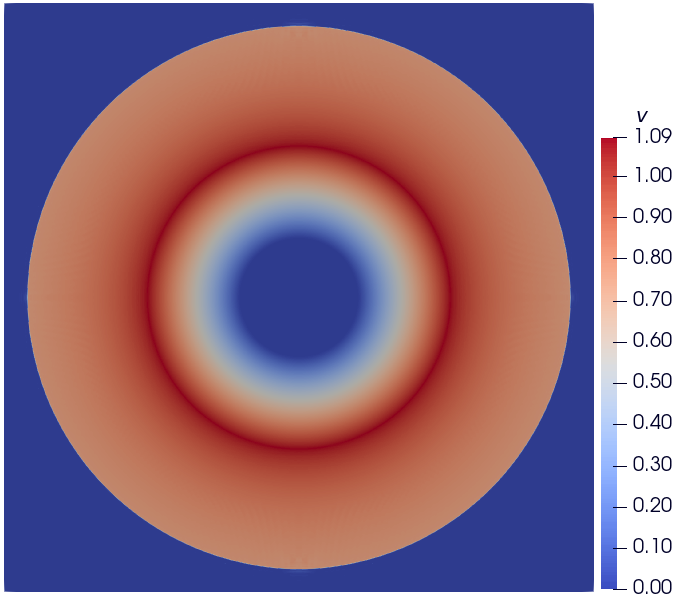}
\includegraphics[width=0.245\textwidth]{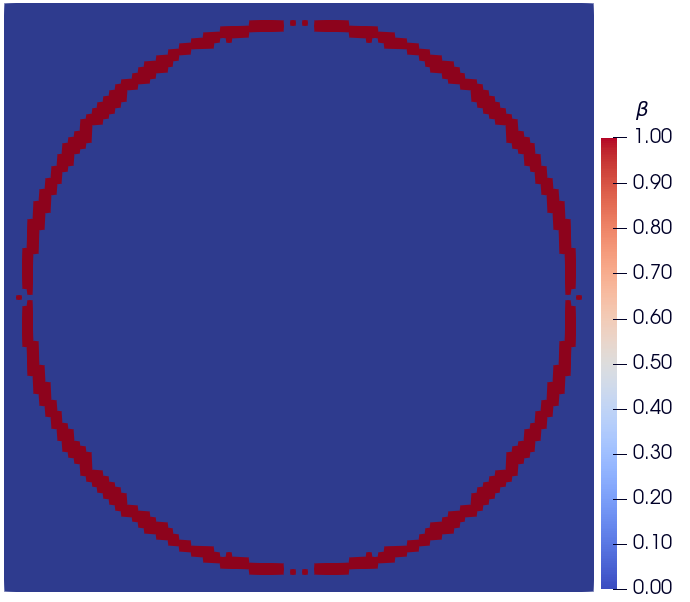}
\caption{\label{fig:sod_2d}
Numerical solution of the two-dimensional Sod explosion problem (a detailed statement of the problem is presented in the text)
obtained using the ADER-DG-$\mathbb{P}_{2}$ (top) and ADER-DG-$\mathbb{P}_{9}$ (bottom) methods on mesh with $101 \times 101$ cells.
The graphs show the coordinate dependencies of the subcells finite-volume representation of density $\rho$, pressure $p$, flow velocity magnitude $v$ 
and troubled cells indicator $\beta$ (from left to right) at the final time $t_{\rm final} = 0.25$.
}\vspace{10mm}
\includegraphics[width=0.33\textwidth]{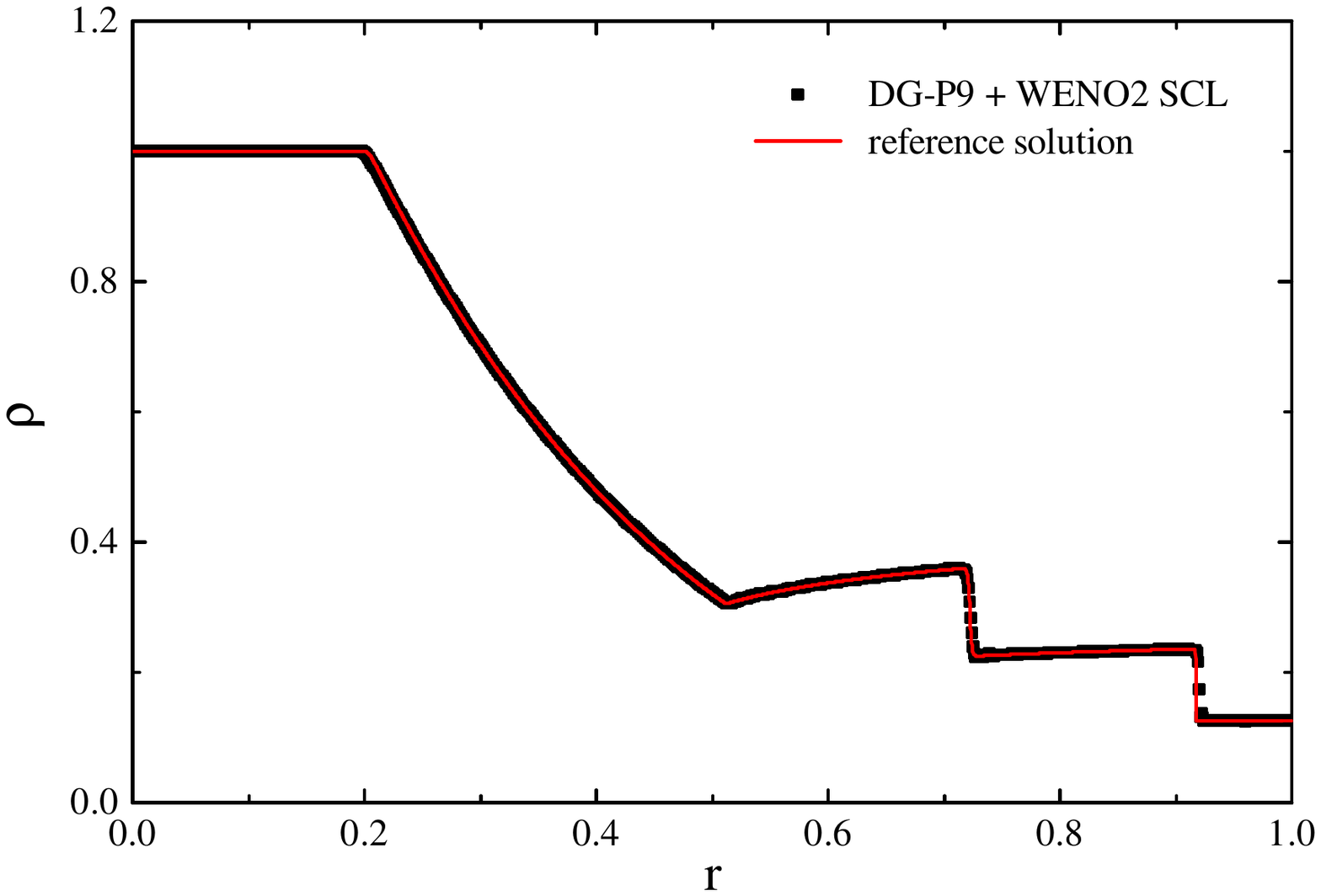}
\includegraphics[width=0.33\textwidth]{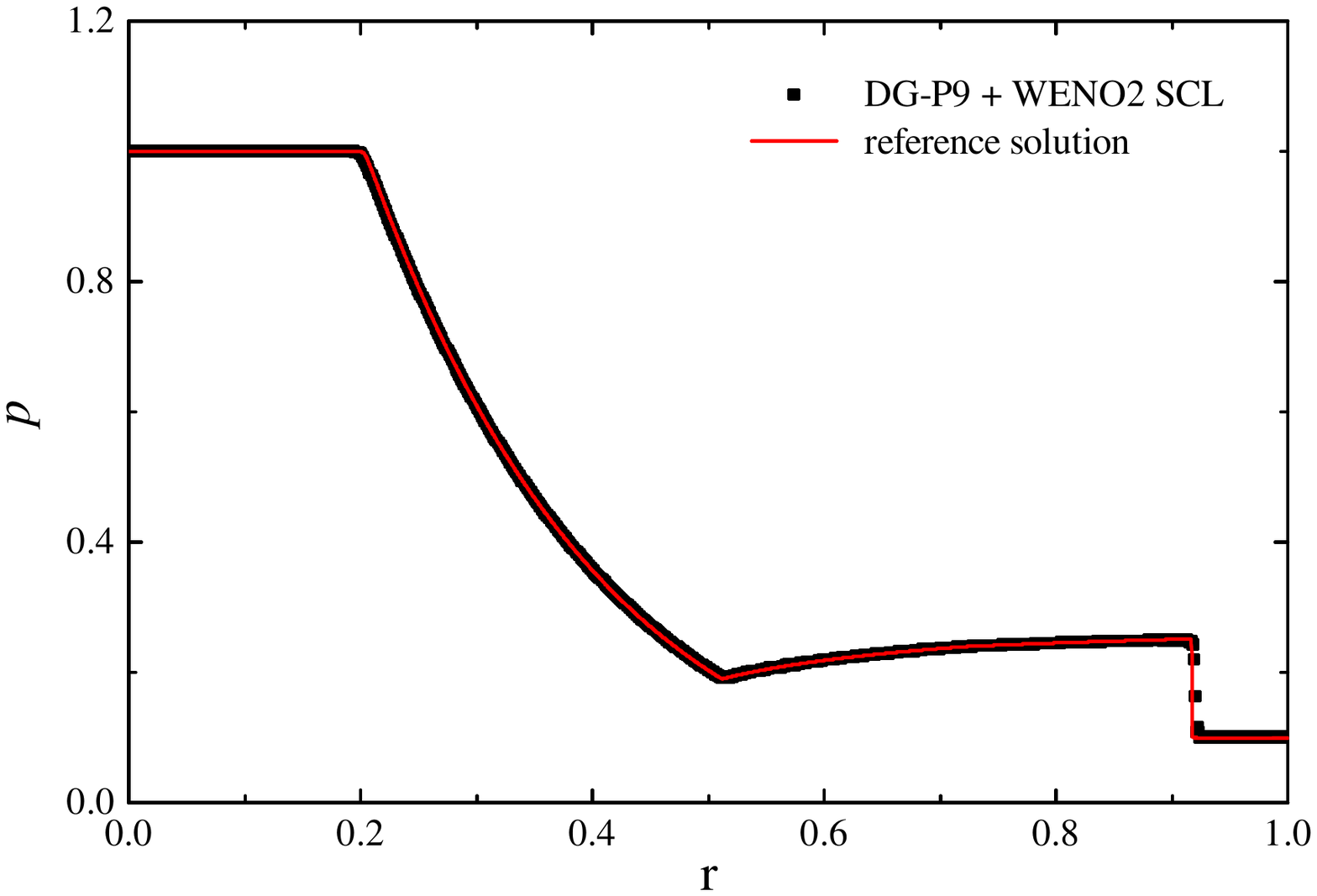}
\includegraphics[width=0.33\textwidth]{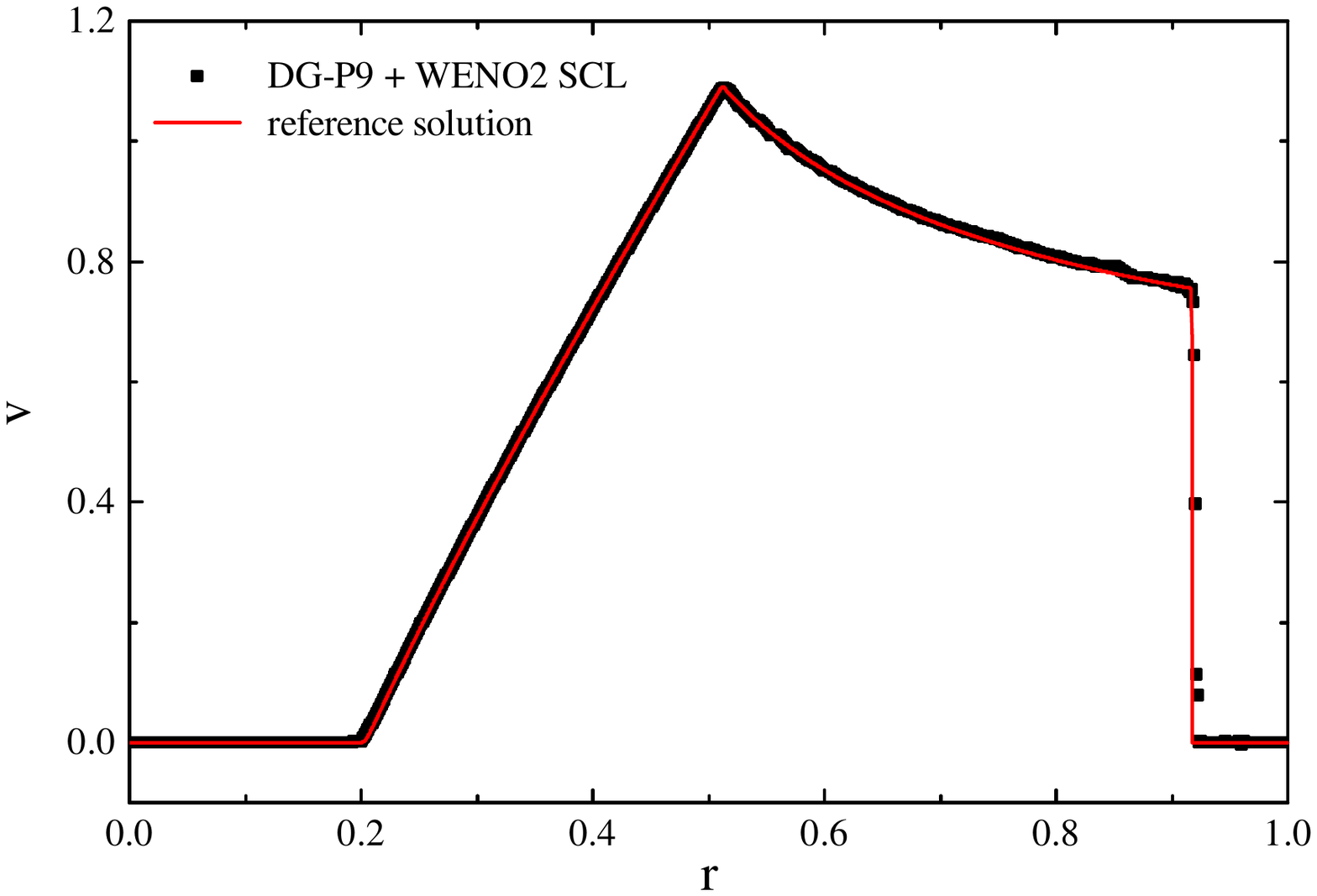}
\caption{\label{fig:sod_2d_slice}
One dimensional cuts of the numerical solution for two-dimensional Sod explosion problem in Figure~\ref{fig:sod_2d},
obtained using the ADER-DG-$\mathbb{P}_{9}$ method on mesh with $101 \times 101$ cells.
The graphs show the coordinate dependence of density $\rho$, pressure $p$ and flow velocity magnitude $v$ (from left to right) 
on the distance $r$ to the point $(0, 0)$ along the direction $(0, 1)$.
The black square symbols represent the subcells finite-volume representation of the numerical solution; 
the red solid lines represents the reference solution of the problem.
}
\end{minipage}
\end{figure*}

\paragraph{Cylindrical explosion problem}
Numerical solution to the cylindrical explosion problem was obtained on a spatial mesh with size $101 \times 101$. This size, in this case and further in this work, was used for obtaining of the one-dimensional cut of a numerical solution along one of the coordinate directions, in order to compare it with the reference solution. The main calculations were performed using the ADER-DG-$\mathbb{P}_{9}$ method with ADER-WENO2 finite volume method used as a posteriori limiter. The main results obtained for cylindrical explosion problem are presented in Figure~\ref{fig:sod_2d}. To quantify the accuracy of the numerical solution, the results in the case of using $N = 2$ are also presented for comparison.

\begin{figure*}[h!]
\begin{minipage}{1.0\textwidth}
\centering
\includegraphics[width=0.33\textwidth]{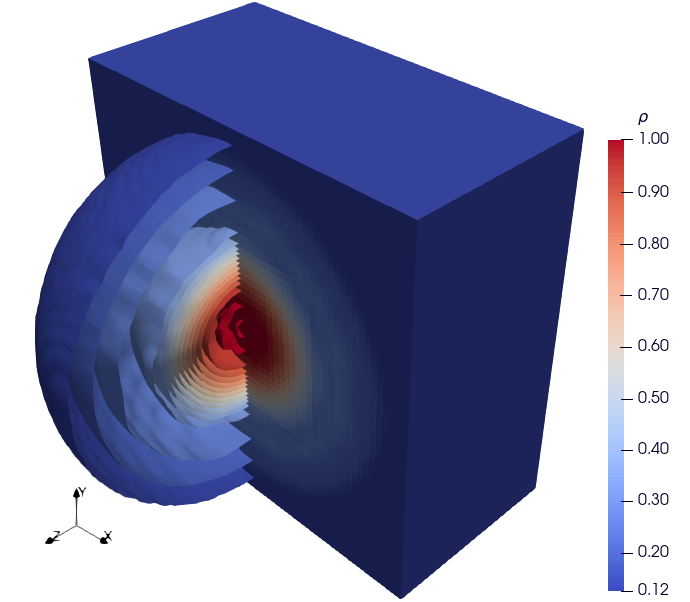}
\includegraphics[width=0.33\textwidth]{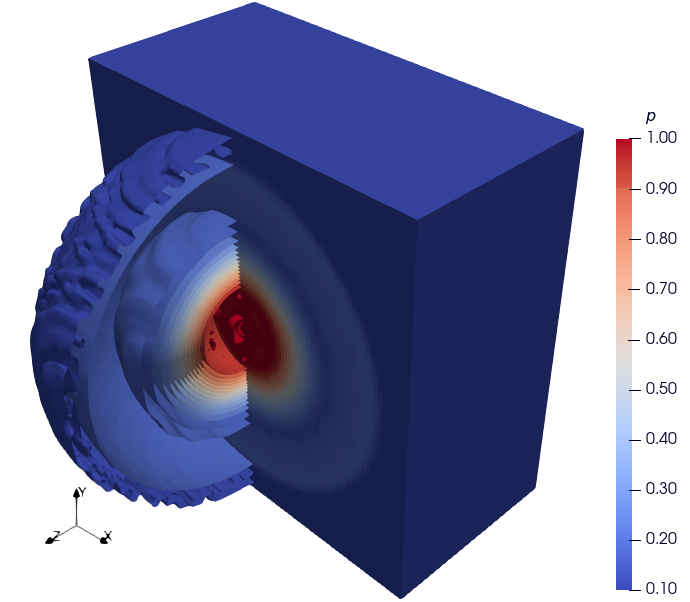}
\includegraphics[width=0.33\textwidth]{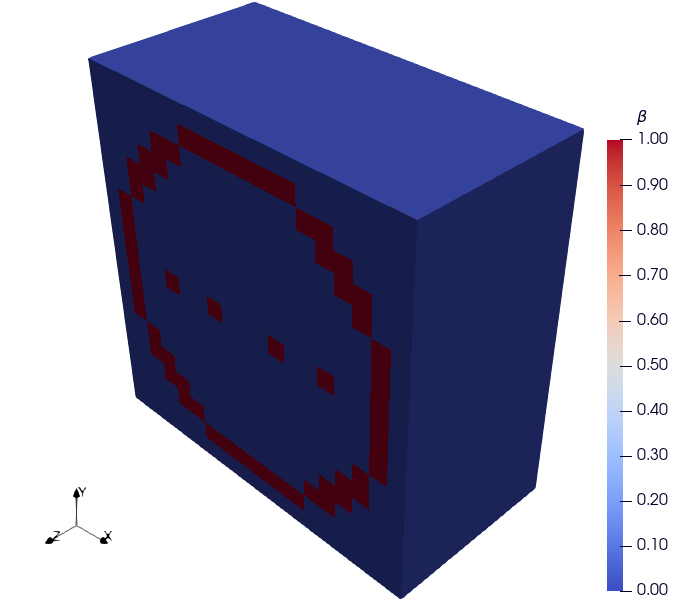}\\
\includegraphics[width=0.33\textwidth]{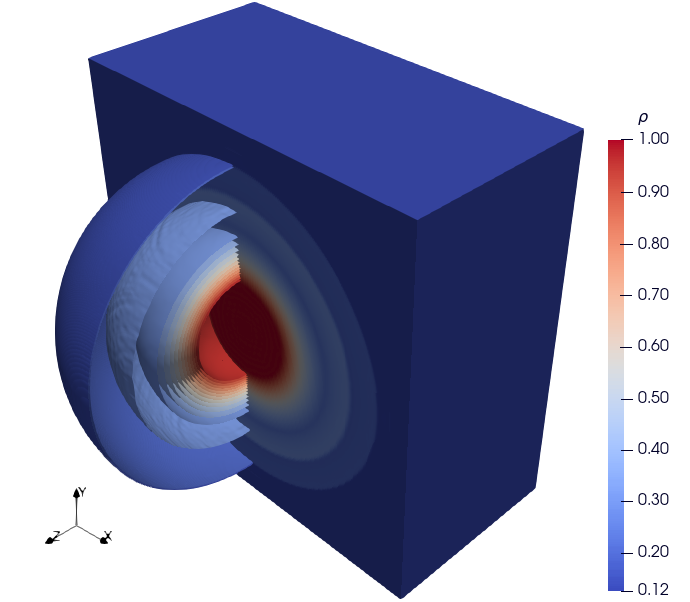}
\includegraphics[width=0.33\textwidth]{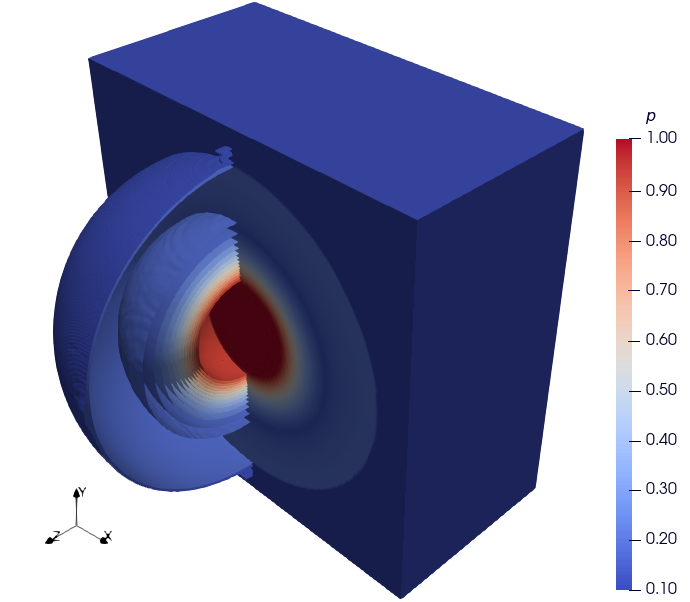}
\includegraphics[width=0.33\textwidth]{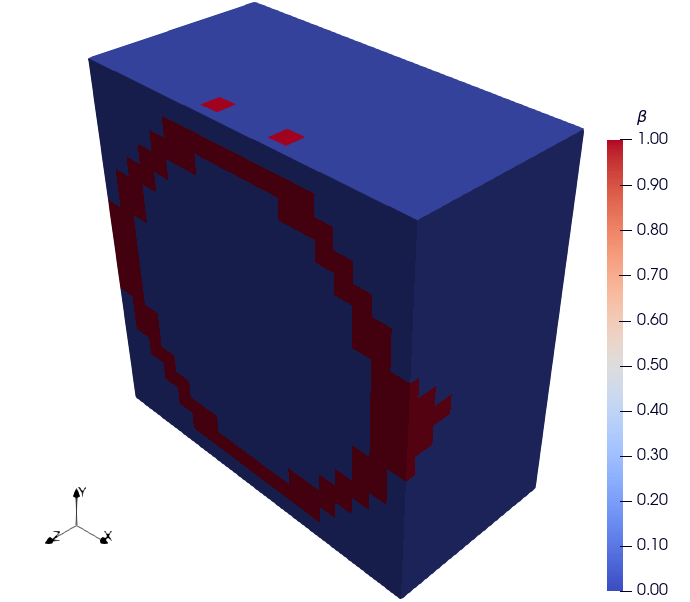}
\caption{\label{fig:sod_3d}
Numerical solution of the three-dimensional Sod explosion problem (a detailed statement of the problem is presented in the text)
obtained using the ADER-DG-$\mathbb{P}_{2}$ (top) and ADER-DG-$\mathbb{P}_{9}$ (bottom) methods on mesh with $19 \times 19$ cells.
The graphs show the coordinate dependencies of the subcells finite-volume representation of density $\rho$, pressure $p$
and troubled cells indicator $\beta$ (from left to right) at the final time $t_{\rm final} = 0.25$.
The opaque fill represents the coordinate domain clip $\Omega_{\rm clip} = \left\{\mathbf{r}\, |\, \mathbf{r} \in [-1, +1]^{3} \land z \geqslant 0 \right\}$; 
the left and center columns also represent density and pressure isosurfaces, uniformly distributed between the boundary values.
}\vspace{10mm}
\includegraphics[width=0.33\textwidth]{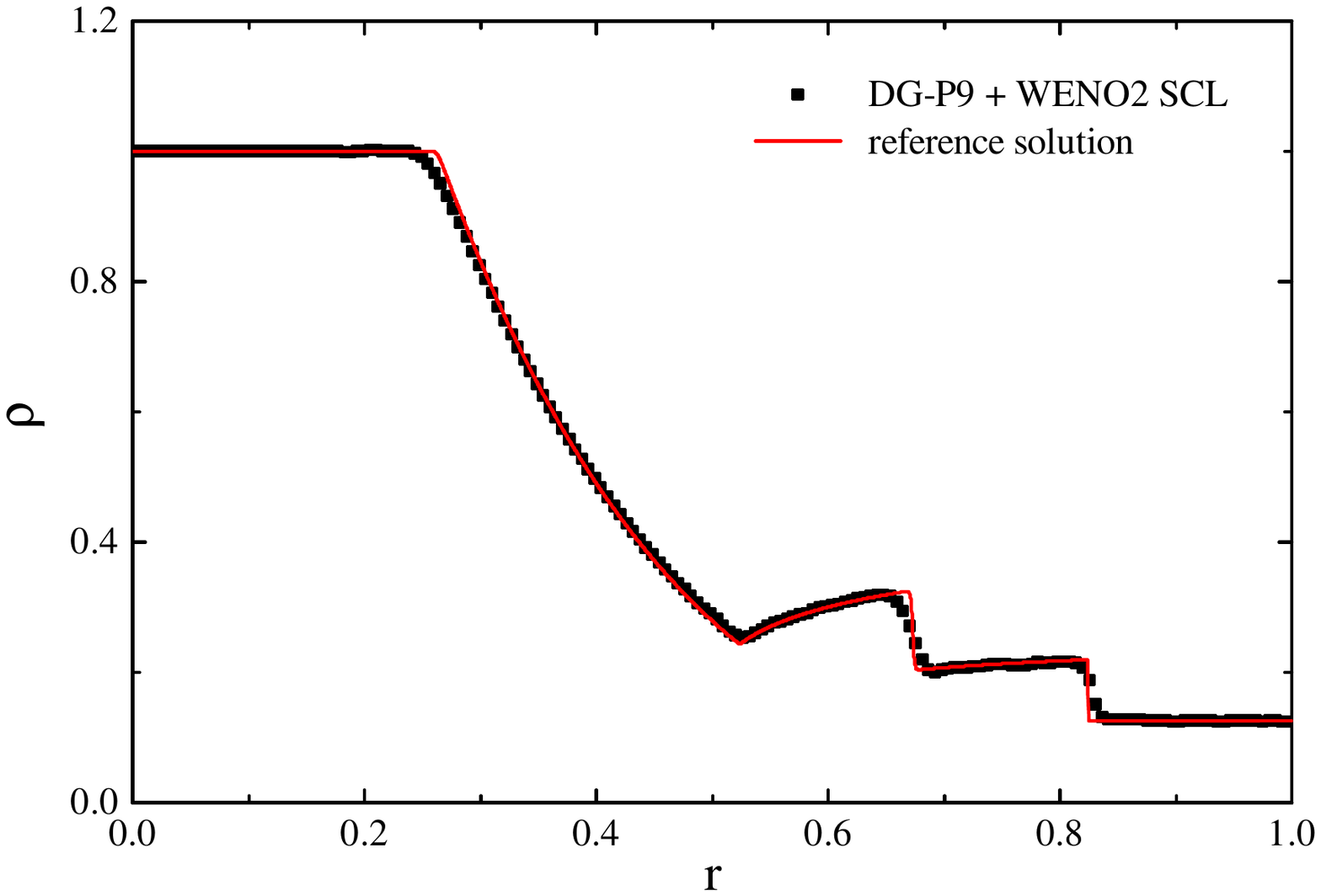}
\includegraphics[width=0.33\textwidth]{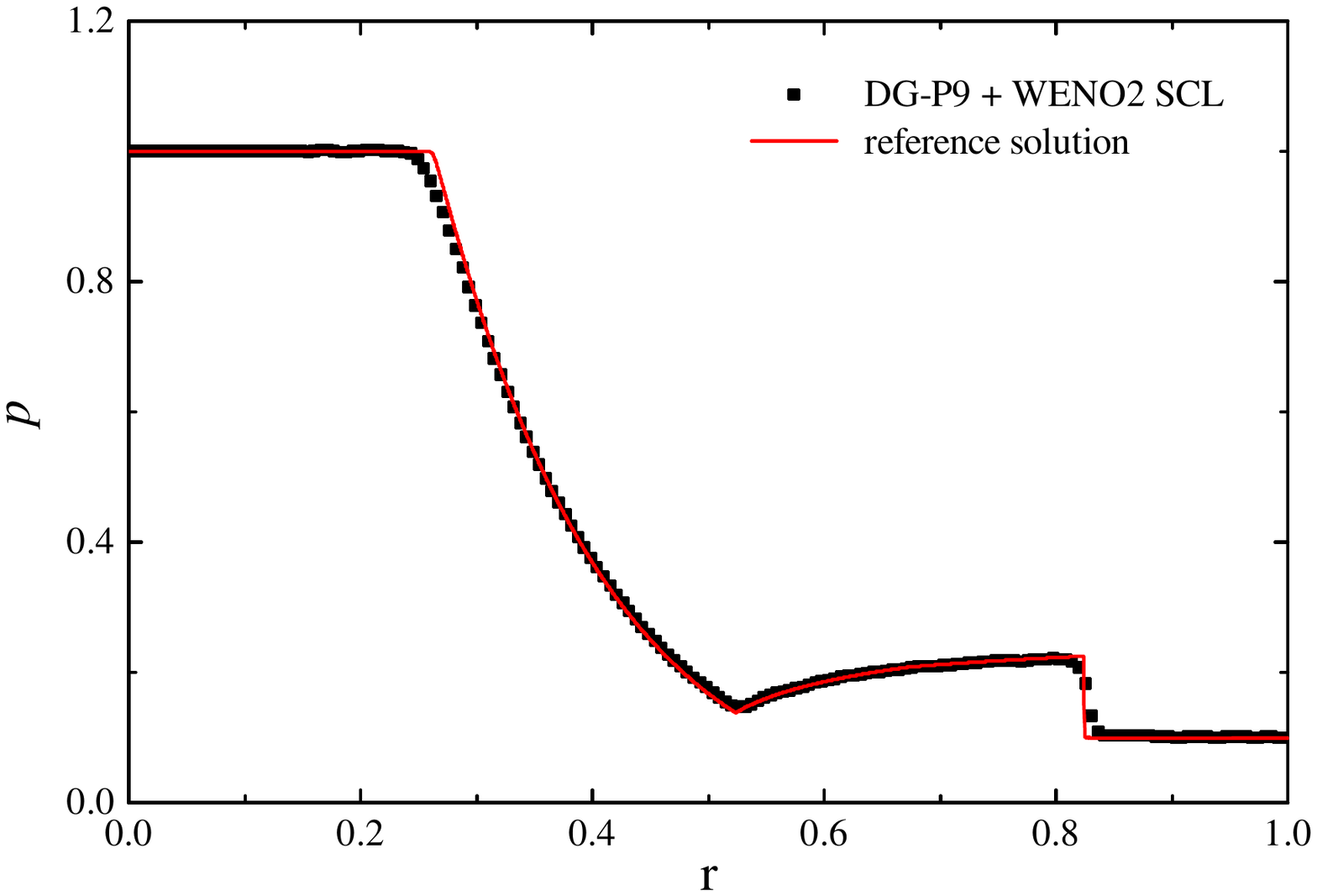}
\includegraphics[width=0.33\textwidth]{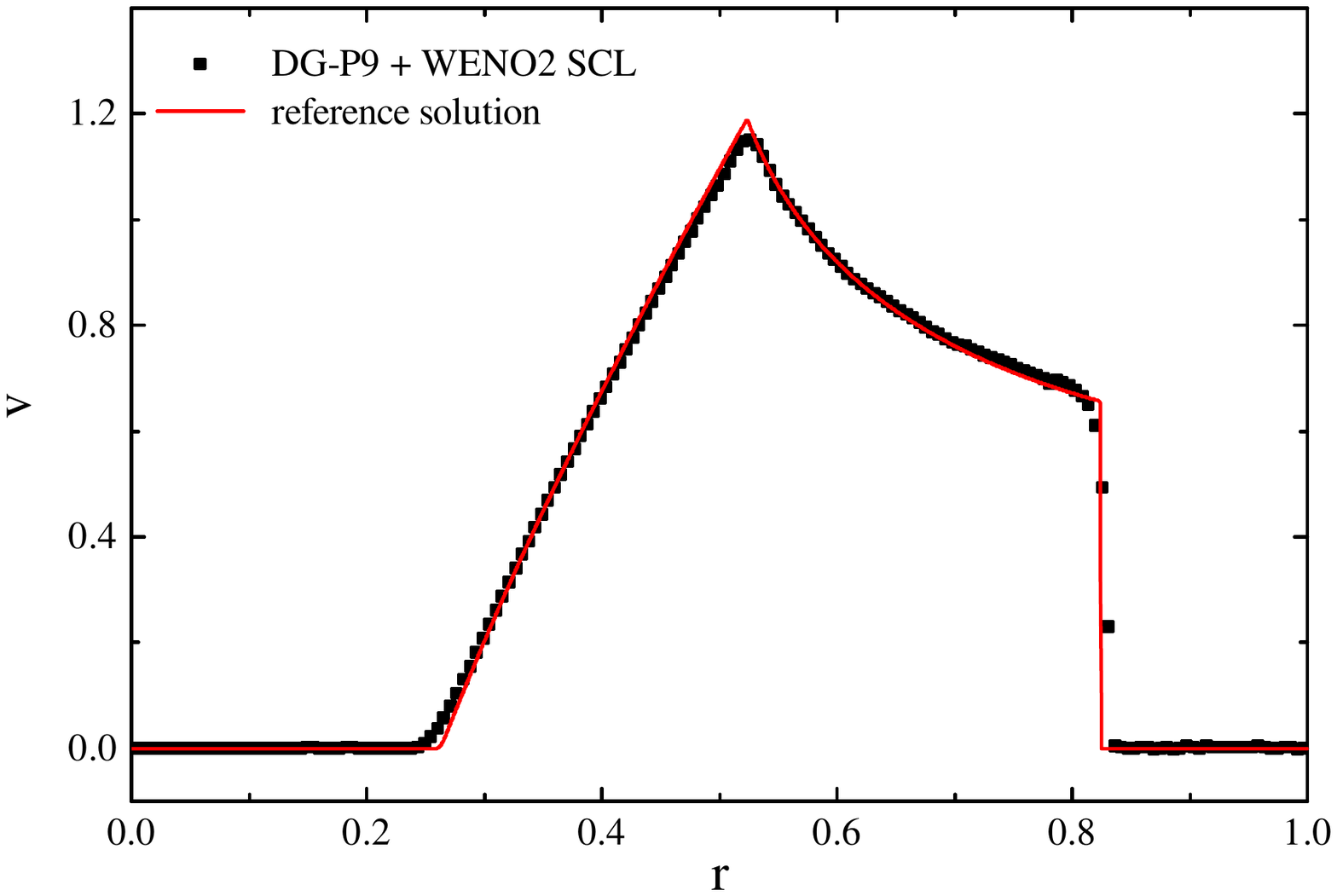}
\caption{\label{fig:sod_3d_slice}
One dimensional cuts of the numerical solution for three-dimensional Sod explosion problem in Figure~\ref{fig:sod_3d},
obtained using the ADER-DG-$\mathbb{P}_{9}$ method on mesh with $19 \times 19$ cells.
The graphs show the coordinate dependence of density $\rho$, pressure $p$ and flow velocity magnitude $v$ (from left to right)
on the distance $r$ to the point $(0, 0, 0)$ along the direction $(0, 0, 1)$.
The black square symbols represent the subcells finite-volume representation of the numerical solution; 
the red solid lines represents the reference solution of the problem.
}
\end{minipage}
\end{figure*}

The presented results show that the ADER-DG-$\mathbb{P}_{9}$ method with ADER-WENO2 finite volume a posteriori limiter correctly resolves the expanding shock wave and contact discontinuity and the converging rarefaction wave in a cylindrical explosive flow. Two-dimensional coordinate dependencies of pressure $p$, density $\rho$ and absolute velocity $\mathrm{v}$ are axially symmetric -- the numerical method correctly preserves the axial symmetry properties of the flow. In the case of using the ADER-DG-$\mathbb{P}_{9}$ method with ADER-WENO2 finite volume a posteriori limiter, it is clear that troubled cells arise only in the vicinity of the shock wave front; the thickness of the ``front'' of troubled cells ranges from one to three cells, while in the vicinity of the axial directions small domains are observed along which troubled cells do not appear. In the case of using numerical method ADER-DG-$\mathbb{P}_{2}$, the results for which are presented for comparison, it is clear that the numerical solution is more diffuse compared to method ADER-DG-$\mathbb{P}_{9}$. However, the symmetry of the numerical solution is preserved. In the case of method ADER-DG-$\mathbb{P}_{2}$, the number of troubled cells is slightly larger than in the case of method ADER-DG-$\mathbb{P}_{9}$, and troubled cells arise not only in the spatial vicinity of the shock wave front, but also in the areas of the contact discontinuity and the boundary characteristics of the rarefaction wave.

The comparison of one dimensional cuts of the numerical solution for two-dimensional Sod explosion problem, obtained using the ADER-DG-$\mathbb{P}_{9}$ method, and reference solution, obtained using one-dimensional problem with a geometric source term, is presented in Figure~\ref{fig:sod_2d_slice}. From the presented comparison it is clear that strong discontinuous components of the flow -- the shock wave and the contact discontinuity, are resolved by a numerical method with subgrid resolution -- the width of the shock wave front is $6$-$8$ subcells, the width of the contact discontinuity front does not exceed $9$-$11$ cells, while the cell size in one direction is $2N+1 = 19$ subcells for degree $N = 9$ polynomials in DG representation. The existence of a contact discontinuity does not manifest itself in any way in the coordinate dependencies of pressure $p$ and velocity $\mathrm{v}$. The rarefaction wave compares very well with the reference solution. It can be concluded that the ADER-DG-$\mathbb{P}_{N}$ method with ADER-WENO finite volume a posteriori limiter and its software implementation make it possible to obtain a numerical solution to the cylindrical explosion problem with very high accuracy, which corresponds to the results of the basic work~\cite{ader_dg_dev_1}.

\paragraph{Spherical explosion problem}
Numerical solution to the spherical explosion problem was obtained on a spatial mesh with size $19 \times 19 \times 19$. The main calculations were performed using the ADER-DG-$\mathbb{P}_{9}$ method with ADER-WENO2 finite volume method used as a posteriori limiter. The main results obtained for spherical explosion problem are presented in Figure~\ref{fig:sod_3d}. To quantify the accuracy of the numerical solution, the results in the case of using $N = 2$ are also presented for comparison.

The presented results show that the ADER-DG-$\mathbb{P}_{9}$ method with ADER-WENO2 finite volume a posteriori limiter correctly resolves the expanding shock wave and contact discontinuity and the converging rarefaction wave in a spherical explosive flow. Three-dimensional coordinate dependencies of pressure $p$, density $\rho$ and absolute velocity $\mathrm{v}$ are axially symmetric -- the numerical method correctly preserves the spherical symmetry properties of the flow. The density $\rho$ and pressure $p$ isosurfaces have well-defined spherical symmetry, despite the rather coarse mesh. In the case of using the ADER-DG-$\mathbb{P}_{9}$ method with ADER-WENO2 finite volume a posteriori limiter, it is clear that troubled cells arise only in the vicinity of the shock wave front; the thickness of the ``front'' of troubled cells ranges from one to three cells, while in the vicinity of the axial directions small domains are observed along which troubled cells do not appear. In the case of using numerical method ADER-DG-$\mathbb{P}_{2}$, the results for which are presented for comparison, it is clear that the numerical solution is more diffuse compared to method ADER-DG-$\mathbb{P}_{9}$. The density $\rho$ and pressure $p$ isosurfaces in the case of method ADER-DG-$\mathbb{P}_{2}$ become more rough compared to the results of method ADER-DG-$\mathbb{P}_{9}$. However, the symmetry of the numerical solution is preserved. In the case of method ADER-DG-$\mathbb{P}_{2}$, the number of troubled cells is slightly larger than in the case of method ADER-DG-$\mathbb{P}_{9}$, and troubled cells arise not only in the spatial vicinity of the shock wave front, but also in the areas of the contact discontinuity and the boundary characteristics of the rarefaction wave.

The comparison of one dimensional cuts of the numerical solution for three-dimensional Sod explosion problem, obtained using the ADER-DG-$\mathbb{P}_{9}$ method, and reference solution, obtained using one-dimensional problem with a geometric source term, is presented in Figure~\ref{fig:sod_3d_slice}. From the presented comparison it is clear that strong discontinuous components of the flow -- the shock wave and the contact discontinuity, are resolved by a numerical method with subgrid resolution -- the width of the shock wave front is $5$-$8$ subcells, the width of the contact discontinuity front does not exceed $9$-$11$ cells, while the cell size in one direction is $2N+1 = 19$ subcells for degree $N = 9$ polynomials in DG representation. The existence of a contact discontinuity does not manifest itself in any way in the coordinate dependencies of pressure $p$ and velocity $\mathrm{v}$. The rarefaction wave compares very well with the reference solution. It can be concluded that ADER-DG-$\mathbb{P}_{9}$ method with ADER-WENO2 finite volume a posteriori limiter and its software implementation make it possible to obtain a numerical solution to the cylindrical explosion problem with very high accuracy, which corresponds to the results of the basic work~\cite{ader_dg_dev_1}.

\subsection{Shock-bubble interaction problem}
\label{sec:apps_cgd_problems:sbi_2d}

\begin{figure}[h!]
\centering
\includegraphics[width=0.49\textwidth]{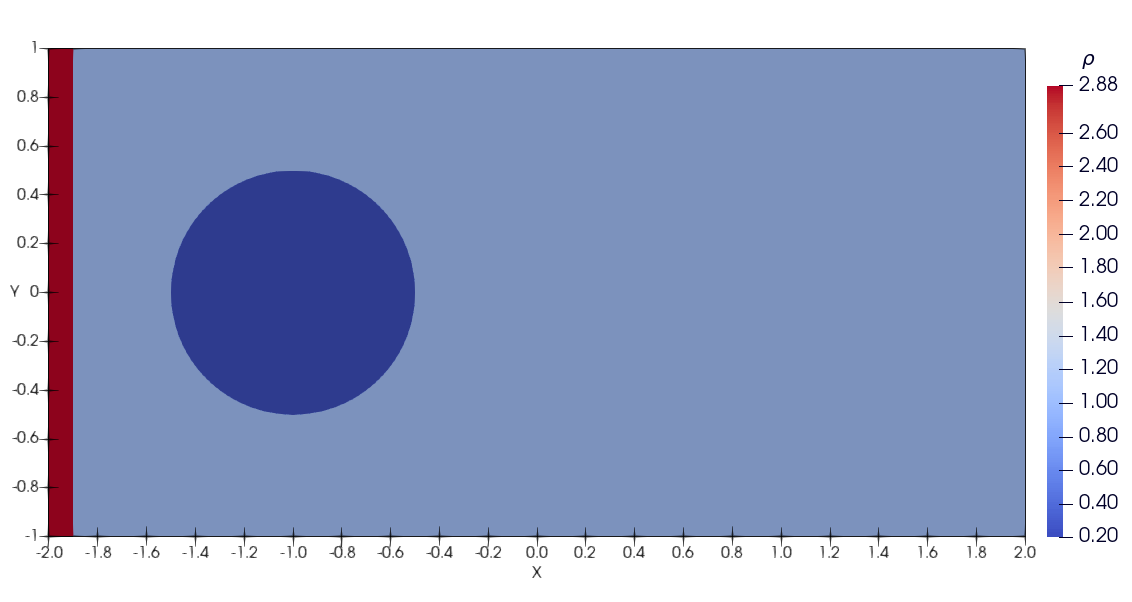}
\caption{\label{fig:sbi_2d_init}
The initial conditions for coordinate dependency of density $\rho$ in the 
two-dimensional shock-bubble interaction problem (a detailed statement of the problem is presented in the text).
}
\end{figure}

The shock-bubble interaction problem was chosen as an interesting, substantially asymmetric problem containing a wide range of spatial and temporal scales. Computational coordinate domain $\Omega = [-2, +2]\times[-1, +1]$. The initial conditions were chosen in the following form:
\begin{equation}
\begin{split}
&\rho = \left\{
\begin{array}{ll}
2.880323, & \mathrm{if}\ x \leqslant -1.9;\\
0.2, & \mathrm{if}\ r \leqslant 0.5;\\
1.0, & \mathrm{if}\ x >\, -1.9 \land r >\, 0.5;
\end{array}
\right.\\
&p = \left\{
\begin{array}{ll}
5.219111, & \mathrm{if}\ x \leqslant -1.9;\\
1.0, & \mathrm{if}\ x >\, -1.9;
\end{array}
\right.\\
&u = \left\{
\begin{array}{ll}
1.659610, & \mathrm{if}\ x \leqslant -1.9;\\
0.0, & \mathrm{if}\ x >\, -1.9;
\end{array}
\right.\\
&v = 0.0;
\end{split}
\end{equation}
where $r^{2} = (x + 1)^{2} + y^{2}$ determines the distance to the point $(-1, 0)$. The selected initial conditions determine a bubble of radius $R = 0.5$ with density $\rho_{0} = 0.2$, center of which is located at the the point $(-1, 0)$, and a shock wave with parameters $(\rho, u, p) = (2.880323, 1.659610, 5.219111)$ behind the shock front, which is located on the line $x = -1.9$. This shock front appears in the solution of a one-dimensional Riemann problem with a pressure difference of 10 times. The configuration of the initial conditions is presented in Figure~\ref{fig:sbi_2d_init}. The boundary conditions were chosen as follows: on the left boundary is the exact solution for the shock wave, on the right boundary are the conditions of free outflow, periodic boundary conditions were set on top and bottom boundaries, which under the symmetry of initial conditions of the problem are equivalent to the solid wall boundary conditions. The final time has been chosen $t_{\rm final} = 1.25$. The adiabatic index $\gamma = 1.4$. The Courant number $\mathtt{CFL} = 0.4$.

\begin{figure*}[h!]
\centering
\includegraphics[width=0.45\textwidth]{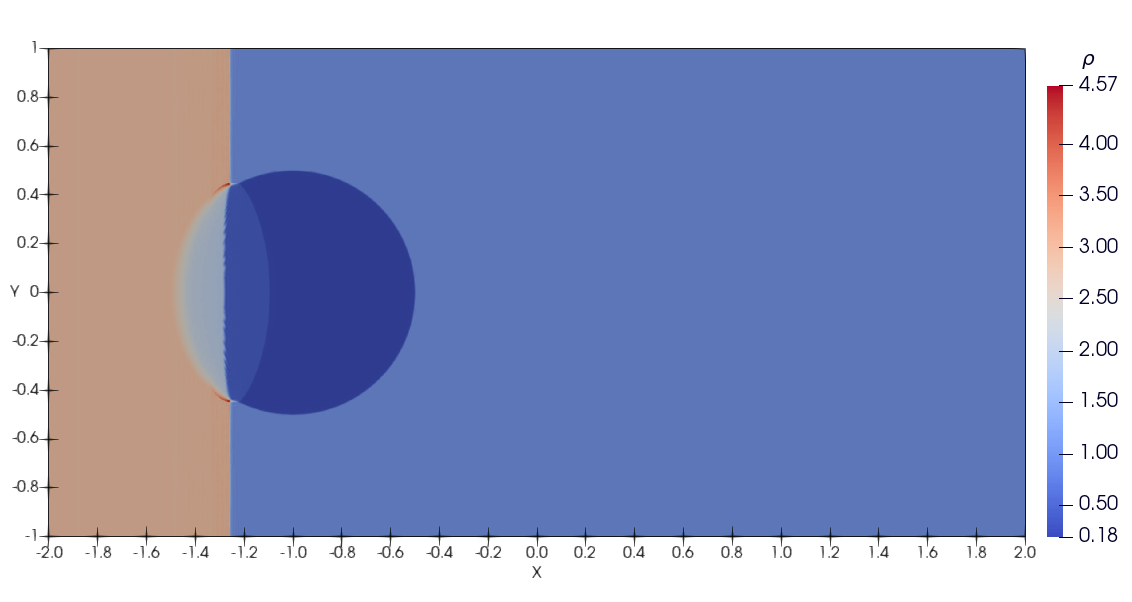}\hspace{10mm}
\includegraphics[width=0.45\textwidth]{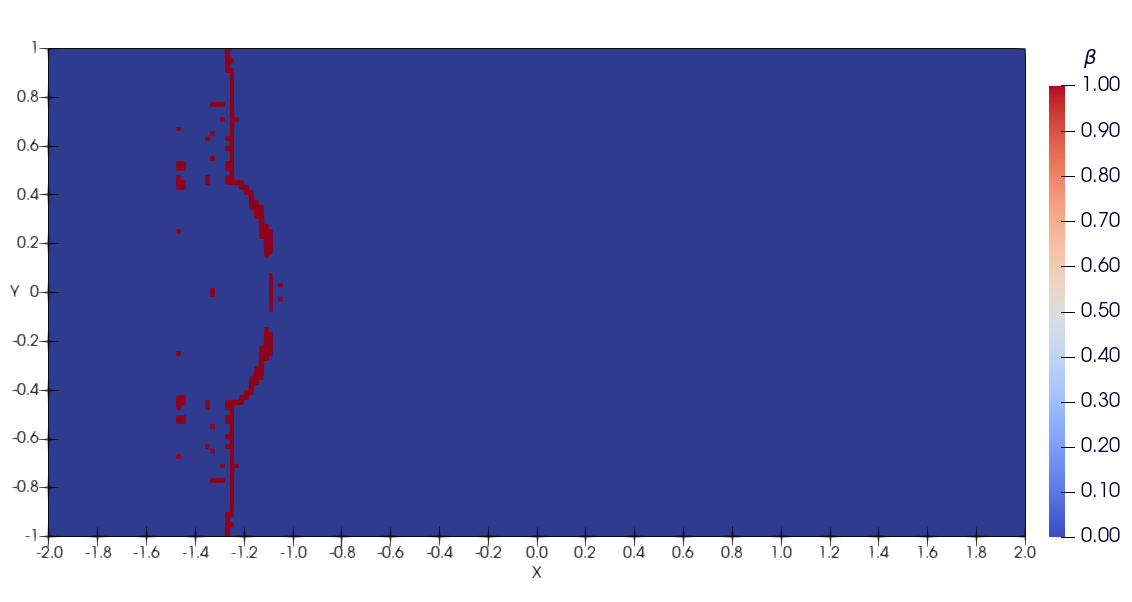}\\
\includegraphics[width=0.45\textwidth]{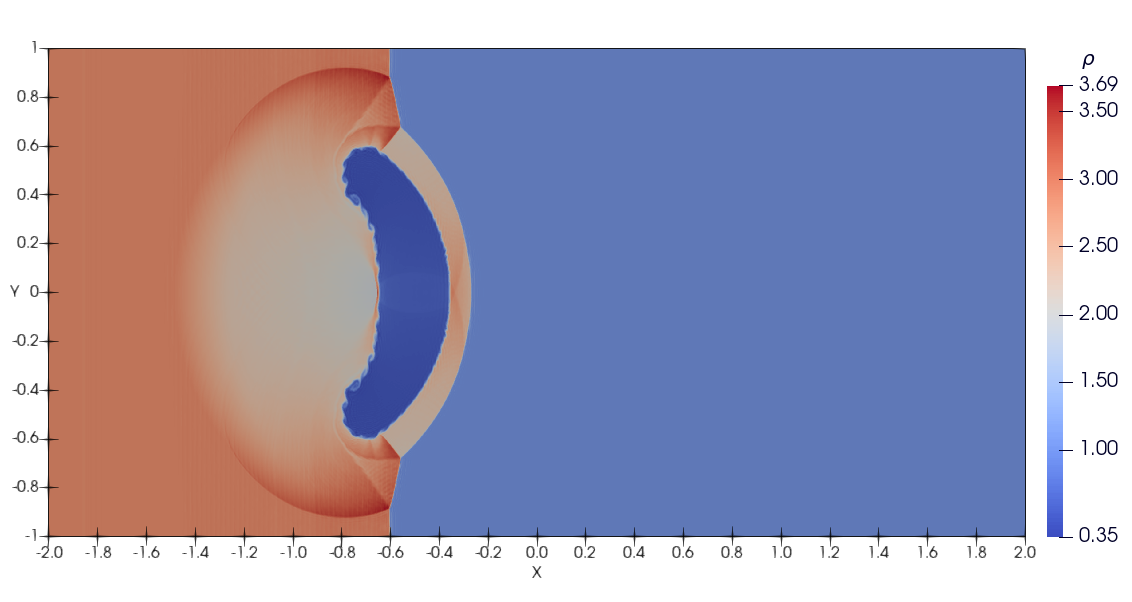}\hspace{10mm}
\includegraphics[width=0.45\textwidth]{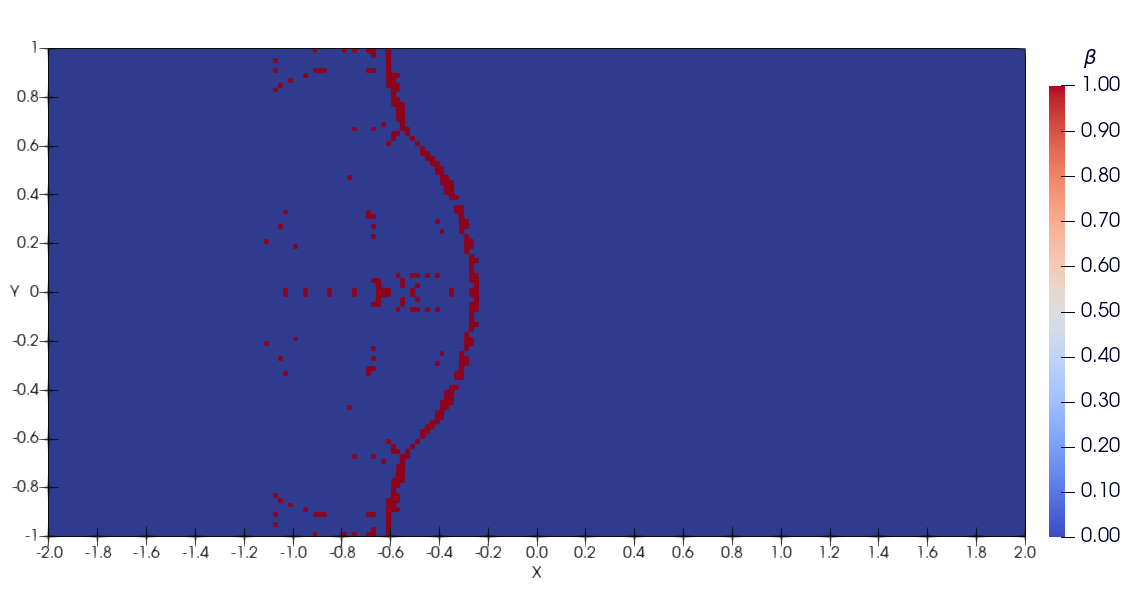}\\
\includegraphics[width=0.45\textwidth]{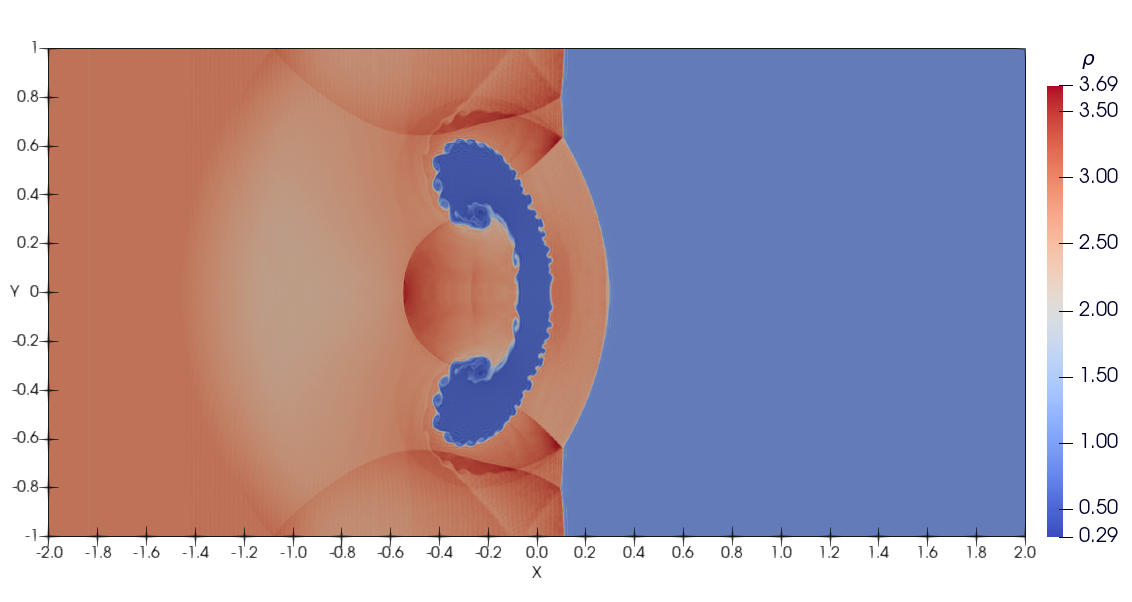}\hspace{10mm}
\includegraphics[width=0.45\textwidth]{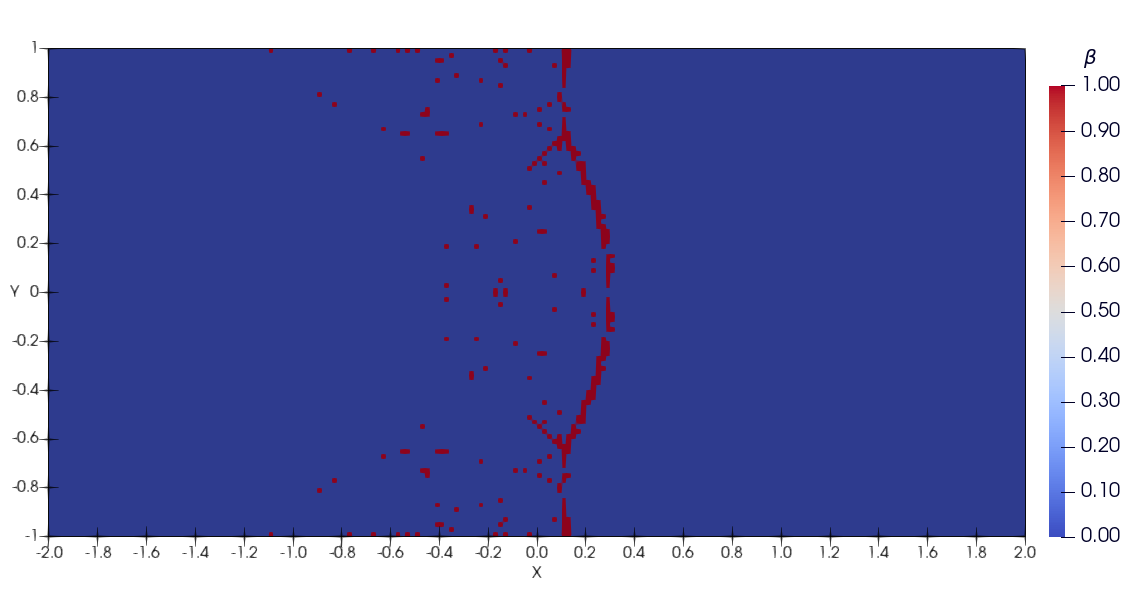}\\
\includegraphics[width=0.45\textwidth]{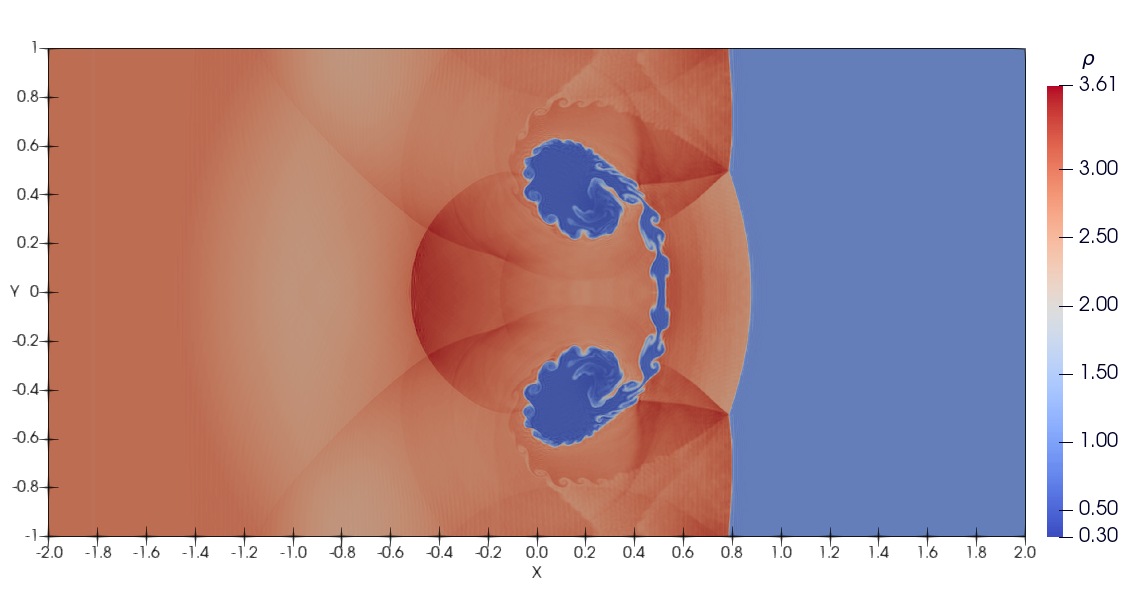}\hspace{10mm}
\includegraphics[width=0.45\textwidth]{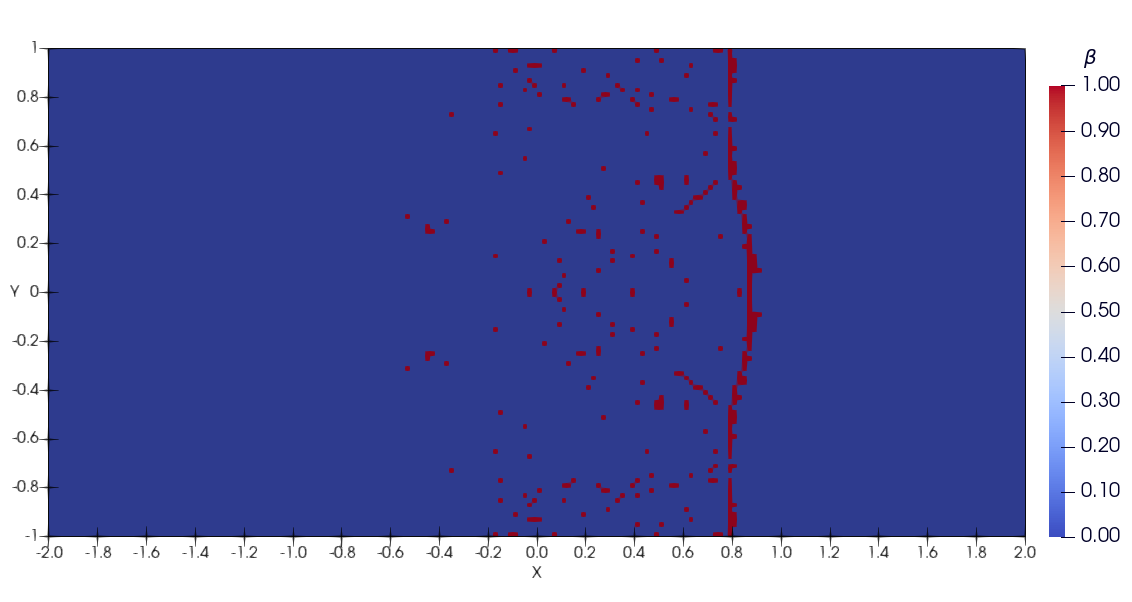}\\
\includegraphics[width=0.45\textwidth]{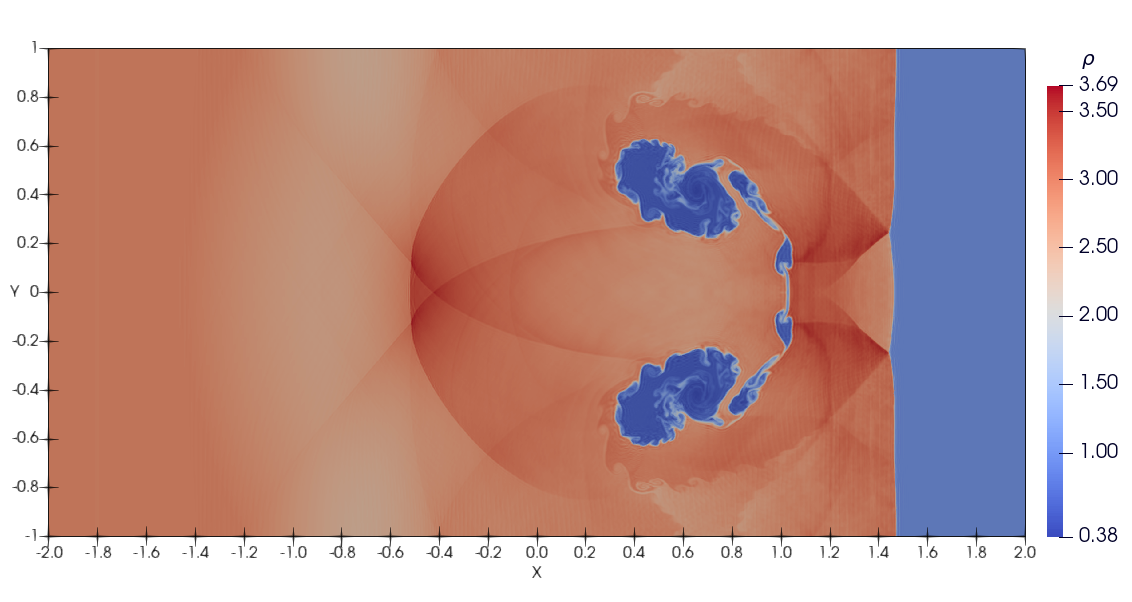}\hspace{10mm}
\includegraphics[width=0.45\textwidth]{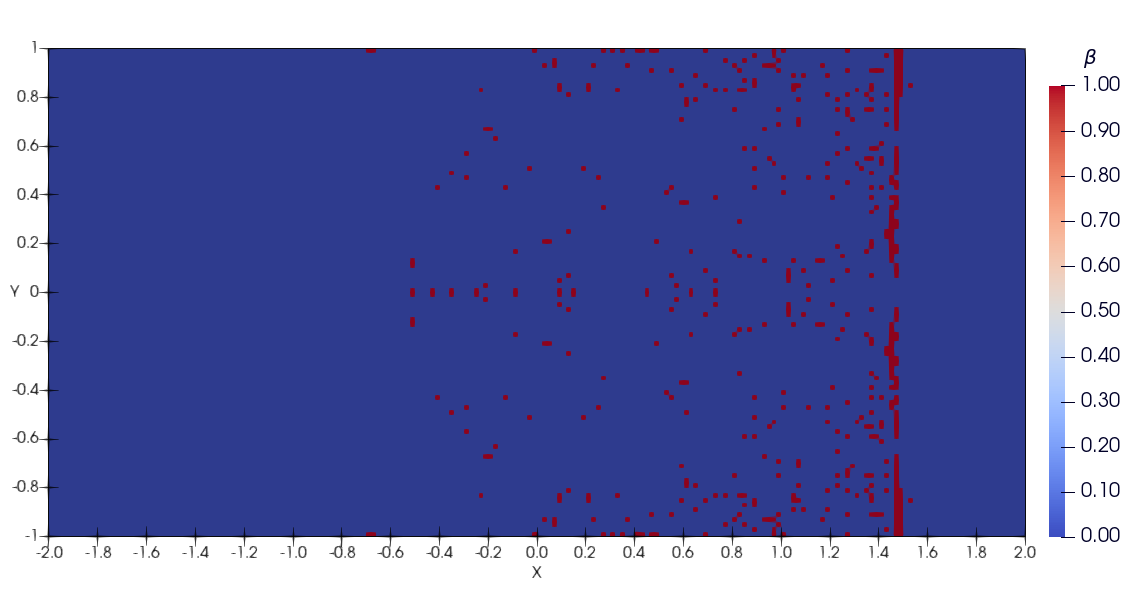}
\caption{\label{fig:sbi_2d}
Numerical solution of the two-dimensional shock-bubble interaction problem (a detailed statement of the problem is presented in the text)
obtained using the ADER-DG-$\mathbb{P}_{5}$ method with a posteriori limitation of the solution by a ADER-WENO2 finite volume limiter 
on mesh with $200 \times 100$ cells at the times $t = 0.25$, $0.50$, $0.75$, $1.00$ and $1.25$ (from top to bottom).
The graphs show the coordinate dependencies of the subcells finite-volume representation of density $\rho$ (left) and troubled cells indicator $\beta$ (right).
}
\end{figure*}

The numerical solution was obtained using ADER-DG-$\mathbb{P}_{5}$ method with ADER-WENO2 finite volume a posteriori limiter on a spatial mesh with sizes $200\times100$. The numerical solution is presented in Figure~\ref{fig:sbi_2d} at several times $t = 0.25$, $0.50$, $0.75$, $1.00$ and $1.25$ to demonstrate the flow dynamics and the arising non-stationary processes. It should be noted that the number of troubled cells on the mesh never exceeded $3.1\%$, and the average number of troubled cells in this test was $\sim 1.7\%$. In this case, of course, the number of troubled cells was determined by the emerging features and structures in the solution.

The presented results demonstrate a classical dynamics of the processes of interaction between a shock wave and a bubble, the gas density in which is lower than the density of the initially surrounding gas. At the moment of time $t = 0.25$, the process of initial interaction of the shock wave with the bubble is observed, which is accompanied by the reflection of the rarefaction wave back, the transmitted of the shock wave deep into the gas in the bubble and its compression at the front of the passing wave; in this case, the inhomogeneities of the contact surface that arise behind the shock wave front are currently very small and are not visually observed. The shock wave in the bubble moves at a speed greater than in the gas surrounding the bubble -- the sound speed in the gas inside the bubble is greater than in the surrounding gas; therefore, refraction of the shock front occurs. At the next presented moment of time $t = 0.50$ shown, the shock wave front emerges from the deformed bubble into the surrounding gas. In this case, a vortex flow is formed inside the bubble, and a vortex street is formed along the side and rear surfaces, the formation process of which is equivalent to the Richthayer-Meshkov instability. The process of movement of the deformed bubble to the right is also visually observed. At the subsequent presented moment of time $t = 0.75$, further movement of the initial shock front is observed, as well as the shock fronts that arose as a result of diffraction on the bubble and reflected from the front boundary of the bubble, where a positive change in acoustic rigidity was observed, and the boundary of the coordinate domain. A vortex street is formed along the entire length of the boundary of the deformed bubble. A complex pattern of movement of contact discontinuities is also observed, in particular, tangential discontinuities arising at two triple points of the shock wave front propagating forward form vortex streets, which are picked up by the return flow behind the deformed vortex. At the subsequent presented moments of time $t = 1.00$ and $1.25$, a consistent fragmentation of the vortex flow inside the deformed bubble is observed, and vortex structures of opposite vorticity arise. At the final moment of time $t = 1.25$, a shock front of a complex structure has moved forward, formed as a result of the merger of several shock waves. The gas that initially formed the bubble fragmented into two symmetrical vortex structures possessing a multi-scale set of vortices, as well as a ``bridge'' formed by the structures with smaller ``bridges'' between them.

The presented numerical solution contains all the main hydrodynamic structures that are characteristic of this problem, and also correctly describes the flow dynamics. From the presented results and their analysis, we can conclude that the ADER-DG-$\mathbb{P}_{N}$ method with ADER-WENO2 finite volume a posteriori limiter and its software implementation can be used to simulate complex multi-scale hydrodynamic flows of compressible media.

\section{Detonation waves simulation}
\label{sec:detonation_waves}

Direct numerical simulation of detonation processes associated with the formation, propagation and dynamic evolution of detonation waves is a rather complex problem~\cite{Oran_Boris_2005, chem_kin_hrs_rev_1, chem_kin_hrs_rev_2}. This is primarily due to a significant difference in the rates of processes occurring in the medium -- the characteristic times of change in hydrodynamic variables are determined by the spatial scale of the problem and the speed of sound, while the characteristic times of exothermic reactions occurring in the medium can be several orders of magnitude higher. In a certain sense, this problem can be characterized as stiffness, problems with which are well known in the field of numerical solution of differential equations. Therefore, the simple use of hydrodynamic numerical methods is associated with obtaining often incorrect numerical solutions, which is very well illustrated in the work~\cite{frac_steps_detwave_sim_2000}. It is known that these features can be explained~\cite{correct_det_wave_speed_2017} by the formation in the numerical solution of a weak detonation front, which propagates ahead of the shock front, which leads to the separation of the processes of energy and momentum transfer in the medium and forms a non-physical process of propagation of the detonation front. 

The use of classical numerical methods of computational fluid dynamics in the case of studying reacting flows with fast reaction processes usually requires the inclusion in the numerical method of some method of splitting into physical processes~\cite{frac_steps_detwave_sim_2000}, similar to the Marchuk-Strang splitting, when hydrodynamic phenomena and the kinetics of reaction processes are calculated at individual steps of the splitting method. This approach is one of the main ones in applied methods for the numerical study of detonation processes. However, it is well known that the use of splitting methods by physical processes leads to a decrease in the order of approximation and the order of convergence of the numerical method, usually to the first order; possible symmetrization of the splitting method to increase the effective order of approximation for use in conjunction with computational fluid dynamics methods is usually too complex and not applied, and can also lead to other, possibly more complex and unexpected, artifacts of the numerical solution~\cite{chem_kin_hrs_weno}. Even in the case of using high-order numerical methods, one often has to resort to the idea of using methods of splitting into physical processes~\cite{chem_kin_hrs_weno}. 

In the work~\cite{popov_j_sci_comp_2023}, the space-time adaptive ADER-DG method with LST-DG predictor and a posteriori sub-cell ADER-WENO finite-volume limiting was used for simulation of the formation and propagation of a one-dimensional detonation wave, where it was shown that this method is very well suited for high-precision modeling of detonation phenomena. In this case, no methods of splitting into physical processes were used. This work uses a slightly different modification of the original numerical method, which is not associated with the introduction of an adaptive time step. It is clear that the use of a global adaptive change in the time step for all mesh cells significantly increases computational costs, while this leads to an effective change in the Courant number, which means it can affect the accuracy of the representation of the smooth components of the numerical solution. The finite-volume ADER-WENO method with LST-DG predictor was used in the work~\cite{ader_stiff_2} to solve a similar one-dimensional problem of a detonation front.

\subsection{Description of the reactive flow model}
\label{sec:detonation_waves:reactive_flow_model}

In this work, the simulation of a detonation wave is considered within the framework of a concept similar to that presented in the works~\cite{frac_steps_detwave_sim_2000, ader_stiff_2, popov_j_sci_comp_2023}. A two-component medium with monomolecular kinetics $A \rightarrow B$ is considered, where the component $A$ is a reaction reagent, and the component $B$ is a reaction product. The reaction rate is determined by the law of mass action.
The discrete ignition temperature kinetics model was represented by the reaction rate constant in the following form:
\begin{eqnarray}\label{rate_constant}
k(T) = \left\{
\begin{array}{cl}
\frac{1}{\tau_{0}},& \mathrm{if}\, T \geqslant T_{\rm ign},\\[2mm]
0,& \mathrm{if}\, T <\, T_{\rm ign},
\end{array}
\right.
\end{eqnarray}
where $T$ is the temperature, $T_{\rm ign}$ is the ignition temperature, and $\tau_{0}$ is the time scale of the chemical reaction. The ignition temperature was chosen $T_{\rm ign} = 0.26$. The temperature $T$ is determined from the Mendeleev-Clapeyron thermal equation of state. Dimensional scales and parameters of medium (such as molar mass $\mu$) are chosen such that $T = p/\rho$. The specific energy yield was chosen $q_{0} = 1$.

The original system of Euler equations (\ref{eq:system_of_equations}), extended by the system of convection-reaction equations (\ref{eq:convection_reaction_equations}), taking into account the chosen reaction mechanism in a two-component medium, takes the following form:
\begin{equation}\label{eq:eqs_two_comps}
\frac{\partial}{\partial t}\left[
\begin{array}{c}
\rho\\
\rho\mathbf{v}\\
\varepsilon\\
\rho c_{1}\\
\rho c_{2}
\end{array}
\right] + 
\nabla\cdot\left[
\begin{array}{c}
\rho\mathbf{v}\\
\rho\mathbf{v}\otimes\mathbf{v} + p\mathbf{I}\\
(\varepsilon + p) \mathbf{v}\\
\rho c_{1}\mathbf{v}\\
\rho c_{2}\mathbf{v}
\end{array}
\right] = 
\left[
\begin{array}{c}
0\\
\mathbf{0}\\
\rho\omega q_{0}\\
-\rho\omega\\
+\rho\omega
\end{array}
\right];
\end{equation}
where $c_{1}$ is the mass concentration of the reaction reagent, $c_{2}$ is the mass concentration of the reaction product, $\omega$ is the reaction rate, which is determined by the expression $\omega = k(T) \cdot c_{1}$.

Two main demonstration limiting cases were chosen: ``slow'' reaction kinetics with $\tau_{0} = 10^{-1}$ and ``fast'' reaction kinetics with $\tau_{0} = 4\cdot10^{-3}$, which correspond to weak and strong stiffness occurring in the system. This choice is based on the results presented in the works~\cite{frac_steps_detwave_sim_2000, ader_stiff_2, popov_j_sci_comp_2023}, and allows us to correctly determine the capabilities of the numerical method under study in a wide range of parameters of the reacting medium. It can be said that in the case of weak stiffness, the use of special methods for directly simulating a detonation wave is not required -- classical numerical methods of computational fluid dynamics, adapted for solving problems of weak stiffness, will allow one to obtain a correct numerical solution. In the case of strong stiffness, it is necessary to be quite strict in the choice of the numerical method. 

The resulting structure of a stationary detonation wave in the case of a ``fast'' reaction should have the classical form of ZND (Zel'dovich, von Neumann, and D\"{o}ring) detonation. A feature of the ZND detonation wave structure is the formation of a chemical Zel'dovich peak, which has the form of a moving extremely sharp change in the pressure and density of the medium, in the spatial region of which the complete burnout of the reagent and the release of the main energy of the reactions occur. The use of standard numerical methods of computational fluid dynamics~\cite{correct_det_wave_speed_2017} and methods based on the idea of splitting into physical processes~\cite{frac_steps_detwave_sim_2000, chem_kin_hrs_weno} leads to difficulty in resolving the chemical peak, and as a consequence, this inaccuracy in resolving the underlying structure of the detonation wave leads to simulation artifacts. In the work~\cite{ader_stiff_2}, it was possible to quite accurately resolve the structure of a one-dimensional ZND detonation wave using the finite-volume ADER-WENO method with LST-DG predictor. In the work~\cite{popov_j_sci_comp_2023}, it was possible to quite accurately resolve the structure of a one-dimensional ZND detonation wave using the space-time adaptive ADER-DG method with LST-DG predictor and a posteriori sub-cell ADER-WENO finite-volume limiting.

Below in this work the results of simulating the formation and propagation of detonation waves in plane, cylindrical and spherical cases are presented, as well as the results of simulating the interaction of detonation waves with inert inhomogeneities of the medium in which detonation propagates.

\subsection{LST-DG predictor for stiffness reactive flows}
\label{sec:detonation_waves:lst_dg_predictor}

The presence in the source terms of terms related to the kinetics of reactions in a multicomponent reacting medium leads to the emergence of high and anomalously strong stiffness. In this case, you can use adaptive time step correction or adaptive mesh refinement (AMR), an additional refinement criterion that will be the relative rate of reactions occurring in the reacting medium. In this work, a new approach was proposed based on modification of the predictor, which allows one to obtain a conditional local discrete space-time solution without using adaptive time step correction. The use of adaptive time step correction, or adaptive change in the time step, was used in the work~\cite{popov_j_sci_comp_2023} to simulate one-dimensional flows with detonation waves, while the expression for the adaptive factor was chosen in a form adapted from the expression proposed in the work~\cite{chem_kin_hrs_weno}. The main idea is to introduce an estimate for the characteristic time of reaction processes $\tau^{k}_{R}$ in a reactive flow based on the expression:
\begin{equation}
\left(\tau^{k}_{R}\right)^{-1} = \frac{[\mathbf{S}]_{k}}{[\mathbf{U}]_{k}},
\end{equation}
where $[\mathbf{U}]_{k}$ and $[\mathbf{S}]_{k}$ are the $k$-th components of the vectors of conserved variables $\mathbf{U}$ and source terms $\mathbf{S}$. In terms of dimension, this expression determines time, and is a certain evaluation expression that allows one to determine, with an accuracy of an order of magnitude, the relative characteristic time scale of processes associated with source terms. In the case of zero source terms $\mathbf{S} \equiv 0$, all values of the times $\tau^{k}_{R} = 0$. In the problems considered in this work, based on the use of a system of equations (\ref{eq:system_of_equations}), the source terms contain only components associated with reactions in a multicomponent reacting medium -- source terms $\mathbf{S}_{r}$ that determine the rates of reactions occurring in the medium, and source terms $\mathrm{S}_{e}$ that determine the release of energy from these reactions. Therefore, in this case, only the ratios of the total reaction rates $\mathbf{S}_{r}$ to the densities $\rho\mathbf{c}$ of individual components and the rate of energy release $\mathrm{S}_{e}$ to the energy density $\varepsilon$ are of interest. The adaptive change in the time step was based on determining the smallest time value $\tau^{min}_{R}$ that is responsible for the strongest stiff components of the solution, and using the adaptive factor $\alpha$, which is determined by the expression:
\begin{equation}\label{adaptive_factor}
\alpha\left(\frac{\tau^{n}_{DG}}{\tau^{min}_{R}}\right) = 
	\left(N_{R} - 1\right)\left[1 - \exp\left(-A \frac{\tau^{n}_{DG}}{\tau^{min}_{R}}\right)\right] + 1,
\end{equation}
where the parameter $N_{R}$ defines the limit value for reducing the time step; the parameter $A > 0$ is chosen for reasons of stability -- the higher the parameter, the faster the step decreases with an increase in the rate of processes associated with the source terms, such as reactions and their energy yield; typical value $A = 10^{1} - 10^{3}$ in work~\cite{chem_kin_hrs_weno} and $A = 1.0$ in work~\cite{popov_j_sci_comp_2023}. The main contribution to the creation of anomalously strong stiffness in flows with detonation waves usually comes from regions in a small vicinity of the detonation wave, where almost complete combustion of reactants occurs and the main part of the reaction energy is released, while in the rest of the flow region such strong local stiffness does not occur. Adaptive correction of the time step leads to a decrease in the time step $\Delta t^{n}$ in the entire computational domain $\Omega$, which, on the one hand, increases the computation time and computational costs, and on the other hand leads to an effective decrease in the value of the Courant number \texttt{CFL}, which can negatively affect the accuracy of the numerical solution in the area of its smoothness. Therefore, in this work, an approach similar to adaptive time step change was applied locally at the scale of individual cells in which rapidly occurring processes associated with source terms are detected.

The source terms associated with the reactions occurring in the reacting medium have only a local influence on the solution. Changes were made only to the process of obtaining a local discrete space-time solution $\mathbf{q}_{h}(\tau, \boldsymbol{\xi})$ -- the LST-DG predictor. The time step was divided into $[\alpha]$ time steps -- an integer part of the adaptive factor $\alpha$. At each local time step $s$, a local solution $\mathbf{q}^{l}_{h}(\tau, \boldsymbol{\xi})$ was calculated, where the initial condition at the first local step $s = 0$ was the solution $\mathbf{u}_{h}(\mathbf{r}, t^{n})$ at the previous time step $t^{n}$, and at subsequent local time steps $s > 0$, the initial condition was the local space-time solution from the previous local time step $\mathbf{q}^{s-1}_{h}(\tau, \boldsymbol{\xi})$. As a result, in each cell where the process of adaptive change in the time step was ``activated'' (the condition for the adaptive factor $\alpha > 1$), $[\alpha]$ local space-time solutions $\mathbf{q}^{s}_{h}(\tau, \boldsymbol{\xi})$, where $0 \leqslant s < [\alpha]$ were calculated. The final local space-time solution $\mathbf{q}_{h}(\tau, \boldsymbol{\xi})$, which was prepared for the output of the LST-DG predictor, was chosen in the form of approximation of solutions $\mathbf{q}^{s}_{h}(\tau, \boldsymbol{\xi})$ using the least squares method based on the functional basis $\{\Theta_{\mathbf{p}}(\tau, \boldsymbol{\xi})\}$ of the final solution. The formulaic apparatus of this local adaptive change in time step procedure is represented based on the representation coefficients of the local space-time solution in the basis $\{\Theta_{\mathbf{p}}(\tau, \boldsymbol{\xi})\}$:
\begin{equation}
\begin{split}
&\mathbf{q}_{h}(\tau, \boldsymbol{\xi}) = \sum_{\mathbf{p}} \hat{\mathbf{q}}_{p} \Theta_{\mathbf{p}}(\tau, \boldsymbol{\xi});\\
&\mathbf{q}^{s}_{h}(\tau^{s}, \boldsymbol{\xi}) = \sum_{\mathbf{p}} \hat{\mathbf{q}}^{s}_{p} \Theta_{\mathbf{p}}(\tau^{s}, \boldsymbol{\xi});\\
\end{split}
\end{equation}
where the argument $\tau\in[0, 1]$ is defined for the entire time step $[t^{n}, t^{n+1}]$ corresponding to the time step value $\Delta t^{n}$, and the argument $\tau^{s}\in[0, 1]$ is defined only for the $s$-th local time step corresponding to the time step value $\Delta t^{n}/[\alpha]$. The approximation was built on the basis of inverting the procedure for calculating the conserved $L_{2}$-projection of the solution $\mathbf{q}_{h}(\tau, \boldsymbol{\xi})$ onto the solution $\mathbf{q}^{s}_{h}(\tau^{s}, \boldsymbol{\xi})$ at the $s$-th local time step:
\begin{equation}\label{eq:q_lts_trans_gen}
\begin{split}
\int\limits_{\omega_{4}} d\tau^{s}d\boldsymbol{\xi} \cdot \Theta_{\mathbf{p}}&(\tau^{s}, \boldsymbol{\xi}) \mathbf{q}^{s}_{h}(\tau^{s}, \boldsymbol{\xi}) \\
&= \int\limits_{\omega_{4}} d\tau^{s}d\boldsymbol{\xi} \cdot \Theta_{\mathbf{p}}(\tau^{s}, \boldsymbol{\xi}) 
	\mathbf{q}_{h}\left(\frac{s + \tau^{s}}{[\alpha]}, \boldsymbol{\xi}\right);
\end{split}
\end{equation}
where the reference space-time element time $\tau$ on the scale of the full time step $\Delta t^{n}$ and the time $\tau^{s}$ on the scale of the local time step $\Delta t^{n}/[\alpha]$ were related by the relation $\tau = (s + \tau^{s})/[\alpha]$. The resulting expression has the form of systems of linear algebraic equations with respect to the coefficients of representation $\hat{\mathbf{q}}_{p}$, however, using the definition of basis functions $\{\Theta_{\mathbf{p}}(\tau, \boldsymbol{\xi})\}$ in the form of tensor products of basis functions $\varphi_{k}(\xi)$ and the property of orthogonality of the basis $\varphi_{k}(\xi)$, the expression immediately reduces to an explicit expression for the matrix of the linear projection operator. \newcoloringtext{%
The expression (\ref{eq:q_lts_trans_gen}) can be rewritten in the following matrix form:
\begin{equation}\label{eq:q_lts_trans_matrices}
\begin{split}
\sum_{\mathbf{q}} \Upsilon^{s}_{\mathbf{p}\mathbf{q}} \hat{\mathbf{q}}_{\mathbf{q}} =
\sum_{\mathbf{q}} \mathbb{M}_{\mathbf{p}\mathbf{q}} \hat{\mathbf{q}}^{s}_{\mathbf{q}},
\end{split}
\end{equation}
where $0 \leqslant s < [\alpha]$ is the number of the local time layer, $\mathbf{p} = (p_{0}, p_{1}, p_{2}, p_{3})$ and $\mathbf{q} = (q_{0}, q_{1}, q_{2}, q_{3})$ are classic multi-indexes: $0 \leqslant p_{0}, p_{1}, p_{2}, p_{3}, q_{0}, q_{1}, q_{2}, q_{3} < N+1$, and the matrices $\mathbb{M} = ||\mathbb{M}_{\mathbf{p}\mathbf{q}}||$ and $\Upsilon^{s} = ||\Upsilon^{s}_{\mathbf{p}\mathbf{q}}||$, taking into account the expression for the basis functions $\Theta_{\mathbf{p}}(\tau, \boldsymbol{\xi}) = \varphi_{p_{0}}(\tau)\varphi_{p_{1}}(\xi)\varphi_{p_{2}}(\eta)\varphi_{p_{3}}(\zeta)$ and the orthogonality property of the nodal basis functions $\varphi_{p}(\xi)$, can be represented in the form of Kronecker products:
\begin{equation}
\begin{split}
\mathbb{M}_{\mathbf{p}\mathbf{q}} = m \otimes m \otimes m \otimes m;\\
\Upsilon^{s}_{\mathbf{p}\mathbf{q}} = \tilde{\varsigma}^{s} \otimes m \otimes m \otimes m;\\
\end{split}
\end{equation}
where $m = \mathrm{diag}(m_{1}, \ldots, m_{N})$ is the matrix of masses of the nodal basis functions $\varphi_{p}(\xi)$, which has a diagonal form due to the $L_{2}$-orthogonality of the functions $\varphi_{p}(\xi)$, and matrix $\tilde{\varsigma}^{s} $ has the following form:
\begin{equation}
\begin{split}
\tilde{\varsigma}^{s}_{pq} = \int\limits_{0}^{1} d\tau^{s} \cdot \varphi_{p}(\tau^{s})\varphi\left(\frac{s + \tau^{s}}{[\alpha]}\right),
\end{split}
\end{equation}
then the equation (\ref{eq:q_lts_trans_matrices}) can be represented in the following form:
\begin{equation}
\begin{split}\label{eq:q_lts_trans_src_slae}
\sum_{\mathbf{q}} \Psi^{s}_{\mathbf{p}\mathbf{q}} \hat{\mathbf{q}}_{\mathbf{q}} = \hat{\mathbf{q}}^{s}_{\mathbf{p}},
\end{split}
\end{equation}
where matrix $\Psi^{s} = ||\Psi^{s}_{\mathbf{p}\mathbf{q}}||$ is defined by the following expression
\begin{equation}
\begin{split}
\Psi^{s} = \varsigma^{s} \otimes I \otimes I \otimes I,
\end{split}
\end{equation}
where matrix $\varsigma^{s} = m^{-1}\tilde{\varsigma}^{s}$, and $I$ is the $(N+1)\times(N+1)$ identity matrix. As a result of introducing an additional multi-index $\mathtt{r} = (s, \mathbf{p})$, which can be represented in continuous form $\mathtt{r} = s \cdot (N+1)^{4} + \mathbf{p}$, where the continuous form for the multi-index $\mathbf{p}$ can be represented by an expression $\mathbf{p} = p_{0}\cdot(N+1)^{3} + p_{1}\cdot(N+1)^{2}+p_{2}\cdot(N+1)+p_{3}$, the expressions for the matrix $\Psi^{s}_{\mathbf{p}\mathbf{q}}$ and ``vector'' $\hat{\mathbf{q}}^{s}_{\mathbf{p}}$ can be represented in a convenient two-index and single-index notations: $\Psi^{s}_{\mathbf{p}\mathbf{q}} \mapsto \Psi_{\mathtt{r}\mathbf{q}}$, $\hat{\mathbf{q}}^{s}_{\mathbf{p}} \mapsto \hat{\mathbf{q}}_{\mathtt{r}}$, and as a result, the expression (\ref{eq:q_lts_trans_src_slae}) becomes the classical form of overdetermined system of linear algebraic equations with respect to $\hat{\mathbf{q}}_{\mathbf{q}}$.} The inverse transformation obtained using a pseudo-inverse matrix determines the procedure for obtaining the final solution $\mathbf{q}_{h}(\tau, \boldsymbol{\xi})$ from solutions at local steps $\mathbf{q}^{s}_{h}(\tau^{s}, \boldsymbol{\xi})$ in the form of approximating solutions using the least squares method based on the basis of the final solution.
From an implementation perspective, this local adaptive change in time step procedure reduces to ordinary matrix-matrix operations.

In this work, the local adaptive change in time step procedure was implemented directly -- the time step was divided into $[\alpha]$ local time steps. The value $[\alpha]$ is limited from above by the value $N_{R}$. In the calculations that were carried out in this work, in the case of problems with strong stiffness, the value $N_{s} = 10$ was used, while the value $A = 1.0$ was chosen. Local adaptive change in time step of the time step was called in no more than $2$-$4$ cells of the spatial mesh in one time step in one-dimensional problems. In the case of problems with weak stiffness, the procedure used the value of $N_{s} = 4$, while the value of $A = 1.0$ was selected, however, as the calculations showed, the time step was called in no more than $1$-$2$ cells of the spatial mesh per time step in one-dimensional problems. The deactivation off the local adaptive change in time step procedure led to an increase in the number of cell troubles, an increase in the width of the detonation waves fronts and, in some cases, the appearance of $\texttt{nan}$ in the numerical solution in the case of problems with strong stiffness. The activation of the local adaptive change in time step procedure led to results that were significantly better compared to the results of the work~\cite{popov_j_sci_comp_2023}. A detailed description of the results obtained related to the simulation of detonation waves is presented later in the text in this work.

The proposed local adaptive change in time step procedure is conceptually similar to the procedure for coarsening cells within the framework of adaptive mesh refinement (AMR)~\cite{ader_dg_ideal_flows, ader_dg_diss_flows}. With AMR, calculations can be performed on a more fined mesh, and then cells at a higher refinement level can be coarsened into fewer cells at a lower refinement level. Mathematically, the implementation of cell coarsening within the framework of AMR is usually carried out by the operator inverse to the mesh refinement operator~\cite{ader_dg_diss_flows, exahype}, which also represents multiplication by a pseudo-inverse matrix. As part of the local adaptive change in time step procedure, a similar coarsening of the numerical solution is performed, only not by the spatial mesh, but in time $t$. As part of AMR, local time stepping (LTS)~\cite{ader_dg_diss_flows, exahype} is also usually used; however, this procedure is in no way equivalent to the proposed local adaptive change in time step procedure.

It is necessary to note some important features associated with the proposed local adaptive change in the time step procedure. The use of this procedure is not necessary to solve problems of flows of multicomponent media with detonation waves. Problem simulation results can be obtained by decreasing and selecting in sufficient detail a small value of the Courant number \texttt{CFL}. In case of problems with weak stiffness, we can limit ourselves to the values of the Courant number $\mathtt{CFL} \leqslant 0.1$. Calculations with such a value of the Courant number will be accompanied by a decrease in the accuracy of the numerical solution and a significant increase in the number of troubled cells in the spatial mesh. A similar phenomenon is observed in work~\cite{popov_j_sci_comp_2023}. Perhaps a similar phenomenon was observed in the work~\cite{chem_kin_hrs_weno}, but the use of the method of splitting into physical processes already had a significant impact on the results obtained using the high-precision WENO scheme. In case of problems with strong stiffness, it is necessary to use an even smaller value of the Courant number $\mathtt{CFL} \leqslant 0.01$-$0.05$, however, this does not guarantee obtaining an admissive numerical solution -- it is necessary to carry out very ``fine'' tuning of the parameters of the numerical method. It should be noted that a decrease in the value of the Courant number immediately leads to a multiple increase in computation time and computational costs -- more than $10$ times in the case of problems with weak stiffness and more than $20$-$100$ times in the case of problems with strong stiffness. It should also be noted that effective decreasing in the Courant number, as well as the use of adaptive correction in the time step throughout the entire spatial domain, does not allow obtaining a high-accuracy solution for the ZND-detonation wave in the case of a ``fast'' reaction in a medium on coarse meshes -- meshes with $200$-$400$ cells~\cite{popov_j_sci_comp_2023}, while in this work the solution was obtained in the case of $100$ cells. In the case of using the procedure of local adaptive change in the time step, calculations are carried out without decreasing the time step $\Delta t^{n}$ in the entire computational domain $\Omega$. All simulation results presented below for flow problems with detonation waves were obtained with a Courant number $\mathtt{CFL} = 0.9$. Therefore, it was concluded that the local adaptive change in the time step procedure is not necessary to obtain simulated results in principle; however, its use significantly decrease computational costs, allows the use of coarse meshes, decrease the number of troubled cell and increases the accuracy of the numerical solution, in the sense that it allows extending the empirical stability limit significantly closer to $\mathtt{CFL} = 1.0$.

It is also necessary to note an important detail from the perspective of algorithmic and software implementation. In this work, the local adaptive change in the time step procedure was implemented directly. The matrix elements were pre-computed to values of $[\alpha] \leqslant N_{R}$ and $N_{R} \leqslant 12$ for all polynomial degrees $N$ used. The calculations were implemented by simple matrix-matrix multiplication. However, it would be possible to achieve an increase in the computational efficiency of the procedure -- instead of coarsening the solution in time as a result of direct multiplication by a matrix, a sequence of coarsening the solution can be organized; for example, instead of coarsening $8$ local time steps, carry out $3$ sequential coarsening stages of $2$ time steps each. It is clear that in the case of large values of the adaptive factor $[\alpha]$, this approach can lead to a decrease in computational costs if the local adaptive change in the time step procedure turns out to be a significant bottleneck in the software implementation. However, in the case of using the local adaptive change in the time step procedure, the first priority may be the question of the load balancing multithreaded and multiprocess execution -- with a uniform load of threads at the stage of calculating a local space-time solution $\mathbf{q}_{h}$, the procedure can be a bottleneck simply because it is called only for a small fraction of mesh cells and requires at least $[\alpha]$ times more computational costs than simply calling the LST-DG predictor once at a time step. Load balancing taking this feature into account can remove the procedure from the list of significant bottlenecks. However, in this work the procedure was implemented directly.

\subsection{Plane ZND-detonation waves}
\label{sec:detonation_waves:cjdw_1d}

\paragraph{Formulation of the problem}
This subsection considers the one-dimensional problem of the formation and propagation of a plane detonation wave. The computational coordinate domain was chosen in the form of a range $\Omega = [0, 1]$. The initial conditions were chosen in the form of the Chapman-Jouguet (CJ) conditions, which define a stationary detonation wave in the instantaneous detonation approximation, which were calculated in the work~\cite{frac_steps_detwave_sim_2000}:
\begin{equation}\label{eq:cjdw_1d_init}
\begin{split}
&\rho(x, t = 0) = \left\{
\begin{array}{ll}
1.4, & \mathrm{if}\ x \leqslant 0.5; \\
0.887565, & \mathrm{if}\ x >\, 0.5; \\
\end{array}
\right.\\
&u(x, t = 0) = \left\{
\begin{array}{ll}
0.0, & \mathrm{if}\ x \leqslant 0.5; \\
-0.577350, & \mathrm{if}\ x >\, 0.5; \\
\end{array}
\right.\\
&p(x, t = 0) = \left\{
\begin{array}{ll}
1.0, & \mathrm{if}\ x \leqslant 0.5; \\
0.191709, & \mathrm{if}\ x >\, 0.5; \\
\end{array}
\right.\\
&c_{1}(x, t = 0) = \left\{
\begin{array}{ll}
1.0, & \mathrm{if}\ x \leqslant 0.5; \\
10^{-14}, & \mathrm{if}\ x >\, 0.5; \\
\end{array}
\right.\\
&c_{2}(x, t = 0) = \left\{
\begin{array}{ll}
10^{-14}, & \mathrm{if}\ x \leqslant 0.5; \\
1.0, & \mathrm{if}\ x >\, 0.5; \\
\end{array}
\right.
\end{split}
\end{equation}
where it is assumed that unburned gas flows onto the burnt gas, which leads to the formation of a detonation wave propagating through the unburnt gas to the right. A small value $10^{-14}$ of mass concentrations $c_{1}$ and $c_{2}$, instead of strictly $0$, was chosen to prevent the occurrence of negative concentrations immediately at the start of the calculation process, which could lead to a meaninglessly large increase in the number of troubled cells in the solution. From the point of view of energy release and flow energy balance, these values of reagent concentration do not have any significant effect.

The approximate speed of the detonation wave $D = 1$ for these initial conditions in the case of a ``fast'' reaction, when the classical ZND detonation structure is formed, taking into account the speed of the hydrodynamic flow~\cite{frac_steps_detwave_sim_2000}; the exact value $D = 1$ occurs in the case of instantaneous detonation.

The reference solution in cases of “slow” and “fast” reactions was obtained using a finite-volume ADER-WENO2 method with LST-DG predictor on a spatial mesh of finite-volume $5000$ cells. The work~\cite{ader_stiff_2} showed that the finite-volume ADER-WENO method allows one to obtain a correct numerical solution without using additional procedures for recalculating the solution, so this method was chosen as the method for obtaining a reference solution.

The boundary conditions were specified in the form of free outflow conditions. The cases of “slow” and “fast” reactions, corresponding to the cases of weak and strong stiffness of the system of equations, are considered. The final time has been chosen $t_{\rm final} = 0.4$. The Courant number was chosen $\mathtt{CFL} = 0.9$.

\begin{figure*}[h!]
\centering
\includegraphics[width=0.245\textwidth]{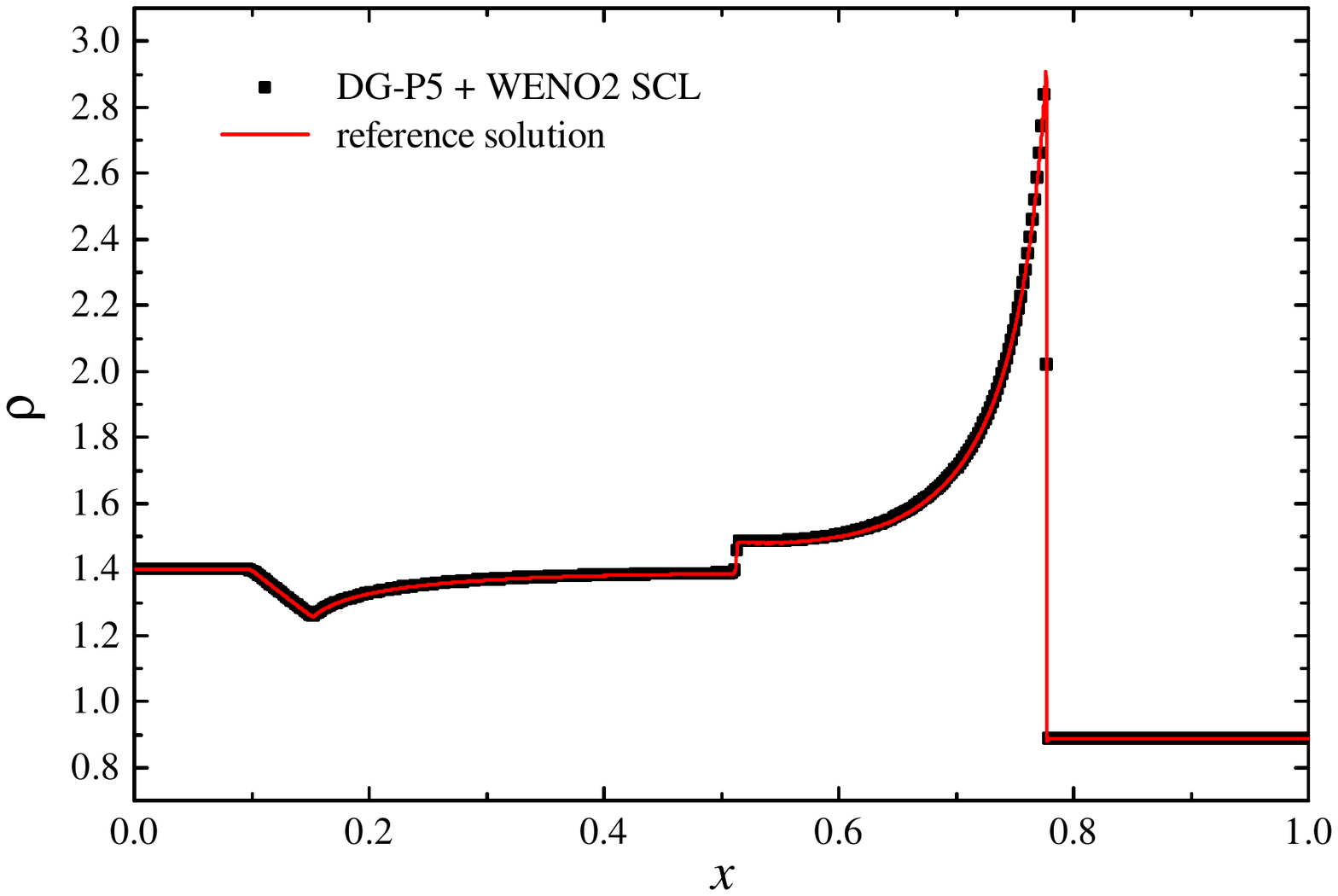}
\includegraphics[width=0.245\textwidth]{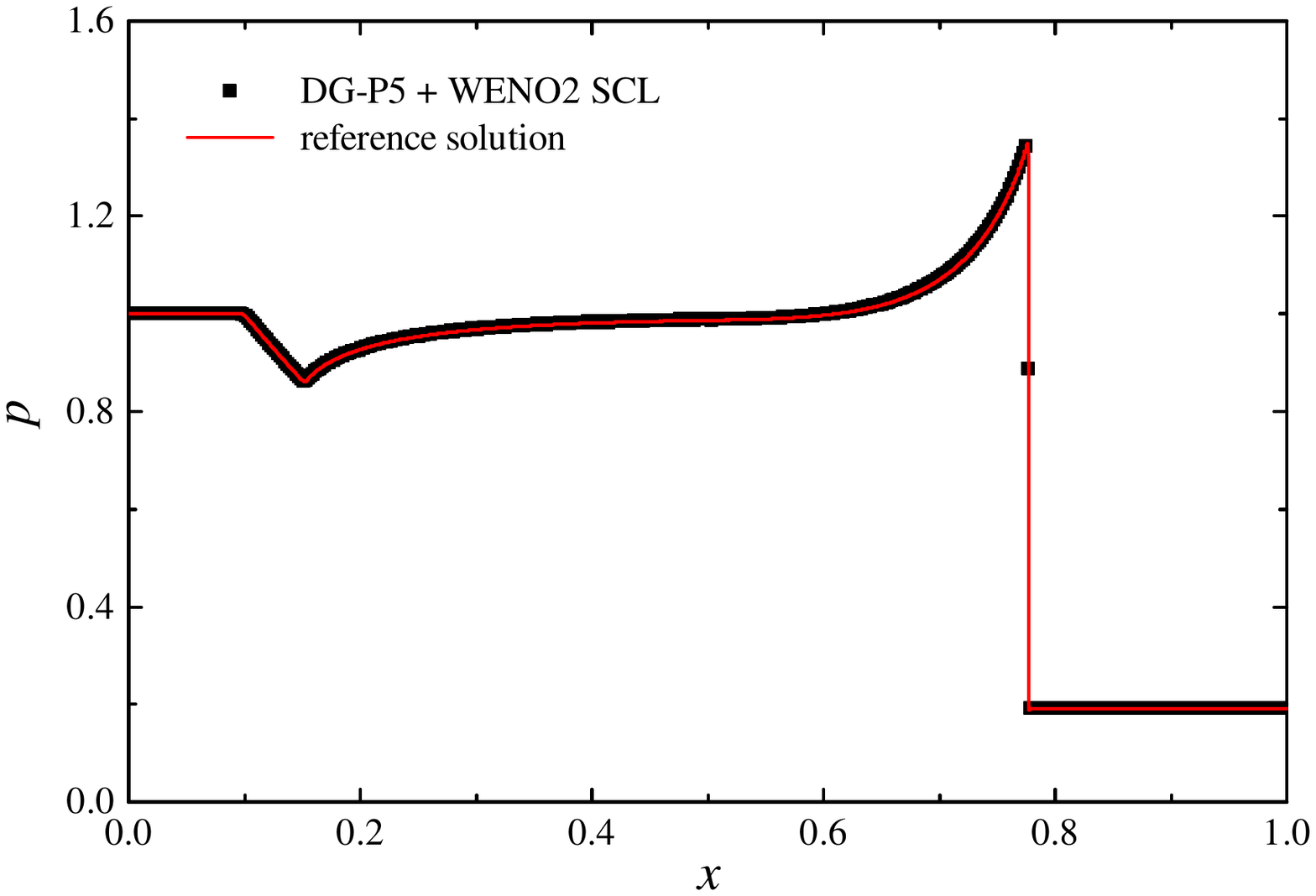}
\includegraphics[width=0.245\textwidth]{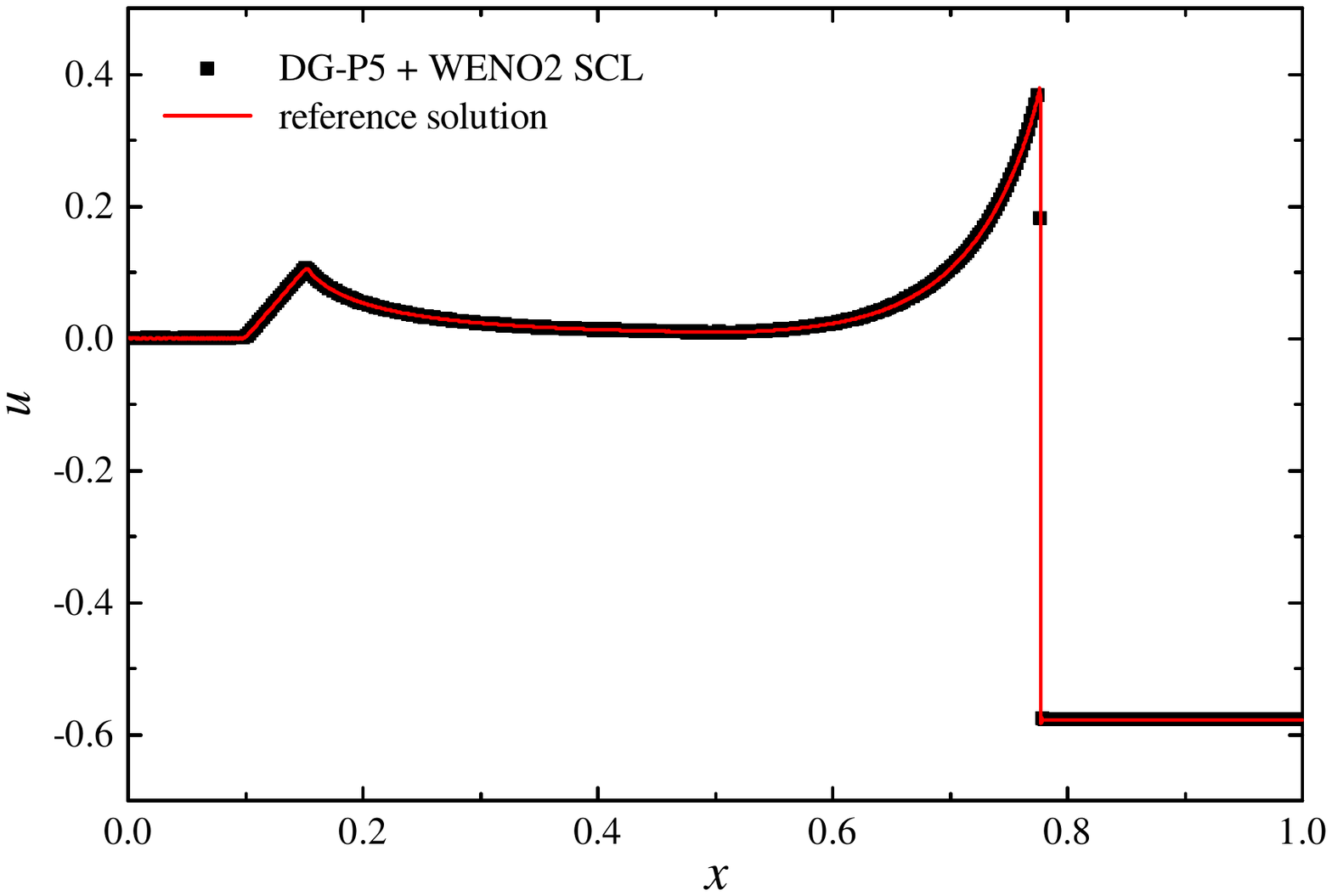}
\includegraphics[width=0.245\textwidth]{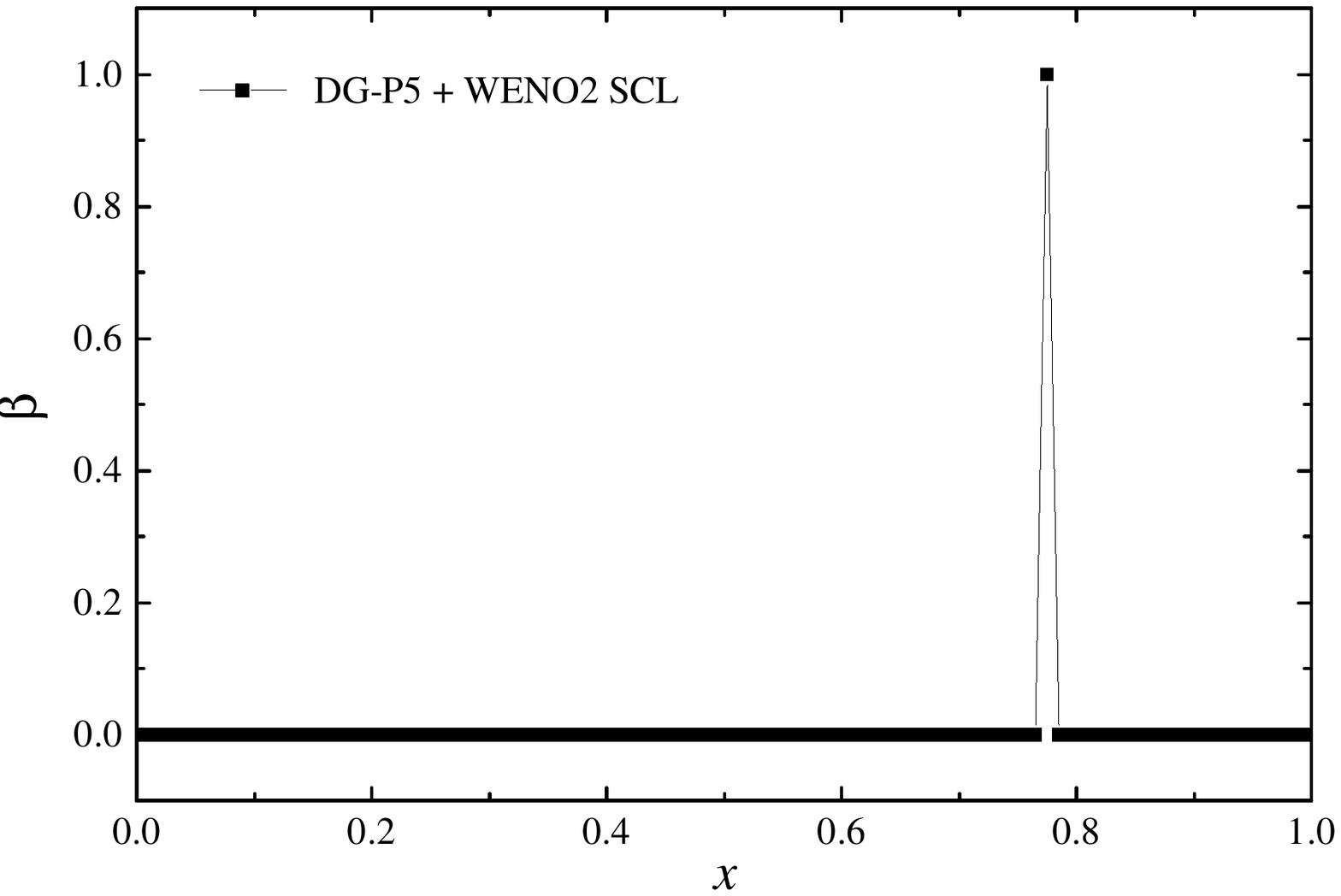}\\
\includegraphics[width=0.245\textwidth]{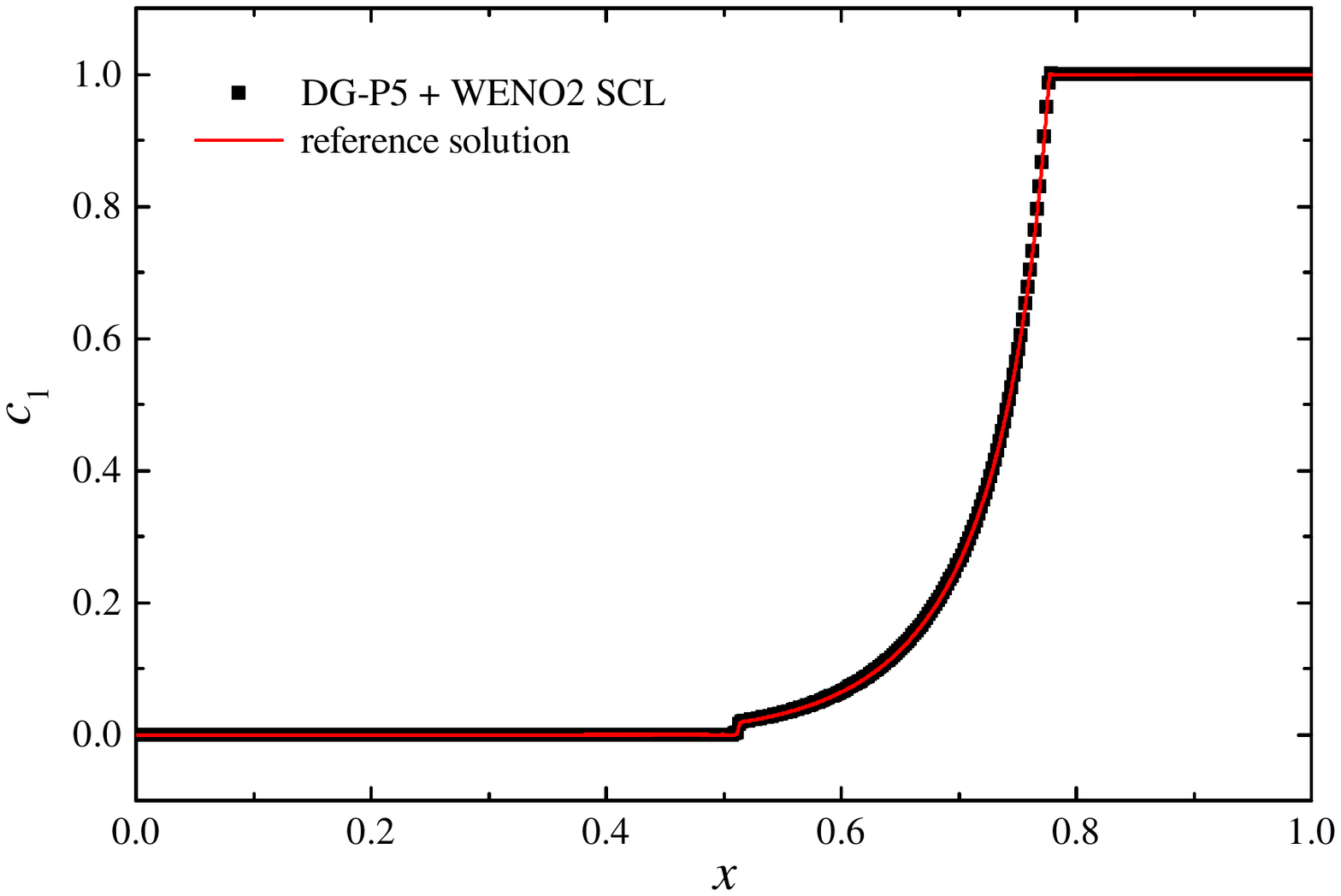}
\includegraphics[width=0.245\textwidth]{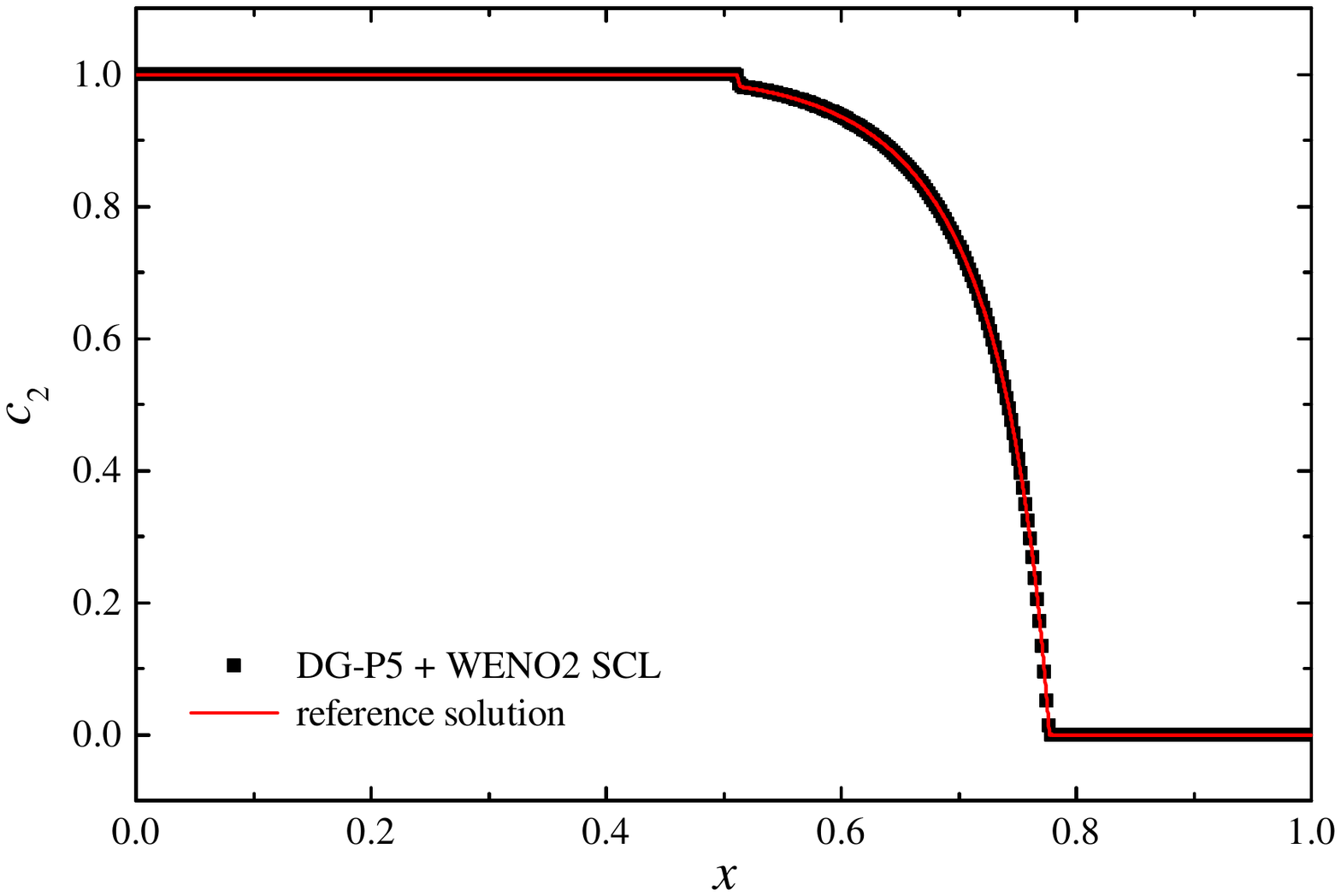}
\includegraphics[width=0.245\textwidth]{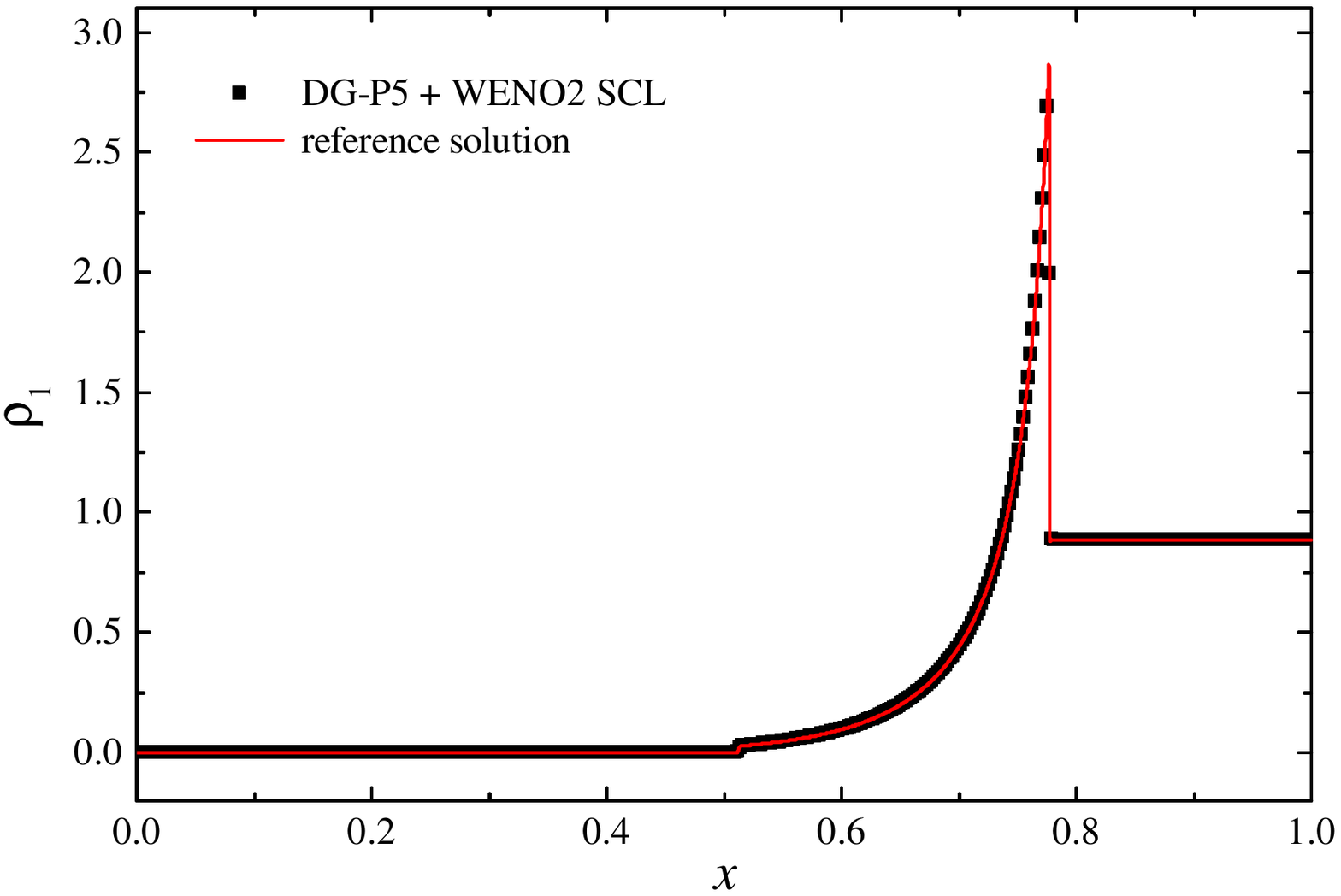}
\includegraphics[width=0.245\textwidth]{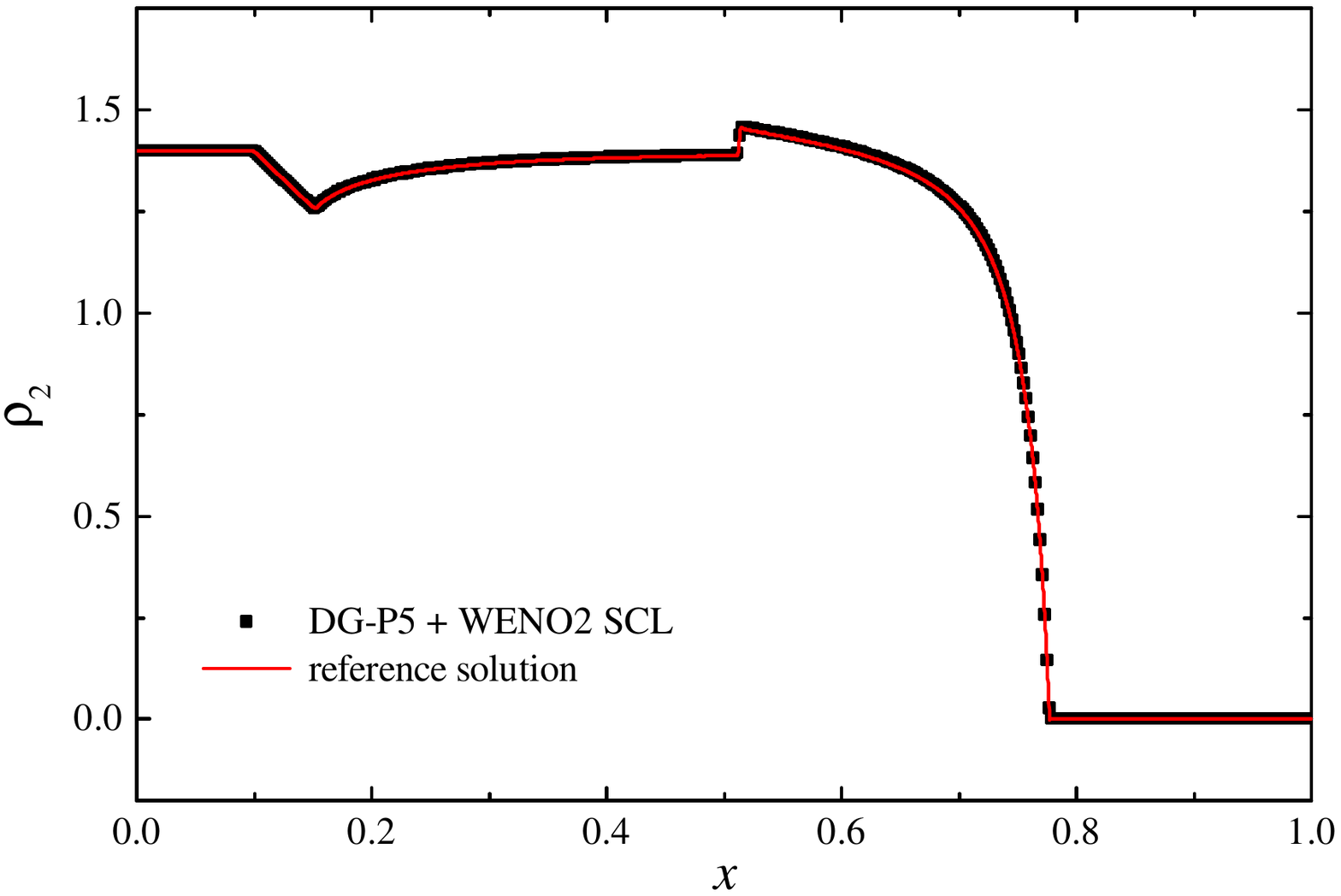}
\caption{\label{fig:cjdws_1d_dg_5}
Numerical solution of the one-dimensional problem of the plane detonation wave formation in a two-component medium with a ``slow'' reaction
(weak stiff case, a detailed statement of the problem is presented in the text), obtained using the ADER-DG-$\mathbb{P}_{5}$ method 
on mesh with $100$ cells at the final time $t_{\rm final} = 0.4$.
The graphs show the coordinate dependence of density $\rho$, pressure $p$, flow velocity $u$,
troubled cells indicator $\beta$, mass concentrations $c_{k}$ and partial densities $\rho_{k} = \rho c_{k}$ 
of the $1$st and $2$nd individual components of the reacting medium.
The black square symbols represent the subcells finite-volume representation of the numerical solution; 
the red solid lines represents the reference solution of the problem.
}
\end{figure*}

\begin{figure*}[h!]
\centering
\includegraphics[width=0.245\textwidth]{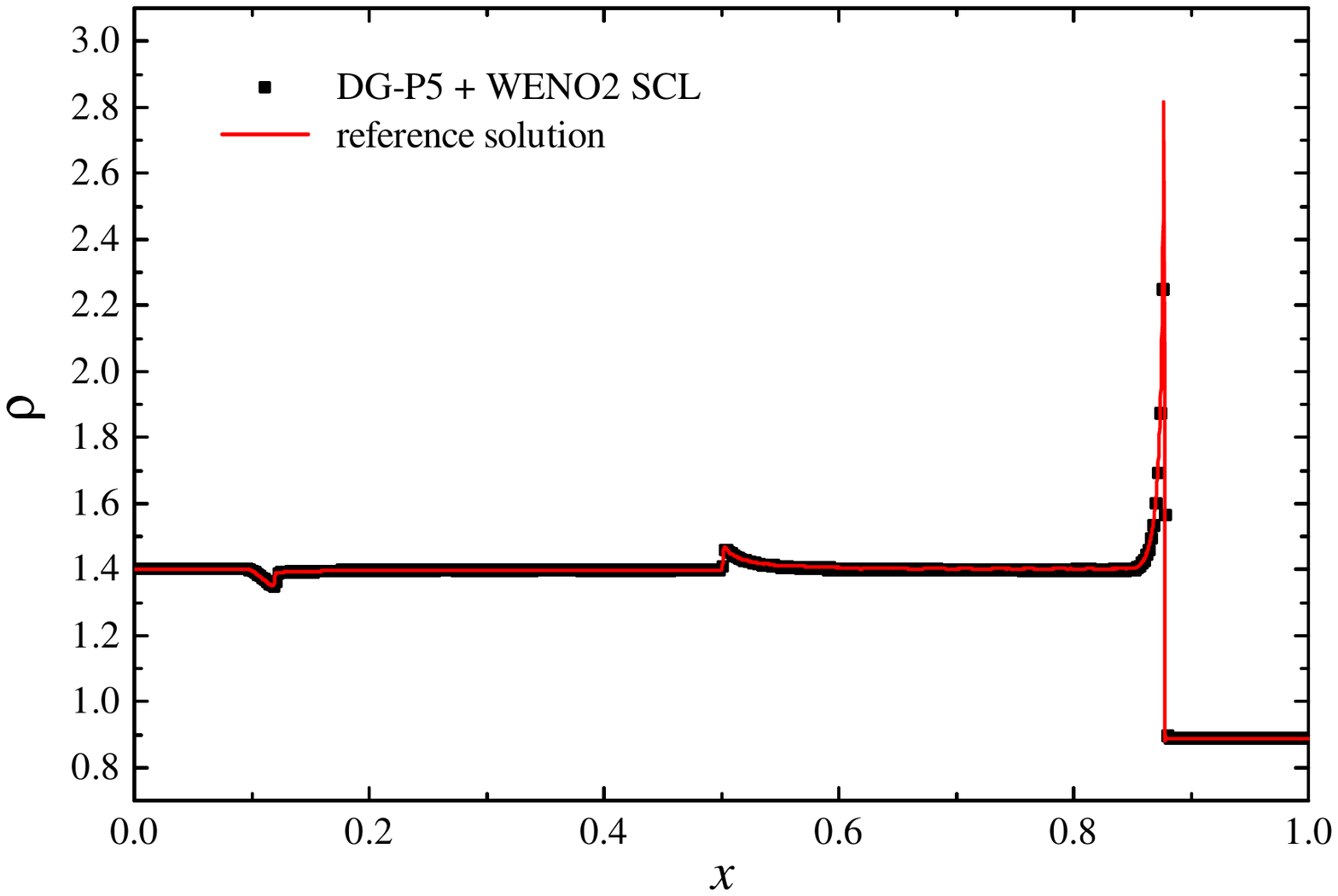}
\includegraphics[width=0.245\textwidth]{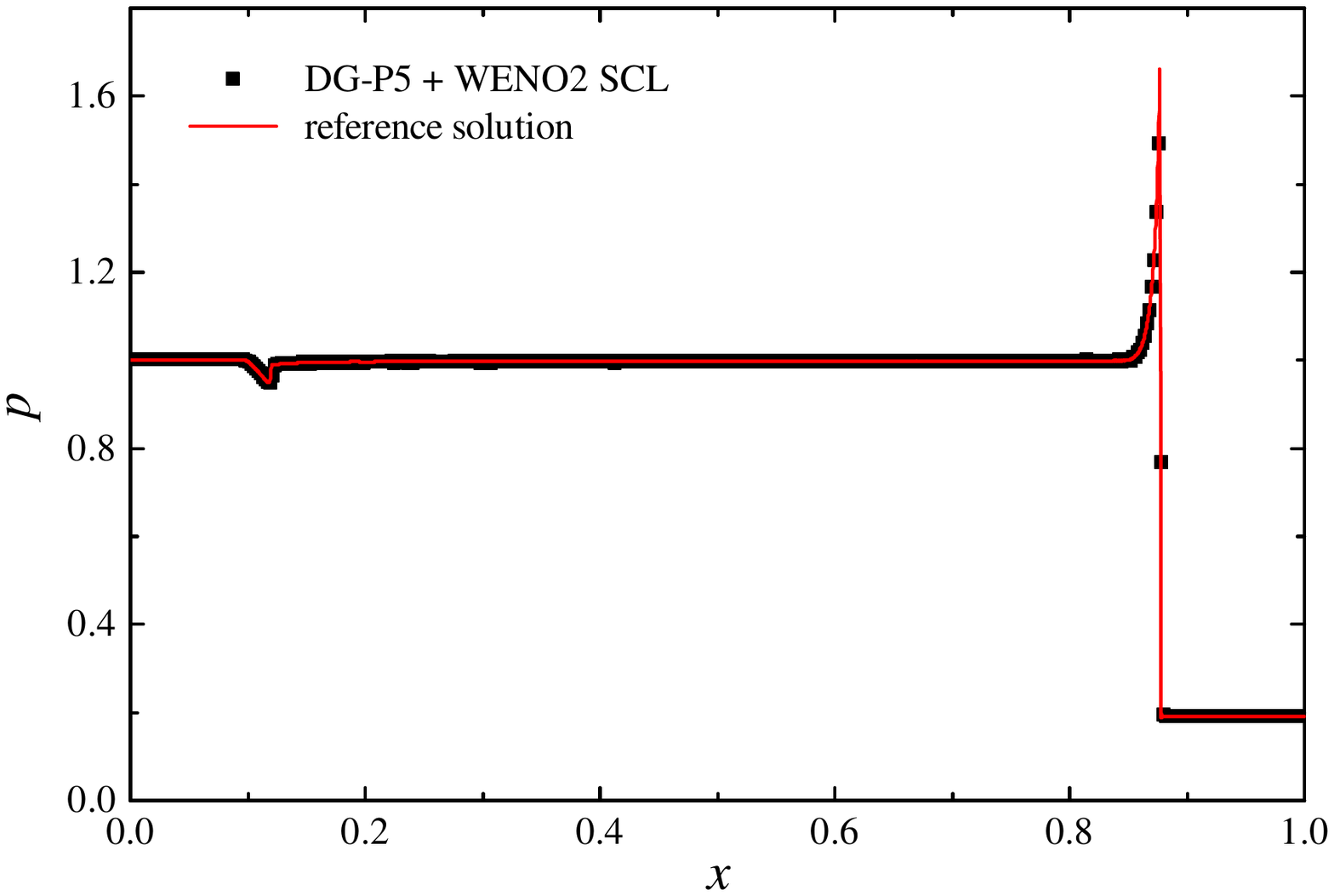}
\includegraphics[width=0.245\textwidth]{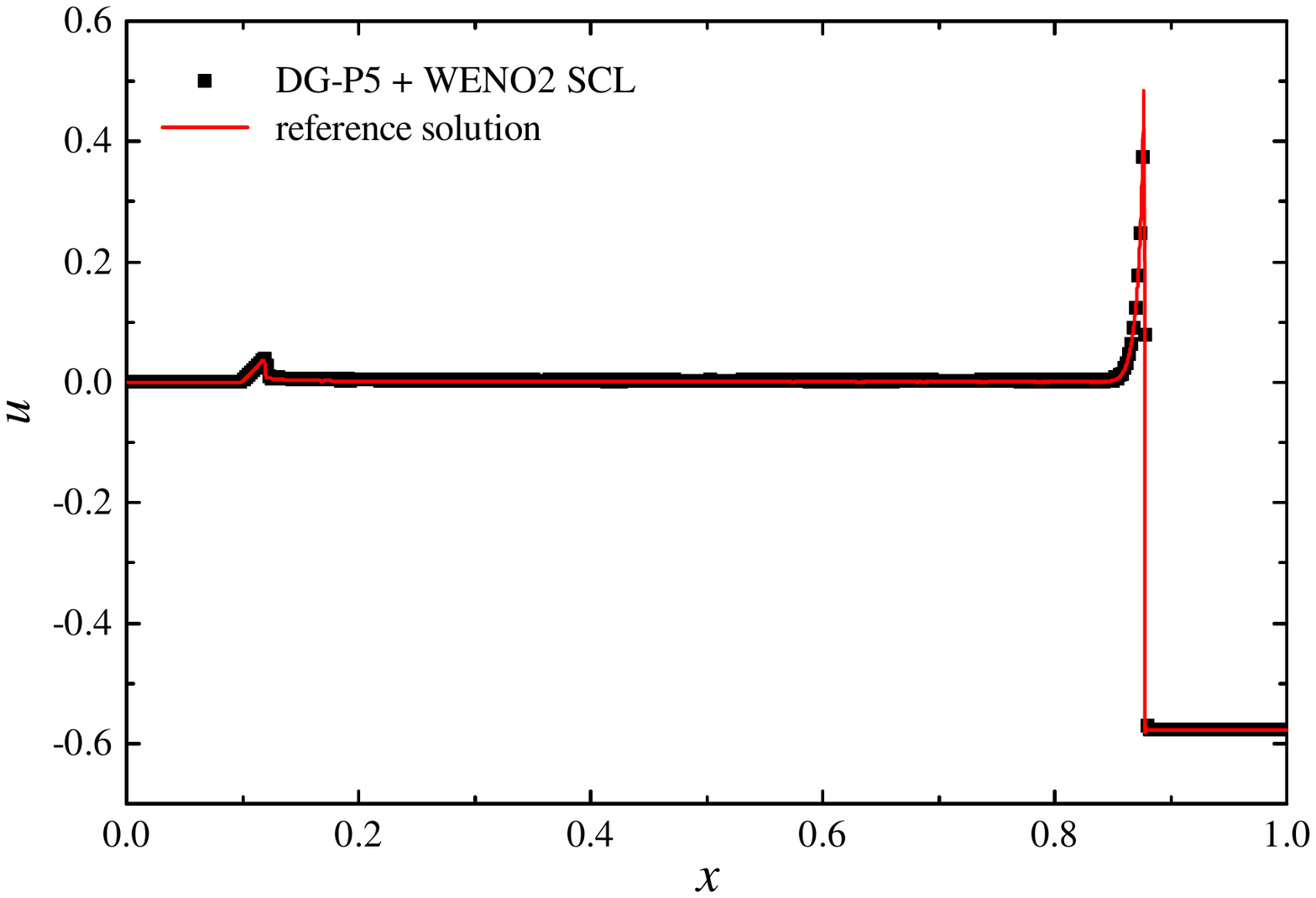}
\includegraphics[width=0.245\textwidth]{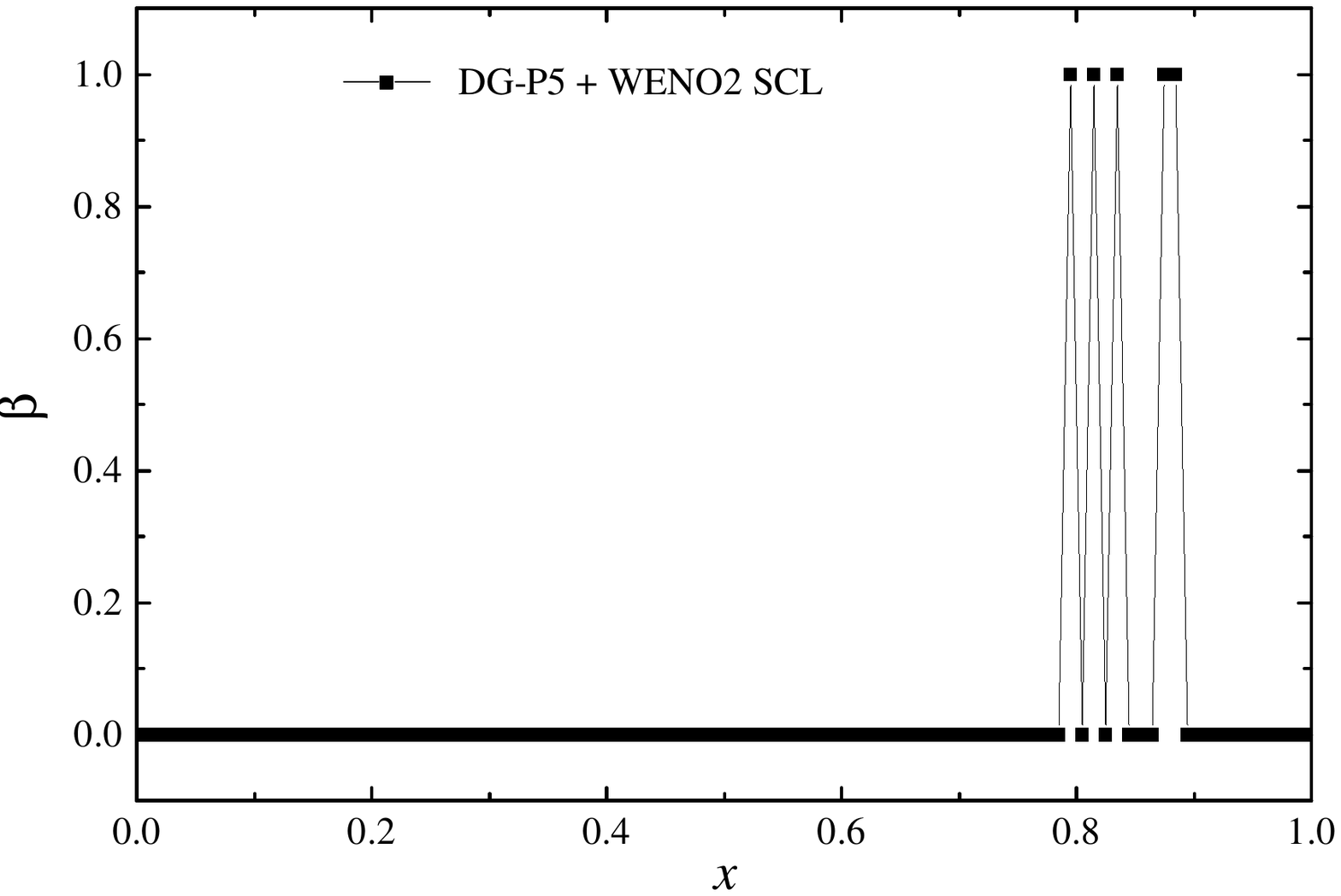}\\
\includegraphics[width=0.245\textwidth]{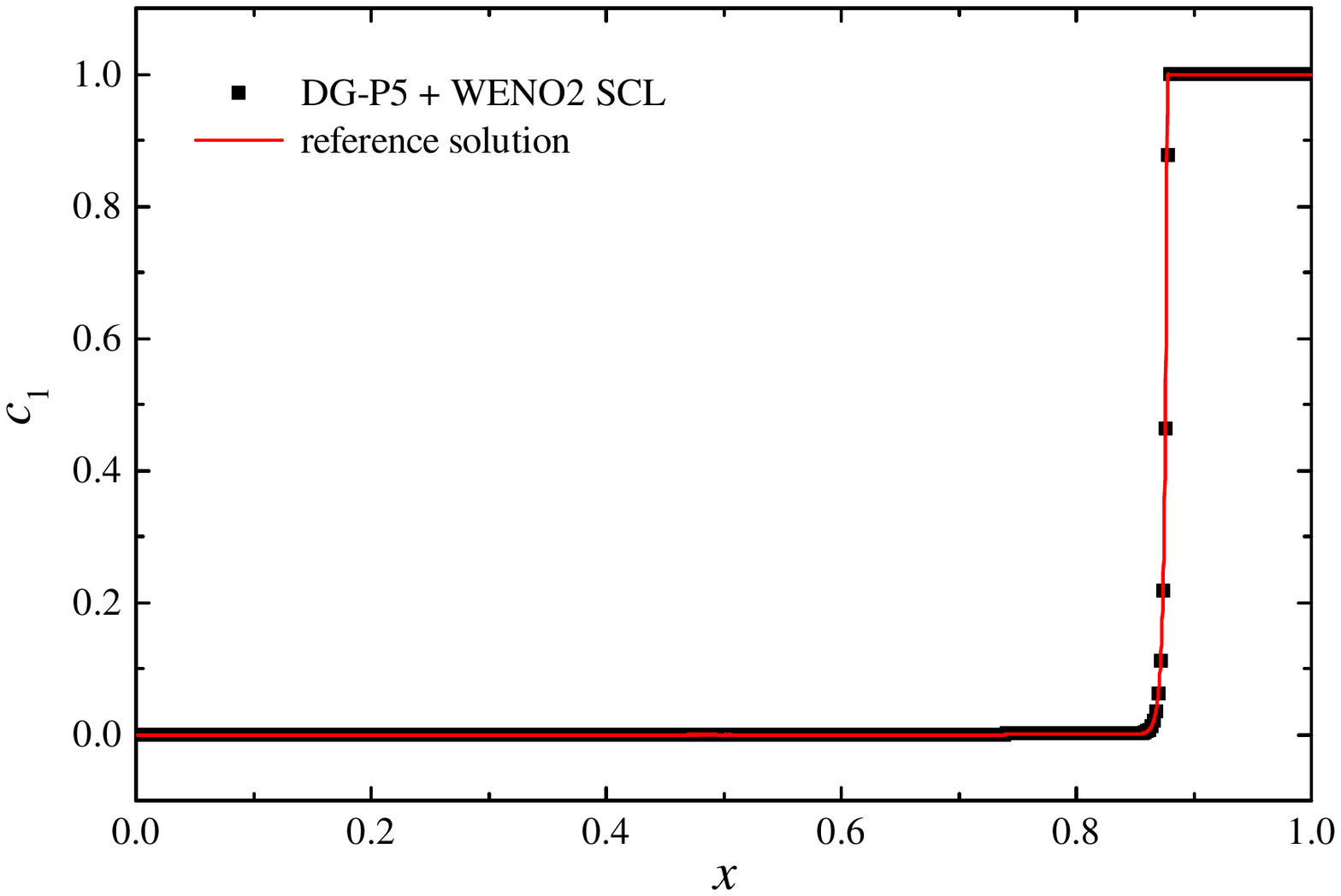}
\includegraphics[width=0.245\textwidth]{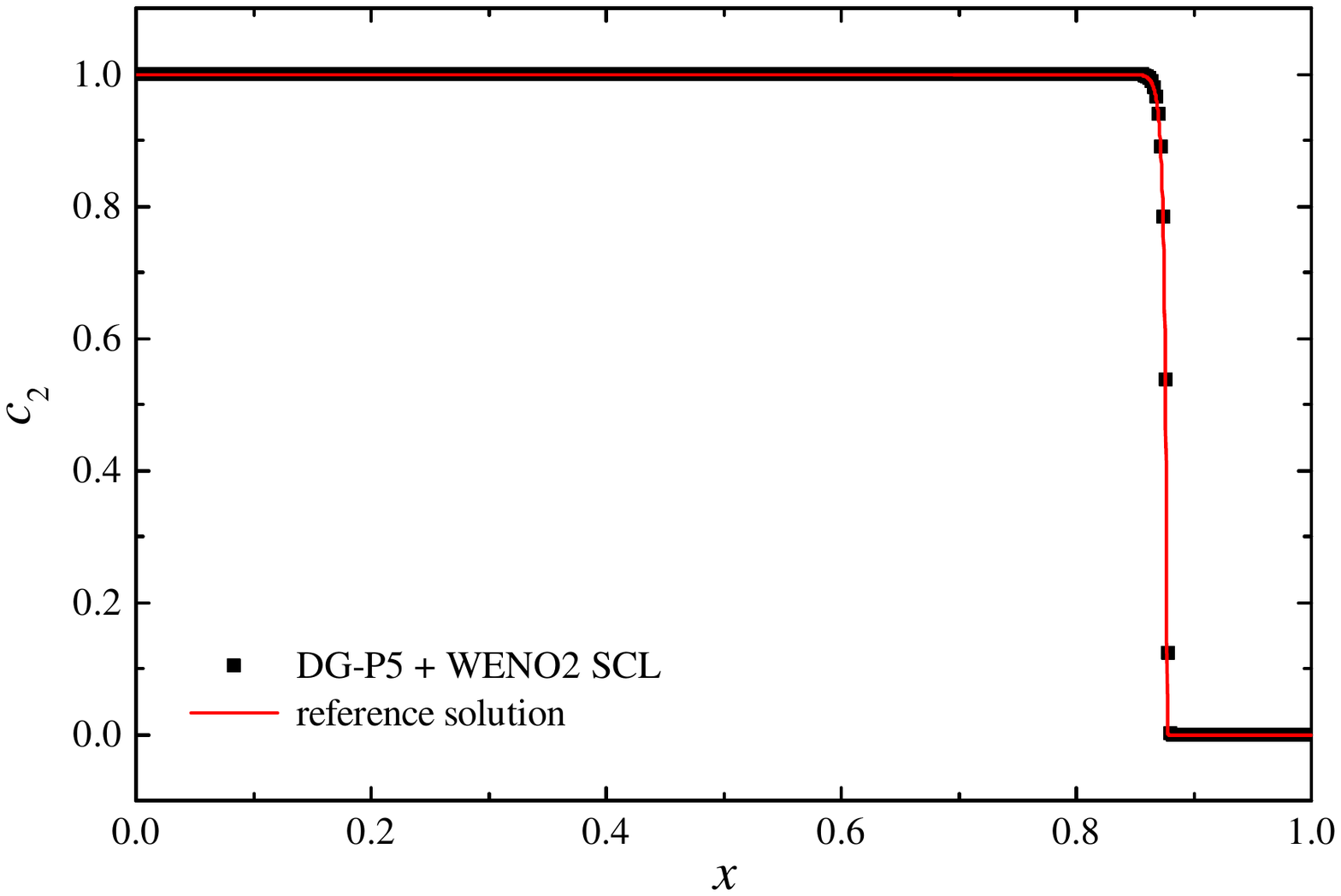}
\includegraphics[width=0.245\textwidth]{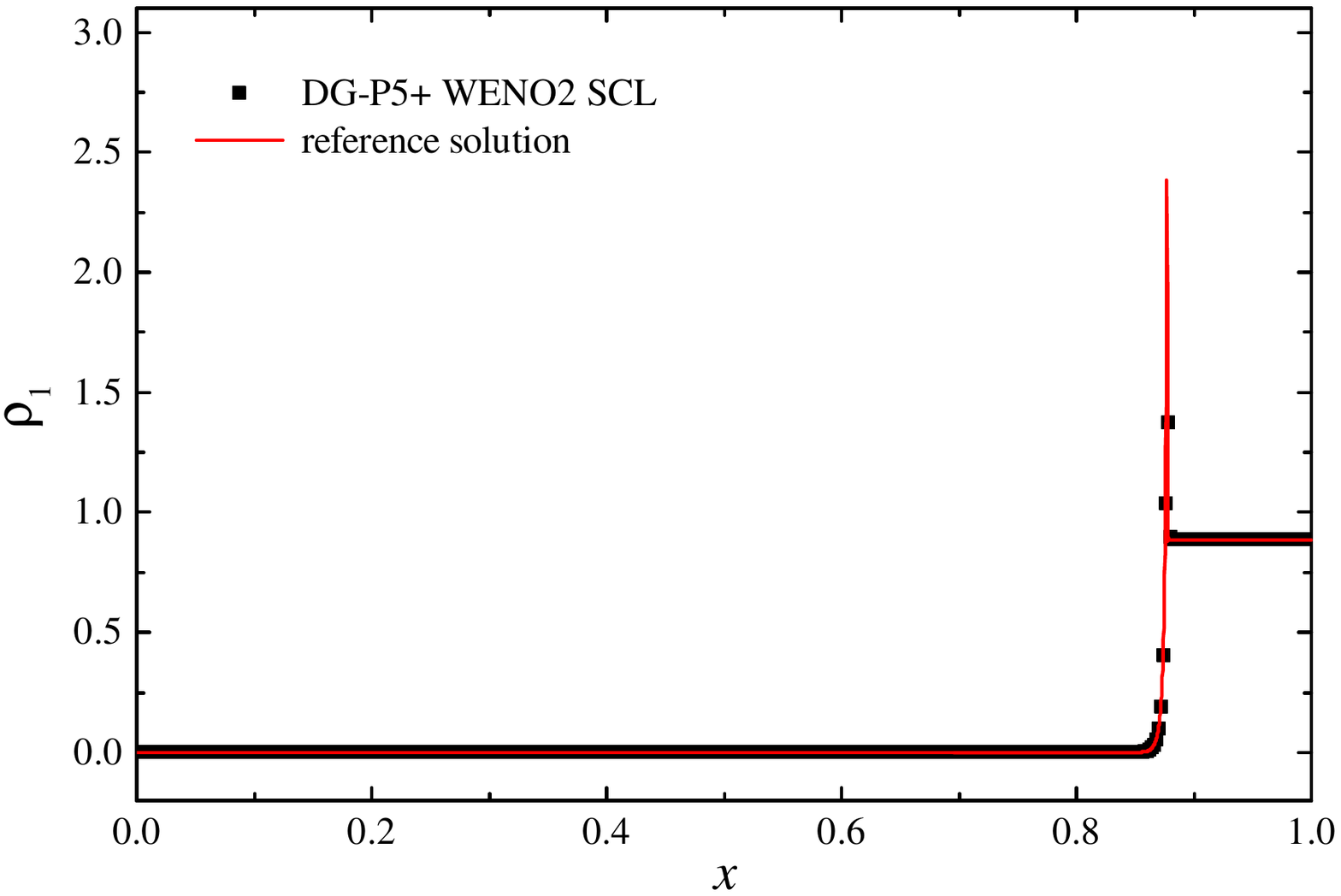}
\includegraphics[width=0.245\textwidth]{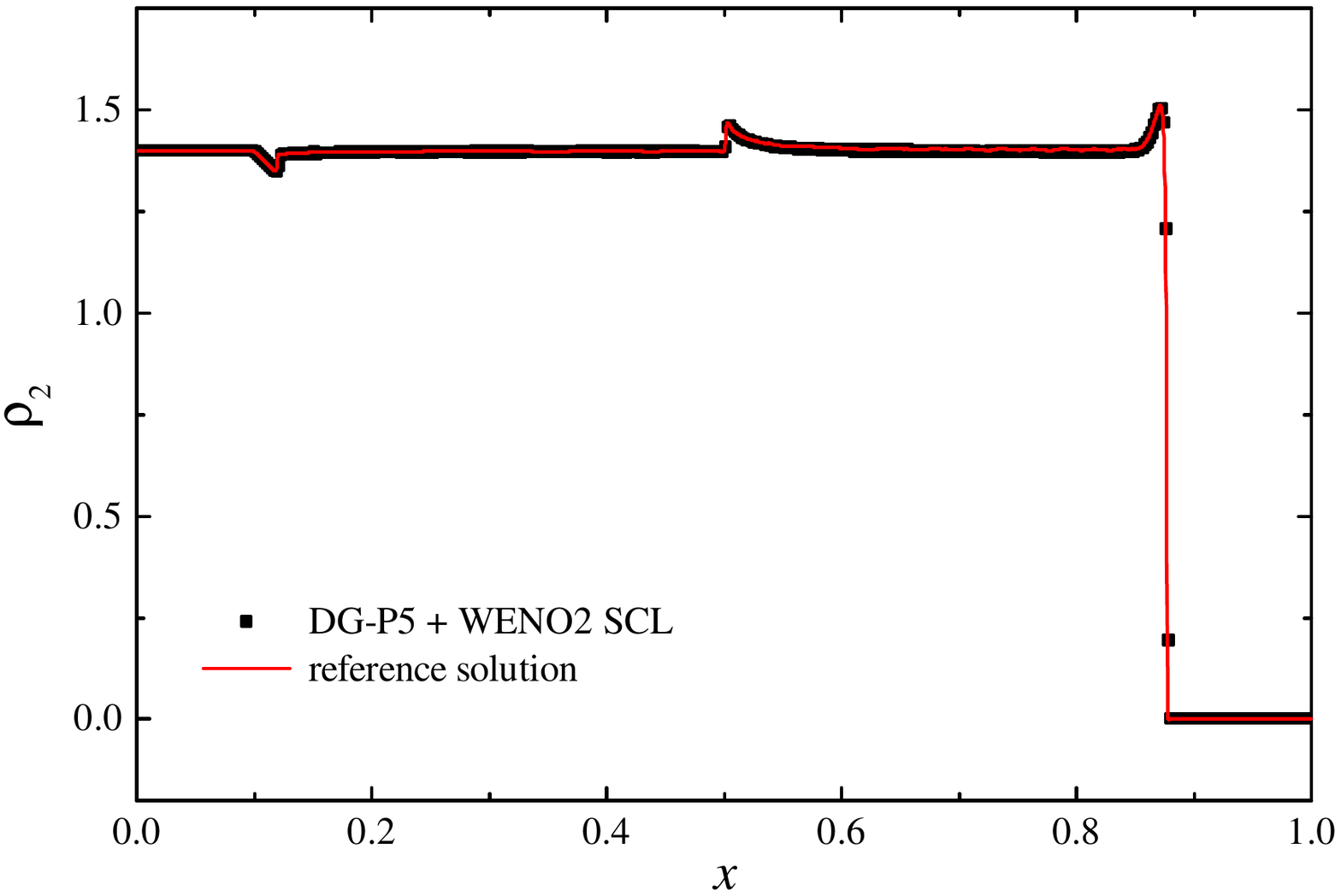}
\caption{\label{fig:cjdwf_1d_dg_5}
Numerical solution of the one-dimensional problem of the plane detonation wave formation in a two-component medium with a ``fast'' reaction
(strong stiff case, a detailed statement of the problem is presented in the text), obtained using the ADER-DG-$\mathbb{P}_{5}$ method
on mesh with $100$ cells  at the final time $t_{\rm final} = 0.4$.
The graphs show the coordinate dependence of density $\rho$, pressure $p$, flow velocity $u$,
troubled cells indicator $\beta$, mass concentrations $c_{k}$ and partial densities $\rho_{k} = \rho c_{k}$ 
of the $1$st and $2$nd individual components of the reacting medium.
The black square symbols represent the subcells finite-volume representation of the numerical solution; 
the red solid lines represents the reference solution of the problem.
}
\end{figure*}

\paragraph{Weak stiff case}
The results of simulating the propagation of a detonation wave in a flow in the case of weak stiffness are presented in Figure~\ref{fig:cjdws_1d_dg_5}. The main feature of the numerical solution obtained in the case of a ``slow'' reaction using the ADER-DG-$\mathbb{P}_{5}$ method with ADER-WENO2 finite volume a posteriori limiter demonstrates unsurpassed accuracy in resolving discontinuous components of the solution -- the detonation front has a width of only $1$-$2$ \textit{subcells}, while in the work~\cite{popov_j_sci_comp_2023} where a similar numerical method was used using adaptive time step correction, the width of the front of the detonation wave was one mesh \textit{cell} -- in the case of degree $N = 5$, one cell is equivalent to $2N+1=11$ mesh subcells. 

In the numerical solution, only one troubled cell is observed, formed in the vicinity of the detonation wave, as can be seen from the troubled cells indicator $\beta$. It should be noted that in the work~\cite{popov_j_sci_comp_2023}, in the case of simulating detonation wave formation in a two-component medium with a ``slow'' reaction, about $10\%$ troubled cells were observed in the case of degree $N = 1$ and $25-35\%$ troubled cells in the case of degrees $N = 2$-$5$, which is a large relative proportion. In the case of using an ADER-DG-$\mathbb{P}_{N}$ method with ADER-WENO finite volume a posteriori limiter that does not use adaptive time step correction, the number of mesh cell troubles decreased to $1$ in all considered degrees $N$ in the case of simulating detonation wave formation in a two-component medium with a ``slow'' reaction.

Comparison of the numerical solution obtained on a spatial mesh of $100$ cells with the reference solution shows excellent point-by-point agreement. The structure of the shock wave and the spatially distributed reaction domain are expressed in the numerical solution correctly and quite accurately, especially considering that only $100$ cells were selected in the spatial mesh. The numerical solution demonstrates the smooth combustion of the reagent behind the detonation wave front, while the conservation law $c_{1} + c_{2} = 1$ (or $\rho_{1} + \rho_{2} = \rho$) for the reagent and product is satisfied with very high accuracy $\sim 10^{-8}$-$10^{-10}$; in spatial domains of constancy and smoothness of flow, the error is even smaller $\sim 10^{-120}$, however, this is expected in this case. Basic hydrodynamic flow structures that were resolved by the reference solution to the problem are also correctly resolved by the numerical solution.

\paragraph{Strong stiff case}
The results of simulating the propagation of a detonation wave in a flow in the case of strong stiffness are presented in Figure~\ref{fig:cjdwf_1d_dg_5}. The main feature of the numerical solution obtained in the case of a ``fast'' reaction using the ADER-DG-$\mathbb{P}_{5}$ method with ADER-WENO2 finite volume a posteriori limiter demonstrates unsurpassed accuracy in resolving discontinuous components of the solution -- the detonation front has a width of only $2$-$4$ \textit{subcells}, while in the work~\cite{popov_j_sci_comp_2023} where a similar numerical method was used using adaptive time step correction, the width of the front of the detonation wave was $1$-$2$ mesh \textit{cells} -- in the case of degree $N = 5$, one cell is equivalent to $11$-$22$ mesh subcells. 

In the numerical solution, only $5$ troubled cells is observed, formed in the vicinity of the detonation wave front and in some spatial region behind the detonation front, as can be seen from the troubled cells indicator $\beta$. It should be noted that in the work~\cite{popov_j_sci_comp_2023}, in the case of simulating detonation wave formation in a two-component medium with a ``fast'' reaction, about $8$-$10\%$ troubled cells were observed in the case of degree $N = 1$ and $\sim 35\%$ troubled cells in the case of degrees $N = 2$-$5$, which is a large relative proportion. In the case of using a ADER-DG-$\mathbb{P}_{N}$ method with ADER-WENO finite volume a posteriori limiter that does not use adaptive time step correction, the number of mesh cell troubles decreased to $3$-$5$ in all considered degrees $N$ in the case of simulating detonation wave formation in a two-component medium with a ``fast'' reaction. 

A well-resolved structure of the detonation wave ZND is observed. The Zel'dovich chemical peak is strictly resolved in the numerical solution. Non-physical artifacts of the numerical solution, characteristic of ``standard'' numerical methods~\cite{frac_steps_detwave_sim_2000, chem_kin_hrs_weno, correct_det_wave_speed_2017} of computational fluid dynamics, are not observed in the numerical solution -- the ZND detonation front has a clearly defined spatial connectivity, forming a single structure in which complete combustion of the reagent into the reaction product occurs. The structure of the shock wave and the spatially distributed reaction domain are expressed in the numerical solution correctly and quite accurately, especially considering that only $100$ cells were selected in the spatial mesh. The numerical solution demonstrates the sharp combustion of the reagent behind the front of the detonation wave, which is a characteristic feature of a ``fast'' reaction in a two-component medium. The conservation law $c_{1} + c_{2} = 1$ (or $\rho_{1} + \rho_{2} = \rho$) for the reagent and product is satisfied with very high accuracy $\sim 10^{-9}$-$10^{-11}$; in spatial domains of constancy and smoothness of flow, the error is even smaller $\sim 10^{-132}$, however, this is expected in this case. Comparison of the numerical solution obtained on a spatial mesh of $100$ cells with the reference solution shows excellent point-by-point agreement. Basic hydrodynamic flow structures that were resolved by the reference solution to the problem are also correctly resolved by the numerical solution.

\subsection{Cylindrical and spherical ZND-detonation waves}
\label{sec:detonation_waves:cjdw_md}

\paragraph{Formulation of the problem}
Cylindrical and spherical ZND-detonation waves problems occupy an interesting place among two-dimensional and three-dimensional problems. On the one hand, these are conceptually one-dimensional problems, where the solution depends on only one spatial coordinate -- the distance $r$ to the center of the explosion, so these problems may seem quite simple. On the other hand, the use of two-dimensional and three-dimensional computational codes to solve these problems can reveal certain problems of numerical methods and their implementations associated with maintaining the spatial symmetry of the original problem -- axial in the two-dimensional case and spherical in the three-dimensional case.

\begin{figure*}[h!]
\centering
\includegraphics[width=0.245\textwidth]{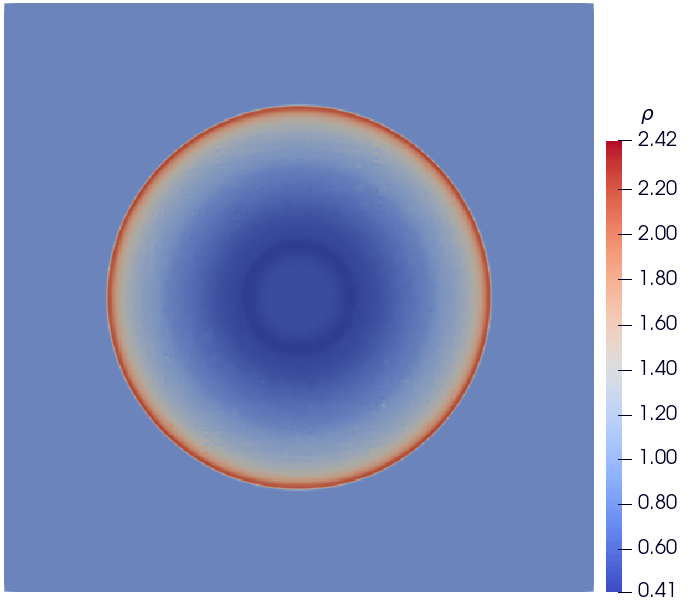}
\includegraphics[width=0.245\textwidth]{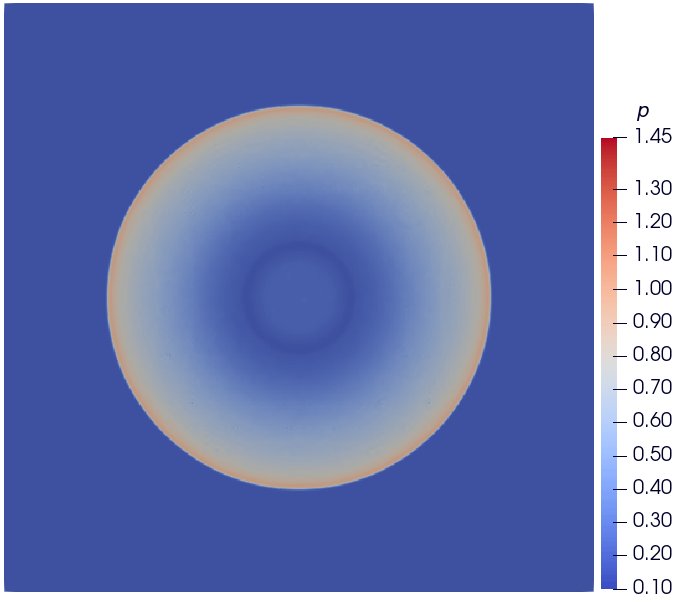}
\includegraphics[width=0.245\textwidth]{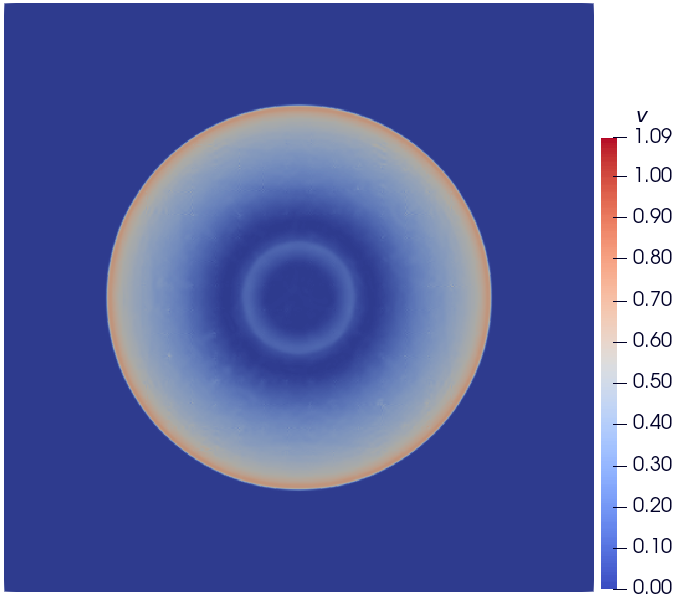}
\includegraphics[width=0.245\textwidth]{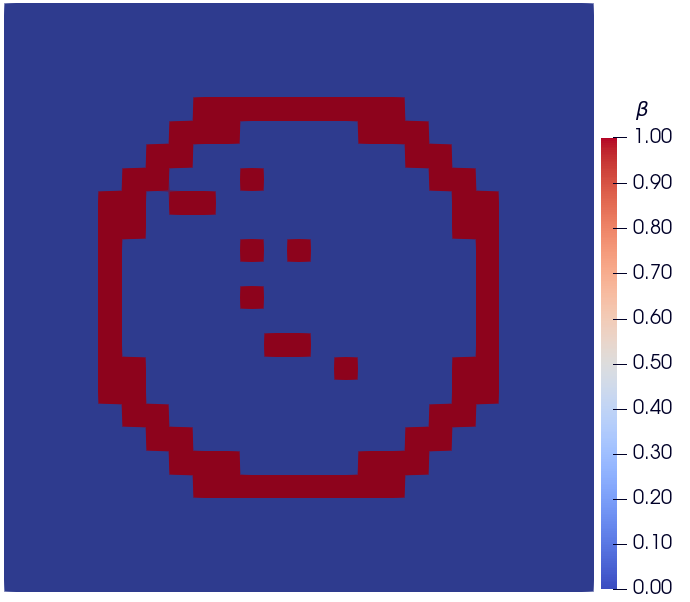}\\
\includegraphics[width=0.245\textwidth]{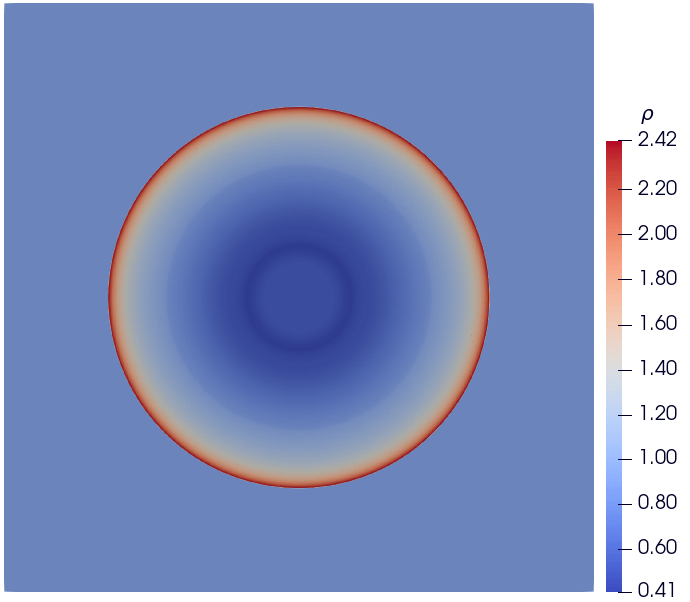}
\includegraphics[width=0.245\textwidth]{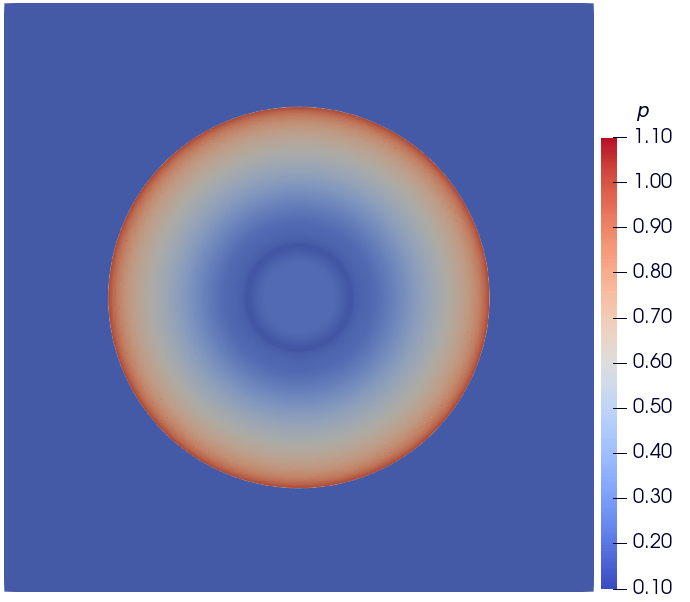}
\includegraphics[width=0.245\textwidth]{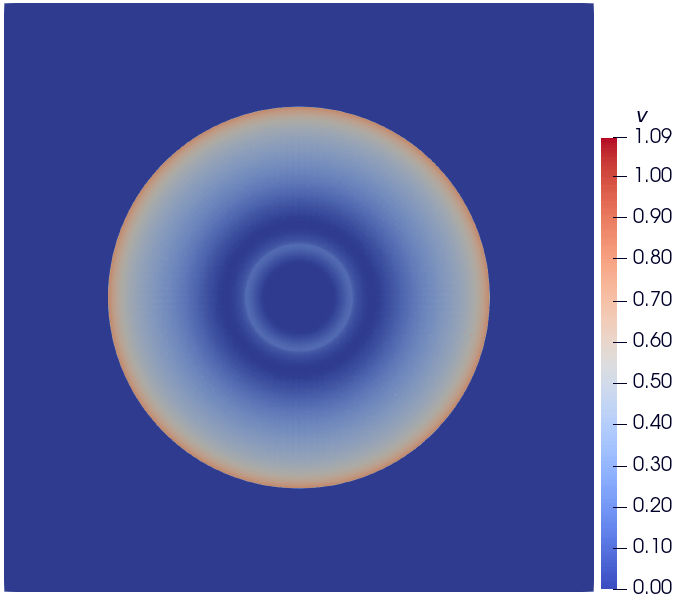}
\includegraphics[width=0.245\textwidth]{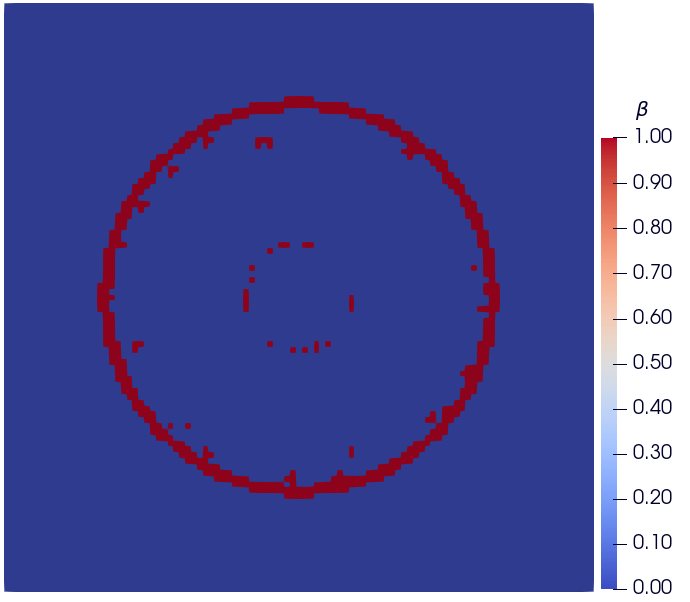}
\caption{\label{fig:cjdws_2d}
Numerical solution of the two-dimensional problem of the cylindrical detonation wave formation in a two-component medium with a ``slow'' reaction
(weak stiff case, a detailed statement of the problem is presented in the text), obtained using the ADER-DG-$\mathbb{P}_{9}$ method 
at the final time $t_{\rm final} = 0.3$ on meshes $25 \times 25$ (top) and $101 \times 101$ (bottom) cells.
The graphs show the coordinate dependence of the subcells finite-volume representation 
of density $\rho$, pressure $p$, flow velocity magnitude $v$ and troubled cells indicator $\beta$.
}
\end{figure*}

\begin{figure}[h!]
\centering
\includegraphics[width=0.239\textwidth]{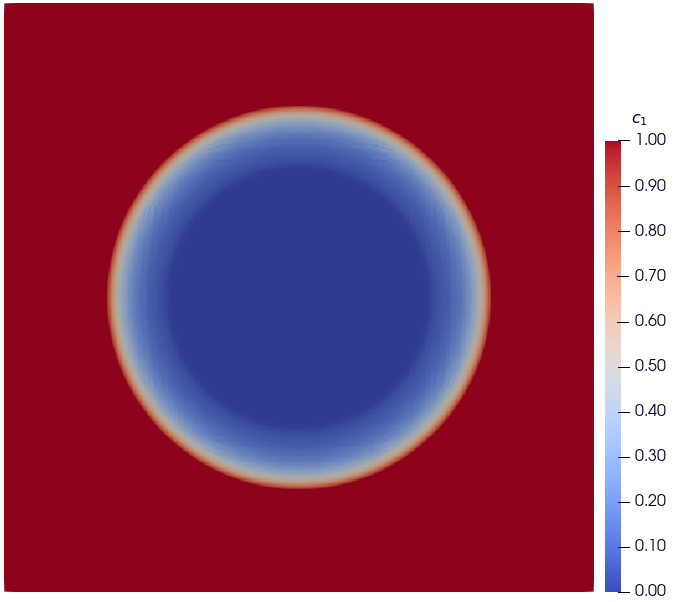}
\includegraphics[width=0.239\textwidth]{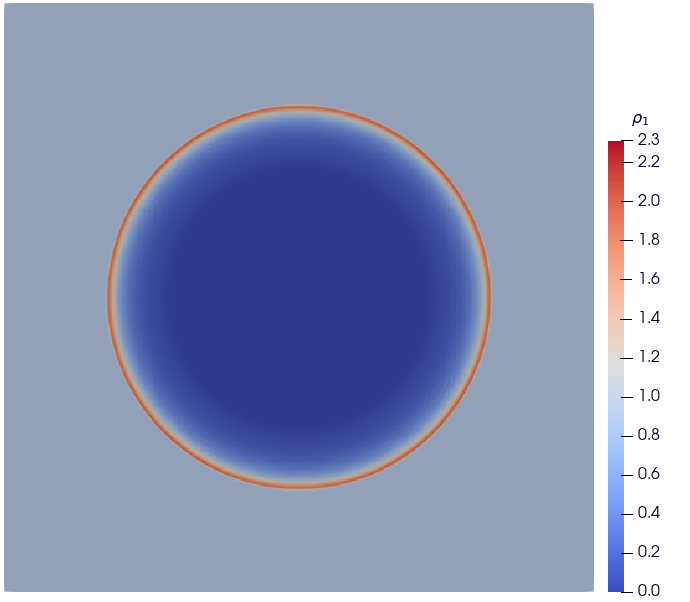}\\
\includegraphics[width=0.239\textwidth]{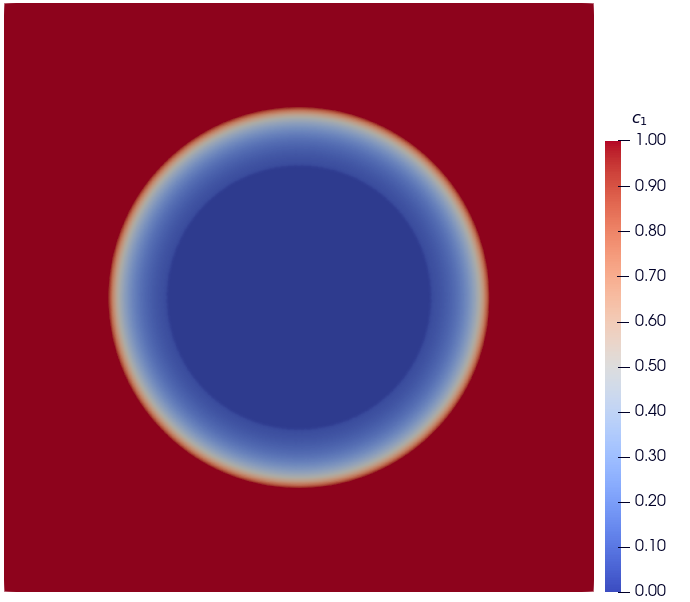}
\includegraphics[width=0.239\textwidth]{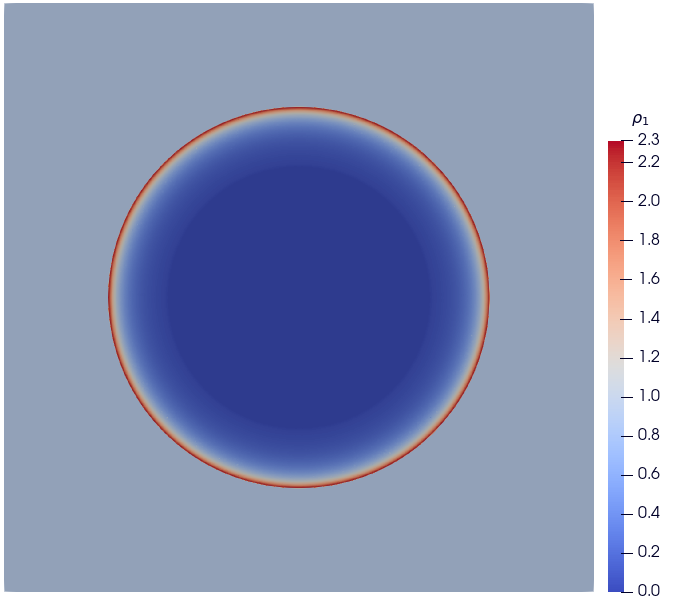}
\caption{\label{fig:cjdws_2d_comps}
The coordinate dependence of the subcells finite-volume representation 
of mass concentration $c_{1}$ and partial density $\rho_{1} = \rho c_{1}$ of the reaction reagent for 
numerical solution of the cylindrical detonation wave formation in a two-component medium with a ``slow'' reaction, 
which is presented in Figure~\ref{fig:cjdws_2d}
on meshes $25 \times 25$ (top) and $101 \times 101$ (bottom) cells.
}
\end{figure}

\begin{figure*}[h!]
\centering
\includegraphics[width=0.245\textwidth]{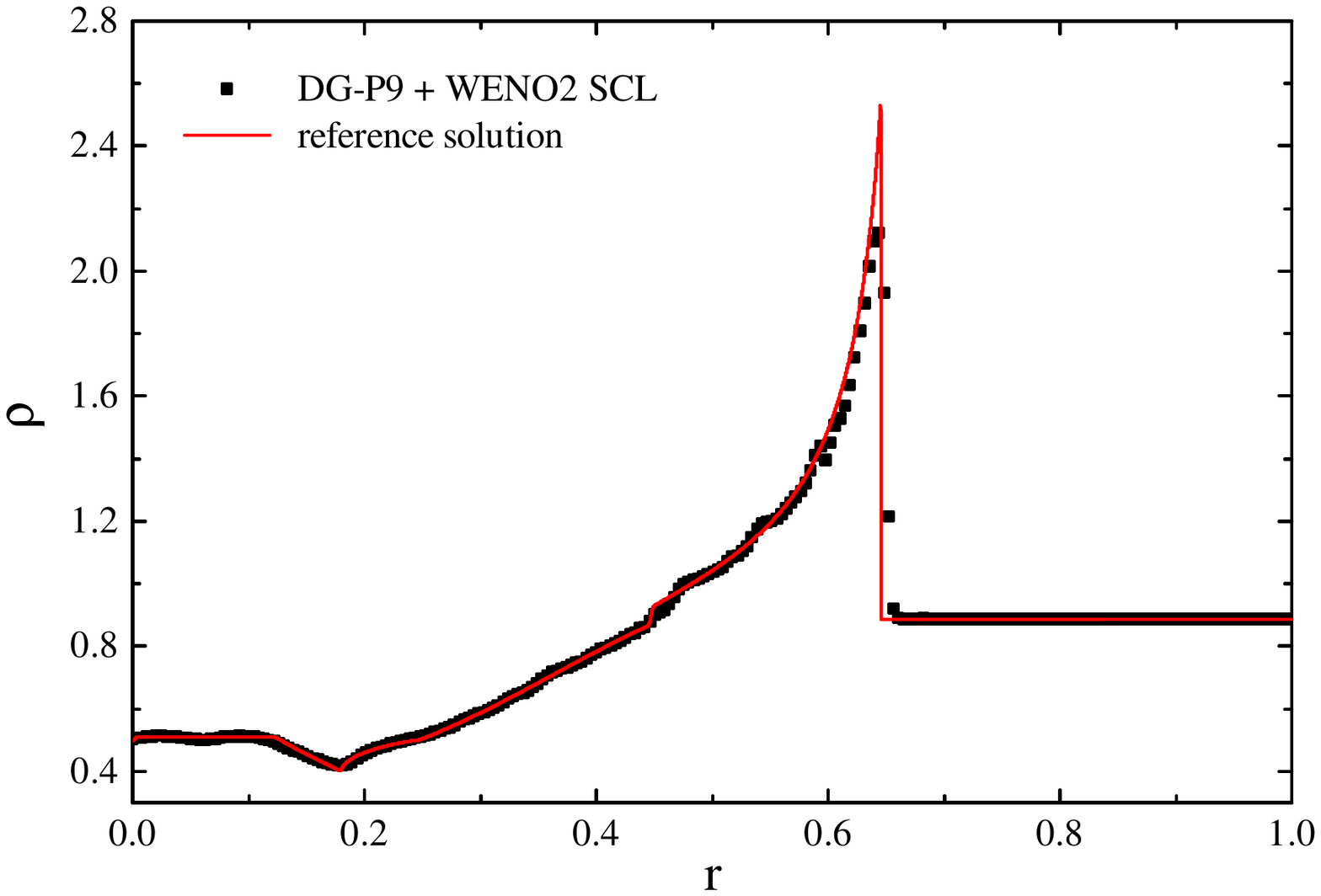}
\includegraphics[width=0.245\textwidth]{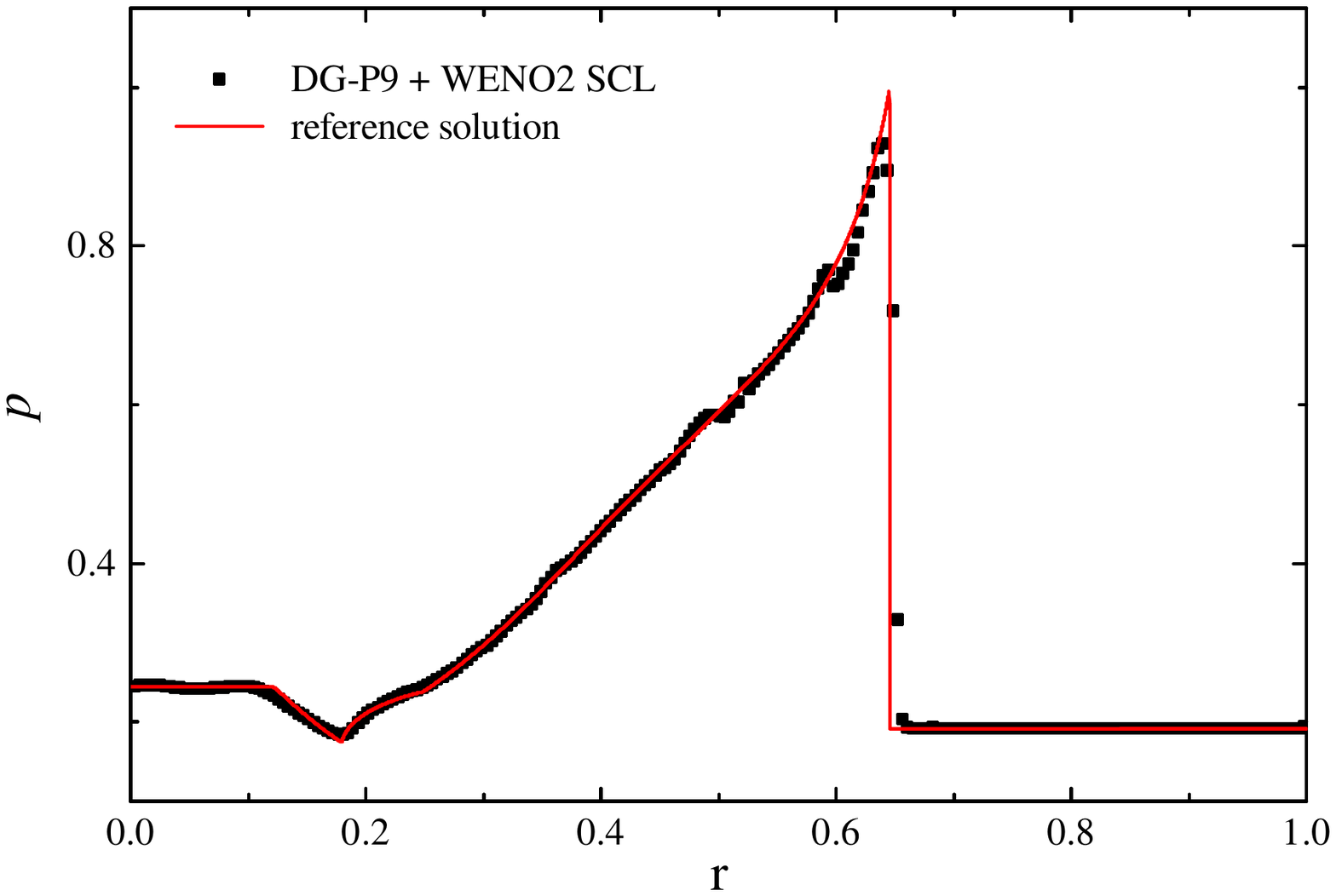}
\includegraphics[width=0.245\textwidth]{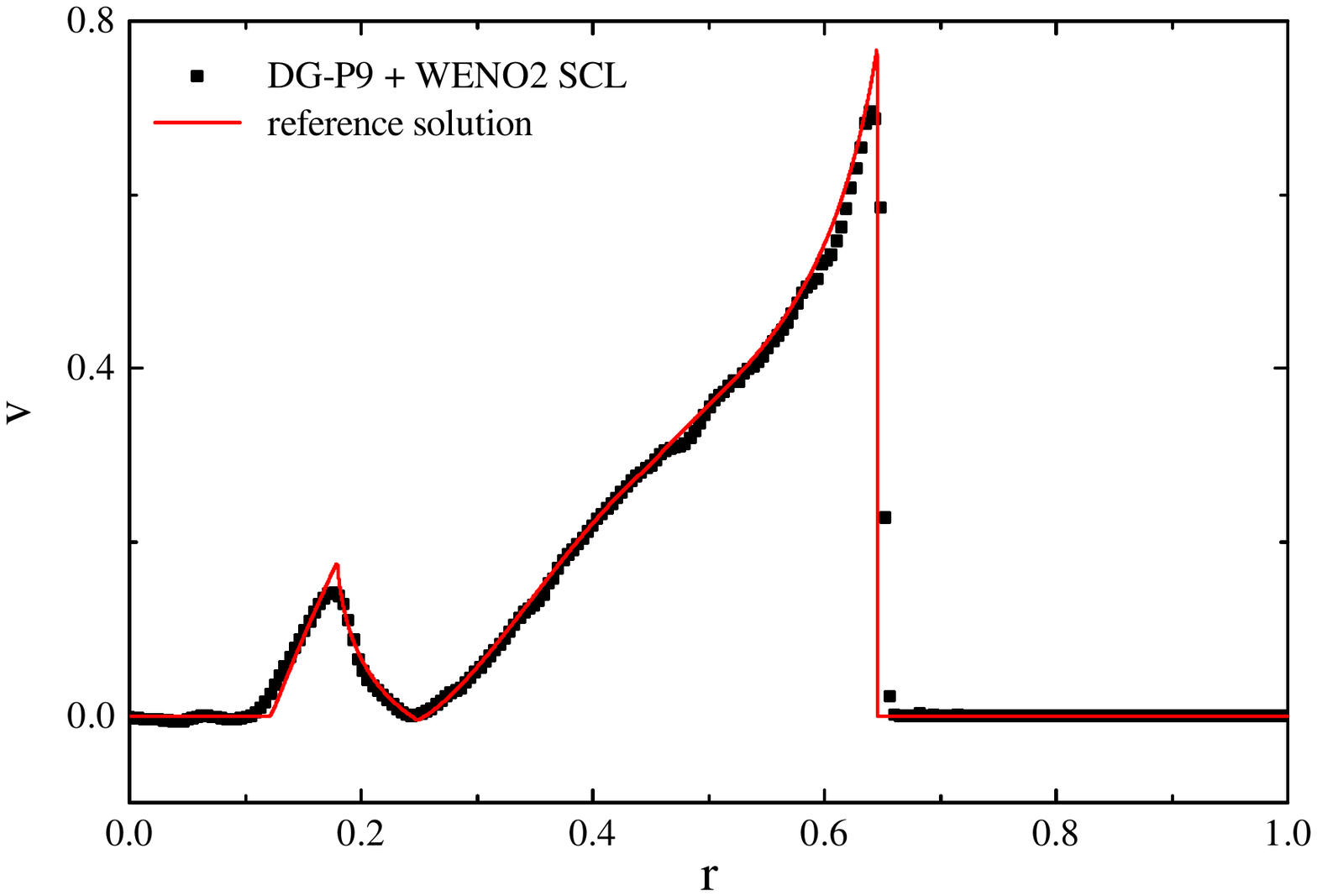}
\includegraphics[width=0.245\textwidth]{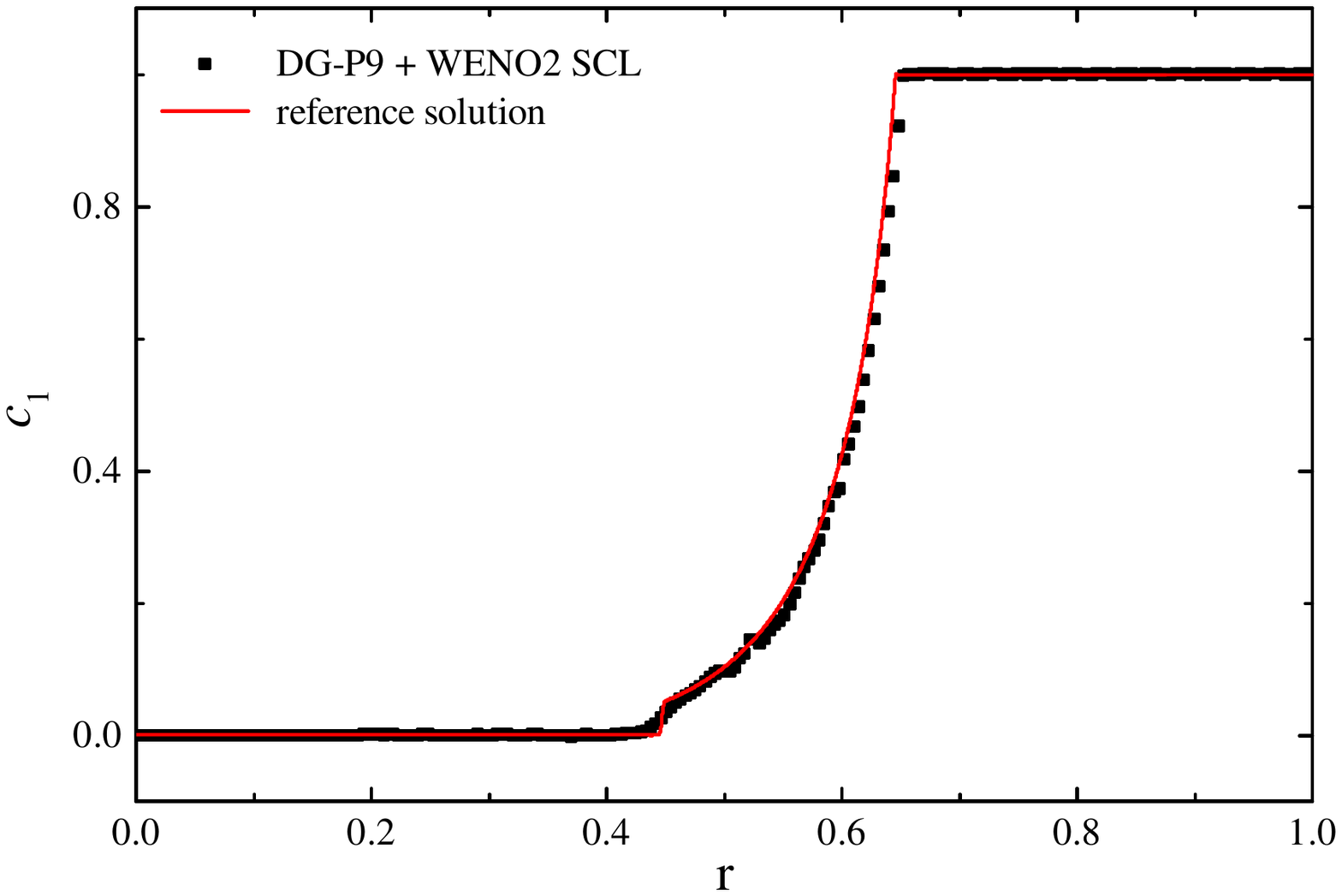}\\
\includegraphics[width=0.245\textwidth]{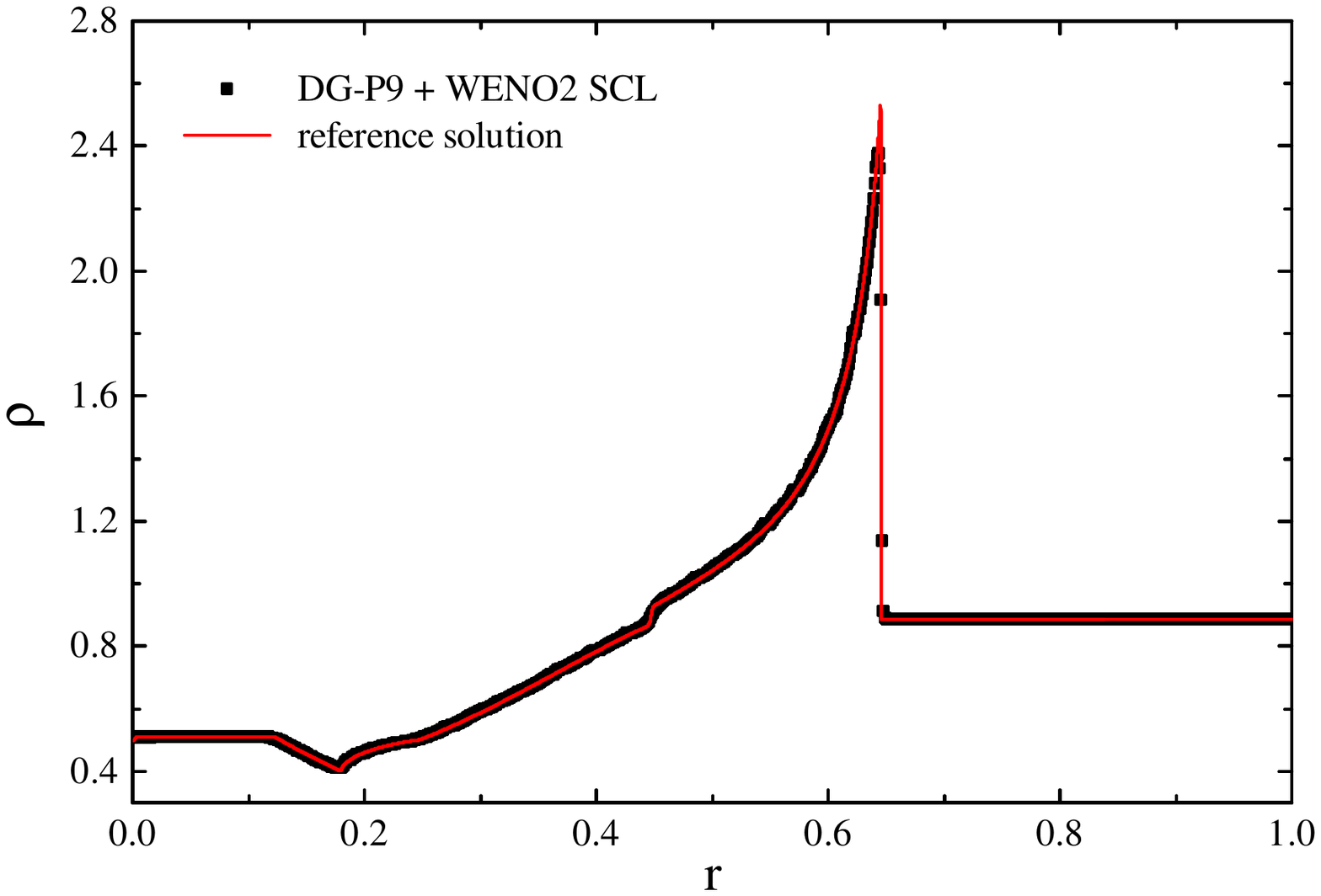}
\includegraphics[width=0.245\textwidth]{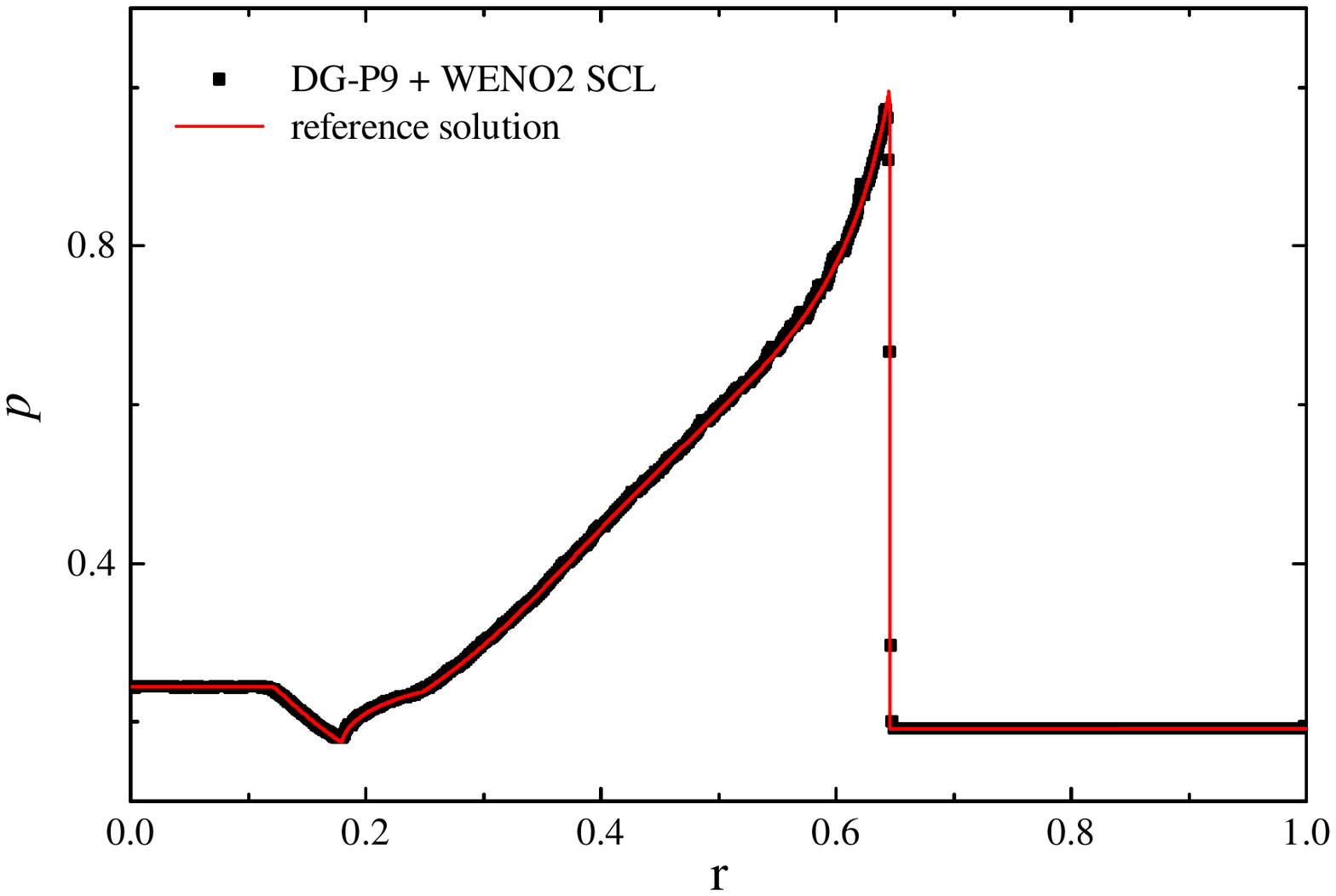}
\includegraphics[width=0.245\textwidth]{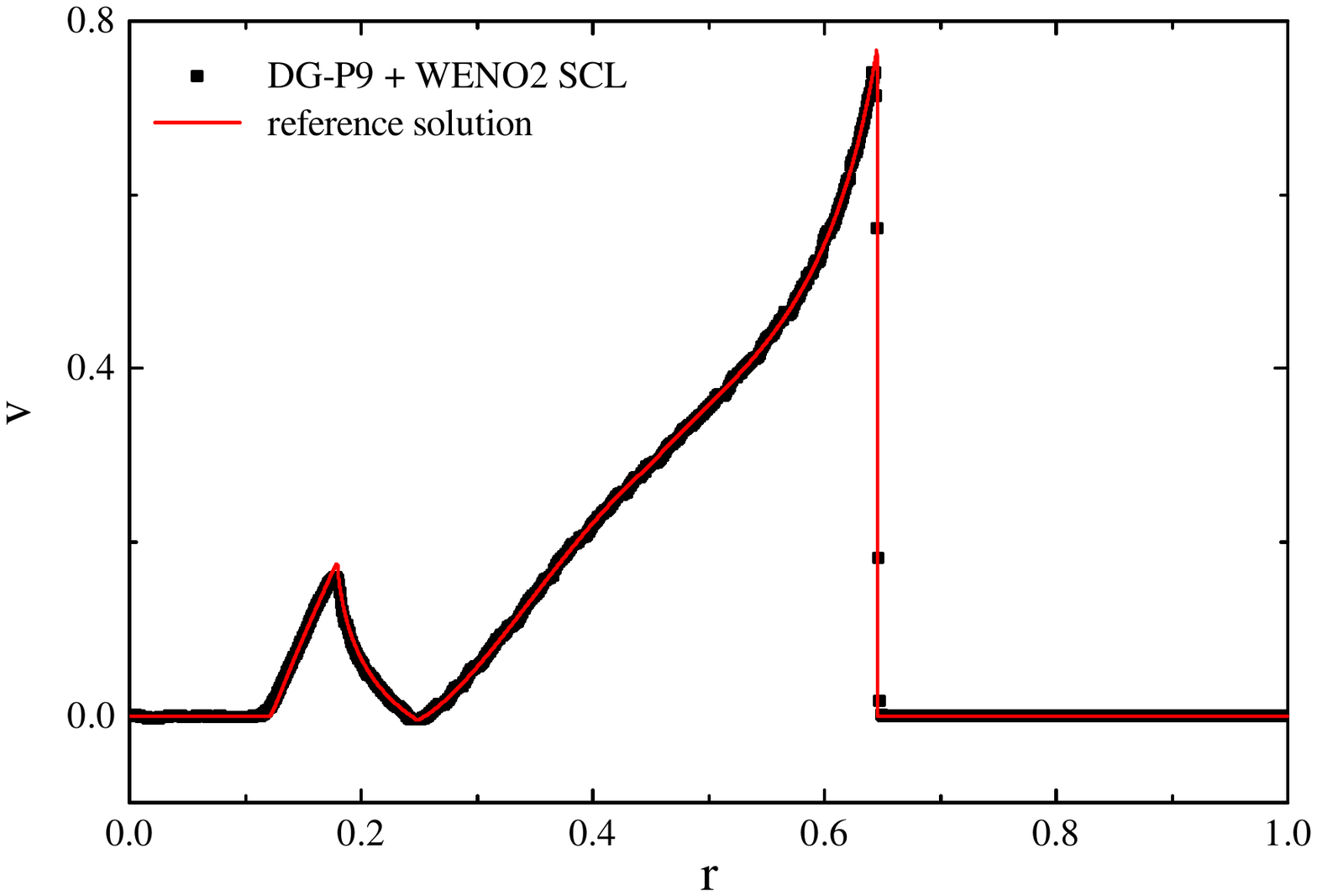}
\includegraphics[width=0.245\textwidth]{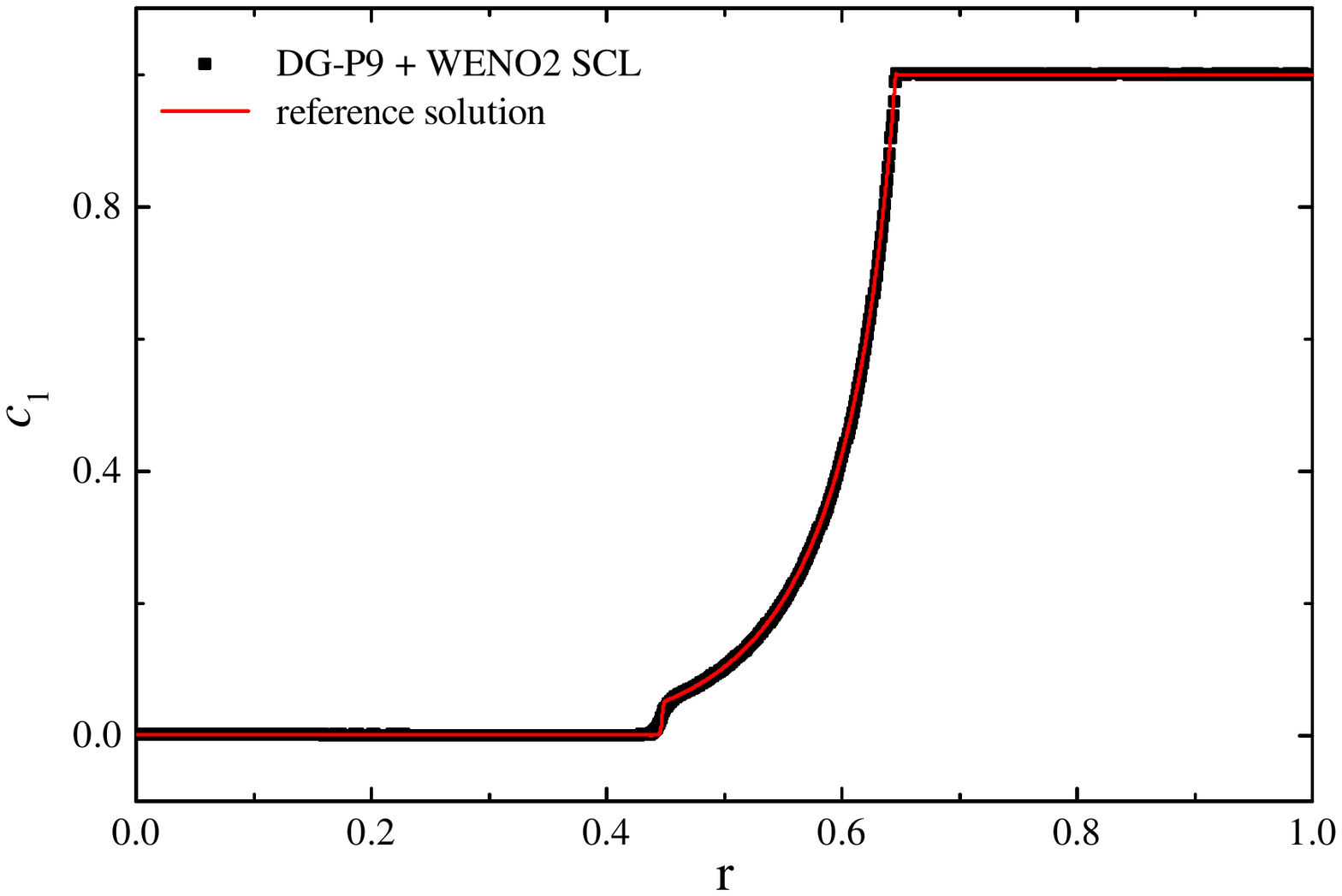}
\caption{\label{fig:cjdws_2d_slice}
One dimensional cuts of the numerical solution for two-dimensional problem of the cylindrical detonation wave formation 
in a two-component medium with a ``slow'' reaction, which is presented in Figure~\ref{fig:cjdws_2d},
obtained using the ADER-DG-$\mathbb{P}_{9}$ method on meshes $25 \times 25$ (top) and $101 \times 101$ (bottom) cells.
The graphs show the coordinate dependence of density $\rho$, pressure $p$, flow velocity $u$ and 
mass concentration $c_{1}$ of the reaction reagent (from left to right)
on the distance $r$ to the point $(0, 0)$ along the direction $(0, 1)$.
The black square symbols represent the subcells finite-volume representation of the numerical solution; 
the red solid lines represents the reference solution of the problem.
}
\end{figure*}

\begin{figure*}[h!]
\centering
\includegraphics[width=0.245\textwidth]{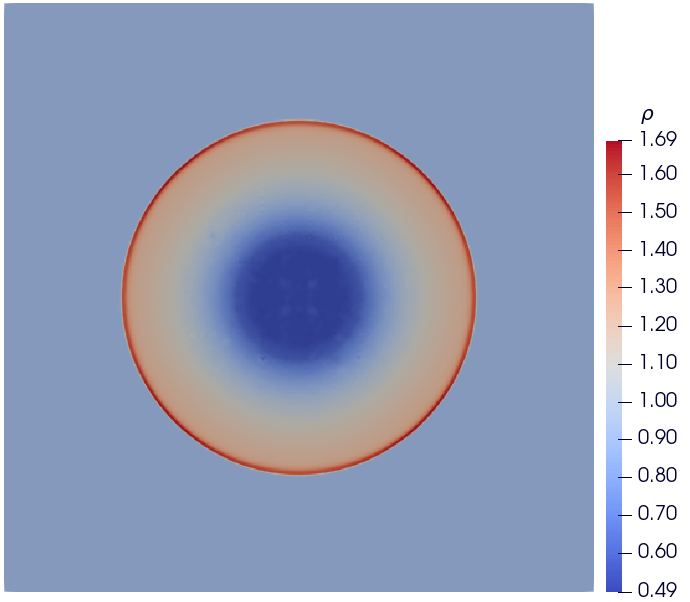}
\includegraphics[width=0.245\textwidth]{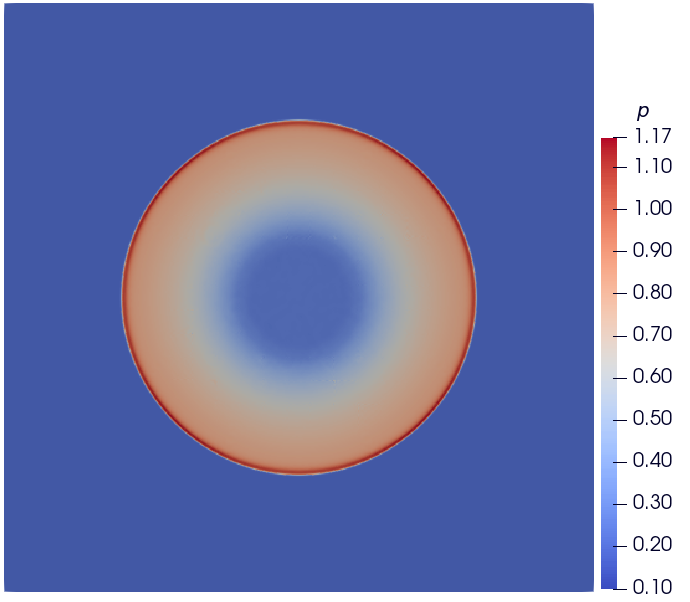}
\includegraphics[width=0.245\textwidth]{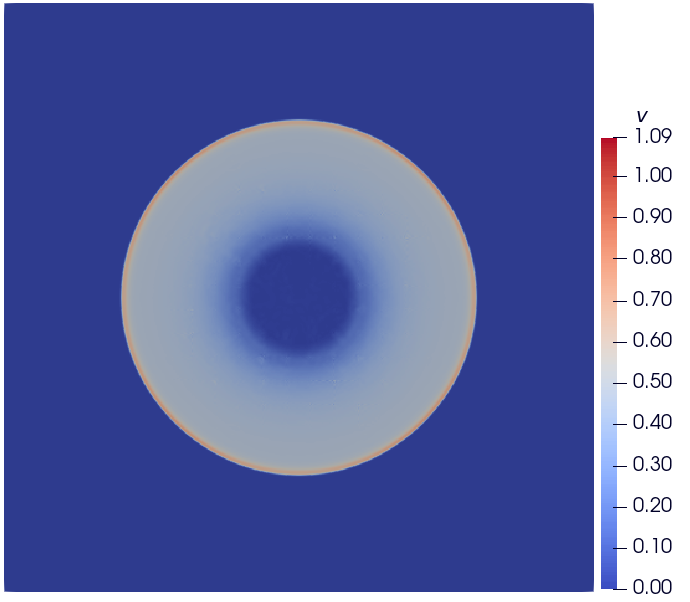}
\includegraphics[width=0.245\textwidth]{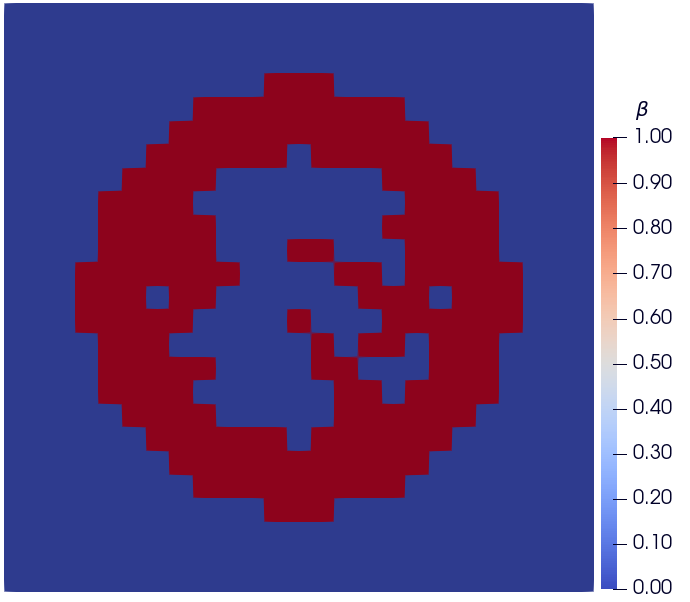}\\
\includegraphics[width=0.245\textwidth]{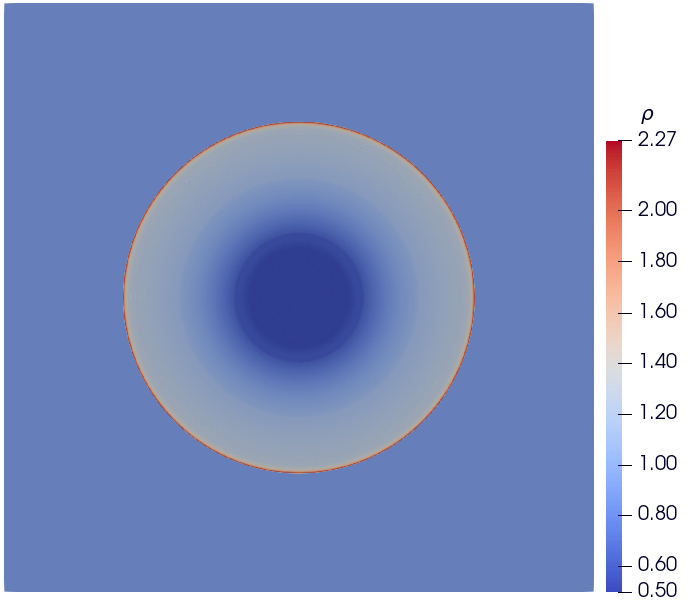}
\includegraphics[width=0.245\textwidth]{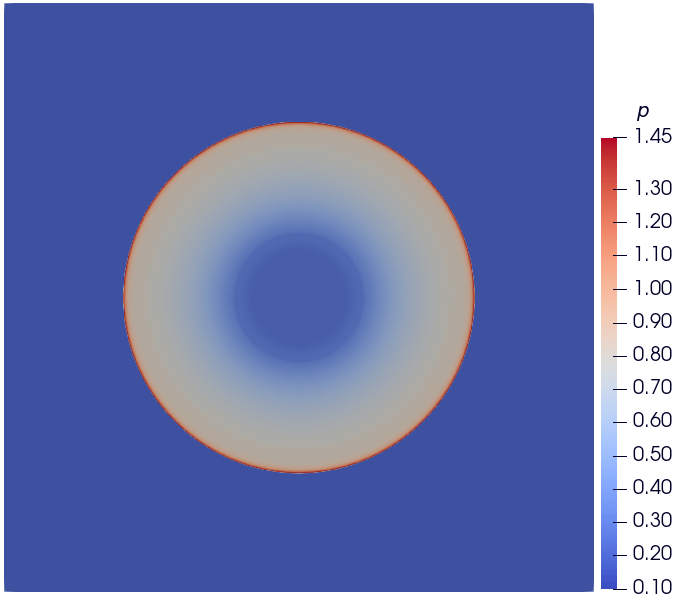}
\includegraphics[width=0.245\textwidth]{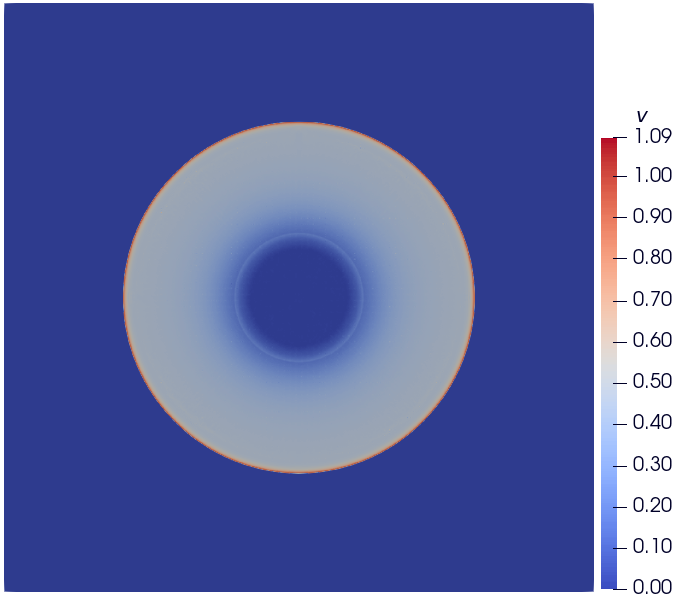}
\includegraphics[width=0.245\textwidth]{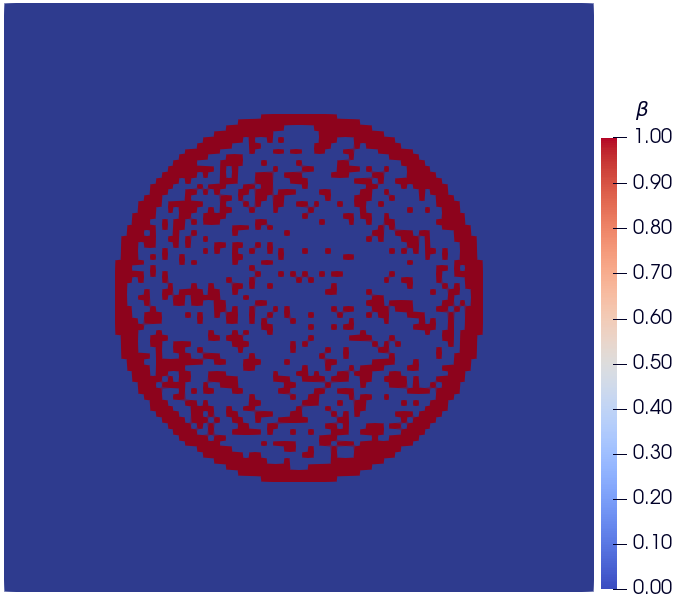}
\caption{\label{fig:cjdwf_2d}
Numerical solution of the two-dimensional problem of the cylindrical detonation wave formation in a two-component medium with a ``fast'' reaction
(strong stiff case, a detailed statement of the problem is presented in the text), obtained using the ADER-DG-$\mathbb{P}_{9}$ method 
at the final time $t_{\rm final} = 0.2$ on meshes $25 \times 25$ (top) and $101 \times 101$ (bottom) cells.
The graphs show the coordinate dependence of the subcells finite-volume representation 
of density $\rho$, pressure $p$, flow velocity magnitude $v$ and troubled cells indicator $\beta$.
}
\end{figure*}

\begin{figure}[h!]
\centering
\includegraphics[width=0.239\textwidth]{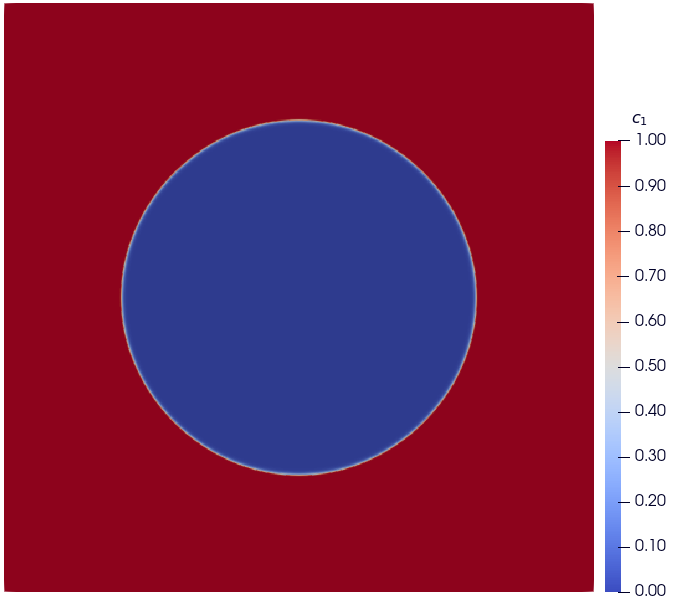}
\includegraphics[width=0.239\textwidth]{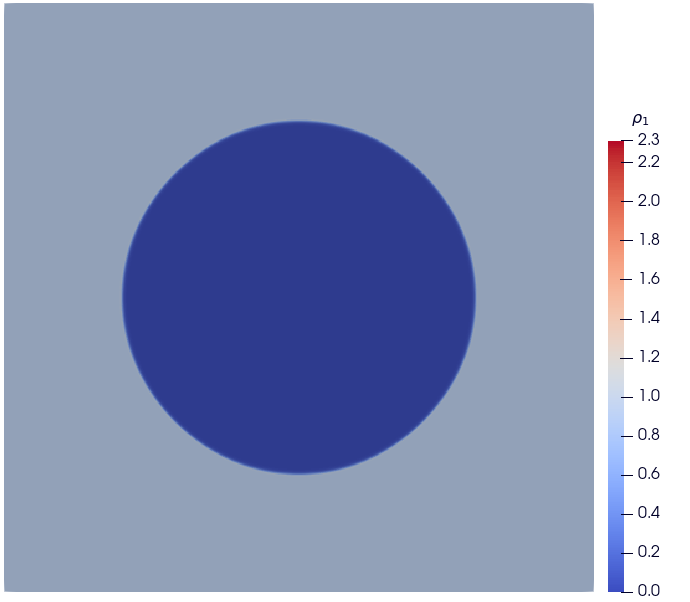}\\
\includegraphics[width=0.239\textwidth]{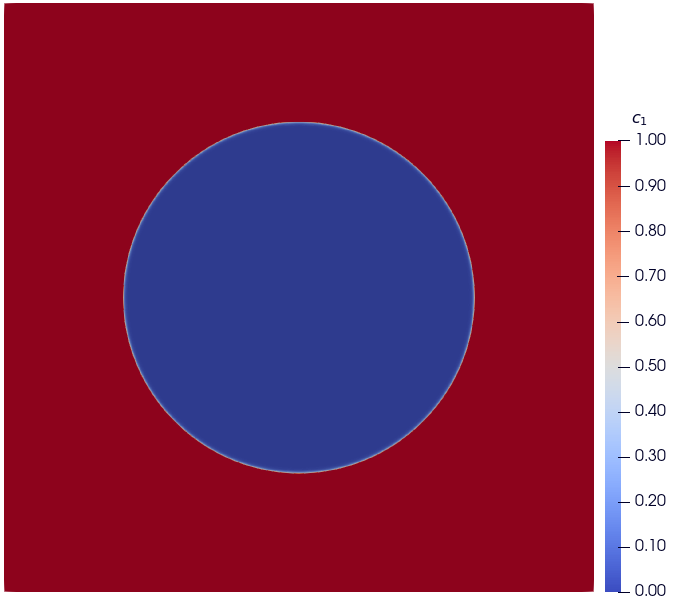}
\includegraphics[width=0.239\textwidth]{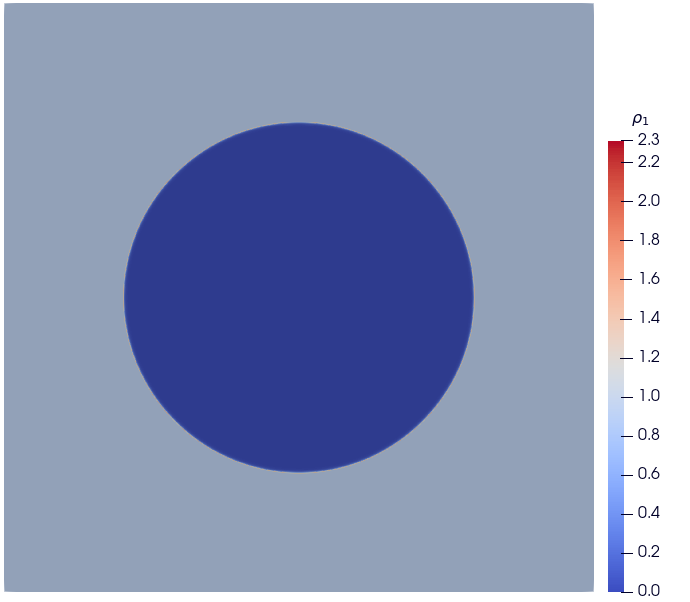}
\caption{\label{fig:cjdwf_2d_comps}
The coordinate dependence of the subcells finite-volume representation 
of mass concentration $c_{1}$ and partial density $\rho_{1} = \rho c_{1}$ of the reaction reagent for 
numerical solution of the cylindrical detonation wave formation in a two-component medium with a ``fast'' reaction, 
which is presented in Figure~\ref{fig:cjdwf_2d},
on meshes $25 \times 25$ (top) and $101 \times 101$ (bottom) cells.
}
\end{figure}

The computational coordinate domain was chosen in the form of a $d$-dimensional cube $\Omega = [-1, +1]^{d}$, where $d = 2$ in two-dimensional case and $d = 3$ in three-dimensional case. The boundary conditions were specified in the form of free outflow conditions. The final time has been chosen $t_{\rm final} = 0.3$ for cylindrical and  $t_{\rm final} = 0.2$ for spherical ZND-detonation waves problems. The adiabatic index $\gamma = 1.4$. The initial conditions were chosen in the form of the Chapman-Jouguet (CJ) conditions, which define a stationary detonation wave in the instantaneous detonation approximation. However, in contrast to the formulation of the problem chosen in the work~\cite{frac_steps_detwave_sim_2000}, the direction of the flow velocity was chosen in the opposite direction:
\begin{equation}\label{eq:cjdw_md_init}
\begin{split}
&\rho(\mathbf{r}, t = 0) = \left\{
\begin{array}{ll}
1.4, & \mathrm{if}\ r \leqslant 0.3; \\
0.887565, & \mathrm{if}\ r >\, 0.3; \\
\end{array}
\right.\\
&\mathrm{v}(\mathbf{r}, t = 0) = \left\{
\begin{array}{ll}
0.577350, & \mathrm{if}\ r \leqslant 0.3; \\
0.0, & \mathrm{if}\ r >\, 0.3; \\
\end{array}
\right.\\
&p(\mathbf{r}, t = 0) = \left\{
\begin{array}{ll}
1.0, & \mathrm{if}\ r \leqslant 0.3; \\
0.191709, & \mathrm{if}\ r >\, 0.3; \\
\end{array}
\right.
\end{split}
\end{equation}
\begin{equation}
\begin{split}
&c_{1}(\mathbf{r}, t = 0) = \left\{
\begin{array}{ll}
1.0, & \mathrm{if}\ r \leqslant 0.3; \\
10^{-14}, & \mathrm{if}\ r >\, 0.3; \\
\end{array}
\right.\\
&c_{2}(\mathbf{r}, t = 0) = \left\{
\begin{array}{ll}
10^{-14}, & \mathrm{if}\ r \leqslant 0.3; \\
1.0, & \mathrm{if}\ r >\, 0.3; \\
\end{array}
\right.
\end{split}
\end{equation}
where $r$ is the distance to the center of the coordinate system, $\mathrm{v} = |\mathbf{v}|$ is the absolute value of flow velocity: $r^{2} = x^{2} + y^{2}$ and $\mathrm{v}^{2} = u^{2} + v^{2}$ in two-dimensional case, $r^{2} = x^{2} + y^{2} + z^{2}$ and $\mathrm{v}^{2} = u^{2} + v^{2} + w^{2}$ in three-dimensional case. The velocity projections $(u, v)$ and $(u, v, w)$ were chosen based on the radial direction of the velocity vector $\mathbf{v}$:
\begin{equation}
u = \mathrm{v}\cdot\cos\gamma_{x};\quad
v = \mathrm{v}\cdot\cos\gamma_{y};
\end{equation}
in two-dimensional case and
\begin{equation}
u = \mathrm{v}\cdot\cos\gamma_{x};\quad
v = \mathrm{v}\cdot\cos\gamma_{y};\quad
z = \mathrm{v}\cdot\cos\gamma_{z};
\end{equation}
in three-dimensional case, where $\cos\gamma_{x}$, $\cos\gamma_{y}$ and $\cos\gamma_{z}$ are the direction cosines of the corresponding coordinate axes. The choice of the direction of velocity $\mathbf{v}$ was associated with the following feature of the cylindrical and spherical formulations of the problem -- in the case of a one-dimensional flow, the choice of flow velocity $u$ towards the origin of coordinates does not have a predominant emphasis, however, in the cylindrical and spherical cases, the choice of the radial direction of the flow velocity towards the origin of coordinates leads to the filling of the coordinate domain $\Omega$ by the mass of the substance, if the boundary conditions are specified in the form of conditions of free outflow. In this case, during the simulation, a large amount of new substance flowing through the boundary will accumulate in the coordinate domain $\Omega$ -- a smooth density profile of the substance will be formed in front of the detonation wave, decreasing towards the boundary. In this case, the detonation front will propagate along a nonuniform radial density profile. Therefore, the choice was made in favor of the radial direction of velocity from the origin. From the position of the speed of the detonation wave, in the reference frame associated with a stationary coordinate domain, it will increase. Thus, the burned medium was represented as a radial ``piston'' for the unburned medium. A small value $10^{-14}$ of mass concentrations $c_{1}$ and $c_{2}$, instead of strictly $0$, was chosen to prevent the occurrence of negative concentrations immediately at the start of the calculation process, which could lead to a meaninglessly large increase in the number of troubled cells in the solution.

The boundary conditions were chosen to exact solution of the problem outside the detonation front, which corresponds to the initial conditions at the boundary of the coordinate domain $\Omega$ -- the perturbations arising in the solution, by the time the simulation ends, do not have time to reach the boundary points and perturb the hydrodynamic flow on the boundary, therefore these boundary conditions with exact solution are equivalent to the conditions of free outflow, in this particular case.

To obtain a reference solution, a one-dimensional problem with a geometric source term was used:
\begin{equation}
\begin{split}
\frac{\partial}{\partial t}\left[
\begin{array}{c}
\rho\\
\rho \mathrm{v}\\
\varepsilon\\
\rho c_{1}\\
\rho c_{2}
\end{array}
\right] +& 
\frac{\partial}{\partial r}\left[
\begin{array}{c}
\rho \mathrm{v}\\
\rho \mathrm{v}^{2} + p\\
(\varepsilon + p) \mathrm{v}\\
\rho c_{1}\mathrm{v}\\
\rho c_{2}\mathrm{v}
\end{array}
\right]\\
=&\left[
\begin{array}{c}
0\\
0\\
\rho\omega q_{0}\\
-\rho\omega\\
+\rho\omega
\end{array}
\right] -
\frac{d - 1}{r} \left[
\begin{array}{c}
\rho \mathrm{v}\\
\rho \mathrm{v}^{2}\\
(\varepsilon + p) \mathrm{v}\\
\rho c_{1}\mathrm{v}\\
\rho c_{2}\mathrm{v}
\end{array}
\right];
\end{split}
\end{equation}
where $\mathrm{v} = |\mathbf{v}|$ is the absolute value of the flow velocity, the remaining designations coincide with those already introduced above. The coordinate domain for obtaining the reference solution was chosen in the form of a range $\Omega_{r} = [0, 1]$. The initial conditions were chosen according to the initial conditions of the original problems. A solid wall condition was specified at the left boundary, and a free outflow condition -- at the right boundary. During the calculations, the point $r = 0$ was not directly involved in the calculations of the source term. The reference solution was obtained for cylindrical and spherical explosion problems using the ADER-WENO2 finite volume method on a mesh with $6000$ finite-volume cells. The work~\cite{ader_stiff_2} showed that the finite-volume ADER-WENO method allows one to obtain a correct numerical solution without using additional procedures for recalculating the solution, so this method was chosen as the method for obtaining a reference solution. The resulting standard solution will be further used in this work for comparison with solutions to the explosion problem in the full two-dimensional and three-dimensional formulations of the problems.

\paragraph{Cylindrical ZND-detonation waves}

The numerical solution to the problem of the formation and propagation of a detonation wave in media with a ``slow'' reaction was obtained using ADER-DG-$\mathbb{P}_{9}$ method with a posteriori ADER-WENO2 finite volume limitation on a spatial mesh of $100 \times 100$ cells. The simulation results are presented in Figure~\ref{fig:cjdws_2d}. Figure~\ref{fig:cjdws_2d_comps} shows the results for the mass concentration $c_{1}$ and density $\rho_{1}$ of the reagent. The numerical solution shows a very high axial symmetry of the solution. The fronts of hydrodynamic quantities are expressed sharply and are clearly visible in the numerical solution. The number of troubled cells does not exceed $15\%$, which is a fairly small value for multidimensional problems. The combustion of the reagent $A$ occurs smoothly, and the structure of the shock front is clearly observed in the coordinate dependence of the density $\rho_{1}$ of the component.

\begin{figure*}[h!]
\centering
\includegraphics[width=0.245\textwidth]{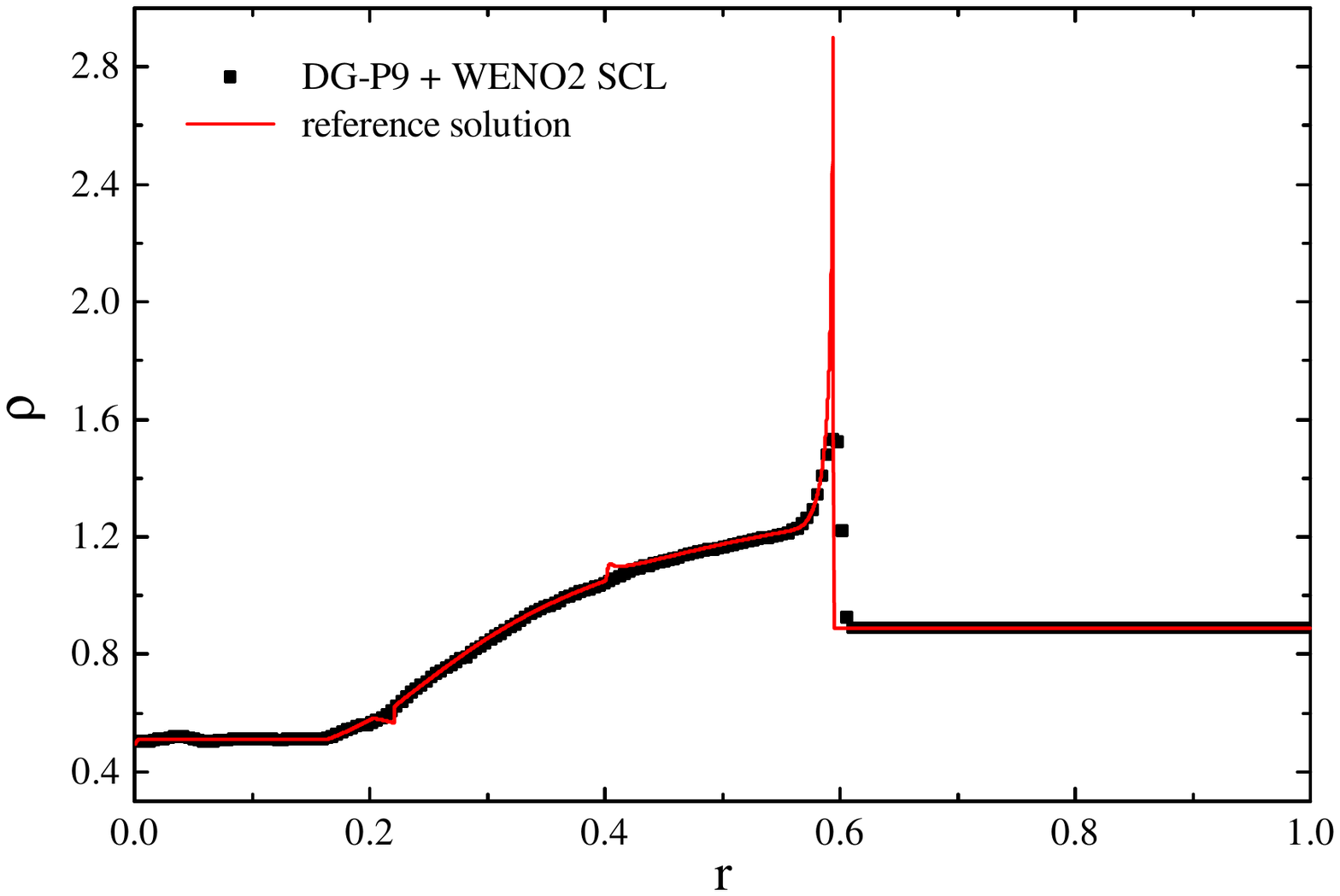}
\includegraphics[width=0.245\textwidth]{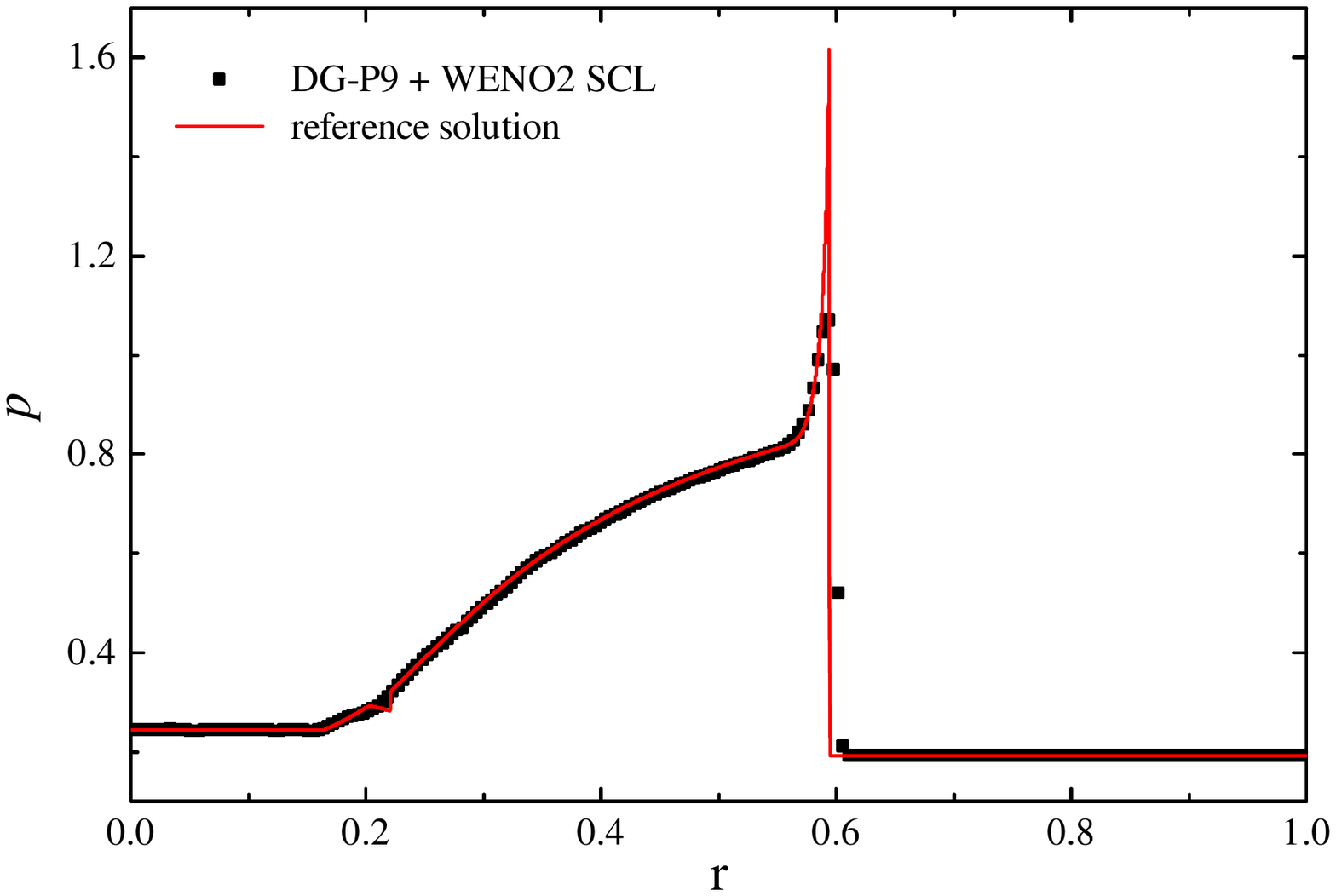}
\includegraphics[width=0.245\textwidth]{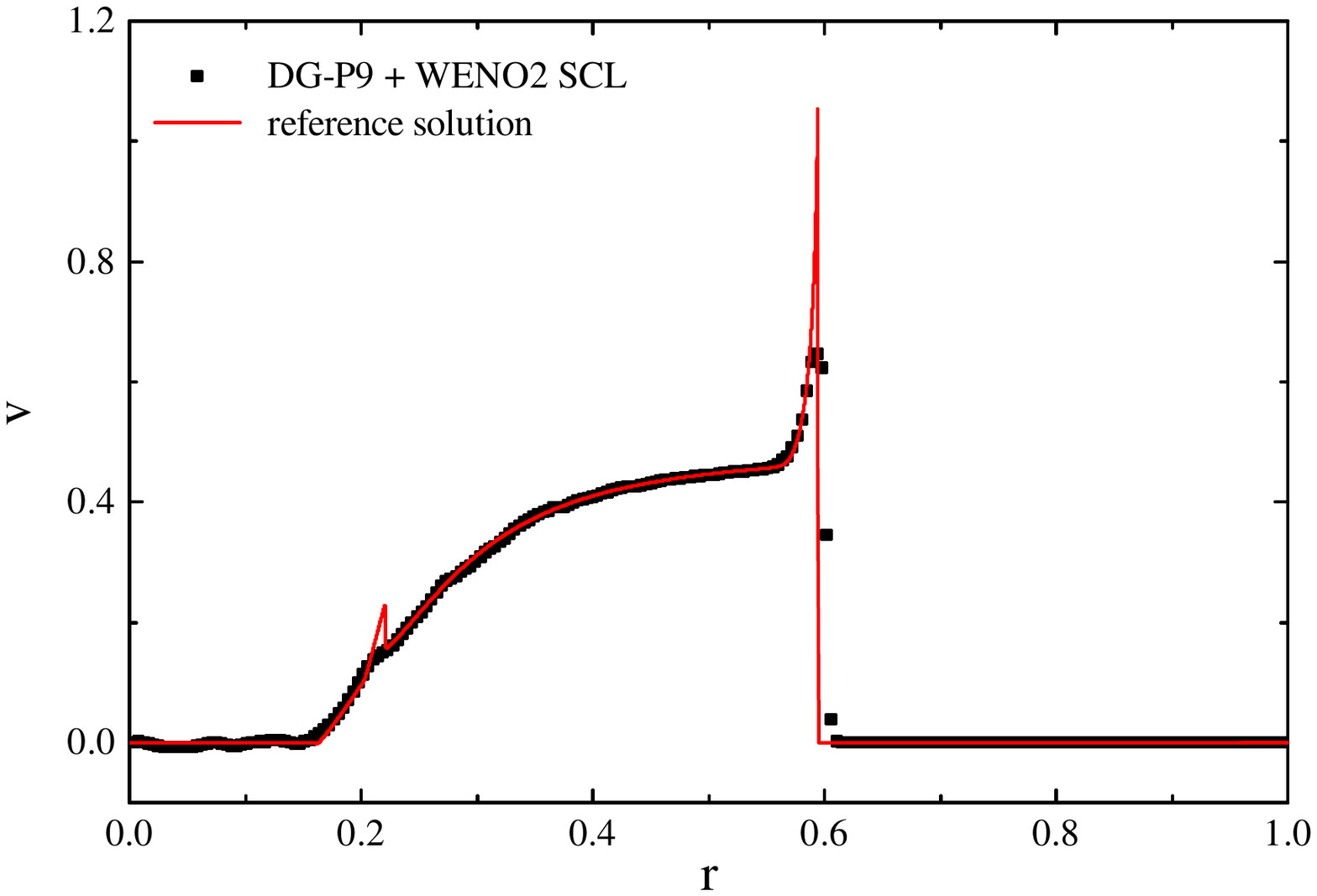}
\includegraphics[width=0.245\textwidth]{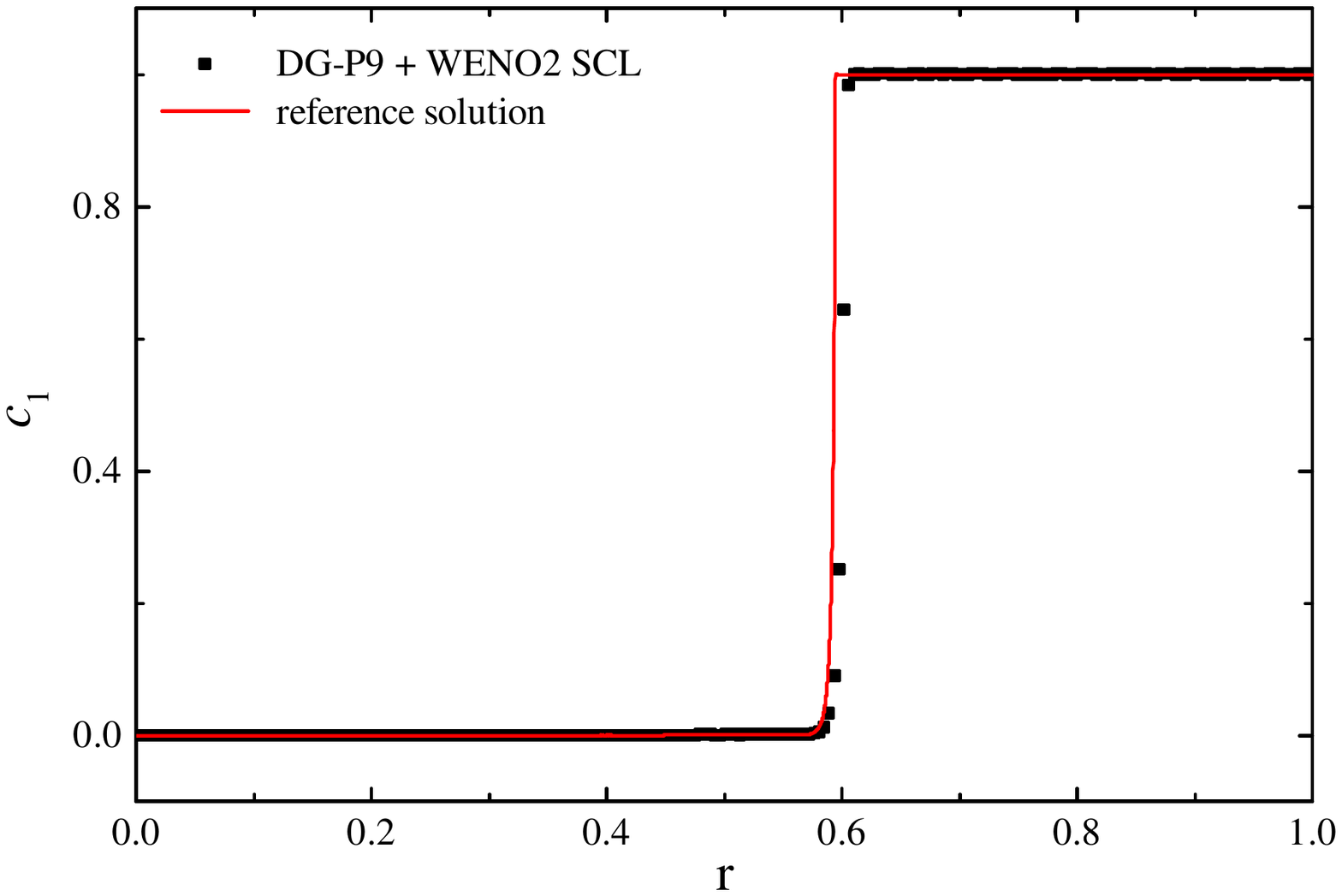}\\
\includegraphics[width=0.245\textwidth]{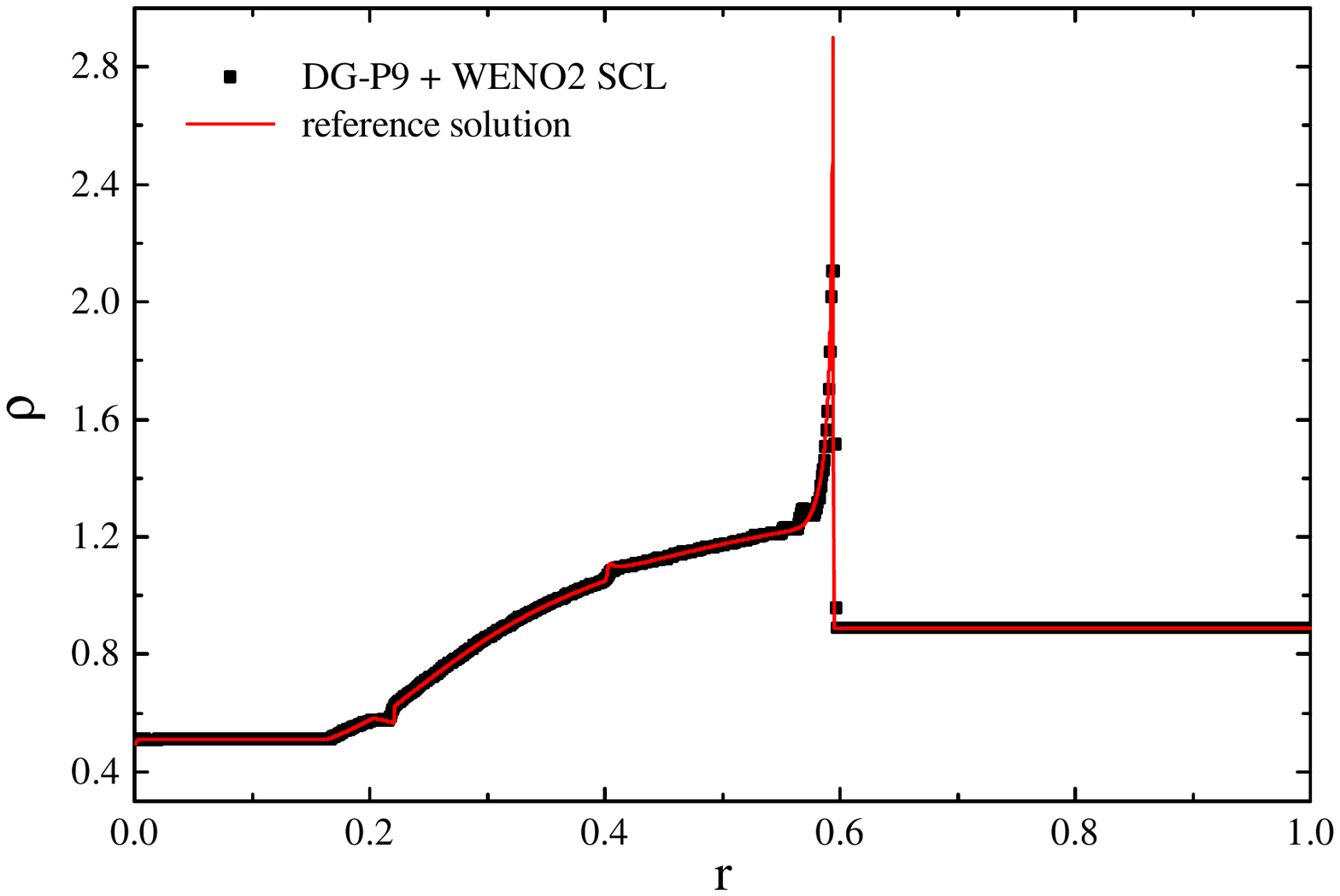}
\includegraphics[width=0.245\textwidth]{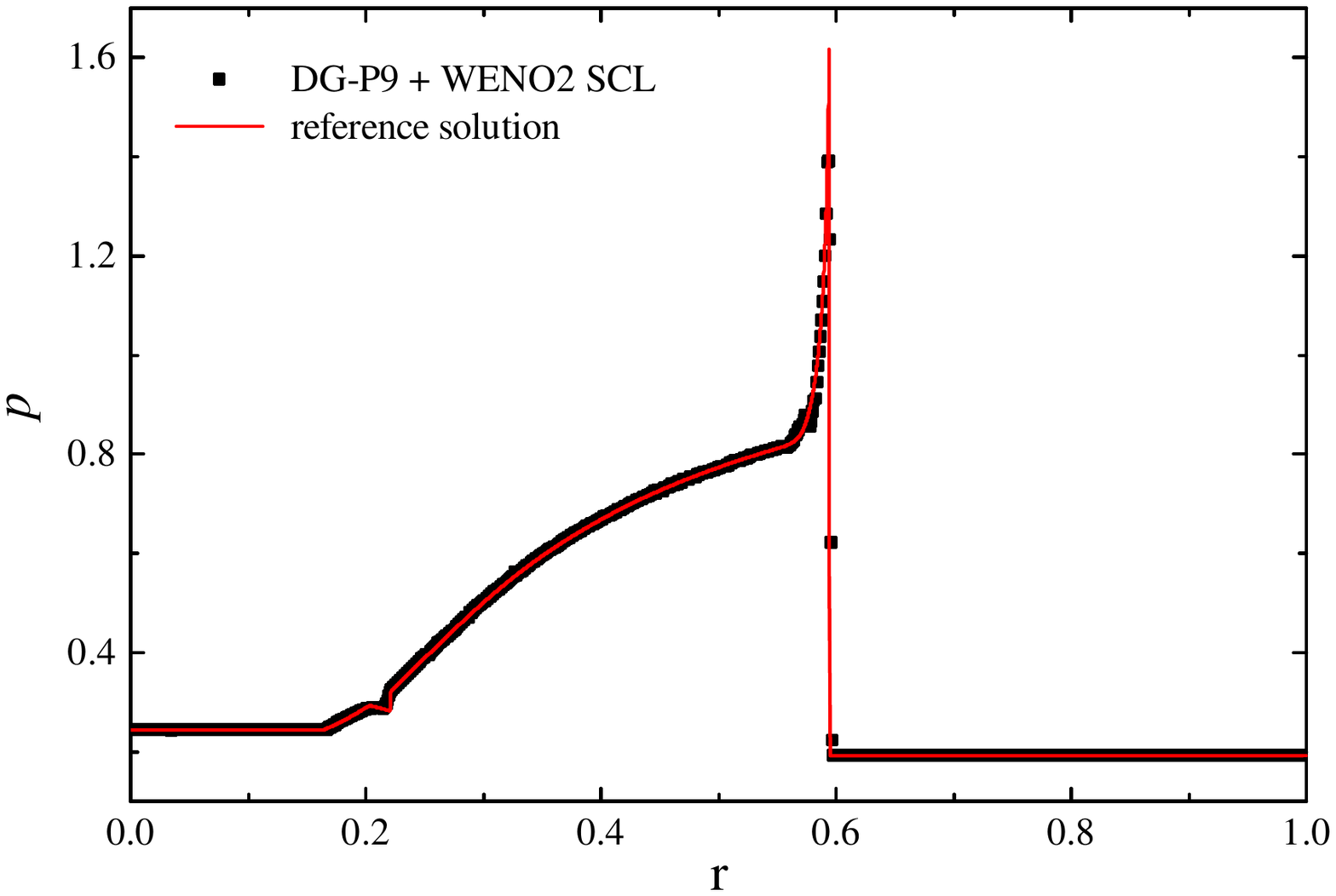}
\includegraphics[width=0.245\textwidth]{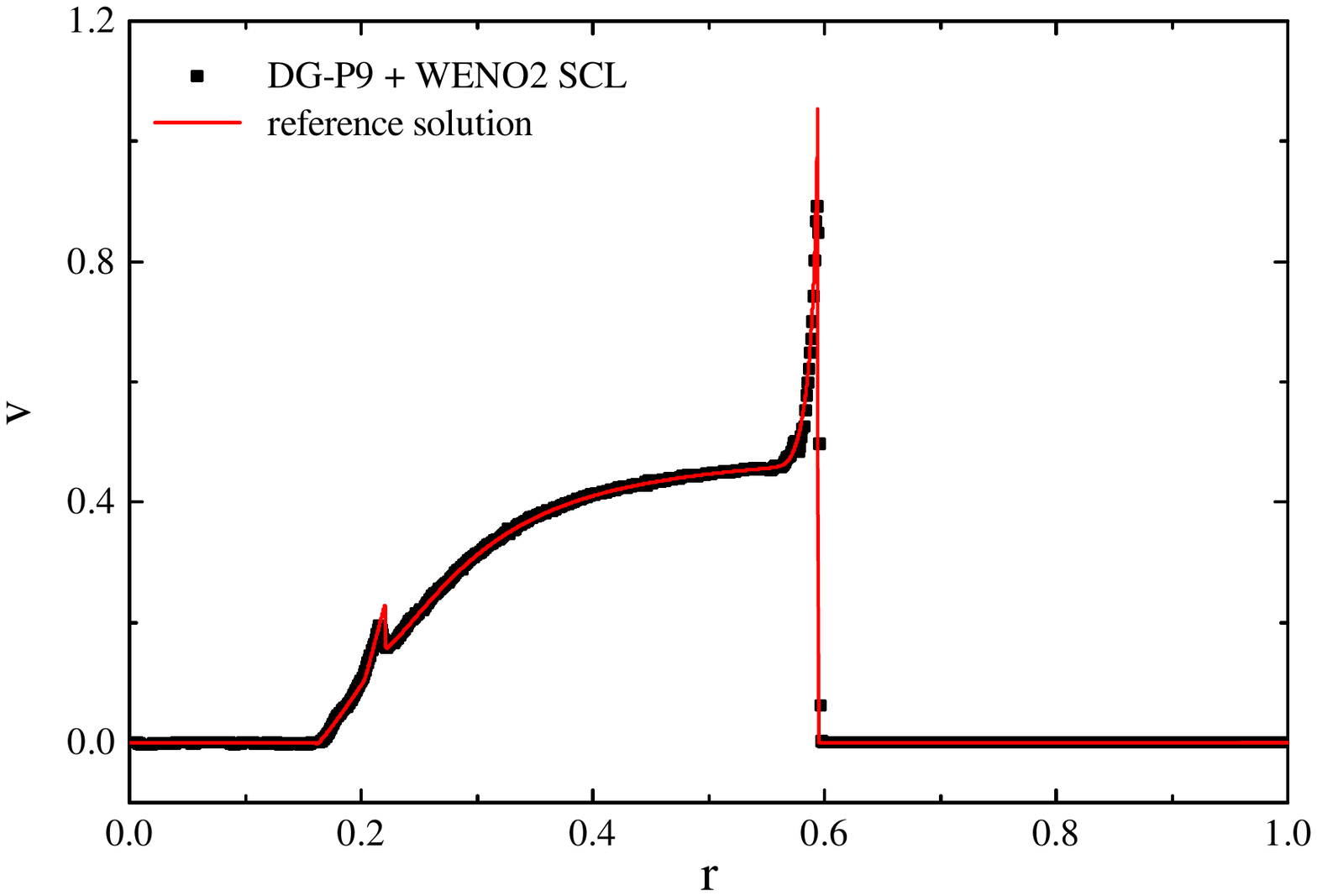}
\includegraphics[width=0.245\textwidth]{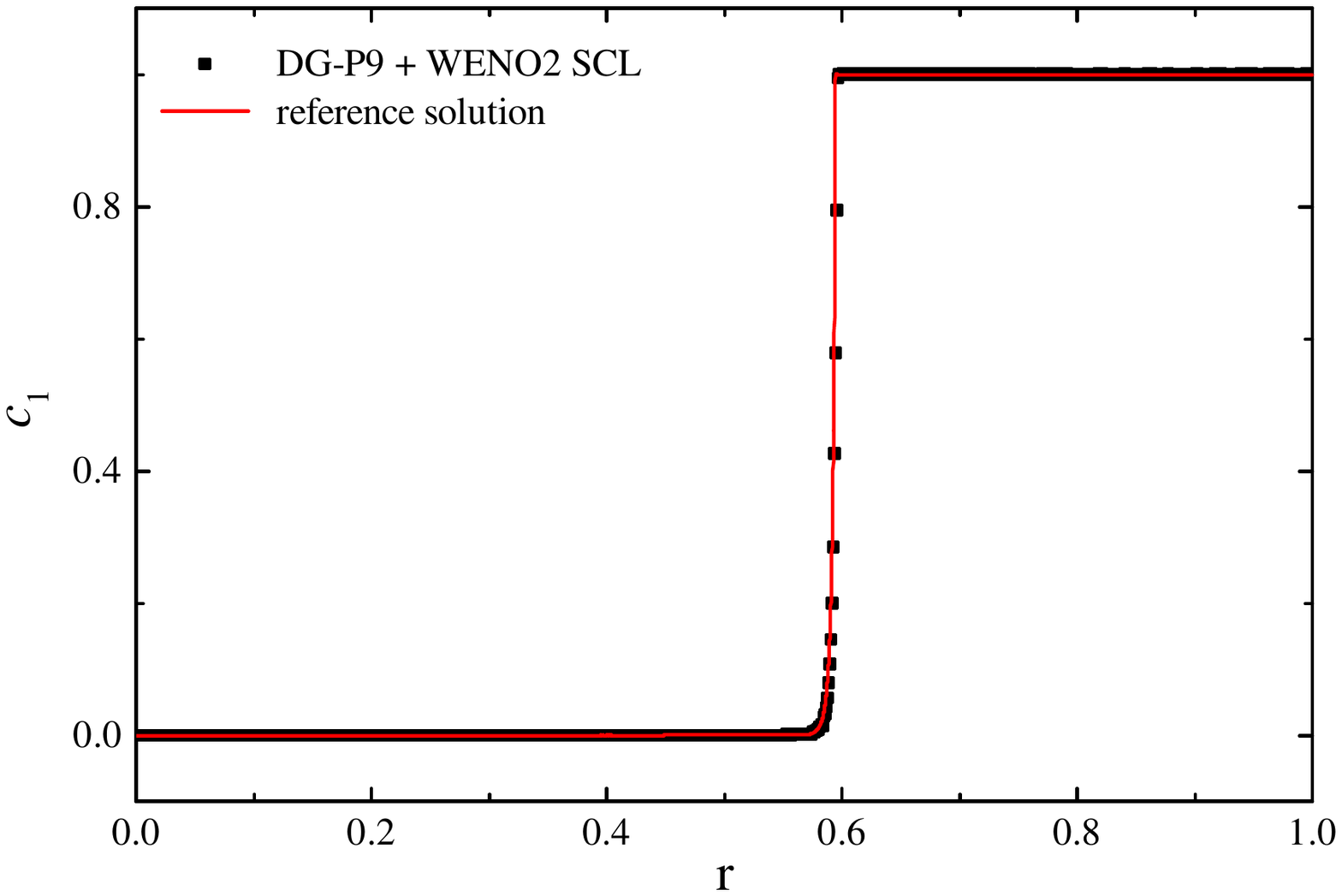}
\caption{\label{fig:cjdwf_2d_slice}
One dimensional cuts of the numerical solution for two-dimensional problem of the cylindrical detonation wave formation 
in a two-component medium with a ``fast'' reaction, which is presented in Figure~\ref{fig:cjdwf_2d},
obtained using the ADER-DG-$\mathbb{P}_{9}$ method on meshes $25 \times 25$ (top) and $101 \times 101$ (bottom) cells.
The graphs show the coordinate dependence of density $\rho$, pressure $p$, flow velocity $u$ and 
mass concentration $c_{1}$ of the reaction reagent (from left to right)
on the distance $r$ to the point $(0, 0)$ along the direction $(0, 1)$.
The black square symbols represent the subcells finite-volume representation of the numerical solution; 
the red solid lines represents the reference solution of the problem.
}
\end{figure*}

\begin{figure*}[h!]
\centering
\includegraphics[width=0.33\textwidth]{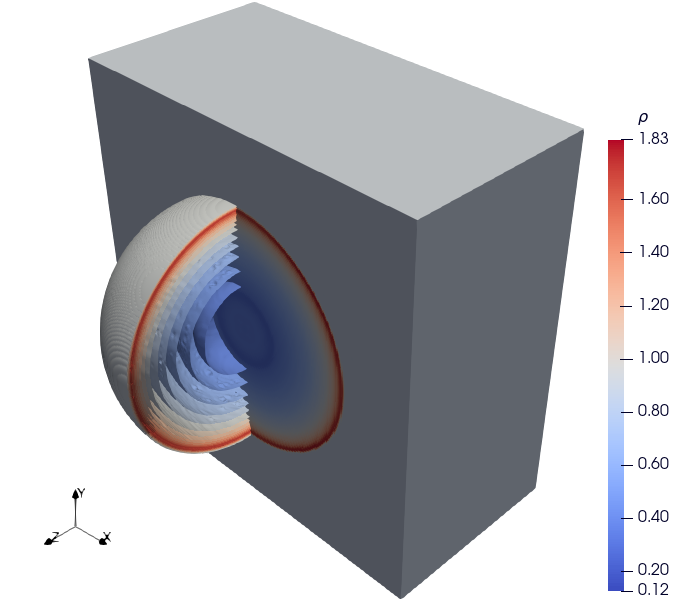}
\includegraphics[width=0.33\textwidth]{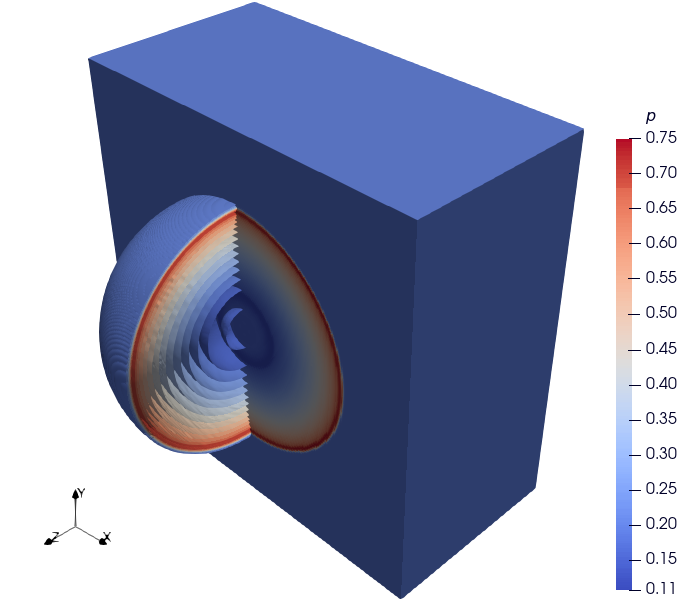}
\includegraphics[width=0.33\textwidth]{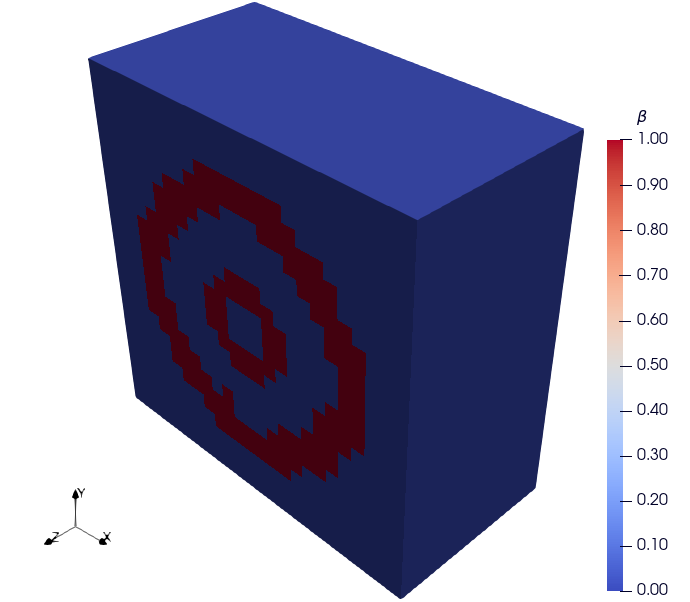}\\
\includegraphics[width=0.33\textwidth]{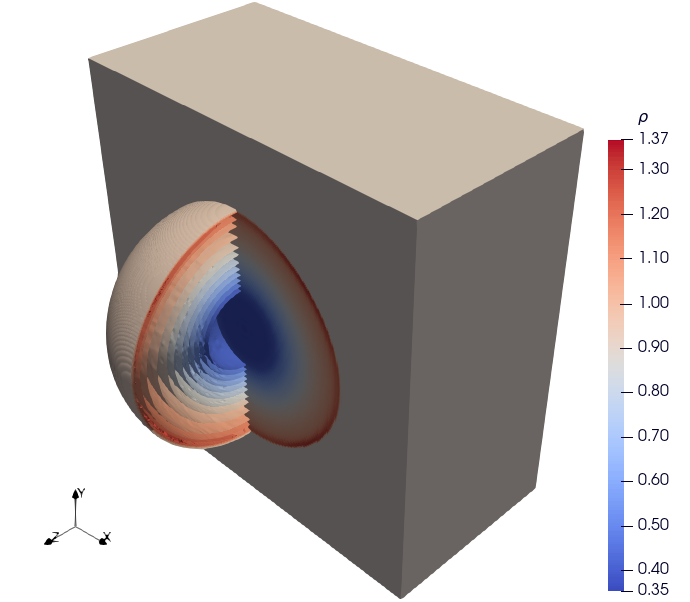}
\includegraphics[width=0.33\textwidth]{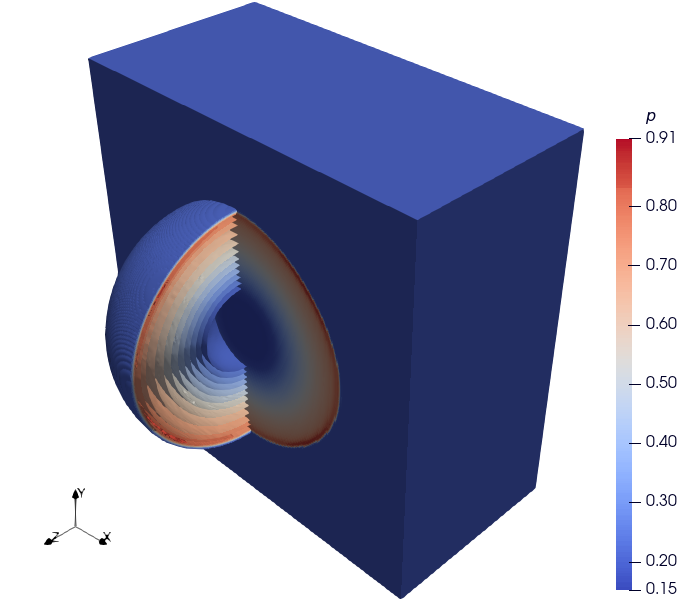}
\includegraphics[width=0.33\textwidth]{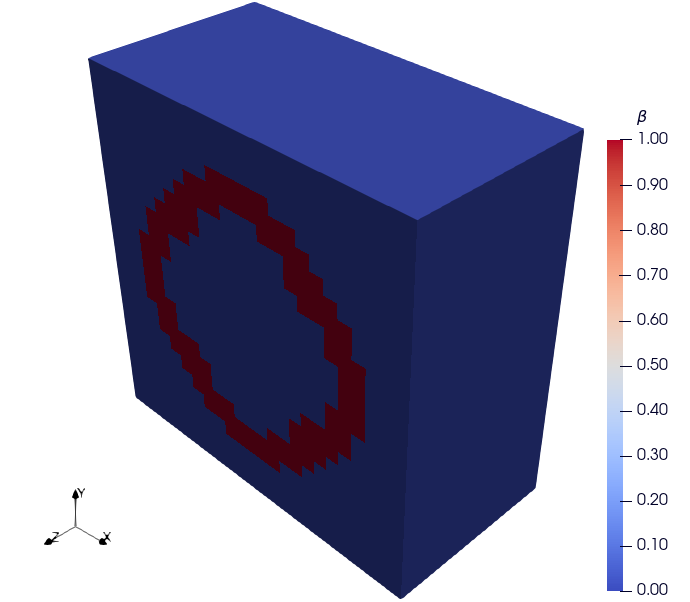}
\caption{\label{fig:cjdw_3d}
Numerical solution of the three-dimensional problem of the spherical detonation wave formation in a two-component medium with a ``slow'' reaction (top)
and a ``fast'' reaction (bottom) (weak and strong stiff cases, respectively; a detailed statement of the problem is presented in the text), 
obtained using the ADER-DG-$\mathbb{P}_{5}$ method at the final time $t_{\rm final} = 0.3$ (top) and $0.2$ (bottom) on meshes $25 \times 25 \times 25$.
The graphs show the coordinate dependence of the subcells finite-volume representation 
of density $\rho$, pressure $p$ and troubled cells indicator $\beta$.
The opaque fill represents the coordinate domain clip $\Omega_{\rm clip} = \left\{\mathbf{r}\, |\, \mathbf{r} \in [-1, +1]^{3} \land z \geqslant 0 \right\}$; 
the left and center columns also represent density and pressure isosurfaces, uniformly distributed between the boundary values.
}
\end{figure*}

\begin{figure}[h!]
\centering
\includegraphics[width=0.239\textwidth]{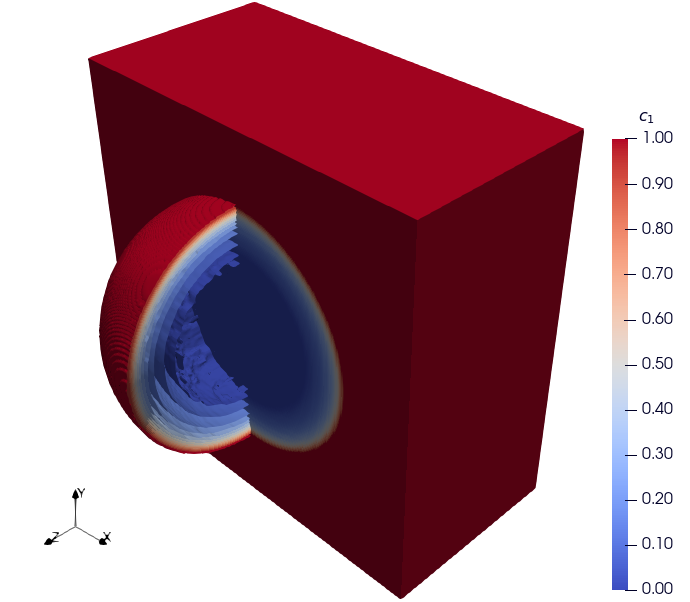}
\includegraphics[width=0.239\textwidth]{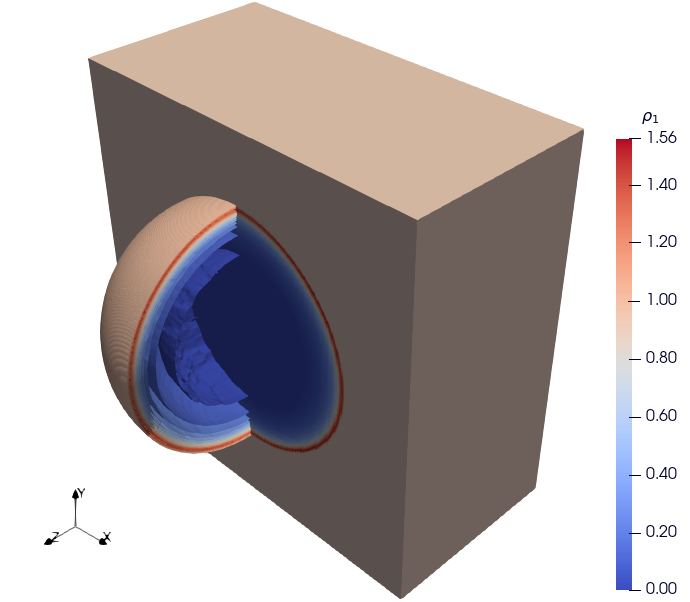}\\
\includegraphics[width=0.239\textwidth]{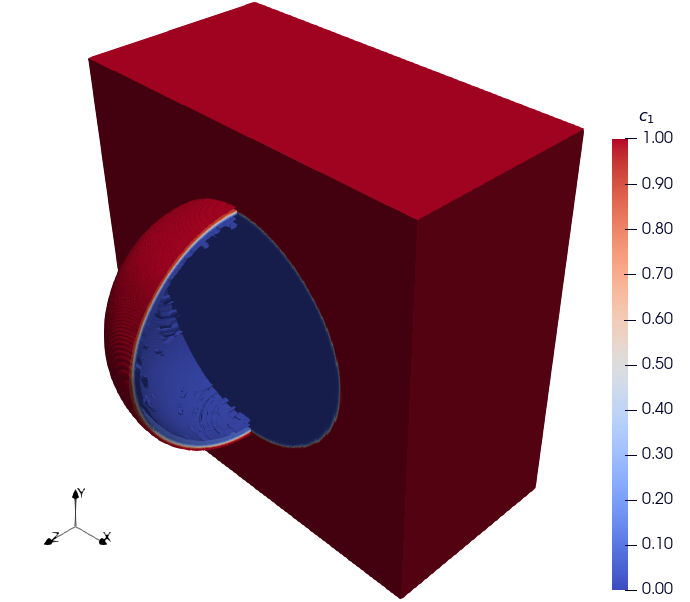}
\includegraphics[width=0.239\textwidth]{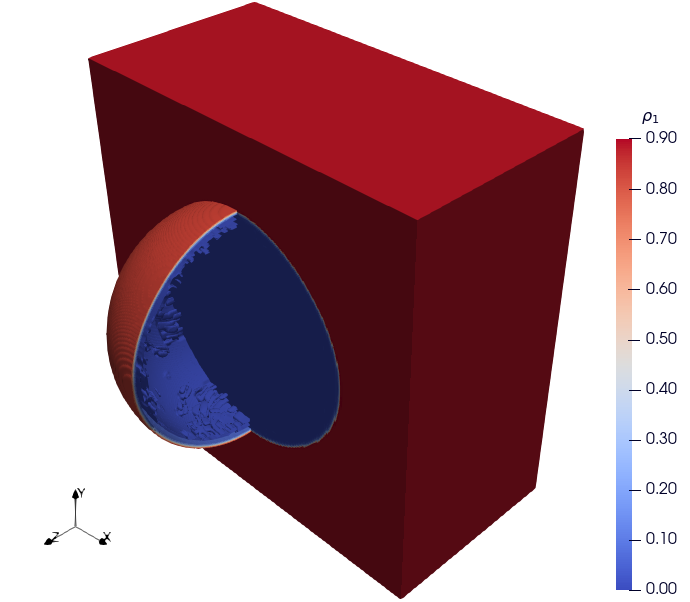}
\caption{\label{fig:cjdw_3d_comps}
The coordinate dependence of the subcells finite-volume representation 
of mass concentration $c_{1}$ and partial density $\rho_{1} = \rho c_{1}$ of the reaction reagent for 
numerical solution of the spherical detonation wave formation in a two-component medium with a ``slow'' reaction (top),
which is presented in Figure~\ref{fig:cjdw_3d},
on mesh $25 \times 25 \times 25$ cells.
}
\end{figure}

\begin{figure*}[h!]
\centering
\includegraphics[width=0.245\textwidth]{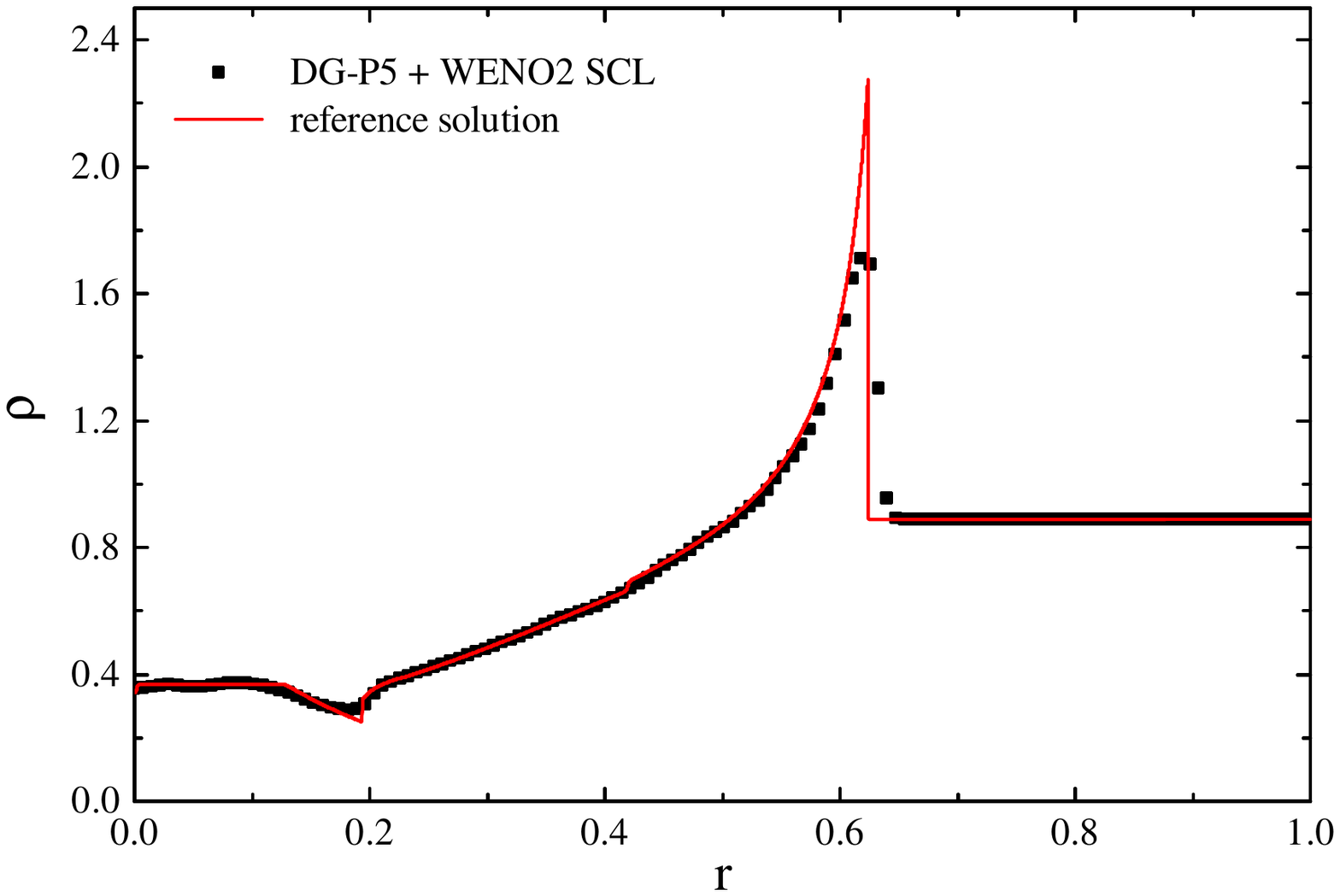}
\includegraphics[width=0.245\textwidth]{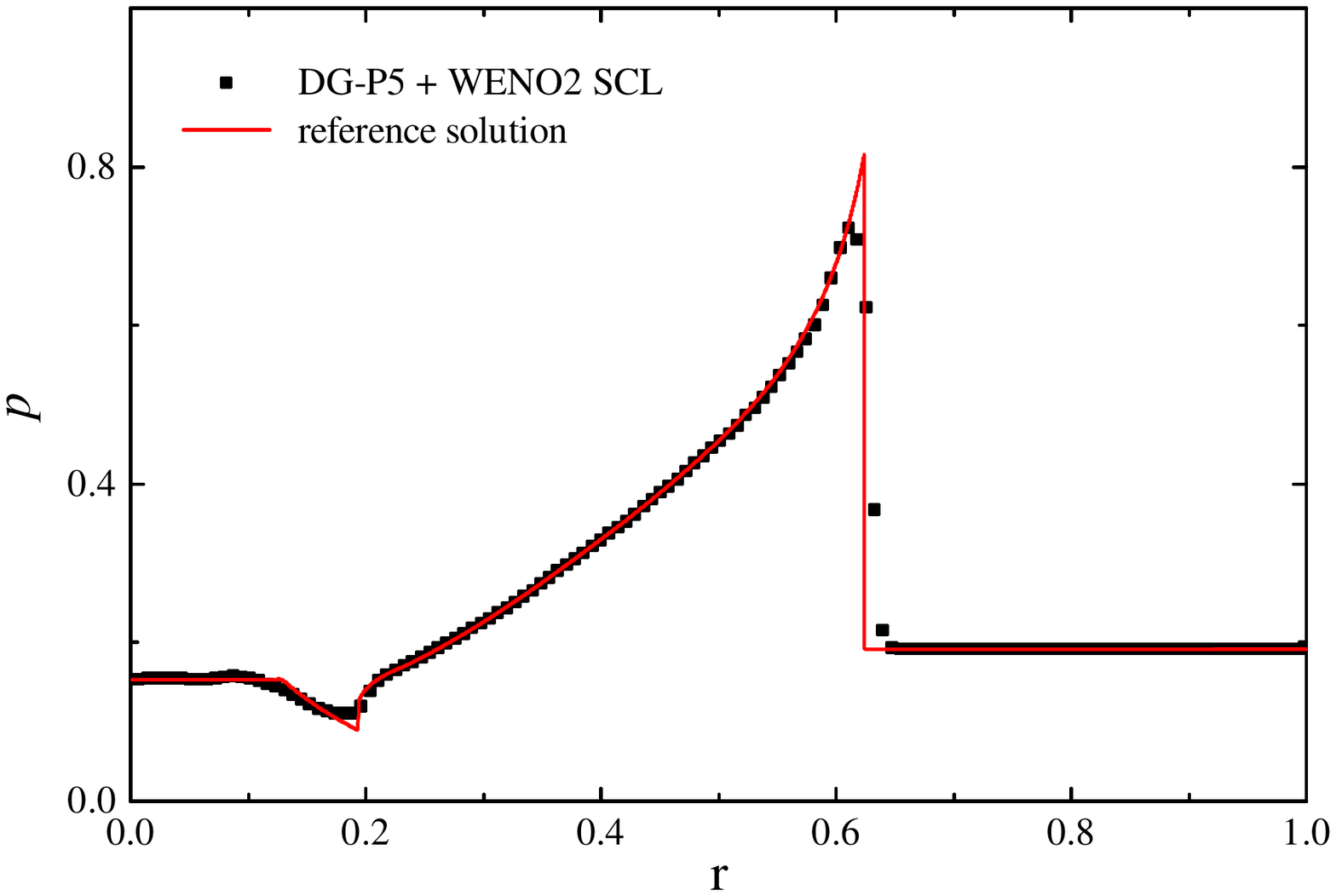}
\includegraphics[width=0.245\textwidth]{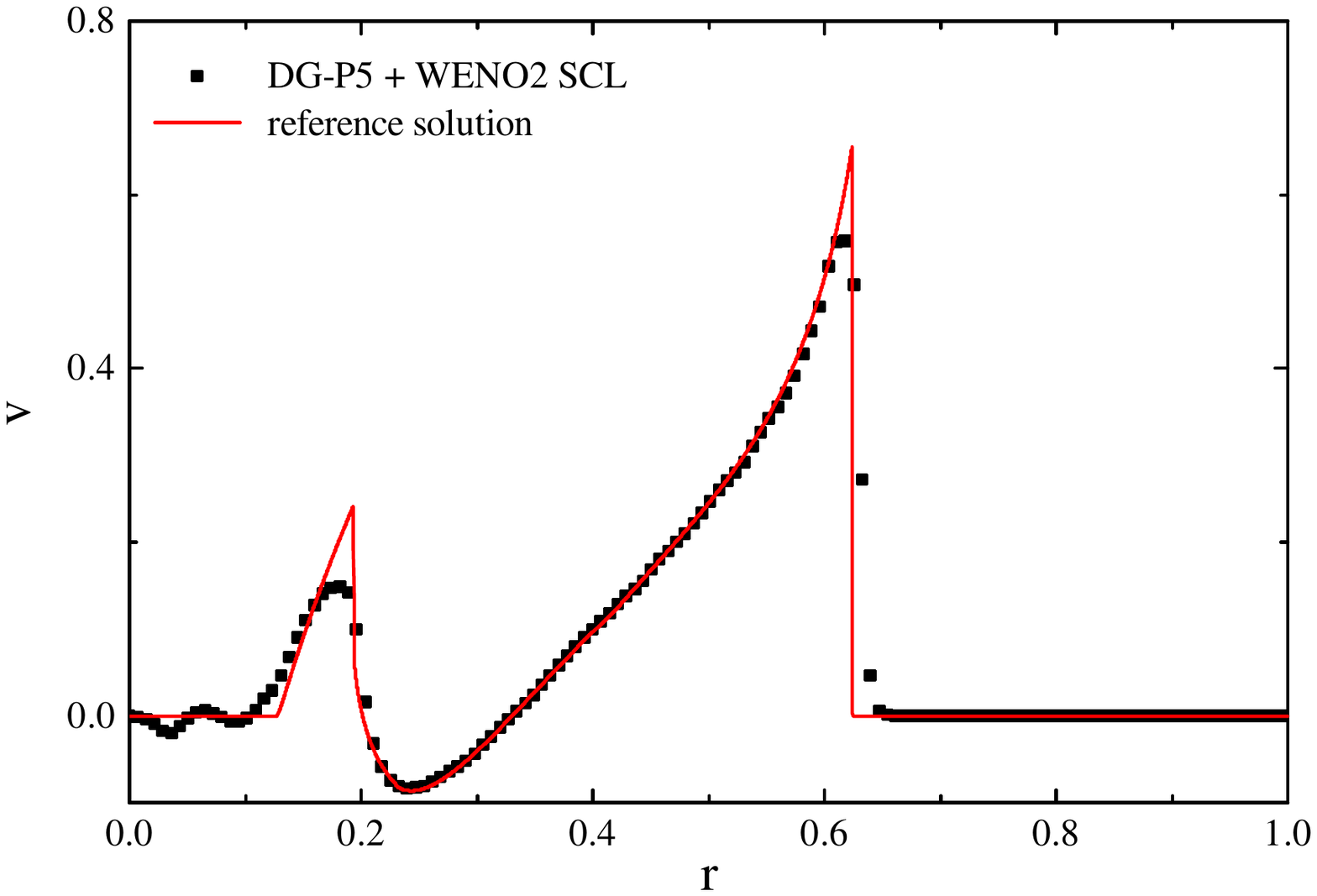}
\includegraphics[width=0.245\textwidth]{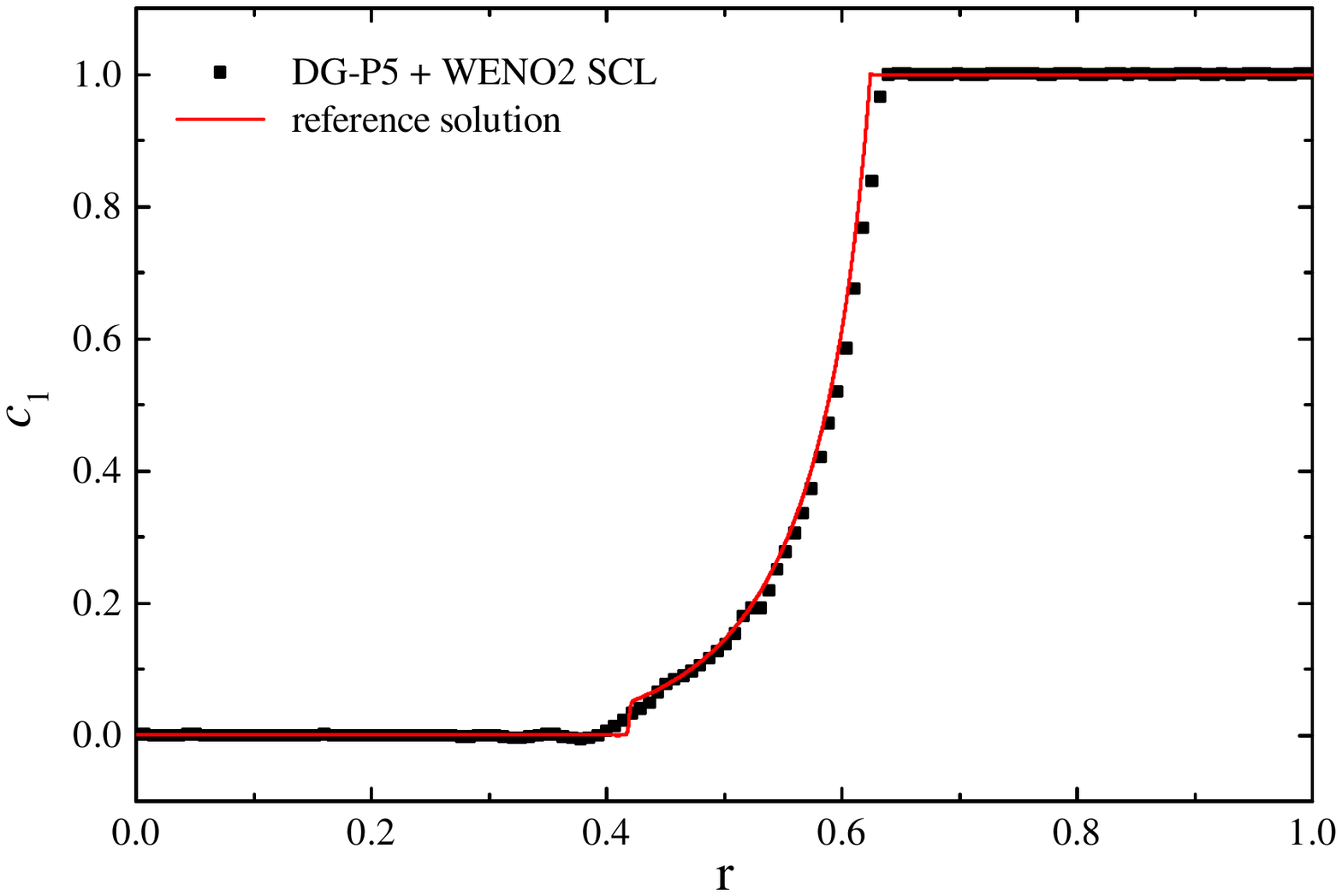}\\
\includegraphics[width=0.245\textwidth]{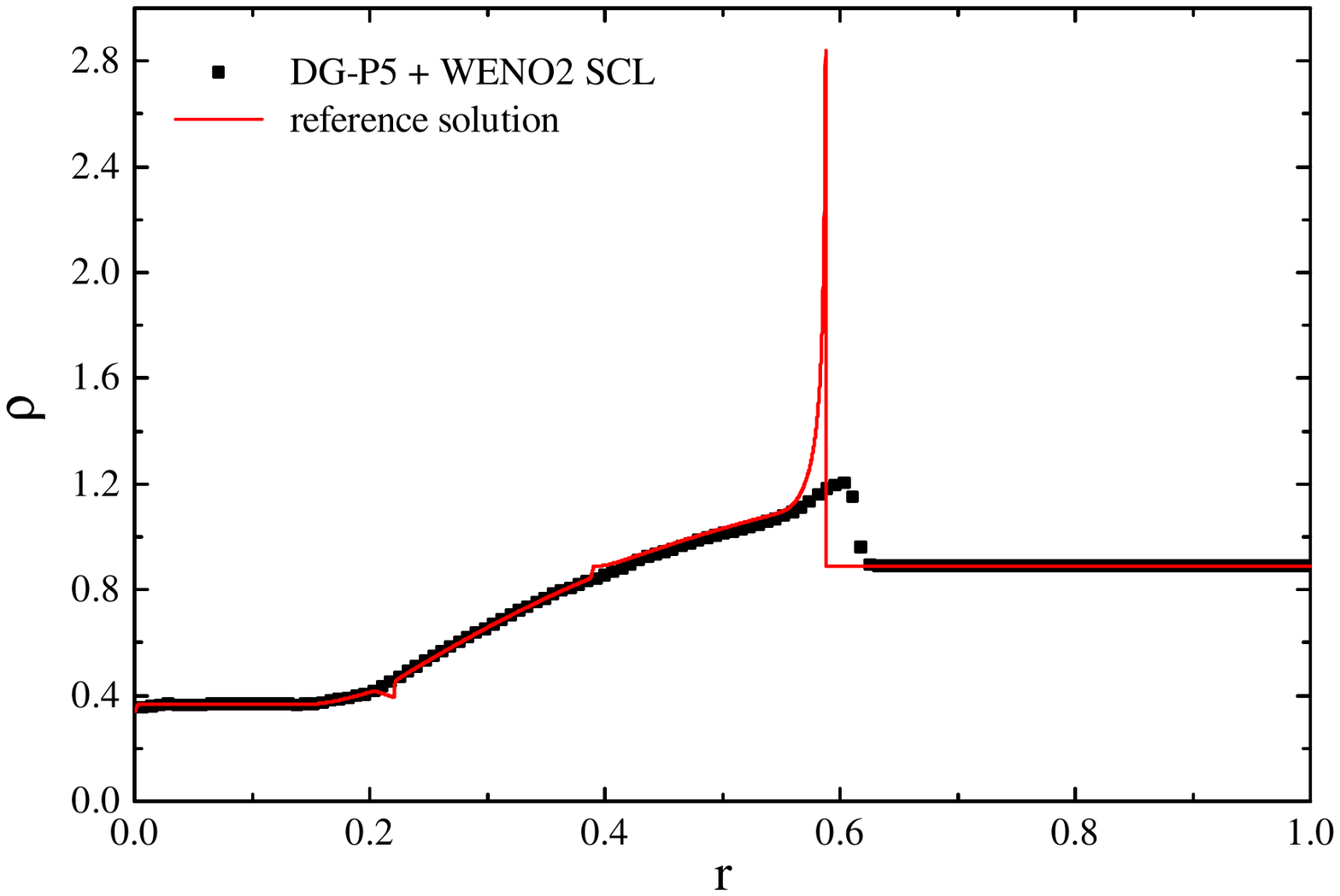}
\includegraphics[width=0.245\textwidth]{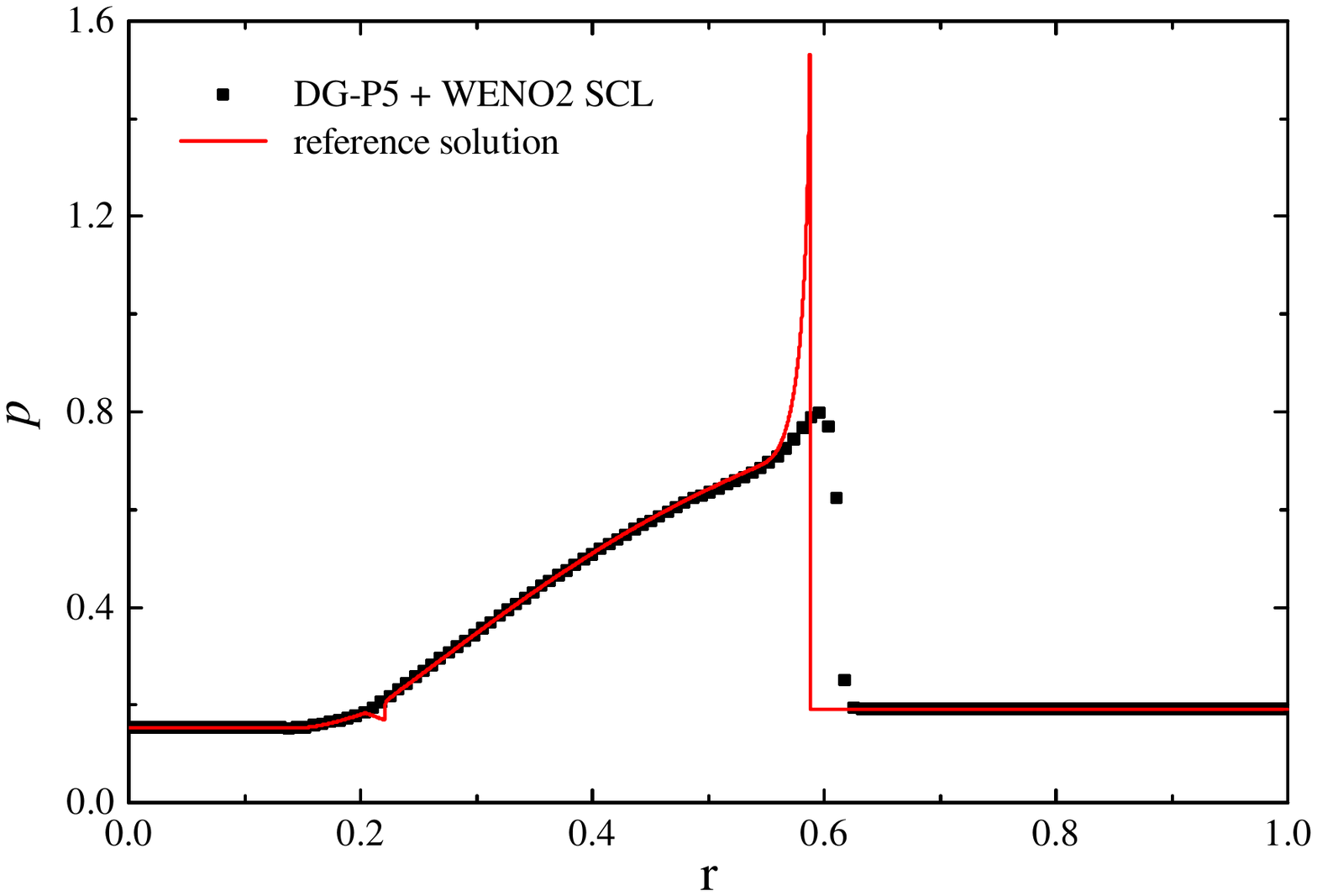}
\includegraphics[width=0.245\textwidth]{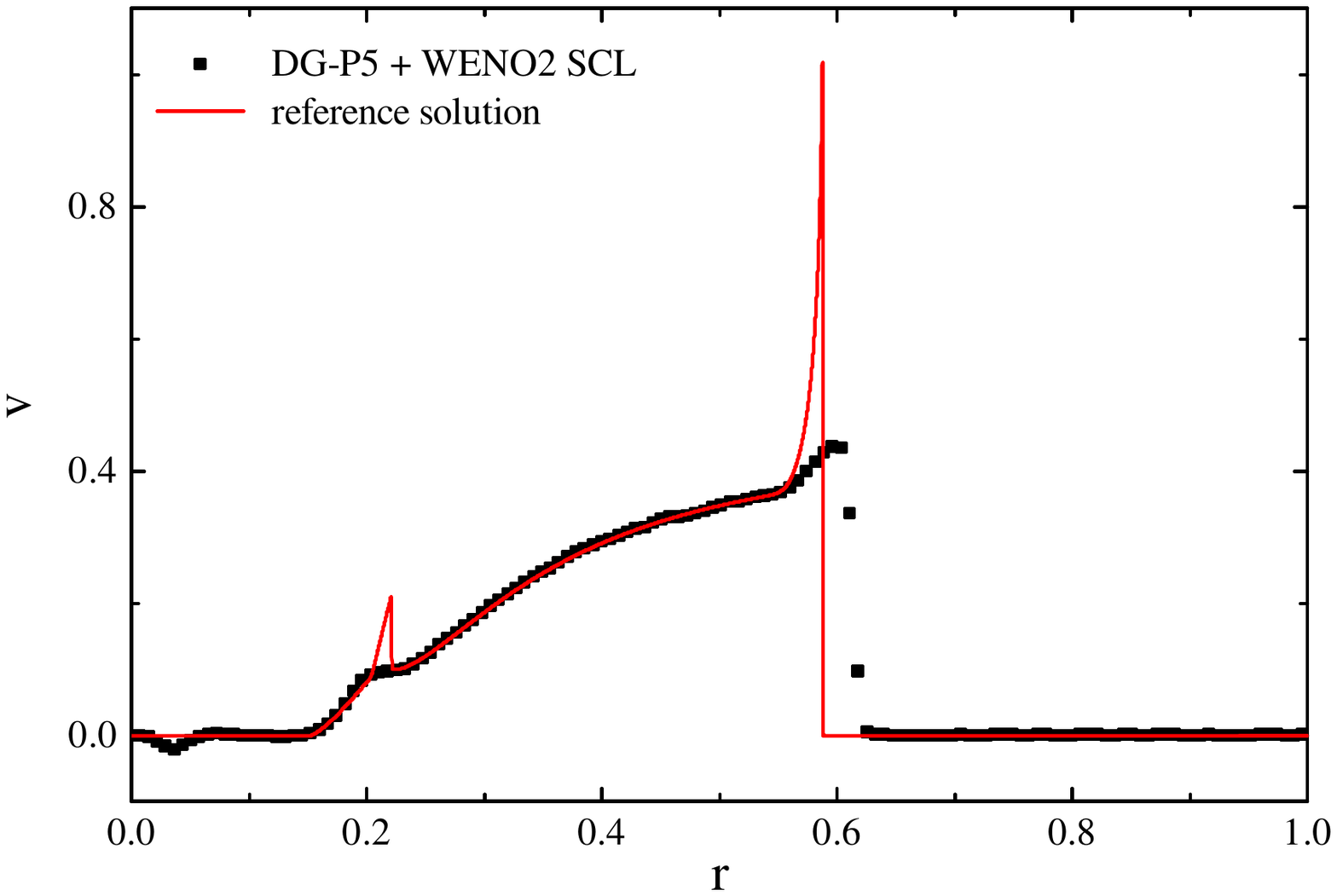}
\includegraphics[width=0.245\textwidth]{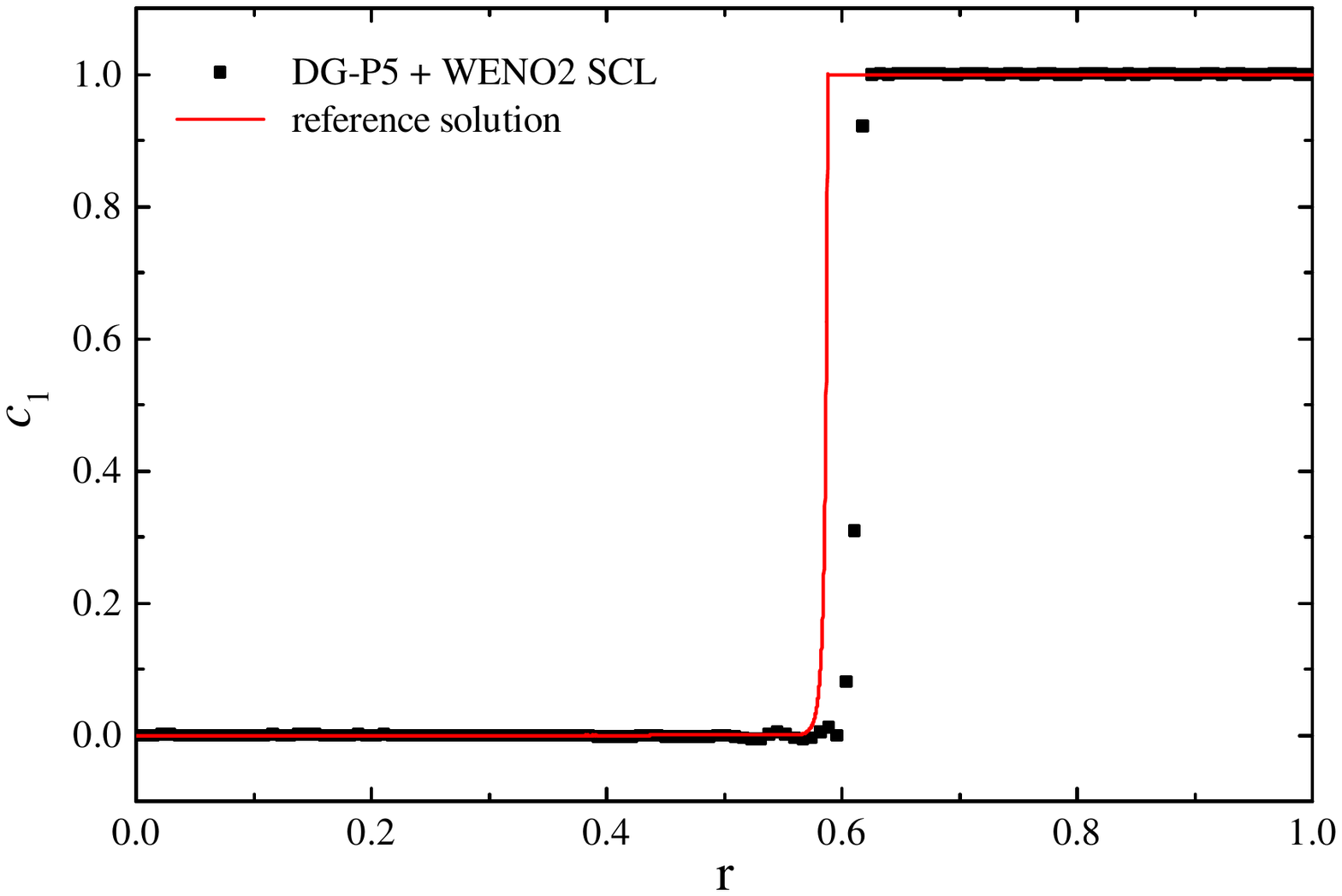}
\caption{\label{fig:cjdw_3d_slice}
One dimensional cuts of the numerical solution for the three-dimensional problem of the spherical detonation wave formation 
in a two-component medium with a ``slow'' (top) and a ``fast'' reaction (bottom), which is presented in Figure~\ref{fig:cjdw_3d},
obtained using the ADER-DG-$\mathbb{P}_{5}$ method on meshes $25 \times 25 \times 25$ cells.
The graphs show the coordinate dependence of density $\rho$, pressure $p$, flow velocity $u$ and 
mass concentration $c_{1}$ of the reaction reagent (from left to right)
on the distance $r$ to the point $(0, 0, 0)$ along the direction $(0, 0, 1)$.
The black square symbols represent the subcells finite-volume representation of the numerical solution; 
the red solid lines represents the reference solution of the problem.
}
\end{figure*}

The numerical solution obtained on a spatial mesh of $25 \times 25$ cells is also presented for comparison. The numerical solution obtained using ADER-DG-$\mathbb{P}_{9}$ method with a posteriori ADER-WENO2 finite volume limitation on a spatial mesh of $25 \times 25$ cells also shows a very high axial symmetry of the solution. However, the number of cell troubles in this case becomes significant and reaches $30\%$. This noticeably affects the numerical solution -- sharp inhomogeneities of density $\rho$, pressure $p$ and velocity $\mathrm{v}$ arise in the coordinate dependencies; in this case, inhomogeneities arise in the spatial region far behind the front of the detonation wave, where the reagent is almost completely burned out, so they are not noticeable on the coordinate dependencies of the mass concentration $c_{1}$ and density $\rho_{1}$ of the reagent.

Comparison of the numerical solution of a two-dimensional problem with a reference solution of a one-dimensional problem with a geometric source terms, which is presented in Figure~\ref{fig:cjdws_2d_slice}, shows that the solution on mesh of $100\times100$ cells compares very well point-wise with the reference solution. The width of the detonation front is impressive -- approximately $4$-$5$ \textit{subcells}, which demonstrates the good subgrid resolution of ADER-DG-$\mathbb{P}_{9}$ method with a posteriori ADER-WENO2 finite volume limitation. It should be noted that the cell size in one coordinate direction in this case is $2N+1=19$ subcells, so the detonation front in a two-dimensional problem ``expands'' by no more than a quarter of one cell. This result is comparable in quality to the results of solving a one-dimensional problem, however, it is quantitatively slightly worse in terms of the width of the detonation front. It should also be noted that all the main structures of the hydrodynamic flow that were resolved in the reference solution are also present and quite well resolved in the one-dimensional cut of the two-dimensional solution.

Comparison of the numerical solution of a two-dimensional problem on a $25\times25$ mesh with a reference solution shows that the width of the detonation front is also $4$-$5$ subcells, however, in coordinate units this is four times larger. Note that sharp inhomogeneities in density $\rho$, pressure $p$ and velocity $\mathrm{v}$, which are observed in two-dimensional coordinate dependencies, are similarly manifested in the one-dimensional cut of the solution.

The numerical solution to the problem of the formation and propagation of a detonation wave in media with a ``fast'' reaction was obtained using ADER-DG-$\mathbb{P}_{9}$ method with a posteriori ADER-WENO2 finite volume limitation on a spatial mesh of $100\times100$ cells. The simulation results are presented in Figure~\ref{fig:cjdwf_2d}. Figure~\ref{fig:cjdwf_2d_comps} shows the results for the mass concentration $c_{1}$ and density $\rho_{1}$ of the reagent. The numerical solution shows a very high axial symmetry of the solution. The fronts of hydrodynamic quantities are expressed sharply and are clearly visible in the numerical solution. The number of troubled cells does not exceed $13\%$, which is a fairly small value for multidimensional problems. The combustion of the reagent $A$ occurs in a very small spatial region behind the front of the shock wave, while a sharp combustion is also observed in the coordinate dependence of the density $\rho_{1}$ -- the reagent practically does not have time to be compressed in the front of the passing shock wave. The structure of the numerical solution is in good agreement with the features of the structure of the detonation wave, while the region of sharp burnout of the reagent coincides with the fronts of the shock wave, which indicates the formation of a single detonation structure. In this case, non-physical artifacts of the numerical solution noted in the works~\cite{chem_kin_hrs_weno, correct_det_wave_speed_2017} do not arise.

The numerical solution obtained on a spatial mesh of $25\times25$ cells is also presented for comparison. The numerical solution obtained using ADER-DG-$\mathbb{P}_{9}$ method with a posteriori ADER-WENO2 finite volume limitation on a spatial $25\times25$ mesh also shows a very high axial symmetry of the solution. However, the number of cell troubles in this case becomes significant and reaches $32\%$. This noticeably affects the numerical solution -- sharp inhomogeneities in density, pressure and velocity arise in the coordinate dependencies. In this case, inhomogeneities arise in the spatial region behind the front of the detonation wave, where the reagent is completely burned out, so they are not noticeable on the coordinate dependencies of the mass concentration $c_{1}$ and density $\rho_{1}$ of the reagent.

Comparison of the numerical solution of a two-dimensional problem with a reference solution of a one-dimensional problem with a geometric source terms, which is presented in Figure~\ref{fig:cjdwf_2d_slice}, shows that the solution on mesh of $100\times100$ cells compares very well point-wise with the reference solution. The width of the detonation front is impressive -- about $3$-$4$ \textit{subcells}, which demonstrates the good subgrid resolution of ADER-DG-$\mathbb{P}_{9}$ method with a posteriori ADER-WENO2 finite volume limitation -- detonation front in a two-dimensional problem ``expands'' by no more than a quarter of one cell. This result is comparable in quality to the results of solving a one-dimensional problem, however, it is quantitatively somewhat worse in terms of the width of the detonation front. However, the amplitude value of the density $\rho$ at the wave front in the numerical solution of two-dimensional problem is approximately $30\%$ lower than in the reference solution; and for pressure $p$ -- by $\sim 15\%$. This is due to the very high sharpness of the change in density $\rho$ and pressure $p$ behind the detonation front, and the subcells spatial step the mesh is not enough to display these dependencies. It should also be noted that all the main structures of the hydrodynamic flow that were resolved in the reference solution are also present and quite well resolved in the one-dimensional cut of the two-dimensional solution.

Comparison of the numerical solution of a two-dimensional problem on a spatial $25\times25$ mesh with a reference solution shows that the width of the detonation front is also $4$-$5$ subcells. However, in this case, the detonation front is approximately $1$ subcell ahead of the detonation front in the reference solution. This inaccuracy is especially clearly observed in the coordinate dependence of the mass of the reagent concentration. We also note that sharp inhomogeneities in density, pressure and velocity, which are observed in two-dimensional coordinate dependencies, are similarly manifested in the one-dimensional cutoff of the solution.

\paragraph{Spherical ZND-detonation waves}

The numerical solution to the three-dimensional problem of the formation and propagation of a detonation wave in a medium with a ``slow'' reaction was obtained using ADER-DG-$\mathbb{P}_{5}$ method with a posteriori ADER-WENO2 finite volume limitation on a spatial mesh of $25\times25\times25$ cells. The simulation results are presented in Figure~\ref{fig:cjdw_3d}. Figure~\ref{fig:cjdw_3d_comps} shows the results for the mass concentration and density of the component. The numerical solution shows a very high spherical symmetry of the solution. The fronts of hydrodynamic quantities are expressed sharply and are clearly visible in the numerical solution. The number of troubled cells does not exceed $16\%$, which is a fairly small value for multidimensional problems. The combustion of the reagent occurs smoothly, and the compression of the unburned reagent in the front of the shock wave ascending into the structure of the detonation front is clearly observed.

Comparison of the numerical solution of a two-dimensional problem with a reference solution of a one-dimensional problem with a geometric source terms, which is presented in Figure~\ref{fig:cjdw_3d_slice}, shows that the solution on mesh of $25\times25\times25$ cells is satisfactorily compared point-wise with the reference solution -- the smooth components of the numerical solution correspond well to the reference solution, the discontinuous components of the numerical solution demonstrate a finite width, however, they are geometrically correctly located. The width of the detonation front is about $5$-$7$ subcells. This result is worse in quality compared to the solution to the two-dimensional cylindrical problem. However, it is necessary to take into account that the mesh size and the degree of polynomials $N$ used in the DG representation in this case are much smaller. It should also be noted that all the main structures of the hydrodynamic flow that were resolved in the reference solution are also present and well resolved in the one-dimensional corner of the three-dimensional solution. Some non-physical velocity oscillations are observed in the central region of the flow.

The numerical solution to the three-dimensional problem of the formation and propagation of a detonation wave in media with a ``fast'' reaction was obtained using ADER-DG-$\mathbb{P}_{5}$ method with a posteriori ADER-WENO2 finite volume limitation on a spatial mesh of $25\times25\times25$ cells. The simulation results are presented in Figure~\ref{fig:cjdw_3d}. Figure~\ref{fig:cjdw_3d_comps} shows the results for the mass concentration and density of the component. The numerical solution shows a very high spherical symmetry of the solution. The fronts of hydrodynamic quantities are expressed sharply and are clearly visible in the numerical solution. The number of troubled cells does not exceed $14\%$, which is a fairly small value for multidimensional problems. The combustion of the reagent occurs abruptly, and sharp combustion is also observed in the coordinate dependence of the density of the component -- the reagent practically does not have time to compress in the front of the passing shock wave. The structure of the numerical solution is in good agreement with the features of the structure of the ZND detonation wave, while the region of sharp burnout of the reagent coincides with the fronts of the shock wave, which indicates the formation of a single detonation structure. In this case, non-physical artifacts of the numerical solution noted in the works~\cite{chem_kin_hrs_weno, correct_det_wave_speed_2017} do not arise.

Comparison of the numerical solution of a two-dimensional problem with a reference solution of a one-dimensional problem with a geometric source terms, which is presented in Figure~\ref{fig:cjdw_3d_slice}, shows that the solution on a mesh of $25\times25\times25$ cells is satisfactorily compared point-wise with the reference solution -- the smooth components of the numerical solution correspond well to the reference solution. The width of the detonation front is about $5$-$7$ subcells, so the detonation front in a three-dimensional problem ``expands'' by no more than one cell. It must be emphasized that the detonation front in the numerical solution is approximately $4$ subcells ahead of the detonation front in the reference solution. This result is significantly worse in quality compared to the solution to the two-dimensional cylindrical problem. However, it is necessary to take into account that the grid size $25\times25\times25$ vs $100\times100$ and the degree of polynomials $N = 5$ vs $9$ used in the DG representation in this case. The amplitude value of the density $\rho$ and pressure $p$ at the wave front in the numerical solution of the fully three-dimensional problem is lower than in the reference solution. This is due to the very sharp change in density $\rho$ and pressure $p$ behind the detonation front, and the subcell spatial step of the mesh is not enough to display these dependencies. It should also be noted that not all the main structures of the hydrodynamic flow that were resolved in the reference solution are also present and well resolved in the one-dimensional cut of the three-dimensional solution. These features are associated, first of all, with the small size $25\times25\times25$ of the mesh and the degree of polynomials $N=5$ used in the DG representation.

\subsection{Detonation waves in a non-uniform medium}
\label{sec:detonation_waves:dwbi}

Demonstration of the method's capabilities for solving problems of the formation and dynamic evolution of multidimensional detonation waves was carried out using the example of two specific problems: interaction of a detonation wave with an inert bubble and interaction of a detonation wave with several inert bubbles. These problems are conceptually similar to the already considered in Subsection~\ref{sec:apps_cgd_problems:sbi_2d} problem of shock-bubble interaction.

\paragraph{Interaction of a detonation wave with an inert bubble}

\begin{figure}[h!]
\centering
\includegraphics[width=0.49\textwidth]{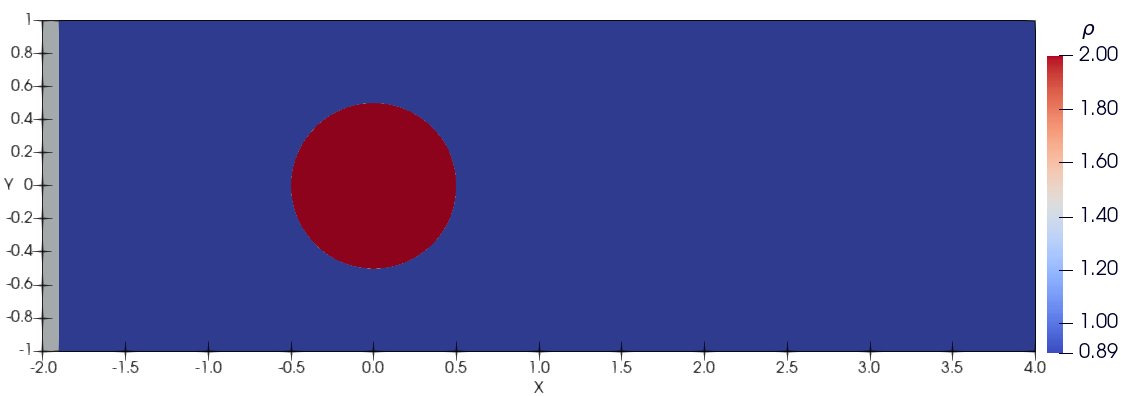}
\includegraphics[width=0.49\textwidth]{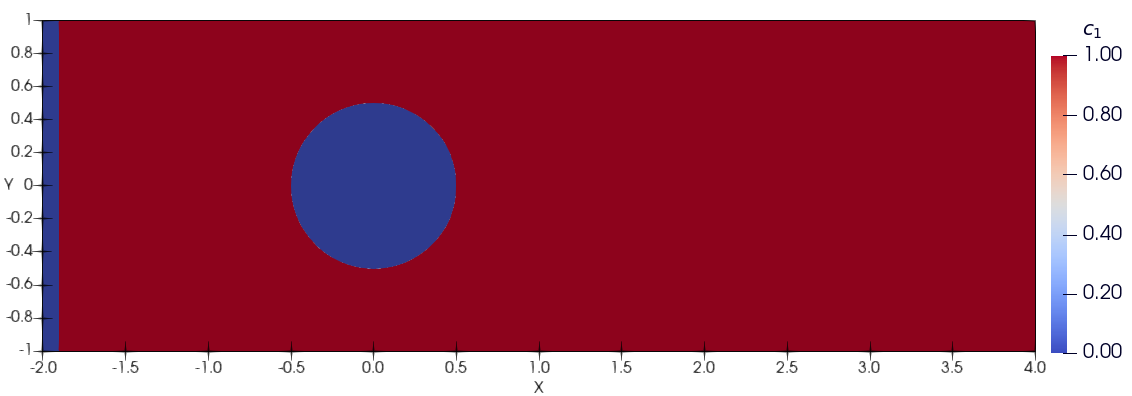}
\caption{\label{fig:dwbi_2d_init}
The initial conditions for coordinate dependency of density $\rho$ (top) and mass concentration $c_{1}$ of the reaction reagent (bottom)
in the two-dimensional problem of interaction between a detonation wave and a reaction-inert bubble (a detailed statement of the problem is presented in the text).
}
\end{figure}

A numerical solution to the problem of the interaction between a detonation wave and an inert bubble made it possible to quantitatively determine the capabilities of the ADER-DG-$\mathbb{P}_{N}$ method with ADER-WENO2 finite volume a posteriori limiter for simulation complex multidimensional problems of the formation and movement of detonation waves in situations where the geometric and physical properties of the medium in which detonation processes are considered imply the occurrence of refraction and diffraction phenomena of detonation waves, the formation of multiple reflected and interacting shock waves, contact discontinuities, rarefaction waves, as well as vortex generation domains and vortex streets.

Computational coordinate domain $\Omega = [-2, +4]\times[-1, +1]$. The initial conditions were chosen in the following form:
\begin{equation}
\begin{split}
&\rho = \left\{
\begin{array}{ll}
1.4, & \mathrm{if}\ x \leqslant -1.9;\\
2.0, & \mathrm{if}\ r \leqslant 0.5;\\
0.887565, & \mathrm{if}\ x >\, -1.9 \land r >\, 0.5;
\end{array}
\right.\\
&p = \left\{
\begin{array}{ll}
1.0, & \mathrm{if}\ x \leqslant -1.9;\\
0.191709, & \mathrm{if}\ x >\, -1.9;
\end{array}
\right.\\
&u = \left\{
\begin{array}{ll}
0.577350, & \mathrm{if}\ x \leqslant -1.9;\\
0.0, & \mathrm{if}\ x >\, -1.9;
\end{array}
\right.\\
&v = 0.0;\\
&c_{1} = \left\{
\begin{array}{ll}
10^{-14}, & \mathrm{if}\ x \leqslant -1.9 \lor r \leqslant 0.5;\\
1.0, & \mathrm{if}\ x >\, -1.9 \land r >\, 0.5;
\end{array}
\right.\\
&c_{2} = \left\{
\begin{array}{ll}
1.0, & \mathrm{if}\ x \leqslant -1.9 \lor r \leqslant 0.5;\\
10^{-14}, & \mathrm{if}\ x >\, -1.9 \land r >\, 0.5;
\end{array}
\right.\\
\end{split}
\end{equation}
where $r^{2} = x^{2} + y^{2}$ determines the distance to the point $(0, 0)$. A small value $10^{-14}$ of mass concentrations $c_{1}$ and $c_{2}$, instead of strictly $0$, was chosen to prevent the occurrence of negative concentrations immediately at the start of the calculation process, which could lead to a meaninglessly large increase in the number of troubled cells in the solution. From the point of view of energy release and flow energy balance, these values of reagent concentration do not have any significant effect. 

The selected initial conditions determine a bubble of radius $R = 0.5$ with density $\rho_{0} = 2.0$, center of which is located at the point $(0, 0)$, and a CJ detonation wave with parameters that were calculated in Subsection~\ref{sec:detonation_waves:cjdw_1d}, which is located on the line $x = -1.9$. The initial conditions for coordinate dependency of density $\rho$ and mass concentration $c_{1}$ of the reaction reagent are presented in Figure~\ref{fig:dwbi_2d_init}. The boundary conditions were chosen as follows: on the left boundary is the exact solution for the \newcoloringtext{detonation front, specified from the CJ conditions}, on the right boundary are the conditions of free outflow, periodic boundary conditions were set on top and bottom boundaries, which under the symmetry of initial conditions of the problem are equivalent to the solid wall boundary conditions. Results were obtained in the cases of a ``slow'' reaction, which corresponds to weak stiffness in the solution, and a ``fast'' reaction, which corresponds to strong stiffness in the solution. The final time has been chosen $t_{\rm final} = 3.5$. The adiabatic index $\gamma = 1.4$. The Courant number $\mathtt{CFL} = 0.9$.

\begin{figure*}[h!]
\centering
\includegraphics[width=0.33\textwidth]{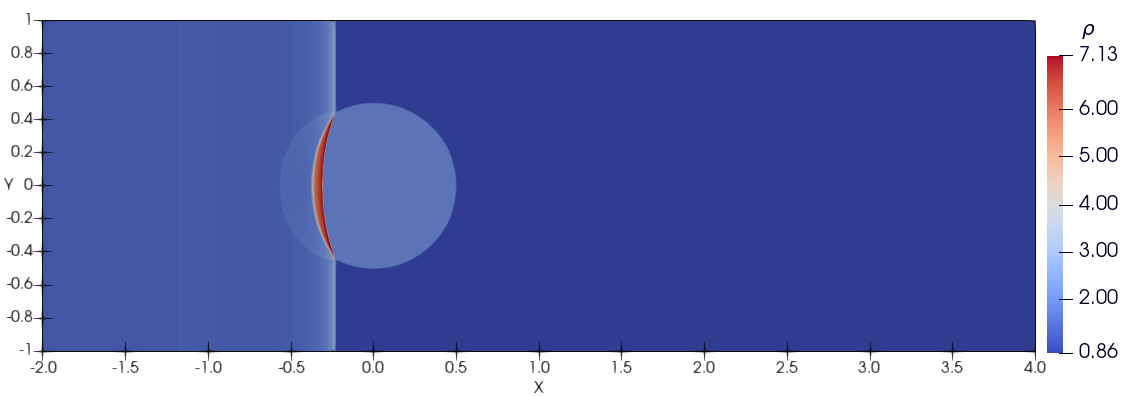}
\includegraphics[width=0.33\textwidth]{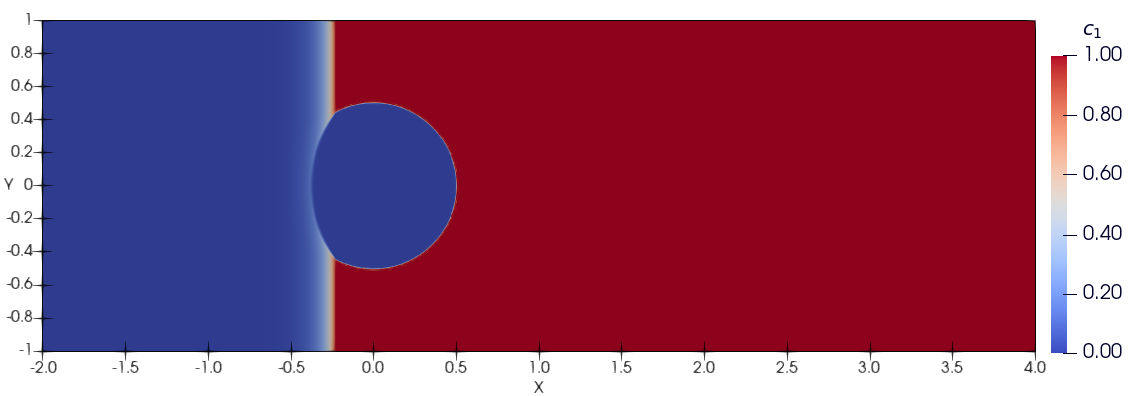}
\includegraphics[width=0.33\textwidth]{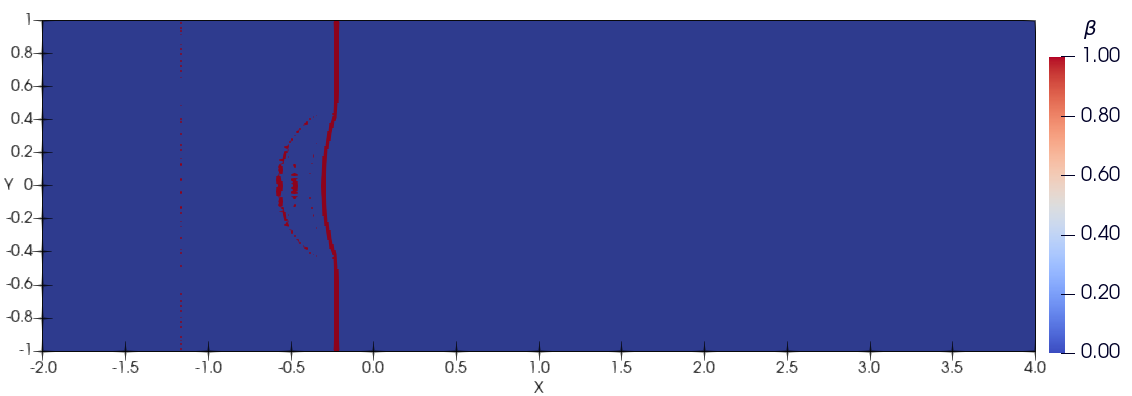}\\
\includegraphics[width=0.33\textwidth]{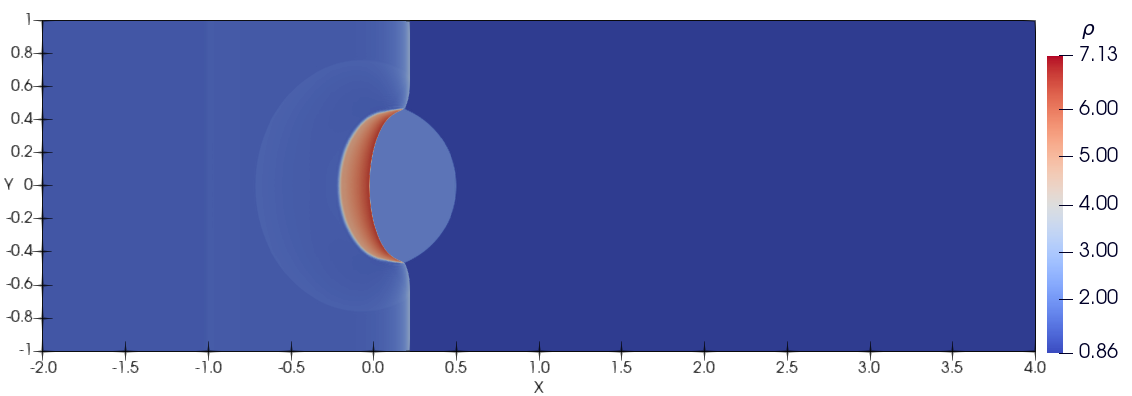}
\includegraphics[width=0.33\textwidth]{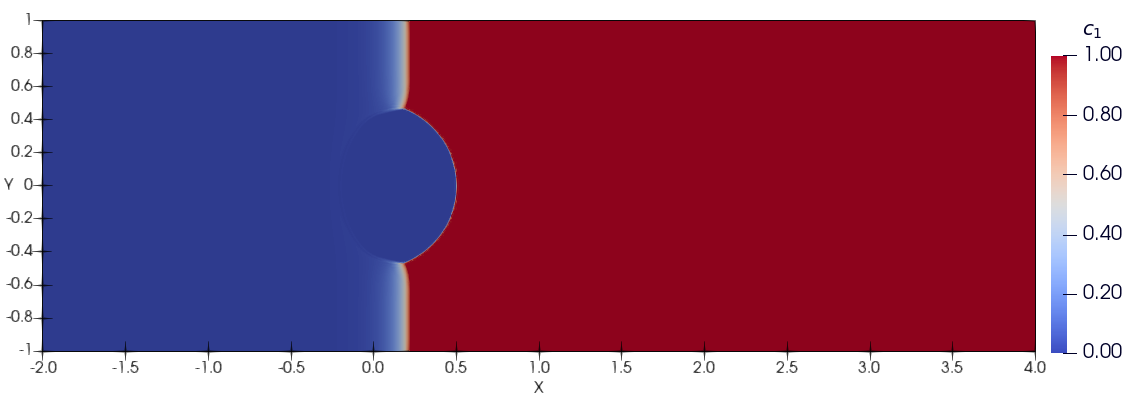}
\includegraphics[width=0.33\textwidth]{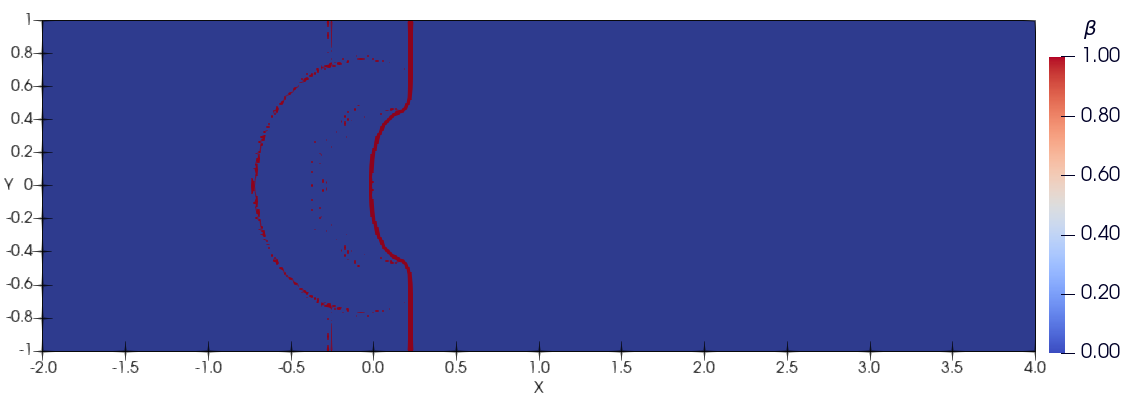}\\
\includegraphics[width=0.33\textwidth]{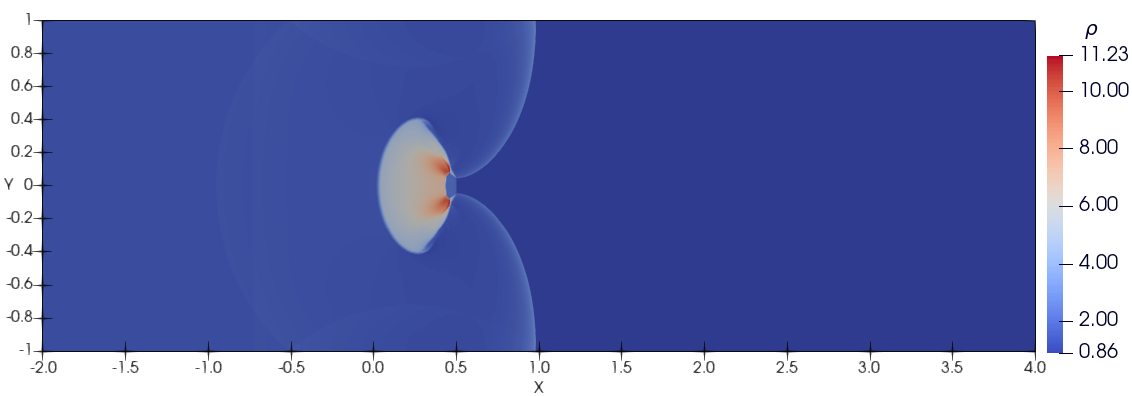}
\includegraphics[width=0.33\textwidth]{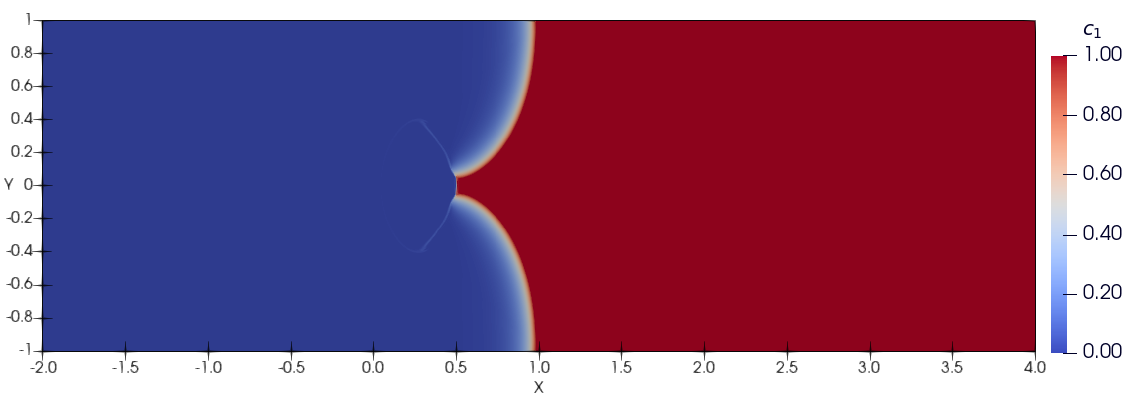}
\includegraphics[width=0.33\textwidth]{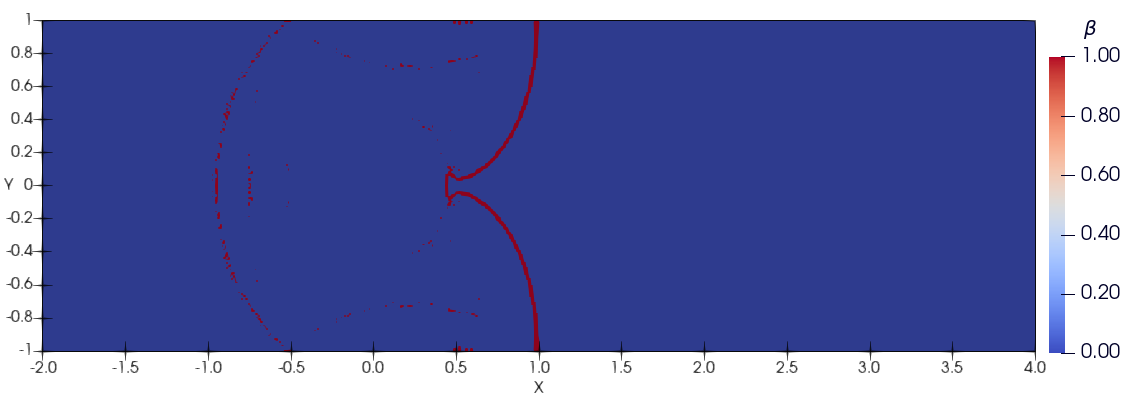}\\
\includegraphics[width=0.33\textwidth]{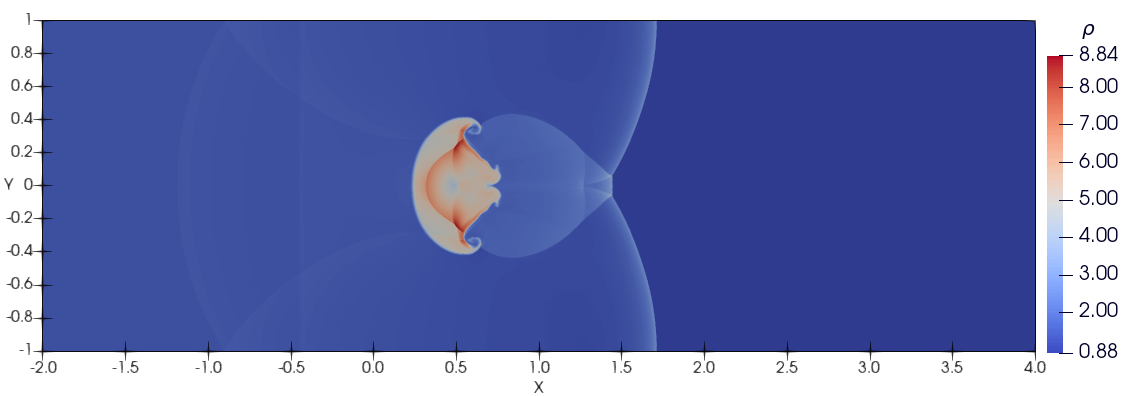}
\includegraphics[width=0.33\textwidth]{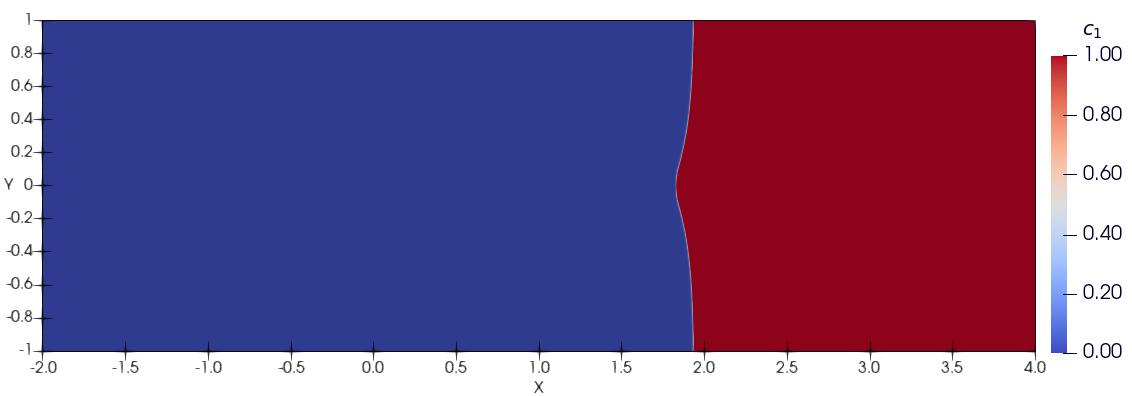}
\includegraphics[width=0.33\textwidth]{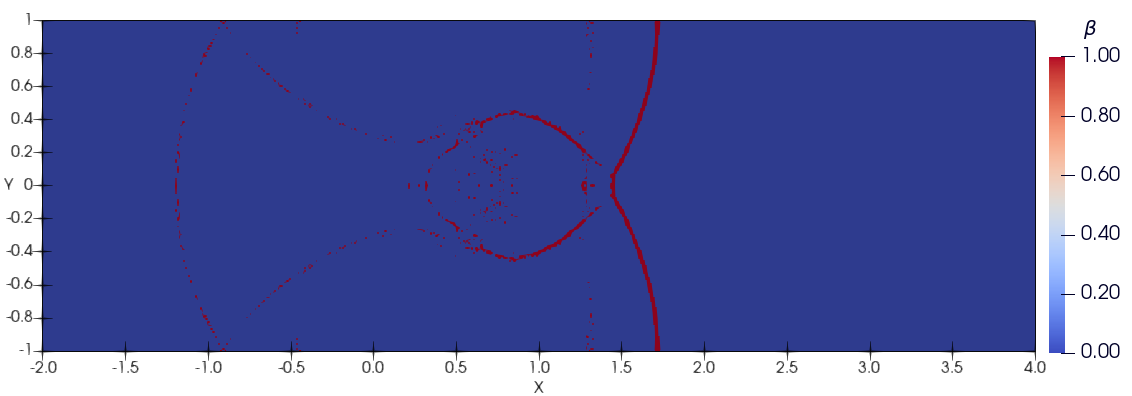}\\
\includegraphics[width=0.33\textwidth]{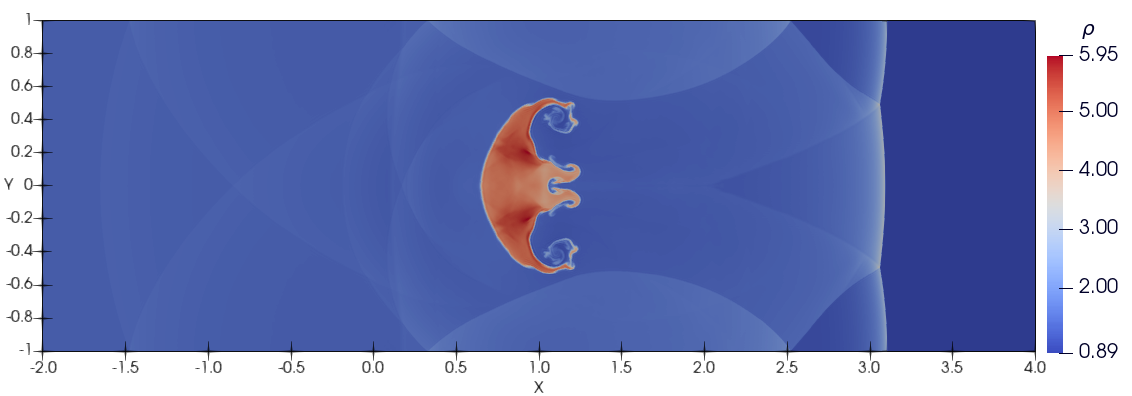}
\includegraphics[width=0.33\textwidth]{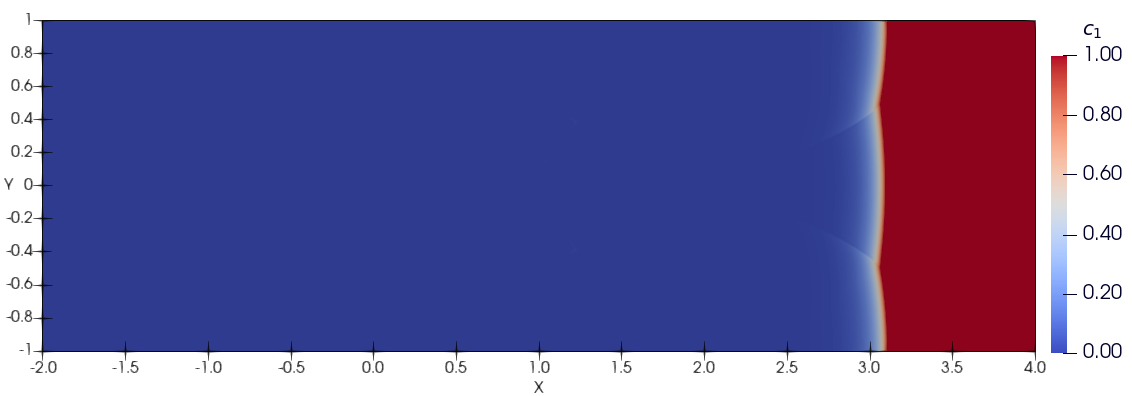}
\includegraphics[width=0.33\textwidth]{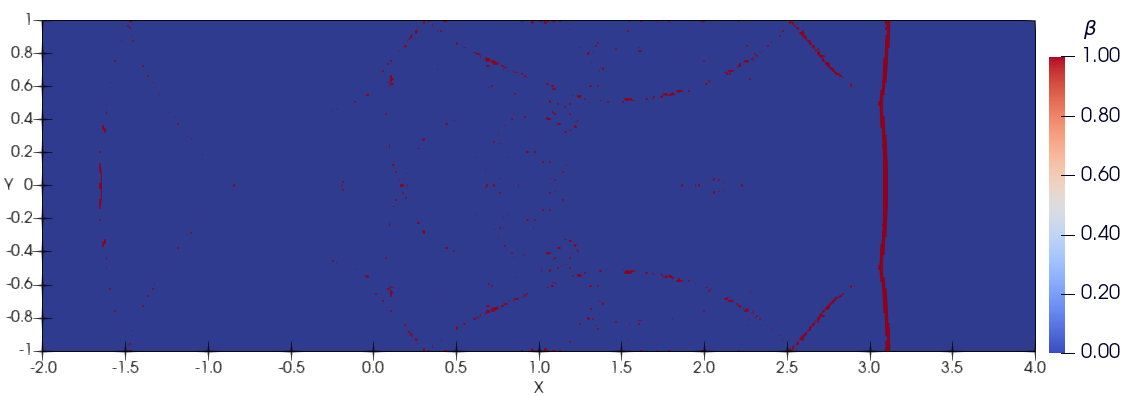}
\caption{\label{fig:dwbis_2d}
Numerical solution of the two-dimensional problem of interaction between a detonation wave and a 
reaction-inert bubble in a two-component medium with a ``slow'' reaction (weak stiff case, a detailed statement of the problem is presented in the text),
obtained using the ADER-DG-$\mathbb{P}_{2}$ method with a posteriori limitation of the solution by a ADER-WENO2 finite volume limiter 
on mesh with $600 \times 200$ cells at the times $t = 1.2$, $1.5$, $2.0$, $2.5$ and $3.5$ (from top to bottom).
The graphs show the coordinate dependencies of the subcells finite-volume representation of density $\rho$ (left), 
mass concentration $c_{1}$ of the reaction reagent (center) and troubled cells indicator $\beta$ (right).
}
\end{figure*}

\begin{figure*}[h!]
\centering
\includegraphics[width=0.33\textwidth]{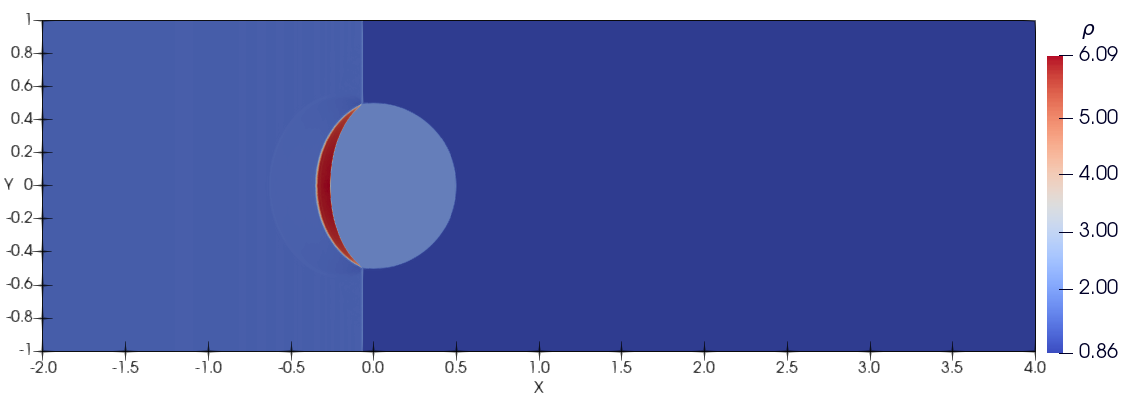}
\includegraphics[width=0.33\textwidth]{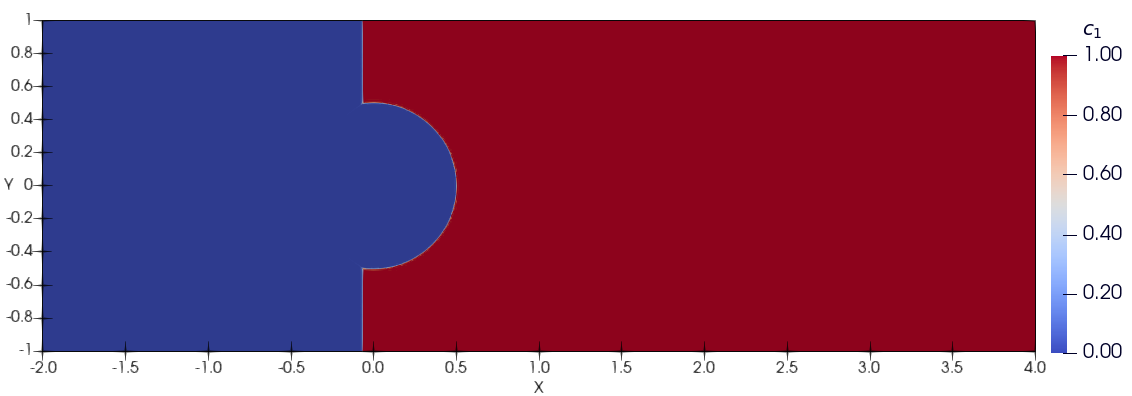}
\includegraphics[width=0.33\textwidth]{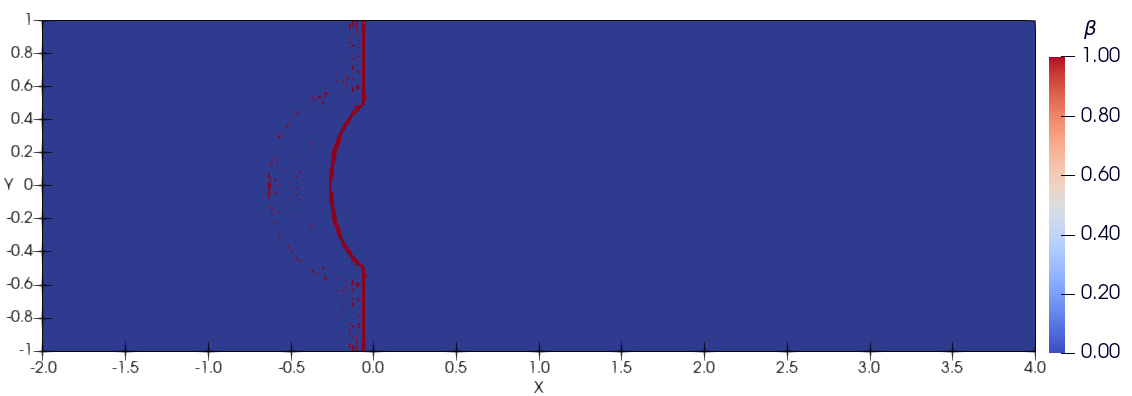}\\
\includegraphics[width=0.33\textwidth]{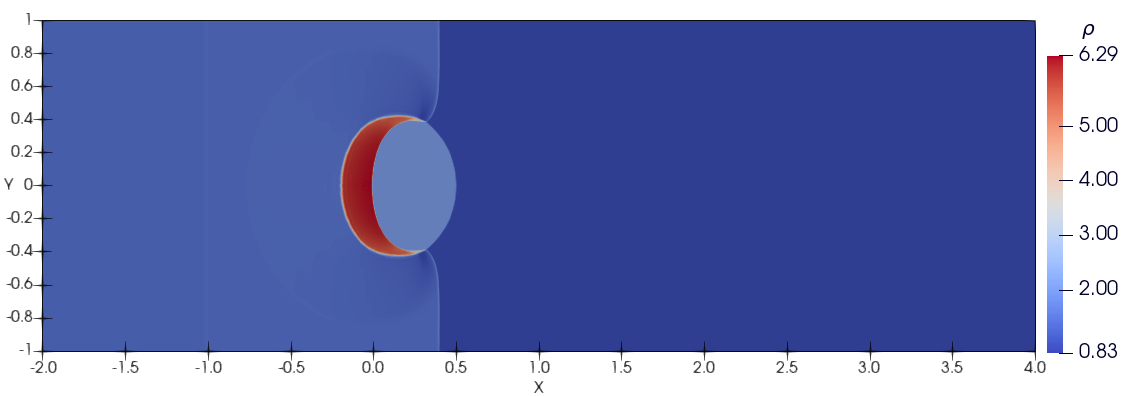}
\includegraphics[width=0.33\textwidth]{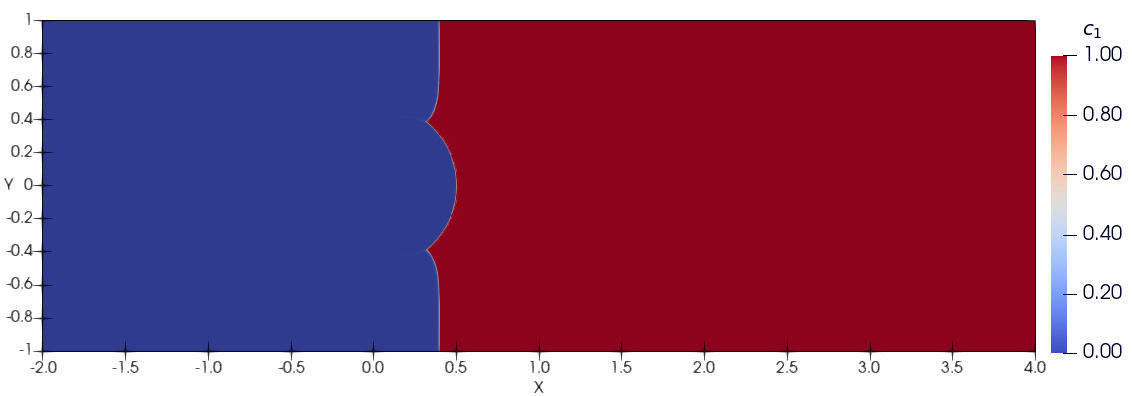}
\includegraphics[width=0.33\textwidth]{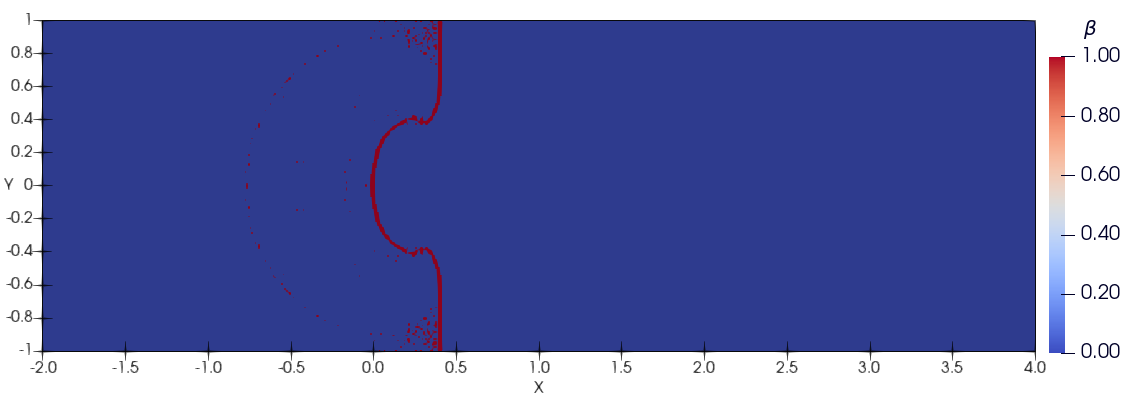}\\
\includegraphics[width=0.33\textwidth]{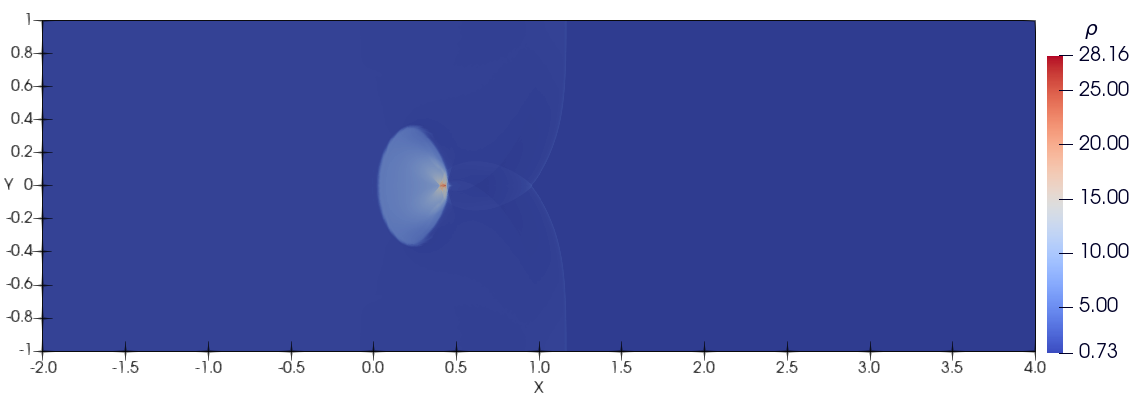}
\includegraphics[width=0.33\textwidth]{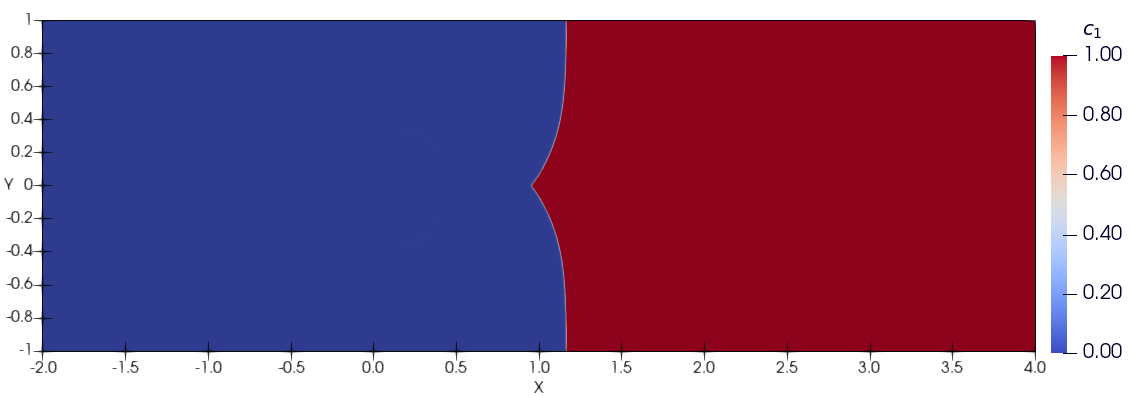}
\includegraphics[width=0.33\textwidth]{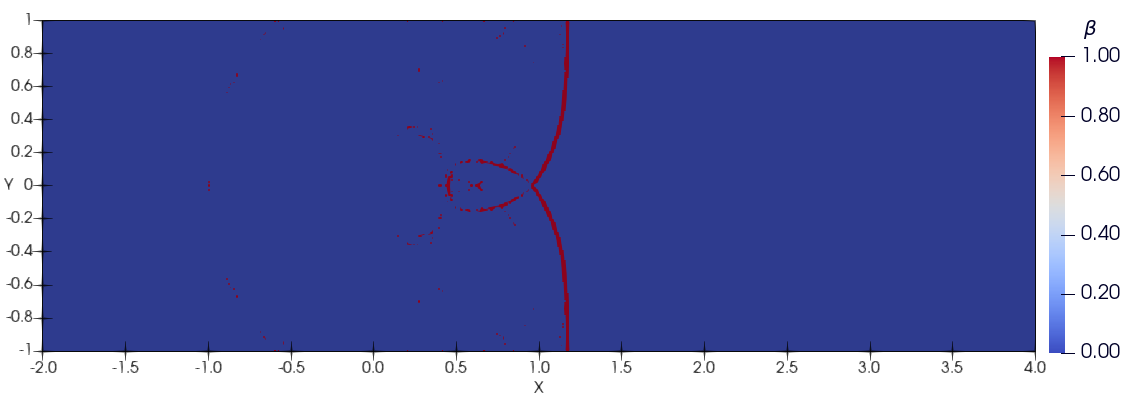}\\
\includegraphics[width=0.33\textwidth]{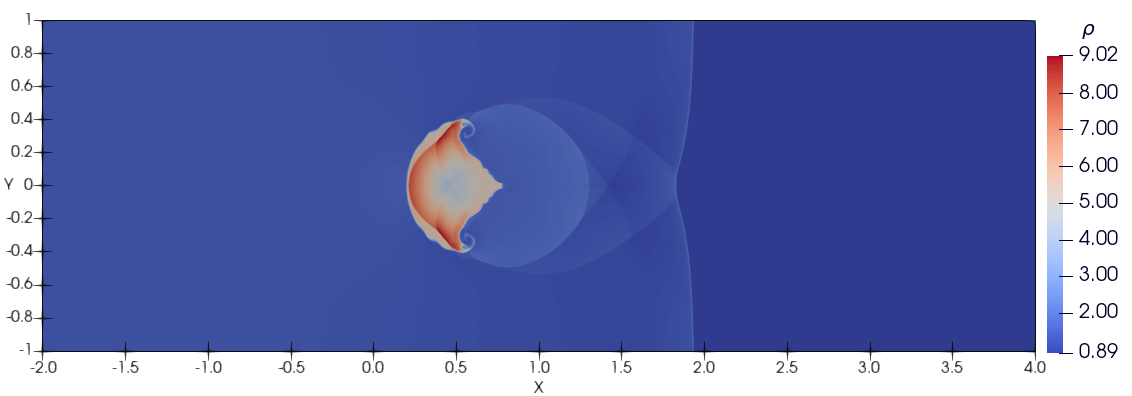}
\includegraphics[width=0.33\textwidth]{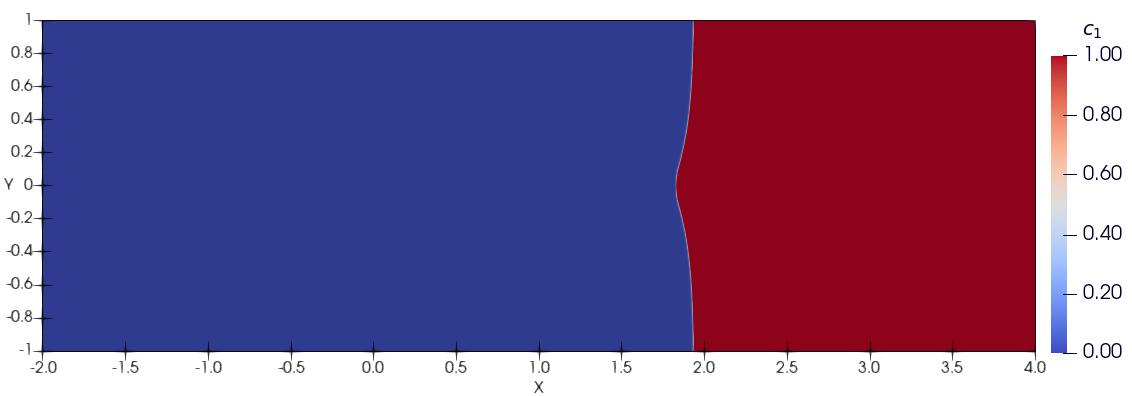}
\includegraphics[width=0.33\textwidth]{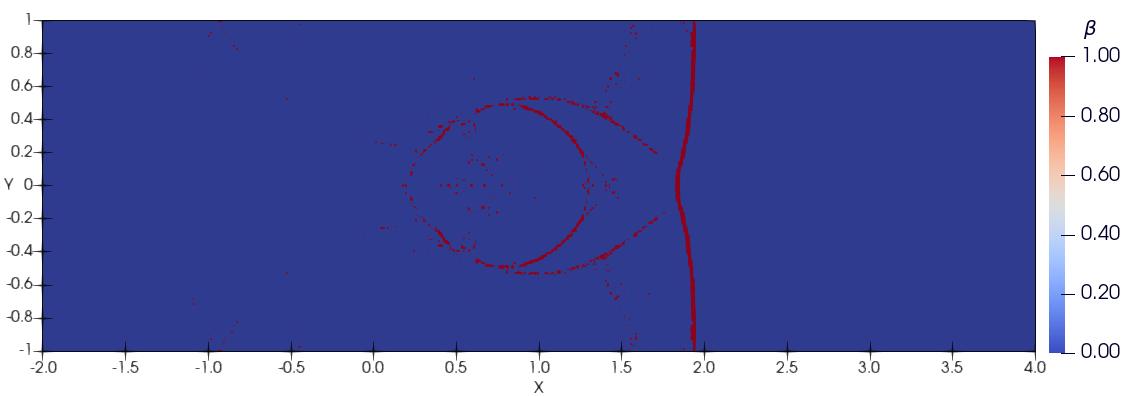}\\
\includegraphics[width=0.33\textwidth]{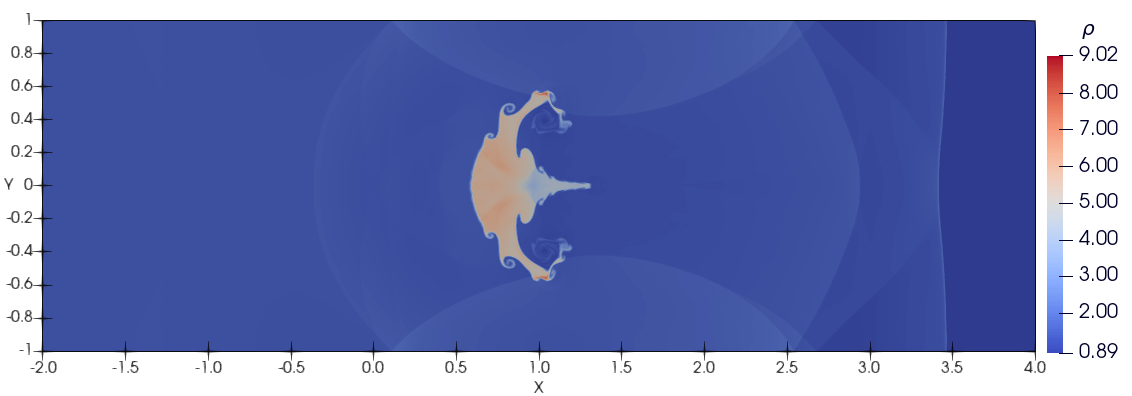}
\includegraphics[width=0.33\textwidth]{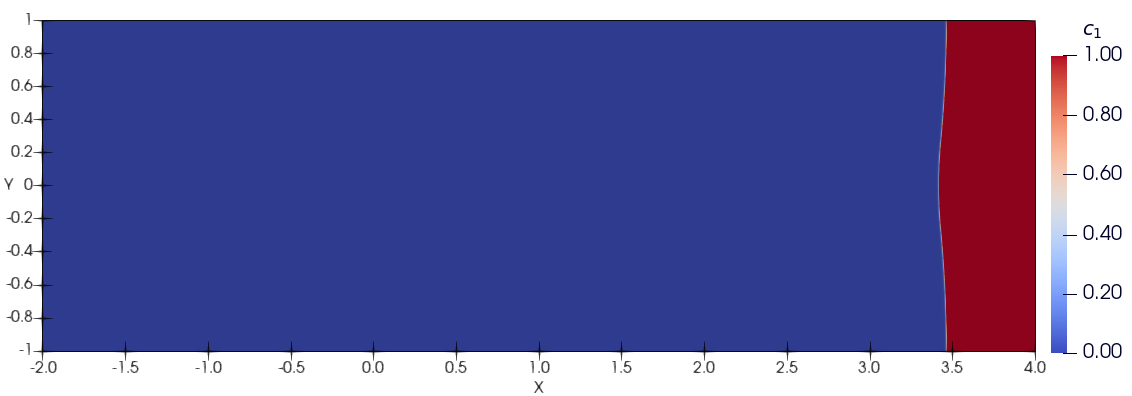}
\includegraphics[width=0.33\textwidth]{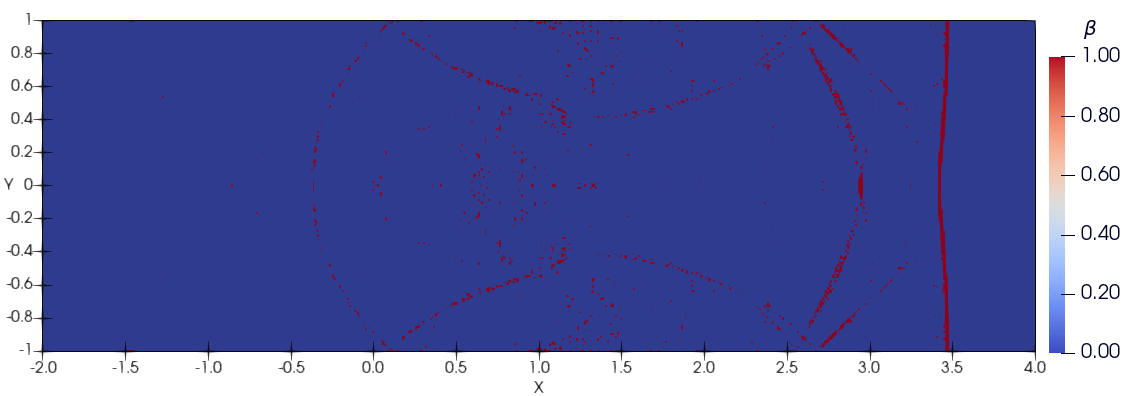}
\caption{\label{fig:dwbif_2d}
Numerical solution of the two-dimensional problem of interaction between a detonation wave and a 
reaction-inert bubble in a two-component medium with a ``fast'' reaction (strong stiff case, a detailed statement of the problem is presented in the text),
obtained using the ADER-DG-$\mathbb{P}_{2}$ method with a posteriori limitation of the solution by a ADER-WENO2 finite volume limiter 
on mesh with $600 \times 200$ cells at the times $t = 1.2$, $1.5$, $2.0$, $2.5$ and $3.5$ (from top to bottom).
The graphs show the coordinate dependencies of the subcells finite-volume representation of density $\rho$ (left), 
mass concentration $c_{1}$ of the reaction reagent (center) and troubled cells indicator $\beta$ (right).
}
\end{figure*}

The numerical solution was obtained using ADER-DG-$\mathbb{P}_{2}$ method with ADER-WENO2 finite volume a posteriori limiter on a spatial mesh with sizes $600\times200$. The numerical solutions are presented in Figure~\ref{fig:dwbis_2d} in the cases of a ``slow'' reaction and in  Figure~\ref{fig:dwbif_2d} in the cases of a ``fast'' reaction. The numerical solutions are presented  at several times $t = 1.2$, $1.5$, $2.0$, $2.5$ and $3.5$ to demonstrate the flow dynamics and the arising non-stationary processes. It should be noted that the number of troubled cells on the mesh never exceeded $1.9\%$ in the cases of a ``slow'' reaction and $2.3\%$ in the cases of a ``fast'' reaction. The average number of troubled cells in this test was $\sim 1.1\%$ in the cases of a ``slow'' reaction and $\sim 1.3\%$ in the cases of a ``fast'' reaction. In this case, of course, the number of troubled cells was determined by the emerging features and structures in the solution.

The numerical solution presented in Figure~\ref{fig:dwbis_2d} in the case of a ``slow'' reaction in a reacting flow demonstrates a classic set of hydrodynamic processes arising as a result of the propagation of a detonation wave in a medium with discontinuous inhomogeneities of the medium. At the initial moment of time $t = 0$, which is presented in Figure~\ref{fig:dwbi_2d_init}, only a detonation wave moving to the right and a stationary inert bubble, the gas density in which exceeds the density of the gas surrounding the bubble, are observed in the flow. At the moment $t = 1.2$ of time presented in Figure~\ref{fig:dwbis_2d}, the detonation wave is already propagating to the distance at which interaction with the inert dense bubble begins. The process of interaction of the detonation wave with a semi-spherical density discontinuity is observed: a shock wave is transmitted forward into the bubble, a reflected shock wave propagates backward, which moves in the already burned gas, and a contact discontinuity is also observed, representing the deformed part of the original bubble. The coordinate dependence of the mass concentration $c_{1}$ of the reagent $A$ demonstrates the process of smooth burnout, characteristic of the case of a ``slow'' reaction. Troubled cells appear only in the vicinity of detonation and shock waves; a contact discontinuity is not detected in them. Note that in the region of coordinate $x \sim -1.15$ there is a small line of cell troubles, which is due to the peculiarities of setting the initial conditions of the problem -- the CJ conditions do not ideally correspond to the one-dimensional Riemann problem in the reacting medium, when only a detonation wave to the right is formed; CJ conditions were obtained for the case of instantaneous detonation, therefore, under the selected conditions they become more accurate with decreasing $\tau_{0}$ (formula (\ref{rate_constant})), therefore, already in the numerical solution for the ``fast'' reaction (which is presented in Figure~\ref{fig:dwbif_2d} and will be discussed further in this text) this anomaly does not arise. At the subsequent moment of time $t = 1.5$ shown in Figure~\ref{fig:dwbis_2d}, the flow does not differ significantly from that presented at the moment of time $t = 1.2$, however, in this case, the initial stage of diffraction of the detonation wave is clearly observed. At the subsequent moment in time $t = 2.0$ presented in Figure~\ref{fig:dwbis_2d}, classical diffraction of a detonation wave on an obstacle is observed -- at the top and bottom, decreasing in size sections of the uncoupled detonation wave are observed, and in the central part, pronounced fronts of diffracted detonation waves are observed, separated by a ``bridge'' of the shock wave passing through the inert bubble substance. The coordinate dependence of the mass concentration $c_{1}$ of the reagent continues to demonstrate the smooth combustion of the reagent behind the front of the detonation wave. Two shock waves also appear, reflected from the upper and lower boundaries of the coordinate domain. The bubble is already significantly deformed at this point in time $t = 2.0$, and a jet flow is formed in the movement of its gas. At the next moment in time $t = 2.5$, merging of diffracted detonation waves is observed, between which a small shock front is observed -- the differences are clearly visible in the coordinate dependencies of the mass concentration $c_{1}$ of the reagent and the cell troubled indicator $\beta$. The formed flow represents a classic sample of the intersection of discontinuity lines with the formation of triple points and points of higher multiplicity. The bubble at this point in time has already been deformed even more significantly, and vortex generation is observed in the movement $t = 2.5$ of its gas. At the last of the presented moments of time $t = 3.5$, a complication of the processes of intersection of discontinuity lines is observed, and many shock waves and contact discontinuities arise. The detonation front looks quite continuous and stable after passing through a dense bubble, which is clearly visible from the coordinate dependence of the mass concentration of the reagent. At this point in time, the bubble continues the process of significant deformation of its shape; vortex generation is observed in the movement of its gas, externally manifested in the form of ``twisting'' of the gas. Finally, it can also be noted that the fronts of detonation and shock waves in the solution are expressed quite sharply, and the contact discontinuities do not expand. The presented numerical solution demonstrates that the ADER-DG-$\mathbb{P}_{N}$ method with ADER-WENO finite volume a posteriori limiter allows one to effectively and accurately simulate the flow of reacting flows of a rather complex geometric configuration.

The numerical solution presented in Figure~\ref{fig:dwbif_2d} in the case of a ``fast'' reaction in a reacting flow also demonstrates a classical set of hydrodynamic processes arising as a result of the propagation of a detonation wave in a medium with discontinuous inhomogeneities of the medium. At the initial moment of time, which is presented in Figure~\ref{fig:dwbi_2d_init}, only a detonation wave moving to the right and a stationary inert bubble are observed in the flow, the gas density in which exceeds the density of the gas surrounding the bubble, however, in this case the speed of the detonation wave is slightly higher than in the ``slow'' reactions case. At the moment of time $t = 1.2$ presented in Figure~\ref{fig:dwbif_2d}, the detonation wave is already propagating to the distance at which interaction with the inert dense bubble begins. The process of interaction of the detonation wave with a semi-spherical density discontinuity is observed: a shock wave is transmitted forward into the bubble, a reflected shock wave propagates backward, which moves in the already burned gas, and a contact discontinuity is also observed, representing the deformed part of the original bubble. The coordinate dependence of the mass concentration $c_{1}$ of the reagent $A$ demonstrates the process of sharp burnout, which is characteristic of the case of a ``fast'' reaction. This feature is characteristic of the classical ZND detonation wave -- the region of localization of combustion of the reagent in the shock wave front is clearly observed, and a pronounced chemical Zel'dovich peak appears. Non-physical effects associated with a weak detonation wave propagating ahead of the shock wave do not arise -- the ZND detonation front is formed. Troubled cells appear only in the vicinity of detonation and shock waves; several troubled cells appear in the area of contact discontinuity. Note that the anomaly associated with the cell line of troubles, which occurs in the case of a ``slow'' reaction, does not appear in this case. At the subsequent moment $t = 1.5$ of time presented in Figure~\ref{fig:dwbif_2d}, the flow does not differ significantly from that presented at the moment of time, however, in this case, the initial stage of diffraction of the detonation wave is clearly observed. At the subsequent presented moments $t = 2.0$ and $2.5$ of time, a merging of diffracted detonation waves is observed -- these phenomena are clearly visible on the coordinate dependencies of the mass concentration $c_{1}$ of the reagent and the indicator of troubled cells $\beta$. The formed flow represents a classic example of the intersection of discontinuity lines. At these times, the bubble is already significantly deformed, the formation of a jet flow and obvious vortex generation are observed in the movement of its gas, and a vortex street is formed at the boundary of the deformed bubble. At the last of the presented moments of time $t = 3.5$, a complication of the processes of intersection of discontinuity lines is observed, and many shock waves and contact discontinuities arise. The detonation front looks quite stable after passing through a dense bubble, which is clearly visible from the coordinate dependence of the mass concentration $c_{1}$ of the reagent. At this finish time, the bubble continues the process of significant deformation of its shape, vortex generation is observed in the movement of its gas, externally manifested in the form of ``twisting'' of the gas, and a jet flow is sharply distinguished. Finally, it can also be noted that the fronts of detonation and shock waves in the solution are expressed quite sharply, and the contact discontinuities do not expand. Non-physical effects characteristic of modeling detonation processes using classical numerical methods~\cite{frac_steps_detwave_sim_2000, chem_kin_hrs_weno} do not arise -- a non-physical weak detonation front is not formed~\cite{correct_det_wave_speed_2017}. The coordinate dependence of the mass concentration $c_{1}$ of the reagent demonstrates the process of sharp burnout, which is characteristic of ZND detonation. The presented numerical solution demonstrates that the ADER-DG-$\mathbb{P}_{N}$ method with ADER-WENO finite volume a posteriori limiter allows one to effectively and quite accurately simulate the flows of reacting flows with strong stiffness due to reaction kinetics in areas of rather complex geometric configuration.

\paragraph{Interaction of a detonation wave with lattice of inert bubbles}

\begin{figure}[h!]
\centering
\includegraphics[width=0.49\textwidth]{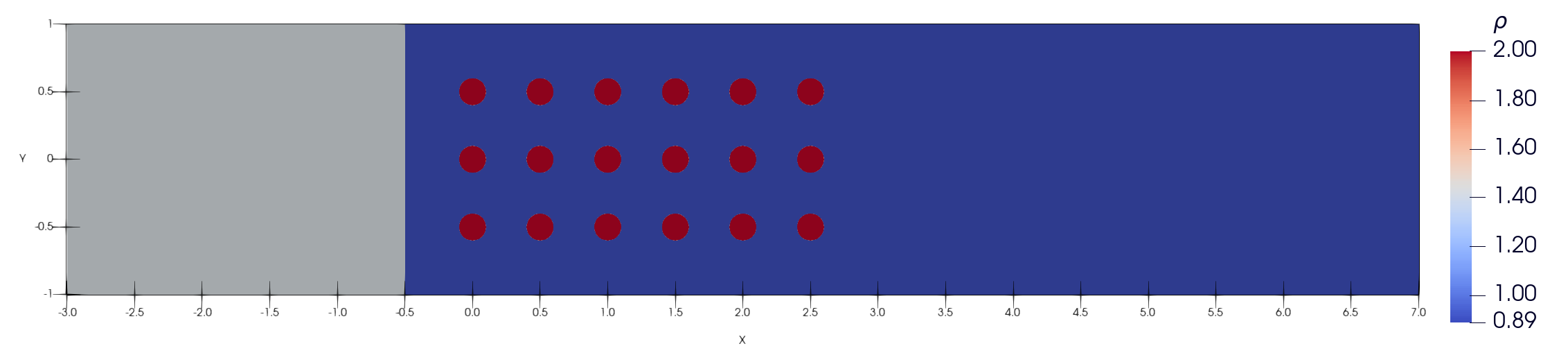}
\includegraphics[width=0.49\textwidth]{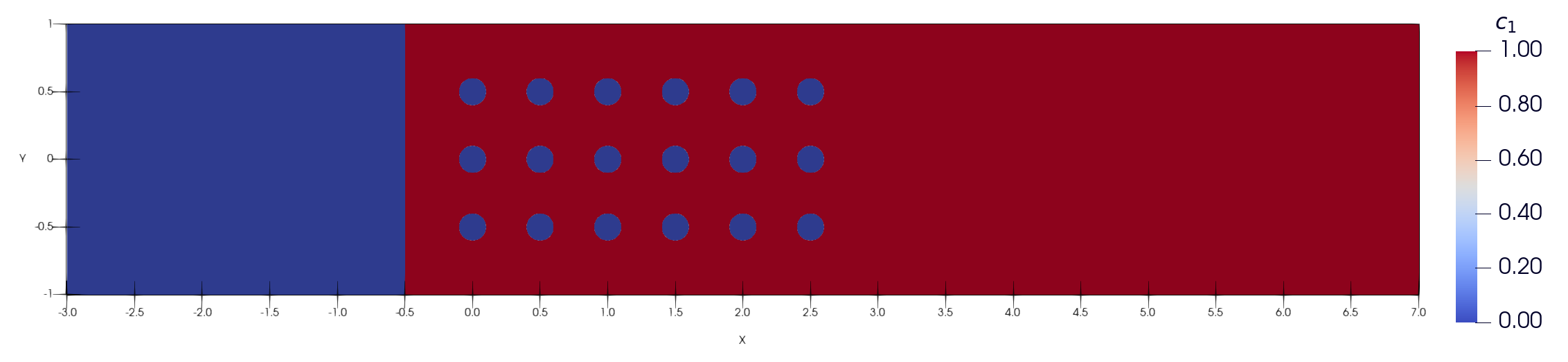}
\caption{\label{fig:dwmbi_3x6_2d_init}
The initial conditions for coordinate dependency of density $\rho$ (top) and mass concentration $c_{1}$ of the reaction reagent (bottom)
in the two-dimensional problem of interaction between a detonation wave and $18$ ($6 \times 3$) reaction-inert bubbles
(a detailed statement of the problem is presented in the text).
}
\end{figure}

\begin{figure*}[h!]
\centering
\includegraphics[width=0.33\textwidth]{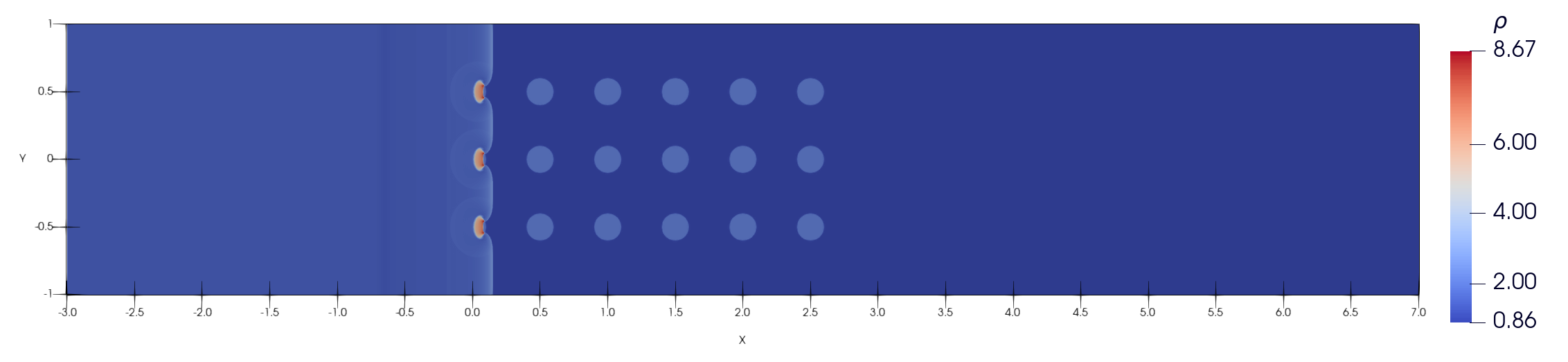}
\includegraphics[width=0.33\textwidth]{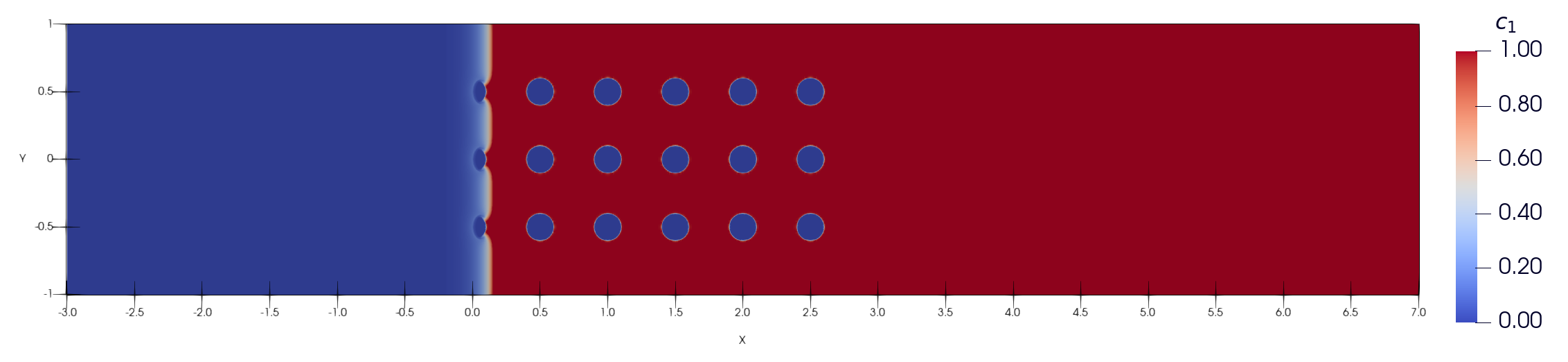}
\includegraphics[width=0.33\textwidth]{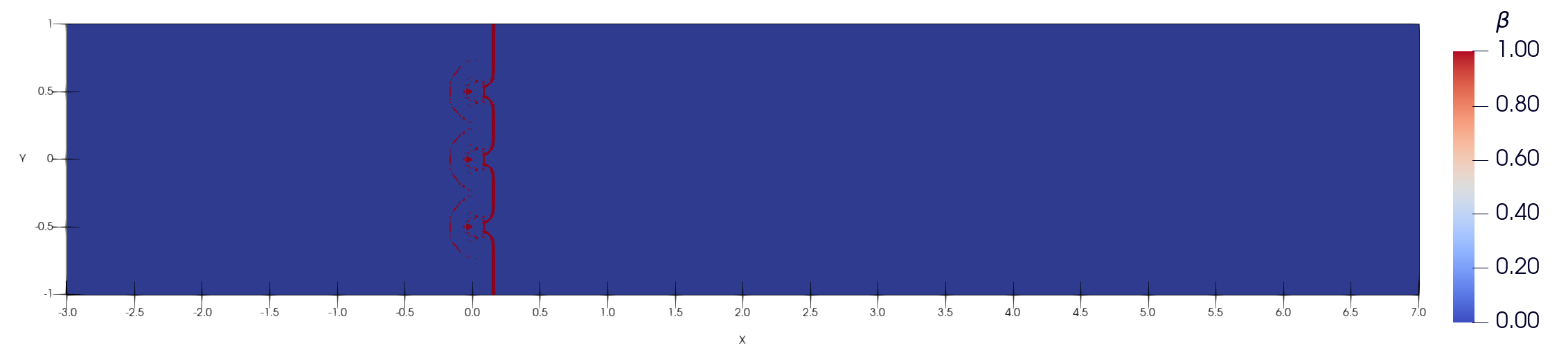}\\
\includegraphics[width=0.33\textwidth]{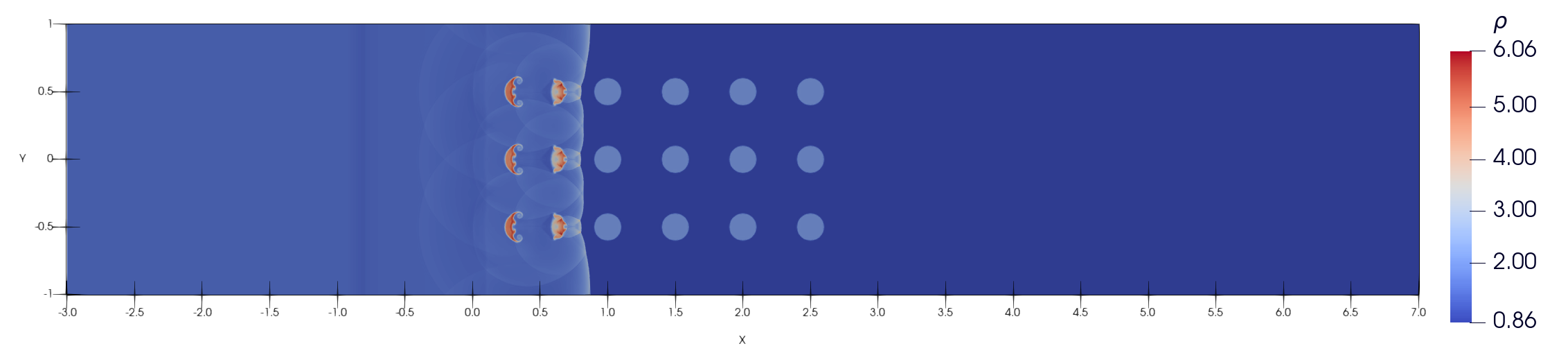}
\includegraphics[width=0.33\textwidth]{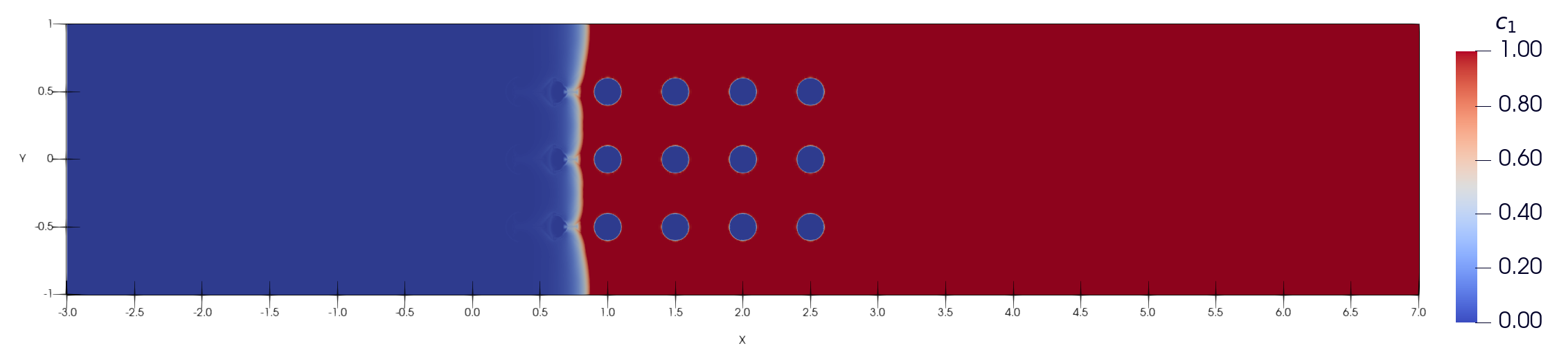}
\includegraphics[width=0.33\textwidth]{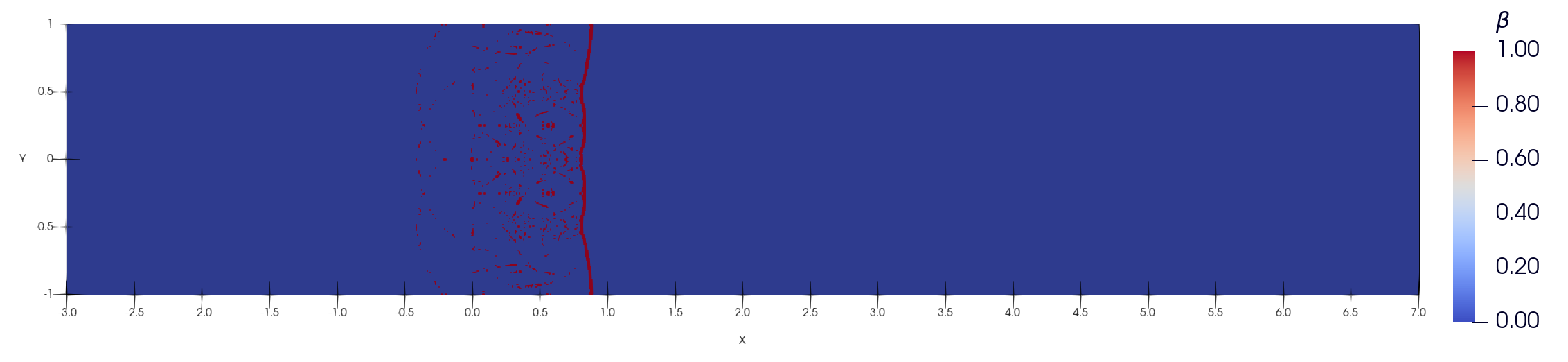}\\
\includegraphics[width=0.33\textwidth]{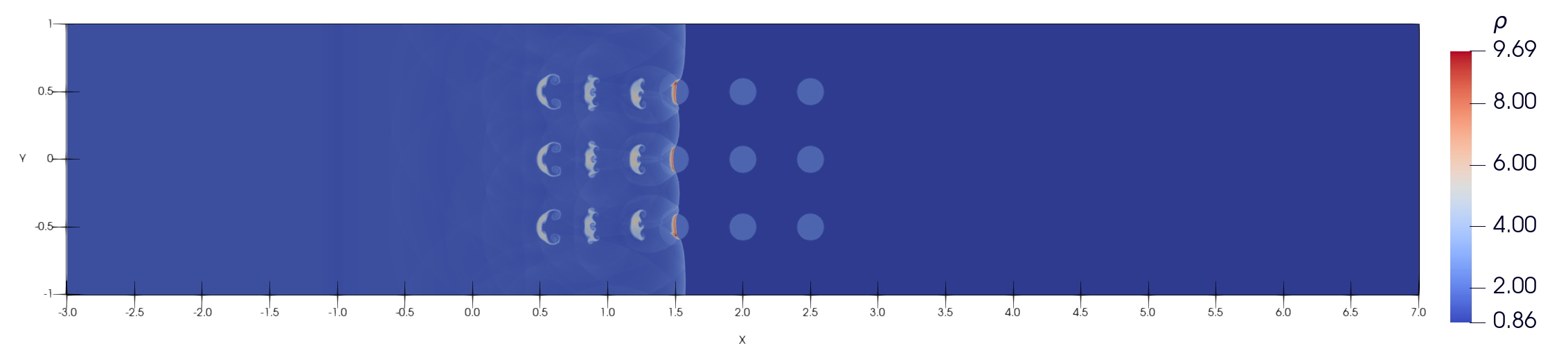}
\includegraphics[width=0.33\textwidth]{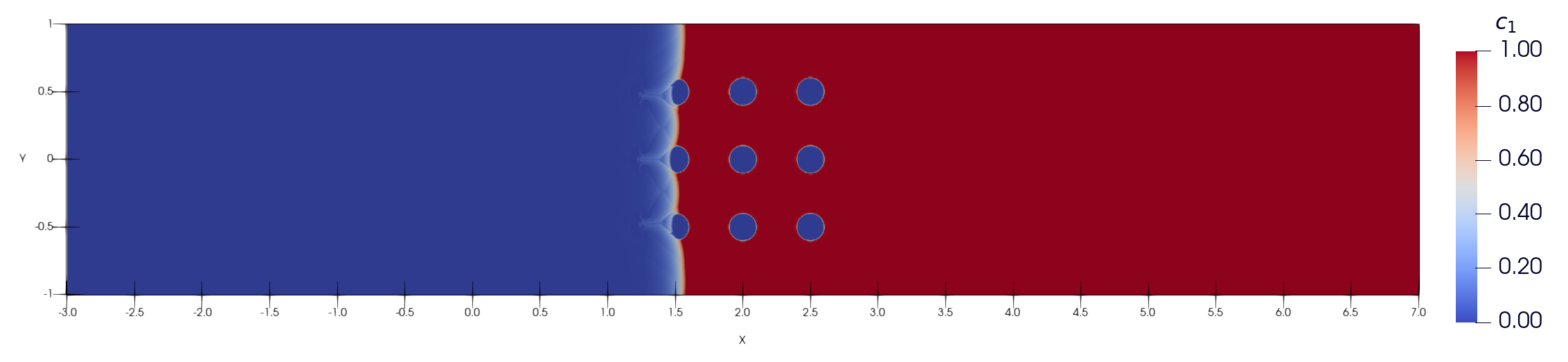}
\includegraphics[width=0.33\textwidth]{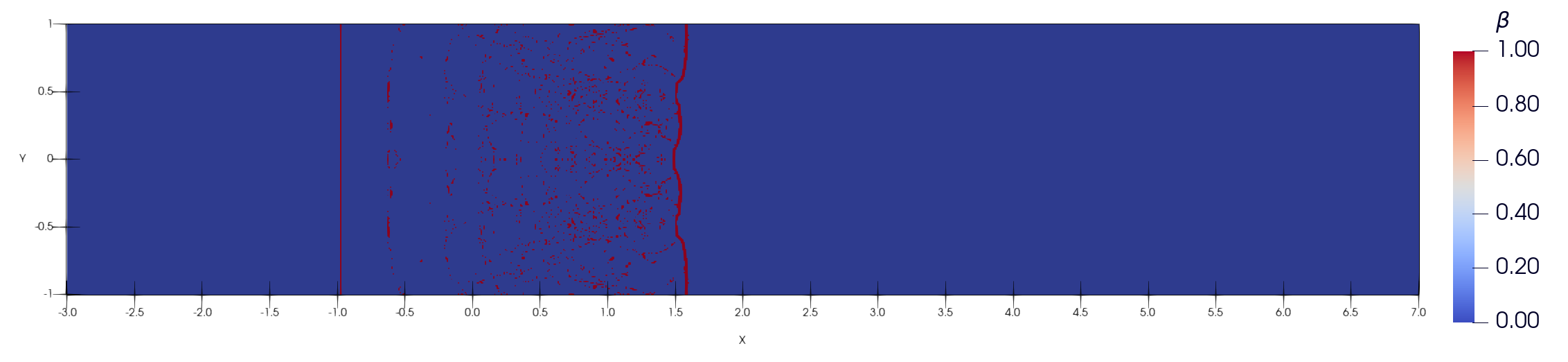}\\
\includegraphics[width=0.33\textwidth]{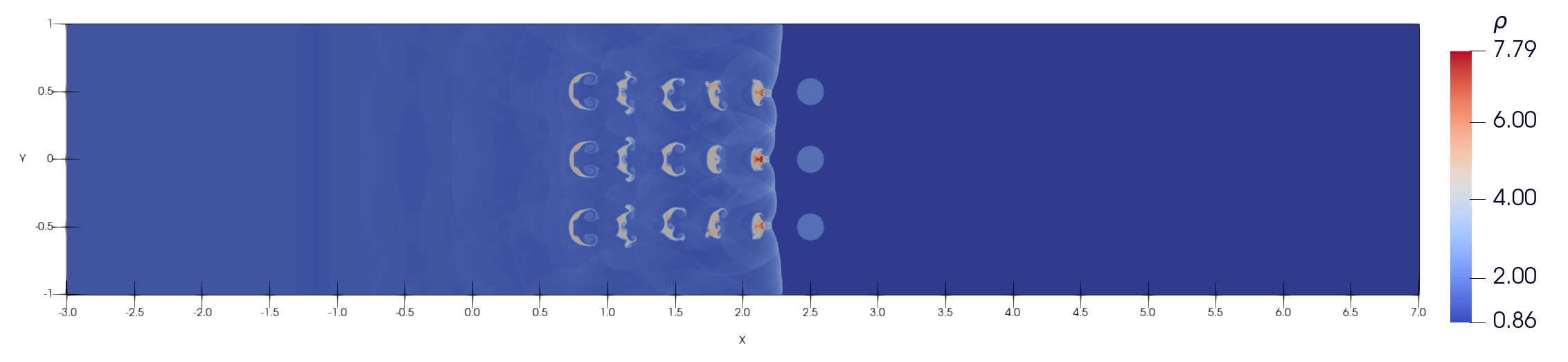}
\includegraphics[width=0.33\textwidth]{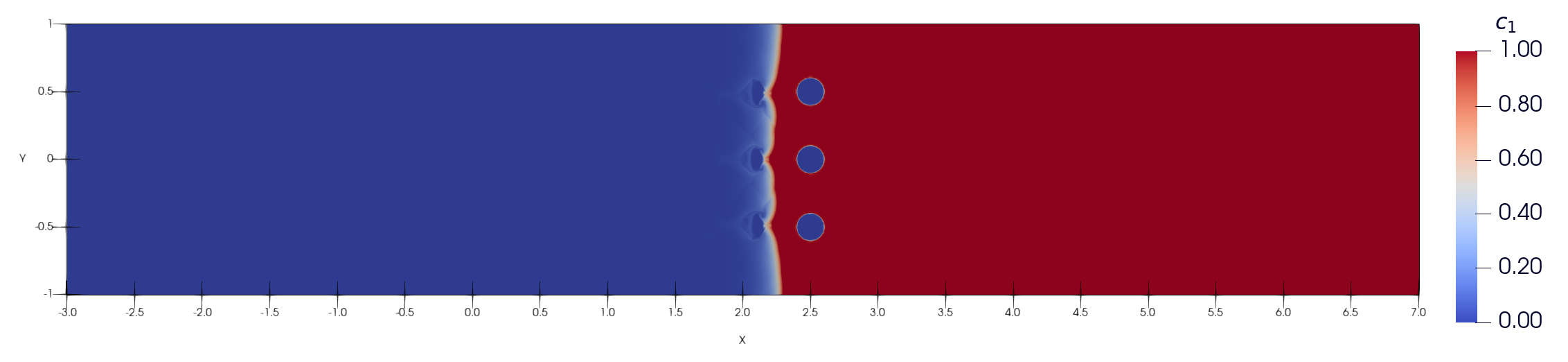}
\includegraphics[width=0.33\textwidth]{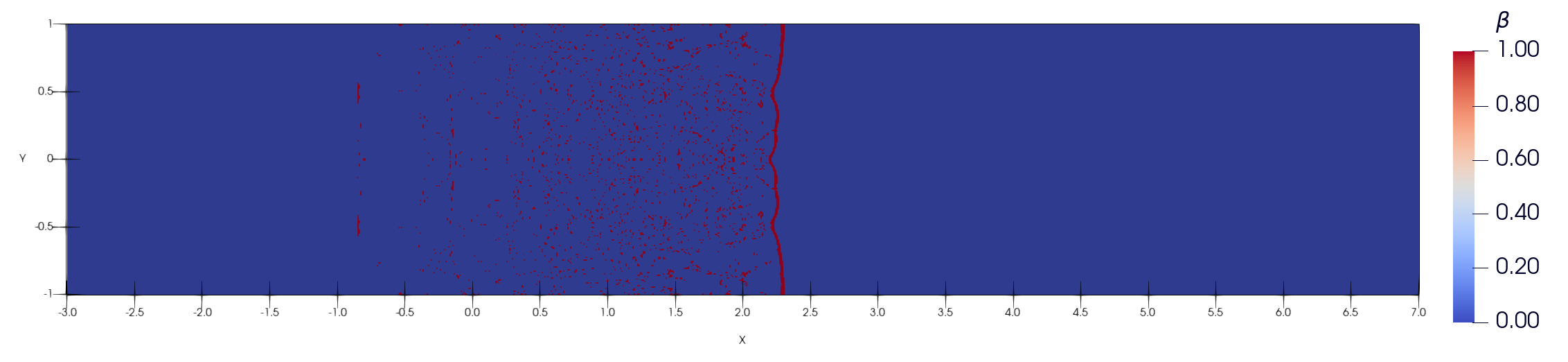}\\
\includegraphics[width=0.33\textwidth]{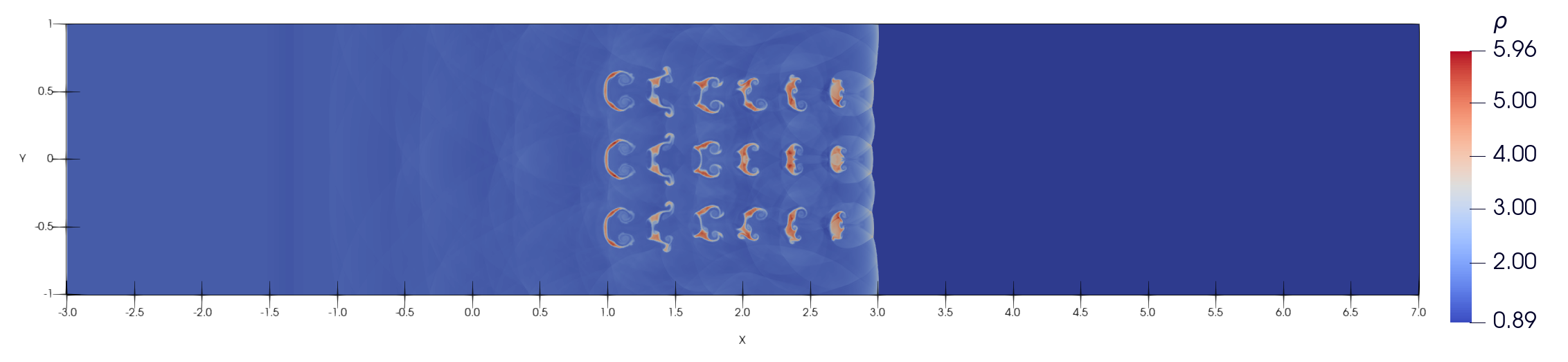}
\includegraphics[width=0.33\textwidth]{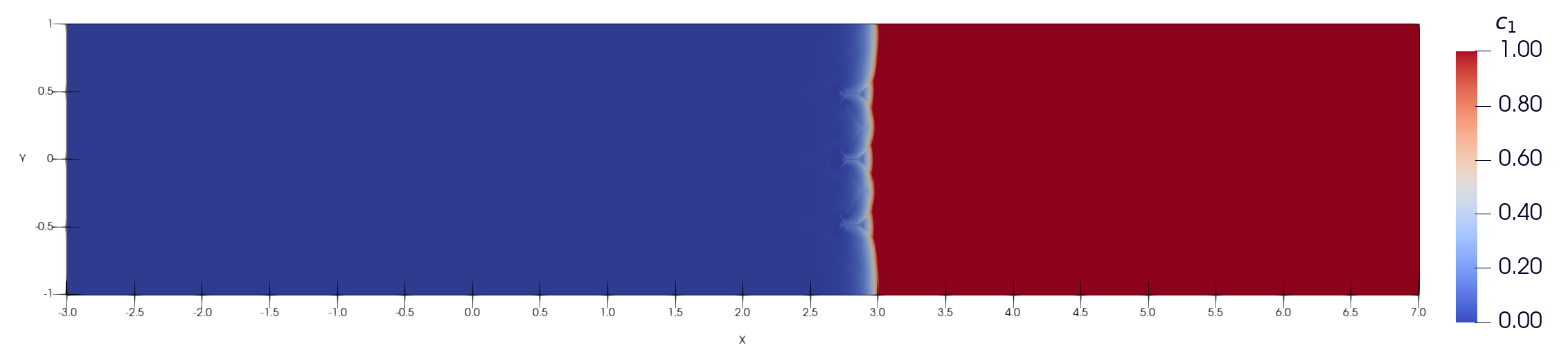}
\includegraphics[width=0.33\textwidth]{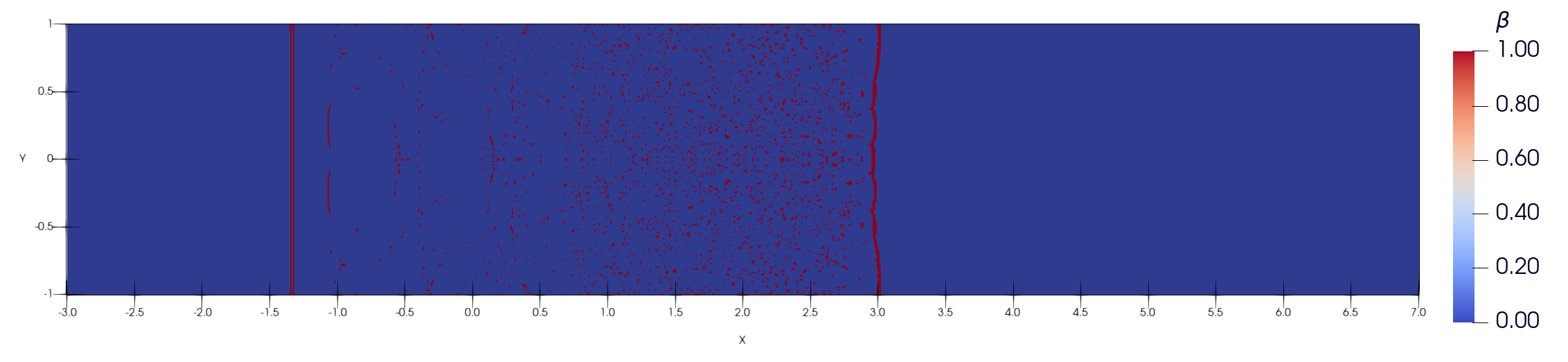}\\
\includegraphics[width=0.33\textwidth]{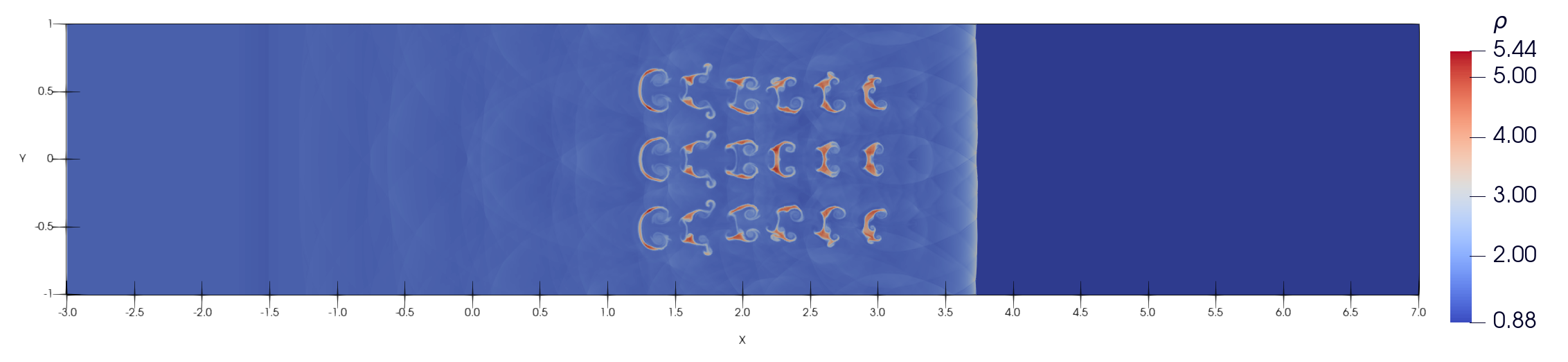}
\includegraphics[width=0.33\textwidth]{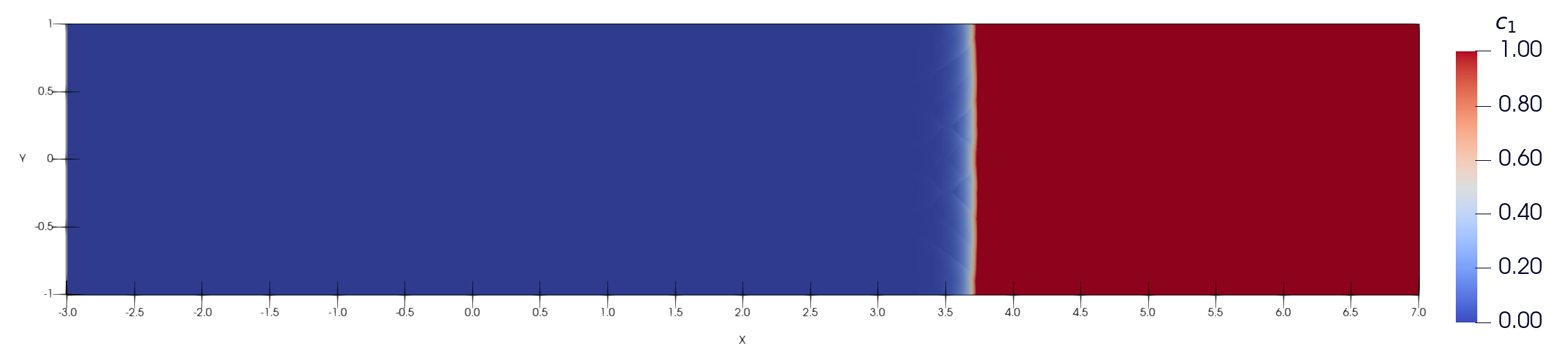}
\includegraphics[width=0.33\textwidth]{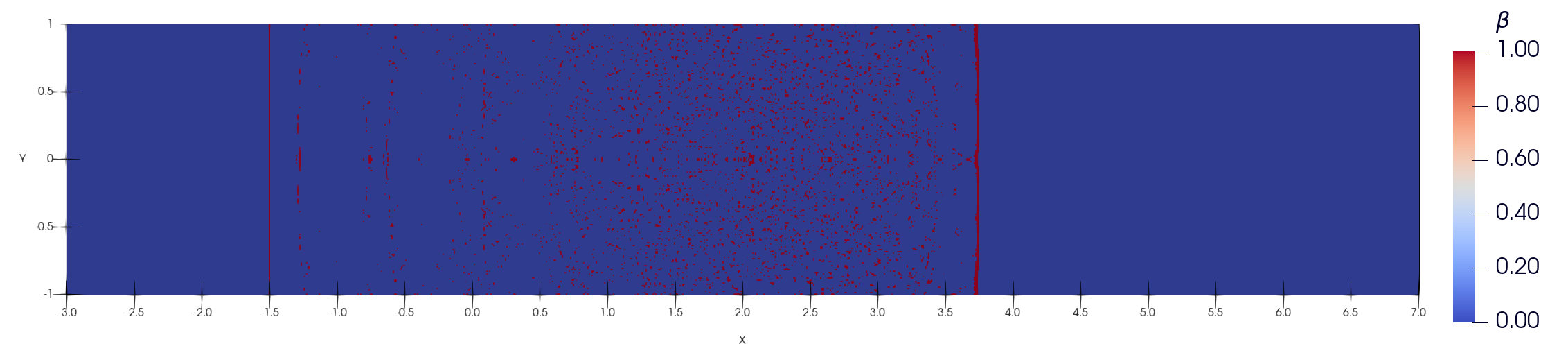}\\
\includegraphics[width=0.33\textwidth]{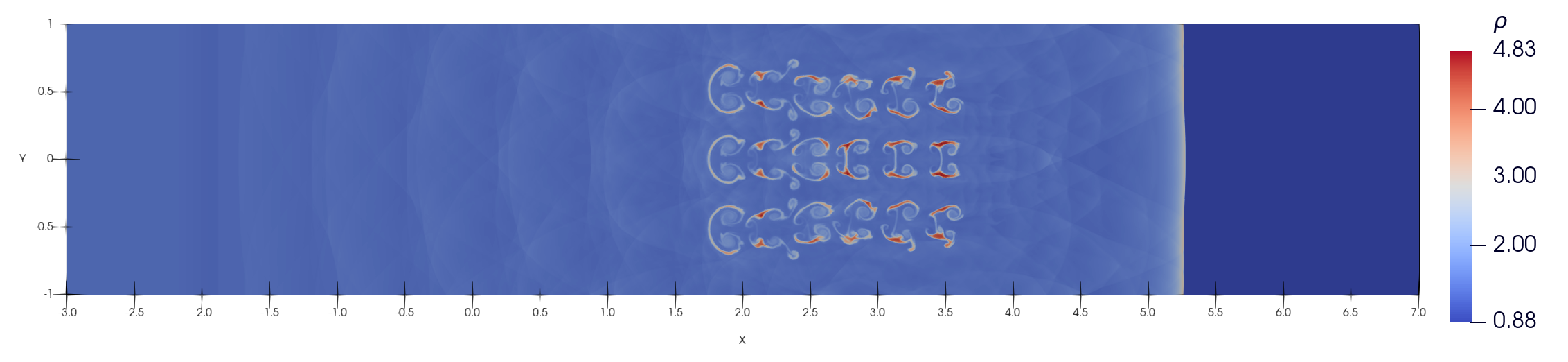}
\includegraphics[width=0.33\textwidth]{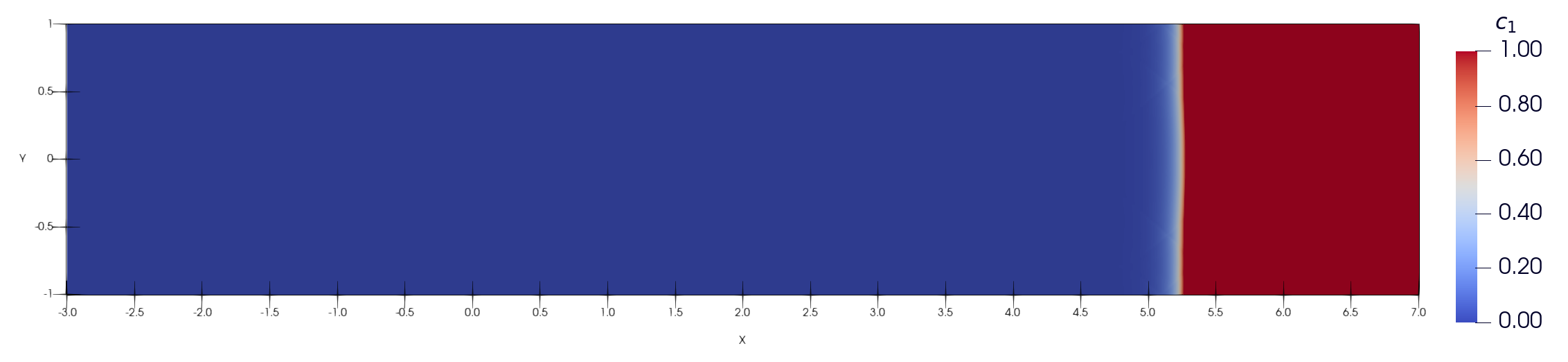}
\includegraphics[width=0.33\textwidth]{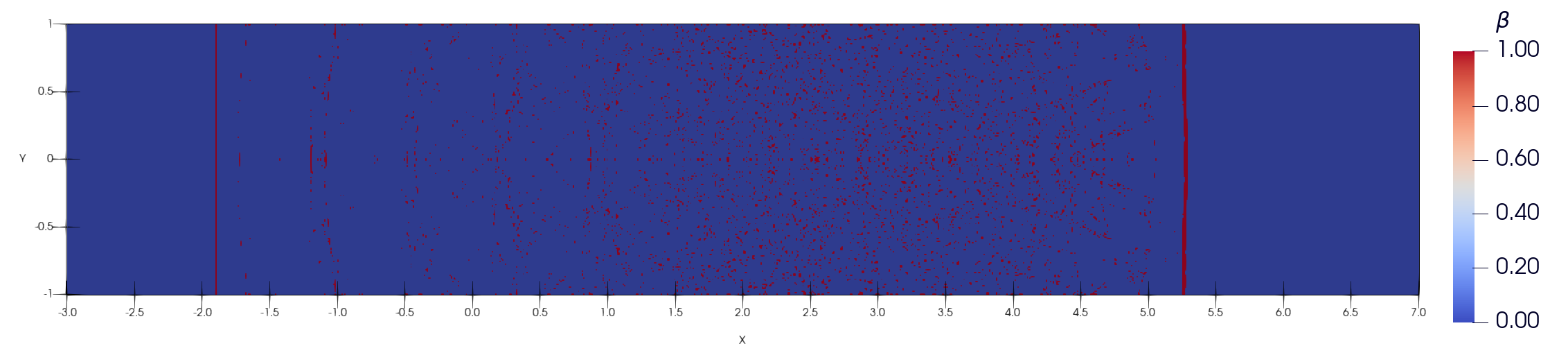}\\
\includegraphics[width=0.33\textwidth]{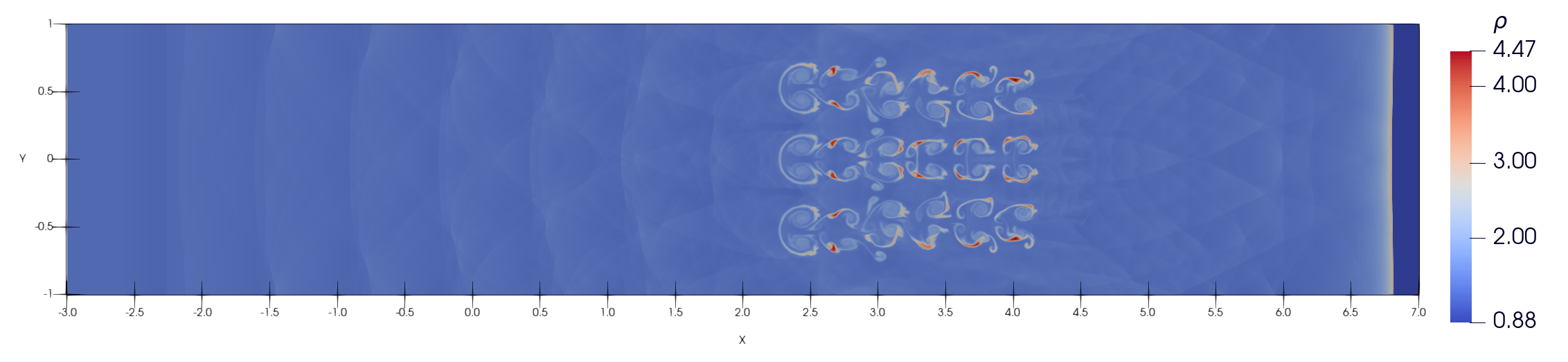}
\includegraphics[width=0.33\textwidth]{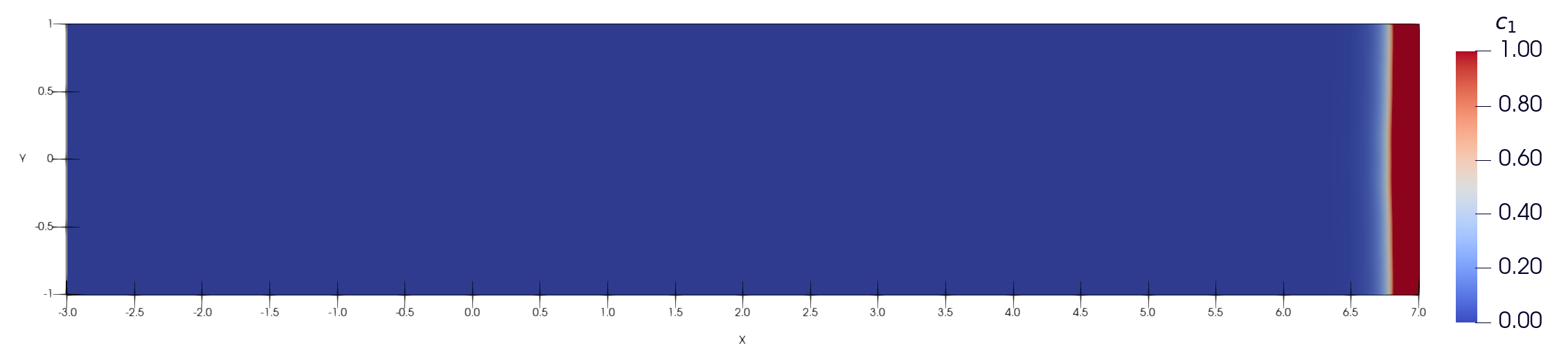}
\includegraphics[width=0.33\textwidth]{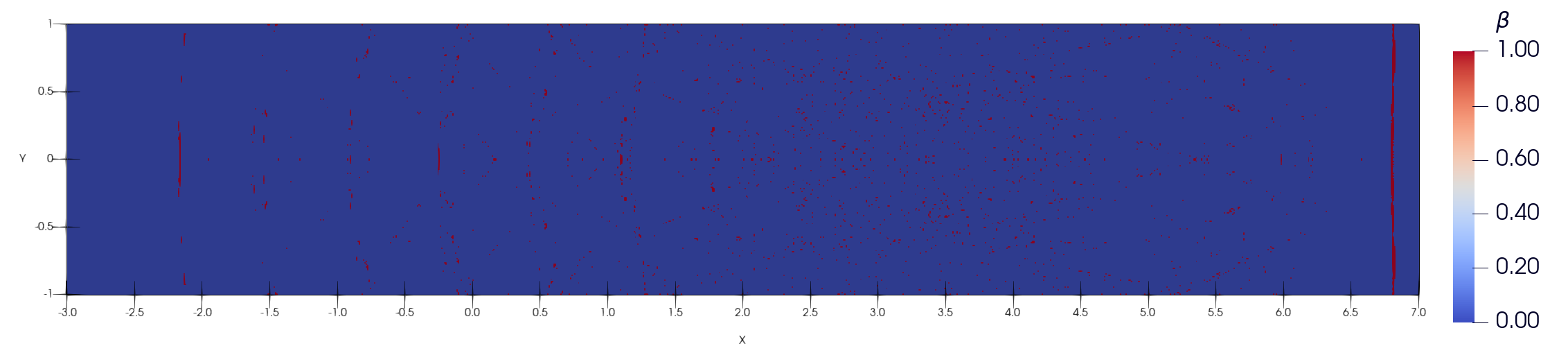}
\caption{\label{fig:dwmbis_3x6_2d}
Numerical solution of the two-dimensional problem of interaction between a detonation wave and $18$ ($6 \times 3$) reaction-inert bubbles
in a two-component medium with a ``slow'' reaction (weak stiff case, a detailed statement of the problem is presented in the text),
obtained using the ADER-DG-$\mathbb{P}_{2}$ method with a posteriori limitation of the solution by a ADER-WENO2 finite volume limiter 
on mesh with $1000 \times 200$ cells at the times $t = 0.5$, $1.0$, $1.5$, $2.0$, $2.5$, $3.0$, $4.0$ and $5.0$ (from top to bottom).
The graphs show the coordinate dependencies of the subcells finite-volume representation of density $\rho$ (left), 
mass concentration $c_{1}$ of the reaction reagent (center) and troubled cells indicator $\beta$ (right).
}
\end{figure*}

\begin{figure*}[h!]
\centering
\includegraphics[width=0.33\textwidth]{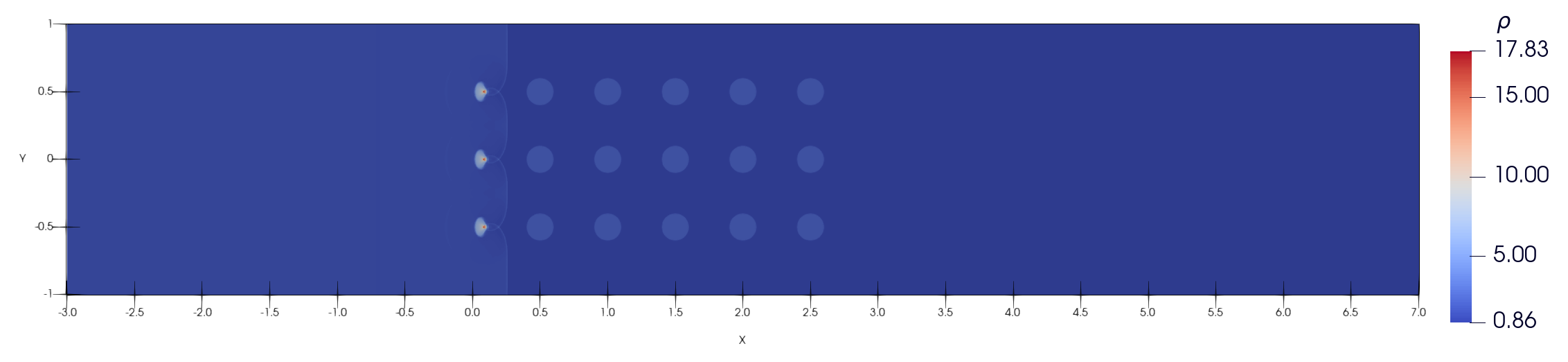}
\includegraphics[width=0.33\textwidth]{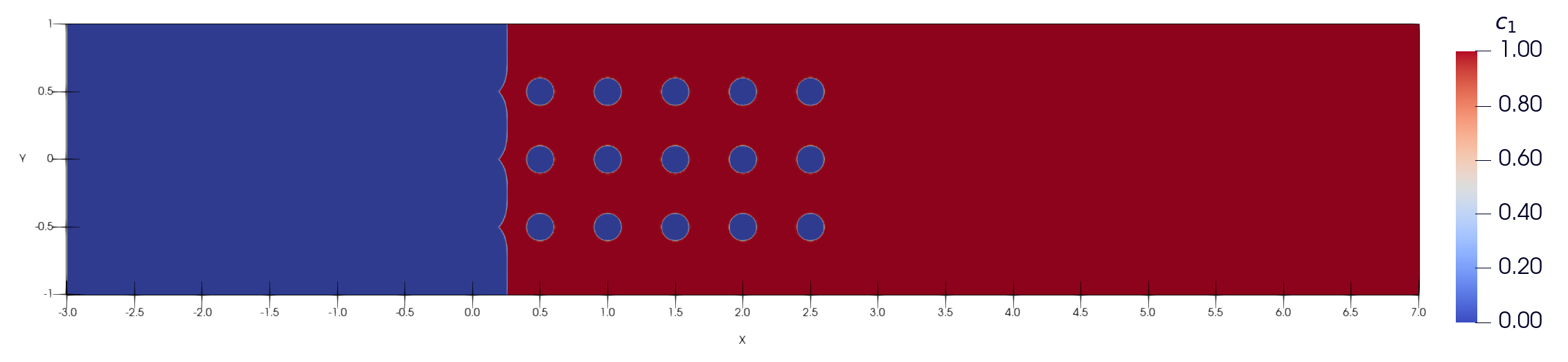}
\includegraphics[width=0.33\textwidth]{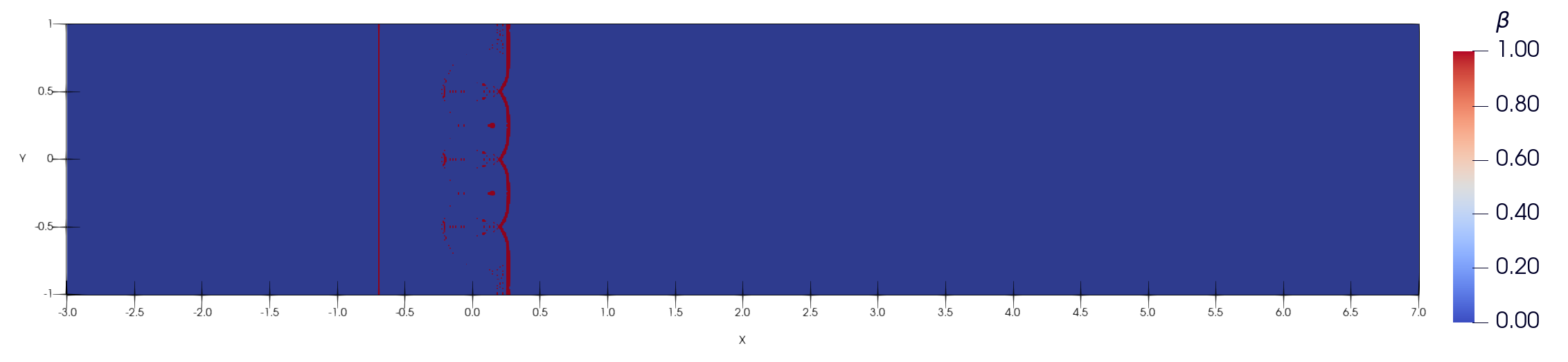}\\
\includegraphics[width=0.33\textwidth]{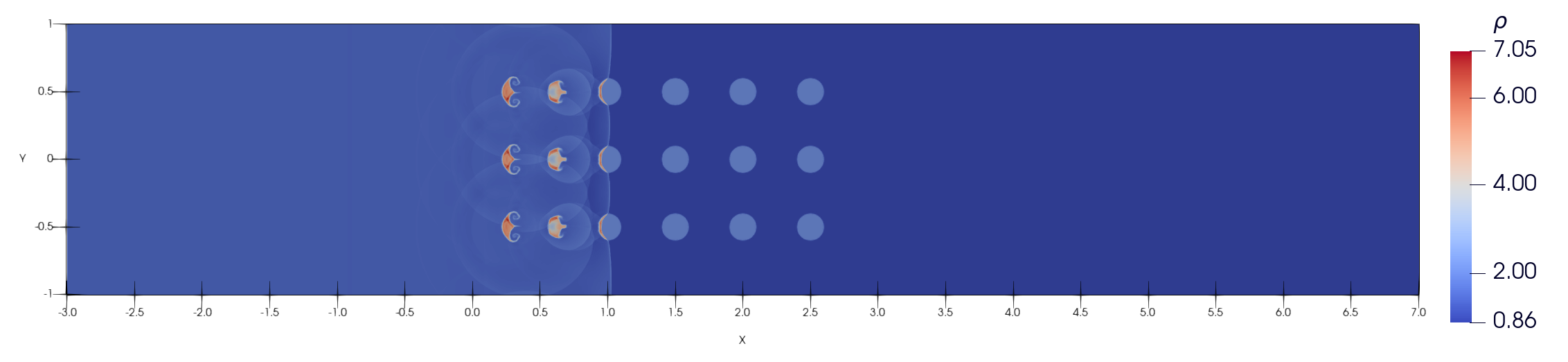}
\includegraphics[width=0.33\textwidth]{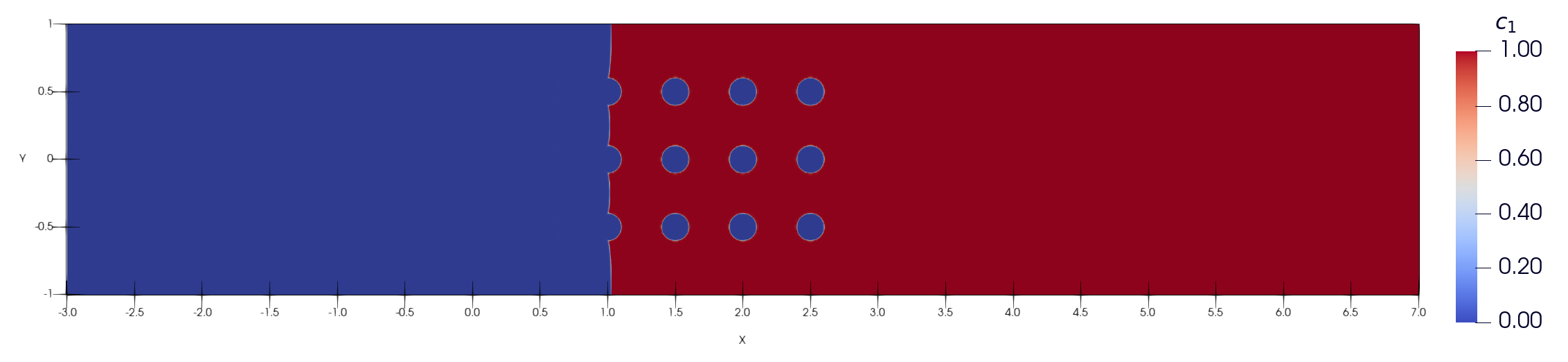}
\includegraphics[width=0.33\textwidth]{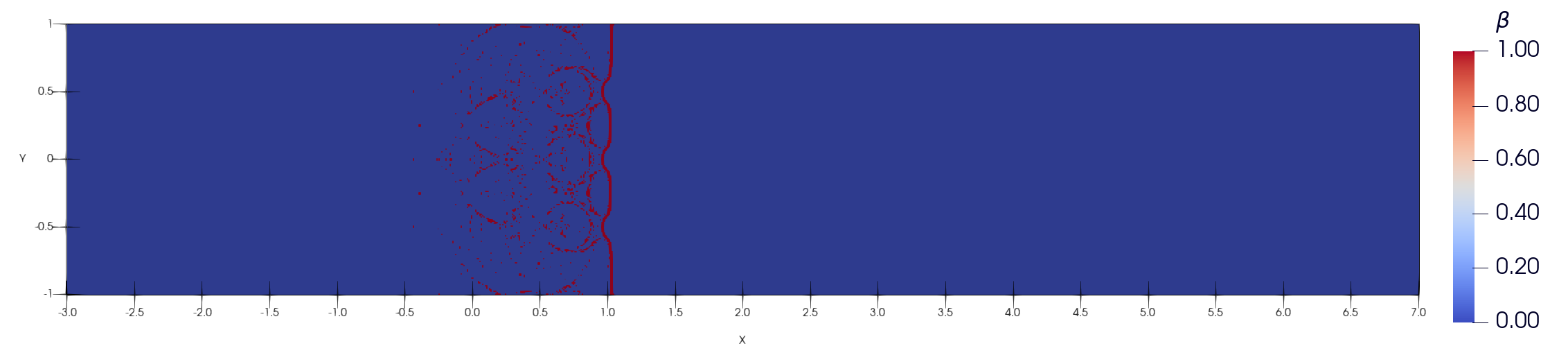}\\
\includegraphics[width=0.33\textwidth]{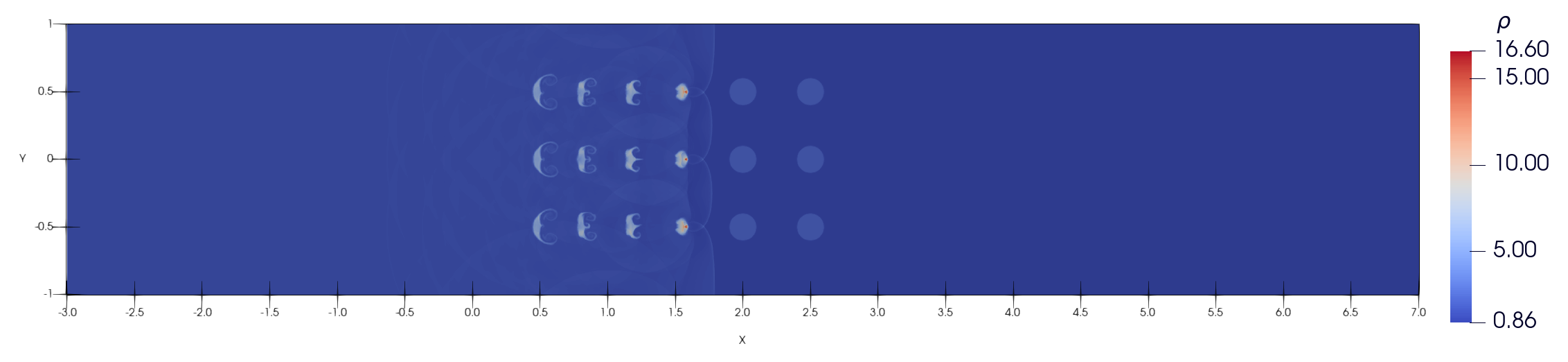}
\includegraphics[width=0.33\textwidth]{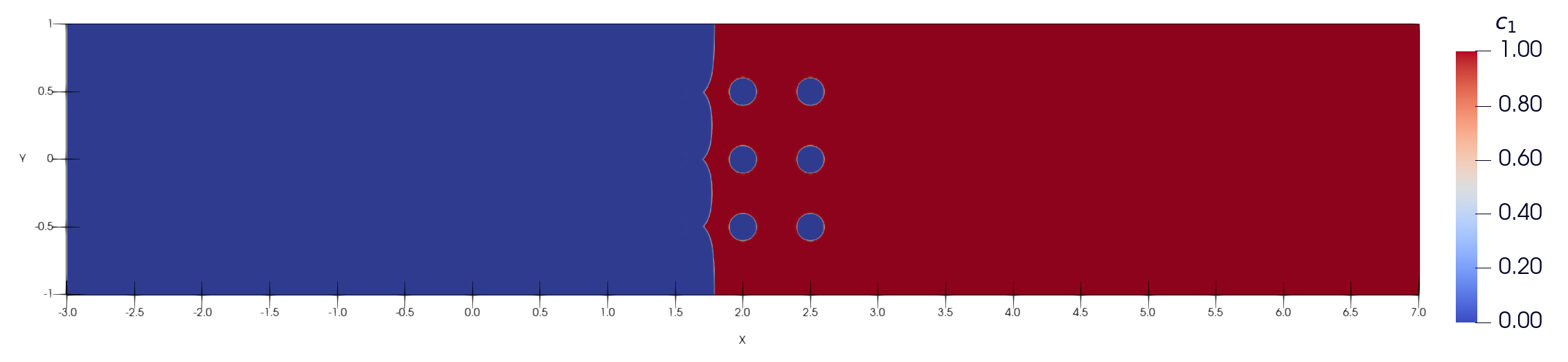}
\includegraphics[width=0.33\textwidth]{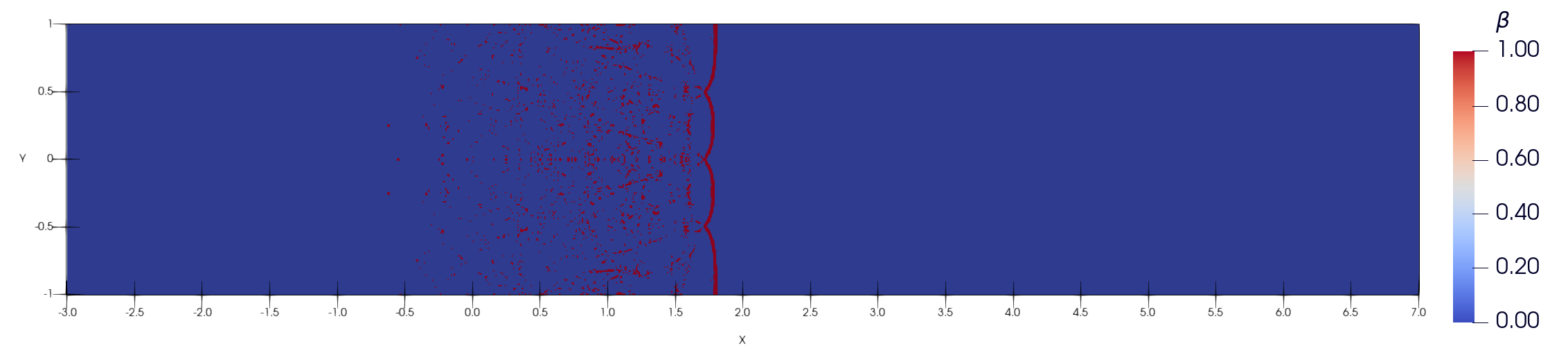}\\
\includegraphics[width=0.33\textwidth]{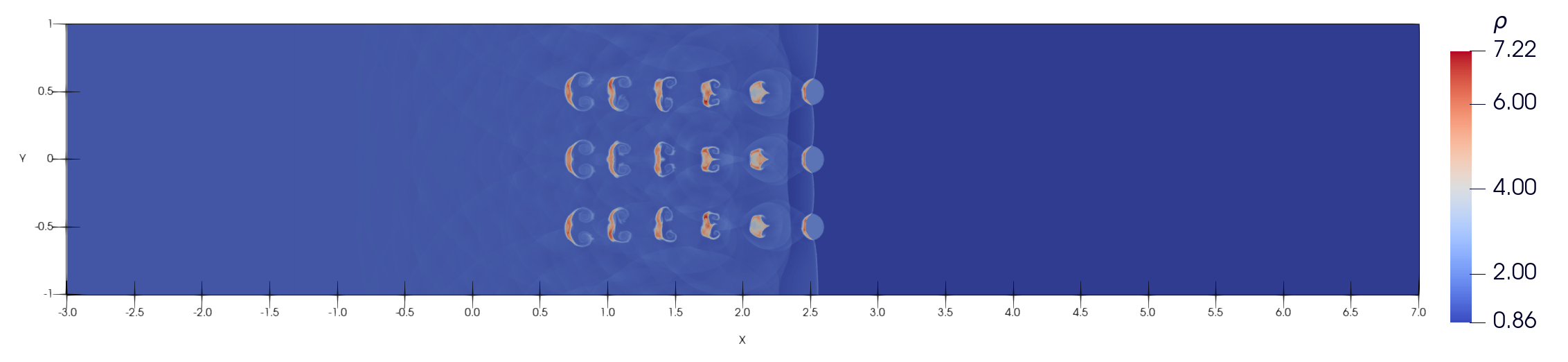}
\includegraphics[width=0.33\textwidth]{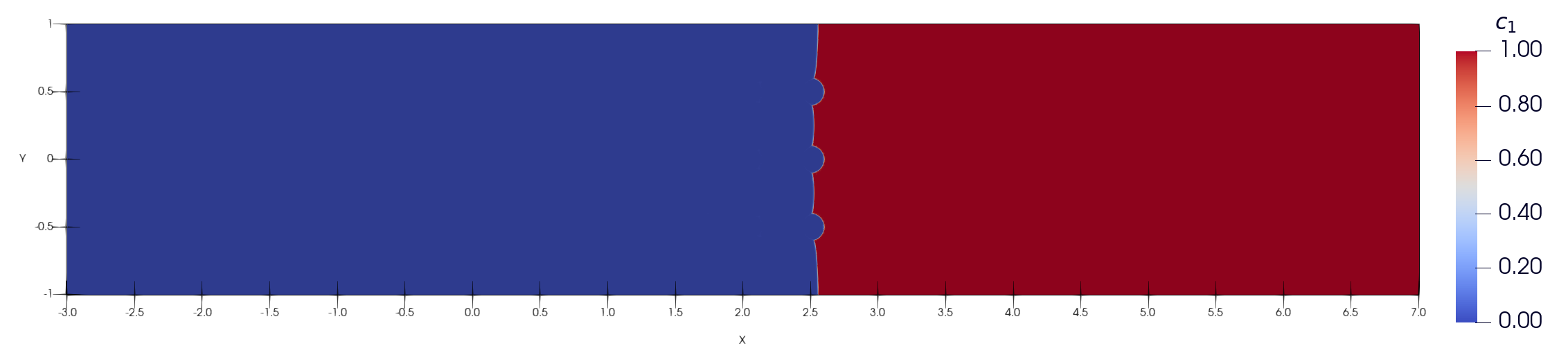}
\includegraphics[width=0.33\textwidth]{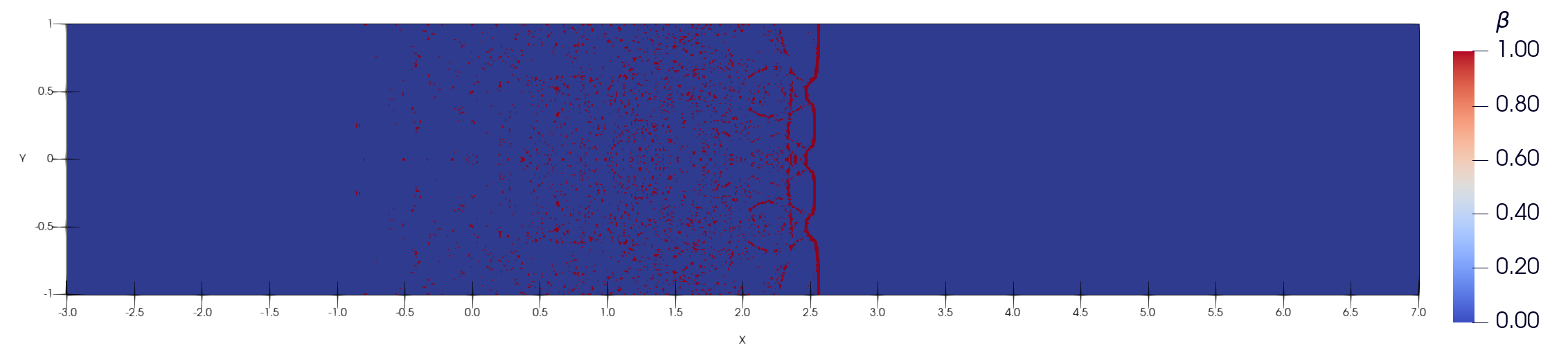}\\
\includegraphics[width=0.33\textwidth]{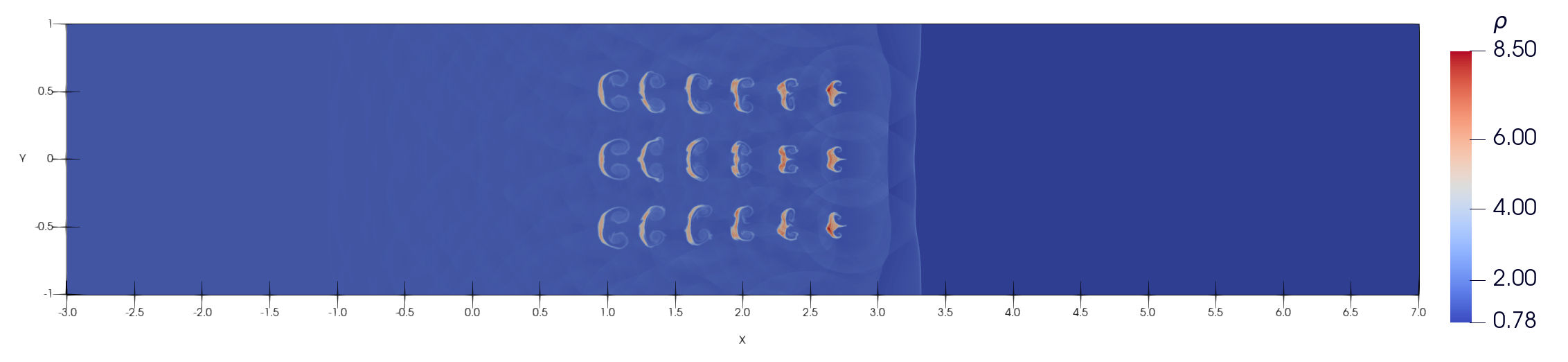}
\includegraphics[width=0.33\textwidth]{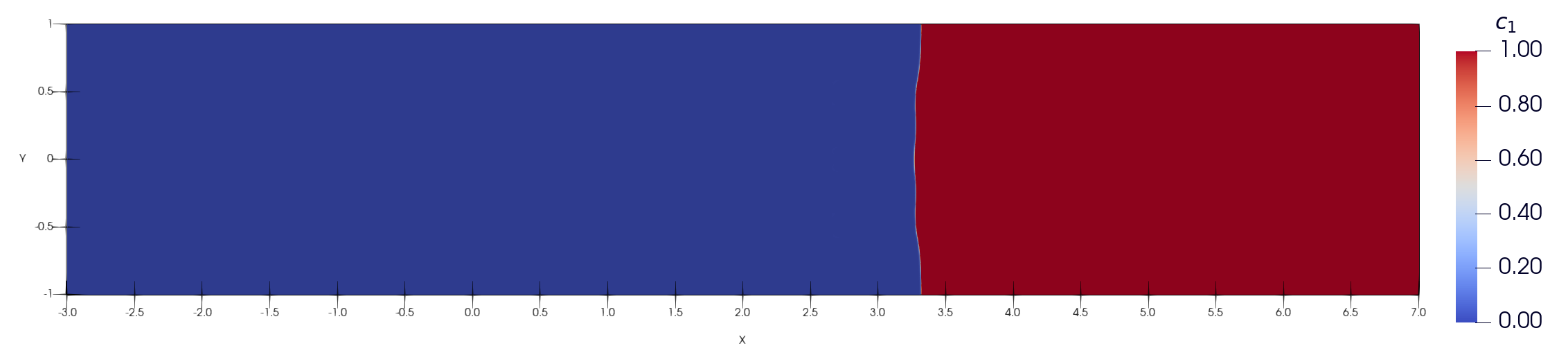}
\includegraphics[width=0.33\textwidth]{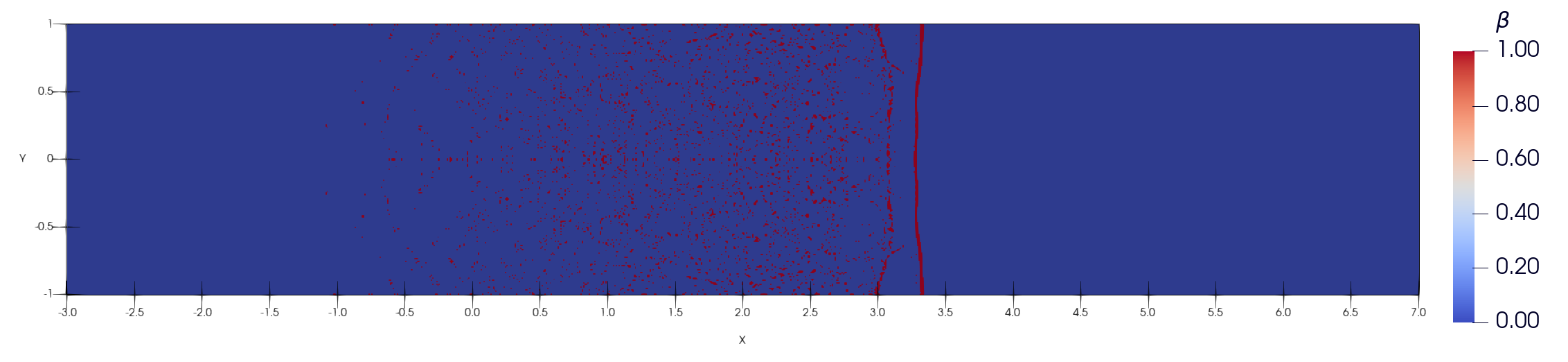}\\
\includegraphics[width=0.33\textwidth]{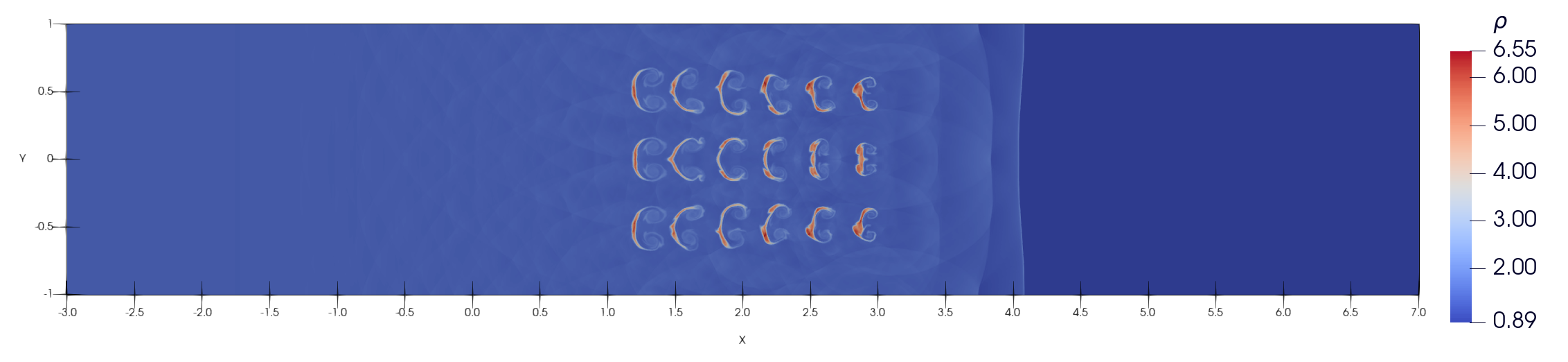}
\includegraphics[width=0.33\textwidth]{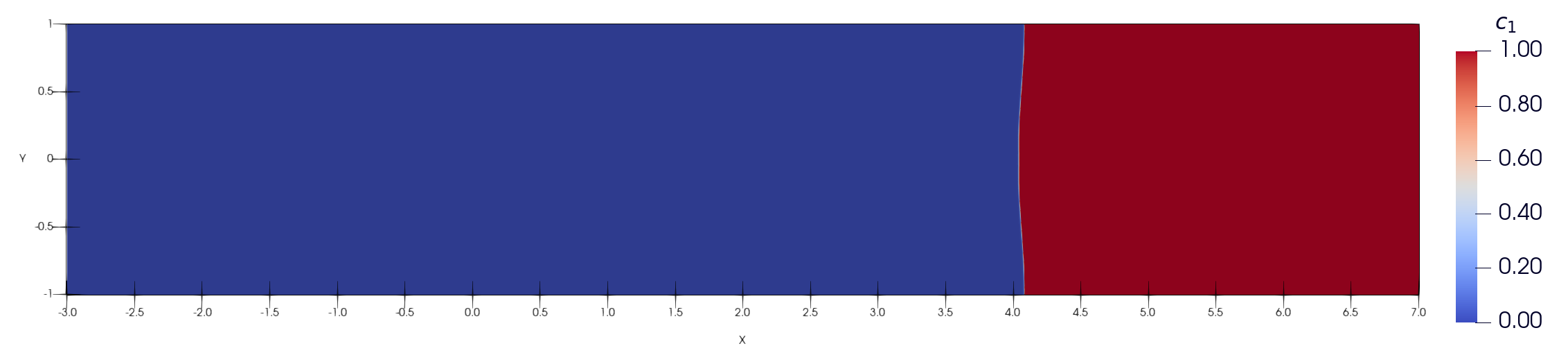}
\includegraphics[width=0.33\textwidth]{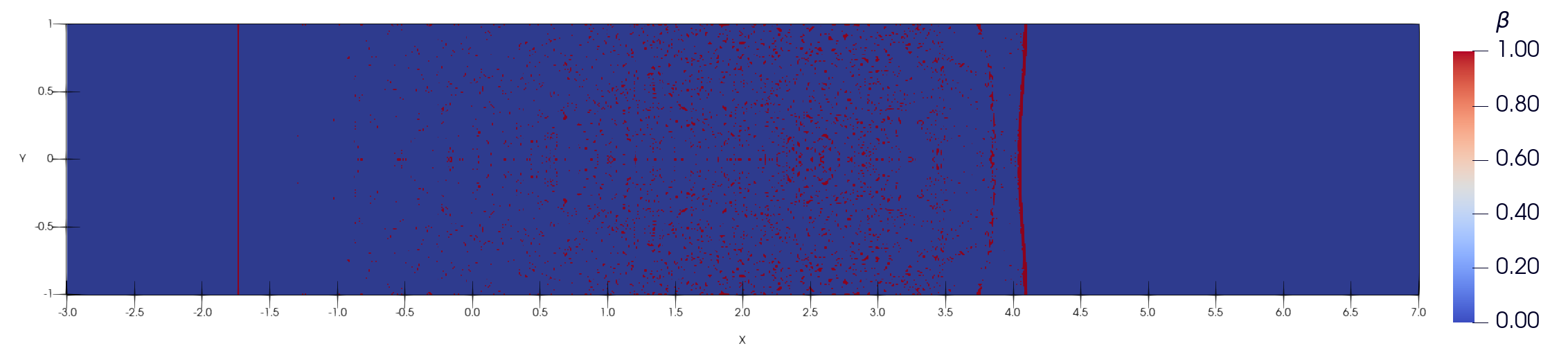}\\
\includegraphics[width=0.33\textwidth]{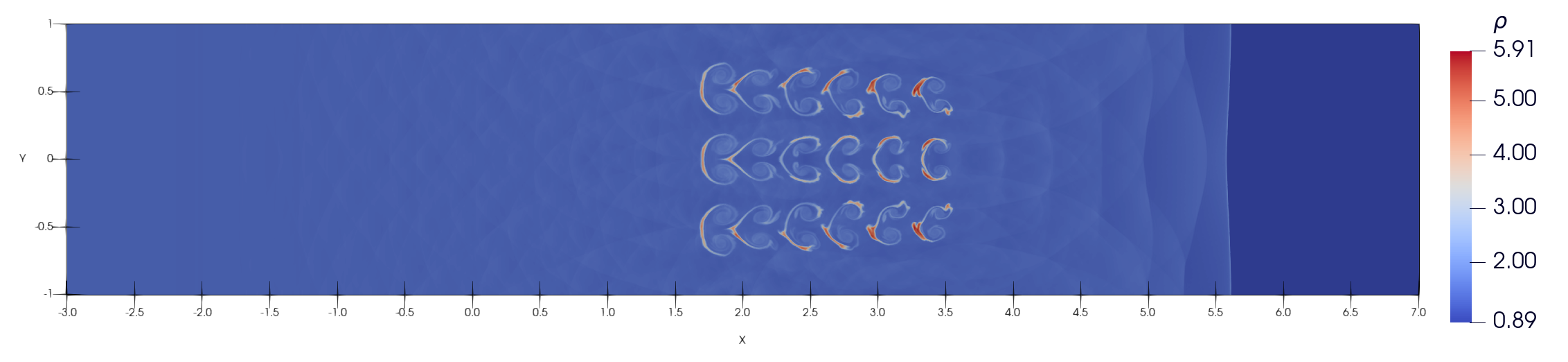}
\includegraphics[width=0.33\textwidth]{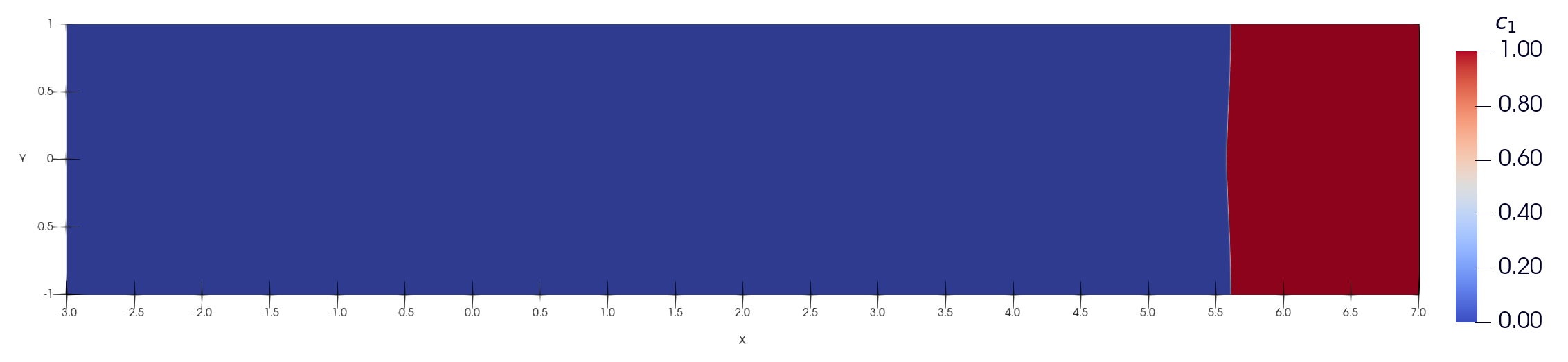}
\includegraphics[width=0.33\textwidth]{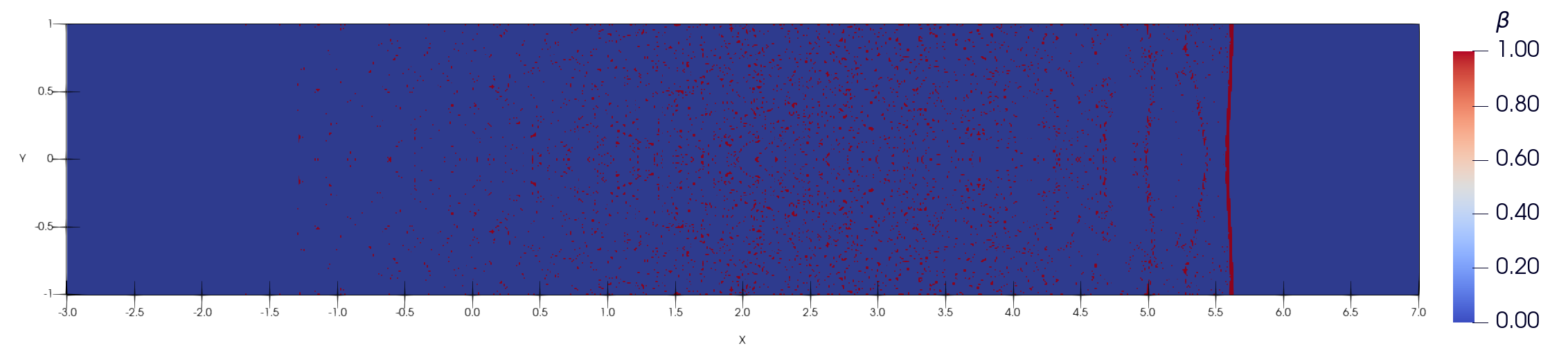}\\
\includegraphics[width=0.33\textwidth]{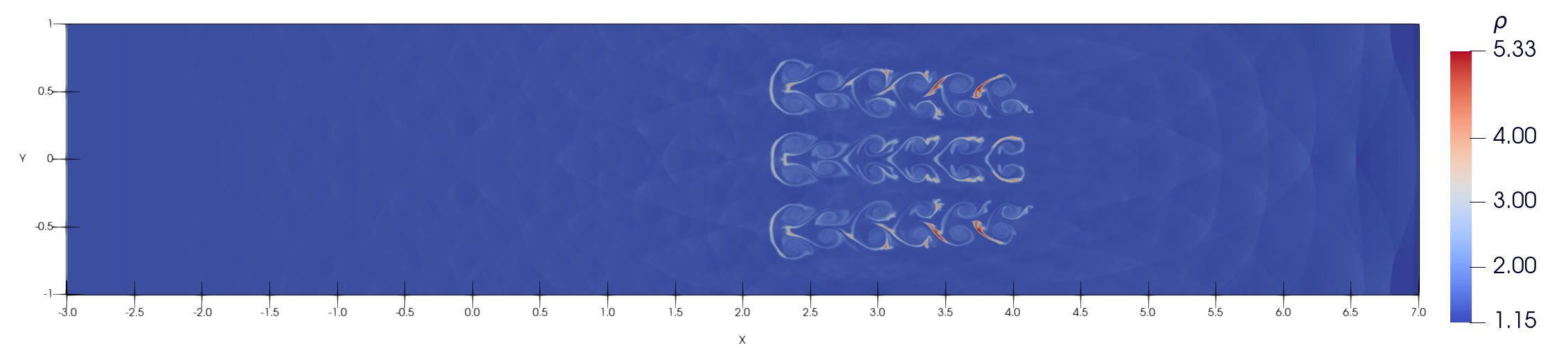}
\includegraphics[width=0.33\textwidth]{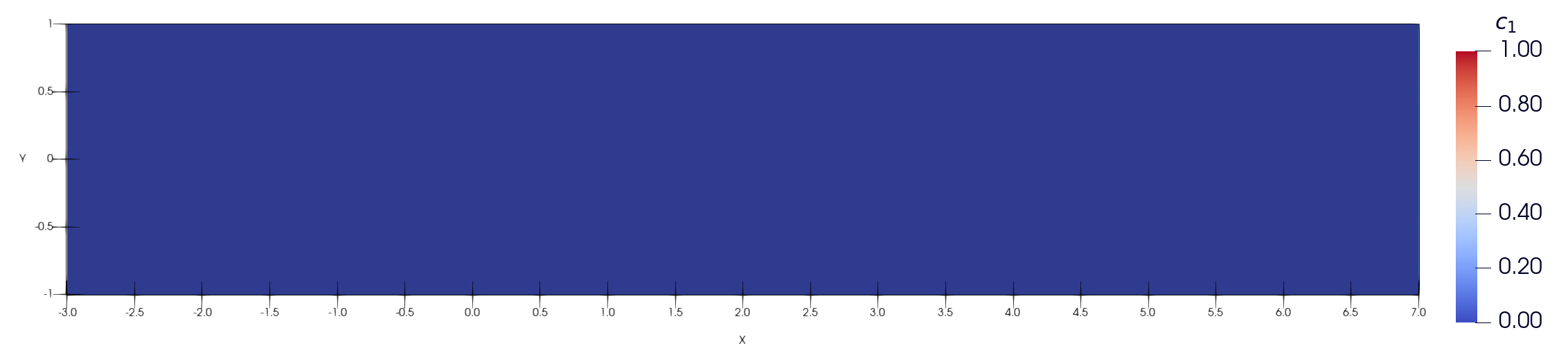}
\includegraphics[width=0.33\textwidth]{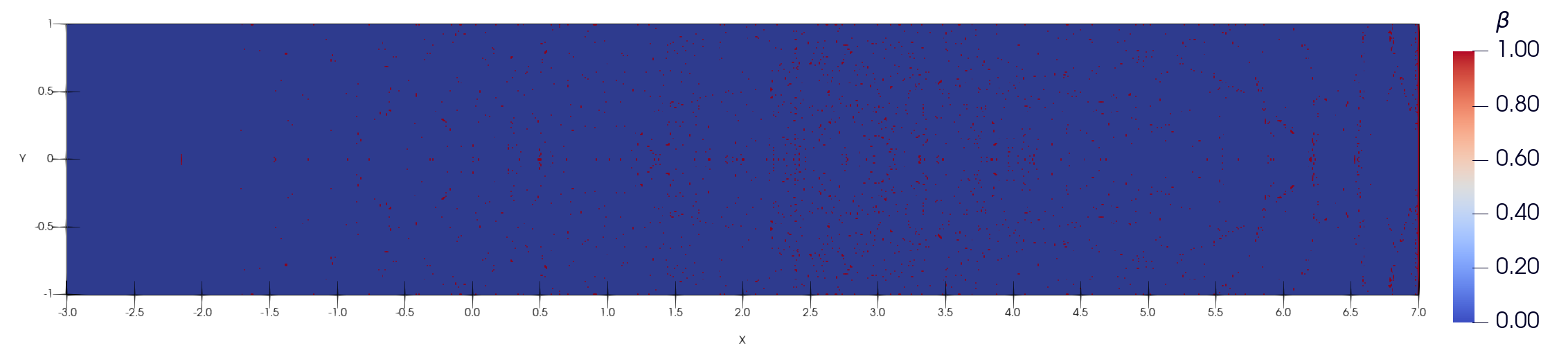}
\caption{\label{fig:dwmbif_3x6_2d}
Numerical solution of the two-dimensional problem of interaction between a detonation wave and $18$ ($6 \times 3$) reaction-inert bubbles
in a two-component medium with a ``fast'' reaction (strong stiff case, a detailed statement of the problem is presented in the text),
obtained using the ADER-DG-$\mathbb{P}_{2}$ method with a posteriori limitation of the solution by a ADER-WENO2 finite volume limiter 
on mesh with $1000 \times 200$ cells at the times $t = 0.5$, $1.0$, $1.5$, $2.0$, $2.5$, $3.0$, $4.0$ and $5.0$ (from top to bottom).
The graphs show the coordinate dependencies of the subcells finite-volume representation of density $\rho$ (left), 
mass concentration $c_{1}$ of the reaction reagent (center) and troubled cells indicator $\beta$ (right).
}
\end{figure*}

A numerical solution to the problem of the interaction between a detonation wave and a regular lattice $6\times3$ of inert bubbles made it possible to quantify the capabilities of the numerical method for solving complex multidimensional problems of the formation and movement of detonation waves in situations where the geometric and physical properties of the medium in which detonation processes are considered suggest the occurrence of multiple refraction phenomena and multiple diffraction of detonation waves on obstacles of complex and moving geometry, the formation of a more complex structure than in the case of interaction with a single bubble (see Subsection~\ref{sec:detonation_waves:dwbi}), multiple reflected and interacting shock waves, contact discontinuities, rarefaction waves, as well as vortex generation domains and vortex streets.

Computational coordinate domain $\Omega = [-3, +7]\times[-1, +1]$. The initial conditions were chosen in the following form:
\begin{equation}
\begin{split}
&\rho = \left\{
\begin{array}{ll}
1.4, & \mathrm{if}\ x \leqslant -0.5;\\
2.0, & \mathrm{if}\ \mathtt{d} \leqslant 0.1;\\
0.887565, & \mathrm{if}\ x >\, -0.5 \land \mathtt{d} >\, 0.1;
\end{array}
\right.\\
&p = \left\{
\begin{array}{ll}
1.0, & \mathrm{if}\ x \leqslant -0.5;\\
0.191709, & \mathrm{if}\ x >\, -0.5;
\end{array}
\right.\\
&u = \left\{
\begin{array}{ll}
0.577350, & \mathrm{if}\ x \leqslant -0.5;\\
0.0, & \mathrm{if}\ x >\, -0.5;
\end{array}
\right.\\
&v = 0.0;\\
&c_{1} = \left\{
\begin{array}{ll}
10^{-14}, & \mathrm{if}\ x \leqslant -0.5 \lor \mathtt{d} \leqslant 0.1;\\
1.0, & \mathrm{if}\ x >\, -0.5 \land \mathtt{d} >\, 0.1;
\end{array}
\right.\\
&c_{2} = \left\{
\begin{array}{ll}
1.0, & \mathrm{if}\ x \leqslant -0.5 \lor \mathtt{d} \leqslant 0.1;\\
10^{-14}, & \mathrm{if}\ x >\, -0.5 \land \mathtt{d} >\, 0.1;
\end{array}
\right.
\end{split}
\end{equation}
where $\mathtt{d} = \min_{ij} r_{ij}$ is the distance to the center of the bubble nearest to the point $(x, y)$, $r_{ij}^{2} = (x - x_{i})^{2} + (y - y_{j})^{2}$ determines the distance to the center of bubble with center in the point $(x_{i}, y_{j})$. A small value $10^{-14}$ of mass concentrations $c_{1}$ and $c_{2}$, instead of strictly $0$, was chosen to prevent the occurrence of negative concentrations immediately at the start of the calculation process, which could lead to a meaninglessly large increase in the number of troubled cells in the solution. From the point of view of energy release and flow energy balance, these values of reagent concentration do not have any significant effect. 

The initial conditions define a regular lattice $6\times3$ of inert bubbles of radius $R = 0.1$ with density $\rho_{0} = 2.0$, the centers of which are located at the nodes of a regular square lattice $(x_{i}, y_{j}) = \{x_{i}\}\times\{y_{j}\}$ of size $6\times3$. The nodes coordinates $\{x_{i}\} = \{0.0,\, 0.5,\, 1.0,\, 1.5,\, 2.0,\, 2.5\}$ and the nodes coordinates $\{y_{j}\} = \{-0.5,\, 0.0,\, +0.5\}$. The initial conditions also define a CJ detonation wave with parameters that were calculated in Subsection~\ref{sec:detonation_waves:cjdw_1d}, which is located on the line $x = -0.5$. The initial conditions for coordinate dependency of density $\rho$ and mass concentration $c_{1}$ of the reaction reagent is presented in Figure~\ref{fig:dwmbi_3x6_2d_init}. The boundary conditions were chosen as follows: on the left boundary is the exact solution for \newcoloringtext{the detonation front, specified from the CJ conditions}, on the right boundary are the conditions of free outflow, periodic boundary conditions were set on top and bottom boundaries, which under the symmetry of initial conditions of the problem are equivalent to the solid wall boundary conditions. Results were obtained in the cases of a ``slow'' reaction, which corresponds to weak stiffness in the solution, and a ``fast'' reaction, which corresponds to strong stiffness in the solution. The final time has been chosen $t_{\rm final} = 5.0$. The adiabatic index $\gamma = 1.4$. The Courant number $\mathtt{CFL} = 0.9$.

The numerical solution was obtained using ADER-DG-$\mathbb{P}_{2}$ method with ADER-WENO2 finite volume a posteriori limiter on a spatial mesh with sizes $1000\times200$. The numerical solutions are presented in Figure~\ref{fig:dwbis_2d} in the cases of a ``slow'' reaction and in Figure~\ref{fig:dwbif_2d} in the cases of a ``fast'' reaction. The numerical solutions are presented  at several times $t = 0.5$, $1.0$, $1.5$, $2.0$, $2.5$, $3.0$, $4.0$ and $5.0$ to demonstrate the flow dynamics and the arising non-stationary processes. It should be noted that the number of troubled cells on the mesh never exceeded $3.8\%$. The average number of troubled cells in this test was $\sim 2.1\%$. In this case, of course, the number of troubled cells was determined by the emerging features and structures in the solution.

The numerical solution presented in Figure~\ref{fig:dwmbis_3x6_2d} demonstrates the propagation of a detonation wave in a medium with a ``slow'' reaction through a regular lattice $6\times3$ of dense bubbles. In this case, a classic set of hydrodynamic processes arises associated with the propagation of detonation and shock waves in an inhomogeneous medium, and the number of emerging hydrodynamic flow structures is much greater than was presented in the numerical solution of the problem of the interaction of a detonation wave with one inert bubble, considered in the previous paragraph. At the initial moment of time $t = 0$, the numerical solution for which is presented in Figure~\ref{fig:dwmbi_3x6_2d_init}, the detonation wave does not interact with the bubble lattice. At the moment of time $t = 0.5$ presented in Figure~\ref{fig:dwmbis_3x6_2d}, the detonation wave reaches the first line of bubbles, and the beginning of classical diffraction of the detonation wave on three obstacles is observed -- the uncoupled sections of the detonation front and the diffracted sections enveloping the front semi-sphere of the three bubbles are clearly visible. In this case, the fronts of shock waves reflected back from the bubbles and transmitted into the bubbles of shock waves are clearly observed. The surfaces of the bubbles are significantly deformed, and at this stage of interaction with the detonation front, when there is no influence of reflected shock waves from the next line of bubbles, as well as the influence of the top and bottom boundaries of the coordinate domain $\Omega$, their deformed shapes are almost identical. The spatial dependence of the mass concentration $c_{1}$ of the reagent demonstrates the process of smooth combustion of the reagent, which is typical for the case of a ``slow'' reaction in the medium. Troubled cells appear predominantly in the spatial regions of detonation and shock waves, however, some contact discontinuities are also covered with troubled cells. At the subsequent presented moment of time $t = 1.0$, the general detonation front has already transmitted the second layer of bubbles on the lattice, also generating a set of reflected shock waves, only of a more complex configuration. The influence of these shock waves, as well as the influence of shock waves reflected from the middle bubble of the first line of bubbles, led to a violation of the identity of the deformation shape of the bubbles of the first line. However, it should be noted that the final shape of the surface of the upper and lower bubbles is symmetrical about the $x$ axis, which is ensured by the high accuracy of the ADER-DG-$\mathbb{P}_{2}$ method used. At subsequent presented moments of time $t = 1.5$, $2.0$, $2.5$, $3.0$, $4.0$ and $5.0$, an increasing complication of the multi-wave structure of the numerical solution is observed -- a complex superposition of shock waves and contact discontinuities in the solution is observed. The bubble boundaries exhibit very strong deformation, leading to significant vortex generation. There are no standard shapes of deformed bubble boundaries as presented in the problem of the interaction of a detonation wave with one inert bubble. This is primarily due to the fact that through each subsequent line of bubbles, not a flat detonation front is transmitted, but a detonation front of a complex structure, resulting from diffraction on the previous line of bubbles. The most similar to the standard shape is demonstrated by only three bubbles, which are located on the first line of the lattice. At the last presented moments of time $t = 4.0$ and $5.0$, deformed bubbles of previous lines ``catch up'' with bubbles from next lines, as a result of which the shapes of their boundaries are deformed even more significantly, especially taking into account vortex generation and the formation of vortex streets. However, it should be noted that despite the high complexity and multi-scale nature of the hydrodynamic flow, the final shapes of the surfaces of all bubbles are symmetrical about the $x$ axis, which allows us to draw conclusions about the high accuracy of the ADER-DG-$\mathbb{P}_{2}$ method used. The detonation front, from the point of view of combustion of the reagent, starting from the moment of time, becomes stable and almost flat, which is expected. The presented numerical solution demonstrates that the ADER-DG-$\mathbb{P}_{N}$ method with ADER-WENO finite volume a posteriori limiter allows one to effectively and accurately simulate the flow of reacting flows of a rather complex geometric configuration.

The numerical solution presented in Figure~\ref{fig:dwmbif_3x6_2d} demonstrates the propagation of a detonation wave in a medium with a ``fast'' reaction through a regular lattice of $6\times3$ dense bubbles. In this case, a classic set of hydrodynamic processes arises associated with the propagation of detonation and shock waves in an inhomogeneous medium, and the number of emerging hydrodynamic flow structures is much greater than it was imagined in the numerical solution of the problem of interaction of a detonation wave with one inert bubble, considered in the previous paragraph. At the initial time $t = 0$, the detonation wave does not interact with the bubble lattice. At the presented moment of time $t = 0.5$ the detonation wave propagates beyond the first line of bubbles and classical diffraction of the detonation wave on three obstacles is observed -- the diffracted areas resulting from rounding the front semi-spheres of three bubbles. The shock wave fronts reflected from the bubbles and the shock waves transmitted into the bubbles are clearly observed. The surfaces of the bubbles are significantly deformed, and at this stage of interaction with the detonation front, their deformed shapes are almost identical. The spatial dependence of the mass concentration of the reagent $c_{1}$ demonstrates the process of abrupt combustion of the reagent, characteristic of the case of a ``fast'' reaction in the medium. This is characteristic of the classical ZND detonation wave -- a chemical Zel'dovich peak appears. At the subsequent instant of time $t = 1.0$, the general detonation front has already spread to the second layer of bubbles on the lattice and the beginning of diffraction is observed on the third line of bubbles. In this case, a set of reflected shock waves of a more complex configuration is generated. The influence of these shock waves, as well as the influence of shock waves reflected from the middle bubble of the first line of bubbles, leads to a violation of the identity of the deformation shape of the bubbles of the first line. However, it should be noted that the final shape of the surface of the upper and lower bubbles is symmetrical about the $x$ axis, which is ensured by the high accuracy of the ADER-DG-$\mathbb{P}_{2 }$ method used. It should also be noted that the intensity and speed of propagation of the detonation wave in the case of a ``fast'' reaction is higher than in the case of a ``slow'' reaction in the reacting medium, and deformed bubbles demonstrate the formation of jet flows. At the subsequent presented times $t = 1.5$, $2.0$, $2.5$, $3.0$, $4.0$ and $5.0$, an increasing complexity of the multi-wave structure of the numerical solution is observed. The bubble boundaries exhibit very strong deformation, leading to significant vortex formation. There are no standard forms of deformed bubble boundaries presented in the problem of the interaction of a detonation wave with one inert bubble. These phenomena are conceptually of the same nature as in the previous case of a ``slow'' reaction. The most similar to the standard shape is demonstrated by only three bubbles located on the first line of the lattice. At the last presented moments of time $t = 4.0$ and $5.0$, the deformed bubbles of the previous rows ``catch up'' with the bubbles of the next lines, as a result of which the shapes of their boundaries are deformed even more strongly, especially taking into account vortex regeneration and the formation of vortex streets. However, it should be noted that, despite the high complexity and multi-scale nature of the hydrodynamic flow, the final shapes of the surfaces of all bubbles are symmetrical relative to the $x$ axis, which allows us to draw conclusions about the high accuracy of the calculation. The ADER-DG-$\mathbb{P}_{2}$ method was used. The detonation front from the point of view of combustion of the reagent, starting from the moment in time, becomes stable, as expected. The spatial dependence of the mass concentration of the reagent $c_{1}$ demonstrates the process of abrupt combustion of the reagent, characteristic of the case of a ``fast'' reaction in the medium, for all presented moments of time. In this case, the Zel'dovich chemical peak and the classical structure of the ZND detonation wave are observed. Non-physical effects characteristic of modeling detonation processes using classical numerical methods~\cite{frac_steps_detwave_sim_2000, chem_kin_hrs_weno} do not arise -- a non-physical weak detonation front is not formed~\cite{correct_det_wave_speed_2017}. The presented numerical solution demonstrates that the ADER-DG-$\mathbb{P}_{N}$ method with a finite volume ADER-WENO and an a posteriori limiter allows one to efficiently and accurately simulate the flow of reacting flows of a rather complex geometric configuration.

\paragraph{Detonation cellular structure}

\newtext{
The large-scale structure of a stationary detonation wave, after the detonation wave reaches stationary propagation, does not depend on the small-scale features of the detonation initiation process~\cite{Lee_2008}. However, detonation processes, especially in reacting gas flows, are characterized by the formation of a detonation cellular structure as a result of a disturbing effect on the detonation front~\cite{Oran_Boris_2005, Lee_2008}. In this case, the classical stationary structure of the detonation wave is destroyed, and a stable non-stationary cellular structure is formed, which exists for a sufficiently long period of time, so it cannot be attributed simply to a transient process. Classic ways of perturbing the detonation front are to place an asymmetrical pocket (especially an elliptical pocket) of unburned material immediately behind the front of a planar stationary detonation wave~\cite{Oran_Boris_2005, dcs_oran_1985, dcs_oran_1987}, and to create a perturbed initial initiation, such as a spherical or sinusoidal ``wave'' irregularity at the initial planar detonation front~\cite{dcs_hydrogen_2020}. An important feature of the detonation cellular structure is the significant non-stationary and variability of the spatial structure of both the detonation front and the flow behind the detonation wave front~\cite{Oran_Boris_2005}. 

In this case, a flow with multiple shock waves and tangential discontinuities is formed, which are characterized by the intersection of discontinuity fronts and significant nonlinear multi-wave interaction. The resulting shock waves and rarefaction waves interact not only with each other and with tangential discontinuities, but also with solid walls limiting the detonation flow, and a complex nonlinear superposition arises in the overall wave pattern of the flow. The structure of the detonation wave in this case is characterized by the existence of disturbances transverse to the direction of detonation propagation. In the case of sufficiently symmetric conditions of the initial initiation disturbance and the geometric conditions of the problem, the symmetry of the flow in the numerical solution is expected to be maintained over a long time period. Numerical solutions to the problem of interaction of a detonation wave with a single-bubble inertial inhomogeneity and inert inhomogeneities in the form of a regular lattice of bubbles, obtained using the ADER-DG-$\mathbb{P}_{N}$ method with ADER-WENO2 finite volume a posteriori limiter presented in the previous paragraphs, demonstrated a very high accuracy of the numerical solution and high quality of maintaining the symmetry of the flow over a long time range. Therefore, it is also interesting to carry out a numerical study of the formation of a detonation cellular structure using the ADER-DG-$\mathbb{P}_{N}$ method with ADER-WENO finite volume a posteriori limiter in order to determine the possibilities of reproducing rather complex patterns of the detonation process.

This paragraph presents the results of a study of the formation and development of a detonation cellular structure in a two-component reacting gas media. The reaction properties of the medium were chosen to be the same as in the previous cases considered. The cases of a ``slow'' reaction, which is characterized by weak stiffness, and a “fast” reaction, which is characterized by strong stiffness, were studied. The numerical solution was obtained by a method that is similar to the cases considered in the previous paragraphs. Computational coordinate domain $\Omega = [0, 10]\times[-1, +1]$. 

The formation of a detonation cellular structure requires disturbance of the detonation front, and it is necessary to form shock waves reflected from the solid walls of the coordinate domain or sudden changes in density, which leads to transverse movement of shock waves directly behind the detonation front, which leads to its local curvature in the case of stable detonation and may lead to unstable propagation of detonation or disruption of the detonation regime in the event of unstable detonation. The initial disturbance in this work was chosen in a form  $\sim|\sin(n \pi x)|$ of the initial stationary detonation front, specified from the CJ conditions, that is essentially similar to the selected initiation method in the work~\cite{dcs_hydrogen_2020}. A pressure disturbance was also added to the initial disturbance of the detonation front shape, which made it possible to ``split'' the flow behind the detonation wave front into two ``parts'' by a region of significant vortex generation associated with the formation and localization of the Richtmyer-Meshkov instability. The initial conditions were chosen in the following form:
\begin{equation}
\begin{split}
&\rho = \left\{
\begin{array}{ll}
1.4, & \mathrm{if}\ x \leqslant x_{\rm b}(y);\\
0.887565, & \mathrm{if}\ x > x_{\rm b}(y);
\end{array}
\right.\\
&p = \left\{
\begin{array}{ll}
1.0, & \mathrm{if}\ x \leqslant x_{\rm b}(y) - \Delta x;\\
2.0, & \mathrm{if}\ x_{\rm b}(y) - \Delta x < x \leqslant x_{\rm b}(y);\\
0.191709, & \mathrm{if}\ x > x_{\rm b}(y);
\end{array}
\right.\\
&u = \left\{
\begin{array}{ll}
0.577350, & \mathrm{if}\ x \leqslant x_{\rm b}(y);\\
0.0, & \mathrm{if}\ x > x_{\rm b}(y);
\end{array}
\right.\\
&v = 0.0;\\
&c_{1} = \left\{
\begin{array}{ll}
10^{-14}, & \mathrm{if}\ x > x_{\rm b}(y);\\
1.0, & \mathrm{if}\ x \leqslant x_{\rm b}(y);
\end{array}
\right.\\
&c_{2} = \left\{
\begin{array}{ll}
1.0, & \mathrm{if}\ x \leqslant x_{\rm b}(y);\\
10^{-14}, & \mathrm{if}\ x > x_{\rm b}(y);
\end{array}
\right.\\
\end{split}
\end{equation}
where $x_{\rm b}(y) = 2.5 + 0.5\cdot|\sin(2 \pi x)|$ is the coordinate $x$ at which the initial detonation front is set, which is determined with additional irregularity $0.5\cdot|\sin(2 \pi x)|$, and $\Delta x = 0.25$ is the depth at which additional pressure disturbances are specified. A small value $10^{-14}$ of mass concentrations $c_{1}$ and $c_{2}$, instead of strictly $0$, was chosen to prevent the occurrence of negative concentrations immediately at the start of the calculation process, which could lead to a meaninglessly large increase in the number of troubled cells in the solution. From the point of view of energy release and flow energy balance, these values of reagent concentration do not have any significant effect. The initial conditions for coordinate dependency of density $\rho$ and mass concentration $c_{1}$ of the reaction reagent is presented in Figure~\ref{fig:dwsines_2d_init}. The boundary conditions were chosen as follows: on the left boundary is the exact solution for the detonation front, specified from the CJ conditions, on the right boundary are the conditions of free outflow, periodic boundary conditions were set on top and bottom boundaries, which under the symmetry of initial conditions of the problem are equivalent to the solid wall boundary conditions. The final time has been chosen $t_{\rm final} = 4.0$. The adiabatic index $\gamma = 1.4$. The Courant number $\mathtt{CFL} = 0.9$.

\begin{figure}[h!]
\centering
\includegraphics[width=0.49\textwidth]{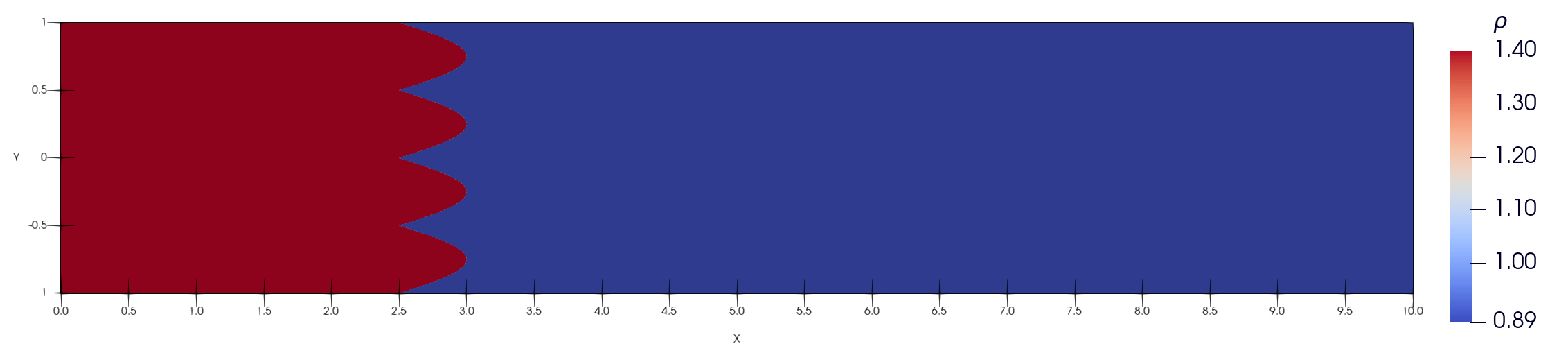}
\includegraphics[width=0.49\textwidth]{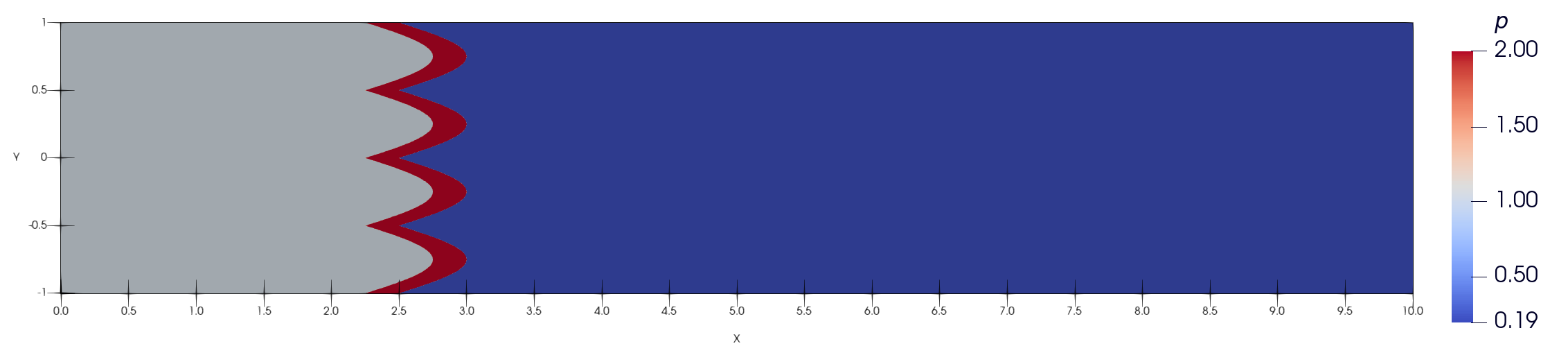}
\includegraphics[width=0.49\textwidth]{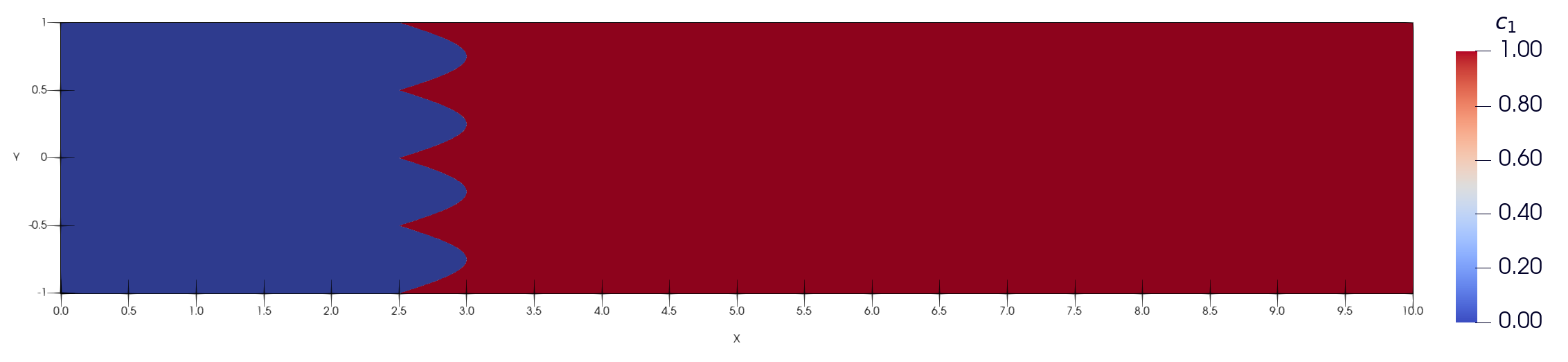}
\caption{\label{fig:dwsines_2d_init}
The initial conditions for coordinate dependency of density $\rho$ (top), 
pressure $p$ (middle) and mass concentration $c_{1}$ of the reaction reagent (bottom)
in the two-dimensional problem of detonation cellular structure (a detailed statement of the problem is presented in the text).
}
\end{figure}

The numerical solution of the two-dimensional problem of detonation cellular structure in a two-component medium was obtained using ADER-DG-$\mathbb{P}_{2}$ method with ADER-WENO2 finite volume a posteriori limiter on a spatial mesh with sizes $1000\times200$. The numerical solutions are presented in Figures~\ref{fig:dwsines_2d_den_p} and~\ref{fig:dwsines_2d_c_1_beta} in the cases of a ``slow'' reaction and in Figures~\ref{fig:dwsinef_2d_den_p} and~\ref{fig:dwsinef_2d_c_1_beta} in the cases of a ``fast'' reaction. Figures~\ref{fig:dwsines_2d_den_p} and~\ref{fig:dwsinef_2d_den_p} presents the coordinate dependencies of the subcells finite-volume representation of density $\rho$ and pressure $p$. Figures~\ref{fig:dwsines_2d_c_1_beta} and~\ref{fig:dwsinef_2d_den_p} presents the coordinate dependencies of the subcells finite-volume representation of mass concentration $c_{1}$ of the reaction reagent and troubled cells indicator $\beta$. The numerical solutions are presented at several times $t = 0.4$, $0.8$, $1.2$, $1.6$, $2.0$, $2.4$, $2.8$, $3.2$, $3.6$ and $4.0$ to demonstrate the flow dynamics and the arising non-stationary processes. It should be noted that the number of troubled cells on the mesh never exceeded $3.4\%$ in the cases of a ``slow'' reaction and $3.1\%$ in the cases of a ``fast'' reaction. The average number of troubled cells in this test was $\sim 2.3\%$ in the cases of a ``slow'' reaction and $\sim 2.1\%$ in the cases of a ``fast'' reaction. In this case, of course, the number of troubled cells was determined by the emerging features and structures in the solution. Figures~\ref{fig:dwsines_2d_2000x400} and~\ref{fig:dwsinef_2d_2000x400} are presented for demonstration of a very high-resolution numerical solution of the two-dimensional problem of detonation cellular structure in a two-component medium obtained using ADER-DG-$\mathbb{P}_{2}$ method with ADER-WENO2 finite volume a posteriori limiter on a spatial mesh with sizes $2000\times400$, for the case of weak and strong stiffness, respectively. The coordinate dependencies of the subcells finite-volume representation of density $\rho$ are presented at several times $t = 0.5$, $2.0$, and $4.0$.

The numerical solution presented in Figures~\ref{fig:dwsines_2d_den_p} and~\ref{fig:dwsines_2d_c_1_beta} demonstrates the development of the detonation process from the initial state in a medium with a ``slow'' reaction. The initial sinusoidal disturbance of the detonation initiation boundary under initial conditions leads to the formation of curved non-stationary detonation fronts, as well as a complex set of shock waves and rarefaction waves, which was expected from this formulation of the problem. The detonation front at times $t = 0.4$ begins to demonstrate a detonation cellular structure, which is realized in the form of a set of shock waves propagating not only in the direction of detonation propagation along the $x$-axis, but also across the front. In this case, several triple points of intersection of the shock wave surfaces are observed with the formation of Mach shock waves. The main detonation front exhibits four distinct ``carbuncles'' associated with the initial disturbance, and three- and four-wave configurations located between them -- four-wave configurations, in reality, are two three-wave configurations with a bow shock wave between them, but in presented Figure~\ref{fig:dwsines_2d_den_p} at time $t = 0.4$ this development of intersection of discontinuities is practically not resolved. The magnitude of the compression of the medium and the pressure $p$ values in the resulting head waves significantly exceed the corresponding values in the coordinate domains surrounding them. The general picture of the propagation and interaction of shock waves and the formation of a detonation front has a highly expressed flow symmetry, corresponding to the symmetry of the initial conditions under periodic boundary conditions along the $y$-axis. In the flow region behind the front of the detonation wave, a classical picture of the development of two-dimensional disintegration of a discontinuity in an already reacted medium is observed. A set of four regions of vorticity formation arises (one ``column'', due to periodic boundary conditions, is divided into two ``half-columns'' -- above and below), which is associated with the development of the Richtmyer-Meshkov instability when shock waves pass at an angle to the contact discontinuity. Also, four shock waves propagate in the opposite direction, having significantly lower intensity compared to the vicinity of the detonation front, in space behind which the areas of intersection of discontinuities and interaction of shock waves are also clearly observed. It should be noted that the classical picture of vortex generation during the development of the Richtmyer-Meshkov instability is usually observed in a problem with one curvilinear boundary; however, in this problem, the vortex generation region also ``collides'' with a denser medium that arose after the passage of shock waves in the opposite direction and formed as a result of the breakup of rarefaction waves, therefore the formation of vortex coherent structures in this case is more complex and interesting. The presented description of the flow is clearly observed based on a comparison of the coordinate dependencies of density $\rho$ and pressure $p$ in Figure~\ref{fig:dwsines_2d_den_p}. The propagation of the detonation wave leads to burnout of the reagent $A$, which is clearly observed in the coordinate dependencies of the mass concentration $c_{1}$ of the reagent presented in Figure~\ref{fig:dwsines_2d_c_1_beta}. By time $t = 0.4$, complete burnout of the reagent is observed deep behind the detonation front. However, in the vicinity of the front, a gradual burnout of the reagent is observed, while the detonation front exhibits coupling. This property fits well into the general ideas about the structure of the detonation front in the case of weak stiffness. It should be noted that at the moment of time, narrow regions of localization of the unburned reagent behind the detonation front are observed, which extend quite deep into the region behind the detonation wave, neatly in the region of zebra vortex formation. This is due to the propagation of rarefaction waves formed as a result of the decay of the initial discontinuity. In this case, the detonation cellular structure is clearly observed in the coordinate dependencies of the mass concentration $c_{1}$ of the reagent. The distribution of troubled cells, as coordinate dependencies of the troubled cells indicator $\beta$, presented in Figure~\ref{fig:dwsines_2d_c_1_beta} shows that the limiter is caused predominantly only in the areas where shock fronts are localized, while the head shock waves and areas of shock wave interaction can be clearly distinguished. In areas of tangential discontinuities, trouble cells practically do not appear (troubled cells appear in the areas of intersection of shock waves and tangential discontinuities).

\begin{figure*}[h!]
\centering
\includegraphics[width=0.49\textwidth]{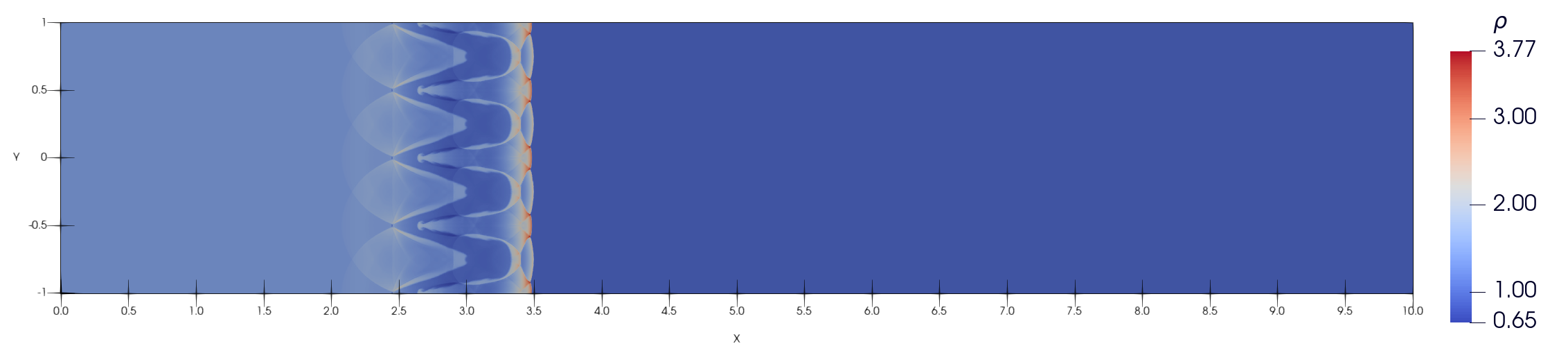}
\includegraphics[width=0.49\textwidth]{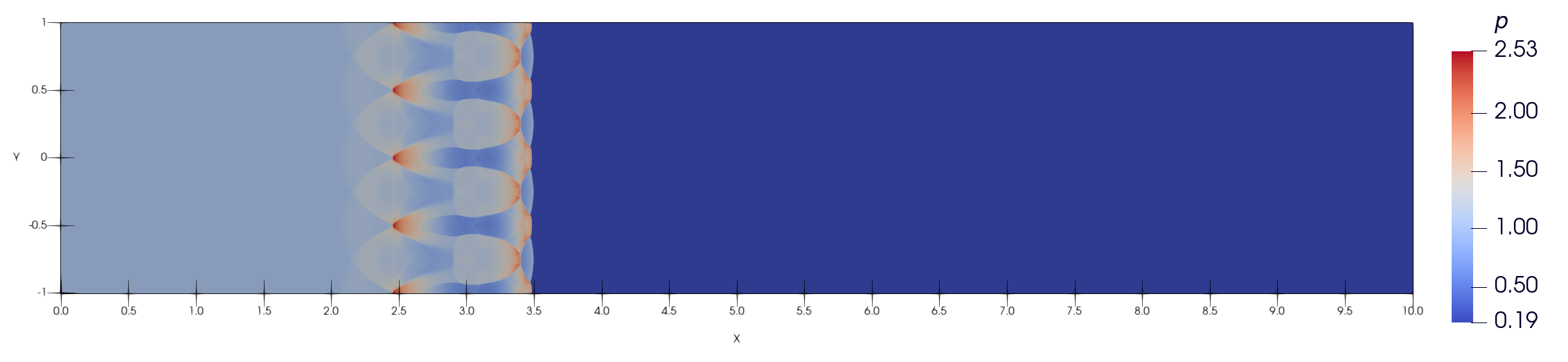}\\
\includegraphics[width=0.49\textwidth]{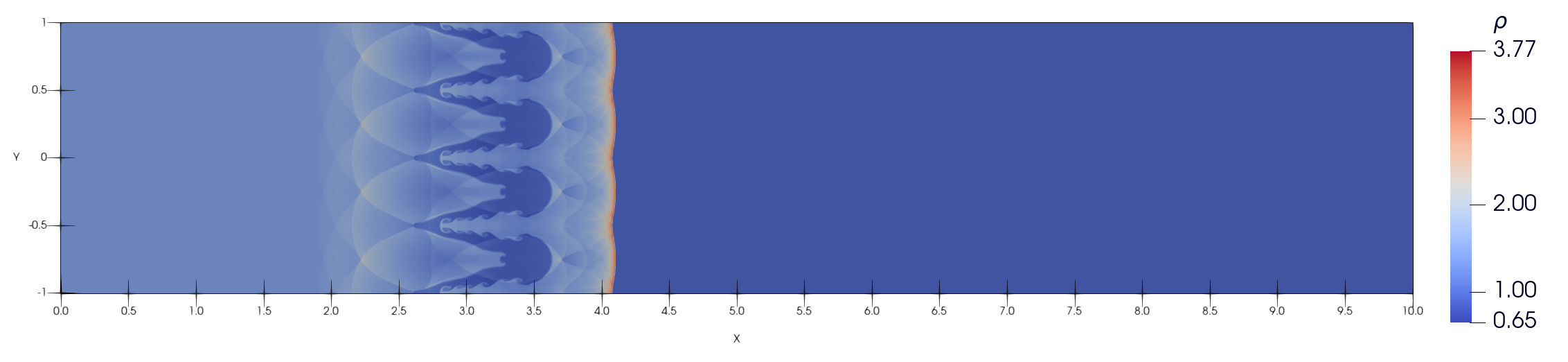}
\includegraphics[width=0.49\textwidth]{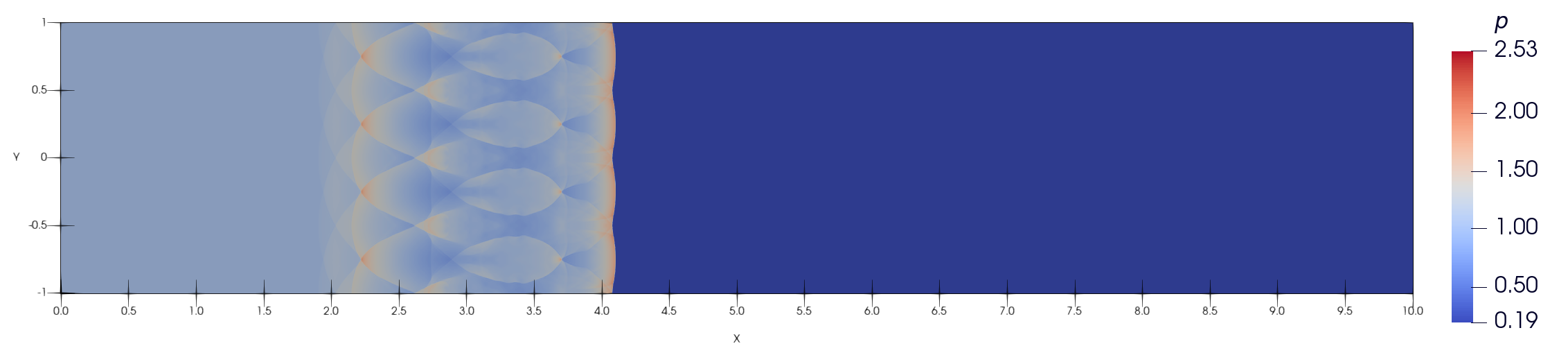}\\
\includegraphics[width=0.49\textwidth]{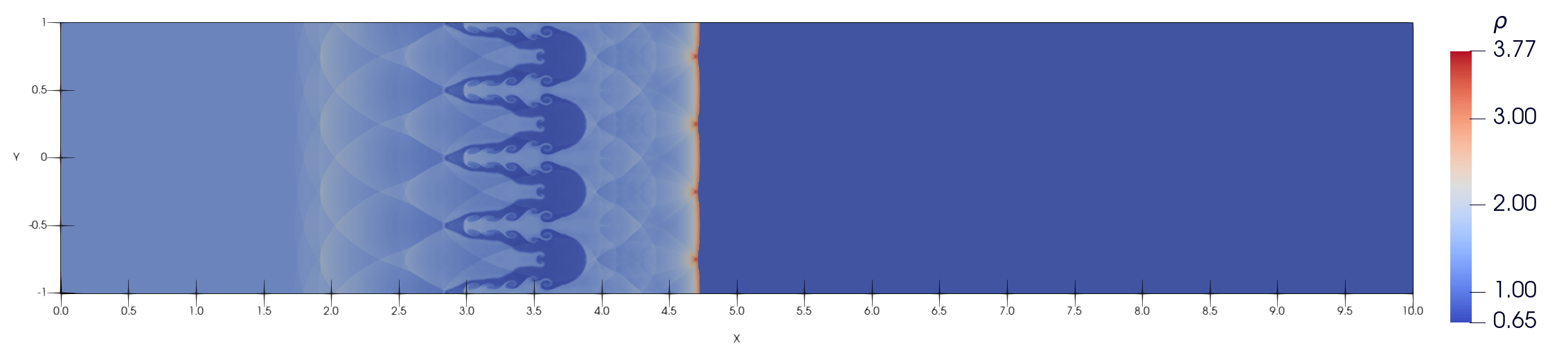}
\includegraphics[width=0.49\textwidth]{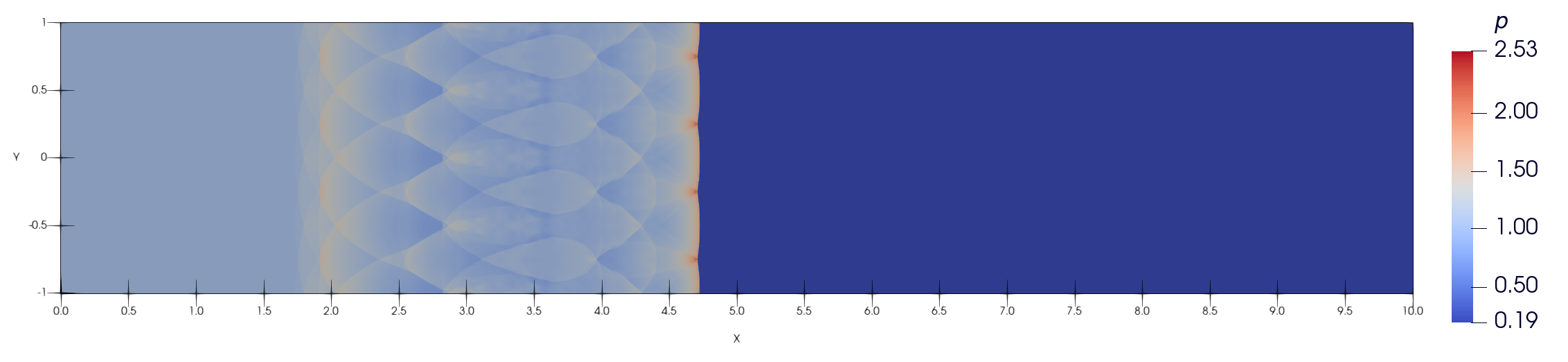}\\
\includegraphics[width=0.49\textwidth]{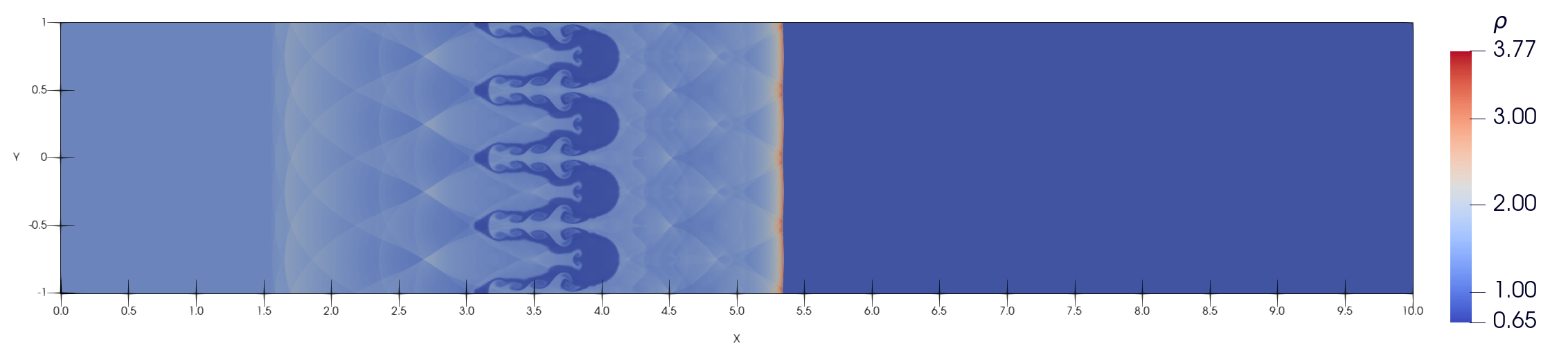}
\includegraphics[width=0.49\textwidth]{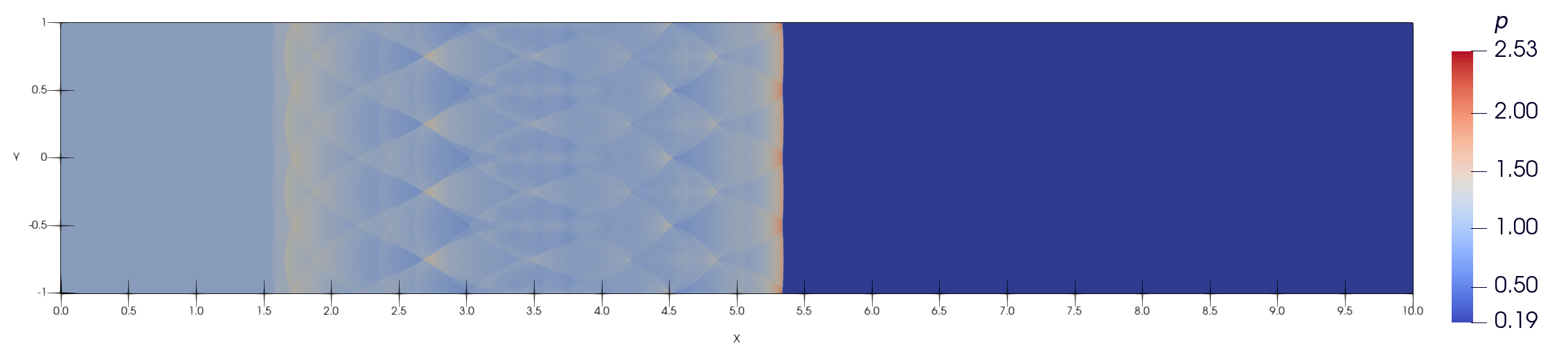}\\
\includegraphics[width=0.49\textwidth]{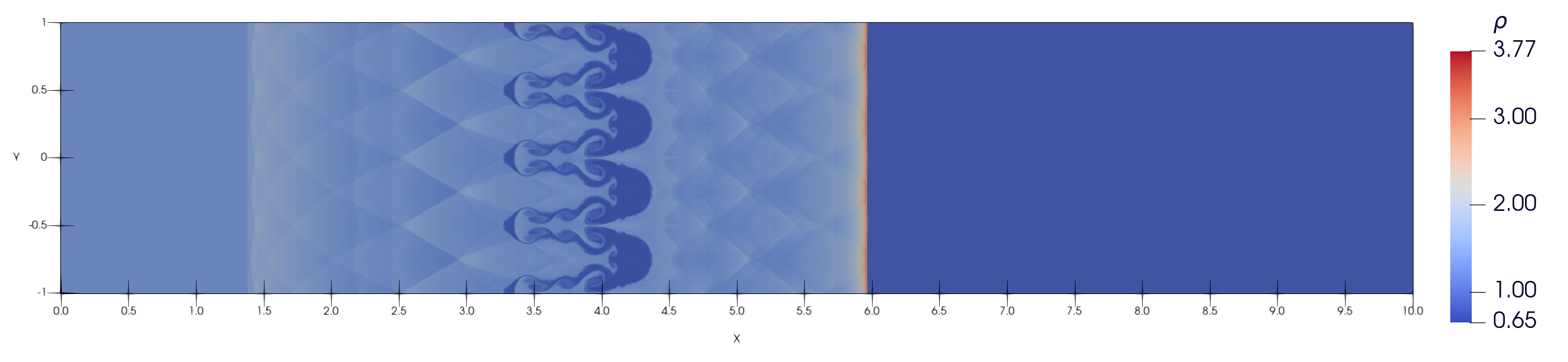}
\includegraphics[width=0.49\textwidth]{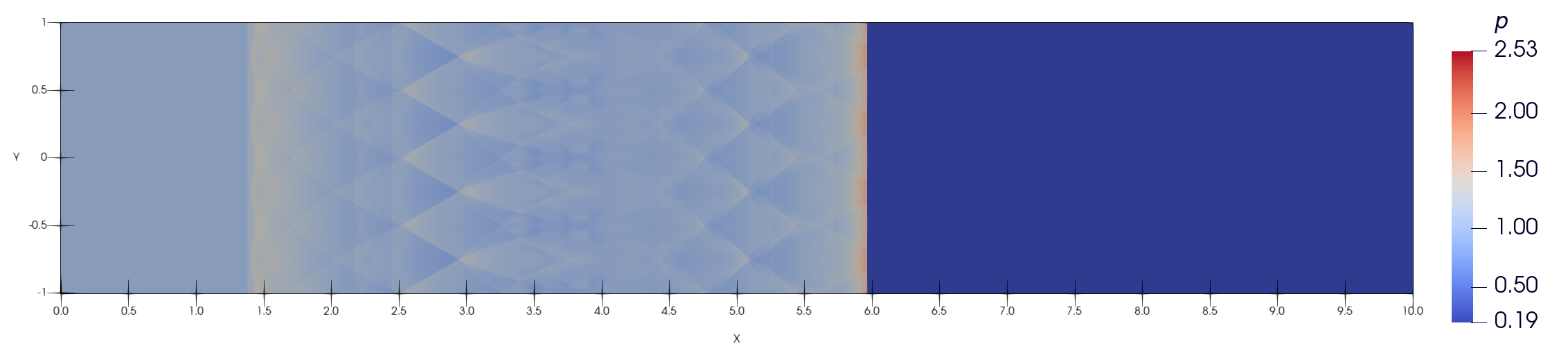}\\
\includegraphics[width=0.49\textwidth]{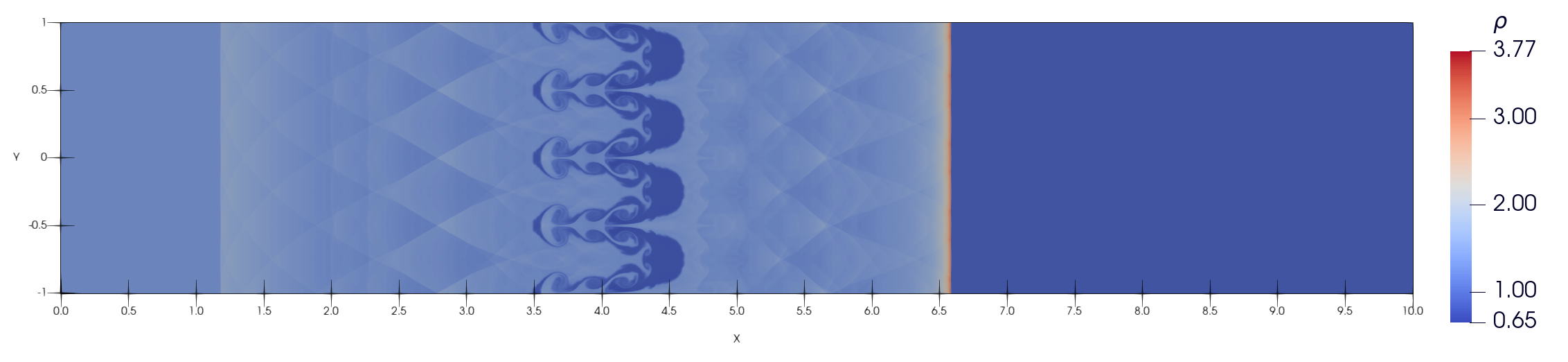}
\includegraphics[width=0.49\textwidth]{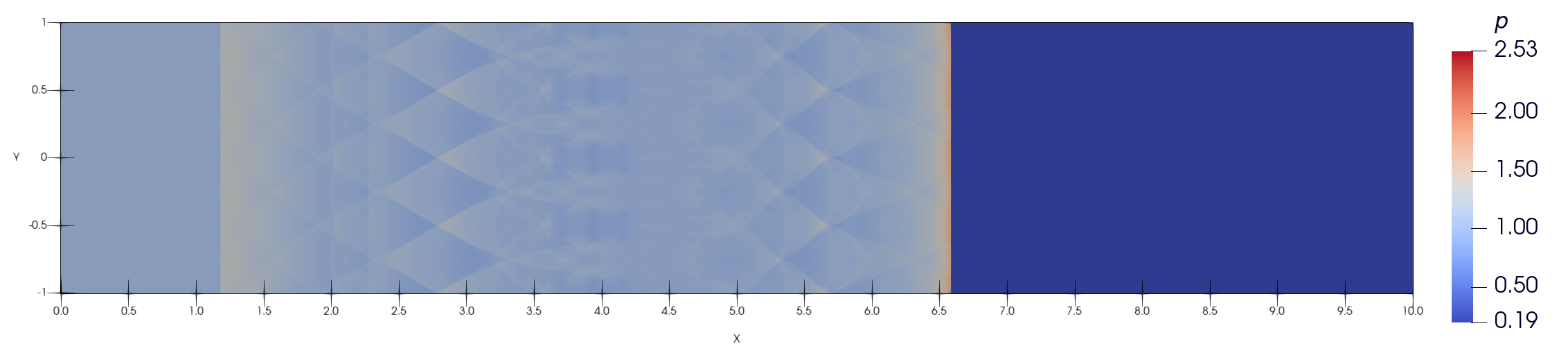}\\
\includegraphics[width=0.49\textwidth]{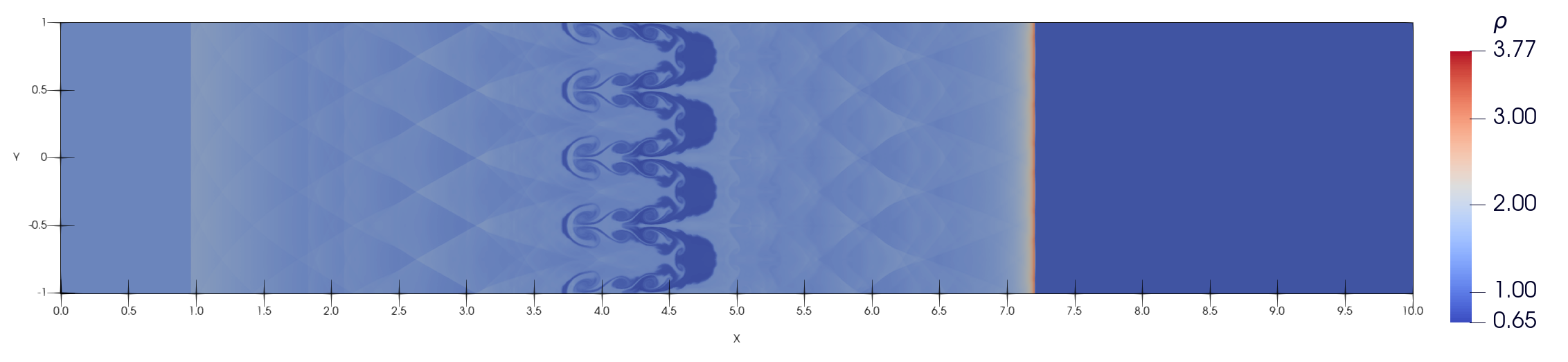}
\includegraphics[width=0.49\textwidth]{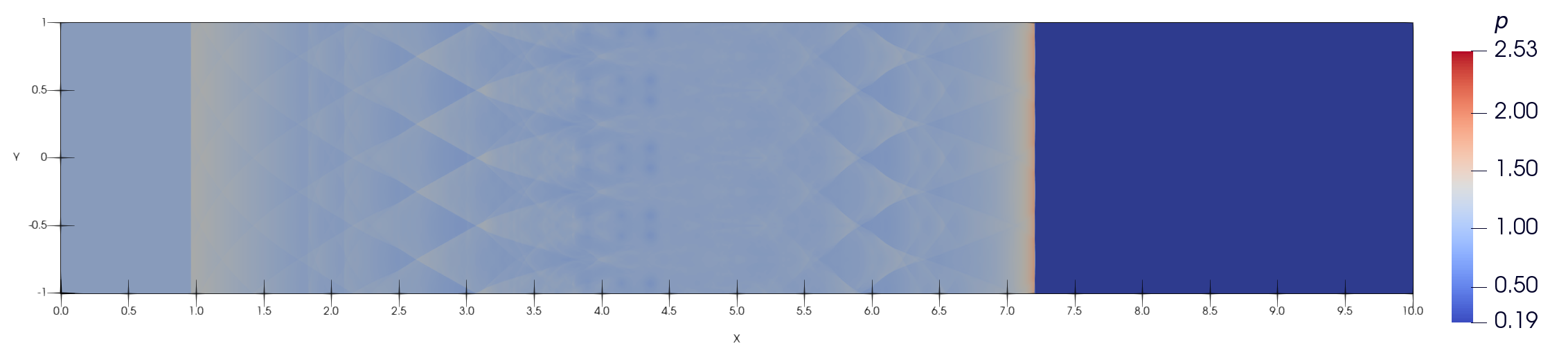}\\
\includegraphics[width=0.49\textwidth]{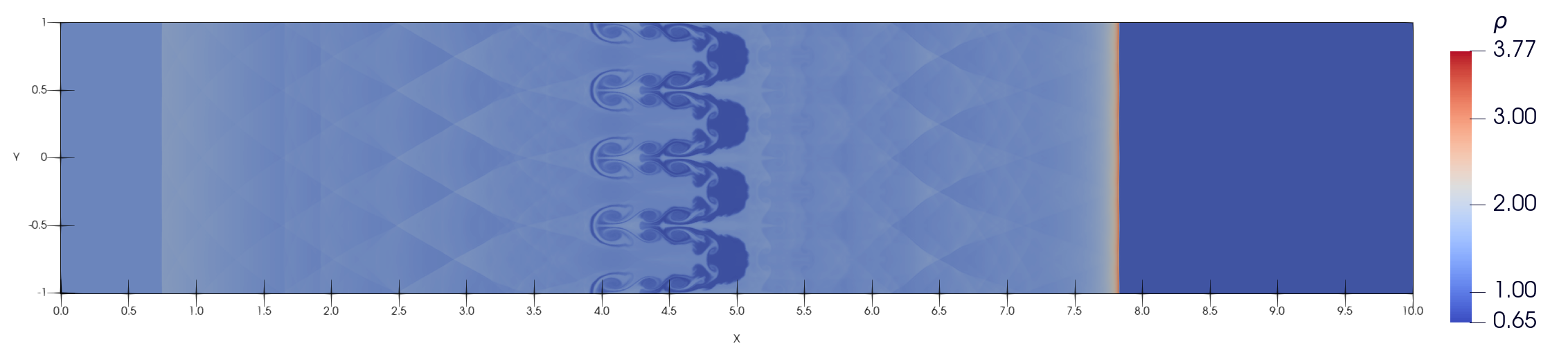}
\includegraphics[width=0.49\textwidth]{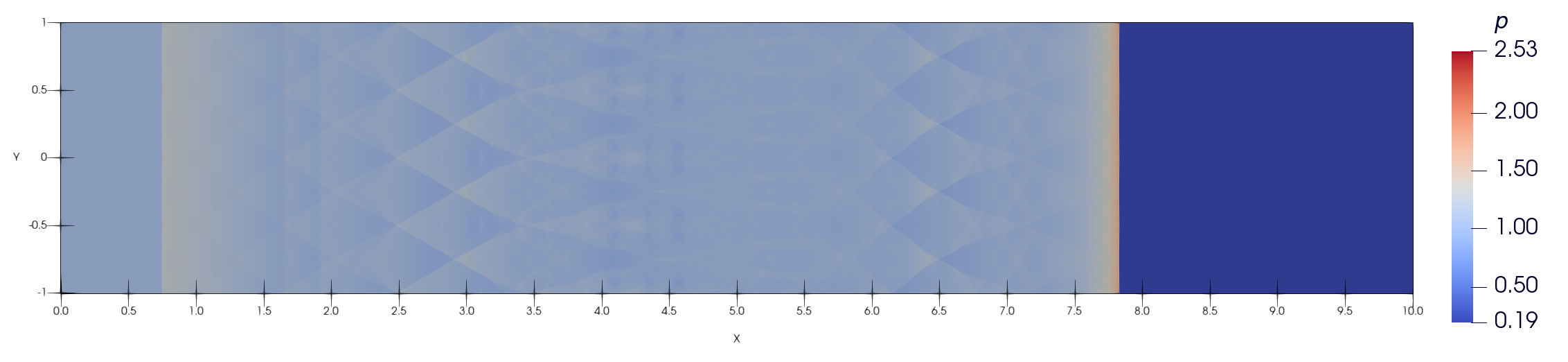}\\
\includegraphics[width=0.49\textwidth]{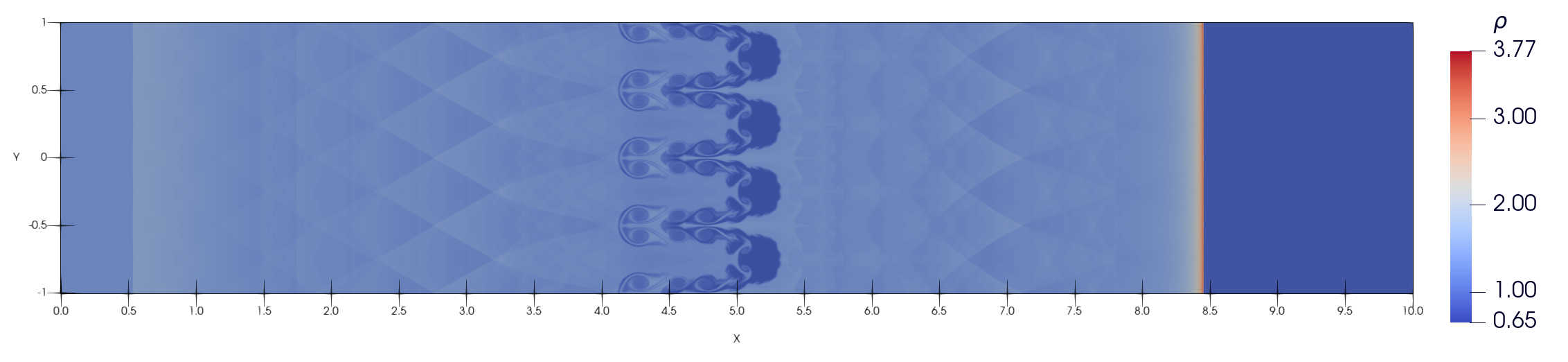}
\includegraphics[width=0.49\textwidth]{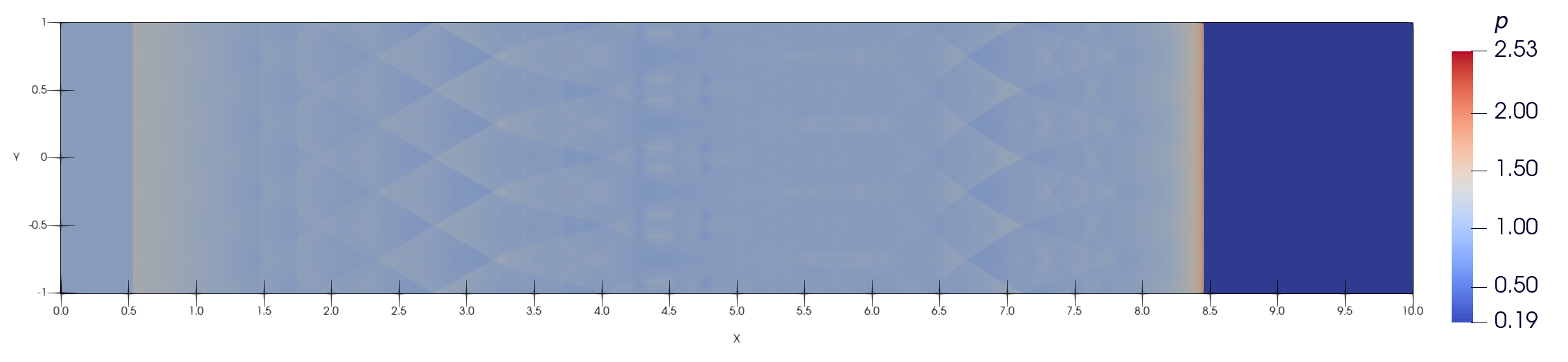}\\
\includegraphics[width=0.49\textwidth]{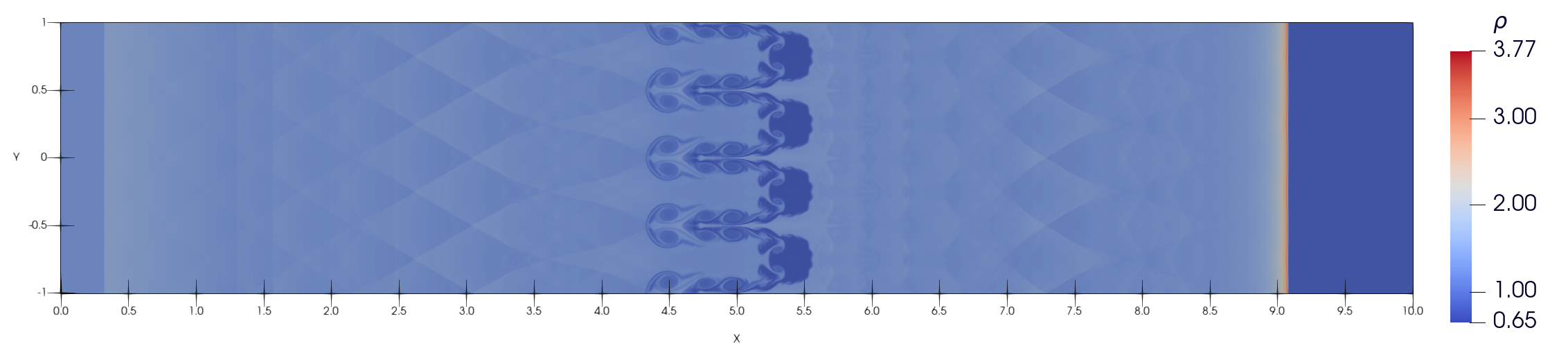}
\includegraphics[width=0.49\textwidth]{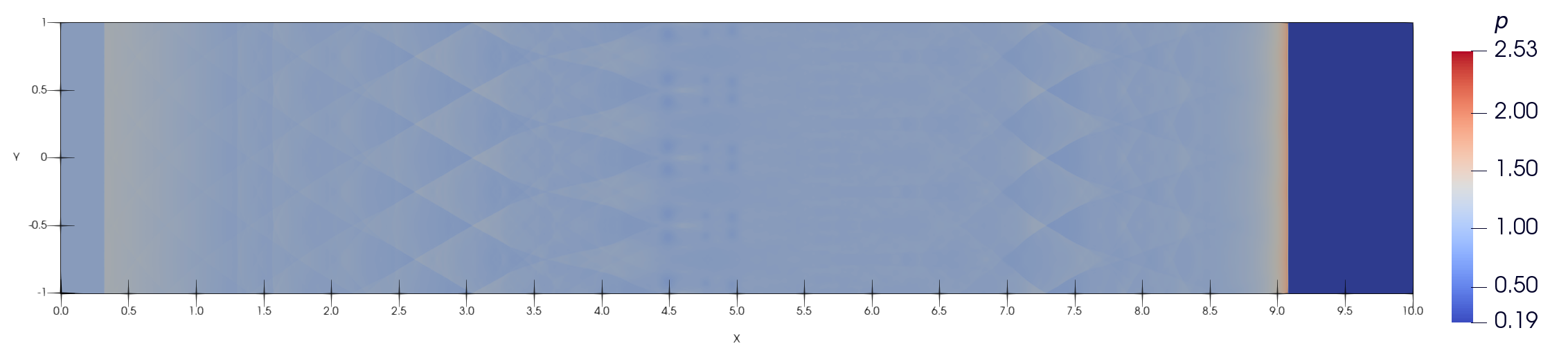}\\
\caption{\label{fig:dwsines_2d_den_p}
Numerical solution of the two-dimensional problem of detonation cellular structure
in a two-component medium with a ``slow'' reaction (weak stiff case, a detailed statement of the problem is presented in the text),
obtained using the ADER-DG-$\mathbb{P}_{2}$ method with a posteriori limitation of the solution by a ADER-WENO2 finite volume limiter 
on mesh with $1000 \times 200$ cells at the times $t = 0.4$, $0.8$, $1.2$, $1.6$, $2.0$, $2.4$, $2.8$, $3.2$, $3.6$ and $4.0$ (from top to bottom).
The graphs show the coordinate dependencies of the subcells finite-volume representation of density $\rho$ (left) and pressure $p$ (right).
}
\end{figure*}

\begin{figure*}[h!]
\centering
\includegraphics[width=0.49\textwidth]{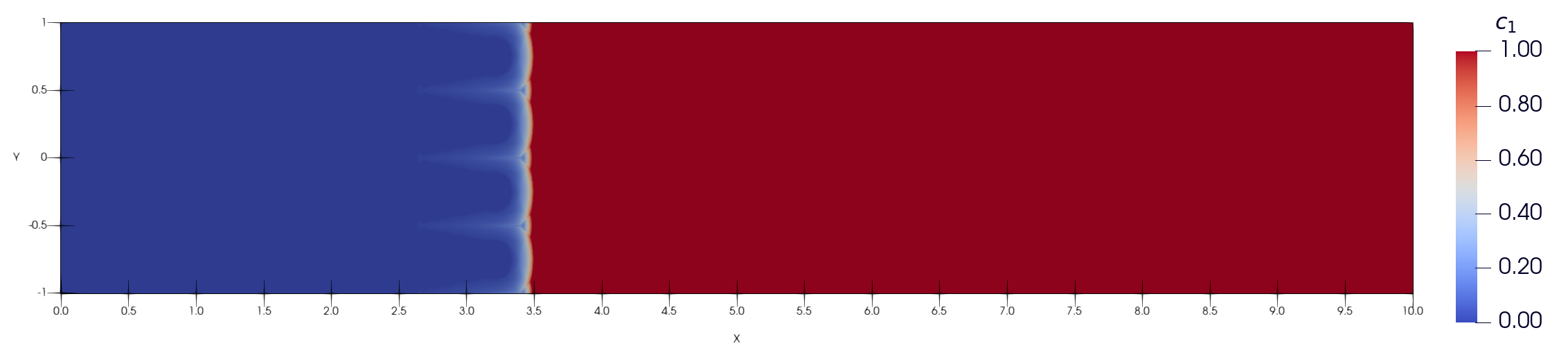}
\includegraphics[width=0.49\textwidth]{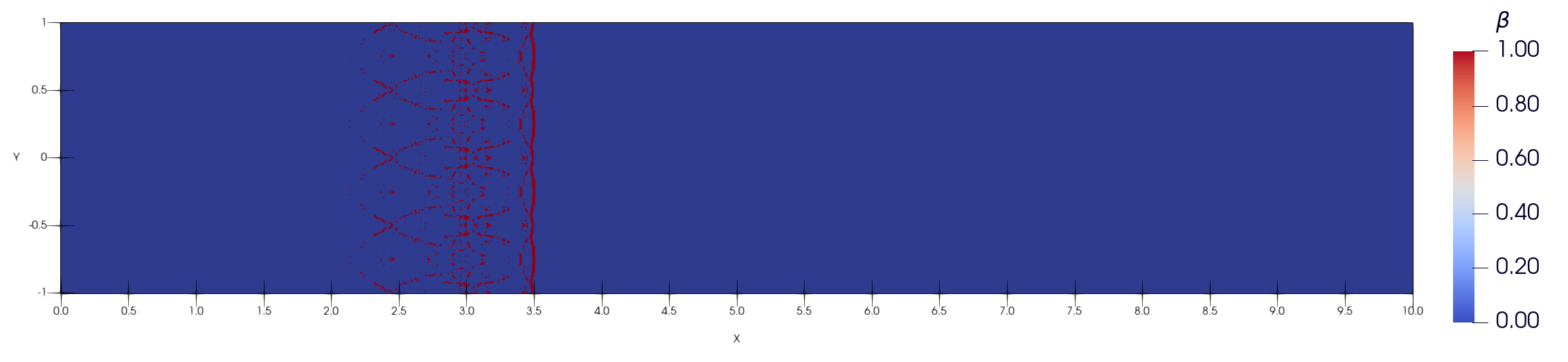}\\
\includegraphics[width=0.49\textwidth]{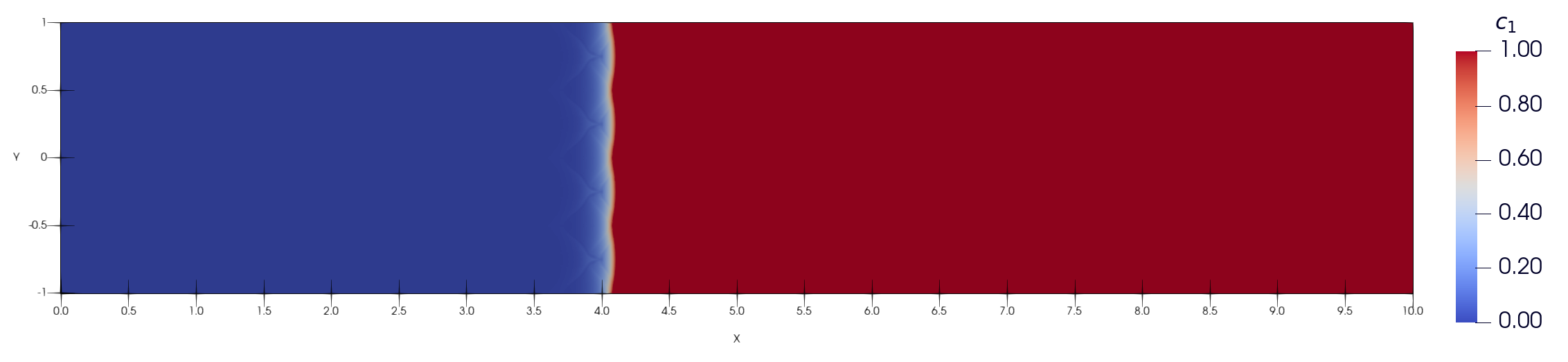}
\includegraphics[width=0.49\textwidth]{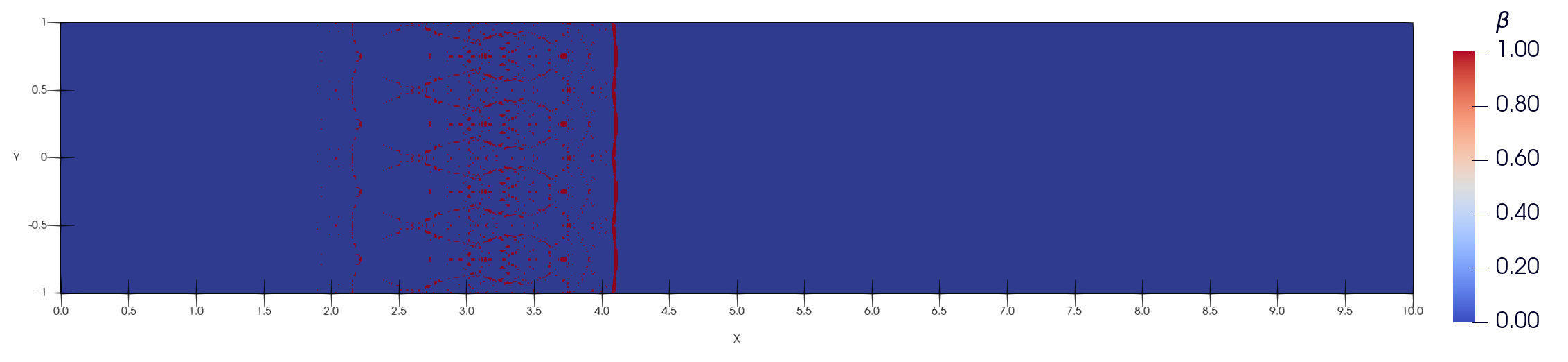}\\
\includegraphics[width=0.49\textwidth]{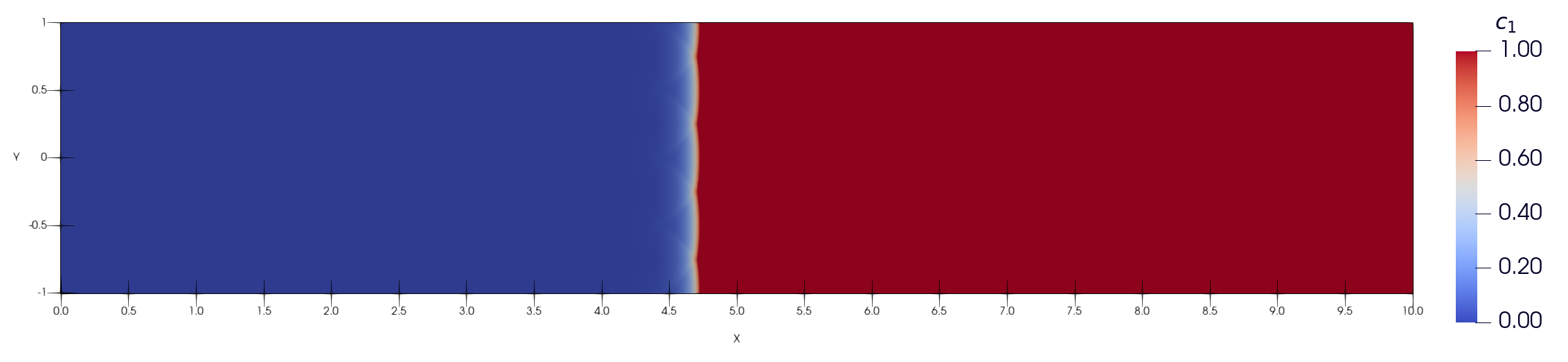}
\includegraphics[width=0.49\textwidth]{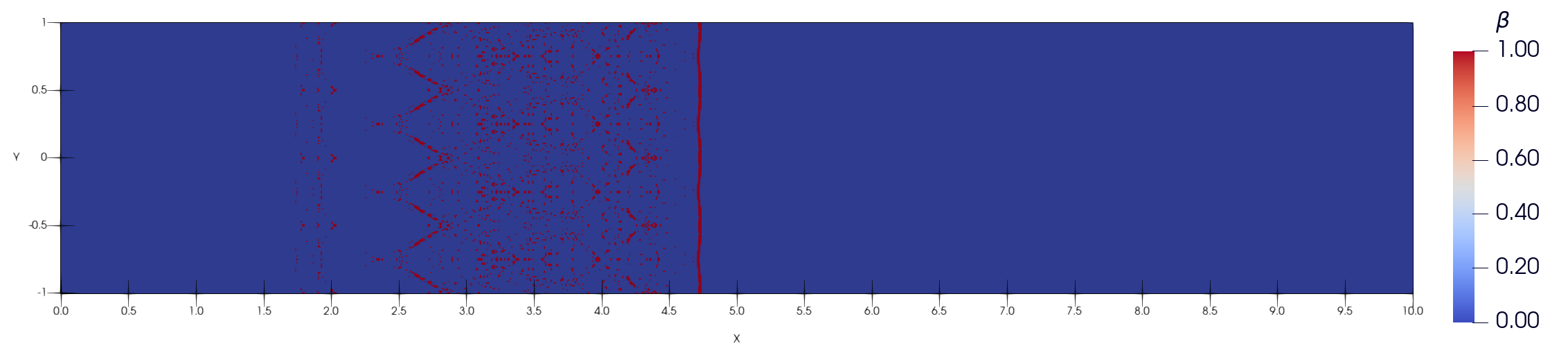}\\
\includegraphics[width=0.49\textwidth]{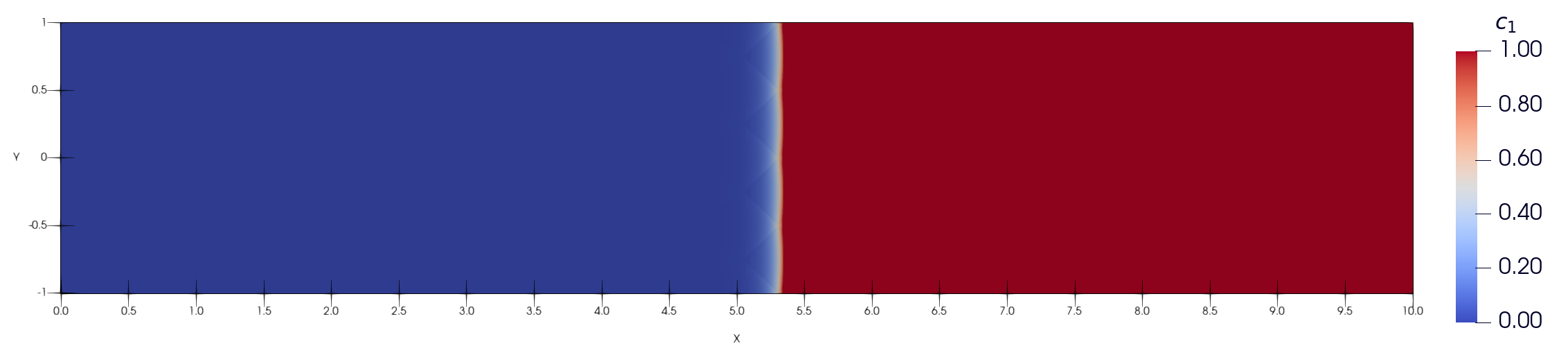}
\includegraphics[width=0.49\textwidth]{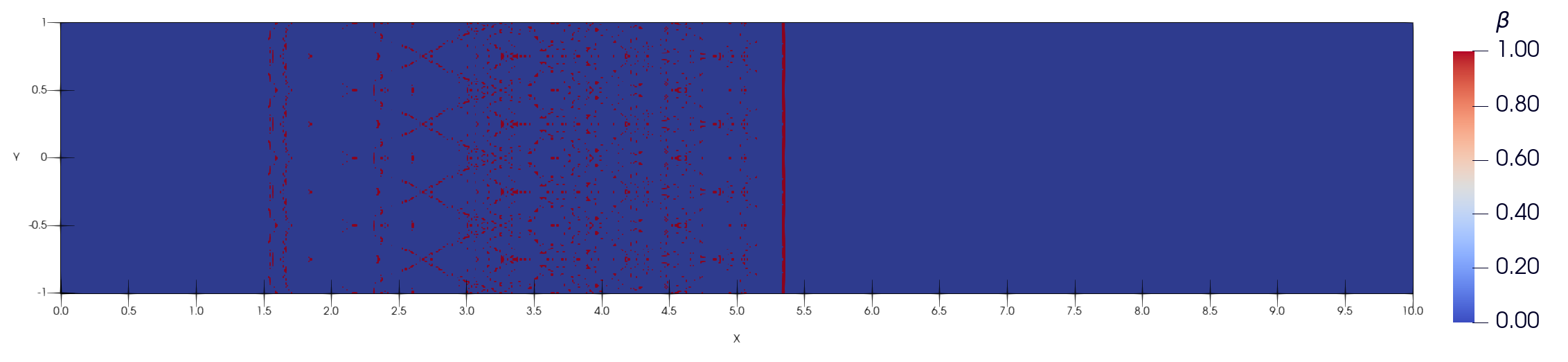}\\
\includegraphics[width=0.49\textwidth]{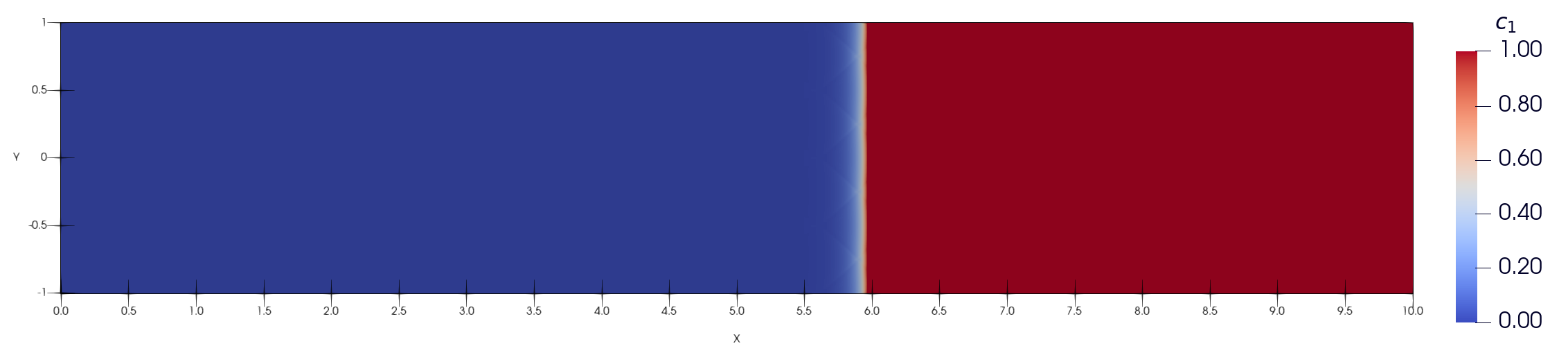}
\includegraphics[width=0.49\textwidth]{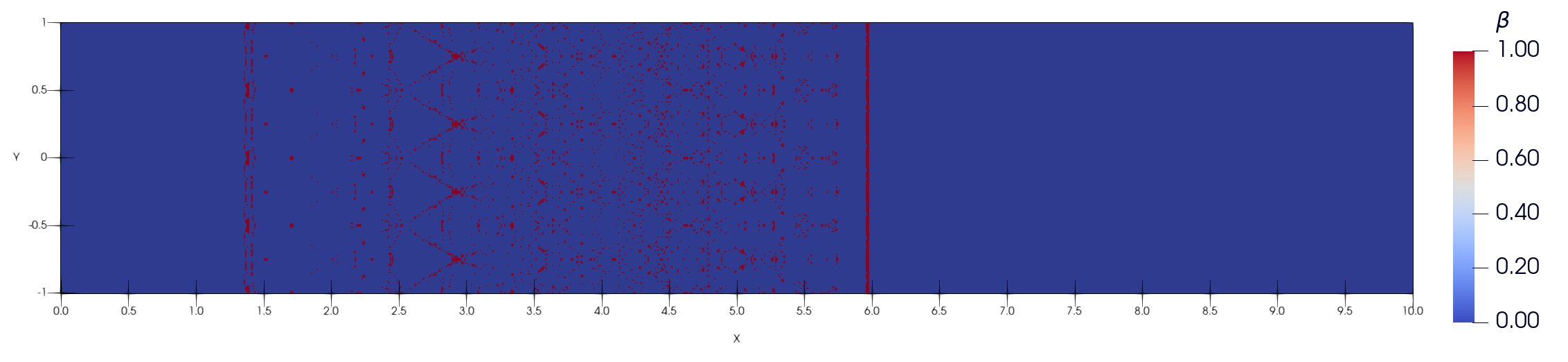}\\
\includegraphics[width=0.49\textwidth]{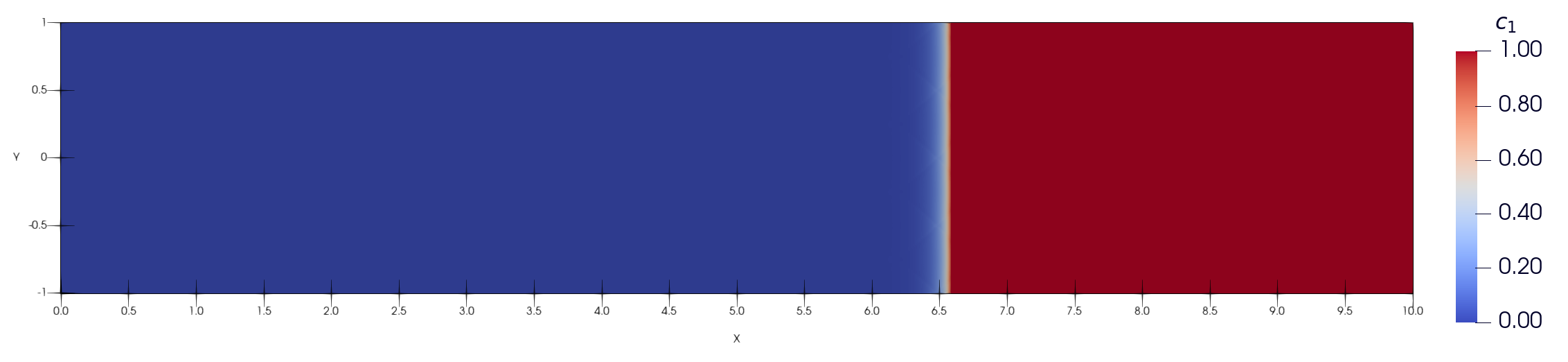}
\includegraphics[width=0.49\textwidth]{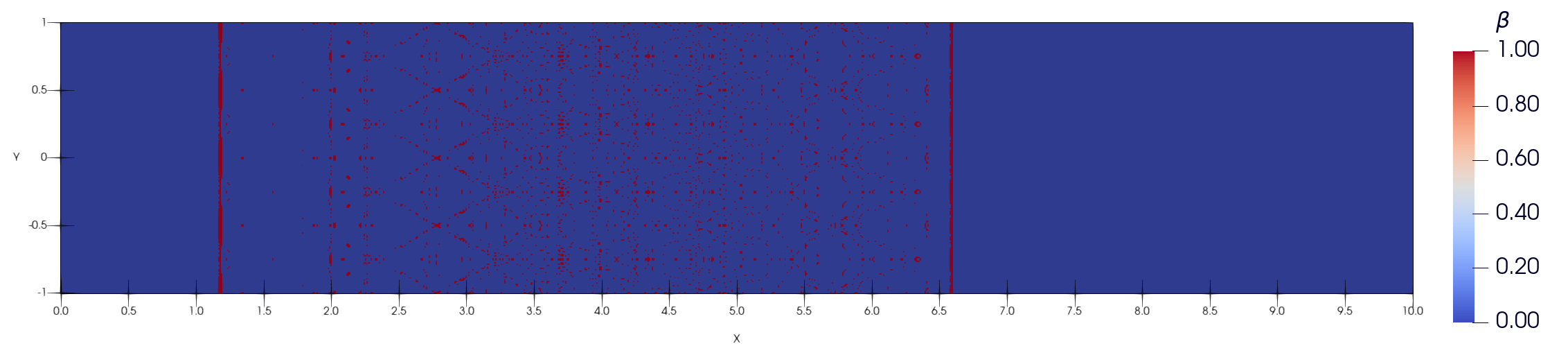}\\
\includegraphics[width=0.49\textwidth]{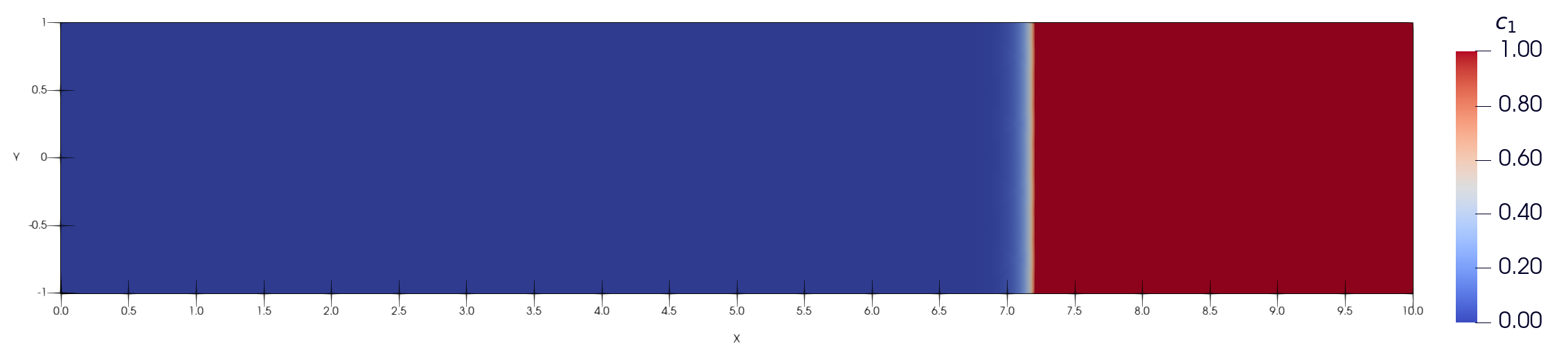}
\includegraphics[width=0.49\textwidth]{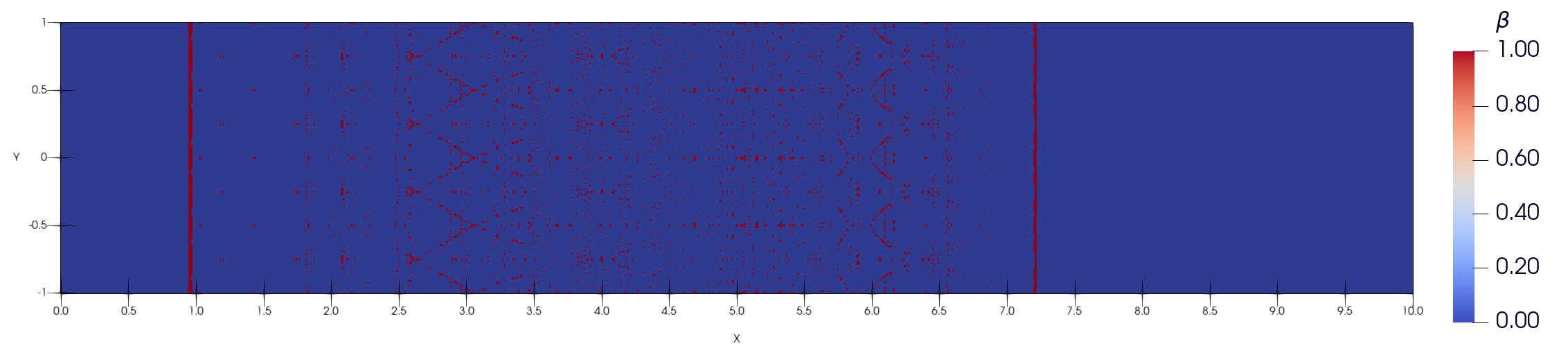}\\
\includegraphics[width=0.49\textwidth]{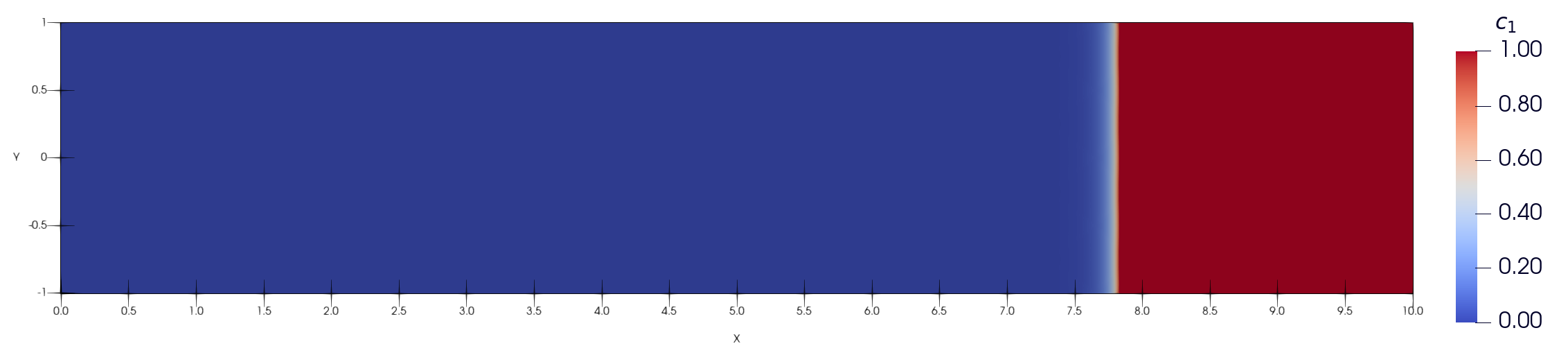}
\includegraphics[width=0.49\textwidth]{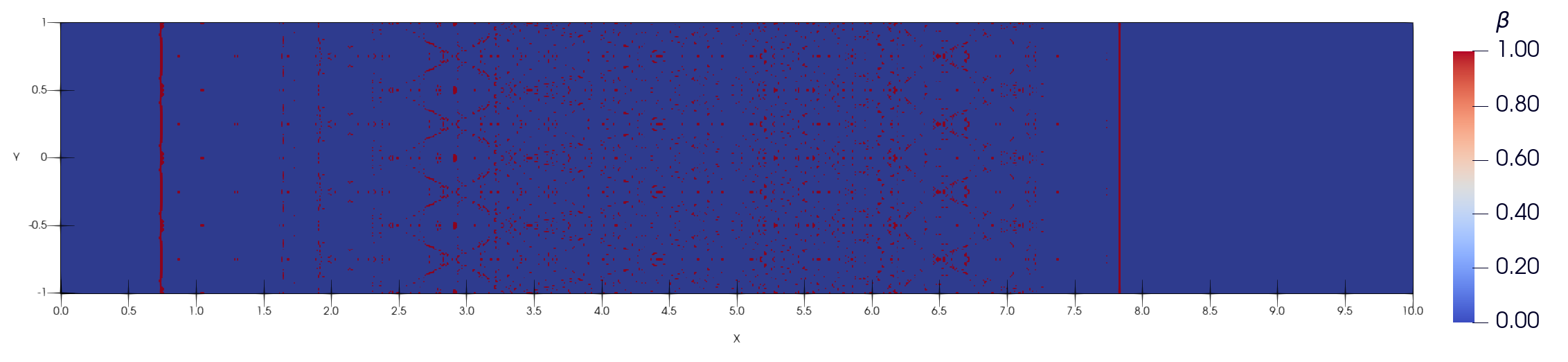}\\
\includegraphics[width=0.49\textwidth]{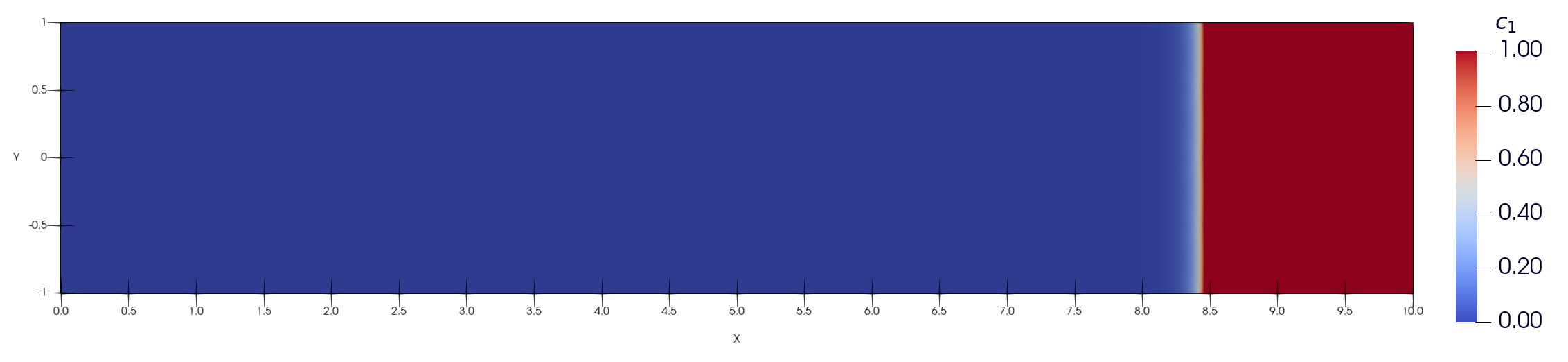}
\includegraphics[width=0.49\textwidth]{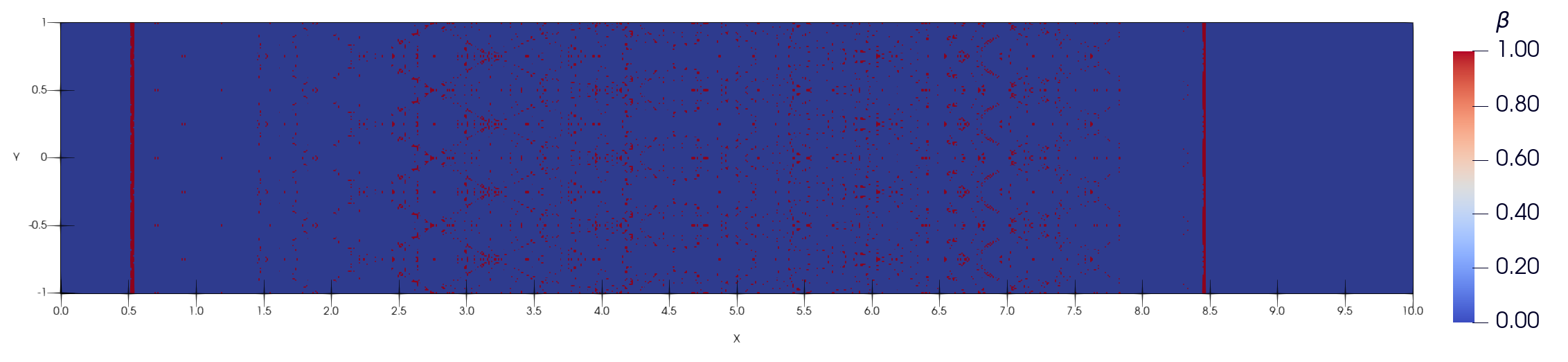}\\
\includegraphics[width=0.49\textwidth]{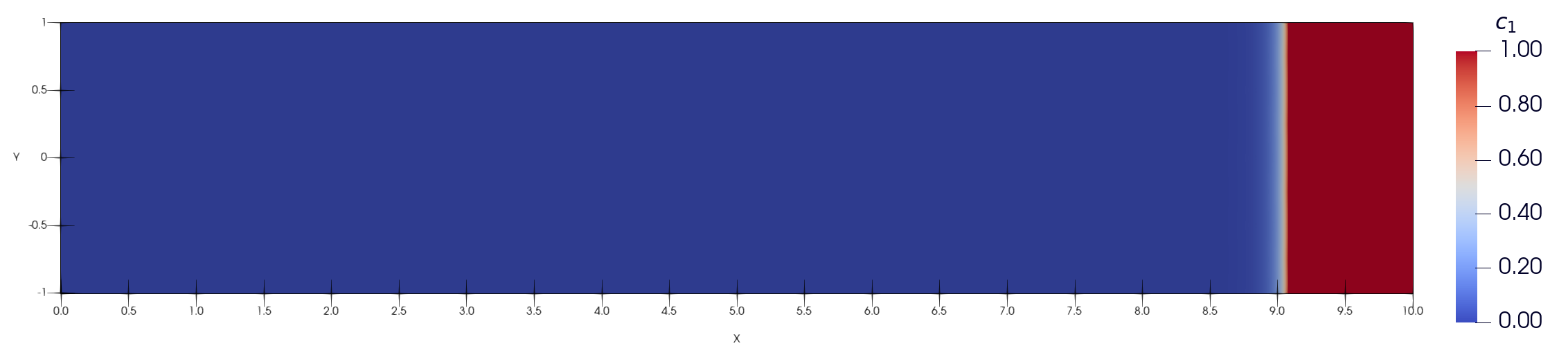}
\includegraphics[width=0.49\textwidth]{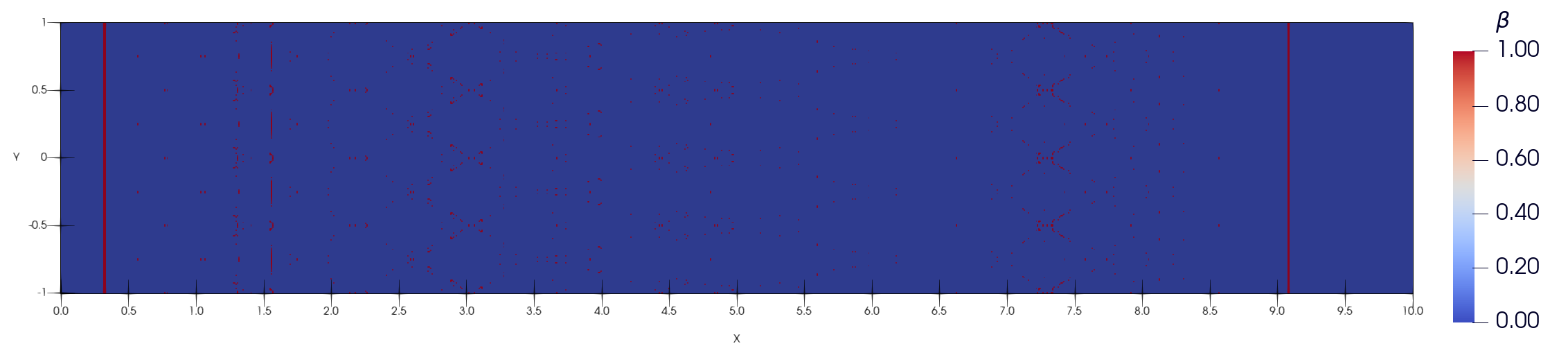}\\
\caption{\label{fig:dwsines_2d_c_1_beta}
Numerical solution of the two-dimensional problem of detonation cellular structure
in a two-component medium with a ``slow'' reaction (weak stiff case, a detailed statement of the problem is presented in the text),
obtained using the ADER-DG-$\mathbb{P}_{2}$ method with a posteriori limitation of the solution by a ADER-WENO2 finite volume limiter 
on mesh with $1000 \times 200$ cells at the times $t = 0.4$, $0.8$, $1.2$, $1.6$, $2.0$, $2.4$, $2.8$, $3.2$, $3.6$ and $4.0$ (from top to bottom).
The graphs show the coordinate dependencies of the subcells finite-volume representation of 
mass concentration $c_{1}$ of the reaction reagent (left) and troubled cells indicator $\beta$ (right).
}
\end{figure*}

At time $t = 0.8$, the coordinate dependencies of density $\rho$ and pressure $p$ in Figure~\ref{fig:dwsines_2d_den_p} clearly demonstrate the formation of a more planar detonation front, behind which a complex superposition of shock waves and tangential discontinuities is clearly observed. The triple points observed on the detonation front and immediately behind it at time $t = 0.4$ had already collapsed by time $t = 0.8$, forming a single stable detonation front, the geometry of which is characterized by cellularity. At the front of the detonation wave, the end points of shock waves are observed, continuing into the region behind the detonation front. The areas of vorticity generation demonstrate stable vortex formation, which led by time $t = 0.8$ to the formation of vortex ``columns'', along which vorticity is formed according to the Kelvin-Helmholtz instability mechanism, as well as within the framework of the classical baroclinity due to transmitted shock waves, which in fact is also complicated picture of the development of the Richtmyer-Meshkov instability. Shock waves propagating behind the detonation fronts are reflected from the upper and lower boundaries of the coordinate domain, propagate across the direction of propagation of the detonation front, and have a significant impact on the pattern of flow development in the region of the burned media. It should be noted that shock waves are reflected not only from the boundaries of the coordinate domain, but also from areas of significant changes in density resulting from the passage of rarefaction waves from the initial decay of the discontinuity -- in this case, reflection occurs not only from inhomogeneities across the $x$-axis, but also from inhomogeneities across the $y$-axis. Regarding the vortex flow and the generation of vorticity behind the detonation wave front, it should also be noted that the initial conditions and the left boundary condition assume the existence of a directed flow to the right along the $x$-axis, therefore the interaction of the general flow with the propagating curvilinear fronts of shock waves and rarefaction waves leads to the formation of additional areas of vortex generation, especially towards the original vortex ``columns'', and also to the right of the initial decay of the discontinuity. These effects are most clearly observed in the coordinate dependencies of very high resolution, which are presented in Figure~\ref{fig:dwsines_2d_2000x400}, and therefore will be discussed further in the text in the appropriate discussion. The coordinate dependencies of the mass concentration $c_{1}$ of the reagent in Figure~\ref{fig:dwsines_2d_c_1_beta} at time $t = 0.8$ demonstrate relatively uniform combustion of the reagent behind the detonation fronts -- narrow regions of localization of the unburnt reagent behind the detonation front are no longer observed. In this case, the cellular structure of the front appears in these dependencies, but not as significantly as at time $t = 0.4$. The distribution of troubled cells presented in Figure~\ref{fig:dwsines_2d_c_1_beta} shows that the limiter is still caused mainly only in the areas where shock fronts are localized, however, due to the complexity of the picture of the interaction of shock waves, it becomes not obvious to track this correspondence. It is clear that over time the detonation front becomes more and more flat, however, this is not a completely monotonic process, since it is associated with nonlinear interference of shock waves behind the detonation front, therefore, on average, the detonation front has become flatter, however, at time $t = 1.2$ an increase in heterogeneity is observed across the front in the coordinate dependencies of density $\rho$ and pressure $p$ in Figure~\ref{fig:dwsines_2d_den_p}. By time $t = 1.6$, there is a significant decrease in the intensity of shock waves behind the detonation wave front due to geometric divergence -- these shock waves have a predominantly two-dimensional quasi-cylindrical structure and at large distances their intensity decreases much faster than the intensity of the shock wave of an almost one-dimensional planar detonation front. In this case, the transverse heterogeneity of the detonation front decreases significantly. By time $t = 2.0$, there is a significant ``separation'' of the detonation front from the region of intense shock wave effects behind the front. In this case, the detonation front becomes almost flat, and the transverse inhomogeneities caused by the passage of shock waves reflected from the upper and lower boundaries of the coordinate domain become very small -- the deviation is no more than $0.05$-$0.1$ from the value for density $\rho$ averaged along the length of the detonation front, and not more than $0.08$-$0.12$ -- for pressure $p$. Further, during the propagation process, density and pressure inhomogeneities are also observed at the detonation front, but they do not exceed the threshold values indicated above. The coordinates of the dependence of the mass concentration $c_{1}$ of the reagent show similar results -- from time $t = 2.0$, burnout inhomogeneities across the detonation front are practically not observed, and from time $t = 3.2$, the representation $c_{1}$ used in Figure~\ref{fig:dwsines_2d_c_1_beta} no longer allows these small inhomogeneities to be resolved across the detonation front. From time points $t = 2.0$-$2.4$, the structure of the burnout front becomes practically one-dimensional. The distribution of troubled cells presented in Figure~\ref{fig:dwsines_2d_c_1_beta} after time $t = 2.0$ allows us to identify the limiter activation regions only with the main structures in the flow -- the detonation front propagating forward, shock wave structures propagating backward, and the strongest shock waves behind the detonation front (and only because these fronts have fairly flat sections, which are clearly visible on the coordinate dependence of the troubled indicator cells). Areas of significant vortex generation associated with the Richtmyer-Meshkov instability are clearly observed for all presented times on the coordinate dependencies of the density $\rho$. Since time $t = 2.0$, a significant interaction of vortex ``columns'' with a region of higher density has been observed, while the dynamics of vorticity development is resolved quite correctly in the numerical solution. 

As a result of the presented analysis, it can be concluded that the ADER-DG-$\mathbb{P}_{N}$ method with a finite volume ADER-WENO and an a posteriori limiter makes it possible to effectively simulate detonation cellular structures, while all the main features of the detonation flow of the reacting medium with a ``slow'' reaction are resolved in a numerical solution.

\begin{figure*}[h!]
\centering
\includegraphics[width=0.49\textwidth]{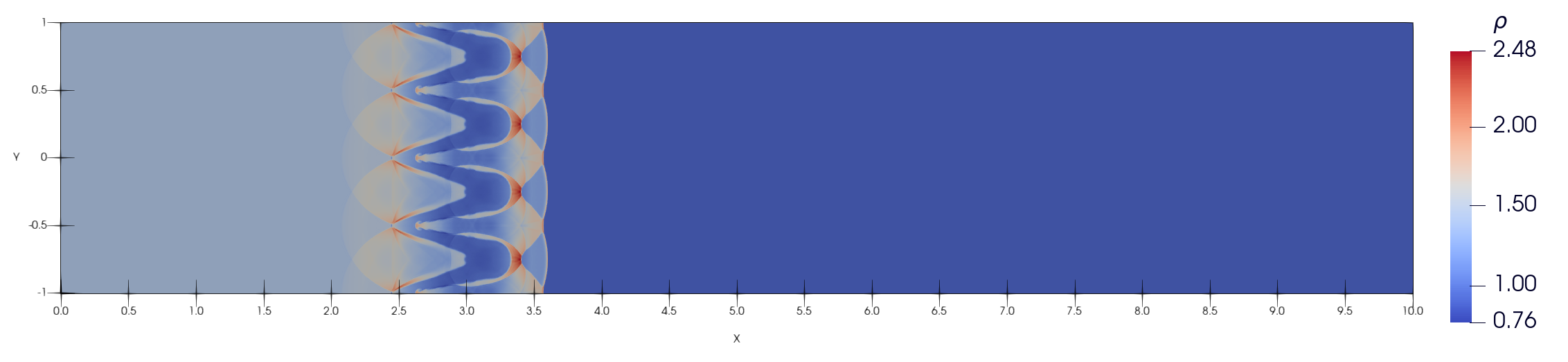}
\includegraphics[width=0.49\textwidth]{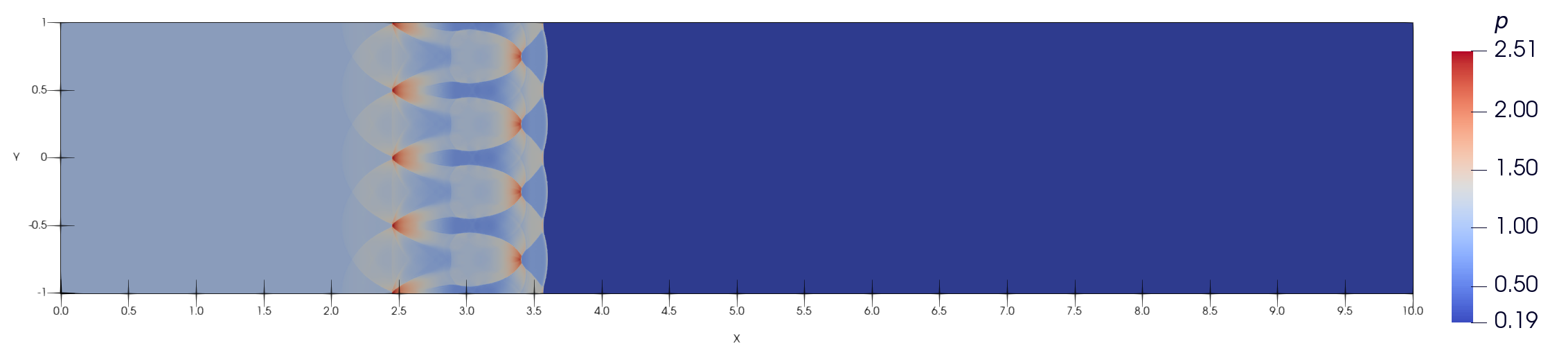}\\
\includegraphics[width=0.49\textwidth]{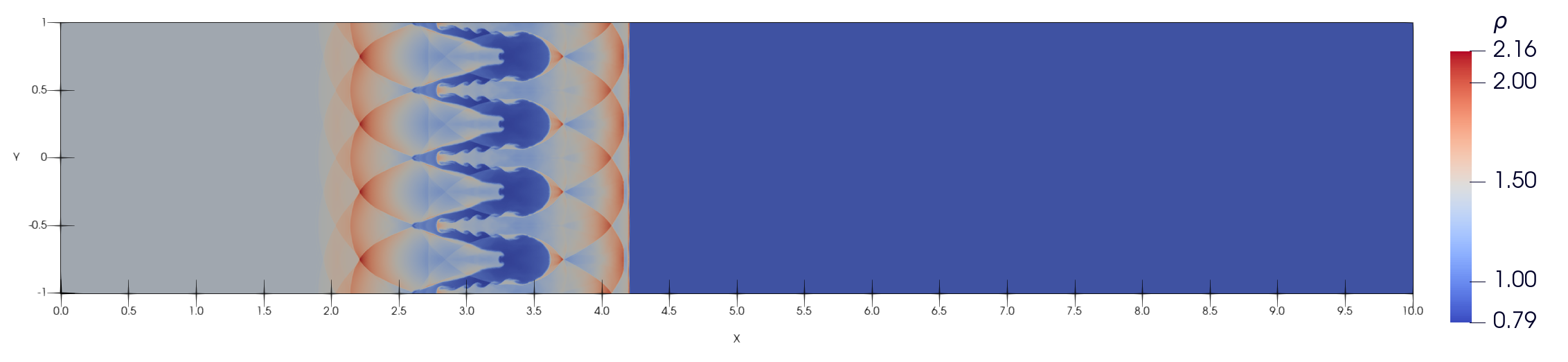}
\includegraphics[width=0.49\textwidth]{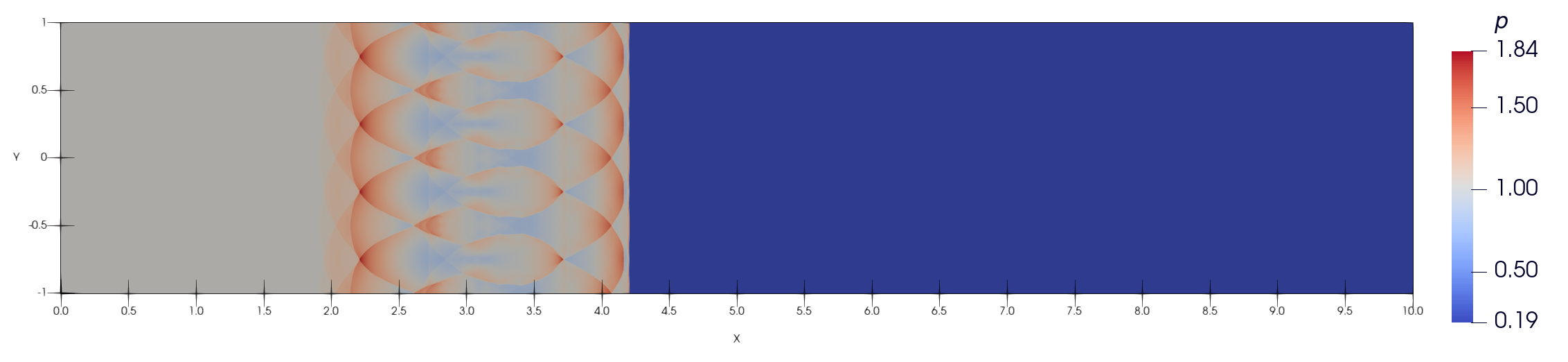}\\
\includegraphics[width=0.49\textwidth]{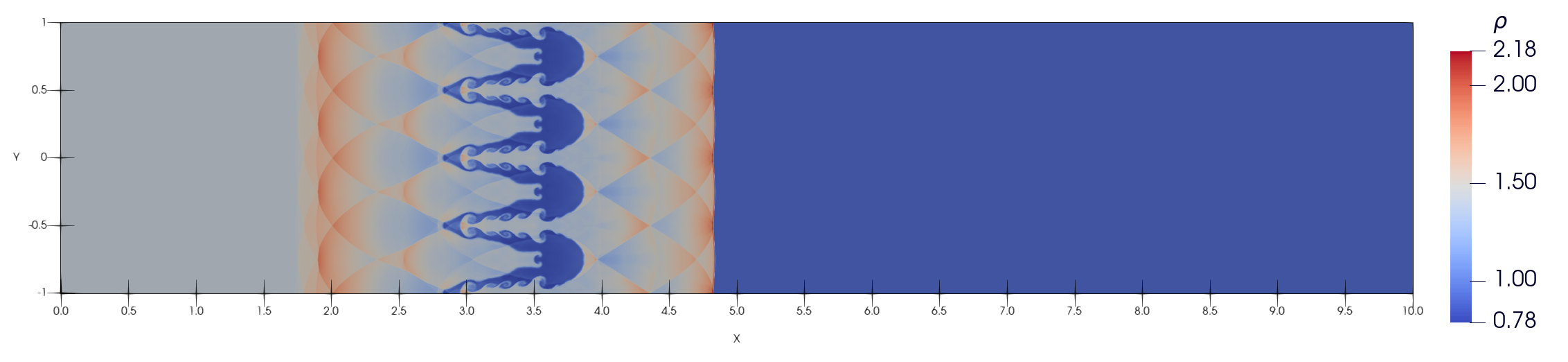}
\includegraphics[width=0.49\textwidth]{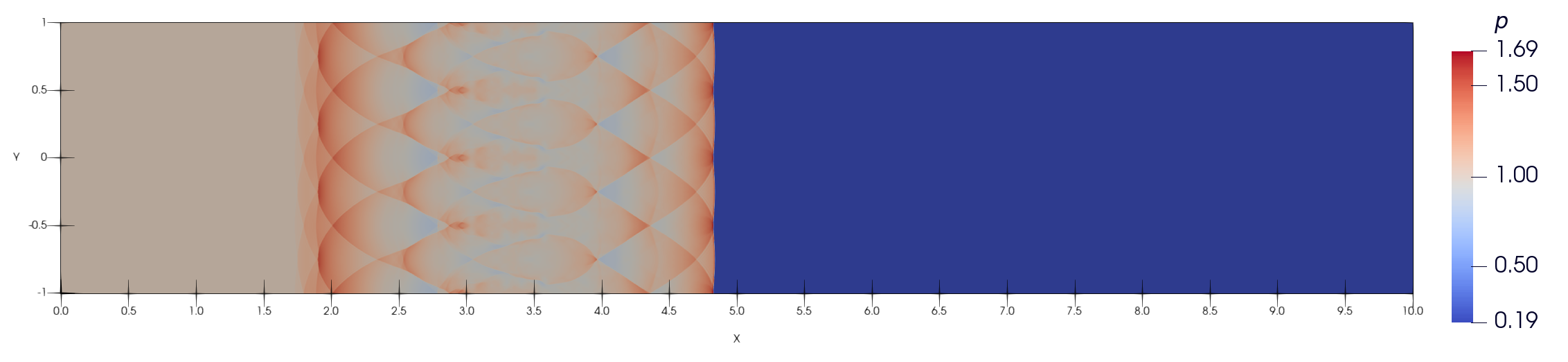}\\
\includegraphics[width=0.49\textwidth]{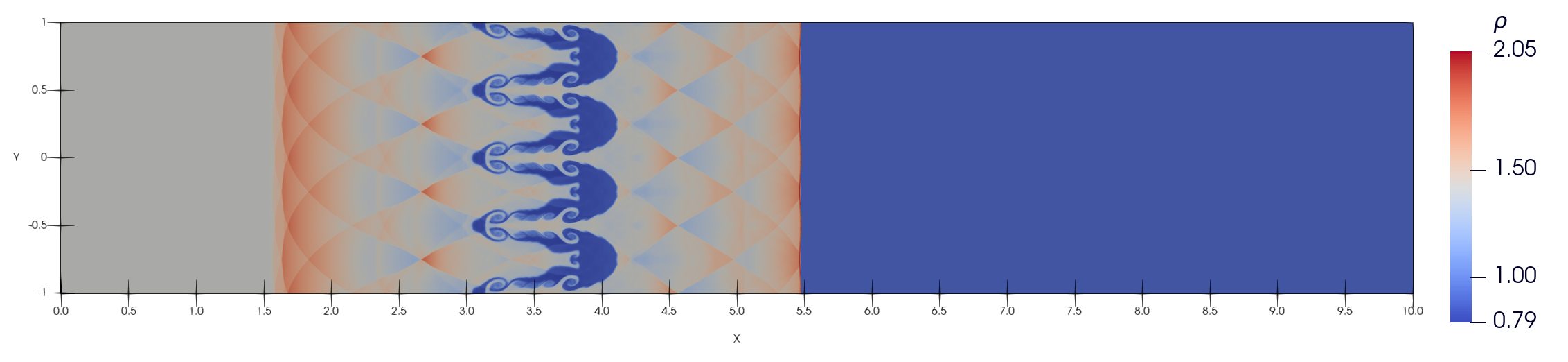}
\includegraphics[width=0.49\textwidth]{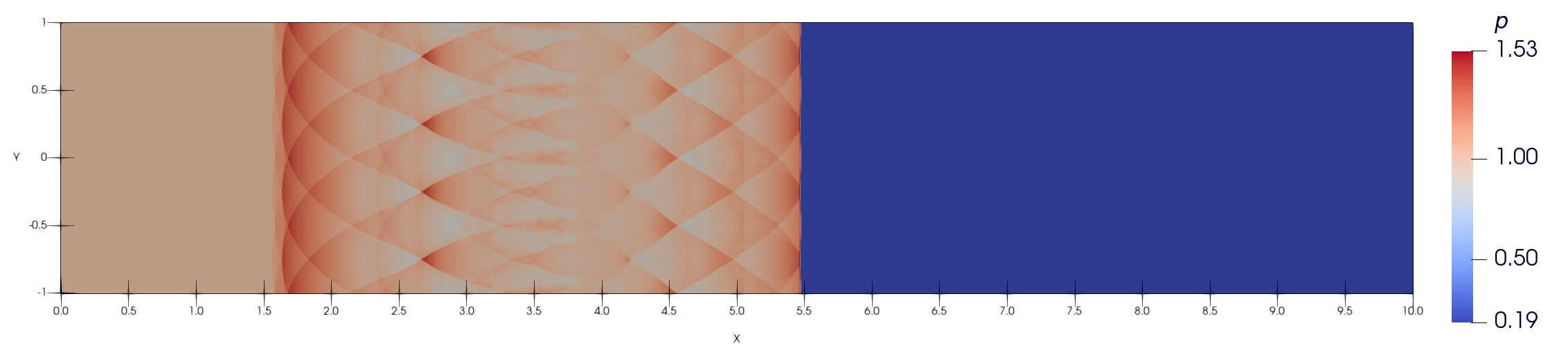}\\
\includegraphics[width=0.49\textwidth]{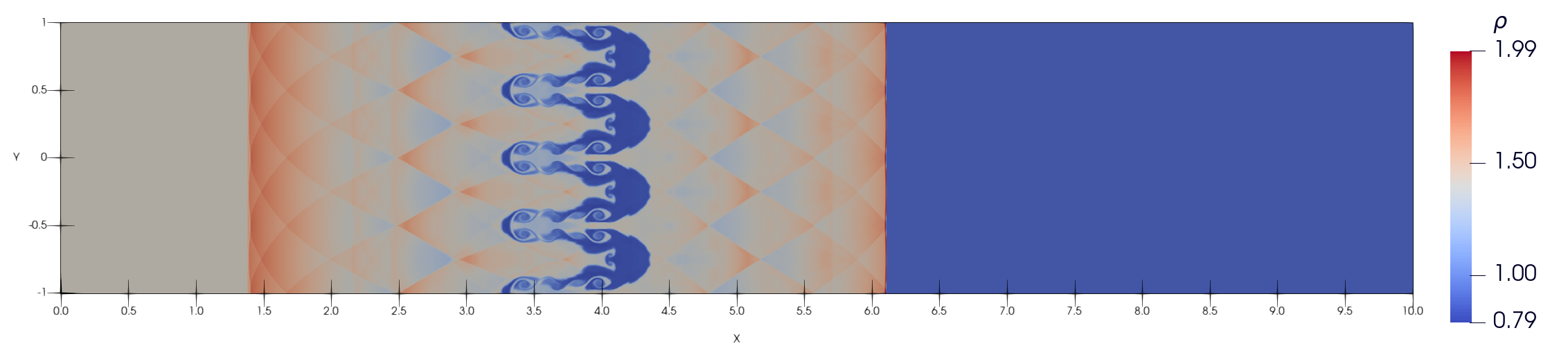}
\includegraphics[width=0.49\textwidth]{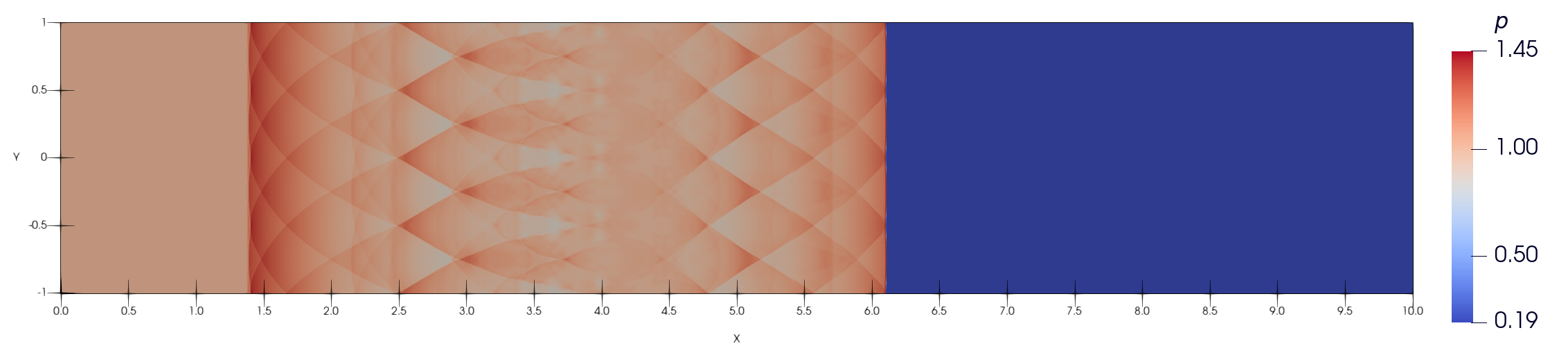}\\
\includegraphics[width=0.49\textwidth]{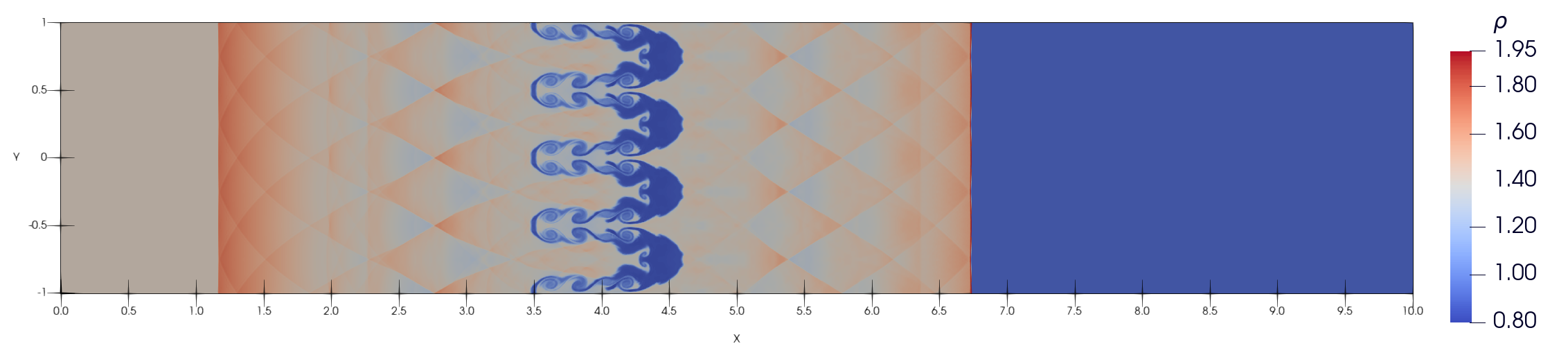}
\includegraphics[width=0.49\textwidth]{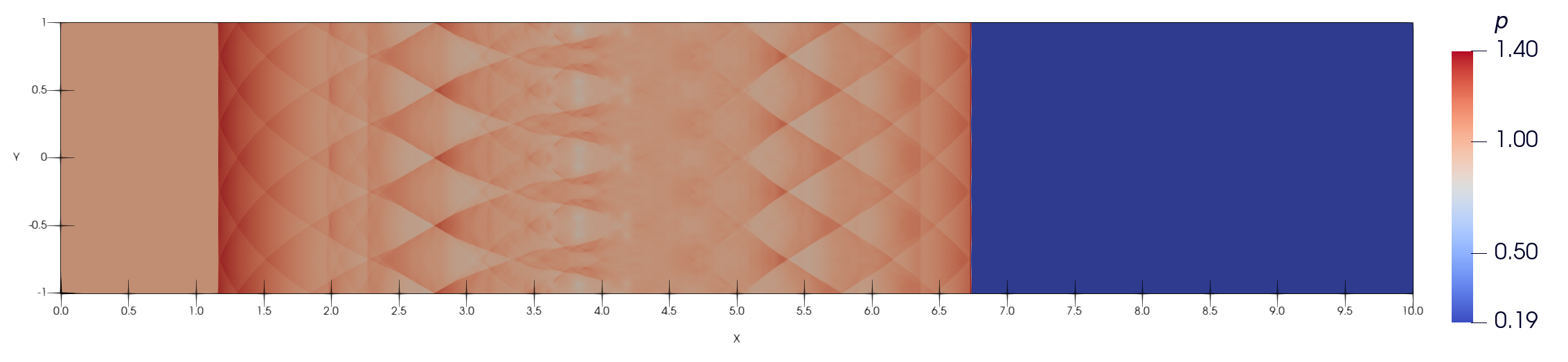}\\
\includegraphics[width=0.49\textwidth]{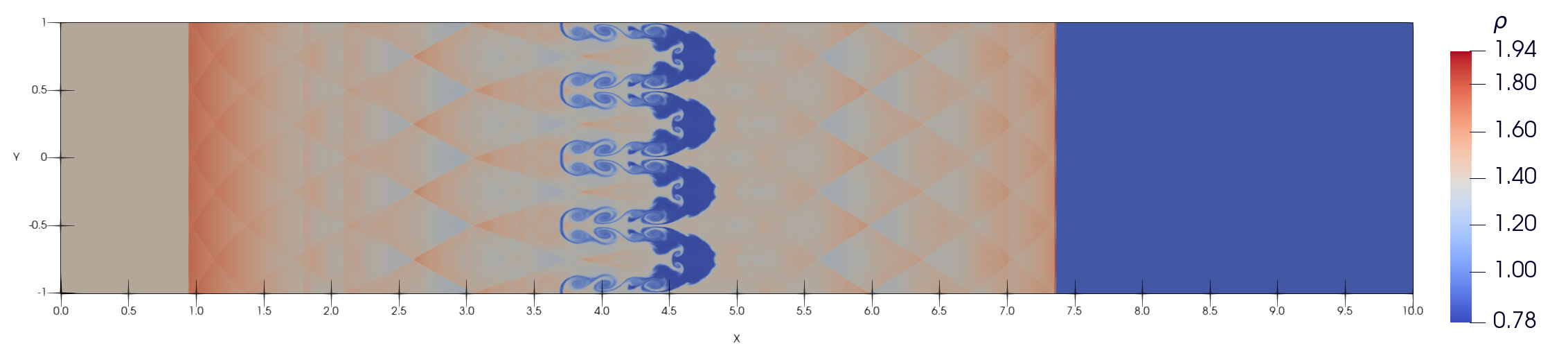}
\includegraphics[width=0.49\textwidth]{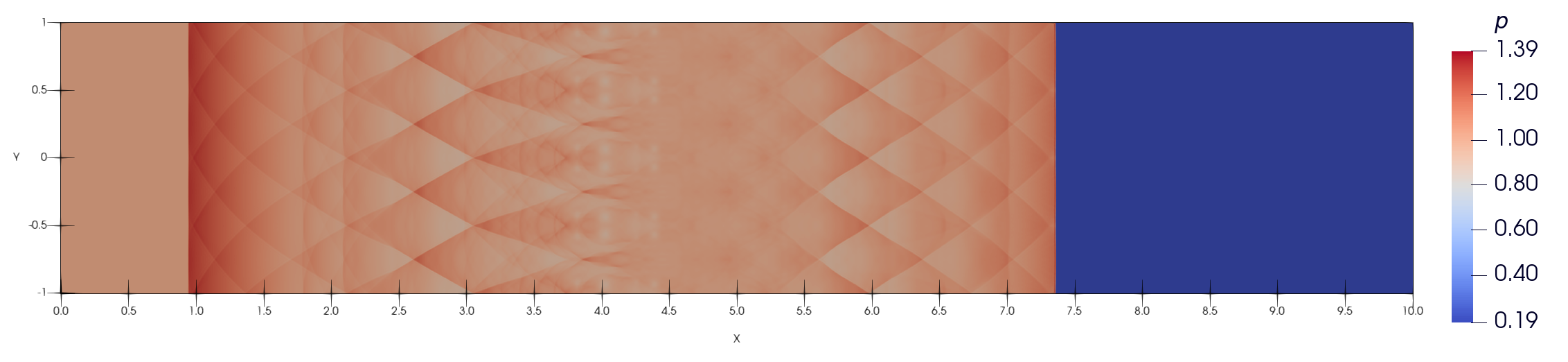}\\
\includegraphics[width=0.49\textwidth]{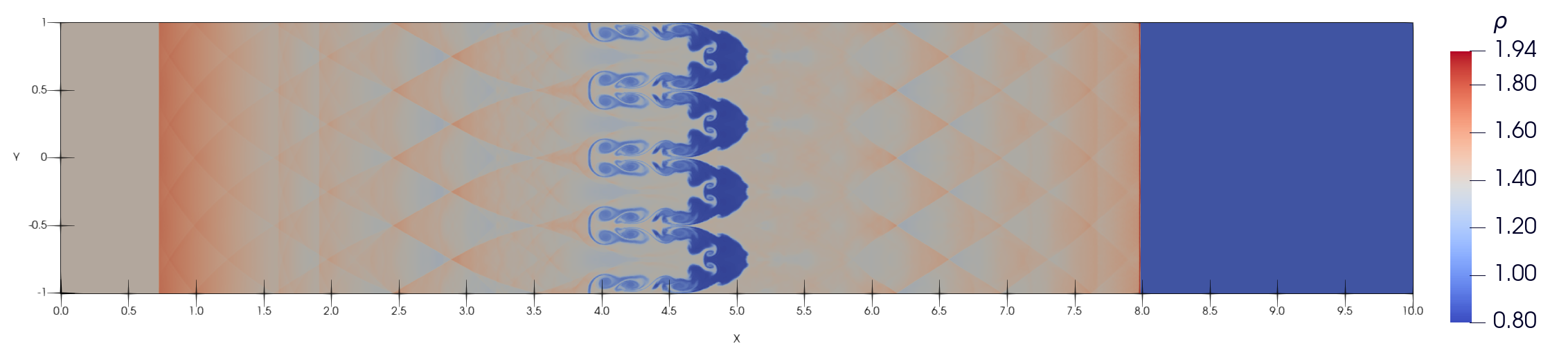}
\includegraphics[width=0.49\textwidth]{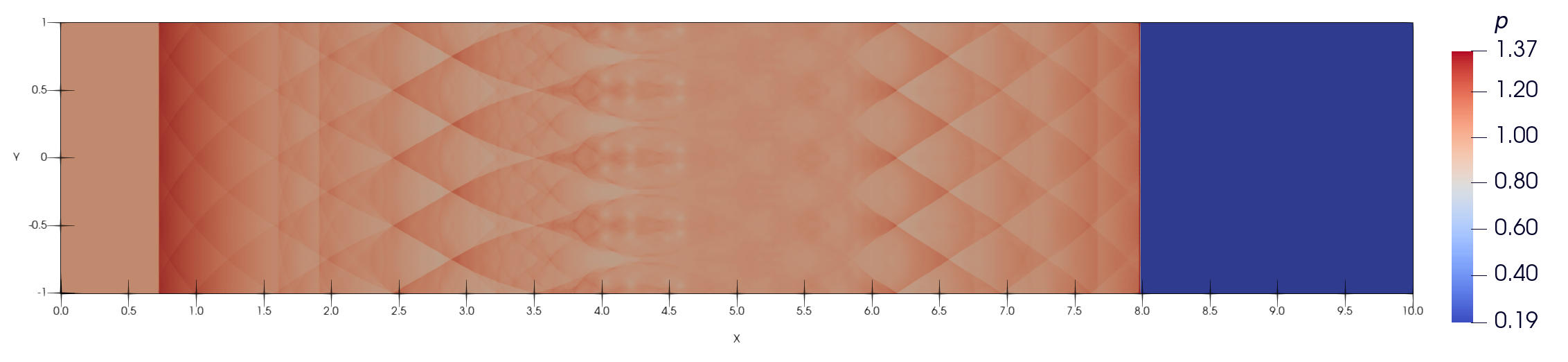}\\
\includegraphics[width=0.49\textwidth]{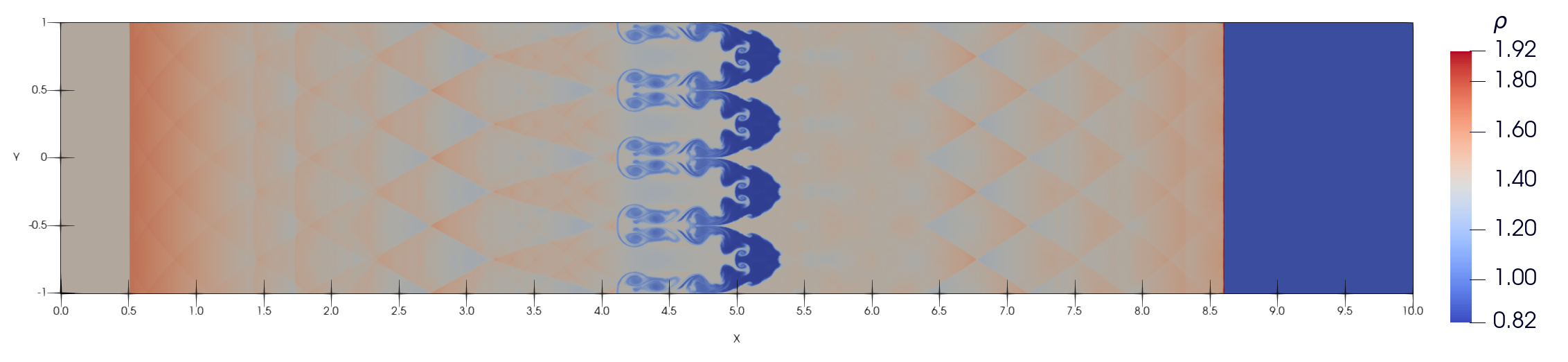}
\includegraphics[width=0.49\textwidth]{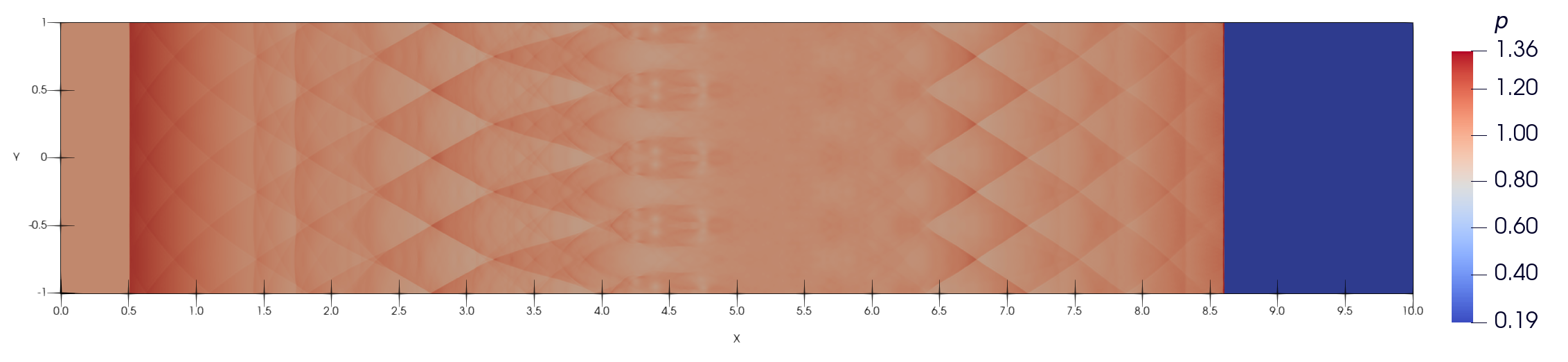}\\
\includegraphics[width=0.49\textwidth]{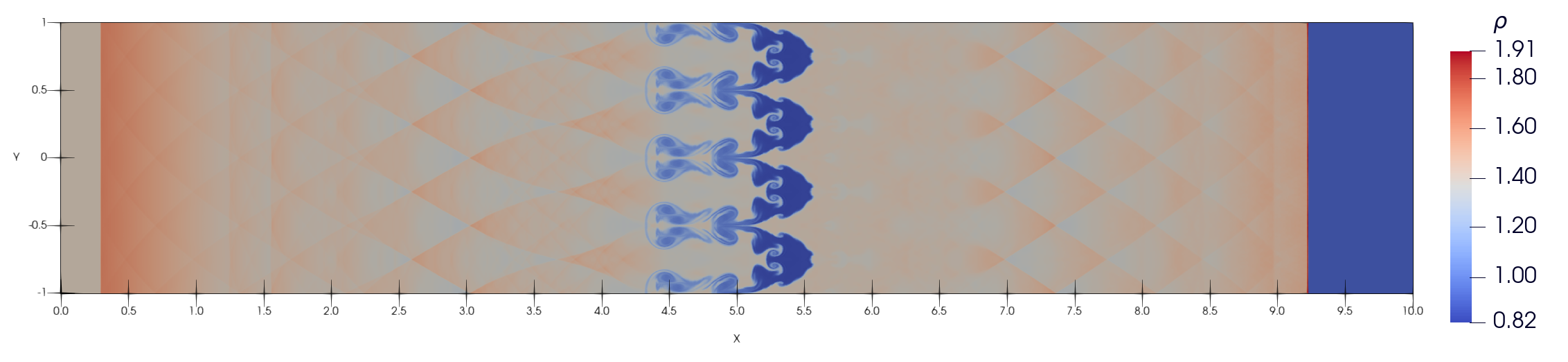}
\includegraphics[width=0.49\textwidth]{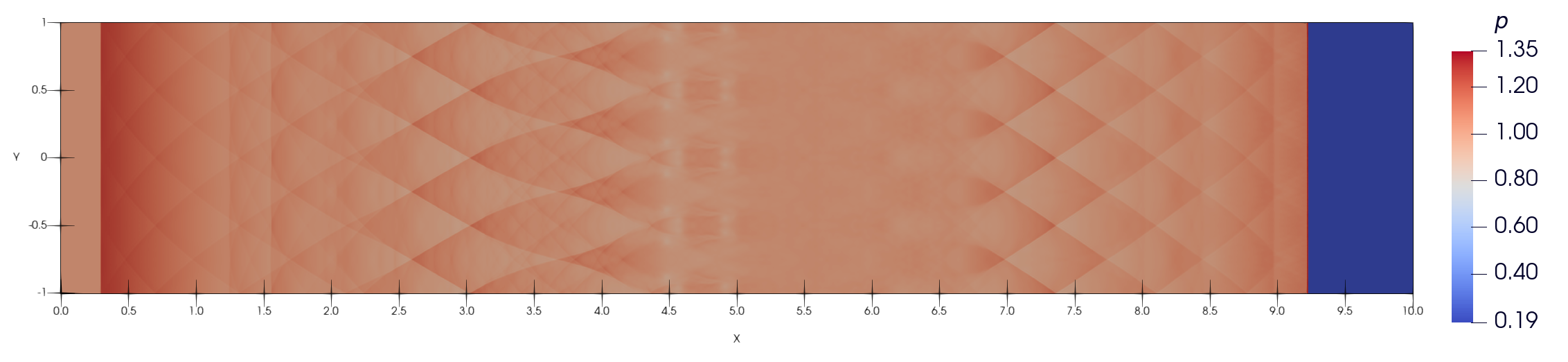}\\
\caption{\label{fig:dwsinef_2d_den_p}
Numerical solution of the two-dimensional problem of detonation cellular structure
in a two-component medium with a ``fast'' reaction (strong stiff case, a detailed statement of the problem is presented in the text),
obtained using the ADER-DG-$\mathbb{P}_{2}$ method with a posteriori limitation of the solution by a ADER-WENO2 finite volume limiter 
on mesh with $1000 \times 200$ cells at the times $t = 0.4$, $0.8$, $1.2$, $1.6$, $2.0$, $2.4$, $2.8$, $3.2$, $3.6$ and $4.0$ (from top to bottom).
The graphs show the coordinate dependencies of the subcells finite-volume representation of density $\rho$ (left) and pressure $p$ (right).
}
\end{figure*}

\begin{figure*}[h!]
\centering
\includegraphics[width=0.49\textwidth]{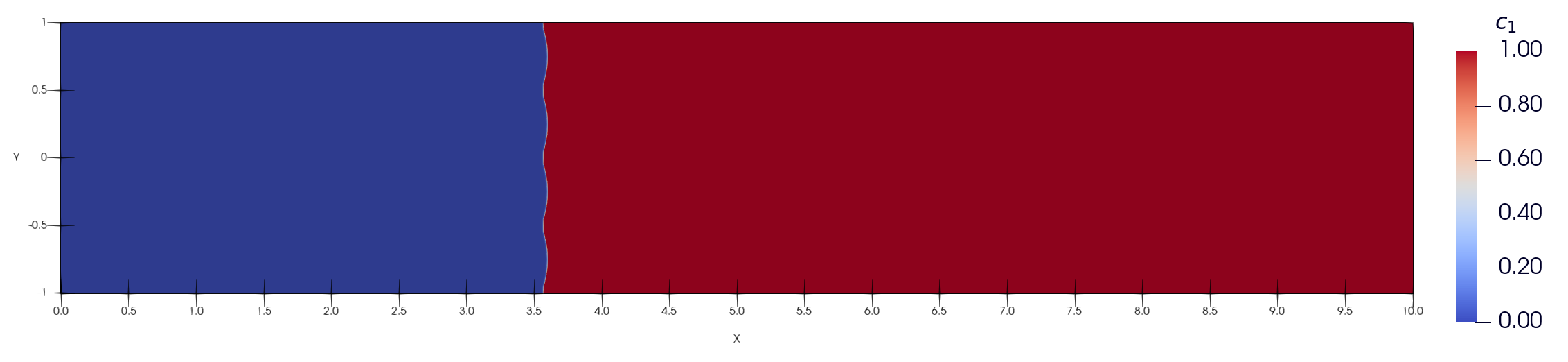}
\includegraphics[width=0.49\textwidth]{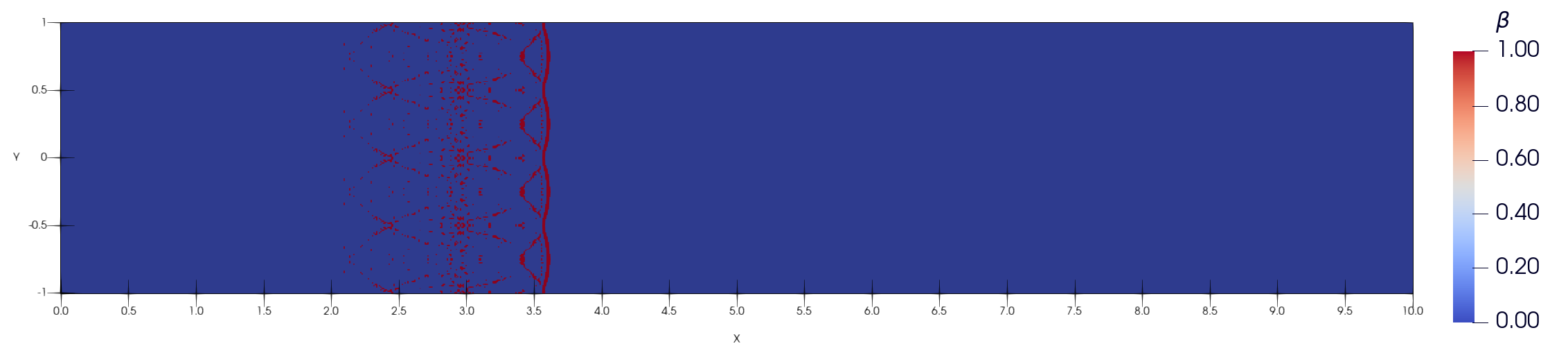}\\
\includegraphics[width=0.49\textwidth]{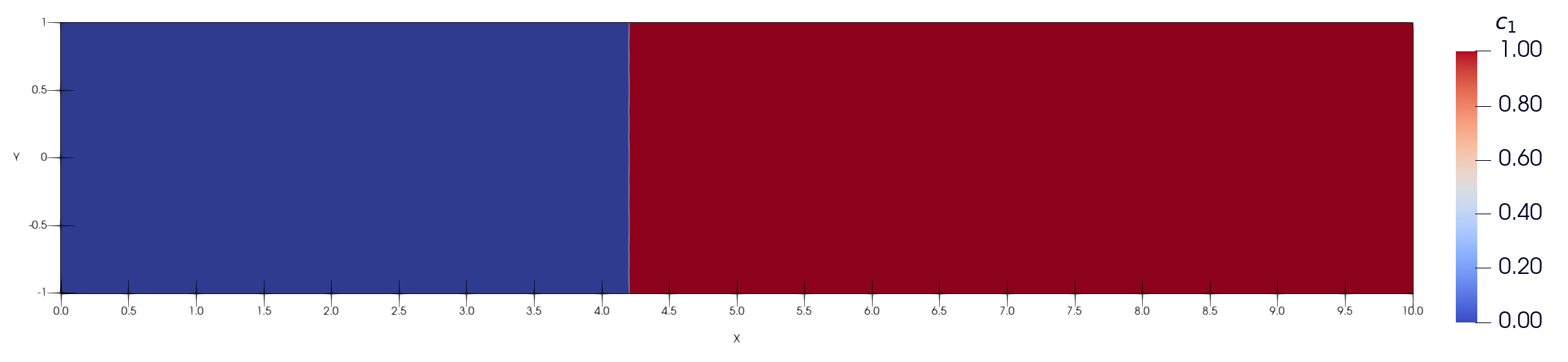}
\includegraphics[width=0.49\textwidth]{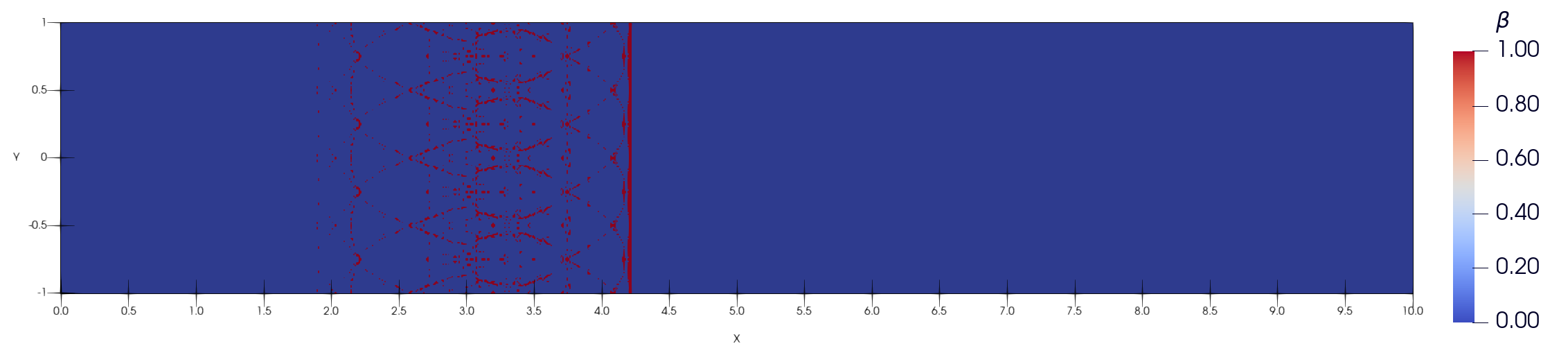}\\
\includegraphics[width=0.49\textwidth]{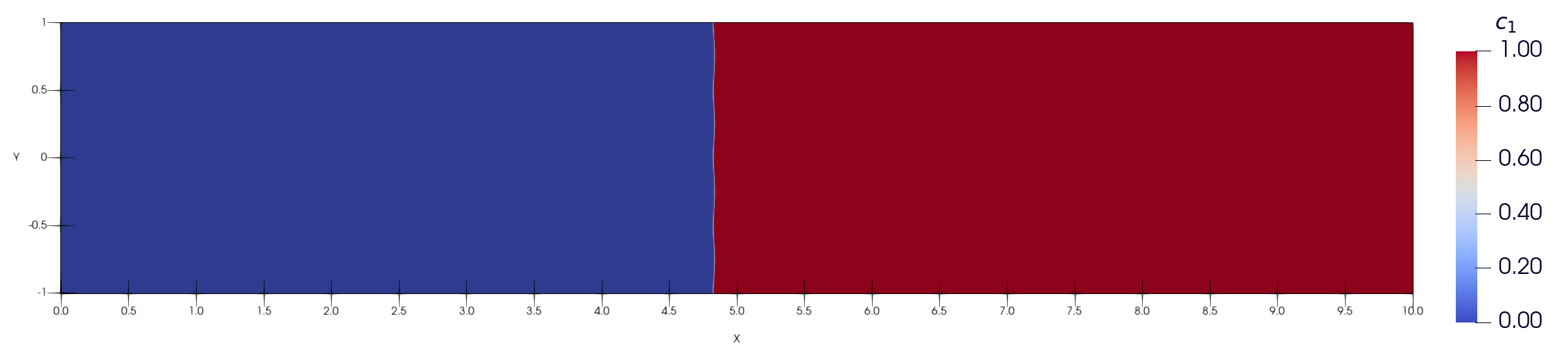}
\includegraphics[width=0.49\textwidth]{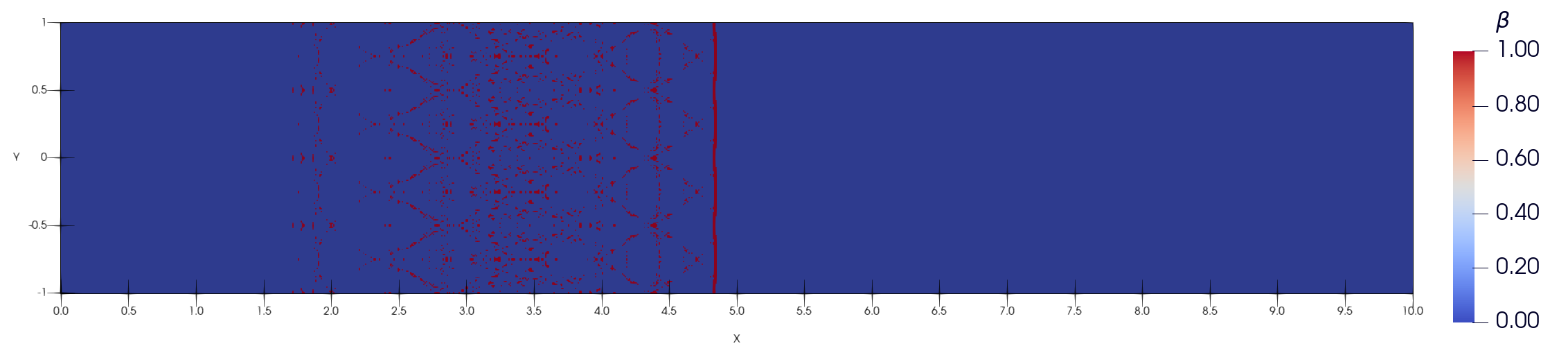}\\
\includegraphics[width=0.49\textwidth]{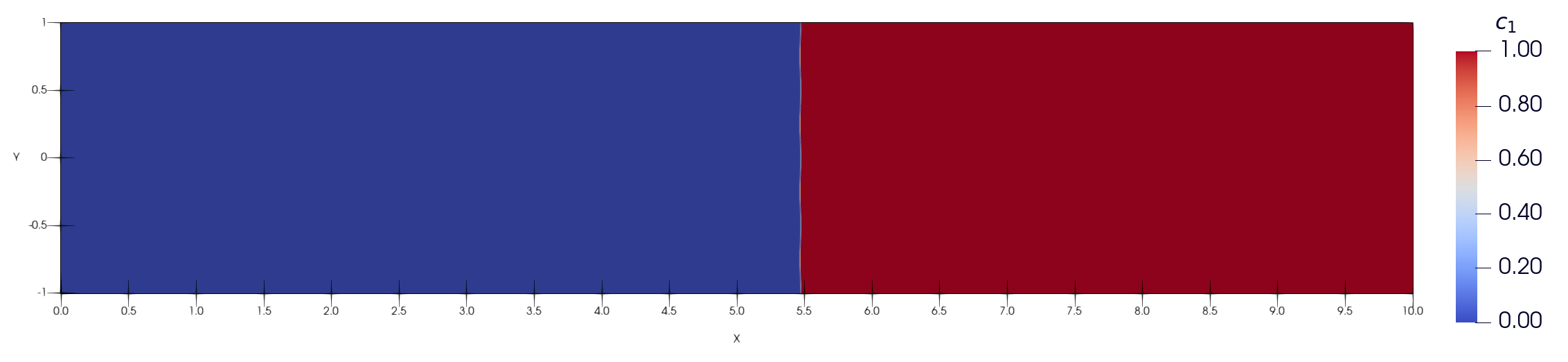}
\includegraphics[width=0.49\textwidth]{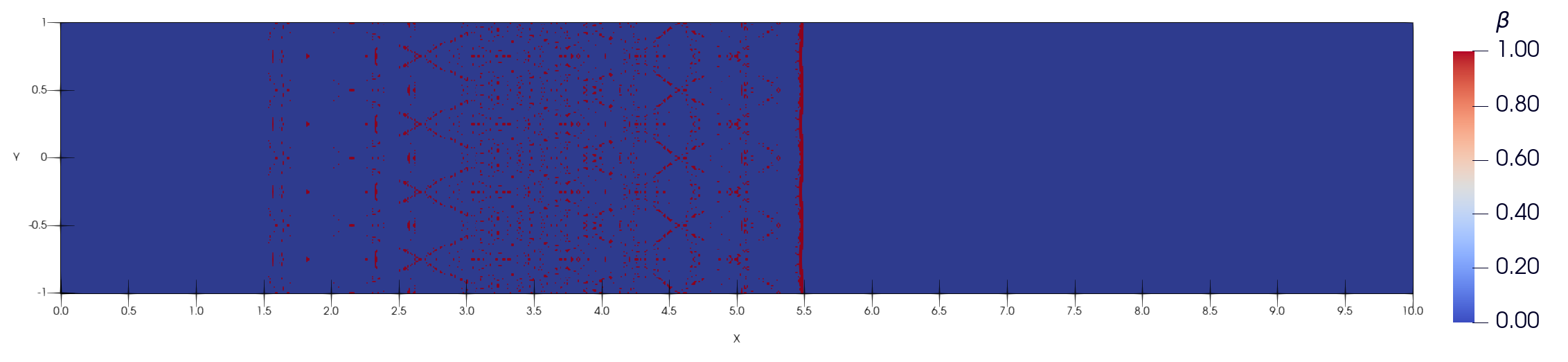}\\
\includegraphics[width=0.49\textwidth]{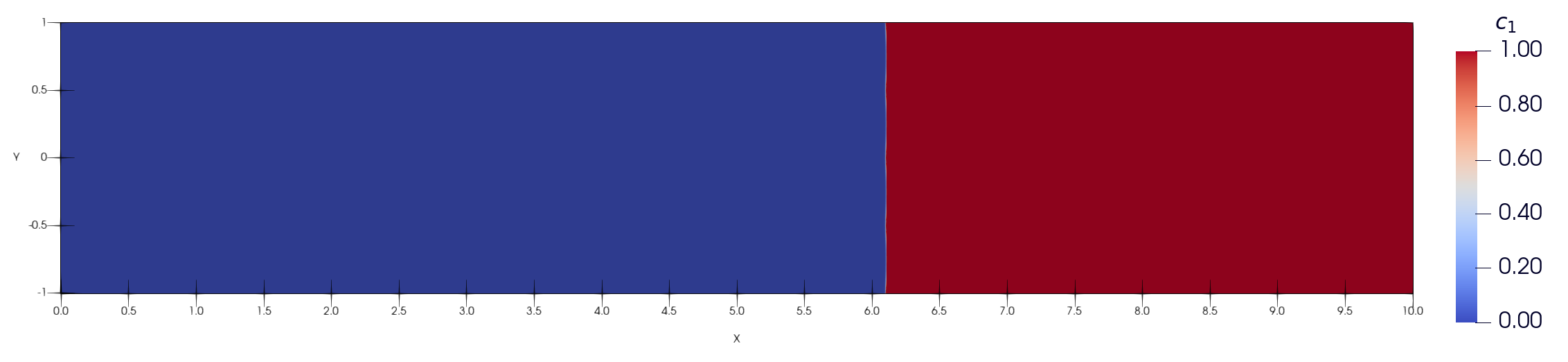}
\includegraphics[width=0.49\textwidth]{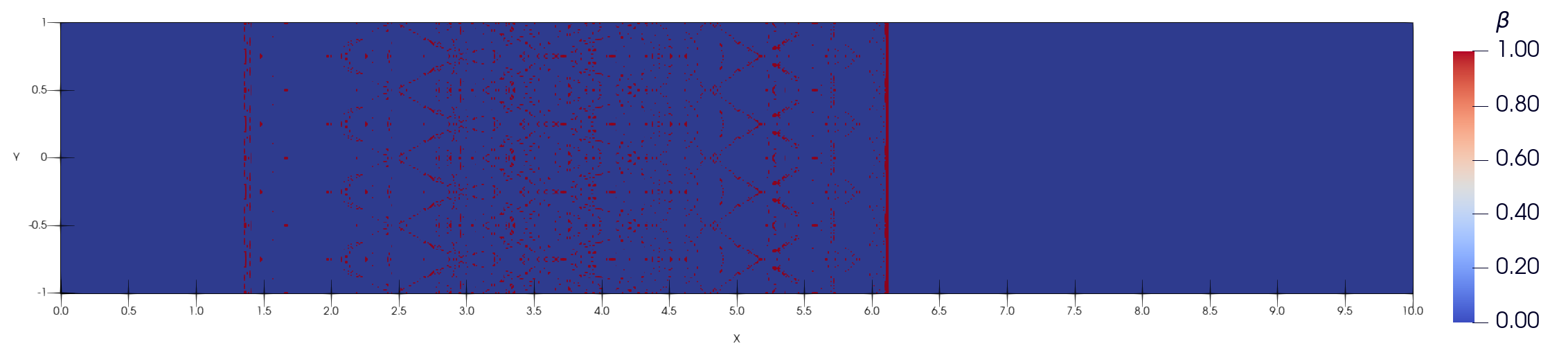}\\
\includegraphics[width=0.49\textwidth]{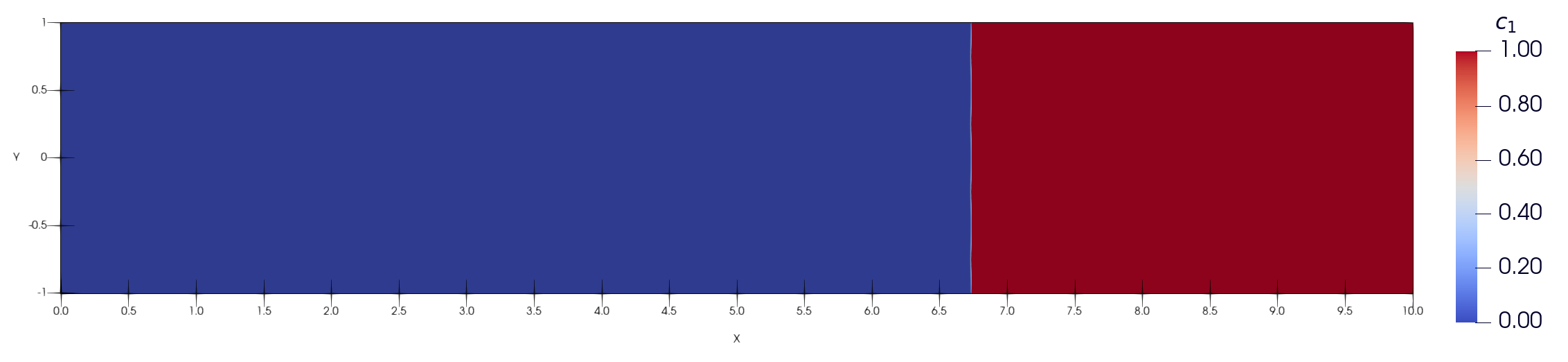}
\includegraphics[width=0.49\textwidth]{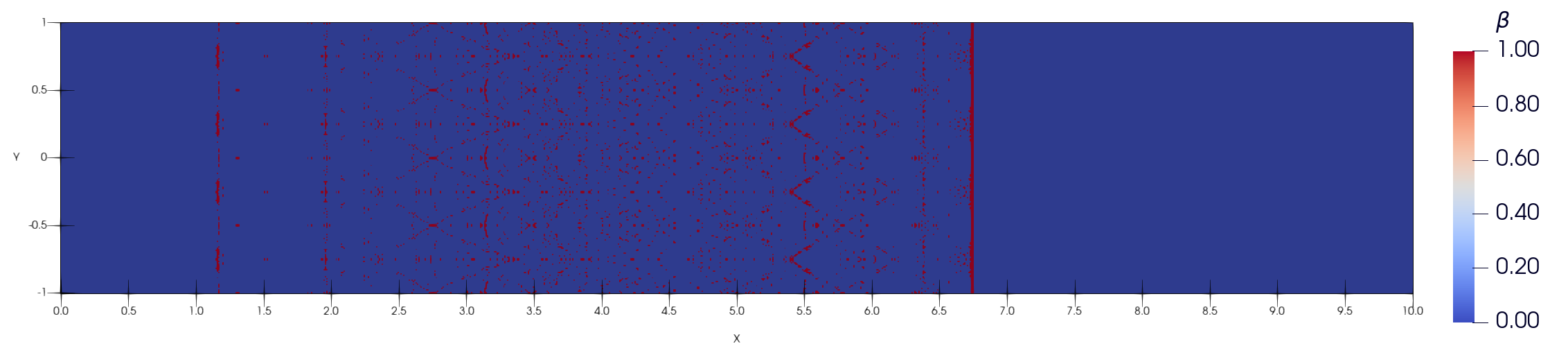}\\
\includegraphics[width=0.49\textwidth]{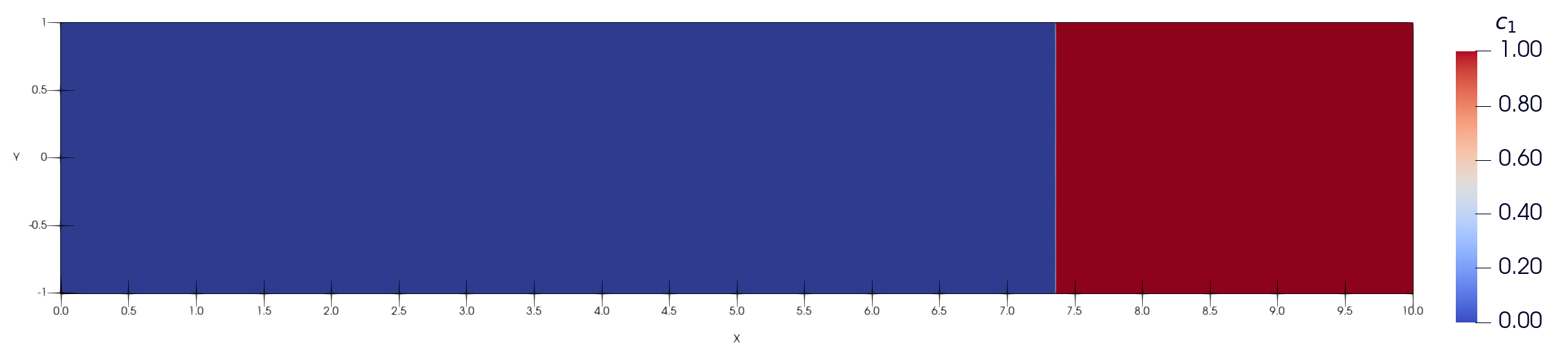}
\includegraphics[width=0.49\textwidth]{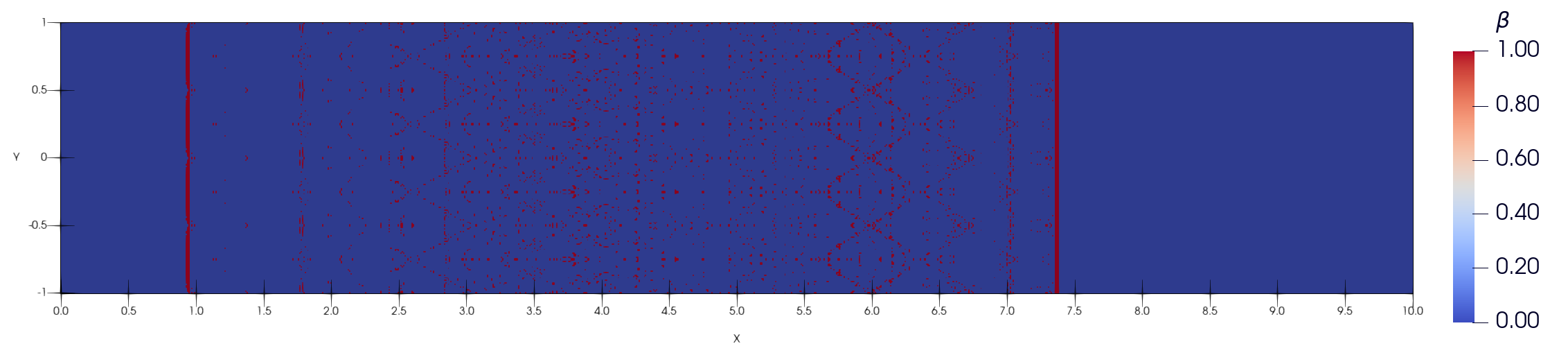}\\
\includegraphics[width=0.49\textwidth]{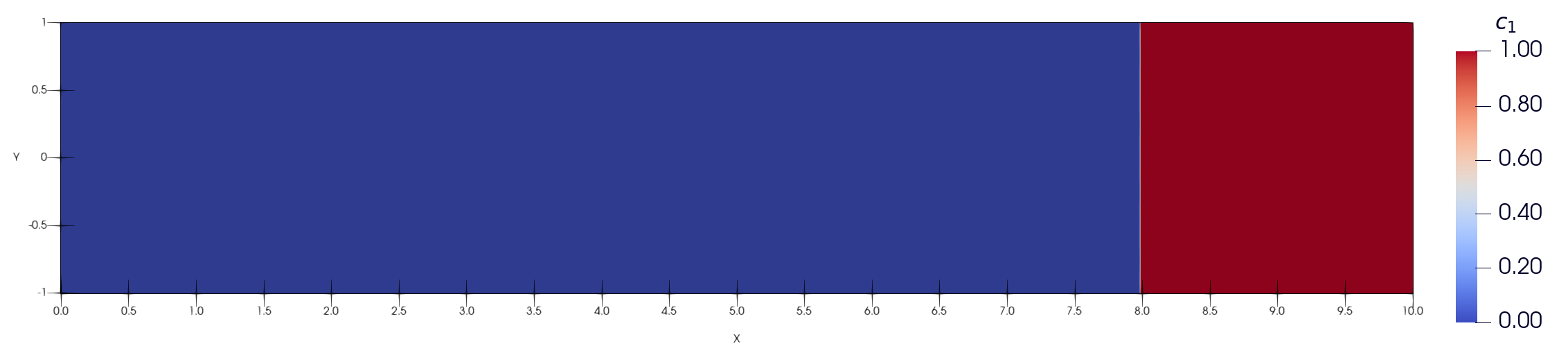}
\includegraphics[width=0.49\textwidth]{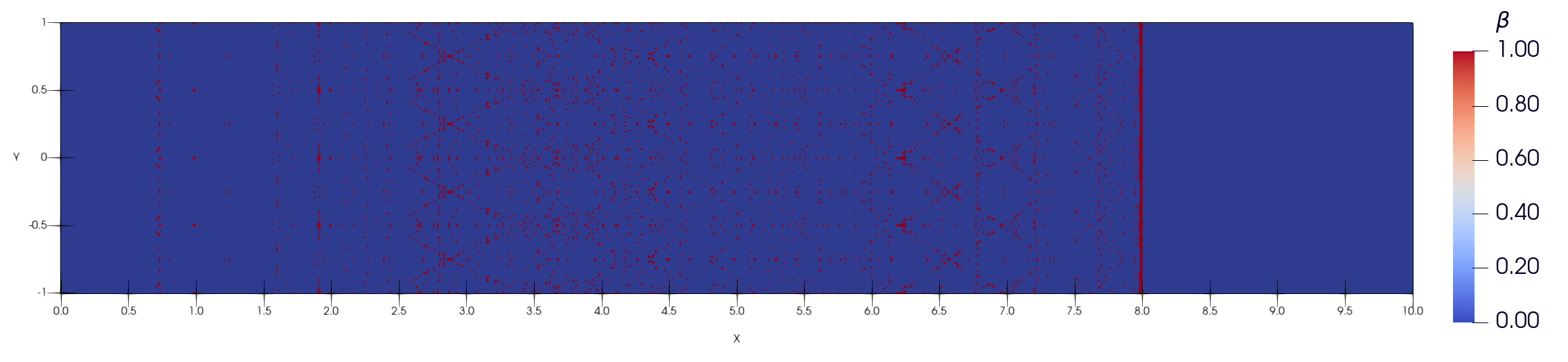}\\
\includegraphics[width=0.49\textwidth]{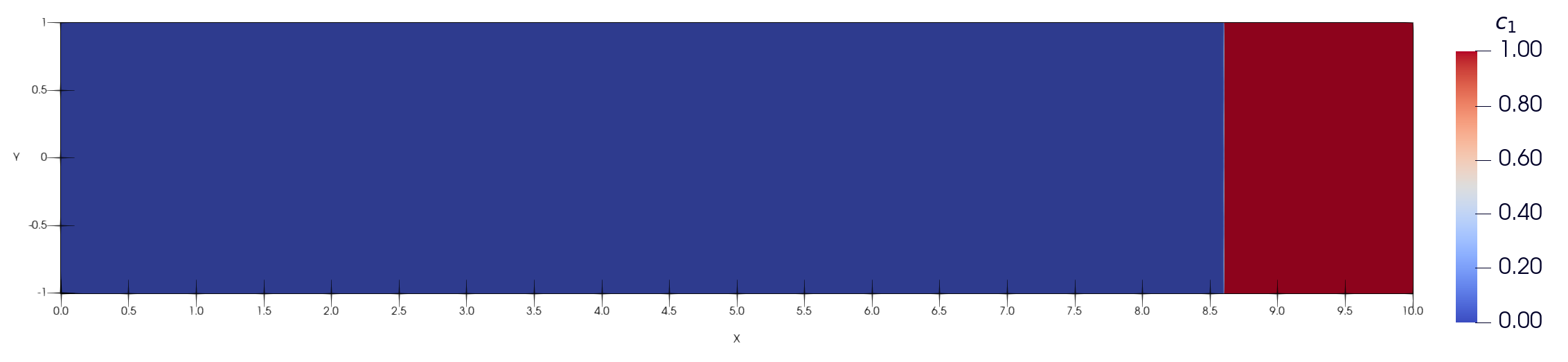}
\includegraphics[width=0.49\textwidth]{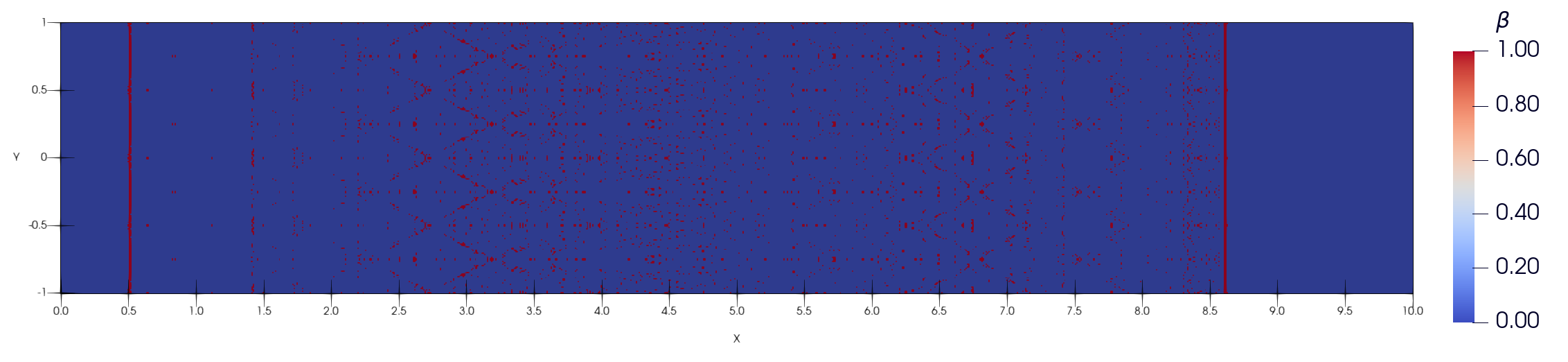}\\
\includegraphics[width=0.49\textwidth]{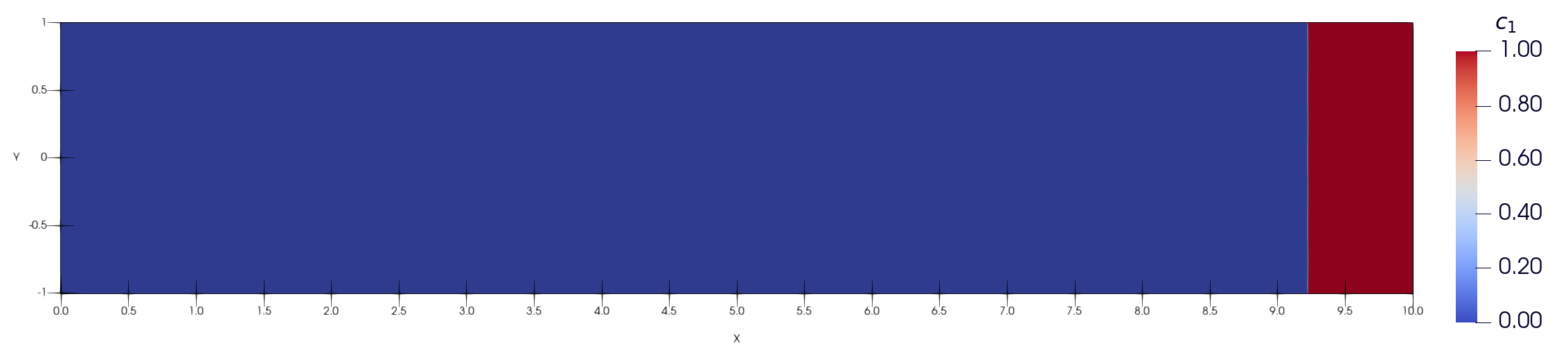}
\includegraphics[width=0.49\textwidth]{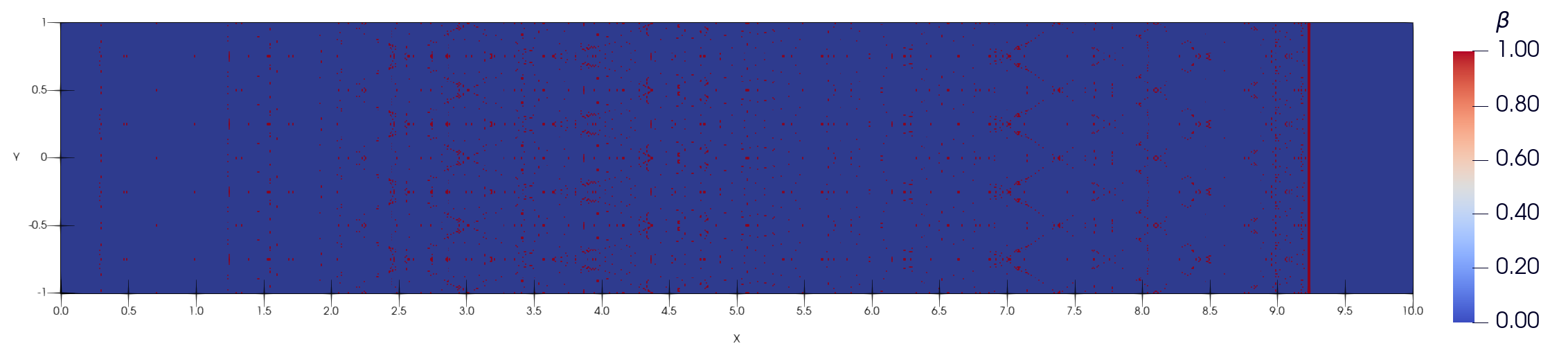}\\
\caption{\label{fig:dwsinef_2d_c_1_beta}
Numerical solution of the two-dimensional problem of detonation cellular structure
in a two-component medium with a ``fast'' reaction (strong stiff case, a detailed statement of the problem is presented in the text),
obtained using the ADER-DG-$\mathbb{P}_{2}$ method with a posteriori limitation of the solution by a ADER-WENO2 finite volume limiter 
on mesh with $1000 \times 200$ cells at the times $t = 0.4$, $0.8$, $1.2$, $1.6$, $2.0$, $2.4$, $2.8$, $3.2$, $3.6$ and $4.0$ (from top to bottom).
The graphs show the coordinate dependencies of the subcells finite-volume representation of 
mass concentration $c_{1}$ of the reaction reagent (left) and troubled cells indicator $\beta$ (right).
}
\end{figure*}

The numerical solution presented in Figures~\ref{fig:dwsinef_2d_den_p} and~\ref{fig:dwsinef_2d_c_1_beta} demonstrates the development of the detonation process from the initial state in an environment with a ``fast'' reaction. Perturbation of the detonation initiation boundary in the initial conditions leads to the formation of curved non-stationary detonation fronts, as well as a complex set of shock waves and rarefaction waves, which was expected from this formulation of the problem. This description of the obtained numerical solution will be limited primarily by the differences between this case and the case of a ``slow'' reaction. The detonation front at times $t = 0.4$ begins to demonstrate a detonation cellular structure, which is realized in the form of a set of shock waves propagating in directions along and across the direction of detonation propagation. Several triple points of intersection of the surfaces of shock waves with the formation of bow shock waves are observed. The main detonation front exhibits pronounced ``carbuncles'' associated with the initial disturbance. 

The general picture of the propagation and interaction of shock waves and the formation of a detonation front has a highly expressed flow symmetry, corresponding to the symmetry of the initial and boundary conditions. In the flow region behind the detonation wave front, a classical picture of the development of two-dimensional discontinuity decay is observed. Regions of vorticity formation arise, associated with the development of the Richtmyer-Meshkov instability. Four shock waves also propagate in the opposite direction, in space behind which the areas of intersection of discontinuities and interaction of shock waves are also clearly observed. The presented description of the flow is clearly observed based on a comparison of the coordinate dependencies of density $\rho$ and pressure $p$ in Figure~\ref{fig:dwsinef_2d_den_p}.

\begin{figure*}[h!]
\centering
\includegraphics[width=0.99\textwidth]{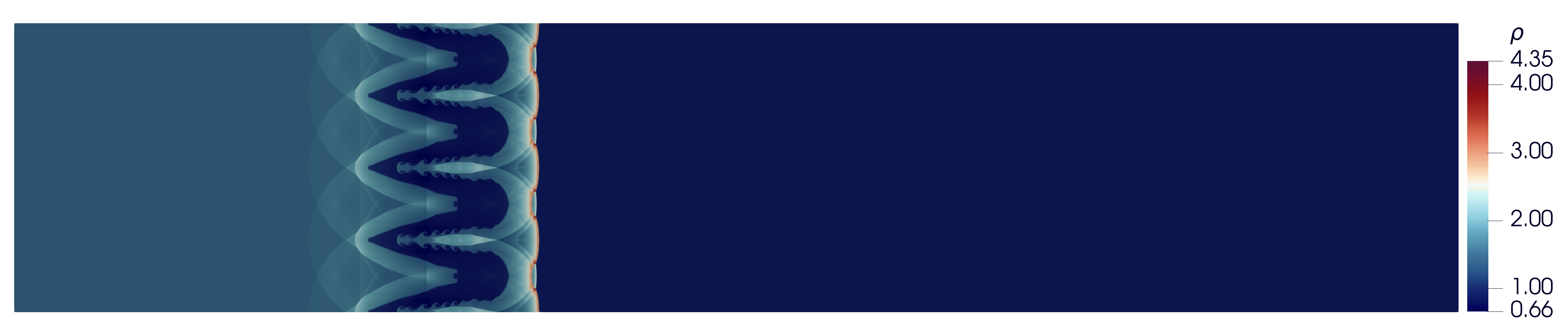}\\
\includegraphics[width=0.99\textwidth]{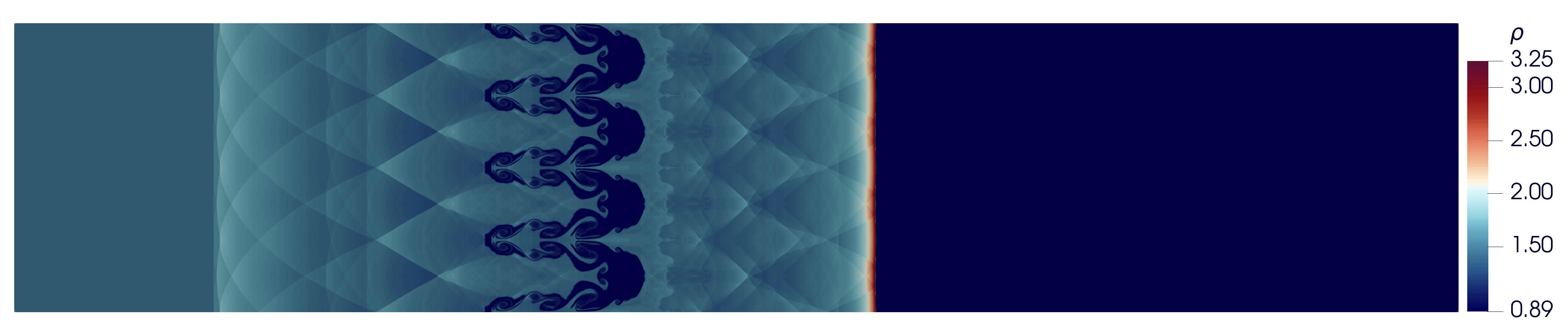}\\
\includegraphics[width=0.99\textwidth]{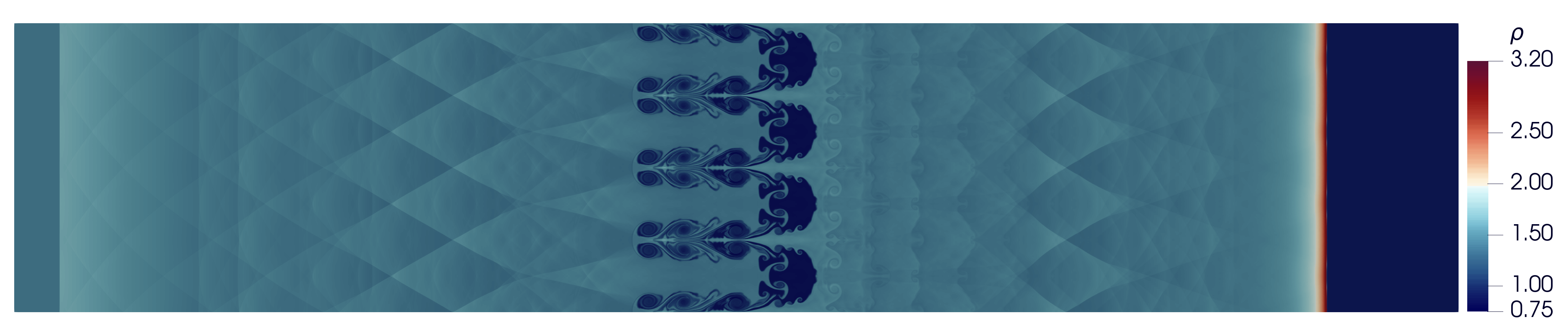}
\caption{\label{fig:dwsines_2d_2000x400}
Numerical solution of the two-dimensional problem of detonation cellular structure
in a two-component medium with a ``slow'' reaction (weak stiff case, a detailed statement of the problem is presented in the text),
obtained using the ADER-DG-$\mathbb{P}_{2}$ method with a posteriori limitation of the solution by a ADER-WENO2 finite volume limiter 
on mesh with $2000 \times 400$ cells at the times $t = 0.5$, $2.0$, and $4.0$ (from top to bottom).
The graphs show the coordinate dependencies of the subcells finite-volume representation of density $\rho$.
}
\end{figure*}

\begin{figure*}[h!]
\centering
\includegraphics[width=0.99\textwidth]{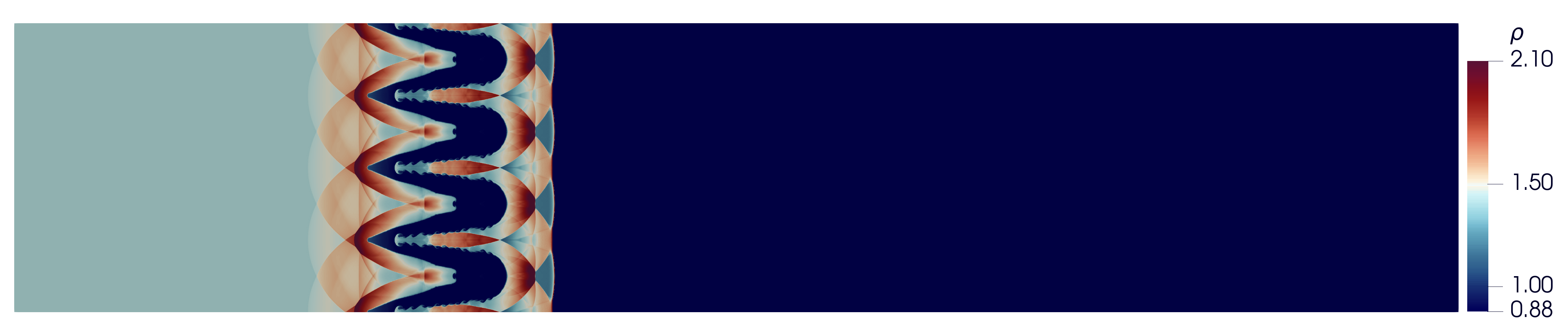}\\
\includegraphics[width=0.99\textwidth]{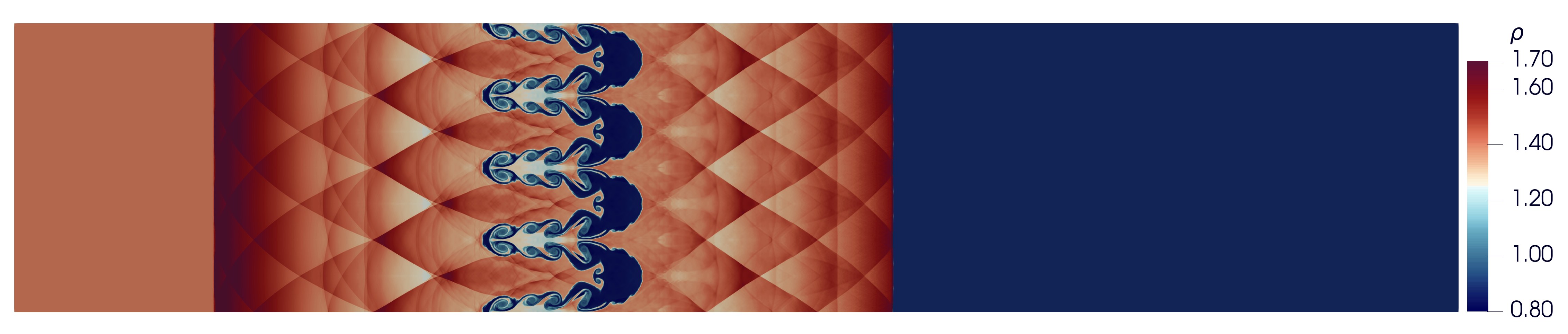}\\
\includegraphics[width=0.99\textwidth]{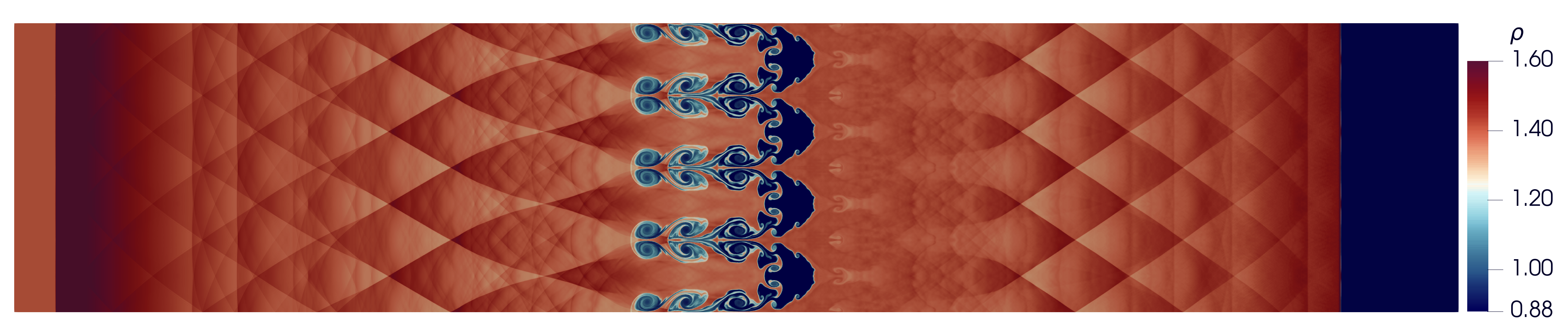}
\caption{\label{fig:dwsinef_2d_2000x400}
Numerical solution of the two-dimensional problem of detonation cellular structure
in a two-component medium with a ``fast'' reaction (strong stiff case, a detailed statement of the problem is presented in the text),
obtained using the ADER-DG-$\mathbb{P}_{2}$ method with a posteriori limitation of the solution by a ADER-WENO2 finite volume limiter 
on mesh with $2000 \times 400$ cells at the times $t = 0.5$, $2.0$, and $4.0$ (from top to bottom).
The graphs show the coordinate dependencies of the subcells finite-volume representation of density $\rho$.
}
\end{figure*}

The propagation of the detonation wave leads to burnout of the reagent, which is clearly observed in the coordinate dependencies of the mass concentration $c_{1}$ of the reagent presented in Figure~\ref{fig:dwsinef_2d_c_1_beta}. In this case of a ``fast'' reaction, a sharp burnout of the reagent is observed in the front of the passing shock wave -- a classic Zel'dovich chemical peak is observed. For all presented moments of time $t$, the burnout region is sharply separated from the region of unburnt reagent, while the boundary of the complete burnout region corresponds well to the detonation front and repeats all its geometric features. This property corresponds well to the general structure of the detonation front in the case of classical ZND detonation. In this case, the detonation front, in the form of a shock wave and a reaction region, is represented in the numerical solution by a coupled structure. It should be noted that, in contrast to the case of a ``slow'' reaction, narrow regions of localization of unburned reagent behind the detonation front are not observed -- this means that the propagation of rarefaction waves from the initial decay of the discontinuity does not lead to disruption of the detonation process. In this case, the detonation cellular structure is clearly observed in the coordinate dependencies of the mass concentration $c_{1}$ of the reagent. At time $t = 0.8$, the coordinate dependencies of density $\rho$ and pressure $p$ in Figure~\ref{fig:dwsinef_2d_den_p} clearly demonstrate the formation of a more planar detonation front, behind which a complex superposition of shock waves and tangential discontinuities is clearly observed. Over time $t$, the detonation front becomes more and more flat, however, this is not a completely monotonic process, since it is associated with nonlinear interference of shock waves behind the detonation front, therefore, on average, the detonation front has become flatter, however, at time $t = 1.2$ there is an increase in heterogeneity across the front in coordinate dependencies of density $\rho$ and pressure $p$ in Figure~\ref{fig:dwsinef_2d_den_p}. By time $t = 1.6$, there is a significant decrease in the intensity of shock waves behind the detonation wave front due to geometric divergence. The transverse heterogeneity of the detonation front decreases significantly. By time $t = 2.0$, there is a significant ``separation'' of the detonation front from the region of intense shock wave effects behind the front. In this case, the detonation front becomes almost flat, and transverse inhomogeneities. Further, during the propagation process, density $\rho$ and pressure $p$ small inhomogeneities are also observed at the detonation front. In this case, the cellular structure of the detonation front manifests itself in the coordinate dependencies of the reagent concentration at times up to $t = 1.2$. At large times $t > 1.2$, the detonation front becomes almost one-dimensional from the position of the region of reagent burnout.

The areas of vorticity generation in Figure~\ref{fig:dwsinef_2d_den_p} demonstrate stable vortex formation, which at time $t = 0.8$ led to the formation of vortex ``columns'', which represents a complicated picture of the development of the Richtmyer-Meshkov instability. Shock waves propagating behind the detonation fronts are reflected from the upper and lower boundaries of the coordinate domain and from areas of significant changes in density $\rho$, propagate across the direction of propagation of the detonation front, and have a significant impact on the pattern of flow development in the region of the burned-out substance. The interaction of the general flow with the propagating curved fronts of shock waves and rarefaction waves leads to the formation of additional areas of vortex generation, especially towards the original vortex ``columns''. These effects are most clearly observed in the coordinate dependencies of very high resolution, which are presented in Figure~\ref{fig:dwsinef_2d_2000x400}, and therefore will be discussed further in the text in the appropriate discussion. Areas of significant vortex generation associated with the Richtmyer-Meshkov instability are clearly observed for all presented times on the coordinate dependencies of the density. Since time $t = 2.0$, a significant interaction of vortex ``columns'' with a region of higher density has been observed, while the dynamics of vorticity development is resolved quite correctly in the numerical solution. 

The distribution of troubled cells, as coordinate dependencies of the troubled cells indicator $\beta$, presented in Figure~\ref{fig:dwsinef_2d_c_1_beta} shows that the limiter is activated predominantly only in the areas where shock fronts are localized, while the head shock waves and areas of shock wave interaction can be clearly distinguished. In areas of tangential discontinuities, trouble cells practically do not appear. The distribution of troubled cells presented after time $t = 2.0$ allows us to identify the limiter activation regions only with the main structures in the flow -- the detonation front propagating forward, shock wave structures propagating backward, and the strongest shock waves behind the detonation front. 

As a result of the presented analysis, it can be concluded that the ADER-DG-$\mathbb{P}_{N}$ method with a finite volume ADER-WENO and an a posteriori limiter makes it possible to effectively simulate detonation cellular structures, while all the main features of the detonation flow of the reacting medium with a ``fast'' reaction are resolved in a numerical solution. In this case, the Zel'dovich chemical peak and the classical structure of the ZND detonation wave are observed. Non-physical effects characteristic of modeling detonation processes using classical numerical methods~\cite{frac_steps_detwave_sim_2000, chem_kin_hrs_weno} do not arise -- a non-physical weak detonation front is not formed~\cite{correct_det_wave_speed_2017}.

Figures~\ref{fig:dwsines_2d_2000x400} and~\ref{fig:dwsinef_2d_2000x400} are presented for demonstration of a very high-resolution numerical solution of the two-dimensional problem of detonation cellular structure in a two-component medium obtained using ADER-DG-$\mathbb{P}_{2}$ method with ADER-WENO2 finite volume a posteriori limiter on a spatial mesh with sizes $2000\times400$, for the case of weak and strong stiffness, respectively. The coordinate dependencies of the subcells finite-volume representation of density $\rho$ are presented at several times $t = 0.5$, $2.0$, and $4.0$.

The coordinate dependence of the density $\rho$ at time $t = 0.5$ presented in Figure~\ref{fig:dwsines_2d_2000x400} makes it possible to clearly clarify the details of the process of initiation of a detonation wave in the case of detonation development in a medium with a ``slow'' reaction. Inhomogeneity of the detonation front, which has characteristic features of a detonation cellular structure, which are clearly related in this case to the structure of the initial conditions of the problem presented in Figure~\ref{fig:dwsines_2d_init}. Intersection of discontinuity surfaces and nonlinear interaction of shock waves occur. As a result of the reflection of shock waves from the solid walls of the coordinate domain and from areas of sharp changes in density, a complex shock-wave flow is formed. The detonation cellular structure is clearly formed by multiple shock waves behind the detonation front. The very high resolution of the numerical solution on mesh with $2000 \times 400$ cells makes it possible to trace in detail the processes of formation of multi-wave structures, details of the processes of interaction of shock waves, rarefaction waves and tangential discontinuities, in particular leading to the formation of complex regions of vortex generation. At least three characteristic areas of vortex generation are observed: classical vortex ``columns'', directly associated with the development of the Richtmyer-Meshkov instability, while in the spatial regions between the vortex columns, in the region of propagation of the rarefaction wave from the initial breakdown of the fracture, the beginning of the process of destruction of the shear layer is also observed with the formation of a vortex street; counter flow in relation to the relative movement of the vortex ``columns'' -- in this case, there is a movement of a dense medium in a much more rarefied medium, so the movement has the opposite character of swirling the flow; Very small classical vortex ``columns'' are also observed, associated with the development of the Richtmyer-Meshkov instability, developing and moving behind the outwardly convex regions of the detonation front. The structure of the boundaries of the vortex flow in the numerical solution is clearly expressed. By time $t = 2.0$ in Figure~\ref{fig:dwsines_2d_2000x400}, the detonation front moves significantly forward along the flow. The detonation cellular structure becomes less pronounced, however, it is quite clearly observed in the numerical solution. Behind the front of the detonation wave, a much more complex configuration of shock waves is observed, which have undergone multiple reflections from the boundaries of the coordinate domain and regions of sharp changes in density. Pronounced triple points of intersection of the discontinuity surfaces are observed, some of which are located on the surface of the detonation front. The detonation cellular structure at time $t = 2.0$ is formed predominantly by multiple reflected and nonlinearly interfered shock waves. The multi-wave structure of the flow behind the front of the detonation wave has created several regularly spaced vortex ``columns'' located on several sections along the flow, on each of which the structure of the vortex-forming regions has a well-conditioned symmetry of the flow -- at least one new section of vortex-forming flows formed at the intersection points of shock waves. The largest-scale structure of vortex ``columns'' has dynamically evolved significantly -- the classical structure of large vortices along the formed shear layer is observed. In this case, a significant part of the length of these vorticity localization regions has significantly penetrated into the region of increased density upstream of the relative flow. In the rear part of the flow formed by the initial disintegration of the discontinuity, multiple shock waves and intersections of their discontinuities are also observed. By time $t = 4.0$, the detonation cellular structure at the front is practically not observed, the detonation front has become almost planar -- the intensity of the shock waves behind the detonation front has decreased significantly, and the detonation front has ``separated'' from the direct influence of this flow region. In this case, shock waves behind the detonation front continue to implement the processes of reflection, scattering and nonlinear interference, and their regular structure is well represented in the numerical solution. The vortex ``columns'' and shear layers induced by these shock waves are represented in the solution as a minimum of three slices across the direction of the general flow. For the largest-scale structure of vortex ``columns'' a stage of developed mixing with the surrounding matter is observed, which is typical for such jet flows. The coordinate dependence of the density $\rho$ presented in Figure~\ref{fig:dwsinef_2d_2000x400} makes it possible to clearly clarify the details of the process of initiation of a detonation wave in the case of detonation development in a medium with a ``fast'' reaction. The main difference from the case of a ``slow'' reaction presented in Figure~\ref{fig:dwsines_2d_2000x400} is the significantly less pronounced heterogeneity of the detonation front at times $t = 2.0$. All main flow structures are presented in a numerical solution; their symmetry and sharpness of identifying the boundaries and surfaces of discontinuities in the numerical solution do not differ significantly in quality from the case considered above. One can note the greater regularity of multiple shock waves behind the detonation fronts at times $t = 2.0$ and $4.0$, while the vicinity of the largest-scale structure of vortex ``columns'' leads to very heterogeneous and irregular scattering and reflection of shock waves, so isolated layers of small areas of vortex formation are not observed downstream. 

As a result of the presented analysis of a very high resolution numerical solution, it can be concluded that the ADER-DG-$\mathbb{P}_{N}$ method with a finite volume ADER-WENO and an a posteriori limiter resolves multiple discontinuous flow components against a background of quite heterogeneous flow. In this case, the main coherent flow structures and their symmetry are well resolved and preserved over fairly long time ranges.
}

\section*{Conclusion and discussion}
\label{sec:conclusion_and_discussion}

\paragraph{Discussion}

\newtext{
In the discussion, it should be noted that in this work, the use of the space-time adaptive ADER-DG-$\mathbb{P}_{N}$ method with LST-DG predictor and a posteriori sub-cell ADER-WENO finite-volume limiting for simulating detonation waves was investigated using the example of a two-component medium with discrete ignition temperature kinetics model of a monomolecular reaction $A \rightarrow B$. It is clear that usually the main interest is in the numerical modeling of detonation processes in real energy-releasing reacting media~\cite{Oran_Boris_2005, Lunev_2017}, such as fuel~\cite{dcs_oran_1985, dcs_oran_1987, dcs_hydrogen_2020}, for example, mixtures of hydrogen, methane and others with oxygen or multicomponent gases, for example, air. Therefore, it should be noted that the presented version of the ADER-DG-$\mathbb{P}_{N}$ method and its software implementation can be used to simulate detonation processes in real multicomponent reacting media.

Expansion of the mechanism of reactions occurring in the medium requires an increase in the dimension of vectors of conserved values $\mathbf{U}$, flux terms $\mathbf{F}$ and source terms $\mathbf{S}$, in accordance with the increase in the number of components $R$ of the reacting media in the system of equations (\ref{eq:system_of_equations}). In the source terms $\mathbf{S}$, it is necessary to add terms related to the rates of reactions occurring in the reacting medium within the framework of the selected reaction mechanism and kinetics. The original form of the system of equations will not change.

A significant change must occur in the equation of state used -- while the thermal equation of state $p = p(\rho, T)$ can be used in the approximation of a perfect or quasi-perfect gas~\cite{Lunev_2017, Nagnibeda_2009}, then the caloric equation of state $e = e(\rho, p)$ must usually be rewritten in a fairly arbitrary form, usually in terms of the enthalpy of the mixture $h = h(\rho, p)$~\cite{Oran_Boris_2005}. It is possible to use the caloric equation of state in a quasi-perfect form $p = (\gamma - 1) e$, with effective adiabatic exponent $\gamma$ (in reality, it will be necessary to introduce several effective exponents $\gamma$ that are not equal to each other), which is often used in the study of flows with high speeds~\cite{Lunev_2017}; however, in the case of modeling detonation waves, this approximation may be incorrect. There is no need for a source term associated with energy release in the system of equations (\ref{eq:eqs_two_comps}) -- energy release will be automatically taken into account in the selected caloric equation of state of a real multicomponent medium. Changing the form of the caloric equation of state of a substance can also lead to a significant complication of the process of calculating the eigenvalues of the Jacobian matrix of a system of equations, as well as to a complication of the procedure for calculating primitive variables from conserved values.

As a result of changing the form of the caloric equation of state, it will be necessary to use Riemann solvers created for equations of state of a fairly arbitrary shape, and taking into account possible thermodynamic features of the media~\cite{rp_real_gases_1, rp_real_gases_2, rp_real_gases_3, rp_real_gases_4, rp_real_gases_5}. The use of numerical methods of the ADER family, and specifically the finite-volume ADER-VENO method, for modeling the flows of real gases with a complex equation of state, is presented in the work~\cite{rp_real_gases_6}, which, in particular, describes the possibilities of using Riemann solvers for complex equations of state of matter, as well as presents an effective method for approximating wide-range equations of state of matter.

The iterative procedure for obtaining a local discrete space-time solution within the framework of solving a system of nonlinear algebraic equations LST-DG-predictor can become quite labor-intensive in terms of computational costs. The use of complex mechanisms of chemical reactions in a medium, taking into account the additional complexity associated with significantly nonlinear dependences of reaction rates on local concentration and temperature fields, can significantly increase computational costs, especially those associated with obtaining a local solution.

Otherwise, significant changes in the ADER-DG-$\mathbb{P}_{N}$ method with a posteriori sub-cell ADER-WENO finite-volume limiting and its software implementation are not expected when it is adapted for studying detonation processes in real multicomponent reacting media. Therefore, it can be concluded that the study of flows of multicomponent flows of real gaseous media will lead to technical complications in the implementation of the numerical method rather than to any fundamental changes in its structure. This study is planned for the future.
}

\paragraph{Conclusion} In conclusion, it is necessary to summarize the results of this work. The space-time adaptive ADER-DG finite element method with LST-DG predictor and a posteriori sub-cell ADER-WENO finite-volume limiting~\cite{ader_dg_ideal_flows, ader_dg_dev_1, ader_dg_dev_2, ader_weno_lstdg_ideal, ader_dg_diss_flows, ader_dg_ale, ader_dg_grmhd, ader_dg_gr_prd, ader_dg_PNPM, PNPM_DG_2009, PNPM_DG_2010, ader_dg_eff_impl, fron_phys, exahype, ader_dg_hpc_impl_1, ader_dg_hpc_impl_2, ader_dg_hpc_impl_3, ader_dg_hpc_impl_4, ader_dg_mod_1, ader_dg_mod_2} was used for the first time to simulate multidimensional reacting flows with detonation waves. Research has mainly been carried out on two-dimensional problems with detonation waves. 

The presented numerical method does not use any ideas of splitting or fractional time steps methods -- the solutions are obtained using a high-order ADER-DG-$\mathbb{P}_{N}$ method with a posteriori limitation of the solution.

The developed software implementation of the numerical method was initially tested on a set of classical gas-dynamic problems to confirm its correctness and performance.

The modification of the LST-DG predictor has been developed that makes it possible not to use the procedure of adaptive change in the time step, which was used in previous work~\cite{popov_j_sci_comp_2023} related to the modeling of one-dimensional detonation waves flows. The proposed modification is based on a local partition of the time step $\Delta t^{n}$ in cells in which strong reaction activity of the medium is observed. Obtaining a discrete space-time solution $\mathbf{q}_{h}^{s}$ was carried out for each individual local step, followed by transformation into the final discrete space-time solution $\mathbf{q}_{h}$ within one space-time element $\Omega_{i}\times[t^{n}, t^{n+1}]$. This approach made it possible to obtain solutions to classical problems of flows with detonation waves having a classical ZND structure, without a decrease of the time step, and the results were obtained at a Courant number of $\mathtt{CFL} = 0.9$.

The results obtained show the very high applicability and efficiency of using the space-time adaptive ADER-DG-$\mathbb{P}_{N}$ method with LST-DG predictor and a posteriori sub-cell ADER-WENO finite-volume limiting for simulation reactive flows with detonation waves. The structure of detonation waves is resolved by this numerical method with subcell resolution even on very coarse spatial meshes. In one-dimensional problems, the results were obtained on a mesh of $100$ cells using the ADER-DG-$\mathbb{P}_{5}$ method with a posteriori sub-cell ADER-WENO2 finite-volume limiting, while the detonation front had a width of only $1$-$2$ subcells in the case of weak stiffness and $2$-$4$ subcells in the case of strong stiffness. In the case of a cylindrical detonation wave, the width of the detonation front of $4$-$5$ subcells on a mesh of $100 \times 100$ cells was obtained using the ADER-DG-$\mathbb{P}_{9}$ method with a posteriori sub-cell ADER-WENO2 finite-volume limiting. The proportion of troubled cells in a spatial mesh usually does not exceed $5\%$ in one-dimensional problems and $10$-$16\%$ in two-dimensional and three-dimensional problems in the case of correctly chosen method parameters and mesh size. The smooth components of the numerical solution, which are represented in the reference problem solutions, are correctly and very accurately reproduced by the numerical method.

It should be noted that non-physical artifacts of the numerical solution\newcoloringtext{, such as formation in the numerical solution of a weak detonation front, which propagates ahead of the shock front,} did not arise in the results obtained. In the case of strongly stiff problems, the numerical solution shows the correct formation and propagation of ZND detonation waves, with a clearly defined Zel'dovich chemical peak. Numerical artifacts noted in a number of works~\cite{frac_steps_detwave_sim_2000, chem_kin_hrs_weno, correct_det_wave_speed_2017} did not arise in this case.

This work also presented the results of simulating rather complex problems associated with the propagation of detonation waves in substantially inhomogeneous domains using a ADER-DG-$\mathbb{P}_{2}$ method with a posteriori sub-cell ADER-WENO2 finite-volume limiting -- interaction of a detonation wave with a single inert bubble and with lattice of inert bubbles\newcoloringtext{, and detonation cellular structures dynamics}. The results obtained show that all the main features of detonation flows are correctly reproduced by space-time adaptive ADER-DG-$\mathbb{P}_{N}$ method with LST-DG predictor and a posteriori sub-cell ADER-WENO finite-volume limiting.

As a result, it can be concluded that the space-time adaptive ADER-DG-$\mathbb{P}_{N}$ method with LST-DG predictor and a posteriori sub-cell ADER-WENO finite-volume limiting is perfectly applicable to simulating fairly complex reacting flows with detonation waves.

\section*{Declarations}

\subsection*{Acknowledgments}
The reported study was supported of the Russian Science Foundation grant No.~21-71-00118 \texttt{https://rscf.ru/en/project/21-71-00118/}.

The author would like to thank the anonymous reviewers for their encouraging comments and remarks that helped to improve the quality and readability of this paper. The author would like to thank Popova A.P. for help in correcting the English text.

\subsection*{Data Availability}
The datasets generated during and/or analysed during the current study are available from the corresponding author on reasonable request.

\subsection*{Conflict of interest}
The author declares that he has no conflict of interest.

\end{document}